\documentclass[11pt]{amsart}
\usepackage[utf8]{inputenc}
\pdfoutput=1
\usepackage{amsmath,amssymb,amsthm,amsfonts}
\usepackage{mathtools}%

\usepackage[english]{babel}%
\usepackage{comment}%
\usepackage[unicode]{hyperref}%
\usepackage{bbm}%
\usepackage{bm}
\usepackage{mathrsfs}%

\usepackage{tikz,graphicx,color}
\usepackage{tikz-cd}%
\usepackage{tikz-3dplot}%
\usetikzlibrary{calc}%
\usetikzlibrary{arrows}%
\usetikzlibrary{shapes}%
\usetikzlibrary{patterns}%
\usetikzlibrary{positioning}%
\usetikzlibrary{arrows.meta}
\usetikzlibrary{decorations.markings}
\usetikzlibrary{knots}

\usepackage{epstopdf}%

\usepackage[arrow]{xy}%
\usepackage{diagbox}%
\usepackage{subfig}%
\usepackage{arcs}%
\usepackage{xcolor}%
\usepackage{xspace}

\usepackage[subtle,tracking=normal,mathdisplays=tight]{savetrees}
\usepackage{microtype}
\usepackage[margin=1in]{geometry}%

\usepackage{enumitem}
\usepackage{letltxmacro}
\usepackage{thmtools,etoolbox}
\makeatletter
\IfFormatAtLeastTF{2024-11-01}{%
 \renewcommand*\@addtoreset[2]{%
 \bgroup
 \edef\aliasctr@@truelist{\aliasctr@follow{#2}}%
 \let\@elt\relax
 \expandafter\@cons\aliasctr@@truelist{{#1}}%
 \egroup
 \expandafter\xdef\csname theH#1\endcsname{%
 \expandafter\noexpand\csname theH#2\endcsname.%
 \noexpand\the\noexpand\value{#1}}%
 }%
}{}
\makeatother

\def\myarabic#1{\normalfont(\roman{#1})}
\newlist{theoremlist}{enumerate}{1}
\setlist[theoremlist]{label=\myarabic{theoremlisti},ref={\myarabic{theoremlisti}},itemindent=0pt,labelindent=0pt,
 leftmargin=*,noitemsep}

\makeatletter
\renewcommand{\p@theoremlisti}{\perh@ps{\thetheorem}}
\protected\def\perh@ps#1#2{\textup{#1#2}}
\newcommand{\itemrefperh@ps}[2]{\textup{#2}}
\newcommand{\itemref}[1]{\begingroup\let\perh@ps\itemrefperh@ps\ref{#1}\endgroup}

\newcommand{\crefi}[2]{%
 \nameCref{#1}~\hyperref[#2]{%
 \ref*{#1}%
 \begingroup\let\perh@ps\itemrefperh@ps\ref*{#2}\endgroup%
 }%
}
\makeatother

\usepackage{crossreftools}

\makeatletter
\newcommand{\optionaldesc}[2]{%
 \phantomsection
 #1\protected@edef\@currentlabel{#1}\label{#2}%
}
\makeatother

\usepackage{nameref}
\usepackage[capitalize,nosort]{cleveref}

\usepackage{scalerel}
\usepackage{stackengine}

\usepackage{xr} %

\makeatletter
\patchcmd{\XR@@}
 {\set@curr@file{#3}}
 {\edef\@curr@file{#3}}
 {}
 {\PackageError{xr-local-fix}
 {Could not patch xr}
 {Check the installed xr version.}}

\def\XRauxsuffix{.aux}
\patchcmd{\XR@@}
 {\@curr@file.aux}
 {\@curr@file\XRauxsuffix}
 {}
 {\PackageError{xr-local-fix}
 {Suffix patch failed}
 {Check the installed xr version.}}

\def\extfile{t_embeddings_v2}
\IfFileExists{\extfile.aux}{%
 \externaldocument[M-]{\extfile}%
}{%
 \def\extfile{t_embeddings_aux}%
 \begingroup
 \def\XRauxsuffix{.tex}%
 \externaldocument[M-]{\extfile}[t_embeddings_v2.pdf]%
 \endgroup
}
\makeatother

\newtheorem{theoremintro}{Theorem}
\newtheorem{theorem}{Theorem}[section]

\newtheorem{lemma}[theorem]{Lemma}
\newtheorem{proposition}[theorem]{Proposition}
\newtheorem{corollary}[theorem]{Corollary}
\newtheorem{conjecture}[theorem]{Conjecture}
\newtheorem{problem}[theorem]{Problem}
\newtheorem{question}[theorem]{Question}
\newtheorem{assumption}{Assumption}

\theoremstyle{definition}
\newtheorem{remark}[theorem]{Remark}
\newtheorem{definition}[theorem]{Definition}
\newtheorem{example}[theorem]{Example}
\newtheorem{notation}[theorem]{Notation}
\newtheorem{algorithm}[theorem]{Algorithm}

\crefname{figure}{Figure}{Figures}
\crefname{equation}{Equation}{Equations}

\def\figref#1(#2){Figure~\hyperref[#1]{\ref*{#1}(#2)}}

\AtBeginEnvironment{theorem}{\crefalias{theoremlisti}{theorem}}
\AtBeginEnvironment{lemma}{\crefalias{theoremlisti}{lemma}}
\AtBeginEnvironment{proposition}{\crefalias{theoremlisti}{proposition}}
\AtBeginEnvironment{corollary}{\crefalias{theoremlisti}{corollary}}

\def\Acal{\mathcal{A}}\def\Bcal{\mathcal{B}}\def\Ccal{\mathcal{C}}\def\Gcal{\mathcal{G}}\def\Hcal{\mathcal{H}}\def\Kcal{\mathcal{K}}\def\Lcal{\mathcal{L}}\def\Mcal{\mathcal{M}}\def\Ncal{\mathcal{N}}
\def\Rcal{\mathcal{R}}

\def\C{{\mathbb{C}}}
\def\R{{\mathbb{R}}}
\def\Z{{\mathbb{Z}}}

\def\<{{\langle}}
\def\>{{\rangle}}
\def\eps{{\epsilon}}

\def\id{\operatorname{id}}
\def\det{\operatorname{det}}
\def\Ker{\operatorname{Ker}}
\def\diag{\operatorname{diag}}
\def\rank{\operatorname{rank}}
\def\Conv{ \operatorname{Conv}}
\def\Span{ \operatorname{Span}}
\def\wt{\operatorname{wt}}
\def\mod{{\operatorname{mod}}}
\def\Hom{\operatorname{Hom}}
\def\RR{{\mathbb R}}
\def\RP{{\RR\mathbb P}}

\def\xrasim{\xrightarrow{\sim}}

\def\GL{\operatorname{GL}}
\def\SL{\operatorname{SL}}
\def\Mat{\operatorname{Mat}}
\def\Mator_#1{\Mat^{\intsup}_{#1}(\R)}
\def\Gr{\operatorname{Gr}}
\def\Grtnn{\Gr_{\ge 0}}
\def\Grtp{\Gr_{>0}}
\def\Pio{\Pi^\circ}
\def\Povtp_#1{\Pi_{#1}^{>0}}
\def\Ptp_#1{\Povtp_{#1}}
\def\Povtnn_#1{\Pi_{#1}^{\geq0}}
\def\BND{\Bcal}
\def\Boundkn{\BND(k,n)}
\def\Boundxx(#1,#2){\BND(#1,#2)}

\numberwithin{equation}{section}

\makeatletter
\@namedef{subjclassname@2020}{\textup{2020} Mathematics Subject Classification}
\makeatother

\def\G{\Gamma}
\def\GD{\Gamma^\ast}
\def\GDp{\Gamma^{\prime\ast}}
\def\vertex{v}
\def\v{\vertex}
\def\vv{u}

\def\fletter{g}
\def\ffletter{f}
\def\fffletter{h}

\def\face{\fletter^\ast}
\def\ff{\ffletter^\ast}
\def\fff{\fffletter^\ast}

\def\f{\face}

\def\fout{\outacc{\fletter}^{\ast}}
\def\ffout{\outacc{\ffletter}^{\ast}}

\def\bdryaccent#1{#1^{\partial}}
\def\bdvletter{u}
\def\bdv{\bdvletter^{\partial}}
\def\bdvp{\bdvletter^{\prime\partial}}
\def\bdvx{\tilde \bdvletter^{\partial}}
\def\bde{\bdryaccent\e}
\def\bdep{\e^{\prime\partial}}
\def\bdeast{\e^{\partial\ast}}
\def\bdf{\ffletter^{\partial\ast}}
\def\bdwx{\tilde \wv^\partial}
\def\ddbdbx{\tilde{\ddot\bv}^\partial}
\def\bdb{\b^{\partial}}
\def\bdbp{\b^{\prime\partial}}
\def\bdbpp{\b^{\prime\prime\partial}}

\def\Verts{\mathbf{V}}
\def\Faces{\Verts^\ast}
\def\dFacesp{\dot\Verts^{\prime\ast}}
\def\Facesout{\outacc{\Verts}^\ast}

\def\intop{\mathtt{int}}
\def\Vint{\Verts_{\intop}}
\def\Vbd{\Verts_\partial}
\def\Fint{\Faces_{\intop}}
\def\Fbd{\Faces_\partial}
\def\bv{b}
\def\wv{w}
\def\e{e}
\def\east{e^\ast}
\def\E{\mathbf{E}}
\def\East{\mathbf{E}^\ast}
\def\Eint{\E_{\intop}}
\def\Ebd{\E_\partial}
\def\alt{\operatorname{alt}}
\def\altp{\alt^\perp}
\def\lalats{\lalatbf_n}
\def\la{\lambda}
\def\lat{\tilde\lambda}
\def\lalat{(\la,\lat)}
\def\lalak{\lalatbf_{k,n}^+}
\def\MPsup{\mathtt{M}+}
\def\lalakMP{\MPkntree}

\def\Matsup{\mathtt{Mat}}
\def\lalakMAT{\lalatbf_{k,n}^{+\Matsup}}
\def\MPkntreeMAT{\Mcal^{\ambsup,\Matsup}_{k,n;L=0}}
\def\lalakMPMAT{\MPkntreeMAT}
\def\lalatsMAT{\lalatbf_n^{\Matsup}}

\def\Meas{\operatorname{Meas}}
\def\BV{\Verts^{\bullet}}
\def\WV{\Verts^{\circ}}
\def\BVint{\BV_{\intop}}
\def\WVint{\WV_{\intop}}
\def\WVintp{\Verts^{\prime\circ}_{\intop}}
\def\WVp{\Verts^{\prime\circ}}
\def\BVbd{\BV_{\partial}}
\def\WVbd{\WV_{\partial}}
\def\Fw{F^\circ}
\def\Fwp{F^{\prime\circ}}
\def\Fb{\tilde F^\bullet}
\def\Fbp{\tilde F^{\prime\bullet}}
\def\I{\mathbf{i}}
\def\gauge{g}
\def\brxx<#1,#2>{\<#1\,#2\>}
\def\brla<#1,#2>{\brxx<#1,#2>_{\la}}
\def\brlabase<#1,#2>{\brxx<#1,#2>_{\labase}}

\def\brlat[#1,#2]{[#1\,#2]_{\lat}}
\def\La{\Lambda}
\def\Lat{\tilde\Lambda}
\def\PhiLL{\Phi_{\La,\Lat}}
\def\PhiLLL{\Phi_{\La,\Lat}^{\funcsub}}

\def\twonondeg{$2^\partial$-nondegenerate\xspace}
\def\Qla{Q_{\la}}
\def\Cast{\C^\times}
\def\Rast{\R^\times}
\def\latp{\lat^\perp}

\def\alphaT{\Tcomp{\alpha}}
\def\alphaO{\Ocomp{\alpha}}
\def\betaT{\Tcomp{\beta}}
\def\betaO{\Ocomp{\beta}}

\def\sumT{\alphaT}
\def\sumO{\alphaO}
\def\sumwT{\alphaT^\circ}
\def\sumbT{\alphaT^\bullet}
\def\sumText{\alphaT^{\mathtt{ext}}}

\def\sumOx{\alphaO_{\xd}}
\def\sumTx{\alphaT_{\xd}}
\def\sumOxout{\alphaO_{\xdout}}
\def\sumTxout{\alphaT_{\xdout}}
\def\sumwTx{\sumwT_{\xd}}
\def\sumbTx{\sumbT_{\xd}}
\def\sumwTxout{\sumwT_{\xdout}}
\def\sumbTxout{\sumbT_{\xdout}}
\def\sumwTy{\sumwT_{\yd}}
\def\sumbTy{\sumbT_{\yd}}

\def\cond(#1|#2){(#1|#2)}
\def\sumwTcond(#1|#2){\sumwT\cond(#1|#2)}
\def\sumbTcond(#1|#2){\sumbT\cond(#1|#2)}
\def\sumTcond(#1|#2){\sumT\cond(#1|#2)}

\def\sumcpT{\alphaT^{\conp}}
\def\sumcmT{\alphaT^{\conm}}
\def\sumcpmT{\alphaT^{\conpm}}
\def\sumxT^#1{\alphaT^{#1}}

\def\Re{\operatorname{Re}}
\def\Im{\operatorname{Im}}
\def\ovlmargin{0.1em}
\def\ovl#1{\hspace{\ovlmargin}\overline{\hspace{-\ovlmargin}#1\hspace{-\ovlmargin}}\hspace{\ovlmargin}}

\def\tdiag{\bt}
\def\epsK{\varepsilon}
\def\pFw{\partial\Fw}
\def\pFb{\partial\Fb}

\def\APMS{\Acal}
\def\APMSGbd(#1){\APMS_{#1}(\G)}
\def\apm{\bm{a}}
\def\by{\bm{y}}
\def\RV#1{{\normalfont(R$#1$)}\xspace} %
\def\MV#1{{\normalfont(M$#1$)}\xspace} %
\def\MVbd{\MV{\bdryaccent{1}}}
\def\wind{\operatorname{wind}}

\makeatletter
\let\save@mathaccent\mathaccent
\newcommand*\if@single[3]{%
 \setbox0\hbox{${\mathaccent"0362{#1}}^H$}%
 \setbox2\hbox{${\mathaccent"0362{\kern0pt#1}}^H$}%
 \ifdim\ht0=\ht2 #3\else #2\fi
 }
\newcommand*\rel@kern[1]{\kern#1\dimexpr\macc@kerna}
\newcommand*\widebar[1]{\@ifnextchar^{{\wide@bar{#1}{0}}}{\wide@bar{#1}{1}}}
\newcommand*\wide@bar[2]{\if@single{#1}{\wide@bar@{#1}{#2}{1}}{\wide@bar@{#1}{#2}{2}}}
\newcommand*\wide@bar@[3]{%
 \begingroup
 \def\mathaccent##1##2{%
 \let\mathaccent\save@mathaccent
 \if#32 \let\macc@nucleus\first@char \fi
 \setbox\z@\hbox{$\macc@style{\macc@nucleus}_{}$}%
 \setbox\tw@\hbox{$\macc@style{\macc@nucleus}{}_{}$}%
 \dimen@\wd\tw@
 \advance\dimen@-\wd\z@
 \divide\dimen@ 3
 \@tempdima\wd\tw@
 \advance\@tempdima-\scriptspace
 \divide\@tempdima 10
 \advance\dimen@-\@tempdima
 \ifdim\dimen@>\z@ \dimen@0pt\fi
 \rel@kern{0.6}\kern-\dimen@
 \if#31
 \overline{\rel@kern{-0.6}\kern\dimen@\macc@nucleus\rel@kern{0.4}\kern\dimen@}%
 \advance\dimen@0.4\dimexpr\macc@kerna
 \let\final@kern#2%
 \ifdim\dimen@<\z@ \let\final@kern1\fi
 \if\final@kern1 \kern-\dimen@\fi
 \else
 \overline{\rel@kern{-0.6}\kern\dimen@#1}%
 \fi
 }%
 \macc@depth\@ne
 \let\math@bgroup\@empty \let\math@egroup\macc@set@skewchar
 \mathsurround\z@ \frozen@everymath{\mathgroup\macc@group\relax}%
 \macc@set@skewchar\relax
 \let\mathaccentV\macc@nested@a
 \if#31
 \macc@nested@a\relax111{#1}%
 \else
 \futurelet\first@char\gobble@till@marker#1\endmarker
 \ifcat\noexpand\first@char A\else
 \fi
 \macc@nested@a\relax111{\first@char}%
 \fi
 \endgroup
}
\makeatother

\makeatletter
\DeclareRobustCommand{\cev}[1]{%
 \mathpalette\do@cev{#1}%
}
\newcommand{\do@cev}[2]{%
 \fix@cev{#1}{+}%
 \reflectbox{$\m@th#1\vec{\reflectbox{$\fix@cev{#1}{-}\m@th#1#2\fix@cev{#1}{+}$}}$}%
 \fix@cev{#1}{-}%
}
\newcommand{\fix@cev}[2]{%
 \ifx#1\displaystyle
 \mkern#23mu
 \else
 \ifx#1\textstyle
 \mkern#23mu
 \else
 \ifx#1\scriptstyle
 \mkern#22mu
 \else
 \mkern#22mu
 \fi
 \fi
 \fi
}

\makeatother

\def\laext{\la^\circ}
\def\latext{\lat^\bullet}
\def\Chat{\widehat C}
\def\Chatext{\widehat C^\circ}
\def\Cext{C^\circ}
\def\Cptext{C^{\perp\bullet}}
\def\labf{\bm{\la}}
\def\latbf{\bm{\lat}}
\def\lalatbf{\bm{\la\!^\perp\!\!\tilde\la}}
\def\LaLatbf{\bm{\La\!\tilde\La}}
\def\lak{\labf_{k,n}^+}
\def\latk{\latbf_{k,n}^+}
\def\lakMAT{\labf_{k,n}^{+\Matsup}}
\def\latkMAT{\latbf_{k,n}^{+\Matsup}}
\def\LaLak{\LaLatbf_{k,n}^+}
\def\LaLat{(\La,\Lat)}

\newlength\arrowheight

\def\Path{\zzpath}
\def\bdryarcs{\partial^{\mathtt{arcs}}}
\def\nbdryarcs#1{|\bdryarcs#1|}
\def\laextsub{\la}
\def\latextsub{\lat}
\def\brlaw<#1,#2>{\brxx<#1,#2>_{\laextsub}}
\def\brlatb[#1,#2]{[#1\,#2]_{\latextsub}}

\def\wtlaK{\Kop_{\laextsub}}
\def\wtlap{\wt_{\laextsub}}
\def\Kop{\operatorname{K}}
\def\wtK{\Kop}
\def\ddwt{\ddot{\wt}}
\def\dwtKrow{\dot{\wtK}_{\row}}
\def\ddwtphan{\smash{\ddwt}\vphantom{\wt}}
\def\ddKop{\ddot{\Kop}}
\def\ddwtlaK{\ddKop_{\laextsub}}
\def\ddwtlap{\ddwtphan_{\laextsub}}
\def\ddG{\ddot{\G}}
\def\dG{\dot{\G}}
\def\Sig{\Sigma}
\def\darkop{\mathtt{dark}}
\def\lightop{\mathtt{light}}
\def\BSig{\Sig^{\darkop}}
\def\WSig{\Sig^{\lightop}}
\def\ddw{\ddot{\wv}}
\def\ddb{\ddot{\bv}}
\def\ddbx{\ddot{\bv}}
\def\dde{\ddot{\e}}
\def\ddface{\ddot\ffletter^*}
\def\ddF{\ddface}
\def\w{\wv}
\def\b{\bv}
\def\triang{\tau}

\def\ddVerts{\ddot{\mathbf{V}}}
\def\ddVbd{\ddVerts_\partial}
\def\dVerts{\dot{\mathbf{V}}}
\def\ddFaces{\ddVerts^\ast}
\def\ddE{\ddot{\mathbf{E}}}

\def\ddBV{\ddVerts^{\bullet}}
\def\ddWV{\ddVerts^{\circ}}
\def\dWV{\dVerts^{\circ}}
\def\ddBVint{\ddBV_{\intop}}
\def\ddWVint{\ddWV_{\intop}}
\def\ddWVintp{\ddVerts^{\prime\circ}_{\intop}}
\def\ddBVp{\ddVerts^{\prime\bullet}}
\def\zzpath{\gamma}
\def\s{s}
\def\dCphan{\smash{\dot{C}}\vphantom{C}}
\def\ddCphan{\smash{\ddot{C}}\vphantom{C}}
\def\ddCext{\ddCphan^\circ}
\def\dCext{\dCphan^\circ}
\def\pddCext{\partial\ddCext}
\def\ss{j}
\def\leq{\leqslant}
\def\geq{\geqslant}
\def\ge{\geqslant}

\def\fC{f_{C}}
\def\fG{f_\G}
\def\Rtp{\R_{>0}}
\def\bzero{\bm{0}}
\def\Id{\bm{1}}

\def\fhat{f^\vee}

\def\Ghat{\G^\vee}
\def\comp#1{#1^{c}}

\def\brn{\bm{[n]}}
\def\brm{\bm{[m]}}
\def\brd{\bm{[d]}}
\def\brx#1{\bm{[#1]}}

\def\Ttaus{\bm{\tau\!T}}
\def\Ttauknij{\Ttaus^{i\ssep j}_{k,n}}
\def\ijsepsup{i\ssep j}
\def\Ttaucoef{c^{\ijsepsup}_{\tau,T}}

\def\brk{\brx{k}}
\def\Pmom{P}
\def\MPop{\Mcal^{\ambsup}}

\def\Gbf_#1{\mathbf{\G}_{#1|\funcsub}}
\def\BCG{\mathbf{\G}^{\BCop}}

\def\brat[#1|{[#1|}
\def\kett|#1]{|#1]}
\def\bralat[#1|{[#1|_{\lat}}
\def\ketlat|#1]{|#1]_{\lat}}
\def\Tcomp#1{\hat{#1}}
\def\Ocomp#1{\check{#1}}
\def\PmomT{\Tcomp\Pmom}
\def\PmomO{\Ocomp\Pmom}
\def\Qmom{Q}
\def\QmomT{\Tcomp\Qmom}
\def\QmomO{\Ocomp\Qmom}
\def\ycomp#1{#1^{\y}}
\def\ytcomp#1{#1^{\yt}}
\def\Pmomy{\ycomp\Pmom}
\def\Pmomyt{\ytcomp{\Pmom}}
\def\Qmomy{\ycomp\Qmom}
\def\Qmomyt{\ytcomp\Qmom}
\def\r{r}
\def\Rmom{R}
\def\RmomT{\Tcomp\Rmom}
\def\RmomO{\Ocomp\Rmom}

\def\PllT{\PbdT_{\la,\lat}}
\def\Pll{\Pbd_{\la,\lat}}

\def\llPll{(\la,\lat)}
\def\llPllL{(\la,\lat;\byL)}
\def\byL{\by_{\bmnL}}
\def\Pcurve{\bm{p}}
\def\PcurveT{\Tcomp{\bm{p}}}
\def\Pbd{\Pcurve^{\partial}}
\def\PbdT{\PcurveT\vphantom{\Pcurve}^{\partial}}

\def\LG{T}
\def\LGp{T_+}
\def\LGpm{T_{\pm}}
\def\dwv{\dot\wv}
\def\dbv{\dot\bv}
\def\row{F}
\def\rowt{\tilde F}
\def\rowext{\row^\circ}
\def\rowtext{\rowt^\bullet}

\def\tauT{(\tau,T)}
\def\Fl{\bm{Fl}}
\def\Ztnn{\Z_{\geq0}}

\def\Rg{R}
\def\Rgout{\outacc{\Rg}}
\def\Rgbipout{\outacc{\Rg}}
\def\RWbipout{\outacc{\Rg}^{\circ}}
\def\RW{\Rg^\circ}
\def\RB{\Rg^\bullet}
\def\Rgc{\comp{\Rg}}

\def\RgEint{\Eint\ind[\Rg]}
\def\RgEintx_#1{\Eint\ind[\Rg_{#1}]}
\def\RgEbd{\Ebd\ind[\Rg]}
\def\RgpEbd{\Ebd\ind[\Rg']}
\def\RgFac{\Facesgr\ind[\Rg]}
\def\RgFacl{\Faces\ind[\Rg]}
\def\RgFaclx_#1{\Faces\ind[\Rg_{#1}]}
\def\RgCyc{\Cyc_{\Rg}}
\def\RgCycx_#1{\Cyc_{\Rg_{#1}}}

\def\Fintgr{\Vint^\times}
\def\Fbdgr{\Vbd^\times}

\def\RgFbd{\Fbdgr\ind[\Rg]}
\def\RgFint{\Fintgr\ind[\Rg]}
\def\RgFintl{\Fint\ind[\Rg]}
\def\RgE{\Edges\ind[\Rg]}

\def\RgWV{\Rg^\circ}
\def\RgBV{\Rg^\bullet}

\def\RgoutFacl{\Fout\ind[\Rgout]}

\def\Rghol{\Rg^{\mathrm{hole}}}
\def\Rgcl{\Sclose{\Rg}}
\def\Faclxout[#1]{\Fout\ind[#1]}

\def\GR{\G\ind[\Rg]}
\def\GRext{\G\ind[\RgCyc]}
\def\GRextx_#1{\G\ind[\RgCycx_{#1}]}

\def\lgg{\Lfunc Grassmannian graph\xspace}
\def\algg{an \Lfunc Grassmannian graph\xspace}
\def\lggs{\Lfunc Grassmannian graphs\xspace}
\def\LggsTITLE{\LfuncTITLE Grassmannian graphs\xspace}

\def\helW{k^\circ_\G}
\def\helB{k^\bullet_\G}
\def\helWsub_#1{k^\circ_{#1}}
\def\helBsub_#1{k^\bullet_{#1}}

\def\helWmin{k^\circ_{\min}}
\def\helBmin{k^\bullet_{\min}}
\def\helmin{k_{\min}}

\def\heldW{\helWsub_{\dG}}
\def\heldB{\helBsub_{\dG}}

\def\gelletter{h}

\def\gelG{\gelletter_\G}
\def\gelW{\gelletter^\circ_\G}
\def\gelB{\gelletter^\bullet_\G}
\def\gelWsub_#1{\gelletter^\circ_{#1}}
\def\gelBsub_#1{\gelletter^\bullet_{#1}}
\def\gelWout{\gelletter^\circ_{\Gout}}
\def\gelBout{\gelletter^\bullet_{\Gout}}
\def\gelWgen{\gelletter^\circ_{\Ggen}}
\def\gelBgen{\gelletter^\bullet_{\Ggen}}

\def\gelcpmout{\gelletter^{\conpm}_{\Gout}}

\def\gelmin{\gelletter_{\min}}
\def\gelWmin{\gelletter^\circ_{\min}}
\def\gelBmin{\gelletter^\bullet_{\min}}

\def\Puncs{\Plocs}
\def\degG{\deg_\G}

\def\xd{\bm{x}}
\def\xdT{\Tcomp{\xd}}
\def\xT{\Tcomp{\xd}}
\def\xO{\Ocomp{\xd}}
\def\xt{\Tcomp{\xd}}
\def\xs{x}
\def\xsT{\Tcomp{x}}
\def\xsO{\Ocomp{x}}
\def\ys{y}
\def\ysT{\Tcomp{y}}
\def\yloc_#1{y_{(#1)}}
\def\yTloc_#1{\Tcomp{y}_{(#1)}}
\def\yOloc_#1{\Ocomp{y}_{(#1)}}

\def\yd{\xd'}
\def\ydout{\xdout'}
\def\yT{\xT'}
\def\yTout{\xTout'}

\def\clique{\Delta}
\def\corner{\nu}
\def\corners{\bm{\nu}}
\def\cornersw{\bm{\nu}^\circ}
\def\cornersb{\bm{\nu}^\bullet}
\def\corsout{\corners_{\GDout}}
\def\cornersGD{\corners_{\GD}}
\def\conven{\oldmathtt{c}}
\def\nL{L}
\def\loopf_#1{\pi^\ast_{(#1)}}

\def\Plocs{\bm{\pi^\ast}_{\bmnL}}
\def\ploc_#1{\loopf_{#1}}

\def\brnLv{\Pfl^{\v}}
\def\brnLcorv{\Pfl^{\corv}}
\def\brnLvo{\Pfl^{\vo}}
\def\brnLvL{\Pflout^{\vL}}
\def\brnLvR{\Pflout^{\vR}}
\def\brnLvout{\Pflout^{\vout}}

\def\vo{v_0}

\def\partFout{\partial_{\Facesout}}

\def\Tcur{\xT_{\eps}}

\def\Vact{\Vint^{\mathtt{flex}}}

\def\Edges{\mathbf{E}}

\def\corfm{\f_{\corner-}}
\def\corfp{\f_{\corner+}}
\def\corfpm{\f_{\corner\pm}}
\def\coream{\east_{\corner-}}
\def\coreap{\east_{\corner+}}
\def\coreapm{\east_{\corner\pm}}

\def\corem{\e_{\corner-}}
\def\corep{\e_{\corner+}}
\def\corepm{\e_{\corner\pm}}
\def\corf{\f_{\corner}}
\def\corfx_#1{\f_{\corner_{#1}}}
\def\corv{\v_{\corner}}
\def\corvx_#1{\v_{#1}}

\def\GDarb{\GD}
\def\FDarb{\Faces}
\def\EDarb{\East}
\def\Tarb{\xT}

\def\brnL{\bm{[\nL]}}
\def\BCGDinitx_#1{\GD_{0|#1}}

\def\Pivots(#1){\partF^{\outsub}(#1)}

\def\BCmvs{\bm{\mu}_{\BCop}^{\BCin}}
\def\BCmvsx_#1{\bm{\mu}_{\BCop}^{#1}}

\def\BCin{\delta}
\def\bridgeop{\mathtt{br}}
\def\uL{u^{\bridgeop}_{+}}
\def\uR{u^{\bridgeop}_{-}}
\def\uLR{u^{\bridgeop}_{\pm}}

\def\vL{\outacc{v}_{+}}
\def\vR{\outacc{v}_{-}}
\def\vLR{\outacc{v}_{\pm}}

\def\Gout{\outacc{\G}}
\def\GDout{\outacc{\G}^{\ast}}
\def\GDpout{\outacc{\G}^{\prime\ast}}

\def\BCGs{\BCG_{k,n;\funcsub}}
\def\BCscl{1}
\def\BCop{\text{\scalebox{\BCscl}{$\mathtt{BCFW}$}}}
\def\kinrealsup{\text{\scalebox{\BCscl}{$\mathtt{BCFW}$}}\ \!\!\intsup}
\def\rBCGs{\mathbf{\G}^{\kinrealsup}_{k,n;\funcsub}}
\def\oBCGs{\rBCGs}
\def\ddoBCGs{\mathrlap{\ddot{\overline{\phantom{\mathbf{\G}}}}}\mathbf{\G}^{\kinrealsup}_{k,n;L}}

\def\CollWsup{}
\def\CollWBsup{}
\def\CollBWsup{'}

\def\CollBCGs{\mathrlap{\overline{\phantom{\mathbf{\G}}}}\mathbf{\G}^{\kinrealsup}_{k,n;L}}

\def\terminal{terminal\xspace}
\def\Terminal{Terminal\xspace}
\def\knL{(k,n;\nL)}
\def\knl{\knL}
\def\bdx{\xs^\partial}
\def\bdxM_#1{M_{\bdx_{#1}}}

\def\Gloop{\Gfunc}
\def\Gloopgr{\G}
\def\Gloopsub_#1{\G_{#1|\funcsub}}
\def\Gloopsubp_#1{\G'_{#1|\funcsub}}
\def\Glinit{\Gloopsub_0}
\def\Glout{\outacc{\G}\subfuncsubG}
\def\GDlout{\outacc{\G}^\ast}%

\def\Gfin{\Gfunc}
\def\Gfingr{\G}
\def\GDfin{\G^\ast}
\def\xdfin{\xd}

\def\MPGfin{\MPop_{\Gfin}}

\def\xdout{\bm{y}}
\def\xTout{\Tcomp{\xdout}}

\def\xdf(#1){\xd(#1)}
\def\xTf(#1){\Tcomp{\xd}(#1)}
\def\Pm{\Pmom_-}
\def\Pp{\Pmom_+}
\def\PmT{\PmomT_-}
\def\PpT{\PmomT_+}
\def\PpO{\PmomO_+}
\def\cc{\corner,\conven}

\def\xrout{\xdr(\ffout)}
\def\xtrout{\xTr(\ffout)}

\def\Rcc{\Rmom_{\cc}}
\def\RccT{\RmomT_{\cc}}
\def\RccO{\RmomO_{\cc}}
\def\open[#1]{[#1]^\intsup}

\def\KT{\Tcomp{\clique}}
\def\KO{\Ocomp{\clique}}
\def\rmax{r_{\maxsup}}
\def\rcrit{\outacc{r}}

\def\Kbig{\clique'}
\def\KTbig{\KT'}
\def\Ksm{\clique}
\def\KTsm{\KT}

\def\xM{M}
\def\ap{a_+}
\def\am{a_-}
\def\CMI{\eta}

\def\lap{\la^\perp}
\def\Qlapp{Q_\la^\vee}
\def\ddC{\ddot{C}}
\def\AA{\Acal^{\ambsup}}
\def\AAkntree{\AA_{k-2,n;L=0}}
\def\AAknL{\AA_{k-2,n;L}}
\def\AAddG{\AA_{\ddGvunc}}

\def\Line_#1{\Lcal_{(#1)}}
\def\Liner{\Lcal_{(\rho)}}
\def\Lineg{\Lcal_{(\gamma)}}

\def\var{\operatorname{var}}
\def\brV[#1,#2,#3,#4]{[#1\,#2\,#3\,#4]_V}
\def\brVL[#1,#2,#3]{[#1\,#2\,(#3)]_{V}}
\def\brLL[#1,#2]{[(#1)\,(#2)]_V}

\def\bdxT{\xsT^{\partial}}
\def\bdxO{\xsO^{\partial}}

\def\MPkntree{\MPop_{k,n;L=0}}
\def\MPG{\MPop_{\Gfunc}}
\def\MPknL{\MPop_{k,n;\nL}}

\def\MPknLFGLS{\Mcal^{\text{\scalebox{\ttscl}{[FGLS]}}}_{k,n;\nL}}
\def\MomLLLprojoFL{\Mcal^{\text{\scalebox{\ttscl}{[FL]}}}_{k,n;L}(\La,\Lat)}

\def\Grsup^#1{\Gr^{(#1)}}
\def\Grsupla{\Grsup^\la}
\def\Grtnnsupla{\Gr_{\geq0}^{(\la)}}
\def\flagsup{\mathtt{flag}}
\def\MAop{\Acal^{\flagsup}}
\def\MAkntree{\MAop_{k-2,n;L=0}}
\def\MAknL{\MAop_{k-2,n;L}}
\def\MAddG{\MAop_{\ddGvunc}}

\def\colB{\bullet}
\def\colW{\circ}
\def\conp{\conven_+}
\def\conm{\conven_-}
\def\conpm{\conven_\pm}
\def\conmp{\conven_\mp}
\def\leftcon{\colB\colW}
\def\rightcon{\colW\colB}

\def\Disk{\mathbb{D}}
\def\cor{\corner}

\def\Skelop{\operatorname{Sk}^1}
\def\SkelGD{\Skelop(\GD)}
\def\Skel{\xd(\SkelGD)}
\def\SkelT{\xT(\SkelGD)}
\def\Skelteh{\xteh(\SkelGD)}
\def\rigop{\mathtt{rig}}
\def\Vrig{\Vint^{\rigop}}
\def\Vrigout{\outacc{\Verts}_{\intop}^{\rigop}}
\def\GDr{\GD_{\threelegs}}
\def\threelegs{
\!\!\scalebox{2}{\begin{tikzpicture}[scale=0.07,baseline=(Z.base)]
\coordinate(Z) at (0,-1);
\node[draw,circle,fill=black,scale=0.05](O) at (0,0) {};
\draw(O)--(0.7,-1);
\draw(O)--(0,-1.2);
\draw(O)--(-0.7,-1);
\end{tikzpicture}}
}

\def\threelegsdash{
\!\!\scalebox{2}{\begin{tikzpicture}[scale=0.07,baseline=(Z.base)]
\coordinate(Z) at (0,-1);
\node[draw,circle,fill=black,scale=0.05](O) at (0,0) {};
\draw[black!20](O)--(0.7,-1);
\draw[black!20](O)--(0,-1.2);
\draw[black!20](O)--(-0.7,-1);
\node[draw,circle,fill=black,scale=0.05](O) at (0,0) {};
\end{tikzpicture}}
}
\def\GDo{\GD_{\threelegsdash}}
\def\Chords{\Ccal_{\,\threelegsdash}}
\def\Chordsiso{\Chords^{\mathtt{iso}}}
\def\Chordsr{\Ccal_{\,\threelegs}}

\def\xdr{\xd_{r}}
\def\xTr{\xT_{r}}
\def\xdrp{\xd_{r'}}
\def\xTrp{\xT_{r'}}

\def\ORATITLE{Origami reconstruction algorithm\xspace}
\def\oraTITLE{origami reconstruction algorithm\xspace}
\def\ORA{origami reconstruction algorithm\xspace}
\def\ora{origami reconstruction algorithm\xspace}
\def\xdrcrit{\xd_{\rcrit}}
\def\xTrcrit{\xT_{\rcrit}}
\def\xdrmax{\xd_{\rmax}}
\def\xTrmax{\xT_{\rmax}}

\def\bendh{\gamma_\eps}
\def\normal{\vec n}
\def\bendhall{\xT_\eps}
\def\hall{\bendhall}

\def\bendhx#1{\bendh^{(#1)}}
\def\bendhr{\bendh^{(r)}}
\def\bendhpm{\bendh^\pm}

\def\convex{convex\xspace}
\def\MCEnoacr{M-\convex weak embedding\xspace}
\def\MCEsnoacr{M-\convex weak embeddings\xspace}
\def\MCEsnoacrTITLE{\MdashTITLE \convex embeddings\xspace}
\def\MCEnoacrTITLE{\MdashTITLE \convex embedding\xspace}

\def\MCE{MCE\xspace}
\def\MCEs{MCEs\xspace}
\def\MCNE{MCE\xspace}
\def\MNNE{M-nonnegative weak embedding\xspace}
\def\MNEsTITLE{\MdashTITLE nonnegative embeddings\xspace}

\def\MCEintro{\MCEnoacrTITLE}
\def\MCEsintro{\MCEsnoacrTITLE}

\def\wemb{weak embedding\xspace}
\def\Wemb{Weak embedding\xspace}
\def\wembs{weak embeddings\xspace}
\def\Wembs{Weak embeddings\xspace}
\def\wtemb{weak t-embedding\xspace}
\def\wtembs{weak t-embeddings\xspace}
\def\Wtemb{Weak t-embedding\xspace}
\def\wtembsTITLE{weak t-embeddings\xspace}
\def\awemb{a weak embedding\xspace}

\def\pMNE{proper \MNE}
\def\pMNNE{proper \MNNE}

\def\pcMNE{properly colored \MNE}
\def\pcMNEs{properly colored \MNEs}
\def\pcMNNE{properly colored \MNNE}

\def\xteh{\Tcomp{\xd}_{\eps}}
\def\xtehint(#1){\xT_{\eps}(#1)^\intsup}
\def\xtehout{\Tcomp{\xdout}_{\eps}}

\def\xTint(#1){\xT(#1)^\intsup}
\def\xToutint(#1){\xTout(#1)^\intsup}
\def\Convint{\Conv^\intsup}
\def\intsup{\diamond}

\def\Convrelint{\Conv_{\mathtt{rel}}^{\intsup}}

\def\Kbigs{\bm{\clique}^{\prime\;\maxsup}_{\ffout}}

\def\ebarast{\bar{e}^\ast}
\def\Ebarast{\bar{\E}^\ast}
\def\Evecast{\vec{\E}^\ast}
\def\evecast{\vec{e}^{\,\ast}}
\def\corevecam{\evecast_{\corner-}}

\def\past{p^\ast}
\def\qast{q^\ast}

\def\RaysDONOTUSE{\Rcal}
\def\Ray{{\mathfrak{R}}}
\def\FRay{\Ray_{\cc}}
\def\Raym{\Ray_{\corfm}}
\def\Rayo{\Ray_{\corf}}
\def\Rayp{\Ray_{\corfp}}
\def\Raypm{\Ray_{\corfpm}}

\def\Rayopen{\Ray^\intsup}
\def\vout{\outacc\v}
\def\vvout{\outacc\vv}
\def\vr{\outacc{\v}_{\,\threelegs}}

\def\outacc{\tilde}
\def\eout{\outacc\e}
\def\eastout{\outacc e^\ast}

\def\partF{\partial_{\Faces}}
\def\partEvecast{\vec\partial_{\Evecast}}
\def\partEast{\partial_{\East}}
\def\partGD{\partial_{\GD}}
\def\partE{\partial_{\Edges}}

\def\iccar{almost-triangular\xspace}

\def\ptrgle{pseudo-triangle\xspace}
\def\ptrgles{pseudo-triangles\xspace}
\def\ptrar{pseudo-triangular\xspace}

\def\ptron{pseudo-triangulation\xspace}
\def\ptrons{pseudo-triangulations\xspace}

\def\Raysffout{\RaysDONOTUSE_{\ffout}}
\def\fmin{\f_{\minsup}}
\def\fmax{\f_{\maxsup}}
\def\bmin{b_{\minsup}}
\def\bmax{b_{\maxsup}}

\def\fall(#1){\preK_{#1}}

\def\preKray{\preK'_\Ray}
\def\Kray{\clique'_\Ray}
\def\KTray{\KT'_\Ray}

\def\Rext{\Ray^{\mathtt{ext}}}
\def\ORst{origami reconstruction step\xspace}
\def\ORsts{origami reconstruction steps\xspace}

\def\sumwTcr{\sumwT}
\def\sumbTcr{\sumbT}

\def\corout{\outacc{\cor}}
\def\BWcon{\leftcon}
\def\WBcon{\rightcon}

\def\Vsact{\Vint^{\text{\scalebox{\ttscl}{$\geq4$}}}}
\def\Vrigsact{\Vint^{\rigop\text{\scalebox{\ttscl}{$\geq4$}}}}
\def\floatsub{\mathtt{float}}
\def\Pfix{\brnL_{\fixsub}}
\def\Pflout{\Pfl}
\def\Pfl{\brnL_{\floatsub}}
\def\Facesgr{\Verts^\times}
\def\GDgr{\G^\times}
\def\partgeom{\partial} %
\def\MCMS{M-\convex moduli space\xspace}
\def\MCMSs{M-\convex moduli spaces\xspace}
\def\MCMSsTITLE{\MdashTITLE \convex moduli spaces\xspace}

\def\MCMSintro{\MdashTITLE \convex moduli space\xspace}

\def\MCM{\Mmce}
\def\ORAop{\mathtt{OR}}
\def\ORmv{\mu_{\ORAop}^{\BCin}}
\def\ORmvx_#1{\mu_{\ORAop}^{#1}}
\def\MCMgen{\MCM^{\BCin}}
\def\MCMgenx_#1{\MCM^{#1}}

\def\Res{\operatorname{Res}}
\def\fpiv{\fout}
\def\rhopiv{\outacc{\rho}}

\def\Vintout{\outacc{\Verts}_{\intop}}
\def\BVintout{\outacc{\Verts}^\bullet_{\intop}}
\def\xdinit{\xd_0}
\def\bic{bicolored\xspace}
\def\unic{unicolored\xspace}

\def\geps{\gamma_\eps}
\def\gzero{\gamma_0}
\def\turnangle{\TURN}
\def\turnanglex_#1{\TURN_{#1}}
\def\winding{winding\xspace}

\def\MNE{M-nonnegative \wemb}
\def\MNEs{M-nonnegative \wembs}

\def\pchord{chord\xspace}
\def\pchords{chords\xspace}
\def\extpchord{external \pchord}
\def\Extpchord{External \pchord}
\def\extpchords{external \pchords}

\def\Extpt{External point\xspace}
\def\Extpts{External points\xspace}
\def\extpt{external point\xspace}
\def\extpts{external points\xspace}

\def\extcycle{external cycle\xspace}
\def\extcycles{external cycles\xspace}
\def\Extcycle{External cycle\xspace}

\def\addable{addable\xspace}

\def\extsup{\mathtt{ext}}

\def\partFext{\partF^{\extsup}}
\def\partFextout{\partFout^{\extsup}}
\def\Cyc{\zeta}
\def\Cycloc_#1{\Cyc_{(#1)}}
\def\CycKmax{\Cyc^{\extsup}_{\clique}}
\def\CycKmaxx{\Cyc^{\extsup}_{\clique_{\xd}}}
\def\CycKsm{\Cyc^{\extsup}_{\Ksm}}

\def\Kmax{\clique}
\def\Kmaxs{\bm{\clique}^{\maxsup}}
\def\Kmaxx{\Kmax_{\xd}}
\def\Kmaxy{\Kmax_{\yd}}
\def\KTmaxx{\KTmax_{\xd}}
\def\KTmaxy{\KTmax_{\yd}}
\def\preK{\nabla}
\def\preKmax{\preK}
\def\preKmaxs{\bm{\preK}^{\max}}

\def\Kmaxp{\clique'}
\def\cliquep{\clique'}
\def\KTmaxp{\KTmax'}
\def\KTp{\KT'}
\def\preKmaxp{\preKmax'}

\def\KTmax{\KT}
\def\KOmax{\KO}
\def\dotKmax{\Kmax'}
\def\dotKTmax{\KTmax'}
\def\Fout{\Facesout}
\def\Fintout{\outacc{\Verts}^\ast_{\intop}}
\def\Eastout{\outacc{\Edges}^\ast}
\def\pspec{$\oplus$-special\xspace}
\def\mspec{$\ominus$-special\xspace}
\def\pmspec{$\opm$-special\xspace}
\def\pandmspec{$\oplus$- and $\ominus$-special\xspace}
\def\pormspec{$\oplus$- or $\ominus$-special\xspace}

\newcommand{\opm}{ 
 \mathbin{
 \mathchoice
 {\buildcirclepm{\displaystyle }{0.14ex}{0.95}{0.05ex}{.7}}
 {\buildcirclepm{\textstyle }{0.14ex}{0.95}{0.05ex}{.7}}
 {\buildcirclepm{\scriptstyle }{0.13ex}{0.955}{0.04ex}{.55}}
 {\buildcirclepm{\scriptscriptstyle}{0.08ex}{0.95}{0.03ex}{.45}}
 } 
}

\newcommand\buildcirclepm[5]{%
 \begin{tikzpicture}[baseline=(X.base), inner sep=-#5, outer sep=-.65]
 \node[draw,circle,line width=#4] (X) {\footnotesize\raisebox{#2}{\scalebox{#3}{$#1\pm$}}};
 \end{tikzpicture}%
}

\def\colop{\operatorname{col}}
\def\colbarop{\overline{\operatorname{col}}}
\def\bperim{\bm{P}}

\def\adatr{an \datr}
\def\datr{algebraic t-realization\xspace}
\def\datrs{algebraic t-realizations\xspace}

\def\Rtpgauge{\Rtp^{|\Faces|-1}}
\def\ddRtpgauge{\Rtp^{|\ddFaces|-1}}
\def\RtpE{\Rtp^{|\Edges|}}
\def\RtpVint{\Rtp^{|\Vint|}}
\def\pmoneVint{\{\pm1\}^{|\Vint|}}
\def\pmoneVintp{\{\pm1\}^{|\Vint'|}}
\def\decacc#1{#1}

\def\talgop{\mathtt{ATR}}
\def\teop{\mathtt{WTE}}
\def\tiop{\mathtt{WTI}}

\def\Mdatr{\decacc\Mcal_{\talgop}}

\def\einjsup{\intsup}

\def\Mdti{\decacc\Mcal_{\tiop}}
\def\Mdte{\decacc\Mcal_{\teop}}
\def\Mdtio{\decacc\Mcal^{\einjsup}_{\tiop}}
\def\Mdteo{\decacc\Mcal^{\einjsup}_{\teop}}
\def\Mdatro{\decacc\Mcal^{\einjsup}_{\talgop}}
\def\MdteMP{\decacc{\Mcal}_{\teop}^{\MPsup}}
\def\MdteoMP{\decacc{\Mcal}_{\teop}^{\intsup,\MPsup}}

\def\gdim{\operatorname{d}_{\MCEop}}
\def\gdimo{\operatorname{d}_0}
\def\gdimatr(#1){\operatorname{d}_{\mathtt{term}}(#1^\ast)}

\def\Neigh{\operatorname{N}}
\def\NeighG{\Neigh_{\G}}
\def\match{\apm}
\def\nvint(#1){|\Sclose{#1}|}
\def\nvintRgp{\Sclose{\Rg}'}
\def\Sclose#1{\widehat{#1}}

\def\MCEop{\mathtt{MCE}}
\def\Mmce{\Mcal_{\MCEop}}

\def\Gbip{\G}
\def\Gbipout{\Gout}

\def\Rdd{\R^{2,2}}
\def\rvalid{\bm{r}}
\def\BCFWid{input datum\xspace}
\def\aBCFWid{an input datum\xspace}
\def\ot{\leftarrow}
\def\algsub{\mathtt{alg}}
\def\dimZ{\dim_{\algsub}}

\def\Resop{\operatorname{Res}}
\def\ResG{\Resop_{\GD}}
\def\ResGx_#1{\Resop_{\GD_#1}}
\def\ResGt{\Resop_{\GD_t}}
\def\ResGlinit{\Resop_{\GDinit}}

\def\MCMalg{\Mcal_{\algsub}}

\def\GDinit{\GD_0}
\def\Tmax{T}

\def\MCMcletter{Y}

\def\MCMcellint{\MCMcletter^\intsup_{\MCEop}}

\def\dMCMcellint{\widehat{\MCMcletter}^\intsup_{\MCEop}}

\def\Kinsupp{Kinematic support\xspace}
\def\kinsupp{kinematic support\xspace}
\def\haskinsupp{has \kinsupp}
\def\havekinsupp{have \kinsupp}

\def\NullGD{\G_{\nullsub}^\ast}
\def\NullG(#1,#2){\Gamma_{\nullsub}^\ast(#2)}
\def\nullsub{\Ncal}
\def\NullE(#1,#2){\Ebarast_{\nullsub}(#2)}
\def\NullED{\Ebarast_{\nullsub}}

\def\ind[#1]{[#1]}
\def\excsub{\mathtt{exc}}
\def\Zexc{Z_{\excsub}(\Glinit)}
\def\ORmvt{\ORmvx_{\BCin_t}}

\def\bending{bending\xspace}
\def\bendings{bendings\xspace}
\def\bent{bent\xspace}
\newcommand\precdot{\mathrel{\ooalign{$\prec$\cr
 \hidewidth\hbox{$\cdot\mkern0.0mu$}\cr}}}
\def\Vbigons{\Vint^{(2)}}
\def\Vtrs{\Vint^{(3)}}
\def\axMCElast{\itemref{MCE5:Mpos_chords}\xspace}
\def\axrangeMCEall{\itemref{MCE1:emb}--\axMCElast}
\def\axrangeMCEchord{\itemref{MCE1:emb}--\itemref{MCE3:M-tnn} and~\itemref{MCE5:Mpos_chords}\xspace}
\def\axrangeMCElasttwo{\itemref{MCE4:properly_colored}--\itemref{MCE5:Mpos_chords}\xspace}
\def\ebendh{\eps}
\def\ebendhx#1{\eps}

\def\btrgle{triangle/bigon\xspace}
\def\btrgles{triangles/bigons\xspace}
\def\btrar{triangular/bigonal\xspace}
\def\ggg{generic Grassmannian graph\xspace}
\def\Ggen{\widehat\G}
\def\Vgen{\widehat\Verts}
\def\Egen{\widehat\Edges}
\def\degGgen{\deg_{\Ggen}}
\def\sconn{simply connected\xspace}
\def\holess{holeless\xspace}
\def\bdvall{\bdv_{\ast}}
\def\nconn{\operatorname{c}}
\def\nconnfl{\operatorname{c}_{\mathtt{float}}}
\def\gg{Grassmannian graph\xspace}
\def\ggs{Grassmannian graphs\xspace}
\def\WKmat{\Kop^{\circ}}
\def\BKmat{\Kop^{\bullet}}
\def\ddWKmat{\ddot{\Kop}\vphantom{\Verts}^\circ}
\def\hasMbd{has M-positive boundary\xspace}
\def\Mbd{M-positive boundary\xspace}
\def\datrQ{(\wt,\epsK,\Fw,\Fb,\xd)}
\def\datrQp{(\wt',\epsK',\Fwp,\Fbp,\xd')}
\def\datrQnox{(\wt,\epsK,\Fw,\Fb)}
\def\datrQL{\bm{T}}
\def\datrQLll{\bm{T}_{\la,\lat}}
\def\quintuple{quintuple\xspace}
\def\wtfloat{\mu}

\def\Cut{\xi}
\def\Grem#1{\G\rem #1}
\def\Giww{\G\rem\{\w_1,\w_2\}}
\def\Gbb{\G\rem\{\b_1,\b_2\}}
\def\Hwspace_#1{\Hcal_{#1}^\circ}
\def\Hbspace_#1{\Hcal_{#1}^\bullet}
\def\Hwbspace_#1{\Hcal_{#1}^{\circ\bullet}}
\def\HHspaceV{\Hwbspace_{\Vspace}\HtripK}
\def\HHspaceC{\Hwbspace_{\C}\HtripK}
\def\HHspaceR{\Hwbspace_{\R}\HtripK}
\def\HHspaceRd{\Hwbspace_{\R^2}\HtripK}
\def\HHspaceRknk{\Hwbspace_{k,n-k}\HtripK}
\def\HHspaceRdnk{\Hwbspace_{2,n-k}\HtripK}
\def\HHspaceRdnd{\Hwbspace_{2,2}\HtripK}

\def\Omegas{\bm{\Omega}}
\def\Om{\Omega}
\def\Oms{\Omegas}

\def\ncycop(#1){|\Cycop(#1)|}
\def\Cycop{\operatorname{Cyc}}
\def\ncycOm{\ncycop(\Om)}

\def\ssep{\|}
\def\OmsG{\Oms_{\G}}
\def\Pathbd{\Path^\partial}
\def\bdbd{boundary-to-boundary\xspace}

\def\bisy{\zeta}
\def\bisyt{\tilde\zeta}
\def\CtoM[#1]{(#1)_{+}}
\def\CtoMt[#1]{(#1)_{-}}
\def\GGsh{\Gcal^+}
\def\GLm{\GL_2^-(\R)}
\def\GLp{\GL_2^+(\R)}
\def\GLpp{\GLp\times\GLp}
\def\GLppar{(\GLp\times\GLp)}
\def\Matddr{\Mat_{2,2}(\R)}
\def\Gror{\vec\Gr}
\def\ambsup{\mathtt{flip}}

\def\bapletter{f}
\def\fap{\bapletter}
\def\einj{edge-injective\xspace}
\def\finj{face-injective\xspace}
\def\wimm{weak immersion\xspace}
\def\wtimm{weak t-immersion\xspace}
\def\wtimms{weak t-immersions\xspace}

\def\justpapone{\cite{origami1}\xspace}
\def\papMone#1{\cite[#1]{origami1}\xspace}
\def\Mref#1{\papMone{\cref*{M-#1}}}
\makeatletter
\newcommand{\Eqref}[1]{\textup{\tagform@{\ref*{#1}}}}
\makeatother
\def\Meqref#1{\papMone{Equation~\Eqref{M-#1}}}

\def\SuppGD{|\GD|}
\def\OACTITLE{Origami-amplituhedron correspondence\xspace}
\def\oac{origami-amplituhedron correspondence\xspace}
\def\Mandash{Mandelstam-}
\def\Mdash{M-}
\def\MdashTITLE{\Mandash}
\def\wdash{$\circ$-}
\def\bdash{$\bullet$-}
\def\wclosed{\wdash closed\xspace}
\def\bclosed{\bdash closed\xspace}
\def\Rgs{\bm{R}}

\def\clsub{\mathtt{cl}}

\def\helWming{\mathrlap{\overline{\phantom{k^{\circ}}}}k^{\circ}_{\min}}
\def\helBming{\mathrlap{\overline{\phantom{k^{\bullet}}}}k^{\bullet}_{\min}}
\def\WNEIg{\mathrlap{\overline{\phantom{\Rgs^{\circ}}}}\Rgs^{\circ}_{\clsub}}
\def\BNEIg{\mathrlap{\overline{\phantom{\Rgs^{\bullet}}}}\Rgs^{\bullet}_{\clsub}}

\def\WNEI{\Rgs^{\circ}_{\clsub}}
\def\BNEI{\Rgs^{\bullet}_{\clsub}}

\def\Woverline#1{\overline{#1}}
\def\WBoverline#1{\overline{\overline{#1}}}
\def\BWoverline#1{\overline{\overline{#1}}}

\def\CollGWB{\WBoverline{\G}\CollWBsup}
\def\CollGWBd{\WBoverline{\G}\CollWBsup\vphantom{\G}^\ast}
\def\CollFWB{\WBoverline{\Verts}\CollWBsup\vphantom{\G}^\ast}
\def\CollpartFWB{\partial_{\CollFWB}}

\def\presub{\mathtt{pre}}
\def\CollGWBpre{\WBoverline{\G}\CollWBsup_{\presub}}
\def\CollGWBpred{\WBoverline{\G}\CollWBsup\vphantom{\G}^\ast_{\presub}}

\def\CollGBWgarb{\BWoverline{\G}\vphantom{\G}\CollBWsup} %
\def\CollGBWpregarb{\BWoverline{\G}\vphantom{\G}_{\presub}\CollBWsup} %
\def\CollGWun{\Woverline{\G}\CollWsup_{\text{\scalebox{\ttscl}{\MVbd}}}}
\def\CollGW{\Woverline{\G}\CollWsup}
\def\CollGDW{\Woverline{\G}\vphantom{\G}^\ast}
\def\CollGWfunc{\Woverline{\G}\subfuncsubG}
\def\CollWFaces{\Woverline{\Verts}^\ast}
\def\CollWRtpgauge{\Rtp^{|\CollWFaces|-1}}

\def\MPCollGWfunc{\MPop_{\CollGWfunc}}

\def\CollRW{\overline{\Rg}}
\def\CollRB{\overline{\Rg}}
\def\CollRWB{\BWoverline{\Rg}}

\def\CollRWxW{\CollRW^\circ}
\def\CollRWxB{\CollRW^\bullet}

\def\kw{k^\circ}
\def\kb{k^\bullet}
\def\rem{\setminus}
\def\KSprim{Kenyon--Smirnov primitive\xspace}
\def\KSprims{Kenyon--Smirnov primitives\xspace}
\def\Vspace{\mathbb{U}}
\def\HtripK{(\G,\wtK)}
\def\HtrippK{(\G',\wtK')}

\def\dturn_#1{\tau_{#1}}
\def\dturnb_#1{\tau^\bullet_{#1}}
\def\dturnw_#1{\tau^\circ_{#1}}
\def\dturnbbd_#1{\dturnb_{#1}}
\def\dturnwbd_#1{\dturnw_{#1}}
\def\dtsumBV{\dturn_{\BVint}}
\def\dtsumWV{\dturn_{\WVint}}
\def\dtsumBF{\dturnb_{\Faces}}
\def\dtsumWF{\dturnw_{\Faces}}
\def\subF{\Faces_0}
\def\dtsumBsubF{\dturnb_{\subF}}
\def\dtsumWsubF{\dturnw_{\subF}}
\def\dtsumBFint{\dturnb_{\Fint}}
\def\Kawangle{Kawasaki angle\xspace}
\def\KawAngle{Kawasaki angle\xspace}
\def\kmone{k-\nconn(\G)}
\def\nkmone{n-k-\nconn(\G)}
\def\dwcor{d^{\circ}}
\def\outsub{\mathtt{out}}
\def\epstra{\delta}

\def\FlnL(#1,#2){\vec{\Fl}(#1,#2;n|L)}
\def\twosep{$2$-separated\xspace}
\def\Twosep{$2$-separated\xspace}
\def\twosepon{$2$-separation\xspace}
\def\pdeg{weakly\xspace}
\def\PLimm{PL immersion\xspace}
\def\PLimms{PL immersions\xspace}
\def\PLemb{PL embedding\xspace}
\def\PLembs{PL embeddings\xspace}
\def\aPLimm{a \PLimm}
\def\aPLemb{a \PLemb}
\def\STRimm{straight-edge immersion\xspace}
\def\STRemb{straight-edge embedding\xspace}
\def\collacc#1{\overline{\overline{#1}}}
\def\Collwt{\collacc{\wt}}
\def\CollepsK{\collacc{\epsK}}
\def\CollFw{\collacc{\Fw}}
\def\CollFb{\collacc{\Fb}}
\def\Collxd{\collacc{\xd}}
\def\ColldatrQ{(\Collwt,\CollepsK,\CollFw,\CollFb,\Collxd)}
\def\ColldatrQL{\collacc{\datrQL}}
\def\ColldatrQLpre{\collacc{\datrQL}_{\presub}}

\def\collWacc#1{\overline{#1}}
\def\CollWdatrQL{\collWacc{\datrQL}}
\def\CollWxd{\collWacc{\xd}}
\def\CollWwt{\collWacc{\wt}}

\def\sign{\operatorname{sign}}
\def\Collv{\collacc{\v}}
\def\CollxT{\Tcomp{\Collxd}}
\def\CollVint{\collacc{\Verts}_{\intop}}
\def\Fcolint[#1]{\Fint\ind[#1]}
\def\Fcolbd[#1]{\Fbd\ind[#1]}
\def\Fcol[#1]{\Faces\ind[#1]}
\def\fixsub{\mathtt{fix}}
\def\Ffix{\Faces_{\fixsub}}

\def\Gbary{\G^{\vartriangle}}
\def\Gbarycol{\Gbary\ind[\Collv]}
\def\Gfloat{\vec \G^{\vartriangle}_1}
\def\Gfloatt{\vec \G^{\vartriangle}_2}

\makeatletter
\providecommand{\Rightarrowfill@}{\arrowfill@\Relbar\Relbar\Rightarrow}

\newcommand{\OverRightarrow}[1]{\mathpalette{\overdoublearrow@\Rightarrowfill@}{#1}}
\def\overdoublearrow@#1#2#3{%
 \vbox{\ialign{##\crcr
 #1#2\crcr
 \noalign{\kern0pt\nointerlineskip}%
 $\m@th\hfil#2#3\hfil$\crcr
 }}%
}
\makeatother

\def\Gfl{\mathrlap{\OverRightarrow{\phantom{\G}}}\G^\ast}
\def\Gflbar{\Gfl_{\pm}}
\def\NeighGfl{\OverRightarrow{\Neigh}^{\outsub}_{\G}}
\def\clNFL(#1){\mathrlap{\OverRightarrow{\phantom{S}}}S^\ast_{#1}\ind[\elline_{#1}]}
\def\preKmaxfl{\OverRightarrow{\preKmax}}

\def\Fflsink{\mathrlap{\OverRightarrow{\phantom{\Verts}}}\Faces_{\mathtt{sink}}}
\def\Fflmid{\mathrlap{\OverRightarrow{\phantom{\Verts}}}\Faces_{\mathtt{mid}}}

\def\GflW{\mathrlap{\overrightarrow{\phantom{\G}}}\G^{\;\ast}}
\def\FflWsink{\mathrlap{\overrightarrow{\phantom{\Verts}}}\Faces_{\mathtt{sink}}}
\def\FflWmid{\mathrlap{\overrightarrow{\phantom{\Verts}}}\Faces_{\mathtt{mid}}}

\def\NeighGbary{\Neigh_{\Gbary}}
\def\Nout{\Neigh^{\outsub}_{\Gfloat}}
\def\Noutt{\Neigh^{\outsub}_{\Gfloatt}}

\def\vbary{g}
\def\ubary{h}
\def\Eastfix{\East_{\fixsub}}

\def\wtfloatt{\mu'}

\def\mtpref{momentum-twistor\xspace}
\def\mta{\mtpref amplituhedron\xspace}

\def\mtas{\mtpref amplituhedra\xspace}

\def\varxxx{\var_{123\ast}}
\def\varx{\var_{1\ast}}
\let\oldmu\mu
\def\mu{\oldmu}

\def\mtiling{$\Rel$-tiling\xspace}
\def\Mtiling{$\Rel$-tiling\xspace}
\def\tMtiling{$\tRel$-tiling\xspace}
\def\tmtilings{$\tRel$-tilings\xspace}

\def\atmtiling{an $\tRel$-tiling\xspace}
\def\atmtilingA{an $\tRelA$-tiling\xspace}
\def\atmtilingB{an $\tRelB$-tiling\xspace}
\def\atmtilingC{an $\tRelC$-tiling\xspace}
\def\atddmtiling{\atmtilingC}

\def\tmtilingsB{$\tRelB$-tilings\xspace}
\def\tmtilingsC{$\tRelC$-tilings\xspace}

\def\projsup{\mathtt{proj}}
\def\Mproj{\Mcal^{\projsup}}
\def\Aproj{\Acal^{\projsup}}
\def\AZprojtree{\Aproj_{k-2,n;L=0}(Z)}
\def\MomLLprojG{\Mproj_{\Gfunc}(\La,\Lat)}
\def\AZprojddG{\Aproj_{\ddCollGWvunc}(Z)}

\def\AZprojLo{\Acal^{\projsup,\geq2}_{k-2,n;L}(Z)}

\def\clAZprojLo{\overline{\AZprojLo}}
\def\MomLLLprojo{\Mcal^{\projsup,\geq2}_{k,n;L}(\La,\Lat)}
\def\clMomLLLprojo{\overline{\MomLLLprojo}}

\def\MomLLLamb{\Mcal^{\ambsup}_{k,n;L}(\La,\Lat)}
\def\clMomLLLamb{\overline{\Mcal^{\ambsup}_{k,n;L}(\La,\Lat)}}
\def\AZambL{\Acal^{\ambsup}_{k-2,n;L}(Z)}

\def\clAZambL{\overline{\Acal^{\ambsup}_{k-2,n;L}(Z)}}
\def\PsiZ{\Phi_Z}
\def\PsiZtree{\PsiZ}
\def\PsiZL{\PsiZ^{\vuncsub}}
\def\LaLaimmnn{\LaLatbf_{k,n}^{\mathtt{imm\geq0}}}

\def\LpuncTITLE{$L$-punctured\xspace}
\def\LfuncTITLE{\texorpdfstring{\Lfunc}{$L$-face-punctured\xspace}}
\def\LvuncTITLE{\texorpdfstring{\Lvunc}{$L$-bivertex-punctured\xspace}}
\def\Lpunc{$L$-punctured\xspace}
\def\Lfunc{$\funcsub$-punctured\xspace}
\def\Lvunc{$\vuncsub$-punctured\xspace}

\def\bmpar#1{\bm{(}#1\bm{)}}

\def\puncsub{\bm{[\nL]}}
\def\bmnL{\bm{[\nL]}}
\def\funcsub{\nL^\ast}
\def\vuncsub{\nL^{\bullet}}

\def\subfuncsubG{_{\!\funcsub}}
\def\subvuncsubG{_{\!\vuncsub}}
\def\Gfuncpair{\Gfunc}
\def\ddGvuncpair{\ddGvunc}

\def\Gfunc{\G\subfuncsubG}
\def\ddGvunc{\ddG\subvuncsubG}

\def\sep{\operatorname{sep}}
\def\fullysep{fully \twosep}
\def\Fullysep{Fully \twosep}
\def\fullysepon{full \twosepon}
\def\Fullysepon{Full \twosepon}

\def\xll{\xd_{\la,\lat}}
\def\O{\xO}
\def\y{\bisy}
\def\yt{\bisyt}

\def\ddbdv{\ddot{\bdvletter}^{\partial}}
\def\dbdv{\dot{\bdvletter}^{\partial}}
\def\dRg{\dot\Rg}
\def\dRB{\dot\Rg^\bullet}
\def\dRW{\dot\Rg^\circ}
\def\Fintup{\Verts^{\ast\uparrow}_{\intop}}
\def\dVint{\dot{\Verts}_{\intop}}
\def\dBV{\dVerts^\bullet}
\def\dBVint{\dVerts^\bullet_{\intop}}
\def\dBVintp{\dVerts^{\prime\bullet}_{\intop}}
\def\dFint{\dot{\Verts}^{\ast}_{\intop}}
\def\dface{\dot\ffletter^\ast}
\def\dFaces{\dot{\Verts}^\ast}
\def\dEint{\dot{\Edges}_{\intop}}
\def\ddbde{\ddot\e^\partial}
\def\dGS{\dG_{\Sig}}
\def\RgSig{T}
\def\ddepsKla{\ddot{\epsK}_{\la}}
\def\ddepsK{\ddot\epsK}
\def\EintSig{\Eint^{\Sig}}
\def\ddbdf{\ddot\ffletter^{\partial\ast}}
\def\ddRg{\ddot{\Rg}}
\def\ddRW{\ddot\Rg^\circ}
\def\CycA{\Cyc_{\lightop}}
\def\CycB{\Cyc_{\partial}}
\def\epsKla{\epsK_{\la}}
\def\GrL(#1,#2){\Gror(#1,#2|\nL)}
\def\repsub{\mathtt{Meas}}

\def\GrtnnLfunc(#1,#2){\Grtnn^{\repsub}(#1,#2|\funcsub)}
\def\GrtnnLfuncx_#1(#2,#3){\Gr_{\geq #1}^{\repsub}(#2,#3|\funcsub)}
\def\GrtnnLvuncprojx_#1(#2,#3){\Gr_{\geq #1}^{\repsub}(#2,#3|\vuncsub)}
\def\GrtnnLvuncambx_#1(#2,#3){\Gr_{\geq #1}(#2,#3|\vuncsub)}
\def\Grtnnflamb{\Grtnn(\ddk,n)\!\underset{\raisebox{3pt}{$\scriptstyle\mathtt{flag}$}}{\times}\!\Grtnn(\ddkpd,n)^\nL}

\def\Linelocs{\Lcal_{\bmnL}}
\def\VLpunc{\mat[V|\Linelocs]}
\def\laVLpunc{\left(\la,\VLpunc\right)}
\def\varVL_#1{\var_{1\ast}[V|\Line_{#1}]}
\def\ML_#1{M_{(#1)}}
\def\MLr{\ML_{\rho}}
\def\matlr[#1]{\left(#1\right)}
\def\knk{(n-k)\times k}
\def\Hknk{H}
\def\Hknkloc_#1{H_{(#1)}}
\def\Hknklocs{H_{\bmnL}}
\def\Hknklocsbd{H_{\brnbd}}

\def\Rknk{\Mat_{n-k,k}(\R)}
\def\Rdnk{\Mat_{n-k,2}(\R)}
\def\Rdnd{\Rdd}

\def\Fwk{\Cext}
\def\Fbnk{\Cptext}

\def\Hdnk{H^{\la}}
\def\Hdnd{H^{\la,\lat}}

\def\Measknk{\Meas}
\def\Measdnk{\Meas^\la}
\def\Measdnd{\Meas^{\la,\lat}}
\def\Amat{A}
\def\Atmat{\tilde A}
\def\TC{T_C}
\def\TLbundle{T^{\puncsub}}
\def\Hloc_#1{H_{(#1)}}
\def\Hdnkloc_#1{\Hdnk_{(#1)}}
\def\Hdndloc_#1{\Hdnd_{(#1)}}
\def\Hdnklocs{\Hdnk_{\bmnL}}
\def\Hdndlocs{\Hdnd_{\bmnL}}
\def\bivertex{bivertex\xspace}
\def\bivertices{bivertices\xspace}
\def\bivletter{\b}
\def\biv{\ddot{\bm{\bivletter}}}
\def\bmbiv{\ddot{\bm{\bivletter}}}
\def\bdbiv_#1{\bmbivloc_{\xbd{#1}}}
\def\bivloc_#1{\biv_{(#1)}}
\def\bmbivloc_#1{\bmbiv_{(#1)}}

\def\Gindependent{$\ddGvunc$-independent\xspace}
\def\CollGindependent{$\ddCollGWvunc$-independent\xspace}
\def\CDindependent{$(\ddC;\DbivlocsL)$-independent\xspace}
\def\oneind{fully $1$-\independent}
\def\dind{fully $d$-\independent}

\def\Measbiv{\Meas}
\def\Dbivloc_#1{\ddot D_{(#1)}}
\def\Dbivlocs_#1{\ddot{D}_{\bmpar{#1}}}
\def\DbivlocsL{\ddot D_{\bmnL}}
\def\Dbivlocext_#1{\ddot D^\circ_{(#1)}}
\def\Dbivlocsext_#1{\ddot D^\circ_{\bmpar{#1}}}
\def\Drestloc_#1{\widehat D_{(#1)}}
\def\Drestlocext_#1{\widehat D^\circ_{(#1)}}
\def\ddCprest{\ddot{U}}
\def\ddCprestw{U^\circ}
\def\ddCprestb{\ddot{U}^\bullet}

\def\munk{\ddCprestw}
\def\Uext{\ddCprestw}
\def\ddFb{\ddot{F}^\bullet}
\def\ddFbloc_#1{\ddFb_{(#1)}}
\def\ddbloc_#1{\ddot\b_{(#1)}}
\def\wloc_#1{\w_{(#1)}}
\def\wlocx(#1,#2){\w\xfun(#1,#2)}
\def\ddlaext{\ddot\la^\bullet}
\def\ddla{\ddot\la}
\def\RgSigloc_#1{\RgSig_{(#1)}}
\def\ddRgloc_#1{\ddRg_{(#1)}}
\def\ddRB{\ddot\Rg^\bullet}
\def\ddRBloc_#1{\ddRB_{(#1)}}
\def\yMloc_#1{\xM_{\yloc_{#1}}}
\def\pLr{\Liner^\perp}
\def\Vlocr{\Vloc_\rho}
\def\VLr{\Vlocr}

\def\bivlocsletter{\bmbiv}
\def\bivlocsUNIletter{\ddot{\bm{B}}}
\def\bivlocs{\bivlocsletter_{\bmnL}}
\def\Bivloc_#1{\bivlocsUNIletter_{\bmpar{#1}}}

\def\Vloc_#1{V_{(#1)}}
\def\Vlocs_#1{V_{\bmpar{#1}}}

\def\turn{\wind}
\def\mat[#1]{(#1)}

\def\GGknL{\bm{\G}_{k,n;\funcsub}}
\def\ddGGknL{\ddot{\bm{\G}}_{k,n;\vuncsub}}
\def\CHL{(C;\Hknklocs)}
\def\ddCDL{(\ddC;\DbivlocsL)}
\def\PtpLfunc_#1{\Ptp_{#1}}
\def\PtpLvunc_#1{\Ptp_{#1}}
\def\PtnnLfunc_#1{\Povtnn_{#1}}
\def\MCMp{\Mcal_{\MCEop}}
\def\dMCMp{\widehat{\Mcal}_{\MCEop}}
\def\MCMpdec{\widehat\Mcal_{\MCEop}}
\def\MCMdec{\widehat\Mcal_{\MCEop}}
\def\dResGlinit{\widehat{\Resop}_{\GDinit}}

\def\CollFaces{\overline{\Verts}^\ast}

\def\Sloc_#1{\mathrlap{\overrightarrow{\phantom{S}}}S^\ast_{(#1)}}
\def\Snoloc_#1{\mathrlap{\overrightarrow{\phantom{S}}}S^\ast_{#1}}
\def\Slocs{\mathrlap{\overrightarrow{\phantom{S}}}S^\ast_{\bmnL}}
\def\dloc_#1{\mathrlap{\overrightarrow{\phantom{d}}}d_{(#1)}}
\def\tloc_#1{\mathrlap{\overrightarrow{\phantom{t}}}t_{(#1)}}

\def\dbloc{\mathrlap{\overrightarrow{\phantom{d}}}d_{\bmnL}}

\def\tbloc{\mathrlap{\overrightarrow{\phantom{t}}}t_{\bmnL}}
\def\Tloc_#1{\mathrlap{\overrightarrow{\phantom{T}}}T^\ast_{(#1)}}
\def\Tnoloc_#1{\mathrlap{\overrightarrow{\phantom{T}}}T^\ast_{#1}}

\def\preKmaxloc_#1{\mathrlap{\overrightarrow{\phantom{\preKmax}}}\preKmax_{(#1)}}
\def\preKmaxnoloc_#1{\mathrlap{\overrightarrow{\phantom{\preKmax}}}\preKmax_{#1}}
\def\Kmaxloc_#1{\mathrlap{\overrightarrow{\phantom{\Kmax}}}\Kmax_{(#1)}}
\def\KTmaxloc_#1{\mathrlap{\overrightarrow{\phantom{\KTmax}}}\KTmax_{(#1)}}

\def\CollGWfin{{\Woverline{\G}}\CollWsup\subfuncsubG}
\def\ddCollGWfin{\mathrlap{\ddot{\phantom{\Woverline{\G}}}}\Woverline{\G}\CollWsup\subvuncsubG}

\def\ddCollGWvunc{\ddCollGWfin}
\def\ddCollGWloc_#1{\ddCollGW_{(#1)}}
\def\ddCollGWlocs_#1{\ddCollGW_{\bmpar{#1}}}
\def\ddCollGW{\mathrlap{\ddot{\phantom{\Woverline{\G}}}}\Woverline{\G}\CollWsup}
\def\ddCollwtlaK{\mathrlap{\ddot{\phantom{\Woverline{\Kop}}}}\Woverline{\Kop}\CollWsup_{\laextsub}}
\def\ccc{\theta} %
\def\bccc{\bm{\ccc}}
\def\bcccloc_#1{\bccc_{(#1)}}
\def\bcccnoloc_#1{\bccc_{#1}}
\def\bcccnL{\bccc_{\bmnL}}

\def\ddbccc{\ddot{\bccc}}
\def\ddbcccloc_#1{\ddbccc_{(#1)}}
\def\ddbcccnL{\ddbccc_{\bmnL}}

\def\Lfuncture{$\funcsub$-puncture\xspace}
\def\Lfunctures{$\funcsub$-punctures\xspace}
\def\Lvuncture{$\vuncsub$-puncture\xspace}
\def\Lvunctures{$\vuncsub$-punctures\xspace}
\def\ddSloc_#1{\ddot{S}^{\bullet}_{(#1)}}
\def\ddSlocs{\ddot{S}^{\bullet}_{\bmnL}}
\def\Simplex{\Sigma^{\intsup}}
\def\Simplexr{\Simplex_{(\rho)}}
\def\Simplexloc_#1{\Simplex_{(#1)}}
\def\clSimplexloc_#1{\Sigma_{(#1)}}
\def\Dimplex{\ddot\Sigma^{\intsup}}
\def\Dimplexr{\Dimplex_{(\rho)}}
\def\SimplexbL{\Simplex_{\bmnL}}

\let\oldmathtt\mathtt
\def\ttscl{0.85}
\def\mathtt#1{\text{\scalebox{\ttscl}{$\oldmathtt{#1}$}}}
\def\xfun(#1,#2){_{(#1)}^{\,#2}}
\def\xfunnoloc(#1,#2){_{#1}^{\,#2}}
\def\bivlocx(#1,#2){\biv\xfun(#1,#2)}
\def\ddblocx(#1,#2){\ddot\b\xfun(#1,#2)}
\def\ffx(#1,#2){\ffletter^{#2\ast}_{(#1)}}
\def\cccx(#1,#2){\ccc\xfun(#1,#2)}
\def\cccxnoloc(#1,#2){\ccc\xfunnoloc(#1,#2)}
\def\ddcccx(#1,#2){\ddot{\ccc}\xfun(#1,#2)}
\def\Hx(#1,#2){H^{\la,#2}_{(#1)}}
\def\laextloc_#1{\laext_{(#1)}}
\def\munkloc_#1{\munk_{(#1)}}
\def\mtconf{momentum-twistor configuration\xspace}
\def\mtconfs{momentum-twistor configurations\xspace}
\def\mtV_#1{V(#1)}
\def\mtL_#1{\Lcal(#1)}

\def\bmtV{V}
\def\bmtL{\Lcal}
\def\mtP_#1{\Pi(#1)}
\def\mtmapT{\Tcomp{\phi}}
\def\mtVrm{V_\r(\east_-)}
\def\mtVom{V_0(\east_-)}
\def\mtLr_#1{\Lcal_\r(#1)}
\def\mtLro_#1{\Lcal_0(#1)}
\def\mtLrcrit_#1{\Lcal_{\rcrit}(#1)}
\def\mceTOwte{\phi^{\MCEop}_{\teop}}
\def\mceTOatr{\mceTOwte}
\def\cseps{\eps}

\def\lamin{\bm{\tau}}
\def\Kine{\Kcal}
\def\labase{\la}

\def\maxsup{\mathtt{max}}
\def\minsup{\mathtt{min}}

\def\twondsup{\mathtt{2^\partial}\text{-}\mathtt{nd}}
\def\Grnd{\Gr^{\twondsup}_{\geq0}}

\def\APMnoacr{almost perfect matching\xspace}
\def\APM{APM\xspace}
\def\APMs{APMs\xspace}
\def\hasnofloat{is boundary-connected\xspace}
\def\havenofloat{are boundary-connected\xspace}
\def\wcp{\pdeg convex polygon\xspace}
\def\ndcp{strictly convex polygon\xspace}
\def\wndcp{$2$-dimensional weakly convex polygon\xspace}
\def\PbdxT{\PbdT_{\xd}}
\def\Pbdx{\Pbd_{\xd}}
\def\Pbdxo{\Pbd_{\xd_0}}
\def\MPknLgen{\Mcal_{k,n;\nL}}
\def\AAknLgen{\Acal_{k-2,n;\nL}}
\def\MPknLgentree{\Mcal_{k,n;\nL=0}}
\def\AAknLgentree{\Acal_{k-2,n;\nL=0}}
\def\AAknLgenONE{\Acal_{k-2,n;\nL=1}}
\def\ddsub{{2\times2}}
\def\bzerodd{\bzero_\ddsub}
\def\Wloc_#1{W_{(#1)}}
\def\basesub{\infty}
\def\Kinebase{\Kine_{\basesub}}
\def\Pine{\mtP}
\def\Piner{\Pine_{\rho}}
\def\TOm{T_{\Om}}
\def\plabicEdgeColor{blue!50!black}
\def\plabicEdgeColordG{orange!80!black}
\def\FwColor{purple}
\def\FbColor{green!70!black}
\def\puncColor{red}

\def\trO{\Ocomp{\xs}}
\def\Arg{\operatorname{Arg}}

\def\TURN{\operatorname{turn}}
\def\Xs{C_{\s}}
\def\Xsi{C_{\s,i}}
\def\Xsx_#1{C_{\s,#1}}
\def\bdesi{\bde_{\s,i}}
\def\bdesx_#1{\bde_{\s,#1}}
\def\bdfsi{\bdf_{\s,i}}
\def\ns{n_\s}
\def\sinds{\ind[\RgCycx_\s]}
\def\Xsext{\Xs^\circ}
\def\wtsm{\wt}
\def\epsKsm{\epsK}
\def\Fwsm{\Fw}
\def\Fbsm{\Fb}
\def\wtKsm{\wtK}
\def\Ebb{\mathbb{E}}
\def\Vardd{\operatorname{Var}_{2,2}}

\def\trpi{p}
\def\supA{1}
\def\supB{2}
\def\supC{3}
\def\supS{\s}

\def\RGbf{\mathbf{\G}}
\def\RG{\G}

\def\RelA{\Rel_{\supA}}
\def\RelB{\Rel_{\supB}}
\def\RelC{\Rel_{\supC}}
\def\RelS{\Rel_{\supS}}

\def\Rclacc{\widetilde}

\def\RXcl{\Rclacc{X}}
\def\RYcl{\Rclacc{Y}}
\def\Wcl{\Rclacc{W}}
\def\RPhicl{\Rclacc{\Phi}}
\def\Relcl{\Rclacc{\Rcal}}
\def\RX{X}
\def\RY{Y}
\def\W{W}
\def\Rel{\Rcal}
\def\RPhi{\Phi}
\def\RXint{\RX^{\intsup}}
\def\RYint{\RY^{\intsup}}
\def\Relint{\Rel^{\intsup}}
\def\Relo_#1{\Relint(#1)}
\def\Wint{\W^{\intsup}}

\def\RXA{\RX_{\supA}}
\def\RXB{\RX_{\supB}}
\def\RXC{\RX_{\supC}}
\def\RXS{\RX_{\supS}}

\def\RXAcl{\RXcl_{\supA}}
\def\RXCcl{\RXcl_{\supC}}
\def\RYAcl{\RYcl_{\supA}}
\def\RYCcl{\RYcl_{\supC}}
\def\WAcl{\Wcl_{\supA}}
\def\WCcl{\Wcl_{\supC}}
\def\RelAcl{\Relcl_{\supA}}
\def\RelCcl{\Relcl_{\supC}}

\def\RYO{\RY_{\supA}}
\def\RYA{\RY_{\supA}}
\def\RYB{\RY_{\supB}}
\def\RYC{\RY_{\supC}}
\def\RYS{\RY_{\supS}}
\def\Rproj{\trpi}

\def\RprojO{\Rproj_{\RYO}}
\def\RprojB{\Rproj_{\RYB}}
\def\RprojC{\Rproj_{\RYC}}
\def\RprojS{\Rproj_{\RYS}}

\def\RprojoO_#1{\Rproj_{\RYoO_{#1}}}

\def\Rphibot{\phi}
\def\Rpi{\pi}
\def\RpiLG{\Rpi_{\LGpm}}
\def\RpiGGsh{\Rpi_{\GGsh}}

\def\ReloO_#1{\tReloA_{#1}}
\def\ReloS_#1{\Relint_{\supS}(#1)}
\def\RXo_#1{\RXint(#1)}
\def\RXocl_#1{\RXcl(#1)}
\def\RYo_#1{\RYint(#1)}
\def\RXoO_#1{\tRXoA_{#1}}
\def\RXoA_#1{\RXint_{\supA}(#1)}
\def\RXoAcl_#1{\RXcl_{\supA}(#1)}
\def\RXoB_#1{\RXint_{\supB}(#1)}
\def\RXoC_#1{\RXint_{\supC}(#1)}
\def\RXoS_#1{\RXint_{\supS}(#1)}
\def\RXoCcl_#1{\RXcl_{\supC}(#1)}
\def\RYoO_#1{\RYoA_{#1}}
\def\RYoA_#1{\RYint_{\supA}(#1)}
\def\RYoB_#1{\RYint_{\supB}(#1)}
\def\RYoC_#1{\RYint_{\supC}(#1)}
\def\RYoS_#1{\RYint_{\supS}(#1)}
\def\Wo_#1{\Wint(#1)}
\def\WoA_#1{\Wint_{\supA}(#1)}
\def\WoC_#1{\Wint_{\supC}(#1)}

\def\tRaccBig{\widehat}
\def\tRaccSm{\widehat}
\def\tRelint{\tRaccBig\Rel^{\intsup}}
\def\tReloA_#1{\tRelint_{\supA}(#1)}
\def\tReloB_#1{\tRelint_{\supB}(#1)}
\def\tReloC_#1{\tRelint_{\supC}(#1)}
\def\tRXoA_#1{\tRXint_{\supA}(#1)}
\def\tRXoB_#1{\tRXint_{\supB}(#1)}
\def\tRXoC_#1{\tRXint_{\supC}(#1)}
\def\tRproj{\tRaccSm\trpi}
\def\tRphitop{\tRaccSm\phi}
\def\tRXB{\tRX_{\supB}}

\def\tRX{\tRaccBig X}
\def\tRel{\tRaccBig\Rcal}
\def\tRelo_#1{\tRelint(#1)}
\def\tRXint{\tRX^{\intsup}}
\def\tRXo_#1{\tRXint(#1)}
\def\tMeas_#1{\Meas_{#1}}
\def\trx{\tRaccSm x}
\def\tRelA{\tRel_{\supA}}
\def\tRelB{\tRel_{\supB}}
\def\tRelC{\tRel_{\supC}}

\def\MCX{\RX_{\MCEop}}
\def\MCW{\W_{\MCEop}}
\def\MCPhi{\RPhi_{\MCEop}}
\def\MCXo_#1{\RXint_{\MCEop}(#1)}
\def\MCWo_#1{\Wint_{\MCEop}(#1)}

\def\amtilingA{an $\RelA$-tiling\xspace}
\def\amtilingB{an $\RelB$-tiling\xspace}
\def\amtilingC{an $\RelC$-tiling\xspace}

\def\WA{\W_{\supA}}
\def\WC{\W_{\supC}}

\def\RGraph{\operatorname{G}}
\def\RPhiA{\RPhi_{\supA}}
\def\RPhiC{\RPhi_{\supC}}
\def\RPhiAcl{\RPhicl_{\supA}}
\def\RPhiCcl{\RPhicl_{\supC}}

\def\supX#1{%
 \ifcase#1\relax
 \or \supA %
 \or \supB %
 \or \supC %
 \fi
}

\def\addmtiling{\amtilingC}

\def\twoind{fully $2$-\independent}
\def\independent{independent\xspace}

\def\bdconn{boundary-connected\xspace}

\def\AAiso{\Rphibot}
\def\AAshift{\overline{\Rphibot}}
\def\AAforg{\Rpi}

\def\nbd{n^\partial}
\def\brnbd{\brx{\nbd}}
\def\brFOURbd{\brx{4^\partial}}
\def\xbd#1{#1^\partial}
\def\ibd{\xbd{i}}
\def\jbd{\xbd{j}}
\def\brnLbd{\brnbd\sqcup\brnL}

\def\brnbdsep{\sep(\brnbd)}
\def\brnLbdsep{\sep(\brnLbd)}
\def\seps{\rho}
\def\sept{\gamma}
\def\sepr{\rho}
\def\bivI{{\bm{\rho}}}
\def\sepst{\{\seps,\sept\}}
\def\afflin{h}
\def\elline{\ell}
\def\PolygT{\Tcomp{\Pcurve}}
\def\PolygTrelint{\PolygT_{\mathtt{rel}}^{\intsup}}
\def\chind[#1]{\chi\ind[#1]}
\def\twosepMCE{$2_{\nullsub}$-separated\xspace}
\def\fullysepMCE{fully \twosepMCE}
\def\twoseponMCE{$2_{\nullsub}$-separation\xspace}
\def\fullyseponMCE{full \twoseponMCE}

\def\bt{\bm{t}}
\def\MCMo{\MCM^{\intsup}}
\def\realizable{realizable\xspace}
\def\MCMto(#1){\MCMcellint(#1)}

\def\tls{(\tlim)}
\def\zls{(0)}
\def\xdtls{\xd^{\tls}}
\def\xdouttls{\xdout^{\tls}}
\def\xdoutzls{\xdout^{\zls}}
\def\xToutzls{\xTout^{\zls}}
\def\tlim{\eps}
\def\rcrittls{\rcrit^{\tls}}
\def\rcritzls{\rcrit^{\zls}}
\def\RMOphi{\phi}
\def\RMOpsi{\psi}
\def\mtwphi{\phi}
\def\bmv{\Rg}
\def\distsep{\operatorname{d}_{\mathtt{sep}}}
\def\distrk{\operatorname{d}_{\mathtt{rank}}}
\def\distcurve{\operatorname{d}_{\nullsub}}
\def\distcurvepre{\distcurve^{\mathtt{pre}}}
\def\Flex{Flexible\xspace}
\def\flex{flexible\xspace}
\def\aflex{a \flex}
\def\tdiv{t'}
\def\Pathbdsub_#1{\Pathbd}

\def\ddk{k-2}
\def\ddkpd{k}
\def\ddkpdd{k-2+2d}
\def\ddkpdI{\klocs_{\bivI}}

\def\deltaw{d^\circ}

\def\tripsup{\mathtt{tripod}}
\def\suptripsup{^{\tripsup}}

\def\epsww_#1{\epsK_{#1}\suptripsup}
\def\epsbb_#1{\epsK_{#1}\suptripsup}
\def\ddepsbb_#1{\ddepsK_{#1}\suptripsup}

\def\ddkvee{\ddot{k}^\vee}
\def\skipddC{\\[3pt]}
\def\brlasepr{\<\sepr\>_{\la}}
\def\brlaseps{\<\seps\>_{\la}}
\def\brlasept{\<\sept\>_{\la}}
\def\Chordsout{\Chords^{\outsub}}
\def\Rgloc_#1{\Rg_{(#1)}}
\def\wrlet{c_{(\rho)}}
\def\wrlai{\wrlet^{\;i}}
\def\wrlaj{\wrlet^{\;j}}
\def\wrlaji{\wrlet^{\;j,i}}
\def\popsub{\mathtt{pop}}
\def\ddGaux{\ddG_{\popsub}}
\def\ddEaux{\ddE_{\popsub}}
\def\ddVintaux{\ddVerts^{\intop}_{\popsub}}
\def\wauxloc_#1{\ddot{\wv}^{\popsub}_{(#1)}}
\def\ddGlocs_#1{\ddG_{\bmpar{#1}}}
\def\ddElocs_#1{\ddE_{\bmpar{#1}}}
\def\ddGloc_#1{\ddG_{(#1)}}
\def\ddEloc_#1{\ddE_{(#1)}}
\def\mloc_#1{m_{(#1)}}
\def\klocs_#1{\ddot k_{\bmpar{#1}}}
\def\blocout_#1{\ddot{S}^{\popsub}_{(#1)}}
\def\ddwtaux{\ddwt_{\popsub}}
\def\ddwtlocs_#1{\ddwt_{\bmpar{#1}}}
\def\ddwtK{\ddKop}
\def\ddwtloc_#1{\ddwt_{(#1)}}
\def\ddwtKloc_#1{\ddKop_{(#1)}}
\def\Rtpgaux{\Rtp^{|\ddEaux|-|\ddVintaux|}}
\def\ddepsKloc_#1{\ddepsK_{(#1)}}
\def\ddepsKlocs_#1{\ddepsK_{\bmpar{#1}}}
\def\ddwtKlocs_#1{\ddwtK_{\bmpar{#1}}}
\def\ddepsKaux{\ddepsK^{\popsub}}
\def\ddWKmatloc_#1{\ddWKmat_{(#1)}}
\def\ddWKmatlocs_#1{\ddWKmat_{\bmpar{#1}}}
\def\Dveeloc_#1{\ddot D^\vee_{(#1)}}
\def\ddWVloc_#1{\ddot{\Verts}^{\circ}_{(#1)}}
\def\ddWVlocs_#1{\ddot{\Verts}^{\circ}_{\bmpar{#1}}}
\def\psitrip{\operatorname{T}}
\def\psitriploc_#1{\operatorname{T}_{(#1)}}
\def\ddeloc_#1{\dde_{(#1)}}
\def\ddlaveeloc_#1{\ddla^\vee_{(#1)}}
\def\rpp{q}
\def\Omij{\Om_{i,j}}
\def\Omxx_#1{\Om_{#1}}
\def\easyred{easily reducible\xspace}
\def\CycRgloc_#1{\Cycloc_{#1}}
\def\nloc_#1{n_{(#1)}}

\makeatletter
\def\part{\@startsection{part}{0}%
 \z@{\linespacing\@plus\linespacing}{.5\linespacing}%
 {\normalfont\Large\bfseries\centering}}
\makeatother

\def\nfloat{\operatorname{float}}
\def\dlog{\operatorname{dlog}}
\def\facewtletter{Y}
\def\facewt(#1){\facewtletter_{#1}}
\def\facewtall{\bm{\facewtletter}}
\def\facewtallg{\facewtall_{\f}}
\def\ellastij{\ell^\ast_{i,j}}
\def\Cpast{(C^\perp)^\ast}
\def\eh{h}
\def\wtoft{\wt_t}
\def\wtKoft{\wtK_t}
\def\Coft{C(t)}
\def\Cofh{C(1+\hbar)}
\def\Cextoft{\Cext_t}
\def\HtripKoft{(\G,\wtKoft)}
\def\subev{\mathtt{ev}}
\def\subodd{\mathtt{odd}}
\def\Sev{S_{\subev}}
\def\Sodd{S_{\subodd}}
\def\OmsGST{\OmsG(S,T)}
\def\apmev{\apm_{\subev}}
\def\apmodd{\apm_{\subodd}}
\def\chicut{\chi_{\Cut}}
\def\EST{\Ebb_{S,T}}
\def\ZSTcut{Z^{\Cut}_{S,T}}

\def\ZST{Z_{S,T}}
\def\inv{\operatorname{inv}}
\def\ehev{\eh_{\subev}}
\def\ehodd{\eh_{\subodd}}
\def\ehdbl{\eh_{\mathtt{dbl}}}
\begin{document} 
 \setlength{\leftmargini}{20pt}
\numberwithin{equation}{section}

\thinmuskip=3mu 

\title{Amplituhedra and origami, II: loop level}
\author{Pavel Galashin}
\address{Department of Mathematics, Cornell University, Ithaca, NY 14850, USA}
\email{{\href{mailto:galashin@cornell.edu}{galashin@cornell.edu}}}
\address{Department of Mathematics, University of California, Los Angeles, CA 90095, USA}
\email{{\href{mailto:galashin@math.ucla.edu}{galashin@math.ucla.edu}}}
\thanks{P.G.\ was supported by the National Science Foundation under Grants No.~DMS-2046915 and No.~DMS-2554159.}
\date{\today}



\begin{abstract}
Building on the recently discovered \oac, we prove 
that the BCFW (Britto--Cachazo--Feng--Witten) cells triangulate the $m=4$ amplituhedron
 in full generality at all loop orders, both in momentum and momentum-twistor space.
Along the way, we develop two natural ``$L$-punctured'' extensions of the positive Grassmannian and relate them via T-duality.
\end{abstract}

\maketitle

\setcounter{tocdepth}{1}
\hypersetup{bookmarksdepth=subsection}
\tableofcontents 

\newpage
\section*{Introduction}\label{sec:intro}

The study of scattering amplitudes in planar $\mathcal{N}=4$ supersymmetric Yang--Mills (SYM) theory has revealed a deep connection between quantum field theory, algebraic geometry, and combinatorics. Central to this development is the \emph{amplituhedron} introduced by Arkani-Hamed and Trnka~\cite{AHT}. A major open problem in this area, known as the \emph{BCFW tiling conjecture}, states that 
 the Britto--Cachazo--Feng--Witten (BCFW) recurrence relations~\cite{BCFW, AHBC_all_loop} manifest geometrically as tilings of the amplituhedron by pairwise non-overlapping subsets.

In the first paper in this series~\justpapone, we introduced the \emph{\oac}, establishing a direct link between the ($m=4$) tree momentum amplituhedron and the space of origami crease patterns~\cite{KLRR,CLR1} planar dual to a fixed bipartite graph embedded in a disk. This correspondence allowed us to give a rigorous proof of the BCFW tiling conjecture at tree level. The approach of~\justpapone yields the result simultaneously for the momentum amplituhedron~\cite{DFLP} and the \mta~\cite{AHT}, linking their tilings through the operation of \emph{T-duality}~\cite{Zono,LPW,PSBW} and
generalizing the BCFW tiling results of~\cite{ELT,ELPTSBW}.
 In this paper, we extend this framework to prove the BCFW tiling conjecture at all loop orders. 

Let $\MPknLgen$ (resp., $\AAknLgen$) denote the $m=4$ amplituhedron in momentum\footnote{Similarly to~\justpapone, we restrict to the class of \emph{Mandelstam-nonnegative} momentum amplituhedra.} (resp., momentum-twistor) space for $n$ particles, helicity $k$, and loop order $L \geq 0$. The BCFW recursion provides a collection of \emph{tiles}, defined as images of positroid cells in the \emph{\Lpunc positive Grassmannian} under specific rational maps. The BCFW tiling conjecture states that these tiles have mutually disjoint interiors and that their closures cover the entire space $\MPknLgen$ (resp., $\AAknLgen$).
This conjecture has been confirmed\footnote{More precisely, making different choices in the BCFW recursion gives rise to many collections of BCFW tiles. The results of~\cite{ELT,Tessler_notes} apply to a single collection of BCFW tiles, while the results of~\cite{ELPTSBW,origami1}---as well as \cref{thm:main_triangulation}---apply to all collections simultaneously.} in~\cite{ELT,ELPTSBW} for $\AAknLgentree$, 
in~\cite{Tessler_notes} for $\AAknLgenONE$,
and independently in~\justpapone for both $\MPknLgentree$ and $\AAknLgentree$.
Our main result resolves the BCFW tiling conjecture in full generality.

\renewcommand{\thetheoremintro}{A}
\begin{theoremintro}\label{thm:main_triangulation}
For all $2\leq k\leq n-2$ and $\nL\geq0$, the BCFW tiles form a tiling of the $m=4$ loop amplituhedra $\MPknLgen$ and $\AAknLgen$.
\end{theoremintro}
\noindent See \cref{lemma:BCLOOP:MPknL_tiling,lemma:BCLOOP:AAknL_tiling,lemma:proj_tiling} for precise statements.

The leap from tree level ($L=0$) to loop level ($L \ge 1$) requires substantial new geometric and combinatorial machinery. We summarize below some of the new ideas developed in the present paper. 

\subsection*{Defining amplituhedra and BCFW tiles}
Establishing rigorous definitions for the spaces in \cref{thm:main_triangulation} has historically been quite challenging. 
The \emph{linear projection} (resp., \emph{sign flip}) definition of the loop \mta $\AAknLgen$ was given in~\cite{AHT} (resp.,~\cite{AHTT}). 
We confirm the equivalence of these two definitions in \crefi{lemma:proj_tiling}{proj_tiling3}.
For the loop momentum amplituhedron $\MPknLgen$, the linear projection (resp., sign flip) definition was given in~\cite{FL} (resp., \cite{FGLS}).\footnote{We thank L.~Ferro, T.~\L{}ukowski, and J.~Stalknecht for bringing the results of~\cite{FGLS} to our attention.} Our definition of $\MPknLgen$ (\cref{ssec:LOOP:ambient_ampl_dfn,ssec:flip_proj}) is new and is directly motivated by the geometry of origami crease patterns. 
We compare our definition to that of~\cite{FGLS} in \cref{rmk:FL_FGLS_compare}. 

To the best of our knowledge, the BCFW tiles inside either $\AAknLgen$ or $\MPknLgen$ 
have not yet been defined at loop level.
 The BCFW recursion for planar bipartite graphs was introduced at loop level in~\cite[Section~4.2]{AHBC_all_loop} (see also~\cite[Section~2.6]{abcgpt}). 
This recursion gives rise to a collection $\BCGs$ of planar bipartite graphs for each $k,n,\nL$. 
However, associating a specific subset of $\MPknLgen$ or $\AAknLgen$ to each of these graphs has been an open problem.

\begin{figure}
\def\scl{1.15}
\begin{tabular}{ccc}
 \includegraphics[scale=\scl]{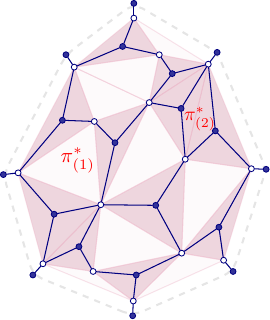}
&
 \includegraphics[scale=\scl]{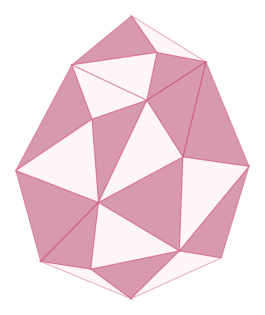}
&
 \includegraphics[scale=\scl]{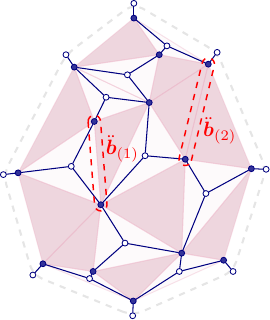}
\\[5pt]
(a) $\Gfunc$ & (b) $\Sig$ & (c) $\ddGvunc$ 
\end{tabular}
 \caption{\label{fig:intro-G-ddG} 
T-duality for \Lpunc planar bipartite graphs; see \crefrange{sec:shift}{sec:LPUNC}.}
\end{figure}

To give a proper definition of loop BCFW tiles, we introduce the \emph{\Lpunc positive Grassmannian} and study the associated boundary measurement map in \cref{sec:LPUNC}, generalizing the results of~\cite{Pos} from the $\nL=0$ case. We consider two incarnations corresponding to momentum and momentum-twistor amplituhedra, respectively: 
{
 \setlength{\leftmargini}{20pt}
\begin{itemize}
\item the \emph{\Lfunc positive Grassmannian} $\GrtnnLfunc(k,n)$ consists of boundary measurements of weighted planar bipartite graphs $(\Gfunc,\wt)$ with $\nL$ marked faces $\ploc_1,\dots,\ploc_\nL$;
\item the \emph{\Lvunc positive Grassmannian} ${\GrtnnLvuncprojx_0(k-2,n)}$ consists of boundary measurements of weighted planar bipartite graphs $(\ddGvunc,\ddwt)$ with $\nL$ marked black \emph{\bivertices} $\bivloc_1,\dots,\bivloc_\nL$, i.e., pairs $\bivloc_\rho=\{\ddbloc_\rho^1,\ddbloc_\rho^2\}$ of black vertices such that $\ddbloc_\rho^1,\ddbloc_\rho^2$ share a face of $\ddG$ for each $1\leq\rho\leq\nL$.
\end{itemize}
}
\noindent See \figref{fig:intro-G-ddG}(a,c).
 We note that the black \bivertices in $\ddGvunc$ are required to be pairwise \emph{non-crossing}; see \cref{dfn:Lvunc_ordinary} for further details.

Recall that the ($\nL=0$) boundary measurement map of~\cite{Pos} associates to each weighted planar bipartite graph $(\ddG,\ddwt)$ a $(k-2)$-plane $\ddC=\Meas(\ddG,\ddwt)$ in the \emph{totally nonnegative Grassmannian} $\Grtnn(k-2,n)$.
 We show in \cref{lemma:DIM:remove_bivertex} that removing a black \bivertex $\bivloc_\rho$ from $\ddG$ and then applying $\Meas$ results in a $k$-plane $\Dbivloc_\rho\in\Grtnn(k,n)$ containing $\ddC$.\footnote{More generally, we show in \cref{lemma:DIM:remove_add} that \emph{popping} a black vertex $\ddbloc_\rho$ of $\ddG$ (i.e., replacing $\ddbloc_\rho$ with a white vertex adjacent to some of the black vertices that share faces with $\ddbloc_\rho$; see \cref{fig:popping}) also gives rise to a $k$-plane $\Dbivloc_\rho\in\Grtnn(k,n)$ containing $\ddC$, and we allow such \emph{generalized \Lvunc graphs} in our definition of $\GrtnnLvuncprojx_0(k-2,n)$.}
Thus, $\GrtnnLvuncprojx_0(k-2,n)$ is contained in the space $\GrtnnLvuncambx_0(k-2,n)$ of tuples $(\ddC;\Dbivloc_1,\dots,\Dbivloc_\nL)\in\Grtnn(k-2,n)\times\Grtnn(k,n)^\nL$ satisfying $\ddC\subset\Dbivloc_\rho$ for each $1\leq \rho\leq \nL$ together with some further positivity conditions (\cref{dfn:LPUNC:Lvunc_Grtnn}). The space $\GrtnnLvuncambx_0(k-2,n)$ is the $\nL$-loop positive Grassmannian originally introduced in~\cite{AHT}. 
Related constructions representing points in $\GrtnnLvuncambx_0(k-2,n)$ using variants of plabic graphs appear in~\cite{BaiHe,BHL}.
Our interpretation of this space in terms of (generalized) \Lvunc planar bipartite graphs is new. 

On the other hand, the space $\GrtnnLfunc(k,n)$ itself is new. It is a subset of the $\nL$-fold tangent bundle $\TLbundle\Gr(k,n)$ of the Grassmannian, and the associated \emph{\Lfunc boundary measurement map} is intimately tied to the \emph{double-dimer model} on the underlying planar bipartite graph $\G$. 

Each BCFW graph $\Gfunc\in\BCGs$ is naturally \Lfunc. 
 The corresponding loop BCFW tile in $\MPknLgen$ is the image of
the \emph{\Lfunc positroid cell} $\PtpLfunc_{\Gfunc}\subset\GrtnnLfunc(k,n)$
 under a ``linear projection'' $\PhiLLL$ similar to the one studied in~\cite{DFLP}.

We apply the \emph{T-duality} operation (see below) to each $\Gfunc\in\BCGs$, obtaining an \Lvunc planar bipartite graph $\ddGvunc$. Applying a positive linear map $Z\in\Grtp(k+2,n)$ to the associated \Lvunc positroid cell $\PtpLvunc_{\ddGvunc}\subset\GrtnnLvuncprojx_0(k-2,n)$, we obtain a loop BCFW tile inside $\AAknLgen$.

\subsection*{T-duality}
A major component of our proof is the \emph{T-duality} map that relates momentum amplituhedra to momentum-twistor amplituhedra. As a combinatorial operation on planar bipartite graphs, it was first introduced in~\cite{Zono} (for unweighted reduced graphs for the top positroid cell in $\Grtnn(k,n)$). It was later generalized in~\cite{BaWe,HSP,crit,LPW,PSBW,CLSBW} to unweighted reduced graphs corresponding to arbitrary positroid cells in $\Grtnn(k,n)$. It was related to the \emph{magic projector} $\Qla$ of~\cite{AHCC} in~\cite{abcgpt} and applied to the $m=2$ amplituhedron in~\cite{LPW,PSBW}. 

In~\Mref{sec:TREE}, we showed that the magic projector $\Qla$ gives rise to an explicit T-duality map between (``ambient'' versions of) tree amplituhedra $\MPknLgentree$ and $\AAknLgentree$ that relates their BCFW tilings. This allowed us to only prove the BCFW tiling result for $\MPknLgentree$ and deduce it for $\AAknLgentree$ as a byproduct. In this paper, we employ a similar strategy. 

The following aspects of T-duality that we develop in \crefrange{sec:shift}{sec:LOOP} are new:
\begin{itemize}
\item extending T-duality to \emph{not necessarily reduced} planar bipartite graphs;
\item lifting it from the combinatorial level to the geometric level, i.e., to \emph{weighted} graphs;\footnote{A simplified version of this construction appeared in the first version of \justpapone. That construction has been moved to \cref{sec:shift} of the present manuscript, extended to the class of not necessarily reduced weighted graphs.}
\item using it to relate \Lfunc and \Lvunc graphs and their boundary measurements;
\item generalizing the magic projector map $\Qla$ between ambient amplituhedra $\MPknLgen$ and $\AAknLgen$ from tree level~\justpapone to loop level.
\end{itemize}
\noindent For example, the \Lfunc graph $\Gfunc$ in \figref{fig:intro-G-ddG}(a) is T-dual to the \Lvunc graph $\ddGvunc$ in \figref{fig:intro-G-ddG}(c), where $\nL=2$. Here, we choose the \bivertex $\bivloc_1$ to 
contain any two vertices of the triangle of $\Sig$ containing $\ploc_1$.
Such choices do not affect the resulting boundary measurements; see \cref{rmk:LPUNC:choices}.
In \cref{fig:BCFW-T-dual} and \cref{ex:BCFW-T-dual}, we compute the T-dual of the \Lfunc BCFW graph from \figref{fig:BCFW-full}(e).

\begin{figure}
 \def\inputscale{1.5}
\begin{tabular}{cc}
\includegraphics[scale=\inputscale]{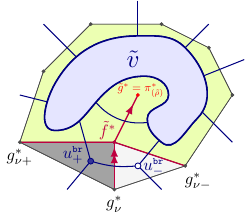}
&
\includegraphics[scale=\inputscale]{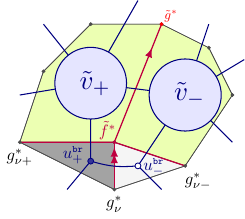}
\end{tabular}
 \caption{\label{fig:BCFW-vertices} Possible generic \ORsts are precisely the planar duals of the loop BCFW recursion steps. 
An example of a non-generic \ORst is shown in \cref{fig:fout-cliques}.
}
\end{figure}

\subsection*{\OACTITLE}
An \emph{origami crease pattern} (also called a \emph{circle pattern} or a \emph{t-embedding} in~\cite{KLRR,CLR1}) is an embedding of the planar dual $\GD$ of a planar bipartite graph $\G$ satisfying the \emph{\Kawangle condition}~\cite{Kawasaki} at each interior vertex $\ff\in\Fint$ of $\GD$: 
 the sum of angles of all black (resp., white) faces of $\GD$ around $\ff$ must be equal to $\pi$. 
In~\justpapone, our proof of the tree BCFW tiling conjecture relied on the \emph{\oac} between origami crease patterns and points in $\MPknLgentree$. 
We established this correspondence for graphs 
 satisfying a certain \emph{surplus condition} $\helmin(\G)\geq2$ (cf. \cref{ssec:BACKGR:surplus}). This includes the reduced graphs of~\cite{Pos}. 

\begin{figure}
 \def\inputscale{1.05}
 \def\inputscalE{1.6}
 \setlength{\tabcolsep}{1pt}
\begin{tabular}{cc|cc}
 \includegraphics[scale=\inputscale]{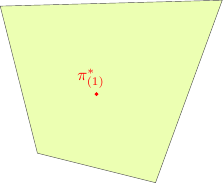}
&
 \includegraphics[scale=\inputscale]{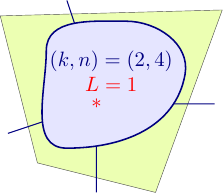}
&
 \includegraphics[scale=\inputscale]{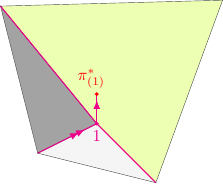}
&
 \includegraphics[scale=\inputscale]{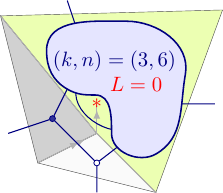}
\\[-15pt]
\multicolumn{2}{c|}{(a) $t=0$} & \multicolumn{2}{c}{(b) $t=1$}
\\\hline
 \includegraphics[scale=\inputscale]{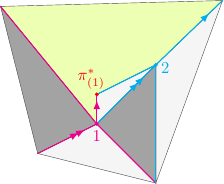}
&
 \includegraphics[scale=\inputscale]{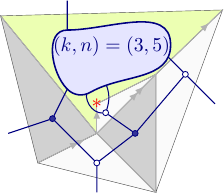}
&
 \includegraphics[scale=\inputscale]{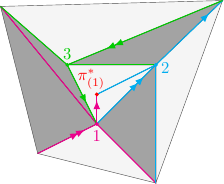}
&
 \includegraphics[scale=\inputscale]{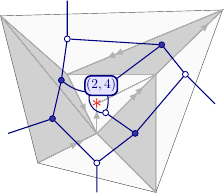}
\\[-15pt]
\multicolumn{2}{c|}{(c) $t=2$} & \multicolumn{2}{c}{(d) $t=3$}
\\[2pt]\hline
\multicolumn{2}{r}{
 \includegraphics[scale=\inputscalE]{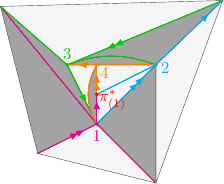}
}
&
\multicolumn{2}{l}{
 \includegraphics[scale=\inputscalE]{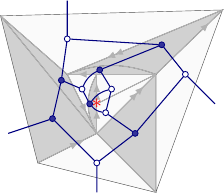}
}
\\[-15pt]
\multicolumn{4}{c}{(e) $t=4$}
\end{tabular}
 \caption{\label{fig:BCFW-full} Applying the \ORA\ /\ loop BCFW recursion.}
\end{figure}

 For the purposes of the loop BCFW tiling conjecture, this is not enough: the (\Lfunc) graphs $\Gloop\in\BCGs$ appearing in the loop BCFW recursion---including the ones with \emph{\kinsupp} (\cref{dfn:BCFW:kinsupp})---only satisfy $\helmin(\G)\geq1$, and the associated origami crease patterns contain degenerate triangular faces; see \figref{fig:BCFW-full}(e) for an example. In \cref{sec:TE}, we generalize the \oac of~\justpapone to a bijection between such \emph{weakly embedded} origami crease patterns for graphs $\G$ satisfying $\helmin(\G)\geq1$ and points in the tree momentum amplituhedron. Having to deal with weak embeddings is a major technical complication in the present paper compared to~\justpapone.

\subsection*{\ORATITLE}
Our proof of the BCFW tiling conjecture for $\MPknLgen$ relies on the \emph{\ORA} developed in \cref{part2}. 
A basic step of the algorithm takes a vertex $\corf$ incident to some face $\corv$ of $\GD$ as an input (where $\cor$ is a \emph{corner} of $\GD$). It creates a new vertex $\ffout$ inside $\corv$, a pair of triangles of opposite color incident to $\ffout$ and $\corf$, and one or several \emph{outgoing edges}, connecting $\ffout$ to other vertices of $\GD$. 
In our figures, the edge from $\corf$ to $\ffout$ is marked with a double arrow and the outgoing edges are marked with single arrows emanating from $\ffout$. An \ORst is called \emph{generic} if it creates exactly one outgoing edge; see \crefrange{fig:BCFW-vertices}{fig:BCFW-full} and \cref{dfn:BCin-generic}. 

We show that any sequence of \ORsts results in a weakly embedded origami crease pattern satisfying the \Kawangle condition at every interior vertex. 
At tree level, the proof of correctness of this algorithm is only a few pages long; see~\Mref{sec:BCFW}. At loop level, while the general idea of the algorithm is roughly the same, the proof of correctness is much more involved. 

Our key tool is the notion of a 
\emph{\MCEintro}
 introduced in \cref{sec:MCE}. 
These objects interpolate between points of $\MPknLgen$ on the one hand and origami crease patterns on the other hand. 
They include \emph{T-graphs} of~\cite{Kenyon_Sheffield,Kenyon_height} and \emph{pointed \ptrons} of~\cite{Streinu,Streinu_ptr,RSS_expansive,RSS} as special cases; see \cref{ssec:pointed}. 

Roughly speaking, a \emph{\MCEintro} of a planar graph $\GD$ (whose planar dual is not necessarily bipartite) is a pair $\xd=(\xT,\xO):\Faces\to\Rdd$ of maps from the vertex set $\Faces$ of $\GD$ to the plane satisfying the following conditions; see \cref{dfn:MCE:MCE} for further details. 
\begin{itemize}
\item \emph{Weak embedding}: $\xT$ is a straight-edge (weak) embedding of $\GD$. 
\item \emph{Mandelstam-nonnegative}: for all $\ff,\f\in\Faces$, we have $|\xT(\ff)-\xT(\f)|\geq|\xO(\ff)-\xO(\f)|$, with equality for all edges $\{\ff,\f\}\in\East$ of $\GD$.
\item \emph{No \pchords}: for all vertices $\ff,\f\in\Faces$ not connected by an edge but satisfying $|\xT(\ff)-\xT(\f)|=|\xO(\ff)-\xO(\f)|$, either the line segment $[\xT(\ff),\xT(\f)]$ intersects some other edge of $\xT(\GD)$, or it violates the \emph{generalized \Kawangle condition} (discussed below) at either $\ff$ or $\f$.
\end{itemize}
\noindent In~\eqref{eq:nearGr:angles_exist}, 
 we define the black and white angle sums $\sumbT(\ff),\sumwT(\ff)$ for arbitrary maps $\xd:\Faces\to\Rdd$ satisfying $|\xT(\ff)-\xT(\f)|=|\xO(\ff)-\xO(\f)|$ for all edges $\{\ff,\f\}\in\East$. When $\xd$ is a Mandelstam-nonnegative \wemb, we automatically have $\sumbT(\ff),\sumwT(\ff)\in\{0,\pi,2\pi\}$ with $\sumbT(\ff)+\sumwT(\ff)=2\pi$, and we impose the generalized \Kawangle condition $\sumbT(\ff)=\sumwT(\ff)=\pi$ for each $\ff\in\Fint$; see~\eqref{eq:MCE:sumwT_sumbT=pi}.

We show that \ORsts preserve the class of \MCEsintro. 
For example, \MCEsintro appear at each step in \cref{fig:BCFW-full}, where we use the following convention. 
 We only depict $\xT(\GD)$, and for a face $\v$ of $\GD$, if the faces $\xT(\v)$ and $\xO(\v)$ are isometric then $\xT(\v)$ is colored either white or black depending on whether the isometry between $\xT(\v)$ and $\xO(\v)$ is orientation-preserving or orientation-reversing. Such faces are called \emph{rigid}. All other faces of $\xT(\GD)$ are called \emph{\flex} and colored green. 
For example, in \figref{fig:BCFW-full}(d), the face with vertices $\ffout_1,\ffout_2,\ffout_3,\ploc_1$ (where for $i=1,2,3,4$, $\ffout_i$ is labeled by $i$ in the figure) is rigid white. 
In particular, $|\xT(\ffout_3)-\xT(\ploc_1)|=|\xO(\ffout_3)-\xO(\ploc_1)|$, even though the vertices $\ffout_3$ and $\ploc_1$ are not connected by an edge. In this case, the edge connecting them would violate the generalized \Kawangle condition at $\ploc_1$; see \cref{ssec:PMNEs}.

The input \MCEintro (\figref{fig:BCFW-full}(a)) of the \ora at time $t=0$
 is an $n$-gon in $\Rdd$ with null sides and $\nL$ isolated vertices inside (with respect to the $\xd\mapsto\xT$ projection), satisfying certain \Mandash positivity and winding number conditions. The moduli space of such \MCEsintro at $t=0$ is the loop momentum amplituhedron $\MPknLgen$. On the other hand, the output \MCEintro (\figref{fig:BCFW-full}(e)) is a weakly embedded origami crease pattern with $\nL$ marked vertices. Its planar dual is among the \Lfunc planar bipartite graphs in $\BCGs$, and the associated \MCMSintro projects 
(under the forgetful map that only remembers the boundary null polygon and the $\nL$ marked points in $\Rdd$)
to a BCFW tile inside $\MPknLgen$.
 The loop BCFW recursion (or, dually, the \ORA) gives a way to reconstruct an origami crease pattern from its image under this forgetful map. 
 In the notation of \cref{fig:BCFW-vertices}, a step of the \ORA consists of determining the location of the point $\xd(\ffout)$ from the geometry of the 
face of $\xT(\GD)$ containing it. 
\MCEsintro are the objects appearing during the intermediate steps of this algorithm.

\subsection*{Outline} 
The general theory of loop amplituhedra and T-duality is developed in \cref{part1}. \Cref{part2} is devoted to the \ORA and its proof of correctness. \Cref{part3} combines the previous results to complete the proof of \cref{thm:main_triangulation}. 
While \cref{part1} may be viewed as an extension of~\justpapone, \cref{part2} is logically independent from both \cref{part1} and~\justpapone.

\subsection*{Acknowledgments}
I am indebted to Nima Arkani-Hamed for suggesting that the results of~\justpapone could potentially be extended to loop amplituhedra. 
I thank Thomas Lam and Lauren Williams for their feedback on a preliminary version of this manuscript. 
I also thank Sasha Goncharov, Rick Kenyon, Tsviqa Lakrec, Matteo Parisi, and Jara Trnka for 
their valuable comments on some of the results presented here. 

\vspace{0.05in}
{\small AI was used for proofreading this document and for creating the interactive applets~\cite{applets}.\par}

\section{Preliminaries}

\subsection{Minkowski space, decorated null polygons, and positive kinematic space}
We review some background on the Minkowski space $\Rdd$.
For $\xs=(\xsT,\xsO)\in\Rdd\cong\C^2$, we set $\xs^2 = |\xsT|^2 - |\xsO|^2$. 

We identify the space $\R^{2,2}$ with the space $\Matddr$ of $2\times 2$ matrices, as follows. 
Given $\xs=(\xsT,\xsO)\in\R^{2,2}$, we define a matrix $\xM_{\xs}$ by 
\begin{equation}\label{eq:x_vs_ap_vs_am}
\xM_{\xs}:= \begin{pmatrix}
\Re(\ap) & \Im(\am) \\ -\Im(\ap) & \Re(\am)
\end{pmatrix},\quad\text{where}\quad
\ap:=\frac12(\xsT + \xsO) \quad\text{and}\quad \am:=\frac12(\xsT-\xsO);\quad\text{thus,}
\end{equation}
\begin{equation}\label{eq:det_xM_vs_Pmom^2}
 \det \xM_{\xs} = \Re(\ap)\Re(\am) + \Im(\ap)\Im(\am) = \frac14 \left(|\xsT|^2 - |\xsO|^2\right) = \frac14 \xs^2.
\end{equation}

\begin{definition}\label{dfn:BACKGR:bispinor}
Suppose that $\Pmom\in\Rdd$ is a nonzero \emph{null vector}, i.e., $\Pmom\neq0$ and $\Pmom^2=0$. We say that a pair $(\bisy,\bisyt)\in(\Cast)^2$ is a \emph{bispinor representation} of $\Pmom$ if $\PmomT = \bisy\bisyt$ and $\PmomO = \ovl{\bisy}\bisyt$. 
\end{definition}
\noindent The pair $(\bisy,\bisyt)$ is determined by $\Pmom$ up to the \emph{little group action} $(\bisy,\bisyt)\mapsto(t\bisy,t^{-1}\bisyt)$ for $t\in\Rast$. By~\eqref{eq:x_vs_ap_vs_am},
\begin{equation}\label{eq:TE:bisy_bisyt_to_M}
\xM_\Pmom = \CtoMt[\bisyt] \cdot \CtoM[\bisy]^T,
\quad\text{where}\quad
 \CtoM[\bisy]:=\begin{pmatrix}
 \Re\bisy\\\Im\bisy
 \end{pmatrix}
 \quad\text{and}\quad
 \CtoMt[\bisyt]:=\begin{pmatrix}
 \Re\bisyt\\-\Im\bisyt
 \end{pmatrix}.
\end{equation}
The Minkowski scalar product of $\Pmom,\Qmom\in\Rdd$ is given by $\Pmom\cdot \Qmom:=\Re(\PmomT\ovl{\QmomT} - \PmomO\ovl{\QmomO})$, and if $\Pmom,\Qmom$ are nonzero null vectors with bispinor representations $(\Pmomy,\Pmomyt)$, $(\Qmomy,\Qmomyt)$ then their scalar product satisfies
\begin{equation}\label{eq:BACKGR:brla_brlat}
 (\Pmom+\Qmom)^2 = 2\Pmom\cdot\Qmom = 4\<\Pmom\,\Qmom\>[\Pmom\,\Qmom], 
\quad\text{where}\quad
\<\Pmom\,\Qmom\> := \det\mat[\Pmomy|\Qmomy], \quad
 [\Pmom\,\Qmom] := -\det\mat[\Pmomyt|\Qmomyt].
\end{equation}
Here, for $z_1,z_2\in\C$, we denote $\det\mat[z_1|z_2]:=\Re(z_1)\Im(z_2) - \Im(z_1)\Re(z_2)$. 

For integers $a,b\geq0$, let $\Mator_{a,b}$ be the set of full rank $a\times b$ matrices over $\R$. 
Let $\Id_r$ denote the $r\times r$ identity matrix and $\bzero_{a\times b}$ denote the $a\times b$ zero matrix.
Given $\lalat\in\Mator_{2,n}^2$ and $i,j\in\brn:=\{1,2,\dots,n\}$, we set $\brla<i,j>:=\det\mat[\la_i|\la_j]$ and $\brlat[i,j]:=\det\mat[\lat_i|\lat_j]$. For a fixed integer $k$ satisfying $2\leq k \leq n-2$, we extend the columns $\la_i$ and $\lat_i$ periodically to $i\in\Z$ by setting $\la_{i+n}:=(-1)^{k-1}\la_i$ and $\lat_{i+n}:=(-1)^{k-1}\lat_i$ for all $i\in\Z$; cf. \cref{notn:BACKGR:cs} below.
When the columns of a $2\times n$ matrix $\la$ are all nonzero and $\la_i$ is not antiparallel to $\la_{i+1}$ (i.e., $\la_{i+1}\notin\R_{<0}\cdot \la_i$) for all $i\in\brn$, we define %
\begin{equation}\label{eq:intro:wind}
\wind(\la) := \sum_{i=1}^n \Arg_{(-\pi,\pi]}(\la_i,\la_{i+1})
\end{equation}
to be the total turning angle of the column vectors of $\la$ around the origin in the counterclockwise direction, where $\Arg_{(-\pi,\pi]}(\la_i,\la_{i+1})$ denotes the angle between $\la_i$ and $\la_{i+1}$. 
 Since $\la_{n+1}=(-1)^{k-1}\la_1$, $\wind(\la)$ equals $(k-1)\pi$ modulo $2\pi$. 
 We set 
\begin{align}
\lalatsMAT&:=\{\lalat\in\Mator_{2,n}^2\mid \lat\cdot \la^T = \bzerodd\},\\
\label{eq:TE:LALAK}
 \lalakMAT &:= \left\{\lalat\in\lalatsMAT\middle| 
\text{\begin{tabular}{l}
 $\brla<i,i+1> >0$ and $\brlat[i,i+1]>0$ for all $i\in\brn$,\\
 $\wind(\la) = (k-1)\pi$, and $\wind(\lat) = (k+1)\pi$
 \end{tabular}
 }\right\}.
\end{align}
Let $\GLp\subset\GL_2(\R)$ be the group of $2\times2$ matrices with positive determinant. We denote by $\Gror(2,n):=\GLp\backslash\Mator_{2,n}$ the \emph{oriented Grassmannian} consisting of oriented $2$-planes in $\R^n$. It is clear that the $\GLp$-action preserves the conditions $\brla<i,i+1> >0$ and $\turn(\la)=(k-1)\pi$. In other words, the (free) $\GLpp$-action preserves the subsets $\lalatsMAT$ and $\lalakMAT$, and we denote the corresponding $\GLpp$-quotients by $\lalats,\lalak\subset\Gror(2,n)^2$. 
We let
 \begin{align}
\label{eq:lak_latk}
 \lak &:= \{\la\in\Gror(2,n)\mid \text{$\brla<i,i+1> > 0$ for all $i\in\brn$ and $\wind(\la) = (k-1)\pi$}\};
\\
\label{eq:lak_latk2}
 \latk &:= \{\lat\in\Gror(2,n)\mid \text{$\brlat[i,i+1] > 0$ for all $i\in\brn$ and $\wind(\lat) = (k+1)\pi$} \}.
\end{align}
\noindent We similarly define $\lakMAT,\latkMAT\subset\Mator_{2,n}$.

\begin{definition}\label{dfn:TODO:null_polygons}
A \emph{null polygon} in $\Rdd$ is a collection $\Pcurve=(\xs_1,\xs_2,\dots,\xs_n)$ of points in $\Rdd$
such that $\Pmom_i:=\xs_i - \xs_{i-1}\neq0$ is a nonzero null vector for each $i\in\brn$ (with subscript $i-1$ taken modulo $n$). 
 We say that $\Pcurve$ is \emph{Mandelstam-positive} (or \emph{\Mdash positive} for short) if $(\xs_i - \xs_j)^2>0$ for all $i\in\brn$ and all $j\notin\{i-1,i,i+1\}$ modulo $n$. We say that $\Pcurve$ is \emph{in normal form} if $\xs_1=0$. We say that $\PcurveT$ is a \emph{simple polygon} if the closed polygonal chain with vertices $(\xsT_1,\xsT_2,\dots,\xsT_n)$ is non-self-intersecting.
\end{definition}

\begin{lemma}[{\Mref{lemma:Mbd=>simple}%
}]\label{lemma:TOP:Mpos=>simple}
Suppose that $\Pcurve=(\xs_1,\xs_2,\dots,\xs_n)$ is an \Mdash positive null polygon in $\Rdd$.
 Then the polygon $\PcurveT$ is simple. 
Furthermore, if $\ys\in\Rdd$ satisfies $(\ys-\xs_i)^2>0$ for all $i\in\brn$ then the point $\ysT$ 
is located either strictly inside or strictly outside $\PcurveT$.
\end{lemma}

\begin{definition}\label{dfn:TREE:lalat_vs_xbd}
Given $\lalat\in\lalatsMAT$ satisfying $\la_i,\lat_i\neq0$ for all $i\in\brn$, 
introduce a null polygon $\Pll=(\bdx_1=0,\bdx_2,\dots,\bdx_n)$ in normal form with sides $\Pmom_i:=\bdx_i - \bdx_{i-1}$ given by 
\begin{equation}\label{eq:TE:decor_dfn}%
\xM_{\Pmom_i} = \lat_i \cdot \la_i^T \quad\text{for all $i\in\brn$}.
\end{equation}
\end{definition}
\noindent 
In other words, for $\bisy_i,\bisyt_i\in\Cast$ defined by $\CtoM[\bisy_i]=\la_i$ and $\CtoMt[\bisyt_i]=\lat_i$ as in~\eqref{eq:TE:bisy_bisyt_to_M}, $(\bisy_i,\bisyt_i)$ is a bispinor representation of $\Pmom_i$ in the sense of \cref{dfn:BACKGR:bispinor}.

\begin{definition}\label{dfn:TREE:M_pos}
For $\lalat\in\lalatsMAT$ and integers $i+2\leq j\leq i+n-2$, consider the \emph{Mandelstam variable}
\begin{equation}\label{eq:TE:Mand_dfn}
\frac14(\bdx_{i} - \bdx_{j})^2 = \sum_{i< p<q\leq j} \brla<p,q>\brlat[p,q];
\end{equation}
cf.~\eqref{eq:BACKGR:brla_brlat} and~\eqref{eq:TE:decor_dfn}. We say that $\lalat$ is \emph{\Mdash positive} if $(\bdx_{i} - \bdx_{j})^2>0$ for all $i+2\leq j\leq i+n-2$.
\end{definition}

\begin{definition}[Ambient tree momentum amplituhedron~{\cite{HZ_notes,DFLP}}]\label{dfn:TREE:MPkn}
For $2\leq k\leq n-2$, let
\begin{align*}%
 \MPkntree&:=\{\lalat\in\lalak\mid \lalat\text{ is \Mdash positive}\};\\
 \MPkntreeMAT&:=\{\lalat\in\lalakMAT\mid \lalat\text{ is \Mdash positive}\}.
\end{align*}
\end{definition}

\begin{definition}\label{dfn:BACKGR:decoration}
A \emph{decoration} of a null polygon $\Pcurve$ is a pair $(\la,\lat)\in\lalatsMAT$ such that 
$\Pcurve$ coincides with $\Pll$ (cf. \cref{dfn:TREE:lalat_vs_xbd}) up to translation in $\Rdd$. When $\Pcurve$ is \Mdash positive, we say that a decoration $\lalat$ is \emph{positive} if $\lalat\in\lalakMAT$ (equivalently, $\lalat\in\MPkntreeMAT$) for some $2\leq k\leq n-2$. 
\end{definition}

\begin{definition}\label{dfn:little_group}
We let $\LG\subset\GL_n(\R)$ be the subgroup of diagonal matrices, and consider the subgroup $\LGp\subset\LG$ of matrices with positive diagonal entries. 
 We let $\LGpm:=\LGp\sqcup(-\LGp)$ be the \emph{sign-constant little group}. The groups $\LGp$ and $\LGpm$ act on $\lalak$ and $\lalakMAT$ by $(\la,\lat)\cdot \tdiag := (\la\cdot \tdiag,\lat\cdot \tdiag^{-1})$ for $\tdiag\in\LGpm$. This action preserves the subsets $\lalakMP\subset\lalak$ and $\lalakMPMAT\subset\lalakMAT$.
\end{definition}

\begin{lemma}[{%
\Mref{lemma:BCFW:null}%
}]\label{lemma:BCFW:null}
Let $\Pcurve = (\xs_1,\xs_2,\dots,\xs_n)$ be an \Mdash positive null polygon such that
\begin{equation}\label{eq:sum_arg_pmom_-2pi}
 \TURN(\PcurveT) := \sum_{i=1}^n \arg_{(-\pi,\pi]}(\PmomT_{i+1} / \PmomT_{i}) = -2\pi.
\end{equation}
Then there exists a positive decoration $\lalat$ of $\Pcurve$, unique up to the action of $\LGpm$. %
\end{lemma}
\noindent Here, $\lalat\in\lalakMPMAT$ for a unique $2\leq k\leq n-2$, and we say that $\Pcurve$ is \emph{of type $(k,n)$}. 

Identifying $\Rdd$ with $\Matddr$ via~\eqref{eq:x_vs_ap_vs_am}, we see that $\GLpp$ acts on $\Matddr$ via left and right multiplication. 
The subgroup $\SL_2(\R)\times\SL_2(\R)\subset\GLpp$ of \emph{Lorentz transformations} preserves the squared Minkowski norm $\det(\cdot)$ on $\Matddr$. Thus, the action of $\GLpp$ preserves the squared Minkowski norm up to multiplication by a positive real number. Furthermore, observe that $\Rdd$ acts on itself by translations.

\begin{definition}\label{dfn:BACKGR:GGsh}
We let $\GGsh:=\GLppar \ltimes \Rdd$ be the group generated by translations, rescalings by positive real numbers, and Lorentz transformations acting on $\Rdd$.
\end{definition}

\subsection{Planar bipartite graphs: notation and basic properties}

Throughout, we assume that $\G$ is a planar bipartite graph embedded in a disk $\Disk$, with $n$ boundary vertices $\bdv_1,\bdv_2,\dots,\bdv_n$ located on the boundary of $\Disk$, each of degree~$1$. 
We denote by $\Vint:=\Verts\setminus\{\bdv_1,\bdv_2,\dots,\bdv_n\}$ the set of interior vertices of $\G$.
 We let $\WV$ and $\BV$ denote the sets of white and black vertices of $\G$ so that $\Verts = \WV\sqcup\BV$, and set $\WVint:=\WV\cap\Vint$ and $\BVint:=\BV\cap\Vint$. We denote by $\Faces$ the set of faces of $\G$, including the $n$ boundary faces $\bdf_1,\bdf_2,\dots,\bdf_n$ (some of which may coincide), where $\bdf_i$ is the face adjacent to the boundary arc between $\bdv_i$ and $\bdv_{i+1}$. We denote by $\Fint:=\Faces\setminus\{\bdf_1,\bdf_2,\dots,\bdf_n\}$ the set of \emph{interior faces} of $\G$. For $i\in\brn$, we let $\bde_i$ be the sole edge connecting $\bdv_i$ to some \emph{next-to-boundary} vertex denoted $\bdvx_i$, and we let $\bdeast_i$ be the edge dual to $\bde_i$. 
We let $\Eint:=\Edges\setminus\{\bde_1,\dots,\bde_n\}$. 
 When using accents to denote a graph (e.g., $\tilde\G$), we implicitly assume that the same accent is used to denote its set of vertices ($\tilde\Verts$), edges ($\tilde\Edges$), etc. 

 For $\Rg\subset\Verts$, let $\GR$ and $\Grem\Rg:=\G\ind[\Verts\setminus \Rg]$ be the induced subgraphs with vertex sets $\Rg$ and $\Verts\setminus \Rg$. 

We let $\nconn(\G)$ be the number of connected components of $\G$. By Euler's formula,
\begin{equation}\label{eq:DIM:Euler}
 |\Verts| - |\Edges| + |\Fint| = 
 |\Vint| - (|\Edges|-n) + |\Fint| = 
\nconn(\G).
\end{equation}
Note that $|\Edges|-n = |\Eint|$ when no edge of $\G$ connects two boundary vertices. 
A connected component of $\G$ is called \emph{floating} if it does not contain any boundary vertices;
we denote the number of such components by $\nconnfl(\G)$. 
We have $|\Faces| = |\Fint| + n + 1 - \nconn(\G) + \nconnfl(\G)$, 
so~\eqref{eq:DIM:Euler} yields
\begin{equation}\label{eq:DIM:Euler_no_float}
 |\Vint| - |\Edges| + |\Faces| = 1 + \nconnfl(\G).
\end{equation}
\noindent 
We say that $\G$ is \emph{boundary-connected} if $\nconnfl(\G) = 0$. 
For \bdconn $\G$, its planar dual graph is denoted $\GD=(\Faces,\East)$. 
The graph $\GD$ may have loop or parallel edges. For a dual edge $\east\in\East$ with endpoints $\ff,\f\in\Faces$, we denote $\ebarast:=\{\ff,\f\}$. We set $\Ebarast:=\{\ebarast\mid\east\in\East\}$.

For $\v\in\Vint$, we denote by $\partF\v\subset\Faces$ the set of vertices of $\GD$ incident to $\v$.
The multiset of edges of $\G$ incident to a face $\ff\in\Faces$ is denoted $\partE\ff$. Here, 
 if $\e\in\Edges$ is incident to $\ff$ on both sides (i.e., the dual edge $\east$ is a loop edge connecting $\ff$ to itself) then $\e$ appears twice in $\partE\ff$. The multiset of dual edges incident to $\v\in\Vint$ is denoted $\partEast\v$. 

\begin{figure}
 \def\inputscale{1.6}
\begin{tabular}{cc}
 \includegraphics[scale=\inputscale]{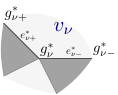}
&
 \includegraphics[scale=\inputscale]{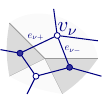}
\\
$\GD$ & \textcolor{\plabicEdgeColor}{$\G$}
\end{tabular}
 \caption{\label{fig:corners}Notation for corners of $\G$.}
\end{figure}

\begin{definition}[Corners]\label{dfn:corners}
A \emph{corner} of $\G$ is a quadruple $\cor=(\corf,\corem,\corep,\corv)\in\Faces\times\Edges^2\times\Vint$ such that the vertex $\corv$ is incident to the edges $\corepm$ and to the face $\corf$, and $\corep,\corv,\corem$ appear on the boundary of $\corf$ in clockwise order; see \cref{fig:corners}. We denote by $\coreapm$ the edges dual to $\corepm$, and let $\corfpm\in\Faces$ be such that $\coreapm$ connects $\corf$ to $\corfpm$. We let $\corners(\G)$ be the set of corners of $\G$ and denote by $\cornersb(\G):=\{\cor\in\corners(\G)\mid \corv\in\BVint\}$ and $\cornersw(\G):=\{\cor\in\corners(\G)\mid \corv\in\WVint\}$ the sets of black and white corners of $\G$, respectively. 
 For $\ff\in\Faces$, we let $\corners(\ff):=\{\cor\in\corners(\G)\mid \corf=\ff\}$, $\cornersb(\ff):=\corners(\ff)\cap\cornersb(\G)$, and $\cornersw(\ff):=\corners(\ff)\cap\cornersw(\G)$. 
For $\v\in\Vint$, we let $\corners(\v):=\{\cor\in\corners(\G)\mid \corv=\v\}$. 
\end{definition}

\begin{figure}
\def\scl{1.22}
\begin{tabular}{c|c|c}
 \includegraphics[scale=\scl]{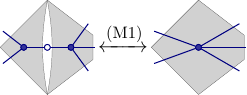}
&
 \includegraphics[scale=\scl]{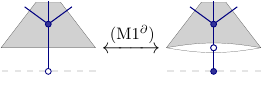}
&
 \includegraphics[scale=\scl]{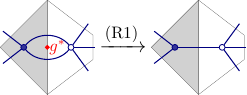}
\end{tabular}
\begin{tabular}{c|c}
\includegraphics[scale=\scl]{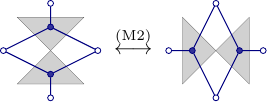} & 
\includegraphics[scale=\scl]{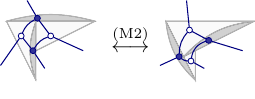}
\end{tabular}
 \caption{\label{fig:moves} Local moves on planar bipartite graphs.}
\end{figure}

We consider the following elementary moves on $\G$; see \cref{fig:moves}. %
Each of these moves induces a natural transformation on the edge weights of $\G$ that preserves its boundary measurements. It also induces a bijection on discrete holomorphic functions (\cref{sec:DIM:KSprim}) as well as \datrs, \wtimms, and \wtembs (\cref{ssec:TE:dfn}) of $\G$; cf.~\cite[Figure~4]{KLRR}.

\begin{definition}[Boundary edge insertion/contraction]\label{dfn:DIM:MVbd}
The move \MVbd on $\G$ consists of declaring $\bdv_i$ (for some $i\in\brn$) to be an interior vertex, introducing a new boundary vertex $\bdvp_i$ of color opposite to that of $\bdv_i$, and connecting it to $\bdv_i$ by a new edge $\bdep_i$. We also denote by \MVbd the reverse move when $\bdv_i$ is connected to an interior next-to-boundary vertex $\bdvx_i$ of degree $2$. 
 The move \MVbd creates or removes a bigonal face at the boundary of $\GD$. Applying such moves, we may arrange that all boundary vertices of $\G$ are black (resp., white). In this case, we say that $\G$ has \emph{black} (resp., \emph{white}) \emph{boundary}.
\end{definition}

\begin{definition}[Degree-$2$ vertex insertion/removal] \label{dfn:DIM:MV1}
Let $\v\in\Vint$ be such that $\degG(\v)=2$ and both of its neighbors are interior non-leaf vertices. The move \MV1 consists of removing $\v$ from $\G$ and identifying its two neighbors. The reverse move is also denoted \MV1. %
 This move creates or removes a bigonal face in $\GD$.
\end{definition}

\begin{definition}[Square/spider move] \label{dfn:DIM:MV2}
The move \MV{2} is shown in \figref{fig:moves}(bottom). 
\end{definition}

\begin{definition}[Parallel edge reduction] \label{dfn:DIM:R1}
Suppose that $\G$ contains a bigonal face incident to a pair $\e_1,\e_2\in\Edges$ of parallel edges connecting vertices $\vv,\v\in\Vint$. The move \RV1 consists of replacing $\e_1,\e_2$ with a single edge connecting $\vv$ to $\v$. 
 This move replaces a degree-$2$ interior vertex of $\GD$ with a single edge. %
\end{definition}

\subsection{Totally nonnegative Grassmannian}\label{ssec:dimer_backgr}
We refer to~\cite{Pos,LamCDM} for full background on the relationship between the dimer model and total positivity. 

Let $0\leq k\leq n$. Let $\Gr(k,n)$ be the space of $k$-dimensional linear subspaces of $\R^n$. The \emph{totally nonnegative Grassmannian}~\cite{Pos,Lus2} is the subset $\Grtnn(k,n)\subset\Gr(k,n)$ where all Pl\"ucker coordinates are nonnegative. Given $C\in\Grtnn(k,n)$, we view it as the row span of a full rank $k\times n$ matrix (also denoted $C$). We consider a linear operator $\alt:\R^n\to\R^n$ sending $(x_1,x_2,\dots,x_n)\mapsto(x_1,-x_2,x_3,\dots,(-1)^{n-1}x_n)$. 
Denoting $\altp(C):=\alt(C^\perp)$, we have the well-known duality
\begin{equation}\label{eq:altp_dfn}
 C\in\Grtnn(k,n)\quad\Longleftrightarrow\quad \altp(C)\in\Grtnn(n-k,n),
 \quad\text{with}\quad \Delta_{\comp{I}}(\altp(C)) = \Delta_I(C)
\end{equation}
for all $I\in{\brn\choose k}:=\{I\subset\brn:|I|=k\}$. Here, we denote $\comp{I}:=\brn\setminus I$. 

\begin{definition}
For a graph $\G$ with $n$ boundary vertices, we have
\begin{equation}\label{eq:DIM:k_dfn}
 k=|\WV| - |\BVint|,\quad n-k = |\BV| - |\WVint|, \quad\text{and}\quad n = |\Verts| - |\Vint|
\end{equation}
for some integer $k$. We say that $\G$ is \emph{of type $(k,n)$}. %
\end{definition}

 We say that edge weights $\wt,\wt'\in\RtpE$ are \emph{gauge equivalent} if there exists a function $\gauge: \Verts\to \Rtp$ such that $\gauge(\bdv_1) = \gauge(\bdv_2) = \dots = \gauge(\bdv_n) = 1$ and $\wt'(\e) = \gauge(\wv)\wt(\e)\gauge(\bv)$ for any edge $\e\in\E$ connecting $\w$ to $\b$. 
We denote by $\Rtpgauge:=\RtpE/\RtpVint$ the corresponding quotient group; cf.~\eqref{eq:DIM:Euler}.
Here, each floating component of $\G$ gives rise to a $1$-parameter subgroup of $\RtpVint$ acting trivially on $\RtpE$. 

\begin{definition}\label{dfn:DIM:apm}
An \emph{\APMnoacr (\APM)} of $\G$ is a subset $\match\subset\E$ of edges of $\G$ covering every interior (resp., boundary) vertex exactly once (resp., at most once). 
The set of \APMs of $\G$ is denoted $\APMS(\G)$. 
For $\apm\in\APMS(\G)$, we let $\partial\apm\subset\brn$ be the set of indices $i$ such that either $\bdv_i$ is black and covered by $\apm$ or $\bdv_i$ is white and not covered by $\apm$.
\end{definition}
\noindent Thus, we have $|\partial\apm| = k$ for any \APM $\apm$ of $\G$, where $k$ is given by~\eqref{eq:DIM:k_dfn}.
When $\G$ admits an \APM, we have $0\leq k \leq n$. 
For $I\in{\brn\choose k}$, we set $\APMSGbd(I):=\{\apm\in\APMS(\G):\ \partial\apm=I\}$ and 
\begin{equation}\label{eq:intro:Delta_G_wt}
 \Delta_I(\G,\wt) := \sum_{\apm\in\APMSGbd(I)} \wt(\apm),
\quad\text{where}\quad \wt(\apm) := \prod_{\e\in\apm} \wt(\e).
\end{equation}
The collection $(\Delta_I(\G,\wt))_{I\in{\brn\choose k}}$ is defined up to common rescaling (in view of the action of $\RtpVint$). When $\G$ admits an \APM, there exists a unique $k$-plane $C\in\Grtnn(k,n)$ whose Pl\"ucker coordinates are given by $(\Delta_I(\G,\wt))_{I\in{\brn\choose k}}$. We denote this $k$-plane by $\Meas(\G,\wt):=C$. 

\begin{notation}[Twisted cyclic symmetry]\label{notn:BACKGR:cs}
We extend the sequence $\mat[X_1|X_2|\dots|X_n]$ of columns of a matrix $X$ to $(X_i)_{i\in\Z}$ by the condition $X_{i+n} = \cseps X_i$ for all $i\in\Z$, where $\cseps\in\{\pm1\}$ is a fixed ``sign twist'' parameter. For $\la,\lat\in\Gr(2,n)$, $C\in\Grtnn(k,n)$, $C^\perp\in\Gr(n-k,n)$, $\ddC\in\Grtnn(k-2,n)$, or $\ddC^\perp\in\Gr(n-k+2,n)$, we set $\cseps = (-1)^{k-1}$. For objects related to $\la,\lat,C$ via the map $\alt$ (such as $\altp(C)$), we set $\cseps = (-1)^{k+n-1}$. 
We write $\bdf_{i+n}=\bdf_i$ and $\bdx_{i+n}=\bdx_i$ for $i\in\Z$. 
\end{notation}

Following~\cite{KLS}, a \emph{bounded affine permutation of type $(k,n)$} is a bijection $\fap:\Z\to\Z$ such that $\fap(i+n) = \fap(i) + n$ and $i\leq \fap(i)\leq i+n$ for all $i\in\Z$, and such that $\frac1n\sum_{i=1}^n(\fap(i)-i) = k$.
The (finite) set of bounded affine permutations of type $(k,n)$ is denoted $\Boundkn$. 
For $C\in\Mator_{k,n}$ or $C\in\Gr(k,n)$, let 
\begin{equation*}%
 \fC(i):=\min\{j\geq i\mid C_i\in\Span(C_{i+1},\dots,C_j)\} \quad\text{for all $i\in\Z$.}
\end{equation*}
It was shown in~\cite{KLS} that $\fC\in\Boundkn$. Thus, we have a \emph{positroid stratification} 
\begin{equation*}%
 \Gr(k,n) = \bigsqcup_{\fap\in\Boundkn} \Pio_\fap, \quad\text{where}\quad
\Pio_\fap:=\{C\in\Gr(k,n)\mid \fC = \fap\}.
\end{equation*}
 For each graph $\G$ of type $(k,n)$ admitting an \APM, there exists $\fG\in\Boundkn$ such that the \emph{positroid cell}
\begin{equation}\label{eq:BACKGR:Ptp_G}
 \Ptp_{\G}:=\{\Meas(\G,\wt)\mid \wt\in\Rtpgauge\} \quad\text{coincides with}\quad \Ptp_{\fG}:=\Pio_{\fG}\cap \Grtnn(k,n).
\end{equation}
\begin{definition}\label{dfn:reduced}
 $\G$ is called \emph{reduced} if the map $\Meas(\G,\cdot):\Rtpgauge\to\Ptp_{\G}$ is a homeomorphism.
\end{definition}
\noindent For reduced $\G$, $\fG$ may be computed from $\G$ using \emph{zig-zag paths}. We will mostly focus on not necessarily reduced graphs $\G$, in which case $\fG$ can only be computed from the boundary measurement map of $\G$.

\begin{remark}\label{rmk:altp_vs_color_swap}
The map $\altp$ from~\eqref{eq:altp_dfn} sends $\Ptp_f$ homeomorphically to $\Ptp_{\fhat}$, where $\fhat\in\Boundxx(n-k,n)$ is given by $\fhat(i) = f^{-1}(i) + n$ for all $i\in\Z$. If $C = \Meas(\G,\wt)$ then $\altp(C) = \Meas(\Ghat,\wt)$, where $\Ghat$ is obtained from $\G$ by changing the colors of all vertices (i.e., swapping the roles of black and white).
\end{remark}

\subsection{Kasteleyn theory and \KSprims}\label{sec:DIM:KSprim}
We discuss the \emph{Kasteleyn sign condition}~\cite{Kasteleyn,AGPR} for planar bipartite graphs and review the \emph{\KSprims} $\xd:\Faces\to\Rdd$ studied originally in~\cite{Kenyon_conf,Smirnov_cluster}. 
See~\Mref{ssec:Kast_KSprims} for further details. 

Assume that $\G$ \hasnofloat. For $\ff\in\Faces$, we denote $\bdryarcs\ff:=\{i\in\brn\mid\bdf_i = \ff\}$ 
and let 
$\dwcor(\ff)$ be the number of white vertices incident to $\ff$, counted with multiplicity. 
This includes the white corners in $\cornersw(\ff)$ introduced in \cref{dfn:corners} together with the boundary white vertices incident to $\ff$. 

\begin{definition}[{\cite{AGPR,SpeyerVariations}}]\label{dfn:DIM:Kast}
We say that $\epsK:\E\to\{\pm1\}$ is a choice of \emph{Kasteleyn signs} for $\G$ if for each face $\ff\in\Faces$, 
\begin{equation}\label{eq:Kast_sign}
 \epsK(\partE\ff):=\prod_{\e\in\partE\ff}\epsK(\e) = (-1)^{\dwcor(\ff)+\nbdryarcs\ff+\epstra(\ff)},\quad\text{where}\quad
\epstra(\ff) = 
\begin{cases}
 1, &\text{if $n\notin\bdryarcs\ff$;}\\
 k+n, &\text{if $n\in\bdryarcs\ff$.}
\end{cases}
\end{equation}
Throughout, we denote by $\wtK:\Edges\to\R$ the \emph{Kasteleyn edge weights} given by $\wtK(\e):=\epsK(\e)\wt(\e)$ for all $\e\in\Edges$. 
\end{definition}
\noindent 
Here, an edge $\e$ incident to $\ff$ on both sides contributes $\epsK(\e)^2=1$ to $\epsK(\partE\ff)$. 

\begin{remark}\label{rmk:Kast_sign_float}
When $\G$ is not necessarily \bdconn, the Kasteleyn sign condition~\eqref{eq:Kast_sign} becomes
$\epsK(\partE\ff)=(-1)^{\dwcor(\ff)+\nbdryarcs\ff+\epstra(\ff)+\nfloat(\ff)}$, where $\nfloat(\ff)$ denotes the number of ``holes'' in the face $\ff$ (i.e., the number of bounded connected components of the complement of $\ff$ in the plane). 
 By convention, isolated white vertices contained in $\ff$ contribute to $\nfloat(\ff)$ but not to $\dwcor(\ff)$.
\end{remark}

\begin{lemma}[{\cite[Lemma~1]{Kenyon_lectures} and~\Mref{lemma:OCP:Kast_even}}]\label{lemma:OCP:Kast_even}
Let $\epsK$ be a choice of Kasteleyn signs for $\G$. Let $\G'$ be a subgraph of $\G$ with the same set of boundary vertices. Then $\epsK$ restricts to a choice of Kasteleyn signs for $\G'$ if and only if each face of $\G'$ that does not contain $\bdf_n$ encloses an even number of vertices in $\Vint\setminus \Vint'$. 
\end{lemma}

Let $\Vspace$ be an $\R$-vector space. 
 A function $\Fw: \WV\to\Vspace$ (resp., $\Fb:\BV\to\Vspace$) is called \emph{\wdash holomorphic} (resp., \emph{\bdash holomorphic}) if for each $\bv\in \BVint$ (resp., $\wv\in\WVint$), 
\begin{equation}\label{eq:OCP:holom}
\sum_{\wv'\sim\bv} \wtK(\wv',\bv) \Fw(\wv') = 0,
\quad\text{resp.,}\quad
\sum_{\bv'\sim\wv} \wtK(\wv,\bv') \Fb(\bv') = 0.
\end{equation} 
Here, the summations are taken over all vertices of $\G$ adjacent to $\bv$ (resp., $\wv$), and $\wtK(\wv',\bv')$ denotes the sum of Kasteleyn weights of the edges connecting $\bv'$ to $\wv'$. 

We refer to \wdash\ and \bdash holomorphic functions collectively as \emph{($\Vspace$-valued) discrete holomorphic functions}. 
We let $\Hwspace_{\Vspace}\HtripK$ (resp., $\Hbspace_{\Vspace}\HtripK$) denote the space of $\Vspace$-valued \wdash holomorphic (resp., \bdash holomorphic) functions 
and set $\HHspaceV:=\Hwspace_{\Vspace}\HtripK\times\Hbspace_{\Vspace}\HtripK$. 
Given $(\Fw,\Fb)\in\HHspaceV$, we let
$\pFw=(\pFw_1,\pFw_2,\dots,\pFw_n)\in\Vspace^n$ (resp., $\pFb=(\pFb_1,\pFb_2,\dots,\pFb_n)\in\Vspace^n$) be given for $i\in\brn$ by
\begin{equation}\label{eq:partial_F_dfn} 
 \pFw_i:=
 \begin{cases}
 -\Fw(\bdv_i), &\text{if $\bdv_i$ is white;}\\
 -\wtK(\bde_i) \Fw(\bdvx_i), &\text{if $\bdv_i$ is black;}
 \end{cases}
 \quad%
 \pFb_i:=
 \begin{cases}
 \Fb(\bdv_i), &\text{if $\bdv_i$ is black;}\\
 -\wtK(\bde_i) \Fb(\bdvx_i), &\text{if $\bdv_i$ is white.}
 \end{cases}
\end{equation}

Given a pair $(\Fw,\Fb)\in\HHspaceC$ of $\C$-valued discrete holomorphic functions, the \emph{\KSprim} $\xd=(\xT,\xO):\Faces\to\Rdd$ is defined up to an overall additive constant by the conditions
\begin{equation}\label{eq:OCP:primitive}
 \xT(\f_2)-\xT(\f_1) = \Fw(\wv) \wtK(\e) \Fb(\bv) 
 \quad\text{and}\quad
 \xO(\f_2)-\xO(\f_1) = \ovl{\Fw(\wv)} \wtK(\e) \Fb(\bv) 
\end{equation}
for all $\e\in\Edges$, where $\bv,\wv$ are the endpoints of $\e$ and $\f_1,\f_2$ are the endpoints of $\east$ such that 
 $\wv$ lies to the left of the oriented edge $\evecast$ pointing from $\f_1$ to $\f_2$. By~\eqref{eq:OCP:holom}, the differences in~\eqref{eq:OCP:primitive} add up to zero around each (black or white) face of $\GD$. Thus, $\xd$ is globally well defined on $\Faces$ up to an additive constant. 
We denote $\bdx_i:=\xd(\bdf_i)$ for $i\in\brn$. 
We note for future reference that by~\eqref{eq:partial_F_dfn}--\eqref{eq:OCP:primitive}, 
\begin{equation}\label{eq:TE:t_imm_bdry_vs_pFw_pFb}
 \bdxT_i - \bdxT_{i-1} = \pFw_i\pFb_i 
 \quad\text{and}\quad
 \bdxO_i - \bdxO_{i-1} = \ovl{\pFw_i}\pFb_i 
\quad\text{for all $i\in\brn$},
\end{equation}
where the index $i-1$ is taken modulo $n$; cf. \cref{notn:BACKGR:cs}.

\begin{remark}\label{rmk:DIM:Kast_gauge_eq}
The action of the gauge group $\RtpVint$ on $\wt\in\RtpE$ extends to $\HHspaceC$: 
for $\gauge\in\RtpVint$, the value $\Fw(\wv)$ (resp., $\Fb(\bv)$) gets divided by $\gauge(\wv)$ (resp., by $\gauge(\bv)$). The \emph{sign gauge group} $\pmoneVint$ acts on $(\epsK,\wtK,\Fw,\Fb)$ by changing the signs at interior vertices. It is well known~\cite[Section~3.2]{Kenyon_lectures} that any two choices $\epsK,\epsK'$ of Kasteleyn signs on $\G$ are related by the $\pmoneVint$-action. By~\eqref{eq:OCP:primitive}, the \KSprim $\xd$ is invariant under the action of the gauge groups $\RtpVint$ and $\pmoneVint$.
\end{remark}

\begin{remark}\label{rmk:DIM:otimes}
For $(\Fw,\Fb)\in\Hwspace_{\Vspace_1}\HtripK\times\Hbspace_{\Vspace_2}\HtripK$, we similarly define the \KSprim $\Hknk:\Faces\to\Vspace_1\otimes_\R\Vspace_2$. For example, we use this construction for $(\Vspace_1,\Vspace_2)=(\R^k,\R^{n-k})$ to define the \emph{\Lfunc boundary measurement map} in \cref{dfn:LPUNC:Meas_func}. For even $m\geq2$, setting $\Vspace_1 = \Vspace_2 = \R^{m/2}$ leads to a starting point for a definition of a \emph{higher-$m$ momentum amplituhedron} which we hope to pursue in future work.
\end{remark}

\begin{definition}\label{dfn:MCE:WKmat}
Assume that $\G$ has white boundary; thus, $\BVint = \BV$. In this case, we denote the $\WV\times\BVint$ Kasteleyn matrix with entries $\wtK(\wv,\bv)$ by $\WKmat$. 
 We assume that the rows of $\WKmat$ are ordered so that the boundary vertices $\bdv_1,\bdv_2,\dots,\bdv_n$ appear first (and in this order). For $I\in{\brn\choose k}$, we let $\Delta_{\WV\setminus I}(\WKmat)$ be the minor of $\WKmat$ with row set $\{\bdv_i\mid i\notin I\}\sqcup\WVint$ and column set $\BVint$. 
\end{definition}

\begin{proposition}[{\cite[Corollary~6.8 and Proposition~6.9]{AGPR}}]%
\label{lemma:MCE:Delta_vs_Kast}
We have $\Delta_I(\G,\wt) = \eps\Delta_{\WV\setminus I}(\WKmat)$ for a fixed $\eps\in\{\pm1\}$ and all $I\in{\brn\choose k}$.
\end{proposition}
\begin{corollary}[{\Mref{lemma:MCE:apm_vs_Kast}}]\ \label{lemma:MCE:apm_vs_Kast}
If $\G$ has white boundary and admits an \APM then
\begin{enumerate}[label=(\arabic*)]
\item\label{apm_vs_Kast1}
 $\WKmat$ is of full rank: $\rank\WKmat = |\BVint|$,
\item\label{apm_vs_Kast2}
 $\dim\Hwspace_{\R}\HtripK = k$, 
\item\label{apm_vs_Kast3} the linear operator $\partial:\Hwspace_{\R}\HtripK\to\R^n$ 
defined in~\eqref{eq:partial_F_dfn}
 is injective, and for $C:=\Meas(\G,\wt)$, %
\begin{equation}\label{eq:MCE:alt(C)_vs_pFw}
 \alt(C) = \{\pFw \mid \Fw\in\Hwspace_{\R}\HtripK\} \quad\text{as elements of $\Gr(k,n)$}.
\end{equation}
\end{enumerate}
\end{corollary}

\begin{remark}\label{rmk:MCE:BKmat}
Similar results hold for the $\BV\times\WVint$ matrix $\BKmat$ when $\G$ has black boundary: there exists $\eps\in\{\pm1\}$ such that for all $I\in{\brn\choose k}$, we have $\Delta_I(\G,\wt) = (-1)^{\sum_{i\in I} i}\eps\Delta_{\{\bdv_i\mid i\in I\}\sqcup\BVint}(\BKmat)$, and if $\G$ admits an \APM then 
\begin{equation}\label{eq:MCE:alt(Cp)_vs_pFb}
\dim\Hbspace_{\R}\HtripK = n-k
\quad\text{and}\quad
 \alt(C^\perp) = \{\pFb \mid \Fb\in\Hbspace_{\R}\HtripK\} \quad\text{as elements of $\Gr(n-k,n)$}.
\end{equation}
\end{remark}

\begin{definition}[Boundary restriction and extension]\label{dfn:OCP:restr_ext}
Assume that $\G$ admits an \APM and let $\Vspace=\R^d$. 
The \emph{boundary restriction} of a discrete holomorphic function $\Fw\in\Hwspace_{\Vspace}\HtripK$ (resp., $\Fb\in\Hbspace_{\Vspace}\HtripK$) is the $d\times n$ matrix $\alt(\pFw)\subset C$ (resp., $\alt(\pFb)\subset C^\perp$). 
In this case, we say that $\Fw$ (resp., $\Fb$) is the \emph{discrete holomorphic extension} of $\alt(\pFw)$ (resp., $\alt(\pFb)$). By \cref{lemma:MCE:apm_vs_Kast}, a discrete holomorphic extension of any $d\times n$ matrix $\Amat\subset C$ (resp., $\Atmat\subset C^\perp$) exists and is unique.
\end{definition}

\begin{proposition}[\Mref{lemma:DIM:remove_bivertex}]\label{lemma:DIM:remove_bivertex}
Suppose that $\w_1,\w_2\in\WVint$ share a face of $\G$. 
Assume that both $\G$ and $\G':=\G\rem\{\w_1,\w_2\}$ admit \APMs and let $\wt':=\wt|_{\Edges'}$.
 Then 
\begin{equation}\label{eq:DIM:remove_bivertex_w}
 C'\subset C,\quad\text{where}\quad
 C':=\Meas(\G',\wt')\in\Grtnn(k-2,n) \quad\text{and}\quad 
C:=\Meas(\G,\wt)\in\Grtnn(k,n).
\end{equation}
Similarly, if $\b_1,\b_2\in\BVint$ share a face of $\G$ and $\G'':=\G\rem\{\b_1,\b_2\}$ admits an \APM then 
\begin{equation}\label{eq:DIM:remove_bivertex_b}
 C\subset C'',\quad\text{where}\quad
C'':=\Meas(\G'',\wt'')\in\Grtnn(k+2,n)
\quad\text{for}\quad
\wt'':=\wt|_{\Edges''}.
\end{equation}
\end{proposition}

\begin{proposition}[Popping a black vertex]
\label{lemma:DIM:remove_add}
Let $(\G,\wt)$ and $(\G',\wt')$ be two weighted planar bipartite graphs that admit \APMs. 
Suppose that $\bv\in\BVint$ and $\wv'\in\WVintp$ are such that the graphs $\G\rem\{\bv\}$ and $\G'\rem\{\wv'\}$ coincide,
$\bv$ and $\wv'$ are located inside the same face of $\G\rem\{\bv\}=\G'\rem\{\wv'\}$, and
 the restrictions of $\wt$ and $\wt'$ to the edges of $\G\rem\{\bv\}=\G'\rem\{\wv'\}$ also coincide. 
Then 
\begin{equation}\label{eq:DIM:remove_add}
 C\subset C', \quad\text{where}\quad
 C:=\Meas(\G,\wt)\in\Grtnn(k,n) \quad\text{and}\quad
 C':=\Meas(\G',\wt')\in\Grtnn(k+2,n). 
\end{equation}
\end{proposition}

\noindent See \cref{fig:popping} and \cref{ex:non_ordinary_Meas_Lvunc}. 

\begin{proof}
Let $\epsK$ and $\epsK'$ be choices of Kasteleyn signs on $\G$ and $\G'$, respectively. 
Let $\ff$ be the face of $\G'':=\G\rem\{\bv\}=\G'\rem\{\wv'\}$ containing $\bv$ and $\wv'$.
By \cref{lemma:OCP:Kast_even}, the restrictions of $\epsK$ and $\epsK'$ to the edges of $\G''$ satisfy the Kasteleyn sign condition in \cref{rmk:Kast_sign_float} for every face of $\G''$ other than $\ff$ (and are violated for $\ff$ if and only if $\ff$ does not contain $\bdf_n$). 
 Thus, the restrictions of $\epsK$ and $\epsK'$ to the edges of $\G''$ are gauge equivalent (cf. \cref{rmk:DIM:Kast_gauge_eq}), and after acting by $\pmoneVint$ and $\pmoneVintp$ on $\epsK$ and $\epsK'$, we may assume that the restrictions of $\epsK$ and $\epsK'$ to $\G''$ coincide. 
Observe that $\WV = \WVp\setminus\{\wv'\}$. 
We have an injection $\Hwspace_{\R}\HtripK\hookrightarrow\Hwspace_{\R}\HtrippK$ which extends $\Fw\in\Hwspace_{\R}\HtripK$ by setting $\Fw(\wv'):=0$. Applying~\eqref{eq:MCE:alt(C)_vs_pFw} to $\HtripK$ and $\HtrippK$, we obtain~\eqref{eq:DIM:remove_add}.
\end{proof}

\begin{remark}\label{rmk:add_remove_vs_remove_two}
\Cref{lemma:DIM:remove_add} implies \cref{lemma:DIM:remove_bivertex}. For example, suppose that $\b_1,\b_2\in\BVint$ share a face of $\G$ and let $\G'':=\G\rem\{\b_1,\b_2\}$ be as in \cref{lemma:DIM:remove_bivertex}. Let $(\G',\wt')$ be obtained from $(\G,\wt)$ by replacing $\b_1$ with a white degree-$1$ vertex connected to $\b_2$ by an edge of weight $1$. Then $\Meas(\G',\wt') = \Meas(\G'',\wt'')$, and
applying~\eqref{eq:DIM:remove_add} to $\G$ and $\G'$, we obtain~\eqref{eq:DIM:remove_bivertex_b}. 
\end{remark}

\subsection{Momentum amplituhedron map and immanant-nonnegativity}\label{ssec:BACKGR:mom_ampl}
We review the construction of the tree momentum amplituhedron map of~\cite{DFLP} and some results of~\justpapone describing when the image of this map is \Mdash nonnegative.

\begin{definition}\label{dfn:BACKGR:nondeg}
We say that $\G$ is \emph{\twonondeg} 
if for any $i\in\brn$, $\G$ admits \APMs $\apm_+,\apm_-$ such that $i,i+1\in\partial\apm_+$ and $i,i+1\notin \partial\apm_-$. We say that $C=[C_1|C_2|\cdots|C_n]\in\Gr(k,n)$ is \emph{\twonondeg} if for all $i\in\Z$, we have 
$\rank[C_i|C_{i+1}] = 2$ and $\rank[C_{i+2}|\dots|C_{i+n-1}] = k$. 
It follows from standard properties of $\Meas$ that $\G$ is \twonondeg if and only if for some (equivalently, any) $\wt\in\Rtpgauge$, $C:=\Meas(\G,\wt)$ is \twonondeg. We let $\Grnd(k,n):=\{C\in\Grtnn(k,n)\mid C\text{ is \twonondeg}\}$.
\end{definition}

\begin{definition}
For $2\leq k\leq n-2$, let
\begin{equation}\label{eq:intro:LaLat}
\LaLak := \alt(\Grtp(n-k+2,n)) \times \Grtp(k+2,n).
\end{equation}
For a fixed pair $\LaLat\in\LaLak$, define the \emph{momentum amplituhedron map}~\cite{DFLP} by
\begin{equation}\label{eq:intro:PhiLL_dfn}
 \PhiLL: \Grtnn(k,n)\to \lalats,\quad C \mapsto (C\cap \La,C^\perp\cap \Lat).
\end{equation} 
\end{definition}

\begin{lemma}[{\cite{DFLP} and~\Mref{prop:momLL_basic}}]\label{lemma:OAC:PhiLL_in_lalak}
For any $\LaLat\in\LaLak$ and $C\in\Grtnn(k,n)$, the intersections $\la := C\cap \La$ and $\lat := C^\perp\cap \Lat$ are $2$-dimensional. Furthermore, if $C\in\Grnd(k,n)$ is \twonondeg then $\PhiLL(C)\in\lalak$. %
\end{lemma}

\begin{remark}\label{rmk:orientation_inherits}
In~\eqref{eq:intro:PhiLL_dfn}, we treat $C,\La,\Lat$ as points in the respective \emph{oriented} Grassmannians where the Pl\"ucker coordinates have prescribed signs. This endows the $2$-planes $\la,\lat$ with canonical orientations. In practice, when $C\in\Grnd(k,n)$, we always orient $\la,\lat$ so that $\brla<i,i+1>,\brlat[i,i+1]>0$ for all $i\in\brn$. 
\end{remark}

 In~\Mref{dfn:immanants_LaLat}, we introduced \emph{immanants} $\{\Ttaucoef(\La,\Lat)\mid i+2\leq j\leq i+n-2,\ \ \tauT\in\Ttauknij\}$ of $\LaLat\in\LaLak$. Each function $\Ttaucoef(\La,\Lat)$ is a polynomial in the Pl\"ucker coordinates of $\La$ and $\Lat$. 
\begin{definition}[\Mref{dfn:Ttauknij}]\label{dfn:BACKGR:immp}
We say that $\LaLat\in\LaLak$ is \emph{immanant-nonnegative} if $\Ttaucoef(\La,\Lat)\geq0$ for all $i+2\leq j\leq i+n-2$ and $\tauT\in\Ttauknij$. We denote 
\begin{equation*}%
 \LaLaimmnn:=\{\LaLat\in\LaLak\mid \LaLat\text{ is immanant-nonnegative}\}.
\end{equation*}
\end{definition}

\begin{proposition}[\Mref{lemma:LaLaimmp_nonempty_Z_dense}]\label{lemma:LaLaimmp_nonempty_Z_dense}
The subset $\LaLaimmnn\subset\LaLak$ is Zariski dense.
\end{proposition}

\subsection{The magic projector \texorpdfstring{$\Qla$}{Q\_lambda}}
Let $\la\in\Gror(2,n)$ be such that $\brla<i,i+1>\neq0$ for all $i\in\brn$. 
Consider the following linear operator $\Qla:\R^n\to\R^n$ introduced in~\cite{AHCC}; see also~\cite[Equation~(8.23)]{abcgpt}. For $C=\mat[C_1|C_2|\cdots|C_n]\in\Gr(k,n)$, the matrix $C\cdot \Qla$ has columns
\begin{equation}\label{eq:intro:Qla}
 (C\cdot \Qla)_i = \frac{1}{\brla<i-1,i>\brla<i,i+1>} 
\left(
C_{i-1}\brla<i,i+1> + C_i\brla<i+1,i-1> + C_{i+1}\brla<i-1,i>
\right) 
\quad\text{for $i\in\brn$;}
\end{equation}
cf. \cref{notn:BACKGR:cs}. 
 It is well known that $\Qla^T = \Qla$ and $\Ker\Qla = \la$. 

We denote $\Gr(k-2,\lap):=\{\ddC\in\Gr(k-2,n)\mid \ddC\subset\lap\}$ and $\Grsupla(k,n):=\{C\in\Gr(k,n)\mid \la\subset C\}$. We set $\Grtnn(k-2,\lap):=\Grtnn(k-2,n)\cap\Gr(k-2,\lap)$ and $\Grtnnsupla(k,n):=\Grsupla(k,n)\cap \Grtnn(k,n)$. 
Since $\Ker\Qla = \la$, for $C\in\Grsupla(k,n)$, we have $C\cdot \Qla\in\Gr(k-2,\lap)$. 
By~\Mref{lemma:TREE:magic_homeo}, $\Qla:\Grsupla(k,n)\xrasim\Gr(k-2,\lap)$ is a homeomorphism. We denote its inverse by $\Qlapp:\Gr(k-2,\lap)\xrasim\Grsupla(k,n)$. By~\Mref{prop:magic_homeo}, when $\la\in\lak$, $\Qla$ and $\Qlapp$ restrict to 
 homeomorphisms 
\begin{equation}\label{eq:Qla_Qlapp_preserve_TNN}
 \Qla:\Grtnnsupla(k,n)\xrasim\Grtnn(k-2,\lap) \quad\text{and}\quad \Qlapp:\Grtnn(k-2,\lap)\xrasim\Grtnnsupla(k,n).
\end{equation}

\part{\OACTITLE and T-duality}\label{part1}

\section{\OACTITLE for \wtembsTITLE}\label{sec:TE}
In~\Mref{thm:intro:t_imm_vs_triples}, we established the \emph{\oac} between \emph{t-immersions} of connected graphs $\G$ satisfying $\helmin(\G)\geq2$ (see \cref{ssec:BACKGR:surplus}) and triples $\la\subset C\subset\latp$ satisfying $C=\Meas(\G,\wt)$ and $\lalat\in\lalak$. In this section, we extend this correspondence (see \cref{thm:TE:OAC}) to graphs $\G$ satisfying $\helmin(\G)\geq0$ (i.e., admitting \APMs), with t-immersions replaced by \emph{\wtimms}. When $\helmin(\G)\geq1$, we show in \cref{thm:TOP:weak_imm=>imm} that \wtimms are \emph{weak immersions}, i.e., limits of immersions. 

\subsection{Defining \wtembsTITLE}\label{ssec:TE:dfn}
It was shown in~\cite[Section~3.2]{KLRR} and \cite[Sections~2 and~3]{CLR1} (see~\Mref{prop:t_imm=>holom}) that every t-immersion arises from a \KSprim of a pair of discrete holomorphic functions. To that end, we make the following definition.

\begin{definition}\label{dfn:TE:datr}
An \emph{\datr} of $\G$ is a \quintuple $\datrQL:=\datrQ$, where $\wt\in\Rtpgauge$, $\epsK$ is a choice of Kasteleyn signs for $\G$, $(\Fw,\Fb)\in\HHspaceC$ is a pair of $\C$-valued discrete holomorphic functions, 
 and $\xd:\Faces\to\Rdd$ is the \KSprim of $(\Fw,\Fb)$; cf. \cref{sec:DIM:KSprim}. The set of \datrs of $\G$ is denoted $\Mdatr(\G)$. 
The subset of \datrs with fixed $\wt$ and $\epsK$ is denoted by $\Mdatr(\G,\wt,\epsK)$. We often omit $\epsK$ from the notation (cf. \cref{rmk:DIM:Kast_gauge_eq}) and write $\Mdatr(\G,\wt)$ instead. 
We view elements of $\Mdatr(\G)$ and $\Mdatr(\G,\wt)$ up to the action of gauge groups $\RtpVint$ and $\{\pm1\}^{\Vint}$ on $\datrQnox$; cf. \cref{rmk:DIM:Kast_gauge_eq}. 
\end{definition}

In view of~\eqref{eq:MCE:alt(C)_vs_pFw} and \cref{notn:BACKGR:cs}, we set $\pFw_{n+1}:=(-1)^{k+n-1}\pFw_1$ and $\pFb_{n+1}:=(-1)^{k+n-1}\pFb_1$.
\begin{definition}[\Wtemb]\label{dfn:TE:tembending}
Assume that $\G$ admits an \APM.
 We say that $\datrQL=\datrQ\in\Mdatr(\Gbip)$ is a \emph{\wtimm} if it satisfies the following conditions.
 \setlength{\leftmargini}{47pt}
\begin{enumerate}[label=(WTE\arabic*)]
\item\label{TE:bdry_angle} \emph{Boundary angle condition:} we have %
\begin{equation}\label{eq:TE:weak_t_imm_bdry}
 \det\mat[\pFw_i|\pFw_{i+1}]<0 \quad\text{and}\quad
 \det\mat[\pFb_i|\pFb_{i+1}]>0 \quad\text{for each $i\in\brn$}.
\end{equation}
In other words, denoting $\arg:=\arg_{[0,2\pi)}$ and
\begin{equation}\label{eq:TE:dfn}
\def\qupad{\quad}
 \sumbT_i:=\arg(-\pFw_{i+1}/\pFw_i), \qupad%
 \sumwT_i:=\arg(\pFb_{i+1}/\pFb_i), 
\qupad\text{we have}\qupad
 0<\sumbT_i,\sumwT_i<\pi \qupad\text{for $i\in\brn$.}
\end{equation}
\item\label{TE:bdry_winding} \emph{Boundary \winding condition:} the angles $\sumbT_i,\sumwT_i$ defined in~\eqref{eq:TE:dfn} satisfy
\begin{equation}\label{eq:TE:bdry_winding}
 \sum_{i=1}^n \sumbT_i = \pi(k-1) \quad\text{and}\quad
 \sum_{i=1}^n \sumwT_i = \pi(n-k-1).
\end{equation}
\item\label{TE:t_imm} \emph{Weak immersion condition:} for each $\b\in\BVint$ (resp., $\w\in\WVint$) connected by edges $\e_1,\e_2,\dots,\e_d$ to vertices $\w_1,\w_2,\dots,\w_d$ (resp., $\b_1,\b_2,\dots,\b_d$) in clockwise order, 
\begin{equation}\label{eq:TE:weak_t_imm_dfn}
 \epsK(\e_\s)\epsK(\e_{\s+1})\det\mat[\Fw(\w_{\s})|\Fw(\w_{\s+1})]\leq0, \quad\text{resp.,}\quad
 \epsK(\e_\s)\epsK(\e_{\s+1})\det\mat[\Fb(\b_{\s})|\Fb(\b_{\s+1})]\leq0
\end{equation}
for all $\s\in\brd$. 
\end{enumerate}
We say that a \wtimm $\datrQL$ is a \emph{\wtemb} if the polygon $\PbdT_{\xd}:=(\bdxT_1,\bdxT_2,\dots,\bdxT_n)$
 is simple. 
We denote the set of \wtimms (resp., \wtembs) of $\G$ by $\Mdti(\G)$ (resp., $\Mdte(\G)$). 
We denote by $\Mdti(\G,\wt)$ (resp., $\Mdte(\G,\wt)$) the subset with fixed $\wt\in\Rtpgauge$.
\end{definition}

We will be particularly interested in the class of \emph{\einj} \wtimms.
\begin{definition}\label{dfn:TE:generic}
 We say that a map $\xd:\Faces\to\Rdd$ is \emph{\einj} if $\xd(\ff) \neq \xd(\f)$ 
when $\ff,\f\in\Faces$ are connected by an edge of $\GD$. 
 We say that $\datrQL=\datrQ\in\Mdatr(\G)$ is \emph{\einj} if $\xd$ is \einj. 
The set of \einj \datrs (resp., \wtimms, \wtembs) is denoted by $\Mdatro(\G)$ (resp., $\Mdtio(\G)$, $\Mdteo(\G)$). 
\end{definition}
\begin{remark}\label{rmk:TE:einj_nonvanishing}
Equivalently, by~\eqref{eq:OCP:primitive}, $\datrQL\in\Mdatr(\G)$ is \einj if and only if for each $\wv\in\WV$ (resp., $\bv\in\BV$) we have $\Fw(\wv)\neq0$ (resp., $\Fb(\bv)\neq0$). In particular, $\xT(\ff)\neq\xT(\f)$ and $\xO(\ff)\neq\xO(\f)$ for all $\ff,\f$ connected by an edge of $\GD$ when $\datrQL$ is \einj.
\end{remark}

\begin{figure}
 \def\inputscale{1.5}
\includegraphics[scale=\inputscale]{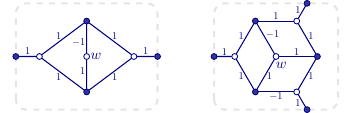}
 \caption{\label{fig:non-einj} Examples of weighted graphs $(\G,\wt)$ (with Kasteleyn edge weights shown) satisfying $\helmin(\G)=1$ but admitting no \einj \wtimms.}
\end{figure}

\begin{example}\label{ex:non-einj}
Two weighted graphs $(\G,\wt)$ are shown in \cref{fig:non-einj}. %
It is easy to check that any \wdash holomorphic function $\Fw\in\Hwspace_{\C}\HtripK$ satisfies $\Fw(\w)=0$ for the vertex $\w$ indicated in the figure. Thus, no \wtimm of $(\G,\wt)$ is \einj.\footnote{
Any weighted graph with outer face of degree $4$ containing one of the graphs in \cref{fig:non-einj} as an induced subgraph gives a counterexample to \cite[Theorem~2]{KLRR}. 
The statement of that theorem only holds for t-embeddings when $\helmin(\G)\geq2$ (\Mref{thm:intro:t_imm_vs_triples}) or for \einj \wtembs when $\helmin(\G)\geq1$ and the edge weights $\wt$ are generic (\crefi{lemma:TE:brla_nonzero_if_apm}{lak_implies4}). 
In the former case, the argument in~\cite{KLRR} does not immediately apply since the condition $\helmin(\G)\geq2$ does not propagate under the moves in \cref{fig:moves}. 
In the latter case, the proof in~\cite{KLRR} and especially the square/spider moves~\MV2 in \cite[Lemma~4]{KLRR} need to be extended to not necessarily convex and not necessarily embedded quadrilaterals such as the one in \figref{fig:moves}(bottom right). 
This can be done by noting that~\eqref{eq:TE:weak_t_imm_dfn} propagates under the moves in \cref{fig:moves}.
We thank R.~Kenyon and M.~Russkikh for discussions related to this issue.
} For different choices of edge weights, two \einj \wtimms of the second graph are shown in \cref{fig:corner-0-pi} below. They can be related by a continuous deformation, and at some point during that deformation, the middle white triangle degenerates into a point, giving rise to a non-\einj \wtimm.
\end{example}

\subsection{Surplus}\label{ssec:BACKGR:surplus}

Given a subset $\Rg\subset\Verts$, we let $\RgWV:=\Rg\cap\WV$ and $\RgBV:=\Rg\cap\BV$. 
We say that $\Rg\subset\Verts$ is \emph{\wclosed} if $\RgBV\subset\BVint$ and $\NeighG(\RgBV)\subset\RgWV$, where $\NeighG(\RgBV)\subset\WV$ denotes the neighborhood of $\RgBV$ in $\G$. We let $\WNEIg(\G)$ be the set of \wclosed subsets in $\G$. 
For $\Rg\in\WNEIg(\G)$, we set $\helW(\Rg):=|\RgWV| - |\RgBV|$. We set 
$\WNEI(\G):=\{\Rg\in\WNEIg(\G)\mid \RgBV\neq\emptyset\}$. 
We similarly define the set $\BNEIg(\G)$ of \emph{\bclosed} subsets and set $\helB(\Rg):=|\RgBV| - |\RgWV|$ for $\Rg\in\BNEIg(\G)$
and $\BNEI(\G):=\{\Rg\in\BNEIg(\G)\mid \RgWV\neq\emptyset\}$. 
We set 
\begin{equation}\label{eq:DIM:helWmin_dfn}
 \helWmin(\G) := \min\{\helW(\Rg)\mid \Rg\in\WNEI(\G)\}
 \quad\text{and}\quad
 \helBmin(\G) := \min\{\helB(\Rg)\mid \Rg\in\BNEI(\G)\}.
\end{equation}
We let $\helmin(\G):=\min(\helWmin(\G),\helBmin(\G))$ be the \emph{surplus} of $\G$; see~\cite[Section~1.3]{Lovasz_Plummer} and~\cite[Section~4.1]{Kenyon_Sheffield} for closely related analysis. We similarly define
\begin{equation}\label{eq:DIM:helWming_dfn}
 \helWming(\G) := \min\{\helW(\Rg)\mid \emptyset\neq\Rg\in\WNEIg(\G)\},
 \quad
 \helBming(\G) := \min\{\helB(\Rg)\mid \emptyset\neq\Rg\in\BNEIg(\G)\}.
\end{equation}

The following result is a variant of Hall's theorem.
 We will generalize it in \cref{lemma:OCP:Hakimi}.%
\begin{proposition}[{\cite[Theorem~1.3.1]{Lovasz_Plummer}}]\label{prop:DIM:helWBmin_geq_0=>apm_exists}
$\G$ admits an \APM if and only if $\helmin(\G)\geq0$.%
\end{proposition}

\begin{lemma}[\Mref{lemma:DIM:deleting_hel_verts}]\label{lemma:DIM:deleting_hel_verts}
Fix integers $\kw,\kb\geq0$. The following are equivalent:
\begin{enumerate}[label=(\arabic*)]
\item\label{DIM:del1} $\helWmin(\G)\geq\kw$ and $\helBmin(\G)\geq\kb$;
\item\label{DIM:del2} for any $\RW\subset\WV$ and $\RB\subset\BV$ with $|\RW|\leq \kw$ and $|\RB|\leq\kb$, $\Grem{(\RW\sqcup\RB)}$ 
 admits an \APM.
\end{enumerate}
\end{lemma}

\begin{corollary}[{\Mref{lemma:hel_geq_1}}]\label{lemma:hel_geq_1}
If $\helmin(\G)\geq1$ then every edge of $\G$ appears in an \APM of $\G$.
\end{corollary}

\begin{lemma}
\label{lemma:DIM:from_WNEI_to_WNEIg}
We have
\begin{equation}\label{eq:BACKGR:helWming_vs_helWmin}
 \helWming(\G)=\min(1,\helWmin(\G)) 
\quad\text{and}\quad
 \helBming(\G)=\min(1,\helBmin(\G)).
\end{equation}
\end{lemma}
\begin{proof}
Let $\emptyset\neq\Rg\in\WNEIg(\G)$. 
If $\Rg\in\WNEI(\G)$ then $\helW(\Rg)\geq\helWmin(\G)$ by~\eqref{eq:DIM:helWmin_dfn}. Otherwise, $\RgBV=\emptyset$ and $\RgWV\neq\emptyset$ so $\helW(\Rg)\geq1$. When $\Rg\in\WNEIg(\G)$ consists of a single white vertex, we get $\helW(\Rg)=1$. 
\end{proof}

\begin{lemma}
\label{lemma:DIM:no_floating}
Suppose that $\helWmin(\G)\geq1$ and $\helBmin(\G)\geq0$ (or vice versa). Then
 $\G$ \hasnofloat.
Furthermore, if $\helmin(\G)\geq1$ then for all $\v\in\Verts$, $\G\rem\{\v\}$ \hasnofloat. 
\end{lemma}
\begin{proof}
Suppose that $\Rg\subset\Vint$ is the set of vertices of a floating connected component of $\G$. Thus, $\Rg\in\WNEIg(\G)\cap\BNEIg(\G)$ with $\helW(\Rg)=-\helB(\Rg)$. By \cref{lemma:DIM:from_WNEI_to_WNEIg}, $\helW(\Rg)\geq1$ and $\helB(\Rg)\geq0$, a contradiction.

Similarly, suppose that $\helmin(\G)\geq1$ and let $\Rg\subset\Vint\rem\{\v\}$ be the set of vertices of a floating connected component of $\G\rem\{\v\}$ for some $\v\in\Verts$. Let $\Rg_+:=\Rg\sqcup\{\v\}$. Assume that $\v$ is, say, white. Then $\Rg\in\BNEIg(\G)$ and $\Rg_+\in\WNEIg(\G)$ with $\helW(\Rg_+)=1-\helB(\Rg)$. By \cref{lemma:DIM:from_WNEI_to_WNEIg}, $\helW(\Rg_+),\helB(\Rg)\geq1$, a contradiction.
\end{proof}

\begin{lemma}[\Mref{lemma:DIM:moves_vs_helWmin}]\label{lemma:DIM:moves_vs_helWmin}
Suppose that $\G_1,\G_2$ are related by moves \MVbd or \MV1. Then for all $\kw,\kb\in\{0,1\}$, 
\begin{equation*}%
 \helWmin(\G_1)\geq\kw,\ \helBmin(\G_1)\geq\kb \quad \Longleftrightarrow\quad 
 \helWmin(\G_2)\geq\kw,\ \helBmin(\G_2)\geq\kb.
\end{equation*}
\end{lemma}

We give a way to extend the functions $\helW(\Rg),\helB(\Rg)$ to not necessarily \wclosed and \bclosed subsets $\Rg$. This approach will become important later in \cref{ssec:MCMS:lggs} for \emph{\ggs}. 

\begin{notation}\label{not:BACKGR:RgEbd_RgEint}
Given a subset $\Rg\subset\Vint$ (resp., $\Rg\subset\Verts$), we let $\RgEbd\subset\Edges$ (resp., $\RgEint\subset\Edges$) be the set of edges of $\G$ incident to exactly one vertex (resp., two vertices) in $\Rg$. 
For $\Rg\subset\Vint$, denote $n(\Rg):=|\RgEbd|$ and $\RgE:=\RgEbd\sqcup\RgEint$. 
\end{notation}

\begin{definition}\label{dfn:gelWB_for_bip}
For $\w\in\WVint$, set $\gelW(\w):=1$ and $\gelB(\w):=\degG(\w)-1$. 
For $\b\in\BVint$, set $\gelB(\b):=1$ and $\gelW(\b):=\degG(\b)-1$. 
For $\Rg\subset\Vint$, set
\begin{align}
\label{eq:BACKGR:hel=sum_hel_minus_E}
 \gelW(\Rg)&:=\sum_{\v\in\Rg} \gelW(\v) - |\RgEint|, &
 \gelB(\Rg)&:=\sum_{\v\in\Rg}\gelB(\v) - |\RgEint|, \\
\label{eq:BACKGR:gelWmin_gelBmin_dfn}
 \gelWmin(\G)&:=\min\{\gelW(\Rg)\mid \emptyset\neq\Rg\subset\Vint\}, &
 \gelBmin(\G)&:=\min\{\gelB(\Rg)\mid \emptyset\neq\Rg\subset\Vint\}. 
\end{align}
\end{definition}

\begin{lemma}\label{lemma:BCFW:gelWmin_vs_helWmin}
We have
\begin{equation}\label{eq:BCFW:helWmin=min_over_gelB}
 \helWmin(\G)=\min\{\gelW(\Rg)\mid \Rg\subset\Vint:\ \RB\neq\emptyset\},
 \quad
 \helBmin(\G)=\min\{\gelB(\Rg)\mid \Rg\subset\Vint:\ \RW\neq\emptyset\}. 
\end{equation}
Furthermore, for all $\kw,\kb\in\{0,1\}$,
\begin{equation}\label{eq:BCFW:gelWmin_vs_helWmin}
 \gelWmin(\G)\geq\kw,\ \gelBmin(\G)\geq\kb
 \quad \Longleftrightarrow\quad 
 \helWmin(\G)\geq\kw,\ \helBmin(\G)\geq\kb.
\end{equation}
\end{lemma}
\begin{proof}
First, observe that by \cref{lemma:DIM:from_WNEI_to_WNEIg},~\eqref{eq:BCFW:gelWmin_vs_helWmin} follows from
\begin{equation}\label{eq:BCFW_gelWmin_vs_helWmin_two_sided}
 \helWming(\G)\leq \gelWmin(\G)\leq\helWmin(\G)
 \quad\text{and}\quad
 \helBming(\G)\leq \gelBmin(\G)\leq\helBmin(\G).
\end{equation}

We prove~\eqref{eq:BCFW:helWmin=min_over_gelB} and~\eqref{eq:BCFW_gelWmin_vs_helWmin_two_sided}. 
 Let $\Rg_1\subset\Vint$ be nonempty. Let $\Rg_2\in\WNEIg(\G)$ be obtained from $\Rg_1$ by adding all white vertices incident to some (black) vertex in $\Rg_1$. Let $\deltaw:=|\Rg_2|-|\Rg_1|$ be the number of such white vertices. We have
$\gelW(\Rg_1) = \sum_{\w\in\RW_1} \gelW(\w) + \sum_{\b\in\RB_1} \gelW(\b) - |\RgEintx_1| = |\RW_1|+(|\RgEintx_2| - |\RB_1|) - |\RgEintx_1|$ since $\gelW(\w)=1$, $\gelW(\b) = \degG(\b)-1$, and each edge in $\RgEintx_2$ is incident to exactly one vertex in $\RB_1=\RB_2$. Since $|\RW_1|=|\RW_2|-\deltaw$ and $|\RgEintx_2| - |\RgEintx_1|\geq \deltaw$, we get $\gelW(\Rg_1) \geq |\RW_2|-|\RB_2| = \helW(\Rg_2)$. Thus, $\gelWmin(\G)\geq\helWming(\G)$. Furthermore, if $\RB_1\neq\emptyset$ then $\Rg_2\in\WNEI(\G)$, so we have $\gelW(\Rg_1)\geq\helWmin(\G)$, which shows the $\helWmin(\G)\leq\min\{\gelW(\Rg)\mid \Rg\subset\Vint:\ \RB\neq\emptyset\}$ inequality in~\eqref{eq:BCFW:helWmin=min_over_gelB}.

Suppose now that $\Rg_2\in\WNEI(\G)$. Let $\Rg_1:=\Rg_2\setminus\WVbd$. 
We have $\emptyset\neq\RB_2=\RB_1\subset\BVint$, so $\emptyset\neq\Rg_1\subset\Vint$. Since each boundary vertex in $\G$ has degree $1$, we have $|\RgEintx_1| \geq |\RgEintx_2| - \deltaw$ for $\deltaw:=|\RW_2| - |\RW_1|$. Similarly to the above, we find
$\gelW(\Rg_1) = |\RW_1| +(|\RgEintx_2| - |\RB_1|) - |\RgEintx_1| \leq |\RW_2|-|\RB_2| = \helW(\Rg_2)$. Thus, $\helWmin(\G)\geq\gelWmin(\G)$, and since $\RB_1\neq\emptyset$, we also get $\helWmin(\G)\geq\min\{\gelW(\Rg)\mid \Rg\subset\Vint:\ \RB\neq\emptyset\}$.

We have shown the first equality in~\eqref{eq:BCFW:helWmin=min_over_gelB} and the two inequalities involving $\gelWmin(\G)$ in~\eqref{eq:BCFW_gelWmin_vs_helWmin_two_sided}. The remaining (in)equalities are obtained similarly by swapping the roles of white and black.
\end{proof}

\begin{definition}\label{dfn:BACKGR:holess}
 We say that $\Rg\subset\Verts$ is \emph{\holess} if $\Grem\Rg$ \hasnofloat. 
 For $\Rg\subset\Verts$, we let $\Rghol\subset\Vint\setminus\Rg$ be the set of vertices contained in the floating connected components of $\Grem\Rg$, 
 and we set $\Rgcl:=\Rg\sqcup\Rghol$. Thus, $\Rgcl$ is always \holess.
\end{definition}

\begin{remark}
By \cref{lemma:DIM:no_floating}, 
if $\helWmin(\G)\geq1$ and $\helBmin(\G)\geq0$ (or vice versa) then $\Rg=\emptyset$ is \holess, and 
if $\helmin(\G)\geq1$ then $\Rg=\{\v\}$ is \holess for all $\v\in\Verts$. 
\end{remark}

The following result is straightforward; see also \cref{lemma:BCFW:Rg_vs_Rg'_helW_vs_helB} below.
\begin{lemma}
Suppose that $\Rg\in\WNEIg(\G)$ is not \holess. Then $\Rghol\in\BNEIg(\G)$, $\Rgcl\in\WNEIg(\G)$, and 
\begin{equation}\label{eq:DIM:Rg_Rgcl_Rghol}
 \helW(\Rg) = \helW(\Rgcl) + \helB(\Rghol).
\end{equation}
\end{lemma}

\begin{definition}\label{dfn:BACKGR:sconn}
We say that $\emptyset\neq\Rg\subset\Verts$ is \emph{\sconn} if it is \holess and $\GR$ is connected.
\end{definition}

\begin{lemma}\label{lemma:BCFW:dualizable_holess_sconn}
Suppose that $\G$ \hasnofloat. 
Then $\emptyset\neq\Rg\subset\Vint$ is \sconn if and only if 
$\Rg$ is the set of faces of $\GD$ enclosed by some simple cycle
 $\RgCyc$ in $\GD$. 
\end{lemma}
\noindent Here and below, a \emph{simple} cycle is a cycle passing through each vertex at most once.
\begin{proof}
This follows from the standard \emph{cycle-bond duality} applied to the connected planar graph $\Ggen$ obtained from $\G$ by identifying all boundary vertices into a single vertex; cf. \cref{rmk:BCFW:gluing_bdv_into_bdvall}.
\end{proof}

\begin{lemma}\label{lemma:BCFW:Rg_simply_conn_lower_bound}
Assume that $\G$ is \bdconn. Fix integers $\kw,\kb\geq0$. 
\begin{enumerate}[label=(\arabic*)]
\item\label{sconn_lower1} If for all \sconn $\Rg\in\WNEI(\G)$ (resp., $\Rg\in\BNEI(\G)$), we have $\helW(\Rg)\geq\kw$ (resp., $\helB(\Rg)\geq\kb$) then $\helWmin(\G)\geq\kw$ and $\helBmin(\G)\geq\kb$.
\item\label{sconn_lower2} If for all \sconn $\emptyset\neq\Rg\subset\Vint$, we have $\gelW(\Rg)\geq\kw$ and $\gelB(\Rg)\geq\kb$ then $\gelWmin(\G)\geq\kw$ and $\gelBmin(\G)\geq\kb$.
\end{enumerate}
\end{lemma}
\begin{proof}
We show part~\itemref{sconn_lower1}. 
Let $\Rg\in\WNEI(\G)$ be \holess. 
If it is \sconn, we have $\helW(\Rg)\geq\kw$ by assumption. Otherwise, let $\G\ind[\Rg_1],\G\ind[\Rg_2],\dots,\G\ind[\Rg_d]$ be the connected components of $\GR$. 
Since $\G$ is \bdconn, each $\Rg_i$ is \sconn. Since $\helW(\Rg) =\sum_{i=1}^d\helW(\Rg_i)$, we get $\helW(\Rg)\geq\kw$. 
Similarly, we find $\helB(\Rg)\geq\kb$ for all \holess $\Rg\in\BNEI(\G)$. 
Next, we show $\helW(\Rg)\geq\kw$ for all $\Rg\in\WNEI(\G)$ and 
$\helB(\Rg)\geq\kb$ for all $\Rg\in\BNEI(\G)$ 
 by induction on $\nvint(\Rg)$.
Let $\Rg\in\WNEI(\G)$ and suppose that the result has been shown for all $\Rg'\in\WNEI(\G)\cup\BNEI(\G)$ with $|\nvintRgp|<\nvint(\Rg)$. If $\Rg$ is \holess, we are done. Otherwise, by~\eqref{eq:DIM:Rg_Rgcl_Rghol}, $\helW(\Rg) = \helW(\Rgcl) + \helB(\Rghol)$. Since $\Rgcl$ is \holess, $\helW(\Rgcl)\geq\kw$. 
By the induction hypothesis, $\helB(\Rghol)\geq\kb\geq0$. Thus, $\helW(\Rg)\geq \kw+\kb \geq \kw$. Similarly, we get $\helB(\Rg) = \helB(\Rgcl) + \helW(\Rghol)\geq \kb$ for $\Rg\in\BNEI(\G)$. This completes the proof of part~\itemref{sconn_lower1}. The proof of part~\itemref{sconn_lower2} is obtained analogously using~\eqref{eq:BCFW:Rg_vs_Rg'_helW_vs_helB} below instead of~\eqref{eq:DIM:Rg_Rgcl_Rghol}. 
\end{proof}

\begin{lemma}\label{lemma:TE:excision_datr}
If $\helmin(\G)\geq0$ and $\Mdatro(\G)\neq\emptyset$ then $\helmin(\G)\geq1$ and $\G$ is \bdconn. 
\end{lemma}
\begin{proof}
Suppose that $\Mdatro(\G)\neq\emptyset$.
Assume first that $\G$ is not \bdconn and let $\Rg$ be the vertex set of a floating connected component of $\G$. We have $\Rg\in\WNEI(\G)\cap\BNEI(\G)$ with $\helW(\Rg)=\helB(\Rg)=0$. 
By \cref{lemma:MCE:apm_vs_Kast}, $\Fw(\w)=0$ for all $\w\in\RW$ and $\Fb(\b)=0$ for all $\b\in\RB$ for any \datr $\datrQ\in\Mdatr(\G)$. By~\eqref{eq:OCP:primitive}, the points $\xd(\ff)$ coincide for all faces $\ff$ incident to some vertex in $\Rg$, a contradiction. 
Thus, $\G$ is \bdconn. Suppose now that we have $\helW(\Rg)=0$ (resp., $\helB(\Rg)=0$) for some \sconn $\Rg\in\WNEI(\G)$ (resp., $\Rg\in\BNEI(\G)$). 
Applying \cref{lemma:MCE:apm_vs_Kast} again, we get $\Fw(\w)=0$ for all $\w\in\RW$ (resp., $\Fb(\b)=0$ for all $\b\in\RB$) for any \datr $\datrQ\in\Mdatr(\G)$, a contradiction. 
By \cref{lemma:BCFW:Rg_simply_conn_lower_bound}, $\helmin(\G)\geq1$. 
\end{proof}

\subsection{\KawAngle condition} \label{ssec:Kawangle}

For $\datrQL=\datrQ\in\Mdatro(\G)$ and a corner $\cor\in\corners(\G)$, let
\begin{equation}\label{eq:TE:sumT_corner_dfn}
 \sumT(\corner):=\arg_{[0,2\pi)}\frac{\xT(\corfp) - \xT(\corf)}{\xT(\corfm) - \xT(\corf)}.%
\end{equation}
Since $\datrQL$ is \einj, the numerator and the denominator are both nonzero. By~\eqref{eq:OCP:primitive} and~\eqref{eq:TE:weak_t_imm_dfn}, 
\begin{equation}\label{eq:TE:sumT_corner_0_pi}
 \sumT(\corner)\in[0,\pi] \quad\text{for $\datrQL\in\Mdtio(\G)$ and $\corner\in\corners(\G)$.}
\end{equation}
In fact,~\eqref{eq:TE:weak_t_imm_dfn} is equivalent to~\eqref{eq:TE:sumT_corner_0_pi} when $\datrQL$ is \einj. 
Given $\datrQL\in\Mdatro(\G)$ and $\ff\in\Faces$, we set
\begin{equation}\label{eq:TE:sumbT_sumwT_sumT_dfn}
 \sumbT(\ff):=\sum_{\cor\in\cornersb(\ff)} \sumT(\corner), \quad
 \sumwT(\ff):=\sum_{\cor\in\cornersw(\ff)} \sumT(\corner), \quad\text{and}\quad
\sumT(\ff) := \sumbT(\ff) + \sumwT(\ff).
\end{equation}

\begin{lemma}[Algebraic \Kawangle condition]\label{lemma:TE:Kawangle_alg}
Let $\datrQL=\datrQ\in\Mdatro(\Gbip)$. Then 
\begin{alignat}{4}%
\label{eq:Kawangle_alg_int}
 \sumwT(\ff) &\equiv\pi, &\quad \sumbT(\ff) &\equiv \pi &\ &\pmod{2\pi} 
&\qquad&\text{for all $\ff\in\Fint$, and}\\
\label{eq:Kawangle_alg_bd}
 \sumbT(\bdf_i)&\equiv\sumbT_i, &\quad \sumwT(\bdf_i)&\equiv\sumwT_i &\ &\pmod{2\pi} 
&\qquad&\text{for all $i\in\brn$ when $\Gbip$ is connected,}
\end{alignat}
where $\sumbT_i,\sumwT_i$ were defined in~\eqref{eq:TE:dfn}.
\end{lemma}
\begin{proof}
See~\cite[Section~3.2]{KLRR}, \cite[Sections~2 and~3]{CLR1}, and~\Mref{lemma:angle_sum_arg_rat}. 
\end{proof}

We generalize~\eqref{eq:Kawangle_alg_bd} to the case when $\Gbip$ is not necessarily connected. 
By \cref{lemma:TE:excision_datr}, $\Gbip$ is \bdconn (but some of the boundary vertices of $\GD$ may coincide when $\Gbip$ is not connected). 
Similarly to~\eqref{eq:Kawangle_alg_int} (cf.~\Mref{lemma:angle_sum_arg_rat}), one can show that for $\ff\in\Fbd$, 
\begin{equation}\label{eq:Kawangle_alg_bd_gen}
 \sumbT(\ff) + \sum_{i\in\bdryarcs\ff} (\pi - \sumbT_i) \equiv \pi 
 \quad\text{and}\quad
 \sumwT(\ff) + \sum_{i\in\bdryarcs\ff} (\pi - \sumwT_i) \equiv \pi 
 \quad\pmod{2\pi}.
\end{equation}
Here, $(\pi - \sumbT_i)$ and $(\pi - \sumwT_i)$ are the black and white angles associated to the corner of the outer face of $\GD$ at $\bdf_i$. When $\Gbip$ is connected, $\bdryarcs\ff$ consists of a single element and~\eqref{eq:Kawangle_alg_bd_gen} specializes to~\eqref{eq:Kawangle_alg_bd}.

\begin{proposition}[\KawAngle condition]\label{lemma:TE:angle_cond}
Any \einj \wtimm $\datrQL\in\Mdtio(\G)$ satisfies the \emph{\Kawangle condition}%
\begin{equation}\label{eq:TE:angle_cond}
 \sumbT(\ff) = \sumwT(\ff) = \pi \quad\text{for all $\ff\in\Fint$}.
\end{equation}
Furthermore, it satisfies the \emph{boundary angle condition}
\begin{equation}\label{eq:TE:angle_cond_bdry}
 \sumbT(\ff) + \sum_{i\in\bdryarcs\ff} (\pi - \sumbT_i) = \pi 
 \quad\text{and}\quad
 \sumwT(\ff) + \sum_{i\in\bdryarcs\ff} (\pi - \sumwT_i) = \pi 
 \quad\text{for all $\ff\in\Fbd$}.
\end{equation}
\end{proposition}
\begin{proof}
Let $\datrQL=\datrQ\in\Mdtio(\G)$. Let $\v\in\Vint$ be an interior vertex of degree $m$.
Let $\cor_1,\cor_2,\dots,\cor_m$ be the corners of $\G$ incident to $\v$ in clockwise order, and set $\ff_i:=\corfx_i$ for $i\in\brm$. 
 Note that we have $m\geq1$ since $\G$ admits an \APM (cf. \cref{dfn:TE:tembending}). In fact, we must have $m\geq2$, for otherwise $\GD$ would have a loop edge which would prevent it from having any \einj \wtimms. Consider a closed polygonal chain $\xT(\v)=(\xT(\ff_1),\xT(\ff_2),\dots,\xT(\ff_m))$.
 We consider angles $\sumT(\cor_i)\in[0,\pi]$ introduced in~\eqref{eq:TE:sumT_corner_dfn}. 
We let $\dturn_\v\in\Z$ be given by 
\begin{equation}\label{eq:TE:turning_of_vertex}
 \sum_{i=1}^m \sumT(\cor_i) = (m - 2(\dturn_\v+1)) \pi.
\end{equation}
Since each $\sumT(\cor_i)$ belongs to $[0,\pi]$ and since the polygonal chain $\xT(\v)$ is closed, $\dturn_\v\geq0$.

Next, by~\eqref{eq:Kawangle_alg_int}, for $\ff\in\Fint$, there exist integers $\dturnb_{\ff},\dturnw_{\ff}\in\Z$ such that 
\begin{equation}\label{eq:TE:turning_sumT}
 \sumbT(\ff) = \pi + 2\pi\dturnb_{\ff} \quad\text{and}\quad
 \sumwT(\ff) = \pi + 2\pi\dturnw_{\ff}.
\end{equation}
Since $\sumbT(\ff),\sumwT(\ff)\geq0$, we have $\dturnb_{\ff},\dturnw_{\ff}\geq0$. 
Finally, for $\ff\in\Fbd$, we let $\dturnbbd_{\ff},\dturnwbd_{\ff}$ be such that 
\begin{equation}\label{eq:TE:turning_sumT_bdry}
 \sumbT(\ff) + \sum_{i\in\bdryarcs\ff} (\pi - \sumbT_i) = \pi + 2\pi\dturnbbd_{\ff}
 \quad\text{and}\quad
 \sumwT(\ff) + \sum_{i\in\bdryarcs\ff} (\pi - \sumwT_i) = \pi + 2\pi\dturnwbd_{\ff}.
\end{equation}
By~\eqref{eq:Kawangle_alg_bd_gen}, 
 we have $\dturnbbd_{\ff},\dturnwbd_{\ff}\in\Z$. Since $\sumbT(\ff),\sumwT(\ff)\geq0$ and $\sumbT_i,\sumwT_i\in(0,\pi)$ by~\itemref{TE:bdry_angle}, we find $\dturnbbd_{\ff},\dturnwbd_{\ff}\geq0$. 
Given a subset $\subF\subset\Faces$, we denote 
\begin{equation*}%
 \dtsumBV:=\sum_{\b\in\BVint} \dturn_{\b}, \quad
\dtsumWV:=\sum_{\w\in\WVint} \dturn_{\w}, \quad
\dtsumBsubF:=\sum_{\ff\in\subF} \dturnb_{\ff}, \quad\text{and}\quad
\dtsumWsubF:=\sum_{\ff\in\subF} \dturnw_{\ff}.
\end{equation*}

 We calculate the sum $\sumbT(\G):=\sum_{\cor\in\cornersb(\G)} \sumT(\cor)$ in two different ways. On the one hand, by~\eqref{eq:TE:turning_of_vertex}, 
\begin{equation*}%
 \frac1\pi\sumbT(\G) = \sum_{\b\in\BVint} (\degG(\b) - 2(\dturn_\b+1)) 
= |\Edges| - |\BVbd| - 2|\BVint| - 2\dtsumBV
= |\Edges| - |\BV| - |\BVint| - 2\dtsumBV.
\end{equation*} 
On the other hand, since $\G$ is \bdconn by \cref{lemma:TE:excision_datr}, we have 
$|\Faces| = |\Fint| + n+1-\nconn(\G)$. 
By~\eqref{eq:TE:turning_sumT}--\eqref{eq:TE:turning_sumT_bdry} and~\eqref{eq:TE:bdry_winding}, 
\begin{equation*}%
 \frac1\pi\sumbT(\G) 
= |\Fint|+ 2\dtsumBFint + \frac1\pi\sum_{\ff\in\Fbd}\sumbT(\ff) 
= |\Faces| + 2\dtsumBF -n + \frac1\pi\sum_{i=1}^n\sumbT_i
= |\Fint| + 2\dtsumBF + k-\nconn(\G).
\end{equation*}
Equating the right-hand sides, we get 
\begin{equation}\label{eq:TE:turning_sum1}
 |\Edges| - |\BV| - |\BVint| = |\Fint| + \kmone + 2\dtsumBF + 2\dtsumBV.
\end{equation}
 Swapping the roles of white and black, we similarly obtain 
\begin{equation}\label{eq:TE:turning_sum2}
|\Edges| - |\WV| - |\WVint| = |\Fint| +\nkmone + 2\dtsumWF + 2\dtsumWV. 
\end{equation}
Taking the sum of these two equations, %
 we find
\begin{equation*}%
 2|\Edges| - |\Verts| - |\Vint| = 2|\Fint| + (n-2\nconn(\G)) + 2(\dtsumBV+\dtsumWV+\dtsumBF+\dtsumWF).
\end{equation*}
Rearranging the terms and applying~\eqref{eq:DIM:Euler}, we get
\begin{equation}\label{eq:TE:dtsums=0}
 \dtsumBV+\dtsumWV+\dtsumBF+\dtsumWF = 0, \quad\text{and thus}\quad
\dtsumBV=\dtsumWV=\dtsumBF=\dtsumWF=0
\end{equation}
since $\dtsumBV,\dtsumWV,\dtsumBF,\dtsumWF\geq0$.
Substituting this into~\eqref{eq:TE:turning_sumT}--\eqref{eq:TE:turning_sumT_bdry}, we obtain~\eqref{eq:TE:angle_cond}--\eqref{eq:TE:angle_cond_bdry}.
\end{proof}

\begin{definition}\label{dfn:DIM:degenerate_convex_polygon}
A \emph{\wcp} is a closed polygonal chain $\PcurveT=(\xsT_1,\xsT_2,\dots,\xsT_m)$ in the plane (with $m\geq2$ vertices) such that $\PcurveT$ is \emph{\einj} meaning $\xsT_{i}\neq\xsT_{i-1}$ for all $i\in\brm$, and such that the \emph{boundary turning angles} 
 $\turnanglex_i(\PcurveT):=\arg_{[-\pi,\pi)}(\PmomT_{i+1}/\PmomT_i)$,
where $\PmomT_i:=\xsT_{i}-\xsT_{i-1}$, 
satisfy 
\begin{equation}\label{eq:DIM:turnangle_cond}
 \turnanglex_i(\PcurveT)\in[-\pi,0] \quad\text{for all $i\in\brm$ \quad and}\quad
 \turnangle(\PcurveT)=\sum_{i=1}^m \turnanglex_i(\PcurveT) = -2\pi.
\end{equation}
We say that $\PcurveT$ is a \emph{\ndcp} (resp., a \emph{\wndcp}) if it is a \wcp such that $\turnanglex_i(\PcurveT)\in(-\pi,0)$ (resp., $\turnanglex_i(\PcurveT)\in(-\pi,0]$) for all $i\in\brm$. 
\end{definition}

\begin{notation}\label{notn:TE:Rg_corners_E_WV_BV_sumTcond}
For $\Rg\subset\Vint$, we denote 
$\corners(\Rg):=\bigsqcup_{\v\in\Rg}\corners(\v)$. 
For $\f\in\Faces$, we set 
$\corners\cond(\f|\Rg):=\corners(\f)\cap\corners(\Rg)$, 
$\cornersb\cond(\f|\Rg):=\cornersb(\f)\cap\corners(\Rg)$, 
$\cornersw\cond(\f|\Rg):=\cornersw(\f)\cap\corners(\Rg)$, 
$\sumbTcond(\f|\Rg):=\sum_{\cor\in \cornersb\cond(\f|\Rg)} \sumT(\cor)$, and 
$\sumwTcond(\f|\Rg):=\sum_{\cor\in \cornersw\cond(\f|\Rg)} \sumT(\cor)$. We set $\sumTcond(\f|\Rg):=\sumbTcond(\f|\Rg) + \sumwTcond(\f|\Rg)$. 
\end{notation}

\begin{lemma}\label{lemma:TE:weakly_convex_helWBmin=1}
Let $\datrQL=\datrQ\in\Mdtio(\G)$ and let $\Rg\subset\Vint$ be \sconn. If $\Rg\in\WNEI(\G)$ (resp., $\Rg\in\BNEI(\G)$) satisfies $\helW(\Rg)=1$ (resp., $\helB(\Rg)=1$) then $\xT(\RgCyc)$ is a \wcp.
\end{lemma}
\begin{proof}
Intersecting $\G$ with the area enclosed by $\RgCyc$, we obtain a connected planar bipartite graph $\GRext$ of type $(k',n')$ with $n':=n(\Rg)$ boundary vertices of degree $1$ and $k'=\helW(\Rg)$ (resp., $n'-k'=\helB(\Rg)$) when $\Rg\in\WNEI(\G)$ (resp., $\Rg\in\BNEI(\G)$).\footnote{By convention, $\GRext$ contains a boundary vertex located in the middle of each edge $\e\in\RgEbd$ of color opposite to that of the endpoint of $\e$ contained in $\Rg$.} 
Let $\f_1,\f_2,\dots,\f_{n'}$ be the vertices of $\RgCyc$ listed in clockwise order. 
We apply the same argument as in the proof of \cref{lemma:TE:angle_cond} to the graph $\GRext$, omitting the terms $\dtsumBV,\dtsumWV,\dtsumBF,\dtsumWF$ in view of~\eqref{eq:TE:dtsums=0}. Summing up the black and white angles in two different ways similarly to~\eqref{eq:TE:turning_sum1}--\eqref{eq:TE:turning_sum2} and applying Euler's formula~\eqref{eq:DIM:Euler} to $\GRext$ with $\nconn(\GRext)=1$, %
 we find that for \sconn $\Rg\in\WNEI(\G)\cup\BNEI(\G)$, 
\begin{equation}\label{eq:DIM:reg_sum_cond}
 \sum_{i=1}^{n'} \sumbTcond(\f_i|\Rg) = (k'-1)\pi\quad\text{and}\quad
 \sum_{i=1}^{n'} \sumwTcond(\f_i|\Rg) = (n'-k'-1)\pi.
\end{equation}
It follows that if, say, $\Rg\in\WNEI(\G)$ and $\helW(\Rg)=1$ then $\sum_{i=1}^{n'} \sumbTcond(\f_i|\Rg) = 0$,
 and $\sum_{i=1}^{n'} \sumwTcond(\f_i|\Rg) = (n' - 2)\pi$,
 so we get $\sumbTcond(\f_i|\Rg)=0$ for each $i\in\brx{n'}$. 
 Letting $\PcurveT:=\xT(\RgCyc)$, we see that the boundary turning angles of $\PcurveT$ are given by $\turnanglex_i(\PcurveT) = \sumTcond(\f_i|\Rg) - \pi=\sumwTcond(\f_i|\Rg) - \pi$, which belongs to $[-\pi,0]$ since $\sumwTcond(\f_i|\Rg)\in[0,\pi]$ by~\eqref{eq:TE:angle_cond}--\eqref{eq:TE:angle_cond_bdry}. Furthermore, $\sum_{i=1}^{n'} \sumwTcond(\f_i|\Rg) = (n' - 2)\pi$ implies that $\turnangle(\PcurveT)=-2\pi$. 
\end{proof}

\subsection{\OACTITLE}

Observe that the group $\GGsh$ (cf. \cref{dfn:BACKGR:GGsh}) naturally acts on the space $\Mdti(\G,\wt)$ of \wtimms of a weighted graph $(\G,\wt)$, where the subgroup $\Rdd$ acts by translations.
Recall also that the elements of $\Mdti(\G,\wt)$ are viewed up to $\RtpVint\times\{\pm1\}^{\Vint}$-action on $\datrQnox$.

\begin{theorem}[\OACTITLE]\label{thm:TE:OAC}
Assume that $\G$ admits an \APM. 
 Let $\wt\in\Rtpgauge$ and $C:=\Meas(\G,\wt)$. %
Then we have homeomorphisms
\begin{align}
 \Mdti(\G,\wt)/\Rdd &\xrasim \{\lalat\in\lalakMAT\mid \la\subset C \subset\latp\} \quad\text{and}\\
 \Mdti(\G,\wt)/\GGsh &\xrasim \{\lalat\in\lalak\mid \la\subset C \subset\latp\}.
\end{align}
\end{theorem}

\begin{proof}
 We start by describing the correspondence. 
Let $\datrQL=\datrQ\in\Mdti(\G,\wt)$. 
Instead of working with $(\Fw,\Fb)\in\HHspaceC$, we work with $(\laext,\latext)\in\HHspaceRd$ related to $(\Fw,\Fb)$ by
\begin{equation}\label{eq:TE:y_to_lalat}
 \laext(\w)=\CtoM[\Fw(\w)] \quad\text{and}\quad \latext(\b)=\CtoMt[\Fb(\b)] \quad\text{for all $\w\in\WV$ and $\b\in\BV$}.
\end{equation}
For $\w_1,\w_2\in\WV$ and $\b_1,\b_2\in\BV$, we set
\begin{equation}\label{eq:brlaw_dfn}
 \brlaw<\w_1,\w_2>:=\det\mat[\laext(\w_1)|\laext(\w_2)]\quad\text{and}\quad \brlatb[\b_1,\b_2]:=\det\mat[\latext(\b_1)|\latext(\b_2)].
\end{equation}

In view of~\eqref{eq:MCE:alt(C)_vs_pFw}--\eqref{eq:MCE:alt(Cp)_vs_pFb}, we consider matrices $\la,\lat\in\Mator_{2,n}$ with columns given by
\begin{equation}\label{eq:TE:lalat_vs_pFw_pFb}
 \la_i := (-1)^{i-1}\partial\laext_i
 \quad\text{and}\quad
 \lat_i := (-1)^{i-1}\partial\latext_i
 \quad\text{for $i\in\brn$}
\end{equation}
in the notation of~\eqref{eq:partial_F_dfn}; thus,
 $\la\subset C \subset\latp$. 
 By~\eqref{eq:TE:weak_t_imm_bdry}, we have $\brla<i,i+1>,\brlat[i,i+1]>0$. Summing up both sides of~\eqref{eq:TE:t_imm_bdry_vs_pFw_pFb} for $i\in\brn$, we get $\lat\cdot \la^T = 0$. As explained in 
\Mref{eq:sumw_sumb_pi_k}, 
\eqref{eq:TE:bdry_winding} is equivalent to $\wind(\la) = (k-1)\pi$ and $\wind(\lat) = (k+1)\pi$. Thus, $\lalat\in\lalakMAT$. 

Conversely, given $\lalat\in\lalakMAT$ satisfying $\la\subset C\subset\latp$, 
we let $\epsK$ be any choice of Kasteleyn signs for $\G$ and let $(\laext,\latext)\in\HHspaceRd$ be the discrete holomorphic extensions (\cref{dfn:OCP:restr_ext}) of $\lalat$. 
Let $(\Fw,\Fb)\in\HHspaceC$ be related to $(\laext,\latext)$ via~\eqref{eq:TE:y_to_lalat}. 
 This defines $(\epsK,\Fw,\Fb)$ up to sign gauge equivalence (cf. \cref{rmk:DIM:Kast_gauge_eq}). The \KSprim $\xd$ of $(\Fw,\Fb)$ is defined by~\eqref{eq:OCP:primitive} up to translation. This gives rise to \adatr $\datrQLll=\datrQ\in\Mdatr(\G,\wt)$. 
As explained above, conditions~\itemref{TE:bdry_angle} and~\itemref{TE:bdry_winding} are equivalent to $\brla<i,i+1>,\brlat[i,i+1]>0$ and $\wind(\la) = (k-1)\pi$, $\wind(\lat) = (k+1)\pi$. Condition~\itemref{TE:t_imm} will follow from~\eqref{eq:TE:same_face_and_same_vertex} below.
\end{proof}

We review the results of \justpapone necessary to finish the proof of \cref{thm:TE:OAC}. 

\begin{proposition}[\Mref{prop:from_la_to_La}]\label{prop:from_la_to_La}
Recall the notation $\lak,\latk$ from~\eqref{eq:lak_latk}--\eqref{eq:lak_latk2}.
\begin{itemize}
\item For each $\la\in\lak$, there exists $\La\in\alt(\Grtp(n-k+2,n))$ such that $\la\subset\La$.
\item For each $\lat\in\latk$, there exists $\Lat\in\Grtp(k+2,n)$ such that $\lat\subset\Lat$.
\end{itemize}
\end{proposition}

\begin{definition}[Tripod insertion]\label{dfn:eksbb}
Suppose that $\w_1,\w_2\in\WV$ share a face of $\G$. 
Let $\G'$ be obtained from $\G$ by adding a trivalent black vertex $\b$ adjacent to $\w_1$, $\w_2$, and a white leaf $\w_3$ by edges $\e_1,\e_2,\e_3$ in clockwise order. 
 Let $\epsK'$ be an extension of $\epsK$ to a choice of Kasteleyn signs on $\G'$. 
(Such an extension $\epsK'$ always exists; cf. \cref{lemma:OCP:Kast_even}.)
We set $\epsww_{\w_1,\w_2}:=\epsK'(\e_1)\epsK'(\e_2)$. Similarly, if $\b_1,\b_2\in\BV$ share a face of $\G$, we let $\G'$ be obtained by adding a trivalent white vertex $\w$ adjacent to $\b_1$, $\b_2$, and a black leaf $\b_3$ by edges $\e_1,\e_2,\e_3$ in clockwise order. We set $\epsbb_{\b_1,\b_2}:=\epsK'(\e_1)\epsK'(\e_2)$. 
\end{definition}

\begin{corollary}[\Mref{cor:lak_implies_black_imm}]\label{lemma:TE:brla_nonzero_if_apm}
Assume that $\G$ admits an \APM, and let $\wt\in\Rtpgauge$ and $C:=\Meas(\G,\wt)$. Consider $2$-planes $\lalat\in\lalats$ satisfying $\la\subset C\subset\latp$. Let $(\laext,\latext)\in\HHspaceRd$ be the discrete holomorphic extensions of $\lalat$ to the vertices of $\G$.
\begin{enumerate}[label=(\arabic*)]
\item\label{lak_implies1} If $\la\in\lak$ then for all distinct $\w_1,\w_2\in\WV$ sharing a face of $\G$, we have $\epsww_{\w_1,\w_2}\brlaw<\w_1,\w_2> < 0$ if $\Giww$ admits an \APM and $\brlaw<\w_1,\w_2>=0$ otherwise.
\item\label{lak_implies2} If $\lat\in\latk$ then for all distinct $\b_1,\b_2\in\BV$ sharing a face of $\G$, we have $\epsbb_{\b_1,\b_2}\brlatb[\b_1,\b_2]>0$ if $\Gbb$~admits an \APM and $\brlatb[\b_1,\b_2]=0$ otherwise.
\item\label{lak_implies3} If $\la\in\lak$ (resp., $\lat\in\latk$) and $\b\in\BVint$ (resp., $\w\in\WVint$) is connected to $\w_1,\dots,\w_d$ (resp., $\b_1,\dots,\b_d$) 
by edges $\e_1,\dots,\e_d$ in clockwise order then %
\begin{equation}\label{eq:TE:same_face_and_same_vertex}
 \epsK(\e_{\s})\epsK(\e_{\s+1})\brlaw<\w_{\s},\w_{\s+1}>\leq0,\quad\text{resp.,}\quad
 \epsK(\e_{\s})\epsK(\e_{\s+1})\brlatb[\b_{\s},\b_{\s+1}]\geq0
 \quad\text{for all $\s\in\brd$}.
\end{equation} 
The inequality is strict if and only if the vertices $\w_\s,\w_{\s+1}$ (resp., $\b_\s,\b_{\s+1}$) are distinct and the graph $\G\rem\{\w_\s,\w_{\s+1}\}$ (resp., $\G\rem\{\b_\s,\b_{\s+1}\}$) admits an \APM. 
\item\label{lak_implies4} Suppose that $\helmin(\G)\geq1$ and the edge weights $\wt\in\Rtpgauge$ are generic. If $\la\in\lak$ (resp., $\lat\in\latk$) then $\laext(\w)\neq0$ for all $\w\in\WV$ (resp., $\latext(\b)\neq0$ for all $\b\in\BV$). 
\end{enumerate}
\end{corollary}

The following result extends trivially from t-immersions studied in~\justpapone to \wtimms.
\begin{lemma}[\Mref{lemma:OAC:nondeg}]\label{lemma:OAC:nondeg}
 Assume that $\G$ admits an \APM. 
 If $\G$ is not \twonondeg (\cref{dfn:BACKGR:nondeg}) then it admits no (weak) t-immersions. 
If $C\in\Grtnn(k,n)\setminus\Grnd(k,n)$ is not \twonondeg then the set of $\lalat\in\lalak$ satisfying $\la\subset C\subset\latp$ is empty.
\end{lemma}

\begin{corollary}[Existence of \wtimms]\label{thm:TE:existence}
Assume that $\G$ is \twonondeg and admits an \APM. Let $\wt\in\Rtpgauge$. 
 Then $(\G,\wt)$ admits \wtimms.
Furthermore, if $\helmin(\G)\geq1$ and $\wt\in\Rtpgauge$ is generic then $(\G,\wt)$ admits \einj \wtimms.
\end{corollary}
\begin{proof}
 Let $C:=\Meas(\G,\wt)$, $\LaLat\in\LaLak$, and $\lalat:=\PhiLL(C)$. 
By \cref{lemma:OAC:PhiLL_in_lalak,thm:TE:OAC}, $\datrQLll\in\Mdti(\G,\wt)$ is a \wtimm. 
When $\helmin(\G)\geq1$ and $\wt\in\Rtpgauge$ is generic, 
$\datrQLll$ is \einj by \cref{rmk:TE:einj_nonvanishing} and 
 \crefi{lemma:TE:brla_nonzero_if_apm}{lak_implies4}.
\end{proof}

\begin{remark}\label{rmk:TE:hel>=1_necessary}
By \cref{lemma:TE:excision_datr}, the condition $\helmin(\G)\geq1$ is necessary in order for $\G$ to admit \einj \wtimms. By \cref{ex:non-einj}, the condition that $\wt$ is generic is also necessary. 
\end{remark}

\section{Weak t-embeddings are weak embeddings}

Our next goal is to show that one can obtain \einj \wtimms and \wtembs as limits of immersions and embeddings; see \cref{thm:TOP:weak_imm=>imm}. Throughout, we assume that $\G$ 
is \bdconn, \twonondeg, and admits an \APM. 

Since $\G$ is \bdconn, its planar dual $\GD$ gives rise to a $2$-dimensional cell complex denoted by $\SuppGD$. We denote its $1$-skeleton by $\SkelGD$.
\begin{definition}
A piecewise-linear map $\xT:\SuppGD\to\C$ is called a \emph{\PLimm} if 
every point $p\in\SuppGD$ has an open neighborhood $U\subset\SuppGD$ such that $\xT:U\to\xT(U)$ is an orientation-preserving homeomorphism.
A \PLimm $\xT$ is a \emph{\PLemb} if it is injective on $\SuppGD$.
\end{definition}

We extend any map $\xT:\Faces\to\C$ linearly to each edge, obtaining a map $\xT:\SkelGD\to\C$. We equip the space of piecewise-linear maps $\SkelGD\to\C$ with the uniform topology. 
\begin{definition}\label{dfn:TOP:wtimm}
A map $\xT:\Faces\to\C$ is called a \emph{\wimm} (resp., \emph{\wemb}) if the extension $\xT:\SkelGD\to\C$ may be obtained as an $\eps\to0$ limit in the uniform topology of (restrictions to $\SkelGD$ of) \PLimms (resp., \PLembs) $\Tcur:\SuppGD\to\C$ depending continuously on $\eps$.
\end{definition}

The main objective of this section is to show the following result.

\begin{theorem}\ \label{thm:TOP:weak_imm=>imm}
Assume that $\helmin(\G)\geq1$. 
\begin{enumerate}[label=(\arabic*)]
\item\label{TOP:weak_imm1} For any \einj \wtimm $\datrQL=\datrQ\in\Mdtio(\G)$, $\xT$ is a \wimm.
\item\label{TOP:weak_imm2} For any \einj \wtemb $\datrQL=\datrQ\in\Mdteo(\G)$, $\xT$ is a \wemb.
\end{enumerate}
\end{theorem}

\subsection{Collapsible subsets}\label{ssec:DIM:coll}
The following construction 
is similar in spirit to the \emph{brick} and \emph{brace decompositions} introduced in~\cite{Lovasz_1987}; see also~\cite[Section~4.3]{Kenyon_Sheffield}.

\begin{lemma}
For any $\Rg_1,\Rg_2\in\WNEIg(\G)$, we have 
$\Rg_1\cap\Rg_2,\Rg_1\cup\Rg_2\in\WNEIg(\G)$ and 
\begin{equation}\label{eq:DIM:submodular}
 \helW(\Rg_1\cap\Rg_2) + \helW(\Rg_1\cup\Rg_2) \leq \helW(\Rg_1) + \helW(\Rg_2).
\end{equation}
\end{lemma}
\begin{proof}
It is clear that $\Rg_1\cap\Rg_2$ and $\Rg_1\cup\Rg_2$ are both \wclosed. For~\eqref{eq:DIM:submodular}, see~\cite[Equation~(1.3.4)]{Lovasz_Plummer}.
\end{proof}

\begin{definition}
Suppose that $\helWmin(\G)\geq1$. We say that $\Rg\in\WNEI(\G)$ is \emph{\wdash collapsible} if $\helW(\Rg) = 1$. 
\end{definition}

\begin{lemma}\label{lemma:DIM:collapsible_union}
Suppose that $\helWmin(\G)\geq1$. Let $\Rg_1,\Rg_2\in\WNEI(\G)$ be two \wdash collapsible subsets such that $\Rg_1\cap\Rg_2\neq\emptyset$. 
Then $\Rg_1\cup\Rg_2\in\WNEI(\G)$ is also \wdash collapsible.
\end{lemma}
\begin{proof}
We have $\emptyset\neq\Rg_1\cap\Rg_2\in\WNEIg(\G)$, so by \cref{lemma:DIM:from_WNEI_to_WNEIg}, $\helW(\Rg_1\cap\Rg_2)\geq1$. 
 Since $\helW(\Rg_1)=\helW(\Rg_2)=1$, we see from~\eqref{eq:DIM:submodular} that $\helW(\Rg_1\cup\Rg_2)\leq 1$. Since $\Rg_1\cup\Rg_2\in\WNEI(\G)$, we must have $\helW(\Rg_1\cup\Rg_2)\geq\helWmin(\G)\geq1$. Thus, $\helW(\Rg_1\cup\Rg_2)=1$.
\end{proof}

We will be interested in \emph{maximal} (by inclusion) \wdash collapsible subsets.

\begin{lemma}\label{lemma:DIM:coll_sconn}
Suppose that $\helWmin(\G)\geq1$ and $\helBmin(\G)\geq0$. Let $\Rg\in\WNEI(\G)$ be a maximal \wdash collapsible subset. Then $\Rg$ is \sconn.
\end{lemma}
\begin{proof}
Let $\Rg_1,\Rg_2,\dots,\Rg_d\in\WNEIg(\G)$ denote the vertex sets of connected components of $\GR$. 
 By \cref{lemma:DIM:from_WNEI_to_WNEIg}, $\helW(\Rg_i)\geq1$ for all $i\in\brd$. Since $1=\helW(\Rg) = \sum_{i=1}^d \helW(\Rg_i) \geq d$, we get $d = 1$, so $\GR$ is connected. 
If $\Rg$ is not \holess then we have $\helW(\Rg) = 1$, $\helW(\Rgcl)\geq1$, and $\helB(\Rghol)\geq0$ by \cref{lemma:DIM:from_WNEI_to_WNEIg}. By~\eqref{eq:DIM:Rg_Rgcl_Rghol}, we must have $\helW(\Rgcl)=1$, contradicting the maximality of $\Rg$.
\end{proof}

\begin{figure}
 \def\inputscale{1.8}
 \setlength{\tabcolsep}{1pt}
\begin{tabular}{ccccc}
 \includegraphics[scale=\inputscale]{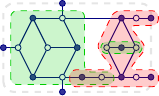}
&
\includegraphics[scale=\inputscale]{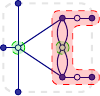}
&
\includegraphics[scale=\inputscale]{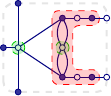}
&
\includegraphics[scale=\inputscale]{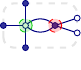}
&
\includegraphics[scale=\inputscale]{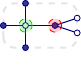}
\\
 (a) $\G$
& (b) $\CollGW$
& (c) $\CollGWun$
& (d) $\CollGWBpre$
& (e) $\CollGWB$
\end{tabular}
 \caption{\label{fig:collapse} Collapsing a graph $\G$; see \cref{dfn:DIM:collapsed_graph,dfn:DIM:CollGBW}. Here, the maximal \wdash collapsible (resp., \bdash collapsible) subsets of $\G$ are circled in green (resp., red).}
\end{figure}

\begin{definition}\label{dfn:DIM:collapsed_graph}
Suppose that $\helWmin(\G)\geq1$, $\helBmin(\G)\geq0$, and that $\G$ has black boundary (cf. \cref{dfn:DIM:MVbd}). Let $\Rg_1,\Rg_2,\dots,\Rg_d\in\WNEI(\G)$ be the maximal \wdash collapsible subsets in $\G$. The \emph{\wdash collapsed graph} $\CollGW$ is obtained from $\G$ by identifying all vertices in $\Rg_i$ into a single white vertex $\w_i$, for each $i\in\brd$.
\end{definition}
\noindent See \cref{fig:collapse}. By \cref{lemma:DIM:coll_sconn}, the graph $\CollGW$ is planar and is obtained from $\G$ by contracting all edges in $\RgEintx_i$ for each $i\in\brd$. By \cref{lemma:DIM:no_floating}, both $\G$ and $\CollGW$ \havenofloat. 

\begin{remark}\label{rmk:DIM:coll_w}
Since each $\Rg_i$ belongs to $\WNEI(\G)$ and since $\G$ has black boundary, $\Rg_i$ does not contain any boundary vertices.
\end{remark}
\begin{lemma}\label{lemma:DIM:CollGW_geq_2}
Suppose that $\helWmin(\G)\geq1$, $\helBmin(\G)\geq0$, and that $\G$ has black boundary. 
 Then $\helWmin(\CollGW)\geq2$ and $\helBmin(\CollGW)\geq\helBmin(\G)$.
\end{lemma}
\begin{proof}
Let $\Rg_1,\Rg_2,\dots,\Rg_d$ and $\w_1,\w_2,\dots,\w_d$ be as in \cref{dfn:DIM:collapsed_graph}. 
For every $\CollRW\in\WNEI(\CollGW)\cup\BNEI(\CollGW)$, let $\Rg_+$ be obtained from $\CollRW$ by replacing each $\w_i\in\CollRW$ with $\Rg_i$. Observe that since $\CollRW\in\WNEI(\CollGW)$ (resp., $\CollRB\in\BNEI(\CollGW)$), we have $\Rg_+\in\WNEI(\G)$ (resp., $\Rg_+\in\BNEI(\G)$). 
Since each $\Rg_i$ satisfies $|\RW_i|-|\RB_i|=1$, we get $\helWsub_{\CollGW}(\CollRW) = \helW(\Rg_+)$ (resp., $\helBsub_{\CollGW}(\CollRB) = \helB(\Rg_+)$). Thus, $\helWmin(\CollGW)\geq\helWmin(\G)\geq1$ and $\helBmin(\CollGW)\geq\helBmin(\G)\geq0$.

To show that we actually have $\helWmin(\CollGW)\geq2$, suppose otherwise that $\CollRW\in\WNEI(\CollGW)$ is such that $\helWsub_{\CollGW}(\CollRW) = \helW(\Rg_+) = 1$. Then $\Rg_+$ must have been contained in a maximal \wdash collapsible subset. On the other hand, since $\CollRW\in\WNEI(\CollGW)$, $\CollRW$ contains a black vertex of $\CollGW$, a contradiction. Thus, $\helWmin(\CollGW)\geq2$. 
\end{proof}

Suppose that $\helmin(\G)\geq1$.
 By \cref{lemma:DIM:CollGW_geq_2}, $\helWmin(\CollGW)\geq2$ and $\helBmin(\CollGW)\geq1$. One can similarly define \bdash collapsible subsets of $\CollGW$ and show that they are pairwise disjoint and \sconn; cf. \cref{lemma:DIM:collapsible_union,lemma:DIM:coll_sconn}. However, unlike in \cref{rmk:DIM:coll_w}, a maximal \bdash collapsible subset of $\CollGW$ may contain one or several (black) boundary vertices. 
To that end, we slightly modify \cref{dfn:DIM:collapsed_graph}. 

\begin{definition}[Fully collapsed graph]\label{dfn:DIM:CollGBW}
Suppose that $\helmin(\G)\geq1$ and that $\G$ has black boundary. 
Let $\Rg'_1,\Rg'_2,\dots,\Rg'_{d'}\in\BNEI(\CollGW)$ be the maximal \bdash collapsible subsets in $\CollGW$. Apply the move \MVbd to each boundary vertex $\bdv_j$ of $\CollGW$ that belongs to some $\Rg'_i$ and denote the resulting graph by $\CollGWun$. 
Let $\CollGWBpre$ be obtained from $\CollGWun$ by replacing each subset $\Rg'_i$ with a single black vertex $\b'_i$. Finally, let $\CollGWB$ be obtained from $\CollGWBpre$ by applying moves \RV1 until no parallel edges are present.
See \cref{fig:collapse}. %
\end{definition}

\begin{lemma}\label{lemma:DIM:CollGBW_geq_2}
Suppose that $\helmin(\G)\geq1$ and that each connected component of $\G$ is incident to at least three boundary vertices. 
Then $\helmin(\CollGWBpre)=\helmin(\CollGWB)\geq2$.
\end{lemma}
\begin{proof}
It is clear that $\helmin(\CollGWBpre) = \helmin(\CollGWB)$. %
Since $\CollGWun$ is obtained from $\CollGW$ by applying \MVbd to some black boundary vertices, the sets $\BNEI(\CollGW)$ and $\BNEI(\CollGWun)$ are in bijection and we have $\helBmin(\CollGWun)=\helBmin(\CollGW)\geq1$ by \cref{lemma:DIM:CollGW_geq_2}. Similarly to \cref{lemma:DIM:CollGW_geq_2}, we find $\helBmin(\CollGWB)\geq2$. 

By \cref{lemma:DIM:CollGW_geq_2}, $\helWmin(\CollGW)\geq2$; however, we may have $\helWmin(\CollGWun)=1$; cf. \cref{lemma:DIM:moves_vs_helWmin}. Let $J\subset\brn$ be the set of $j\in\brn$ such that 
the move \MVbd was applied to $\bdv_j$ when transforming $\CollGW$ into $\CollGWun$. 
 For $j\in J$, let $\bdvx_j$ denote the corresponding (black, interior, degree-$2$) vertex of $\CollGWun$. We denote the (white, degree-$1$) boundary vertex of $\CollGWun$ adjacent to $\bdvx_j$ by $\bdv_j$. 

Suppose for contradiction that $\CollRWB\in\WNEI(\CollGWB)$ satisfies $\helWsub_{\CollGWB}(\CollRWB)=1$.
Let $\CollRW_+\in\WNEI(\CollGWun)$ be obtained by replacing each $\b'_i\in\CollRWB$ with $\Rg'_i$ in the notation of \cref{dfn:DIM:CollGBW}. 
We have $\helWsub_{\CollGWB}(\CollRWB)=\helWsub_{\CollGWun}(\CollRW_+)\geq1$. 
Let $\CollRW_-$ be obtained from $\CollRW_+$ by removing all (black, degree-$2$) vertices $\{\bdvx_j\in\CollRW_+\mid j\in J\}$ together with their (white, boundary) neighbors $\bdv_j$. 
Thus, $\CollRW_-\in\WNEIg(\CollGW)$ satisfies $\helWsub_{\CollGW}(\CollRW_-)=\helWsub_{\CollGWun}(\CollRW_+)\geq1$ and is therefore nonempty. If $\CollRW_-\in\WNEI(\CollGW)$ then by \cref{lemma:DIM:CollGW_geq_2}, $\helWsub_{\CollGW}(\CollRW_-)\geq2$ and we are done. Suppose otherwise that $\CollRWxB_-=\emptyset$ and $\helWsub_{\CollGW}(\CollRW_-)=|\CollRWxW_-|=1$. 
Let $\wv$ be the sole vertex in $\CollRWxW_-$. 
 Since $\CollRWxB_-=\emptyset$, we have $\CollRWxB_+\subset\{\bdvx_j\mid j\in J\}$. Since each connected component of $\G$ is incident to at least three boundary vertices, each (degree-$2$) vertex in $\CollRWxB_+$ is adjacent to $\wv$. 
Since $\CollRW_+\in\WNEI(\CollGWun)$, $\CollRWxB_+\neq\emptyset$. By the definition of $J$, we must have $\{\wv\}\sqcup\CollRWxB_+\subset\Rg'_i$ for some $i$. By construction, we must have $\Rg'_i\subset\CollRW_+$ so $\Rg'_i=\{\wv\}\sqcup\CollRWxB_+$. 
Since $\Rg'_i$ is \bdash collapsible, we get $\helBsub_{\CollGWun}(\Rg'_i) = 1=|\CollRWxB_+|-1$, so $|\CollRWxB_+|=2$. 
Since $\Rg'_i$ is \bdash closed, $\CollRWxB_+=\Neigh_{\CollGWun}(\w)$. 
Thus, $\CollRW_+$ is the vertex set of a connected component of $\CollGWun$ that contains 
 at most two boundary vertices, contradicting the assumption of the lemma.
\end{proof}

\begin{remark}
One can similarly define a fully collapsed graph $\CollGBWgarb$ by swapping the roles of white and black colors in 
\cref{dfn:DIM:collapsed_graph,dfn:DIM:CollGBW}. 
 One can show that the two graphs $\CollGWB\cong\CollGBWgarb$ are isomorphic (however, we need not have $\CollGWBpre\cong\CollGBWpregarb$). 
 We will only use the graph $\CollGWB$ in what follows.
\end{remark}

\subsection{Collapsing \wtimms}
In the following definition, for an \einj map $\xT:\Faces\to\C$, we define the \emph{exterior angles} $\sumText_i\in(0,2\pi]$ similarly to~\eqref{eq:TE:sumT_corner_dfn} by 
$\sumText_i:=\arg_{(0,2\pi]}\frac{\xT(\bdf_{i-1}) - \xT(\bdf_i)}{\xT(\bdf_{i+1}) - \xT(\bdf_i)}$.

\begin{definition}\label{dfn:TOP:STRimm}
A map $\xT:\Faces\to\C$ is called a \emph{\STRimm} if %
\begin{enumerate}[label=(\arabic*)]
\item\label{TOP:STRimm1} $\xT$ is \emph{\einj}: $\xT(\ff)\neq\xT(\f)$ for all $\{\ff,\f\}\in\Ebarast$;
\item\label{TOP:STRimm2} for each $\v\in\Vint$ with $\degG(\v)\geq3$, $\xT(\v)$ is a \wndcp (\cref{dfn:DIM:degenerate_convex_polygon});
\item\label{TOP:STRimm3} $\sumT(\ff)=2\pi$ for all $\ff\in\Fint$ and 
$\sumT(\ff) + \sum_{i\in\bdryarcs\ff} \sumText_i = 2\pi$ for all $\ff\in\Fbd$; cf.~\eqref{eq:TE:sumbT_sumwT_sumT_dfn}--\eqref{eq:Kawangle_alg_bd_gen}.
\end{enumerate}
A \STRimm $\xT:\Faces\to\C$ is called a \emph{\STRemb} if $\PbdT_{\xd}$ is a simple polygon of turning number $-2\pi$; cf.~\eqref{eq:sum_arg_pmom_-2pi}.
\end{definition}
\noindent It is straightforward to see using the argument principle that every \STRimm (resp., \STRemb) $\xT:\Faces\to\C$ extends to a \PLimm (resp., \PLemb) $\xT:\SuppGD\to\C$.
In the terminology of~\justpapone, a \emph{t-immersion} (resp., \emph{t-embedding}) is a \wtimm (resp., \wtemb) that is simultaneously a \STRimm (resp., embedding); cf. \cref{lemma:TE:angle_cond}.

\begin{proposition}\label{lemma:TOP:weak_imm_collapsed}
Assume that $\helmin(\G)\geq2$.
\begin{enumerate}[label=(\arabic*)]
\item\label{TOP:coll1} Every \wtimm of $\G$ is \einj: $\Mdti(\G)=\Mdtio(\G)$.
\item\label{TOP:coll2} If $\datrQL=\datrQ\in\Mdti(\G)$ then $\xT$ is a \STRimm, i.e., a t-immersion.
\item\label{TOP:coll3} If $\datrQL=\datrQ\in\Mdte(\G)$ then $\xT$ is a \STRemb, i.e., a t-embedding.
\end{enumerate}
\end{proposition}
\begin{proof}
Let $\datrQL=\datrQ\in\Mdti(\G)$. By \cref{thm:TE:OAC}, we have $2$-planes $\lalat\in\lalak$ whose discrete holomorphic extensions $(\laext,\latext)\in\HHspaceRd$ are related to $(\Fw,\Fb)$ by~\eqref{eq:TE:y_to_lalat}. 

Let $\w\in\WVint$. Since $\helBmin(\G)\geq2$, we must have $\degG(\w)\geq|\NeighG(\w)|\geq3$. We claim that $\xT(\w)$ is a \wndcp. Let $\NeighG(\w)=\{\b_1,\b_2,\dots,\b_d\}$, listed in clockwise order. Note that $\G$ may contain parallel edges, so let $\e_{\s+}$ (resp., $\e_{\s-}$) be the last (resp., the first) edge of $\G$ in clockwise order connecting $\w$ to $\b_\s$. 
 Since $\helBmin(\G)\geq2$, for $\s\in\brd$, $\G\rem\{\b_\s,\b_{\s+1}\}$ admits an \APM by \cref{lemma:DIM:deleting_hel_verts}. By 
 \crefi{lemma:TE:brla_nonzero_if_apm}{lak_implies3},
 we get $\epsK(\e_{\s+})\epsK(\e_{(\s+1)-})\brlatb[\b_\s,\b_{\s+1}]>0$.
 In particular, $\Fb(\b_\s)\neq0$ for all $\s\in\brd$. Thus, $\Fb(\b)\neq0$ for all $\b\in\BV$, and by a similar argument, $\Fw(\w)\neq0$ for $\w\in\WV$. By \cref{rmk:TE:einj_nonvanishing}, $\xT$ is \einj. 
 Letting $\cor_\s$ be the corner of $\GD$ incident to $\w$ located between $\east_{\s+}$ and $\east_{(\s+1)-}$, we get $0<\sumT(\cor_\s)<\pi$. For all other corners $\cor\in\corners(\w)$ (located between parallel edges emanating from $\w$), we have $\sumT(\cor)=\pi$. Thus, $\xT(\w)$ is a \wndcp in the sense of \cref{dfn:DIM:degenerate_convex_polygon}. \crefi{dfn:TOP:STRimm}{TOP:STRimm3} holds by \cref{lemma:TE:angle_cond}, with $\sumText_i = 2\pi-\sumwT_i-\sumbT_i\in(0,2\pi)$ for each $i\in\brn$. Therefore, $\xT$ is a \STRimm. 
 Furthermore, if $\datrQL\in\Mdte(\G)$ then by \cref{dfn:TOP:STRimm}, $\xT$ is a \STRemb.
\end{proof}

\Cref{lemma:TOP:weak_imm_collapsed} implies \cref{thm:TOP:weak_imm=>imm} in the case $\helmin(\G)\geq2$. We will deduce the more general result in the case $\helmin(\G)\geq1$ by applying \cref{lemma:TOP:weak_imm_collapsed} to the fully collapsed graph $\CollGWB$ (which satisfies $\helmin(\CollGWB)\geq2$ by \cref{lemma:DIM:CollGBW_geq_2}). 
Observe that the graph $\CollGWB$ is obtained from $\G$ by contracting edges, removing loop edges, uncontracting some boundary edges, and identifying parallel edges. Thus, we may naturally view the set of faces of $\CollGWB$ as a subset of the set of faces of $\G$. 
\begin{lemma}\label{lemma:TOP:restr}
Assume that $\helmin(\G)\geq1$.
Let $\datrQL\in\Mdti(\G)$. 
Then $\datrQL$ restricts to 
a \wtimm $\CollWdatrQL\in\Mdti(\CollGW)$.
Furthermore, if each connected component of $\G$ is incident to at least three boundary vertices 
then $\datrQL$ restricts to 
an (\einj) t-immersion $\ColldatrQL\in\Mdtio(\CollGWB)$. 
\end{lemma}
\begin{proof}
Assume that $\G$ has black boundary. 
Let $\Rg_1,\Rg_2,\dots,\Rg_d$ and $\w_1,\w_2,\dots,\w_d$ be as in \cref{dfn:DIM:collapsed_graph}. Set $(\G_0,\wt_0,\epsK_0,\Fw_0,\Fb_0,\xd_0):=(\G,\wt,\epsK,\Fw,\Fb,\xd)$. For each $\s=1,2,\dots,d$, let $\G_\s$ be obtained from $\G_{\s-1}$ by replacing all vertices in $\Rg_\s$ with a single white vertex $\w_\s$. Let $\xd_\s:=\xd|_{\Faces_\s}$ be the restriction of $\xd$ to the faces of $\G_{\s}$. 
Similarly to the proof of \cref{lemma:TE:weakly_convex_helWBmin=1}, consider the graph $\GRextx_\s$ obtained by intersecting $\G$ with the disk enclosed by $\RgCycx_\s$ (where $\Rg_\s$ is \sconn by \cref{lemma:DIM:coll_sconn}). Thus, $\GRextx_\s$ has black boundary, satisfies $\helW(\GRextx_\s)=1$, and is therefore of type $(1,\ns)$ for $\ns:=n(\Rg_\s)$. 

 Since $\helmin(\G)\geq1$, $\GRextx_\s$ admits an \APM. Let $\wt\sinds$ and $\epsK\sinds$ be the restrictions of $\wtsm$ and $\epsKsm$ to the edges of $\GRextx_\s$. After acting by $\{\pm1\}^{\Vint}$, we may assume that $\epsK\sinds$ is a choice of Kasteleyn signs on $\GRextx_\s$. Denote the boundary edges of $\GRextx_\s$ by $\bdesx_1,\dots,\bdesx_{\ns}$. We view these edges also as edges of $\G_\s$ incident to the vertex $\wv_\s$. 

Let $\Xs:=\Meas(\GRextx_\s,\wt\sinds)\in\Grtnn(1,\ns)$. We claim that $\Xs\in\Grtp(1,\ns)$. Otherwise, the $i$-th entry $\Xsi$ of $\Xs=(\Xsx_1:\Xsx_2:\cdots:\Xsx_{\ns})$ is zero for some $i\in\brx{\ns}$. Thus, no \APM of $\GRextx_\s$ uses $\bdesi$. Since every \APM of $\G$ restricts to an \APM of $\GRextx_\s$, we see that $\bdesi$ does not appear in any \APM of $\G$. This contradicts \cref{lemma:hel_geq_1}.

We set $\epsK_\s(\bdesi):=(-1)^{i}$, $\wt_\s(\bdesi):=\Xsx_i$, and $\wtK_\s(\bdesi):=(-1)^i\Xsx_i$ for each $i\in\brx{\ns}$. 
By \cref{lemma:MCE:apm_vs_Kast}, $\dim\Hwspace_{\R}(\GRextx_\s,\wtK\sinds)=1$. 
Thus, there exists $z_\s\in\C$ such that $\Fwsm|_{\RW_\s}=z_\s\Xsext$ is a $z_\s$-multiple of the discrete holomorphic extension $\Xsext$ of $\Xs$ to $\RW_\s$. We set $\Fw_\s(\wv_\s):=z_\s$. 
For all edges $\e$ of $\G_{\s}$ not incident to $\wv_\s$, we set $\wtK_\s(\e):=\wtK_{\s-1}(\e)$, $\wt_\s(\e):=\wt_{\s-1}(\e)$, and $\epsK_{\s}(\e):=\epsK_{\s-1}(\e)$. 
 We set $\Fw_\s(\w):=\Fw_{\s-1}(\w)$ for all $\w\in\WV_\s\setminus\{\w_\s\}$ and $\Fb_\s(\b):=\Fb_{\s-1}(\b)$ for all $\b\in\BV_\s$. 
 Recall also that $\xd_\s:=\xd|_{\Faces_\s}$. 

We claim that the above defined \quintuple $\datrQL_\s:=(\wt_\s,\epsK_\s,\Fw_\s,\Fb_\s,\xd_\s)$ belongs to $\Mdti(\G_\s)$. First, we check that $\epsK_\s$ is a choice of Kasteleyn signs for $\G_\s$. Let $\ff_\s$ be a face of $\G_\s$. We only need to check~\eqref{eq:Kast_sign} in the case when the corresponding face $\ff_{\s-1}$ of $\G_{\s-1}$ contains a boundary face $\bdfsi$ of $\GRextx_\s$. 
Since $\Rg_\s$ is \sconn, $\nbdryarcs\bdfsi = 1$. 
By construction, $\dwcor(\ff_{\s-1}) = \dwcor(\ff_\s) + \dwcor(\bdfsi) - 1$. Since we set $\epsK_\s(\bdesi):=(-1)^{i}$ for all $i\in\brx{\ns}$, we have $\epsK_{\s}(\ff_\s) = \eps_i \epsK_{\s-1}(\ff_{\s-1}) \epsKsm(\bdfsi)$, where $\eps_i=-1$ if $i\neq\ns$ and $\eps_i=(-1)^{1+\ns}$ if $i=\ns$. 
Since $\epsK\sinds$ is a choice of Kasteleyn signs for $\GRextx_\s$, $\epsKsm(\bdfsi)$ is given by~\eqref{eq:Kast_sign} with $\epstra(\bdfsi) = 1$ if $i\neq \ns$ and $\epstra(\bdfsi) = 1+\ns$ if $i=\ns$. It follows that $\epsK_\s$ is indeed a choice of Kasteleyn signs for $\G_\s$. 

Since $\Fbsm|_{\RB_\s}$ is \bdash holomorphic on $\GRextx_\s$, its boundary restriction belongs to $\Xs^\perp$. Thus, $\Fb_\s$ satisfies~\eqref{eq:OCP:holom} at $\w_s$ and is therefore \bdash holomorphic on $\G_\s$. 
Let $\bdvx_{\s,i}$ be the (white) next-to-boundary endpoint of $\bdesi$. Recall that $\Fwsm|_{\RW_\s}=z_\s\Xsext$. By~\eqref{eq:partial_F_dfn} and~\eqref{eq:MCE:alt(C)_vs_pFw}, 
$\wtKsm(\bdesi) \Fwsm(\bdvx_{\s,i}) = -(z_\s\partial\Xsext)_i=(-1)^iz_\s\Xsi = \wtK_\s(\bdesi)\Fw_\s(\w_\s)$. Thus, $\Fw_\s$ is \wdash holomorphic on $\G_\s$ and $\xd_\s$ is the \KSprim of $(\Fw_\s,\Fb_\s)$. 
We have shown that $\datrQL_\s\in\Mdatr(\G_\s)$. Since the boundary restrictions of $\Fw_\s$ and $\Fb_\s$ coincide with those of $\Fw$ and $\Fb$, we see that $\datrQL_\s\in\Mdti(\G_\s)$ by \cref{thm:TE:OAC}. 
In particular, $\CollWdatrQL:=\datrQL_d\in\Mdti(\CollGW)$ since $\G_d=\CollGW$. 

The moves \MVbd and \RV1 preserve the class of \wtimms. 
We repeat the above steps (following the construction in \cref{dfn:DIM:CollGBW}) for all maximal \bdash collapsible subsets of $\CollGWun$, obtaining a \wtimm $\ColldatrQLpre\in\Mdti(\CollGWBpre)$. 
Applying a sequence of moves \RV1 yields the desired \wtimm $\ColldatrQL\in\Mdti(\CollGWB)$. It is an \einj t-immersion by \cref{lemma:DIM:CollGBW_geq_2,lemma:TOP:weak_imm_collapsed}.
\end{proof}

\begin{proposition}[{\Mref{lemma:Mbd=>face_Mnn}}]\label{lemma:Mbd=>face_Mnn}
Assume that $\helmin(\G)\geq2$. 
Suppose that $\datrQL=\datrQ$ is a t-embedding of $\G$ 
such that $\Pbdx$ is \Mdash positive. 
 Then for all $\ff,\f\in\Faces$, we have $(\xd(\ff)-\xd(\f))^2>0$ if $\ff,\f$ do not share a face of $\GD$ and $(\xd(\ff)-\xd(\f))^2=0$ otherwise.
\end{proposition}

\begin{corollary}\label{lemma:TE:hasMbd=>simple_and_MCE3}
Assume that $\helmin(\G)\geq1$. 
 Suppose that $\datrQL=\datrQ\in\Mdti(\Gbip)$ is a \wtimm such that $\Pbdx$ is \Mdash positive. Then $\datrQL$ is a \wtemb and moreover, $\xd$ is \emph{\Mdash nonnegative}, i.e., $(\xd(\ff)-\xd(\f))^2\geq0$ for all $\ff,\f\in\Faces$.
\end{corollary}
\begin{proof}
By \cref{lemma:TOP:Mpos=>simple}, $\PbdxT$ is simple, so $\G$ is connected and $\datrQL$ is a \wtemb. 
By \cref{lemma:DIM:CollGBW_geq_2}, $\helmin(\CollGWB)\geq2$. 
Applying \cref{lemma:Mbd=>face_Mnn} to the t-embedding $\ColldatrQL=\ColldatrQ$ of $\CollGWB$ from \cref{lemma:TOP:restr}, for any two faces $\ff,\f$ of $\CollGWB$, we get $(\xd(\ff)-\xd(\f))^2\geq0$ (with equality if and only if $\ff,\f$ share a face of $\CollGWBd$). More generally, if $y,z\in\Rdd$ are contained in the convex hulls $\Conv\Collxd(\CollpartFWB\vv),\Conv\Collxd(\CollpartFWB\v)$ of faces $\vv,\v$ of $\CollGWBd$ then $\afflin_y(z):=(y-z)^2$ is affine linear as a function of $z$ or of $y$; see \cref{lemma:MCE:clique_affine_linear} below. Since $\afflin_y(z)\geq0$ when both $y$ and $z$ are vertices of the respective convex hulls, it follows by multilinearity that $\afflin_y(z)\geq0$ for all $y\in\Conv\Collxd(\CollpartFWB\vv)$ and $z\in\Conv\Collxd(\CollpartFWB\v)$. Thus, $(\xd(\ff)-\xd(\f))^2\geq0$ for all $\ff,\f\in\Faces$, where we take $\vv$ and $\v$ to be the faces of $\CollGWBd$ containing $\ff$ and $\f$, respectively. 
\end{proof}

\subsection{Proof of Theorem~\ref{thm:TOP:weak_imm=>imm}}\label{ssec:TOP:proof_of_weak_imm=>imm}

Let $\datrQL=\datrQ\in\Mdtio(\G)$ and let $\ColldatrQLpre=\ColldatrQ\in\Mdtio(\CollGWBpre)$ be obtained via the collapsing procedure described in the proof of \cref{lemma:TOP:restr}. %
By \cref{lemma:DIM:CollGBW_geq_2} and \cref{lemma:TOP:weak_imm_collapsed}, $\CollxT$ is a \STRimm, and in particular, every interior face $\Collv\in\CollVint$ of $\CollGWBpred$ maps to a \wndcp $\CollxT(\Collv)$.\footnote{
Since $\helmin(\G)\geq1$, by \cref{lemma:DIM:no_floating}, 
 each connected component $\G_i$ of $\G$ is incident to $n(\G_i)\geq2$ boundary vertices. However, \cref{lemma:DIM:CollGBW_geq_2} only applies when $n(\G_i)\geq3$. Suppose that $n(\G_i)=2$ and 
let $\bdb_j$ be a boundary vertex of $\G_i$ connected to a next-to-boundary vertex $\bdwx_j$. 
Then we can replace $\bdb_j$ with two black boundary vertices $\bdbp_j,\bdbpp_j$ connected to $\bdwx_j$ and extend $\Fb$ to $\bdbp_j,\bdbpp_j$ so that the result is still a \wtimm. 
 The \wimm of the resulting graph produced by our proof can be easily modified into a \wimm of the original graph by ``cutting off'' a white boundary triangle incident to $\bdbp_j,\bdbpp_j$.
}

Each boundary vertex $\bdf_i$ of $\GD$ is also a boundary vertex of $\CollGWBpred$. Each interior vertex $\ff\in\Fint$ of $\GD$ is either an interior vertex of $\CollGWBpred$ or is contained inside a unique face $\Collv\in\CollVint$ of $\CollGWBpred$. 
We denote by $\Ffix\subset\Faces$ (resp., $\Eastfix\subset\East$) the set of vertices (resp., edges) of $\GD$ that are also vertices (resp., edges) of $\CollGWBpred$. 
For $\Collv\in\CollVint$, we let $\Fcolint[\Collv]\subset\Faces\setminus\Ffix$ be the set of vertices $\ff\in\Faces\setminus\Ffix$ such that $\ff$ is contained inside $\Collv$, and let $\Fcol[\Collv]:=\Fcolint[\Collv]\sqcup\Fcolbd[\Collv]$, where $\Fcolbd[\Collv]\subset\Ffix$ is the set of vertices of $\CollGWBpred$ incident to $\Collv$. We consider the induced subgraph $\GD\ind[\Collv]:=\GD\ind[{\Fcol[\Collv]}]$. 

Our goal is to approximate $\xT$ by a family $\Tcur:\SuppGD\to\C$ of \PLimms. By definition, the restriction of $\Tcur$ to $\CollGWBpred$ coincides with $\CollxT$ for all $\eps>0$. This includes the vertices in $\Ffix$ and the edges connecting them. Each of the remaining edges, vertices, and faces of $\GD$ is contained inside $\Collv$ for a unique $\Collv\in\CollVint$. Thus, it suffices to construct, independently for each $\Collv\in\CollVint$, the restriction of $\Tcur$ to the subcomplex $|\GD\ind[\Collv]|$ of $\SuppGD$ 
consisting of the vertices, edges, and faces of $\GD\ind[\Collv]$. 
 Since $\CollxT(\Collv)$ is a \wndcp, we will use the machinery of \emph{convex combination mappings} developed in~\cite{Floater}.

\begin{figure}
 \def\inputscale{2.5}
 \setlength{\tabcolsep}{0.2pt}
\begin{tabular}{ccc}
 \includegraphics[scale=\inputscale]{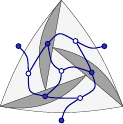}
&
 \includegraphics[scale=\inputscale]{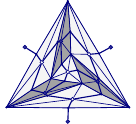}
&
 \includegraphics[scale=\inputscale]{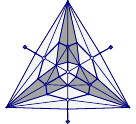}
\\
(a) $\xteh(\GD)$ for $\eps\to0$ & (b) $\xteh(\Gbary)$ for $\eps=0.3$ & (c) $\xteh(\Gbary)$ for $\eps=1$
\end{tabular}
 \caption{\label{fig:bary} 
Embeddings of the barycentric subdivision $\Gbary$ of $\SuppGD$ constructed via convex combination mappings of~\cite{Floater} in the proof of \cref{thm:TOP:weak_imm=>imm}. 
}
\end{figure}

Let $\Gbary$ be the barycentric subdivision of $\SuppGD$. Thus, the vertex set of $\Gbary$ is $\Faces\sqcup\East\sqcup\Vint$,\footnote{In \cref{fig:bary}, we additionally added the boundary vertices of $\G$ to the vertex set of $\Gbary$.} and the edges of $\Gbary$ are $\{\ff,\east\}$ for $\east\in\East$ and $\ff\in\ebarast$, $\{\ff,\v\}$ for $\v\in\Vint$ and $\ff\in\partF\v$, and $\{\east,\v\}$ for $\v\in\Vint$ and $\east\in\partEast\v$. All faces of $\Gbary$ are triangular and correspond to \emph{flags} $(\ff,\east,\v)\in\Faces\times\East\times\Vint$ such that $\east$ is incident to both $\ff$ and $\v$. See \cref{fig:bary}. 
We extend the map $\xT:\Faces\to\C$ to the vertices of $\Gbary$ by linearity. That is, for $\east\in\East$ with $\ebarast=\{\ff,\f\}$, we set $\xT(\east):=\frac12(\xT(\ff)+\xT(\f))$, and for $\v\in\Vint$, we let $\xT(\v)$ be the average of $\xT(\ff)$ over all $\ff\in\partF\v$.

We introduce two orientations $\Gfloat,\Gfloatt$ of $\Gbary$. Each vertex $\vbary$ of $\Gbary$ that belongs to $\Ffix\sqcup\Eastfix$ is a sink in both $\Gfloat$ and $\Gfloatt$, and we set $\Nout(\vbary)=\Noutt(\vbary):=\emptyset$. For every other vertex $\vbary\in(\Faces\setminus\Ffix)\sqcup(\East\setminus\Eastfix)\sqcup\Vint$, let $\Noutt(\vbary):=\NeighGbary(\vbary)$,
 and let 
 $\Nout(\vbary)\subset\NeighGbary(\vbary)$ be such that 
 $\Conv\{\xT(\ubary)\mid\ubary\in\Nout(\vbary)\}$ is the face of $\Conv\{\xT(\ubary)\mid\ubary\in\NeighGbary(\vbary)\}$ that contains $\xT(\vbary)$ in its relative interior. By definition, the digraph $\Gfloat$ (resp., $\Gfloatt$) contains arrows from $\vbary$ to each $\ubary\in\Nout(\vbary)$ (resp., $\ubary\in\Noutt(\vbary)$). 

 For each arrow $\vbary\to\ubary$ of $\Gfloatt$ that is not an arrow of $\Gfloat$,
 set $\wtfloat(\vbary\to\ubary):=0$. For each arrow $\vbary\to\ubary$ of $\Gfloat$, choose a positive real number $\wtfloat(\vbary\to\ubary)>0$ such that for each non-sink vertex $\vbary$ of $\Gfloatt$, %
\begin{equation}\label{eq:TE:Floater}
\sum_{\ubary\in\Noutt(\vbary)} \wtfloat(\vbary\to\ubary) = 1
 \quad\text{and}\quad
 \xT(\vbary) = \sum_{\ubary\in\Noutt(\vbary)} \wtfloat(\vbary\to\ubary) \xT(\ubary).
\end{equation}
 (It is possible to choose such positive coefficients since by construction, 
$\xT(\vbary)$ belongs to the relative interior of $\Conv\{\xT(\ubary)\mid\ubary\in\Nout(\vbary)\}$.) 
Following~\cite{Floater}, we call a map $\xT$ satisfying~\eqref{eq:TE:Floater} for each non-sink vertex $\vbary$ of $\Gfloatt$ a \emph{weak convex combination mapping}. If another map $\xT'$ satisfies
\begin{equation}\label{eq:TE:Floater'}
 \sum_{\ubary\in\Noutt(\vbary)} \wtfloatt(\vbary\to\ubary) = 1 \quad\text{and}\quad
 \xT'(\vbary) = \sum_{\ubary\in\Noutt(\vbary)} \wtfloatt(\vbary\to\ubary) \xT'(\ubary)
\end{equation}
with a different collection of coefficients $\wtfloatt$ that are positive for \emph{every} arrow of $\Gfloatt$ 
 then the map $\xT'$ is called a \emph{convex combination mapping}. We view~\eqref{eq:TE:Floater} (resp.,~\eqref{eq:TE:Floater'}) as a linear system of equations in the variables $\xT(\vbary)$ (resp., $\xT'(\vbary)$) for all non-sink vertices $\vbary$ of $\Gfloatt$. Recall that for each sink vertex $\vbary\in\Ffix\sqcup\Eastfix$, the value of $\xT(\vbary)=\CollxT(\vbary)$ has been fixed.

 Fix $\Collv\in\CollVint$ and let $\Gbarycol$ be the induced subgraph of $\Gbary$
on the vertices, edges, and faces of $\GD\ind[\Collv]$. 
Since $\CollxT(\Collv)$ is a \wndcp, by~\cite[Theorem~4.1 and Corollary~6.2]{Floater}, restricting a convex combination mapping $\xT'$ to $\Gbarycol$ 
 yields an embedding of $\Gbarycol$. Furthermore, as explained in~\cite[Section~3]{Floater}, the system~\eqref{eq:TE:Floater'} has a unique solution by the \emph{discrete maximum principle}: the real (resp., imaginary) part of every solution $\xT'$ to~\eqref{eq:TE:Floater'} must achieve its maximum on some vertex $\vbary\in\Ffix\sqcup\Eastfix$ on the boundary of $\Collv$. 

We claim that the discrete maximum principle still holds for every solution $\xT$ of~\eqref{eq:TE:Floater}. Indeed, 
suppose that the maximum of, say, $\Re(\xT)$ is achieved at some non-sink vertex $\vbary\notin\Ffix\sqcup\Eastfix$. Pick a generic vector $\normal\in\R^2$. 
Since $\datrQL$ is \einj, by~\eqref{eq:TE:sumT_corner_0_pi}, there exists an arrow $\vbary\to\ubary$ in $\Gfloat$ such that $\<\xT(\ubary),\normal\> > \<\xT(\vbary),\normal\>$. By construction, $\wtfloat(\vbary\to\ubary)>0$. Since $\Re(\xT(\vbary))$ is maximal, we must have $\Re(\xT(\ubary))=\Re(\xT(\vbary))$ by~\eqref{eq:TE:Floater}. Set $\ubary_0:=\vbary$ and $\ubary_1:=\ubary$. Continuing in this fashion, we construct a directed path $\ubary_0\to\ubary_1\to\cdots\to\ubary_d$ in $\Gfloat$ such that $\ubary_d$ is a sink. By construction (since all vertices and edges of $\CollGWBpred$ are sinks), the entire directed path must stay inside $\Gbarycol$ for some $\Collv\in\CollVint$, and the sink $\ubary_d$ is located on the boundary of $\Collv$. Thus, the restriction of $\Re(\xT)$ to each convex polygon $\CollxT(\Collv)$ achieves its maximum on the boundary of $\CollxT(\Collv)$ (and similarly for $\Im(\xT)$), and therefore $\xT$ satisfies the discrete maximum principle. 

We conclude that $\xT$ is the unique solution to~\eqref{eq:TE:Floater} satisfying the given boundary conditions on the sink vertices of $\Gfloat$. For each $\eps>0$, pick a collection of positive weights $\wtfloatt_{\eps}$ on the arrows of $\Gfloatt$ converging to $\wtfloat$ as $\eps\to0$, and let $\Tcur$ be the (unique) solution to~\eqref{eq:TE:Floater'} with weights $\wtfloatt_{\eps}$ and boundary conditions $\Tcur(\vbary)=\xT(\vbary)$ for each sink $\vbary$. Since $\xT$ is the unique solution to~\eqref{eq:TE:Floater}, it follows that $\Tcur$ converges to $\xT$ as $\eps\to0$ in the uniform topology. As explained above, each $\Tcur$ restricts to an embedding of $\Gbarycol$ for each $\Collv\in\CollVint$, and therefore $\Tcur$ is \aPLimm of $\GD$. Thus, $\xT$ is a \wimm of $\GD$. This shows part~\itemref{TOP:weak_imm1} of \cref{thm:TOP:weak_imm=>imm}. Part~\itemref{TOP:weak_imm2} follows from part~\itemref{TOP:weak_imm1}: if $\datrQL\in\Mdte(\G)$ then by \cref{dfn:TOP:wtimm}, $\Tcur$ is \aPLemb of $\GD$ for all small $\eps>0$. 
\qed

\section{T-duality for planar bipartite graphs}\label{sec:shift}
The goal of this section is to introduce 
 a local transformation of (not necessarily reduced) weighted planar bipartite graphs called \emph{T-duality}. 

\subsection{Combinatorial shift by $1$}\label{ssec:SHIFT:by1}
Our goal is to extend the construction of the \emph{combinatorial shift by $1$ map} originally introduced in~\cite[Lemma~4.2]{Zono} (see also~\cite{BaWe,HSP,PSBW,CLSBW}) to the case where $\G$ is not necessarily reduced. Our exposition follows that of~\cite{CLSBW}. 

\begin{definition}[{\cite[Definition~7.14]{HSP}}] We say that a planar bipartite graph $\G$ is 
\emph{\bdash trivalent} if all interior black vertices of $\G$ have degree $3$ and all boundary vertices of $\G$ are black (of degree $1$). Similarly, $\G$ is \emph{\wdash trivalent} if all interior white vertices of $\G$ have degree $3$ and all boundary vertices of $\G$ are white (of degree $1$).
\end{definition}

\begin{figure}
 \def\inputscale{2}
 \setlength{\tabcolsep}{10pt}
\begin{tabular}{cc}
 \includegraphics[scale=\inputscale]{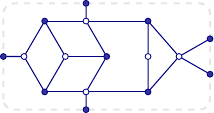}
&
 \includegraphics[scale=\inputscale]{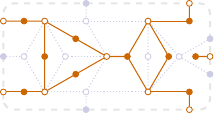}
\\
(a) \textcolor{\plabicEdgeColor}{$\G$} & (b) \textcolor{\plabicEdgeColordG}{$\dG$}
\end{tabular}
 \caption{\label{fig:shift-1} A graph $\G$ (left) and its shift by $1$ (right); see \cref{dfn:shift1}.}
\end{figure}

\begin{definition}[Combinatorial shift by $1$]\label{dfn:shift1}
Assume that $\G$ is \bdash trivalent and \bdconn. 
The \emph{shift by $1$} of $\G$ is a \wdash trivalent graph $\dG$ defined as follows.
For each face $\f\in\Faces$ of $\G$, $\dG$ contains a black vertex $\dbv(\f)$. For each trivalent black vertex $\bv\in\BVint$ incident to faces $\f_1,\f_2,\f_3$ (some of which may coincide), $\dG$ contains a trivalent white vertex $\dwv(\bv)$ adjacent to $\dbv(\f_1),\dbv(\f_2),\dbv(\f_3)$. In addition, for each $i\in\brn$, $\dG$ contains a degree-$1$ white boundary vertex $\dbdv_i$ located between $\bdv_i$ and $\bdv_{i+1}$, incident to $\dbv(\bdf_i)$.
See \cref{fig:shift-1}.
\end{definition}

The following topological result is straightforward.
\begin{lemma}\label{lemma:shift1-top}
Assume that $\G$ is a \bdash trivalent \bdconn planar bipartite graph. 
\begin{enumerate}[label=(\arabic*)]
\item\label{shift1-top1} $\dG$ is a \wdash trivalent \bdconn planar bipartite graph. 
\item\label{shift1-top2} The faces of $\G$ (resp., $\dG$) are in bijection with $\dBV=\dBVint$ (resp., $\WV=\WVint$).
\item\label{shift1-top3} $\G$ (resp., $\dG$) is connected if and only if $\dG$ (resp., $\G$) contains $n$ distinct next-to-boundary vertices.
\item\label{shift1-top4} $\G$ is obtained from $\dG$ by applying the shift in \cref{dfn:shift1} with the roles of black and white swapped.
\end{enumerate}
\end{lemma}
\noindent For $\w\in\WV$, we denote by $\dface(\w)\in\dFaces$ the corresponding face of $\dG$.

Recall from \cref{lemma:DIM:no_floating} that when $\helWmin(\G)\geq1$ and $\helBmin(\G)\geq0$, $\G$ is \bdconn.

\begin{lemma}\label{lemma:SHIFT:by1}
Assume that $\G$ is \bdash trivalent of type $(k,n)$ and satisfies $\helWmin(\G)\geq1$ and $\helBmin(\G)\geq0$.
Then $\dG$ is of type $(k-1,n)$ and satisfies
\begin{equation}\label{eq:SHIFT:by1_helW_helB_ineqs}
 \helWmin(\dG)\geq0 \quad\text{and}\quad
 \helBmin(\dG)\geq\min(\helBmin(\G)+1,2).
\end{equation}
Moreover, when $\G$ contains $n$ distinct next-to-boundary vertices (with $n\geq2$),
\begin{equation}\label{eq:SHIFT:by1_helW_helB_ineqs2}
 \helWmin(\dG)\geq\min(\helWmin(\G)-1,1).
\end{equation}
\end{lemma}
\begin{proof}
By~\eqref{eq:DIM:k_dfn}, $k=k(\G)=|\WV| - |\BVint| = |\WVint| - |\BVint|$ (because $\G$ has black boundary) and $k(\dG) = |\dWV| - |\dBVint| = |\BV|-|\Faces|=|\BVint|-|\Fint|+\nconn(\G)-1$. 
Since $\G$ is \bdash trivalent, $|\Eint| = 3|\BVint|$, so~\eqref{eq:DIM:Euler} yields 
$\nconn(\G) = |\Vint| - |\Eint| + |\Fint| = |\WVint| - 2|\BVint| + |\Fint| = k(\G) - k(\dG)+\nconn(\G)-1$. Thus, $\dG$ is of type $(k-1,n)$. 

We show~\eqref{eq:SHIFT:by1_helW_helB_ineqs}. 
Assume first $\dRg\in\WNEI(\dG)$ is such that $\dRg\subset\dVint$ and $\dG\ind[\dRg]$ is connected. 
Set $\Fintup(\dRB):=\{\ff\in\Fint\mid \dbv(\ff)\in\dRB\}$. Let $\Rg:=\RW\sqcup\RB$, where $\RB:=\{\b\in\BVint\mid \dwv(\b)\in\dRW\}$ and $\RW:=\{\w\in\WVint\mid \w\text{ is incident to some $\ff\in\Fintup(\dRB)$}\}$. 
Let $\ff\in\Fintup(\dRB)$. 
If $\b\in\BVint$ is incident to $\ff$ then $\dwv(\b)$ is adjacent to $\dbv(\ff)\in\dRB$. Since $\dRg$ is \wclosed, we find $\dwv(\b)\in\dRW$ and thus $\b\in\RB$. Thus, for $\ff\in\Fintup(\dRB)$, every vertex (white or black) incident to $\ff$ is contained in $\Rg$.

Recall that $\GR$ has vertex set $\Rg$ and edge set $\RgEint$. Let $\RgFintl$ be the set of interior faces of $\GR$. We have shown that 
 $\Fintup(\dRB)\subset\RgFintl$, and thus $|\dRB| = |\Fintup(\dRB)|\leq|\RgFintl|$. 
 Note that $\Rg$ is not necessarily \wclosed.
Since $\G$ is \bdash trivalent, $\gelW(\b)=2$ for all $\b\in\RB$. Applying~\eqref{eq:BACKGR:hel=sum_hel_minus_E} and~\eqref{eq:DIM:Euler} to $\GR$, we get 
\begin{equation*}%
 \gelW(\Rg) = 2|\RB| + |\RW| - |\RgEint| 
= |\RB| - |\RgFintl| + \nconn(\GR) 
\leq |\dRW| - |\dRB| + \nconn(\GR)
= \heldW(\dRg) + \nconn(\GR).
\end{equation*}
Since $\dG\ind[\dRg]$ is connected, so is $\GR$. Thus, 
 $\heldW(\dRg)\geq\gelW(\Rg)-1$. 
Since $|\RB| = |\dRW| >0$, by~\eqref{eq:BCFW:helWmin=min_over_gelB}, $\gelW(\Rg)\geq\helWmin(\G)$. Thus, 
$\heldW(\dRg)\geq\helWmin(\G)-1\geq0$. 

Now, let $\dRg\in\WNEI(\dG)$ be arbitrary. 
Let $\dRg_-$ be obtained from $\dRg$ by removing all (white) boundary vertices and their (black) neighbors. We see that $\dRg_-\in\WNEIg(\dG)$ and $\heldW(\dRg)\geq\heldW(\dRg_-)$. 
Observe that $\heldW(\dRg_-)$ is the sum of $\heldW(\dRg_i)$ over all connected components $\dG\ind[\dRg_i]$ of $\dG\ind[\dRg_-]$. For each $i$, either $\dRg_i$ is a single isolated white vertex with $\heldW(\dRg_i)=1$ or $\dRg_i\in\WNEI(\dG)$. In the latter case, since $\dRg_i\subset\dVint$ and $\dG\ind[\dRg_i]$ is connected, $\heldW(\dRg_i)\geq\helWmin(\G)-1\geq0$ as we showed above. Thus, $\heldW(\dRg_-)\geq0$. This shows the first inequality in~\eqref{eq:SHIFT:by1_helW_helB_ineqs}.

Suppose that $\dRg\in\BNEI(\dG)$. 
By \cref{lemma:BCFW:Rg_simply_conn_lower_bound}, we may assume that $\dRg$ is \sconn. 
Let $\dFint\ind[\dRg]$ be the set of interior faces of $\dG\ind[\dRg]$. 
Let $\Rg:=\RB\sqcup\RW$ be given by
$\RB:=\{\b\in\BVint\mid \dwv(\b)\in\dRW\}$ and 
$\RW:=\{\w\in\WVint\mid \dface(\w)\in\dFint\ind[\dRg]\}$.
 We have $\heldB(\dRg) = |\dRB| - |\dRW| = |\dRB| - |\RB|$. 
Since $\dRg$ is \sconn, $\dFint\ind[\dRg]\subset\dFint$ and for each $\dface\in\dFint\ind[\dRg]$, every vertex of $\dG$ incident to $\dface$ belongs to $\dRg$. It follows that $\Rg\in\BNEIg(\G)$. 
Since $\dG$ is \wdash trivalent and $\dRg$ is \bclosed, $|\dEint\ind[\dRg]| = 3|\dRW|$. Applying~\eqref{eq:DIM:Euler} to $\dG\ind[\dRg]$, we find 
$1= \nconn(\dG\ind[\dRg]) = |\dRB| - 2|\dRW| + |\dFint\ind[\dRg]| = |\dRB| - |\dRW| - |\RB| + |\RW| = \heldB(\dRg) - \helB(\Rg)$. Thus, $\heldB(\dRg)=\helB(\Rg)+1\geq\helBming(\G)+1=\min(\helBmin(\G)+1,2)$ by~\eqref{eq:BACKGR:helWming_vs_helWmin}. This shows the second inequality in~\eqref{eq:SHIFT:by1_helW_helB_ineqs}. 

We prove~\eqref{eq:SHIFT:by1_helW_helB_ineqs2}. As we showed in the proof of~\eqref{eq:SHIFT:by1_helW_helB_ineqs}, for each $\dRg\in\WNEI(\dG)$ such that $\dRg\subset\dVint$ and $\dG\ind[\dRg]$ is connected, we have $\heldW(\dRg)\geq\helWmin(\G)-1$. Suppose that $\dRg\in\WNEI(\dG)$ is arbitrary 
 and let $\dRg_-$ be defined as above. We already showed that each connected component $\dG\ind[\dRg_i]$ of $\dG\ind[\dRg_-]$ satisfies $\heldW(\dRg_i)\geq\min(\helWmin(\G)-1,1)$. 
Since $\heldW(\dRg_-)$ is the sum of 
$\heldW(\dRg_i)$ for all $i$ (and since $\helWmin(\G)-1\geq0$ by assumption), we get $\heldW(\dRg_-)\geq\min(\helWmin(\G)-1,1)$ unless $\dRg_-=\emptyset$. In this case, $\dRg$ contains only (white) boundary and (black) next-to-boundary vertices. Since $\dRg$ is \wclosed, the only way to not have $\heldW(\dRg)\geq1$ is if every next-to-boundary vertex in $\dRg$ has degree $1$ in $\dG$. Since $\dG$ is connected, it follows that $\dG$ consists of a single boundary edge. But we have assumed $n\geq2$, a contradiction. 
\end{proof}

\begin{corollary}
If $\G$ is \bdash trivalent with $\helWmin(\G)\geq1$ and $\helBmin(\G)\geq0$ then 
$\helWmin(\dG)\geq0$ and $\helBmin(\dG)\geq1$.
In particular, by \cref{prop:DIM:helWBmin_geq_0=>apm_exists}, $\dG$ admits an \APM.
\end{corollary}

\subsection{Combinatorial shift by $2$ (T-duality)}\label{sec:shift:combin}
For the rest of this section, we assume the following. 
\begin{definition}\label{dfn:G:T_dualizable}
$\G$ is called \emph{T-dualizable} if it is \bdash trivalent of type $(k,n)$, satisfies 
 ${\helWmin(\G)\geq2}$ and $\helBmin(\G)\geq0$, and contains $n$ distinct next-to-boundary vertices (with $n\geq2$). 
\end{definition}
\noindent Our goal is to define the \emph{T-duality} map $\G\mapsto\ddG$ which essentially consists of applying the shift by $1$ twice. 

Our assumptions on $\G$ together with \crefrange{lemma:shift1-top}{lemma:SHIFT:by1} imply that $\dG$ is connected and $\helmin(\dG)\geq1$. Since $\helWmin(\dG)\geq1$, each interior black vertex of $\dG$ has degree at least $2$, and $\dG$ contains no cycles in which all black vertices are of degree exactly $2$. This observation allows us to make the following definition. 
\begin{definition}\label{rmk:SFIFT:uncontract_black_trivalent}
Let $\dG'$ be a \bdash trivalent graph obtained from $\dG$ by applying moves \MVbd and \MV1: we apply \MVbd to make all boundary vertices black and then apply \MV1 to contract all degree-$2$ black vertices and uncontract each black vertex of degree $\geq4$ into several degree-$3$ black vertices separated from each other by degree-$2$ white vertices. 
We let $\ddG'$ be the result of applying the shift by $1$ map to $\dG'$ and relabeling the boundary vertices of the resulting graph by $\bdv_i\mapsto \bdv_{i+1}$ for all $i\in\brn$. 
\end{definition}

\begin{remark}\label{rmk:SHIFT:dG_ddG_helW_helB_connected}
By Lemmas~\ref{lemma:DIM:moves_vs_helWmin} and \ref{lemma:shift1-top}--\ref{lemma:SHIFT:by1},
 $\dG'$ is connected and satisfies $\helmin(\dG')\geq1$. 
Thus, by \crefrange{lemma:shift1-top}{lemma:SHIFT:by1} again, $\ddG'$ 
is \bdconn of type $(k-2,n)$, has $n$ distinct next-to-boundary vertices, and satisfies
\begin{equation}\label{eq:dG_ddG_helW_helB}
 \helWmin(\ddG')\geq0 \quad\text{and}\quad\helBmin(\ddG')\geq2. 
\end{equation}
By \cref{prop:DIM:helWBmin_geq_0=>apm_exists}, $\ddG'$ admits an \APM.
\end{remark}

We give a streamlined description of the map $\G\mapsto\ddG'$. 
 First, we claim that each $\f\in\Faces$ is incident to $\dwcor(\f)\geq2$ white vertices (counted with multiplicity). Indeed, if $\f\in\Fint$ is an interior face then $\dwcor(\f)\geq1$, and if $\dwcor(\f)=1$ then $\f$ is located between a pair of parallel edges. Let $\b,\w$ be the vertices of $\G$ connected by such a pair of parallel edges. Since $\b$ is trivalent, letting $\Rg:=\{\b\}\sqcup\NeighG(\b)$, we see that $\Rg\in\WNEI(\G)$ and $\helW(\Rg)\leq1$, a contradiction. If $\f=\bdf_i$ is a boundary face then it is incident to next-to-boundary white vertices $\bdwx_i,\bdwx_{i+1}$ that are distinct by assumption, so again $\dwcor(\f)\geq2$. 

\begin{figure}
\includegraphics[scale=1.2]{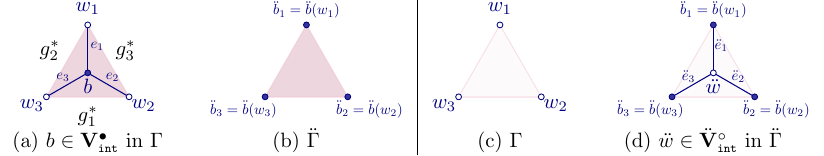}
 \caption{\label{fig:black3} 
Notation around trivalent vertices before and after applying T-duality.
}
\end{figure}

\begin{definition}\label{dfn:SHIFT:Sig}
Let $\Sig$ be the following two-dimensional simplicial complex with vertex set $\WV=\WVint$ and with triangular faces colored \emph{light} and \emph{dark}. The set $\BSig$ of dark faces of $\Sig$ consists of triangles with vertices $\NeighG(\b)=\{\wv_1,\wv_2,\wv_3\}$ (cf. \figref{fig:black3}(a)), one for each trivalent vertex $\bv\in\BVint$ of $\G$. 
For each face $\f\in\Faces$ of $\G$, let $(\wv_1,\wv_2,\dots,\wv_{\dwcor(\f)})$ be the white vertices incident to $\f$ listed with multiplicity in clockwise order. 
By definition, if $\dwcor(\f) = 2$ then $\Sig$ contains an edge $\{\wv_1,\wv_2\}$, and if $\dwcor(\f)\geq3$, 
we choose an arbitrary triangulation of the area inside the closed polygonal chain with vertices $\wv_1,\wv_2,\dots,\wv_{\dwcor(\f)}$ into light triangles and add them to $\WSig$. 
\end{definition}
\noindent Thus, $\Sig$ is a triangulation of the polygon with vertices $(\bdwx_1,\bdwx_2,\dots,\bdwx_n)$; see \figref{fig:intro-G-ddG}(a,b). 

\begin{definition}[T-dual graphs]\label{dfn:SHIFT:ddG}
 For each vertex $\wv\in\WVint$ of $\Sig$, the graph $\ddG$ contains a black interior vertex $\ddbx(\wv)\in\ddBVint$. For each light triangle $\{\wv_1,\wv_2,\wv_3\}\in\WSig$, $\ddG$ contains a trivalent interior white vertex $\ddw\in\ddWVint$ adjacent to the corresponding black vertices $\ddb_1,\ddb_2,\ddb_3$ (with $\ddb_\s:=\ddbx(\wv_\s)$ as in \figref{fig:black3}(d)) by edges $\dde_1,\dde_2,\dde_3$. The boundary vertices $\ddbdv_1,\ddbdv_2,\dots,\ddbdv_n$ of $\ddG$ are all white and degree-$1$. For $i\in\brn$, the $i$-th boundary edge $\ddbde_i$ of $\ddG$ connects $\ddbdv_i$ to $\ddbx(\bdwx_i)$. 
 See \figref{fig:intro-G-ddG}(c).
\end{definition}

\begin{definition}\label{dfn:SHIFT:light_dark_regions}
Consider the planar dual graph $\Sig^\ast$ of $\Sig$, with vertices corresponding to light and dark triangles and two vertices connected by an edge in $\Sig^\ast$ when the two corresponding triangles share an edge of $\Sig$. 
A \emph{light} (resp., \emph{dark}) \emph{region} of $\Sig$ is a maximal by inclusion collection 
 $\RgSig\subset\WSig$ (resp., $\RgSig\subset\BSig$) of triangles such that the induced subgraph $\Sig^\ast\ind[\RgSig]$ is connected. 
\end{definition}

\begin{remark}\label{rmk:SHIFT:choices}
One can make different choices in \cref{rmk:SFIFT:uncontract_black_trivalent} when uncontracting black interior vertices of $\dG$ of degree $\geq4$. 
Similarly, in \cref{dfn:SHIFT:Sig}, one can choose different triangulations of faces $\f\in\Faces$ of $\G$ when $\dwcor(\f)\geq4$. 
By \crefi{lemma:shift1-top}{shift1-top2}, black interior vertices of $\dG$ are in bijection with faces of $\G$. From now on, we assume that the uncontraction choices in \cref{rmk:SFIFT:uncontract_black_trivalent} agree with the corresponding triangulation choices in \cref{dfn:SHIFT:Sig} under this bijection. 
\end{remark}

\begin{lemma}\label{lemma:SHIFT:ddG=ddG'}
The graphs $\ddG$ and $\ddG'$ coincide (assuming compatibility of choices in \cref{rmk:SHIFT:choices}).
\end{lemma}
\begin{proof}
Let $\dGS$ be the ``bipartite planar dual'' of $\Sig$, obtained as follows. Place a trivalent black (resp., white) vertex of $\dGS$ inside each light (resp., dark) triangle of $\Sig$ and connect it by a half-edge to the midpoint of each side of the triangle. If two light (resp., dark) triangles of $\Sig$ share an edge then we place a white (resp., black) degree-$2$ vertex in the middle of this edge. We place a degree-$1$ white (resp., black) boundary vertex in the middle of each boundary edge of $\Sig$ incident to a light (resp., dark) triangle of $\Sig$. 

It follows that the graphs $\dGS$ and $\dG'$ are related by moves \MVbd and \MV1. 
Thus, the face sets $\dFaces_{\Sig}$ and $\dFacesp$ of $\dGS$ and $\dG'$ are both in bijection with the vertex set $\WVint$ of $\Sig$. We therefore have bijections $\ddBV\cong\WVint\cong\dFacesp\cong \ddBVp$ and $\ddWVint\cong\WSig\cong\dBVintp\cong\ddWVintp$. 
 Comparing the edges of $\ddG$ and $\ddG'$, we see that these two graphs indeed coincide. The labelings of the boundary vertices agree in view of the $\bdv_i\mapsto \bdv_{i+1}$ relabeling in \cref{rmk:SFIFT:uncontract_black_trivalent}.
\end{proof}

Combining \cref{lemma:SHIFT:ddG=ddG'} with \cref{rmk:SHIFT:dG_ddG_helW_helB_connected}, we obtain the following.
\begin{corollary}\label{lemma:SHIFT:ddG_admits_apms}
The T-dual graph $\ddG$ \hasnofloat of type $(k-2,n)$, satisfies
 $\helWmin(\ddG)\geq0$ and $\helBmin(\ddG)\geq2$, 
contains $n$ distinct next-to-boundary vertices, 
and admits an \APM. 
\end{corollary}

\subsection{\texorpdfstring{$\la$}{lambda}-Kasteleyn weights}\label{sec:shift:lambda_Kast_weights}
We continue to assume that $\G$ is T-dualizable. 
By \cref{prop:DIM:helWBmin_geq_0=>apm_exists}, $\G$ admits an \APM. 
Let $\wt\in\Rtpgauge$ and $C:=\Meas(\G,\wt)$. Fix a choice $\epsK$ of Kasteleyn signs on $\G$.
 In addition, we fix a $2$-plane $\la\subset C$. For now, we only assume that $\brla<i,i+1>\neq0$ for all $i\in\brn$.
 We explain how to use the $2$-plane $\la$ to recover the Kasteleyn weights $\wtK(\e)=\epsK(\e)\wt(\e)$ up to gauge equivalence. 

After applying row operations to $C$, we may assume that 
$C = \begin{pmatrix}
\la \\
\Chat
\end{pmatrix},$
 where $\Chat\in\Mator_{k-2,n}$. Then $C\cdot \Qla$ has the top two rows equal to zero, and so it equals $\Chat\cdot \Qla$ as an element of $\Gr(k-2,n)$.
 Let $\Cext = \begin{pmatrix}
\laext \\
\Chatext
\end{pmatrix}\in\Hwspace_{\R^k}\HtripK$ be the \wdash holomorphic extension of $C$. %

Given $\wv_1,\wv_2\in\WV$, recall from~\eqref{eq:brlaw_dfn} that we set 
 $\brlaw<\wv_1,\wv_2> := \det\mat[\laext(\wv_1)|\laext(\wv_2)]$.
We now define the \emph{$\laext$-Kasteleyn weights} $\wtlaK$. For each boundary edge $\bde_i$, we set $\wtlaK(\bde_i) := \wtK(\bde_i)$. For each (trivalent) black vertex $\bv\in\BVint$ of $\G$, let $\wv_1,\wv_2,\wv_3$ be its neighbors in clockwise order and let $\e_1,\e_2,\e_3$ be the edges connecting them to $\bv$; see \figref{fig:black3}(a). 
We let
\begin{equation}\label{eq:wtlaK_dfn}
 \wtlaK(\e_\s) := \brlaw<\wv_{\s+1},\wv_{\s+2}> 
 \quad\text{for $\s=1,2,3$.}
\end{equation}
\noindent (Throughout this section, the index $\s$ is always taken modulo $3$.)

 For general $\la\subset C$, the edge weights $\wtlaK(\e)$ are not guaranteed to be nonzero for all $\e\in\E$. The situation changes when $\la\in\lak$: since $\G$ is T-dualizable, $\helWmin(\G)\geq2$ and $\helBmin(\G)\geq0$, so by \cref{lemma:DIM:deleting_hel_verts} and \crefi{lemma:TE:brla_nonzero_if_apm}{lak_implies1}, we get
\begin{equation}\label{eq:la_generic}
 \brlaw<\wv_1,\wv_2>\neq0 \quad
\text{whenever $\la\in\lak$ and $\wv_1,\wv_2$ are connected by an edge of $\Sig$.}
\end{equation}
\noindent In view of~\eqref{eq:la_generic}, we set 
$\epsKla(\e):=\sign(\wtlaK(\e))$ and $\wtlap(\e):=|\wtlaK(\e)|=\epsKla(\e)\wtlaK(\e)$ for all $\e\in\Edges$.

\begin{lemma}\label{lemma:black_gauge_eq}
For $\la\in\lak$, the nonzero edge weights $\wtlaK(\e)$ are \emph{black gauge equivalent} to $\wtK(\e)$: there exists a function $\gauge:\BV\to\Rast$ 
equal to $1$ on $\BVbd=\Vbd$ 
 such that
\begin{equation}\label{eq:black_gauge_equivalent}
 \wtK(\e) = \wtlaK(\e) \gauge(\bv) \quad\text{for all $\b\in\BV$ and all $\e\in\E$ incident to $\b$}.
\end{equation}
In particular, the positive edge weights $\wtlap(\e)$ are black gauge equivalent to $\wt(\e)$.
\end{lemma}
\begin{proof}
Recall that $\G$ has black boundary. 
It is clear that~\eqref{eq:black_gauge_equivalent} holds for boundary edges $\bde_i$ since we have $\wtlaK(\bde_i) = \wtK(\bde_i)$ by definition for $i\in\brn$. If $\bv\in\BVint$ is a trivalent black vertex then in the notation of \figref{fig:black3}(a), no two vectors among $\laext(\wv_1),\laext(\wv_2),\laext(\wv_3)\in\R^2$ are collinear by~\eqref{eq:la_generic}. Since $\laext$ is \wdash holomorphic, the edge weights $\wtK(\e_1),\wtK(\e_2),\wtK(\e_3)$ give the coefficients of a unique (up to scaling) linear dependency between these three vectors. On the other hand, using Cramer's rule, we find 
\begin{equation}\label{eq:la_Cramer}
 \laext(\wv_1)\brlaw<\w_2,\w_3> +
 \laext(\wv_2)\brlaw<\w_3,\w_1> + 
 \laext(\wv_3)\brlaw<\w_1,\w_2> = 0.%
\end{equation}
This shows~\eqref{eq:black_gauge_equivalent}. Taking absolute values, 
 we see that $\wt$ and $\wtlap$ are black gauge equivalent as well.
\end{proof}

\noindent We note that when $\la\subset C$ satisfies $\la\in\lak$, the above lemma implies more generally that
\begin{equation}\label{eq:Cext_Cramer}
 \Cext(\wv_1)\brlaw<\w_2,\w_3> +
 \Cext(\wv_2)\brlaw<\w_3,\w_1> + 
 \Cext(\wv_3)\brlaw<\w_1,\w_2> = 0.
\end{equation}

\subsection{T-dual edge weights}
Let $\G$ be T-dualizable. 
We assume from now on that $\la\subset C$ satisfies $\la\in\lak$. 
We have introduced the T-dual graph $\ddG$ in \cref{dfn:SHIFT:ddG}. 
Recall from \cref{lemma:SHIFT:ddG_admits_apms} that $\ddG$ admits an \APM. 
Our goal is to define certain edge weights $\ddwtlaK:\ddE\to\Rast$ on $\ddG$ using $\laext$.
 We will then show that these edge weights satisfy the Kasteleyn sign condition~\eqref{eq:Kast_sign} for each face of $\ddG$.

Given a trivalent white vertex $\ddw$ of $\ddG$, we consider the corresponding light triangle $\{\wv_1,\wv_2,\wv_3\}\in\WSig$ with $\wv_1,\wv_2,\wv_3$ listed in clockwise order. 
 In the notation of \figref{fig:black3}(d), the Kasteleyn edge weights $\ddwtlaK$ are defined on interior edges of $\ddG$ by
\begin{equation}\label{eq:ddwtlaK_dfn}
 \ddwtlaK(\dde_\s) := \brlaw<\wv_{\s+1},\wv_{\s+2}>, \quad\text{for $\s=1,2,3$.}
\end{equation}
For a boundary edge $\ddbde_i=\{\ddbdv_i,\ddbx(\bdwx_i)\}$ of $\ddG$, we set $\ddwtlaK(\ddbde_i) := -\wtK(\bde_i)$ for $i\in\brn$. By~\eqref{eq:la_generic}, the edge weights $\ddwtlaK$ are nonzero. 
For each $\dde\in\ddE$, we set $\ddepsKla(\dde):=\sign(\ddwtlaK(\dde))$ and
\begin{equation}\label{eq:ddwtlap_dfn}
	\ddwtlap(\dde):=\ddepsKla(\dde)\ddwtlaK(\dde)=|\ddwtlaK(\dde)|.
\end{equation}

\begin{lemma}\ 
Let $\w_1,\w_2,\w_3$ be the vertices of a triangle $\triang$ of $\Sig$, listed in clockwise order. Then 
\begin{equation}\label{eq:SHIFT:sign_triangles}
 \sign\left(\brlaw<\w_1,\w_2>\cdot \brlaw<\w_2,\w_3>\cdot \brlaw<\w_3,\w_1>\right) = 
 \begin{cases}
 +1, &\text{if $\triang$ is light;}\\
 -1, &\text{if $\triang$ is dark.}
 \end{cases}
\end{equation}
More generally, for any clockwise simple cycle $\Cyc=(\w_1,\w_2,\cdots,\w_m)$ in $\Sig$,
\begin{equation}\label{eq:SHIFT:sign_cycles}
 \sign\left(\brlaw<\w_1,\w_2>\cdot \brlaw<\w_2,\w_3>\cdots\brlaw<\w_m,\w_1>\right) 
= (-1)^{|\BSig\ind[\Cyc]| + |\EintSig\ind[\Cyc]|},
\end{equation}
where $\BSig\ind[\Cyc]$ is the set of dark triangles of $\Sig$ contained inside $\Cyc$ and $\EintSig\ind[\Cyc]$ is the set of edges of $\Sig$ contained strictly inside $\Cyc$. 
\end{lemma}
\begin{proof}
Assume that $\triang$ is light. Inserting black tripods at $(\w_1,\w_2)$, $(\w_2,\w_3)$, $(\w_3,\w_1)$ creates an interior face $\ff$ inside $\triang$ with $\dwcor(\ff)=6$. Applying~\eqref{eq:Kast_sign} to $\ff$, we find $\epsww_{\w_1,\w_2}\epsww_{\w_2,\w_3}\epsww_{\w_3,\w_1} = -1$. Thus,~\eqref{eq:SHIFT:sign_triangles} follows from \crefi{lemma:TE:brla_nonzero_if_apm}{lak_implies1} when $\triang$ is light.
For dark $\triang$, \eqref{eq:SHIFT:sign_triangles} follows from~\eqref{eq:TE:same_face_and_same_vertex} and \cref{lemma:black_gauge_eq}.
Finally,~\eqref{eq:SHIFT:sign_cycles} is deduced from~\eqref{eq:SHIFT:sign_triangles} by taking the product over all triangles of $\Sig$ contained inside $\Cyc$ and observing that every edge in $\EintSig\ind[\Cyc]$ contributes to the product twice with opposite signs.
\end{proof}

\begin{lemma}\label{lemma:SHIFT:ddepsKla_is_Kast}
$\ddepsKla$ is a valid choice of Kasteleyn signs for $\ddG$.
\end{lemma}
\begin{proof}
The faces of $\ddG$ correspond to (i) dark regions of $\Sig$ (\cref{dfn:SHIFT:light_dark_regions}), (ii) interior edges of $\Sig$ incident to two light triangles, and (iii) boundary edges of $\Sig$ incident to light triangles. We check that $\ddepsKla$ satisfies~\eqref{eq:Kast_sign} for each face. 

Let $\ddF$ be an (interior, square) face of $\ddG$ of type (ii). Let $\w_1,\w_2,\w_3,\w_4$ be the vertices of the quadrilateral formed by the union of the corresponding two light triangles of $\Sig$, listed in clockwise order. By~\eqref{eq:ddwtlaK_dfn}, $\ddepsKla(\ddF) = \sign\left(\brlaw<\w_1,\w_2>\cdot \brlaw<\w_2,\w_3>\cdot \brlaw<\w_3,\w_4>\cdot \brlaw<\w_4,\w_1>\right)$. By~\eqref{eq:SHIFT:sign_cycles}, $\ddepsKla(\ddF)=-1$.

Let $\ddF$ be a face of $\ddG$ of type (i). 
Let $\RgSig$ be the dark region of $\Sig$ corresponding to $\ddF$, and let $\Cyc$ be the union of all oriented edges of $\Sig$ that are incident to a triangle in $\RgSig$ on the right but not on the left. 
We claim that the union $|\Sig\ind[\Cyc]|=\bigcup_{\triang\in\RgSig}\triang$ of closed triangles is \sconn, $\Cyc=(\w_1,\w_2,\cdots,\w_m)$ is a simple clockwise cycle, and every vertex of $\Sig$ incident to some triangle in $\RgSig$ belongs to $\Cyc$. By \cref{dfn:SHIFT:light_dark_regions}, $|\Sig\ind[\Cyc]|$ is connected. If the boundary $\Cyc$ of $|\Sig\ind[\Cyc]|$ is not connected or if $|\Sig\ind[\Cyc]|$ contains a vertex that does not belong to $\Cyc$ then $\ddG$ contains a floating connected component, contradicting~\cref{lemma:SHIFT:ddG_admits_apms}. If $\Cyc$ passes through some vertex $\w$ twice then since $\RgSig$ is a (connected) dark region, 
there exists a closed curve $\gamma$ passing through $\w$, lying entirely inside $|\Sig\ind[\Cyc]|$, and enclosing a nonempty white region of $\Sig$. Thus, 
$\ddG\rem\{\ddbx(\w)\}$ contains a floating connected component with vertex set $\ddRg$ such that $\ddRW,\ddRB\neq\emptyset$.
 Thus, $\ddRg\in\WNEI(\ddG)$ is \wclosed, and $\ddRg_+:=\ddRg\sqcup\{\ddbx(\w)\}\in\BNEI(\ddG)$ is \bclosed. 
Following \cref{lemma:DIM:no_floating}, we get $\helBsub_{\ddG}(\ddRg_+)=-\helWsub_{\ddG}(\ddRg)+1$, contradicting \cref{lemma:SHIFT:ddG_admits_apms}.

We have thus shown that $|\Sig\ind[\Cyc]|$ is a triangulation of a simple $m$-gon with vertices $\Cyc=(\w_1,\w_2,\cdots,\w_m)$ containing no vertices of $\Sig$ inside. Any such triangulation uses $m-3=|\EintSig\ind[\Cyc]|$ diagonals and $m-2=|\BSig\ind[\Cyc]|$ triangles in the notation of~\eqref{eq:SHIFT:sign_cycles}. 
 Thus, the right-hand side of~\eqref{eq:SHIFT:sign_cycles} is equal to $-1$. 

 Every edge of $\Cyc$ is either an interior edge of $\Sig$ incident to a light triangle or a boundary edge of $\Sig$. Denote the set of such interior (resp., boundary) edges by $\CycA$ (resp., $\CycB$). For every edge $\w_i\to\w_{i+1}$ in $\CycA$, the boundary of $\ddF$ contains a trivalent white vertex $\ddw'_i$ connected to $\ddbx(\w_i),\ddbx(\w_{i+1})$ by edges $\dde_i,\dde_i'$ with $\ddepsKla(\dde_i)\ddepsKla(\dde'_i)=\sign(-\brlaw<\w_i,\w_{i+1}>)$ by~\eqref{eq:ddwtlaK_dfn} and~\eqref{eq:SHIFT:sign_triangles}. For every edge $\w_i\to\w_{i+1}$ in $\CycB$, the boundary of $\ddF$ contains two white boundary vertices $\ddbdv_{j},\ddbdv_{j+1}$ and a boundary arc $j\in\bdryarcs\ddF$ between them. The boundary edge $\w_i\to\w_{i+1}$ of $|\Sig\ind[\Cyc]|$ is incident to a dark triangle $\triang\in\RgSig$. Let $\w'_i$ be the third vertex of $\triang$. The boundary face $\bdf_{j}$ of $\G$ is incident to one boundary arc, two white vertices, and four edges with Kasteleyn signs $\epsK(\bde_j)$, $\epsK(\bde_{j+1})$, $\sign\brlaw<\w_{i+1},\w'_i>$, and $\sign\brlaw<\w'_i,\w_i>$. By~\eqref{eq:Kast_sign}, the product of these four signs is $\epsKla(\bdf_{j})=(-1)^{\epstra(\bdf_j)+1}$. 
By~\eqref{eq:SHIFT:sign_triangles}, 
$\sign(\brlaw<\w_{i+1},\w'_i>\cdot \brlaw<\w'_i,\w_i>)=\sign(-\brlaw<\w_i,\w_{i+1}>)$. 
Since $\ddepsKla(\ddbde_j) = -\epsK(\bde_j)$ and $\ddepsKla(\ddbde_{j+1}) = -\epsK(\bde_{j+1})$, 
we get 
$ \ddepsKla(\ddbde_j)\ddepsKla(\ddbde_{j+1}) = \sign(-\brlaw<\w_i,\w_{i+1}>) (-1)^{\epstra(\bdf_j)+1}$. 
Thus, %
\begin{equation*}%
 \ddepsKla(\ddF) = \prod_{i=1}^m \sign(-\brlaw<\w_i,\w_{i+1}>) \cdot \prod_{j\in\bdryarcs\ddF} (-1)^{\epstra(\bdf_j)+1}
 = (-1)^{m +1 + \sum_{j\in\bdryarcs\ddF}(\epstra(\bdf_j)+1)},
\end{equation*}
where we have used that the right-hand side of~\eqref{eq:SHIFT:sign_cycles} is equal to $-1$ as we showed above. 
On the other hand, $\dwcor(\ddF) = m + \nbdryarcs{\ddF}$ and $\epstra(\ddF)=1$ if $n\notin\bdryarcs\ddF$ and $\epstra(\ddF)=k+n-2$ if $n\in\bdryarcs\ddF$. Similarly, we find $(-1)^{\sum_{j\in\bdryarcs\ddF}(\epstra(\bdf_j)+1)}=+1$ if $n\notin \bdryarcs\ddF$ and $(-1)^{\sum_{j\in\bdryarcs\ddF}(\epstra(\bdf_j)+1)}=(-1)^{k+n-1}$ if $n\in\bdryarcs\ddF$. Thus, in each case, we get $\ddepsKla(\ddF)=(-1)^{\dwcor(\ddF)+\nbdryarcs{\ddF}+\epstra(\ddF)}$, as desired. 

Finally, consider a (boundary) face $\ddbdf_j$ of $\ddG$ of type (iii). Thus, $\ddbdf_j$ is incident to three white vertices, one boundary arc, and four edges so that $\ddepsKla(\ddbdf_j)=\ddepsKla(\ddbde_j)\ddepsKla(\ddbde_{j+1})\sign(-\brlaw<\bdwx_j,\bdwx_{j+1}>)$ by~\eqref{eq:SHIFT:sign_triangles}. 
 Computing $\epsKla(\bdf_j)$ similarly to what we did above, we find $\sign(\brlaw<\bdwx_j,\bdwx_{j+1}>)=\epsKla(\bde_j)\epsKla(\bde_{j+1})(-1)^{\epstra(\bdf_j)+1}$. Since $\ddepsKla(\ddbde_j)=-\epsKla(\bde_j)$ and $\ddepsKla(\ddbde_{j+1})=-\epsKla(\bde_{j+1})$, we see that~\eqref{eq:Kast_sign} is satisfied for $\ddepsKla(\ddbdf_j)$. 
\end{proof}

\subsection{T-dual boundary measurements}
Our next goal is to show that the positive edge weights $\ddwtlap(\dde):=\ddepsKla(\dde)\ddwtlaK(\dde)$ defined above satisfy $\Meas(\ddG,\ddwtlap) = \Chat\cdot \Qla$. 
We will show that the \wdash holomorphic 
 extension $\ddCext\in\Hwspace_{\R^{k-2}}(\ddG,\ddwtlaK)$ of $\Chat\cdot \Qla$ is given explicitly as follows. 
 For a trivalent white vertex $\ddw\in\ddWVint$ adjacent to black vertices $\ddbx(\wv_\s)$ for $\s=1,2,3$ as in \figref{fig:black3}(d), we set
\begin{equation}\label{eq:ddCext_dfn}
 \ddCext(\ddw) := \frac{
\Chatext(\wv_1)\brlaw<\wv_2,\wv_3> +
\Chatext(\wv_2)\brlaw<\wv_3,\wv_1> +
\Chatext(\wv_3)\brlaw<\wv_1,\wv_2> 
}{
\brlaw<\wv_2,\wv_3>\cdot \brlaw<\wv_3,\wv_1>\cdot \brlaw<\wv_1,\wv_2>
}.
\end{equation}
This expression is well defined in view of~\eqref{eq:la_generic}. For a (white) boundary vertex $\ddbdv_i$ of $\ddG$, using~\eqref{eq:partial_F_dfn}, we set 
\begin{equation}\label{eq:ddCext_bdry_dfn}
 \ddCext(\ddbdv_i):=(-1)^{i} (\Chat\cdot \Qla)_i \quad\text{for $i\in\brn$, \quad so that\quad}
\pddCext = \alt(\Chat\cdot \Qla).
\end{equation}

\begin{proposition}\label{lemma:SHIFT:ddCext_in_Hwspace}
We have $\ddCext\in\Hwspace_{\R^{k-2}}(\ddG,\ddwtlaK)$. 
\end{proposition}
\begin{proof}
For a (dark or light) triangle $\triang$ of $\Sig$ with vertices $\wv_1,\wv_2,\wv_3$ in clockwise order, let $\ddCext(\triang)$ be given by the right-hand side of~\eqref{eq:ddCext_dfn}. If $\triang\in\WSig$ and $\ddw\in\ddWVint$ is the corresponding white vertex of $\ddG$ then $\ddCext(\triang) = \ddCext(\ddw)$. By~\eqref{eq:Cext_Cramer}, $\ddCext(\triang) = 0$ for all $\triang\in\BSig$.

We check that $\ddCext$ is \wdash holomorphic. Let $\wv\in\WVint$ and let $\ddb:=\ddbx(\wv)$ be the corresponding black vertex of $\ddG$. We first treat the case where $\ddb$ is not a next-to-boundary vertex of $\ddG$. Let $\triang_1,\triang_2,\dots,\triang_d$ be the triangles of $\Sig$ incident to $\wv$ in clockwise order. Thus, their union contains $\wv$ in its interior. Let $\wv_1,\wv_2,\dots,\wv_d\in\WV$ be such that $\triang_\ss$ has vertices $\wv_\ss,\wv_{\ss+1},\wv$ in clockwise order for each $\ss\in\brd$, where the index $\ss$ is taken modulo $d$ here and below. 
 We claim that 
\begin{equation}\label{eq:sum_triangles_around}
 \sum_{\ss=1}^d \brlaw<\wv_\ss,\wv_{\ss+1}> \ddCext(\triang_\ss) = \bzero\in\R^{k-2}.
\end{equation}
We first explain why this is equivalent to the condition of $\ddCext$ being \wdash holomorphic at $\ddb$. Recall that $\ddCext(\triang_\ss)=0$ for all dark triangles $\triang_\ss$. Each light triangle $\triang_\ss$ contains a trivalent white vertex $\ddw_\ss$ of $\ddG$, and by~\eqref{eq:ddwtlaK_dfn}, 
 $\ddwtlaK(\ddw_\ss,\ddb) = \brlaw<\wv_\ss,\wv_{\ss+1}>$. Thus, the left-hand side of~\eqref{eq:sum_triangles_around} equals $\sum_{\ss:\triang_\ss\in\WSig}\ddwtlaK(\ddw_\ss,\ddb) \ddCext(\ddw_\ss)$, which is the left-hand side of the definition~\eqref{eq:OCP:holom} of a \wdash holomorphic function in $\Hwspace_{\R^{k-2}}(\ddG,\ddwtlaK)$.

Next, we prove~\eqref{eq:sum_triangles_around}. By~\eqref{eq:ddCext_dfn}, the $\ss$-th term in~\eqref{eq:sum_triangles_around} is given by
\begin{equation*}%
 \brlaw<\wv_\ss,\wv_{\ss+1}> \ddCext(\triang_\ss) = 
\frac{
\Chatext(\wv_\ss)
}{
\brlaw<\wv,\wv_\ss>
}
+
\frac{
\Chatext(\wv_{\ss+1})
}{
\brlaw<\wv_{\ss+1},\wv>
}
+
\frac{
\Chatext(\wv)\brlaw<\wv_{\ss},\wv_{\ss+1}>
}{
\brlaw<\wv_{\ss+1},\wv>\brlaw<\wv,\wv_\ss>
}.
\end{equation*}
The first two terms above form a telescoping sum: $\sum_{\ss=1}^d \left(\frac{\Chatext(\wv_\ss)}{\brlaw<\wv,\wv_\ss>}+\frac{\Chatext(\wv_{\ss+1})}{\brlaw<\wv_{\ss+1},\wv>}\right)=0$.
 The third term is proportional to $\Chatext(\wv)$. We prove that the coefficients of $\Chatext(\wv)$ sum up to zero by induction on $d$. For $d=2$, the result is trivial. Let $d>2$. Summing up the coefficients for $\ss=1,2$, we get
$\frac{
\brlaw<\wv_{1},\wv_{2}>
}{
\brlaw<\wv_{2},\wv>\brlaw<\wv,\wv_1>
} +
 \frac{
\brlaw<\wv_{2},\wv_{3}>
}{
\brlaw<\wv_{3},\wv>\brlaw<\wv,\wv_2>
} = 
\frac{
\brlaw<\wv_{1},\wv_{2}>\brlaw<\wv_{3},\wv> + \brlaw<\wv_1,\wv> \brlaw<\wv_2,\wv_3>
}{
\brlaw<\wv_{2},\wv>\brlaw<\wv,\wv_1>\brlaw<\wv_{3},\wv>
}.$
Applying the Pl\"ucker relation $\brlaw<\wv_{1},\wv_{2}>\brlaw<\wv_{3},\wv> + \brlaw<\wv_1,\wv> \brlaw<\wv_2,\wv_3> = \brlaw<\wv_{1},\wv_{3}>\brlaw<\wv_{2},\wv>$ to the numerator, 
this fraction becomes 
$\frac{
\brlaw<\wv_{1},\wv_{3}>
}{
\brlaw<\wv_{3},\wv>\brlaw<\wv,\wv_1>
}.$
The result follows by the induction hypothesis applied to $(\wv_1,\wv_3,\wv_4,\dots,\wv_d)$. 

We have shown that $\ddCext$ is \wdash holomorphic in the interior of $\ddG$. 
The case when $\wv = \bdwx_i$ is a next-to-boundary vertex of $\G$ is handled similarly. Applying the above telescoping sum and induction argument, we transform the left-hand side of~\eqref{eq:sum_triangles_around} into
\begin{equation}\label{eq:sum_triangles_bdry}
\frac
{
\Chatext(\bdwx_{i-1})\brlaw<\bdwx_{i},\bdwx_{i+1}> + 
\Chatext(\bdwx_{i})\brlaw<\bdwx_{i+1},\bdwx_{i-1}> + 
\Chatext(\bdwx_{i+1})\brlaw<\bdwx_{i-1},\bdwx_{i}>
}{
\brlaw<\bdwx_{i-1},\bdwx_{i}>\brlaw<\bdwx_{i},\bdwx_{i+1}>
}.
\end{equation}
By~\eqref{eq:partial_F_dfn} and~\eqref{eq:intro:Qla}, this expression equals $(-1)^{i} \wtK(\bde_i) (\Chat\cdot \Qla)_i$ and since $\ddwtlaK(\ddbde_i) := -\wtK(\bde_i)$, it cancels out with the contribution from $\ddbde_i = \{\ddbdv_i,\ddb\}$; cf.~\eqref{eq:ddCext_bdry_dfn}. It follows that $\ddCext$ is \wdash holomorphic at each next-to-boundary vertex $\ddb$.
\end{proof}

\begin{corollary}
We have 
\begin{equation}\label{eq:Meas_ddG=Chat_Qla}
 \Meas(\ddG,\ddwtlap) = \Chat\cdot \Qla \quad\text{as elements of $\Gr(k-2,n)$.}
\end{equation}
\end{corollary}
\begin{proof}
By \cref{lemma:SHIFT:ddG_admits_apms}, $\ddG$ has white boundary, is of type $(k-2,n)$, and admits an \APM. 
 By \cref{lemma:SHIFT:ddepsKla_is_Kast}, \cref{lemma:SHIFT:ddCext_in_Hwspace}, \cref{lemma:MCE:apm_vs_Kast}, and~\eqref{eq:ddCext_bdry_dfn}, we have $\Chat\cdot\Qla\subset\Meas(\ddG,\ddwtlap)$. Since $\la\subset C$ and $\la$ spans the kernel of $\Qla$, we have $\dim(\Chat\cdot \Qla)=k-2$, which implies~\eqref{eq:Meas_ddG=Chat_Qla}. 
\end{proof}

\begin{remark}[Inverse T-duality]\label{rmk:SHIFT:inverse}
In summary, T-duality transforms an input tuple $(\G,\wt,\la\subset C)$ into an output tuple $(\ddG,\ddwtlap,\la\subset\ddC^\perp)$, where
\begin{itemize}
\item $(\G,\wt)$ is a \bdash trivalent weighted graph of type $(k,n)$ satisfying $\helWmin(\G)\geq2$ and $\helBmin(\G)\geq0$;
\item $(\ddG,\ddwtlap)$ is a \wdash trivalent weighted graph of type $(k-2,n)$ satisfying $\helWmin(\ddG)\geq0$ and $\helBmin(\ddG)\geq2$;
\item $C:=\Meas(\G,\wt)$, $\ddC:=\Meas(\ddG,\ddwtlap)$, and $\la\in\lak$ satisfies $\la\subset C$ and $\la\subset\ddC^\perp$.
\end{itemize}
The inverse of this operation is obtained by simply interchanging the roles of black and white colors; cf. \crefi{lemma:shift1-top}{shift1-top4}. Explicitly, recall from \cref{rmk:altp_vs_color_swap} that for $(\ddG^\vee,\ddwtlap)$ obtained from $(\ddG,\ddwtlap)$ by swapping black and white colors, we have $\Meas(\ddG^\vee,\ddwtlap)=\ddC^\vee:=\altp(\ddC)$. Let $\la^\vee:=\diag(1,-1)\cdot \alt(\la)\in\Gror(2,n)$. We have $\wind(\la^\vee) = n\pi - \wind(\la)$ by~\Mref{eq:wind_alt}, so $\la^\vee\in\labf_{\ddkvee,n}^+$ for $\ddkvee:=n-k+2 = \dim \ddC^\vee$. 
 Applying T-duality to $(\ddG^\vee,\ddwtlap,\la^\vee\subset \ddC^\vee)$, we arrive at $(\G^\vee,\wtlap,\la^\vee\subset (C^\vee)^\perp)$, where
 $C^\vee = \altp(C)=\Meas(\G^\vee,\wtlap)$ as in \cref{rmk:altp_vs_color_swap}.
\end{remark}

\subsection{Shifting edge weights by $1$ twice}\label{sec:shift:by1}

We briefly explain how to transform the edge weights under the shift by $1$ (\cref{ssec:SHIFT:by1}). A closely related construction appears in~\cite[Section~8]{crit}; see also~\cite{Affolter}. 
Up to applying gauge equivalences at the vertices of $\G,\dG,\ddG$, the Kasteleyn edge weight transformation $\wtlaK\mapsto\ddwtlaK$ may be factored through two applications $\wtlaK\mapsto\dwtKrow\mapsto\ddwtlaK$ of the shift by $1$. 

Let $\row\in \la$ be a row of $\la\subset C$ and let $\rowext\in\Hwspace_{\R}\HtripK$ be its \wdash holomorphic extension. We assume that $\row$ is \emph{generic} in the sense that $\rowext(\wv)\neq0$ for all $\wv\in\WVint$. Let $\bv\in\BVint$ be a trivalent black vertex of $\G$.
We set $\dwtKrow(\dwv(\bv),\dbv(\f_{\s})):=\frac{\wtK(\e_\s)}{\rowext(\wv_{\s+1})\rowext(\wv_{\s+2})}$ for $\s=1,2,3$; cf. \figref{fig:black3}(a). 
We also set $\dCext(\dwv(\bv)):= \wtK(\e_\s)^{-1}(
\rowext(\wv_{\s+1})\Cext(\wv_{\s+2}) - \rowext(\wv_{\s+2})\Cext(\wv_{\s+1}))$, where it is easy to check that the right-hand side of this expression does not depend on $\s=1,2,3$. 
Similarly to the proof of \cref{lemma:SHIFT:ddCext_in_Hwspace}, one can show that $\dCext:\dWV\to\R^{k-1}$ is a \wdash holomorphic function on $\dG$ with respect to the intermediate edge weights $\dwtKrow$. 
However, these edge weights do not in general satisfy the Kasteleyn sign condition (unlike in the case of \cref{lemma:SHIFT:ddepsKla_is_Kast}).

\section{\LpuncTITLE positive Grassmannian}\label{sec:LPUNC}
We introduce two \Lpunc generalizations of Postnikov's totally nonnegative Grassmannian~\cite{Pos}. We will eventually relate each generalization to amplituhedra in momentum and momentum-twistor space, respectively, and 
relate them to each other via T-duality. 

\subsection{Double dimers and \KSprims}\label{ssec:LPUNC:Lfunc_graphs}
Our first goal is to relate the \KSprims introduced in~\eqref{eq:OCP:primitive} to the combinatorics of the \emph{double-dimer model} on $\G$. 

\begin{definition}
A \emph{double-dimer configuration} on $\G$ is a multiset $\Om$ of edges of $\G$ that uses every interior (resp., boundary) vertex of $\G$ exactly twice (resp., at most twice). The set of double-dimer configurations on $\G$ is denoted $\OmsG$.
\end{definition}
\noindent Thus, $\Om$ consists of (i) doubled edges, (ii) cycles of even length, and (iii) \emph{\bdbd paths} starting and ending at the boundary of $\G$. Given a weighted graph $(\G,\wt)$, we let $\wt(\Om):=2^{\ncycOm}\prod_{\e\in\Om} \wt(\e)$, where the product is taken with multiplicity and $\ncycOm$ is the number of cycles in $\Om$ (excluding doubled edges).

Following \cref{rmk:DIM:otimes}, we denote $\HHspaceRknk:=\Hwspace_{\R^k}\HtripK\times\Hbspace_{\R^{n-k}}\HtripK$.%

\begin{definition}\label{dfn:LPUNC:Hknk}
Assume that $\G$ admits an \APM. 
Let $\wt\in\Rtpgauge$ and $C:=\Meas(\G,\wt)\in\Grtnn(k,n)$. 
Let $(\Fwk,\Fbnk)\in\HHspaceRknk$ be the discrete holomorphic extensions of $(C,C^\perp)$, and let
$\Hknk:\Faces\to\Rknk$ be the \KSprim of $(\Fwk,\Fbnk)$. That is, for $\f_1,\f_2,\e,\b,\w$ as in~\eqref{eq:OCP:primitive}, $\Hknk$ satisfies
\begin{equation}\label{eq:LPUNC:primitive_knk}
 \Hknk(\f_2)-\Hknk(\f_1) = \wtK(\e)\cdot \Fbnk(\bv) \cdot \Fwk(\wv)^T.
\end{equation}
This defines $\Hknk$ up to an additive constant.
\end{definition}

\begin{remark}\label{rmk:LPUNC:GL_x_GL}
The group $\GL_k(\R)$ (resp., $\GL_{n-k}(\R)$) acts on the columns of $C$ (resp., $C^\perp$) and thus on the vectors $\Fwk(\w)$ (resp., $\Fbnk(\b)$). Consequently, it acts on $\Hknk(\ff)$ by right (resp., left) multiplication. 
\end{remark}

A $\GL_{n-k}(\R)\times\GL_k(\R)$-invariant way to think of $\Hknk(\ff)$ is to view it as an element of $\Hom(C,\Cpast)$. Explicitly, choose row vectors $(\row,\rowt)\in C\times C^\perp$ and consider their discrete holomorphic extensions $(\rowext,\rowtext)\in\HHspaceR$. Let $H_{1\times 1}:\Faces\to\R$ be the \KSprim of $(\rowext,\rowtext)$. Thus, for each face $\ff\in\Faces$, $H_{1\times 1}(\ff):C\times C^\perp\to\R$ is a bilinear map, i.e., an element of $\Hom(C,\Cpast)$, and $\Hknk(\ff)$ defined in~\eqref{eq:LPUNC:primitive_knk} represents this linear map by an explicit $\knk$ matrix. 

\begin{remark}\label{rmk:LPUNC:tangent}
It is well known that the tangent space $\TC\Gr(k,n)$ to $\Gr(k,n)$ at a point $C\in\Gr(k,n)$ is canonically identified with $\Hom(C,\R^n/C)\cong\Hom(C,\Cpast)$. Thus, one may view $(\Hknk(\ff))_{\ff\in\Faces}$ as an arrangement of $|\Faces|$ points in $\TC\Gr(k,n)$, well defined up to shift. 
\end{remark}

\begin{figure}
 \def\inputscale{1.2}
 \setlength{\tabcolsep}{0pt}
\begin{tabular}{cccc}
\hspace{-5pt}
\includegraphics[scale=\inputscale]{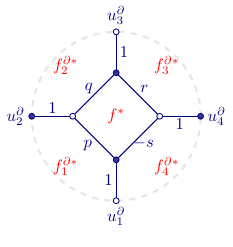}
&
\includegraphics[scale=\inputscale]{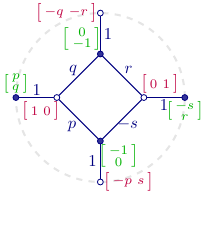}
&
\includegraphics[scale=\inputscale]{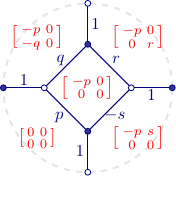}
&
\hspace{3pt}\includegraphics[scale=\inputscale]{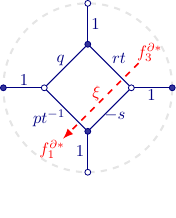}
\\[-13pt]
 (a) $\textcolor{\plabicEdgeColor}{(\G,\wtK)}$ 
& (b) $(\textcolor{\FwColor}{\Fwk},\textcolor{\FbColor}{\Fbnk})$
& (c) $\textcolor{\puncColor}{\Hknk}$
& (d) $\textcolor{\puncColor}{\Cut}$ %
\end{tabular}
 \caption{\label{fig:holom2} Computing the \KSprim $\textcolor{\puncColor}{\Hknk}:\Faces\to\Rknk$.}
\end{figure}

We explain a direct way to compute $\Hknk(\ff)-\Hknk(\f)\in\TC\Gr(k,n)$ for $\ff,\f\in\Faces$. 
Choose a simple path $\Cut$ in $\GD$ from $\ff$ to $\f$. Consider a curve $\Coft:=\Meas(\G,\wtoft)$ in $\Gr(k,n)$, where $\wtoft$ is obtained from $\wt$ by multiplying each edge weight $\wt(\e)$ by $t^{\<\e,\Cut\>}$, with the intersection number $\<\e,\Cut\>\in\{-1,0,1\}$ equal to $+1$ (resp., $-1$) if $\e$ intersects $\Cut$ with the white vertex of $\e$ appearing to the left (resp., to the right) of $\Cut$, and $0$ otherwise; see \figref{fig:holom2}(d). Thus, $C(t=1) = C = \Meas(\G,\wt)$,
 and we are interested in the derivative $C'(t=1)$ of $C(t)$ at $t=1$. 

\begin{proposition}\label{lemma:LPUNC:H_vs_C'(1)}
In the above notation, 
\begin{equation}\label{eq:H_vs_C'(1)_invariant}
	 \Hknk(\ff)-\Hknk(\f) = C^\perp\cdot C'(1)^T.
\end{equation}
Speaking in $\GL_{n-k}(\R)\times\GL_k(\R)$-invariant terms, the derivative $C'(1)\in\TC\Gr(k,n)$ of the curve $C(t)$ at $t=1$ coincides with $\Hknk(\ff)-\Hknk(\f)\in\TC\Gr(k,n)$. 
\end{proposition}
\begin{proof}
Fix an arbitrary function $\eh:\Edges\to\Z$ and consider deformed edge weights $\wtoft(\e):=\wt(e) t^{\eh(e)}$ and $\wtKoft(\e):=\wtK(\e) t^{\eh(\e)}$. Let $\Coft:=\Meas(\G,\wt_t)$ and let $\Cextoft\in\Hwspace_{\R^k}\HtripKoft$ be the \wdash holomorphic extension of $\Coft$. We claim that 
\begin{equation}\label{eq:Cp_C(t)_=_sum_edges}
	C^\perp \cdot \Coft^T = \sum_{\e\in\Edges} (\wtK(\e) - \wtKoft(\e))\cdot \Fbnk(\bv) \cdot \Cextoft(\wv)^T,
\end{equation}
where $\bv,\wv$ are the black and white endpoints of $\e$. Indeed, by~\eqref{eq:OCP:holom}, for each $\bv\in\BVint$, we have 
$\sum_{\e\sim\bv} \wtKoft(\e) \cdot \Fbnk(\bv) \cdot \Cextoft(\wv)^T = \bzero_{(n-k)\times k}$, and for each $\wv\in\WVint$, we have 
$\sum_{\e\sim\wv} \wtK(\e) \cdot \Fbnk(\bv) \cdot \Cextoft(\wv)^T = \bzero_{(n-k)\times k}$. 
Adding up these contributions, we see that the interior vertices contribute zero to the right-hand side of~\eqref{eq:Cp_C(t)_=_sum_edges}, and by~\eqref{eq:MCE:alt(C)_vs_pFw}--\eqref{eq:MCE:alt(Cp)_vs_pFb}, the boundary vertex contributions give precisely $C^\perp \cdot \Coft^T$. 

Given $\ff,\f\in\Faces$, choose a cut $\Cut$ from $\ff$ to $\f$ and set $\eh(\e):=\<\e,\Cut\>$. Differentiating both sides of~\eqref{eq:Cp_C(t)_=_sum_edges} by $t$ and substituting $t=1$, we find
\begin{equation}\label{eq:Cp_C'(1)_=_H_1-H_2}
	C^\perp \cdot C'(1)^T = -\sum_{\e\in\Edges} \<\e,\Cut\>\cdot \wtK(\e)\cdot \Fbnk(\bv) \cdot \Fwk(\wv)^T = \Hknk(\ff)-\Hknk(\f).\qedhere
\end{equation}
\end{proof}

Next, we give a probabilistic interpretation of the entries and, more generally, the minors of $\Hknk(\ff)-\Hknk(\f)$ for $\ff,\f\in\Faces$. First, pick a set $I\in{\brn\choose k}$ such that $\Delta_I(C)>0$ and put $C$ in \emph{reduced row-echelon form}, with an identity submatrix $C|_I = \Id_k$ on column set $I$. Let $C|_{\comp{I}}$ be the submatrix of $C$ on the remaining columns. An explicit representative matrix for $C^\perp$ can be chosen so that $C^\perp|_{\comp{I}} = \Id_{n-k}$ and $C^\perp|_I = -(C|_{\comp{I}})^T$. Choosing such representatives fixes the $\GL_{n-k}(\R)\times\GL_k(\R)$-redundancy.

It is natural to label the rows (resp., columns) of $\Hknk(\ff)-\Hknk(\f)$ by elements of $\comp{I}$ (resp., $I$). Pick subsets $A\subset I$ and $B\subset \comp{I}$ of the same size $d:=|A|=|B|$. Let $S:=A\sqcup B$ and $T:=I\setminus A$. Thus, $|S|=2d$, and we write $S=\{s_1<s_2<\dots<s_{2d}\}$. Let $\Sev:=\{s_2,s_4,\dots,s_{2d}\}$ and $\Sodd:=\{s_1,s_3,\dots,s_{2d-1}\}$. Let $\OmsGST\subset\OmsG$ be the set of double-dimer configurations $\Om=\apmev\cup\apmodd$ obtained by superimposing \APMs $\apmev\in\APMSGbd(T\sqcup\Sev)$ and $\apmodd\in\APMSGbd(T\sqcup\Sodd)$. Thus, each $\Om\in\OmsGST$ contains precisely $d$ \bdbd paths $\Pathbd_1(\Om),\dots,\Pathbd_d(\Om)$, all of which have both endpoints in $S$. (Conversely, any noncrossing pairing of the elements of $S$ consists of arcs connecting $\Sodd$ to $\Sev$.) We orient each \bdbd path $\Pathbd_i(\Om)$ from $\Sev$ to $\Sodd$, and consider its signed intersection number $\<\Pathbd_i(\Om),\Cut\>\in\{-1,0,1\}$ with the cut $\Cut$ from $\ff$ to $\f$. We will be interested in the signed sum
\begin{equation}\label{eq:LPUNC:ZSTcut}
	\ZSTcut = \sum_{\Om\in\OmsGST} \wt(\Om) \chicut(\Om),\quad\text{where}\quad \chicut(\Om):=\prod_{i=1}^d \<\Pathbd_i(\Om),\Cut\>\in\{-1,0,+1\}.
\end{equation}
Observe that $\chicut(\Om)$ is zero unless all \bdbd paths of $\Om$ separate $\ff$ from $\f$. 

\begin{remark}\label{rmk:LPUNC:prob_ZST}
In other words, when $\ZST:= \Delta_{T\sqcup\Sev}(\G,\wt)\Delta_{T\sqcup\Sodd}(\G,\wt) = \sum_{\Om\in\OmsGST}\wt(\Om)$ is nonzero, 
$\EST(\chicut) = \ZSTcut/\ZST$ is the expected value of the random variable $\chicut(\Om)$ 
 on the probability space $\APMSGbd(T\sqcup\Sev)\times\APMSGbd(T\sqcup\Sodd)$, with the probability of $(\apmev,\apmodd)$ proportional to $\wt(\apmev)\cdot \wt(\apmodd)$. 
Here, the multiset union $\Om:=\apmev\cup\apmodd\in\OmsGST$ is viewed as a random double-dimer configuration. 
\end{remark}

The following result gives a generalization of~\Mref{lemma:TE:OmsGf_pos}.
\begin{proposition}\label{lemma:LPUNC:minors_dH}
Assume that $(C,C^\perp)$ are in reduced row-echelon form with respect to $(I,\comp{I})$ 
and let $A\subset I$, $B\subset \comp{I}$ with $d:=|A|=|B|$, $T:=I\setminus A$, $S:=A\sqcup B$, and $\Cut$ be as above. 
The $B\times A$ minor of $\Hknk(\ff)-\Hknk(\f)$ is given by
\begin{equation}\label{eq:LPUNC:minors_dH}
	\Delta_{B,A}(\Hknk(\ff)-\Hknk(\f)) = (-1)^{\inv(I,S)} \frac{\ZSTcut}{\Delta_I(\G,\wt)^2}, 
 \quad\text{where}\quad
 \inv(I,S):=|\{(i,s)\in I\times S\mid i<s\}|.
\end{equation}
\end{proposition}
\begin{proof}
Set $\wtoft(\e):=t^{\<\e,\Cut\>}\wt(\e)$ as in the proof of \cref{lemma:LPUNC:H_vs_C'(1)} and write $S = \{s_1<s_2<\dots<s_{2d}\}$. 
Applying the Cauchy--Binet formula and using~\eqref{eq:altp_dfn}, we get
\begin{equation*}%
 (-1)^{\inv(I,S)}\Delta_{B,A}(C^\perp \cdot \Coft^T)
 = \sum_{R\in{S\choose d}} (-1)^{|R\cap\Sev|} \cdot \Delta_{T\sqcup R}(C(t)) \cdot \Delta_{T\sqcup (S\setminus R)}(C)
 = \frac{\sum_{\Om\in\OmsGST} c_\Om(t)}{\Delta_I(\G,\wt)\Delta_I(\G,\wtoft)}.
\end{equation*}
In the second identity, we have used that 
$\Delta_{J}(C) = \frac{\Delta_J(\G,\wt)}{\Delta_I(\G,\wt)}$ for $J\in{\brn\choose k}$ (and similarly for $C(t)$), 
and that superimposing \APMs 
$(\apm,\apm')\in\APMSGbd(T\sqcup R)\times\APMSGbd({T\sqcup(S\setminus R)})$ 
 produces a double-dimer $\Om=\apm\cup\apm'\in\OmsGST$. Conversely, any such pair $(\apm,\apm')$ is obtained by splitting each cycle and \bdbd path of $\Om$ into alternating sets of edges. 
 Switching the two \APMs on a \bdbd path $\Pathbd_i(\Om)$ reverses the sign $(-1)^{|R\cap\Sev|}$, since the
endpoints of $\Pathbd_i(\Om)$ have opposite parity in $S$; switching them on a cycle preserves
the sign. 

Let us split $\Om=\apmev\cup\apmodd$ with $(\apmev,\apmodd)\in \APMSGbd(T\sqcup\Sev)\times\APMSGbd(T\sqcup\Sodd)$ (in one of $2^{\ncycOm}$-many ways). For each \bdbd path $\Pathbd_i(\Om)$, we set 
$\ehev(\Pathbd_i(\Om)) := \sum_{\e\in\Pathbd_i(\Om)\cap\apmev} \<\e,\Cut\>$ and 
$\ehodd(\Pathbd_i(\Om)) := \sum_{\e\in\Pathbd_i(\Om)\cap\apmodd}\<\e,\Cut\>$,
and similarly for each cycle $\Cyc$ of $\Om$. Thus, 
\begin{equation*}%
	c_\Om(t) = t^{\ehdbl(\Om)}\cdot \prod_{\Cyc\in\Cycop(\Om)}(t^{\ehodd(\Cyc)} + t^{\ehev(\Cyc)}) \cdot 
\prod_{i=1}^d (t^{\ehodd(\Pathbd_i(\Om))} - t^{\ehev(\Pathbd_i(\Om))})\cdot \prod_{\e\in\Om}\wt(\e),
\end{equation*}
where $\ehdbl(\Om)$ is the sum of $\<\e,\Cut\>$ over doubled edges $\e$ of $\Om$.

Write $\hbar:=t-1$, and put 
$M(\hbar):=C^\perp \cdot \Cofh^T$. 
By~\eqref{eq:Cp_C'(1)_=_H_1-H_2}, 
we have $M(\hbar) = {(\Hknk(\ff)-\Hknk(\f)) \hbar} + O(\hbar^2)$, so 
$\Delta_{B,A}(\Hknk(\ff)-\Hknk(\f))$ is the coefficient of $\hbar^d$ in $\Delta_{B,A}(M(\hbar))$. For each $\Om\in\OmsGST$, taking the $2^{\ncycOm}$ factor into account and using 
$t^{\ehodd(\Pathbd_i(\Om))} - t^{\ehev(\Pathbd_i(\Om))} = \hbar\<\Pathbd_i(\Om),\Cut\> + O(\hbar^2)$,
 we get
$c_\Om(1+\hbar) = \wt(\Om)\cdot \chicut(\Om)\cdot \hbar^d + O(\hbar^{d+1})$.
Summing over $\Om\in\OmsGST$, we obtain~\eqref{eq:LPUNC:minors_dH}. 
\end{proof}

\begin{corollary}\label{lemma:LPUNC:dist}
Assume that the edge weights $\wt\in\Rtpgauge$ are generic and let $\Hknk$ be given by~\eqref{eq:LPUNC:primitive_knk}.
For $\ff,\f\in\Faces$, let $\distrk(\ff,\f):=\rank(\Hknk(\ff)-\Hknk(\f))$. 
Let $\distsep(\ff,\f)$ be the maximum over all $\Om\in\OmsG$ of the number of \bdbd paths of $\Om$ separating $\ff$ from $\f$. 
 Then 
\begin{equation}\label{eq:LPUNC:dist}
 \distrk(\ff,\f) = \distsep(\ff,\f)\quad\text{for all $\ff,\f\in\Faces$}. 
\end{equation}
\end{corollary}
\begin{proof}
 By~\eqref{eq:LPUNC:minors_dH}, we have
\begin{equation}\label{eq:distrk<distsep}
	\distrk(\ff,\f) \leq \distsep(\ff,\f) \quad\text{for all $\ff,\f\in\Faces$ and $\wt\in\Rtpgauge$.}
\end{equation}
Conversely, observe that $\distrk(\ff,\f)$ is $\GL_{n-k}(\R)\times\GL_k(\R)$-invariant and thus independent of the choice of $I$ in~\eqref{eq:LPUNC:minors_dH}. Let $\Om$ be a double-dimer with $d=\distsep(\ff,\f)$ \bdbd paths separating $\ff$ from $\f$. Replacing its remaining \bdbd paths with doubled edges if necessary, we assume that each \bdbd path of $\Om$ separates $\ff$ from $\f$. 
 Let us split $\Om=\apmev\cup\apmodd$ as above and let $T:=\partial\apmev\cap\partial\apmodd$, 
 $\Sev:=\partial\apmev\setminus T$, and $\Sodd:=\partial\apmodd\setminus T$.
Applying~\eqref{eq:LPUNC:minors_dH} with $I:=T\sqcup\Sev$, $A:=\Sev$, and $B:=\Sodd$, we see that $\Om$ contributes to the right-hand side of~\eqref{eq:LPUNC:minors_dH}. Since the edge weights $\wt$ are generic, there is no cancellation in~\eqref{eq:LPUNC:ZSTcut}, so $\Delta_{B,A}(\Hknk(\ff)-\Hknk(\f))\neq0$, and thus $\distrk(\ff,\f)\geq\distsep(\ff,\f)$. 
\end{proof}

\begin{conjecture}\label{conj:dist}
For $\ff\neq\f$, let
$\distcurvepre(\ff,\f):=\min_{\Rg} \gelmin(\Rg)$,
where the minimum is taken over all \sconn subsets $\emptyset\neq\Rg\subset\Vint$ such that each of $\ff,\f$ is incident to a vertex of $\Rg$, 
and $\gelmin(\Rg):=\min(\gelW(\Rg),\gelB(\Rg))$.\footnote{We also set $\distcurvepre(\ff,\ff)=0$, and when $\ff\neq\f$ share a \bdbd edge, we set $\distcurvepre(\ff,\f)=1$.} 
Let 
$\distcurve(\ff,\f) := \min_{(\ff=\ff_0,\ff_1,\dots,\ff_m=\f)} \sum_{i=1}^m\distcurvepre(\ff_{i-1},\ff_i)$. 
 Then 
\begin{equation}\label{eq:conj_dist}
 \distsep(\ff,\f) = \distcurve(\ff,\f) \quad\text{for all $\ff,\f\in\Faces$}. 
\end{equation}
\end{conjecture}
\noindent We treat the equivalence $\distsep(\ff,\f)\geq2$ $\Longleftrightarrow$ $\distcurve(\ff,\f) \geq2$ 
geometrically in \cref{lemma:fullysepMCE_vs_fullysep} below.

\begin{example}
Consider a weighted graph $(\G,\wt)$ with Kasteleyn edge weights shown in \figref{fig:holom2}(a), where $p,q,r,s>0$. For $I=\{2,4\}$, we have $C =\Meas(\G,\wt) = \begin{pmatrix}
p & 1 & q & 0\\
-s & 0 & r & 1
\end{pmatrix}$ and $C^\perp = \begin{pmatrix}
1 & -p & 0 & s\\
0 & -q & 1 & -r
\end{pmatrix}$. The discrete holomorphic extensions $(\Fwk,\Fbnk)\in\HHspaceRknk$ are shown in \figref{fig:holom2}(b) and their \KSprim $\Hknk:\Faces\to\Rknk$ is shown in \figref{fig:holom2}(c). 
To compute $\Hknk(\bdf_3)-\Hknk(\bdf_1)$ as in \figref{fig:holom2}(d), we pick a cut $\Cut$ from $\bdf_3$ to $\bdf_1$ crossing the edges of weight $r$ and $p$. The corresponding edge weight transformation is given by $(p,q,r,s)\mapsto(pt^{-1},q,rt,s)$. We get 
$C(t) = \begin{pmatrix}
pt^{-1} & 1 & q & 0\\
-s & 0 & rt & 1
\end{pmatrix}$, 
$C'(1) = \begin{pmatrix}
-p & 0 & 0 & 0\\
0 & 0 & r & 0
\end{pmatrix}$, and 
$C^\perp\cdot C'(1)^T = \begin{pmatrix}
-p & 0\\
0 & r
\end{pmatrix} = \Hknk(\bdf_3)-\Hknk(\bdf_1)$, 
in agreement with~\eqref{eq:H_vs_C'(1)_invariant} and \figref{fig:holom2}(c). We have $T=\emptyset$, $S = \{1,2,3,4\}$, $\inv(I,S)=2$, and $\ZSTcut = -pr = \det(\Hknk(\bdf_3)-\Hknk(\bdf_1))$, in agreement with~\eqref{eq:LPUNC:minors_dH}.
\end{example}

\subsection{\LfuncTITLE positive Grassmannian}\label{sec:LPUNC:Gr}

\begin{definition}[\Twosep faces]\label{dfn:PROP:2sep}
We say that $\ff,\f\in\Faces$ are \emph{\twosep} if $\distsep(\ff,\f)\geq2$ in the notation of \cref{lemma:LPUNC:dist} (that is, if there exists $\Om\in\OmsG$ containing at least two boundary-to-boundary paths separating $\ff$ from $\f$). We denote by $\OmsG(\ff\ssep\f)$ the set of such $\Om$.
\end{definition}

\begin{definition}\label{dfn:LPUNC:Lfunc_graph}
An \emph{\Lfunc} graph $\Gfuncpair$ is a planar bipartite graph $\G$ equipped with a choice of an $\nL$-tuple $\Plocs=(\ploc_1,\ploc_2,\dots,\ploc_L)$ of faces of $\G$. We refer to $\ploc_1,\ploc_2,\dots,\ploc_L$ as \emph{\Lfunctures}.
\end{definition}

\begin{definition}\label{dfn:fullysep}
We consider the alphabet $\brnbd:=\{\xbd1,\xbd2,\dots,\nbd\}$ and denote $\ploc_{\ibd}:=\bdf_i$ for $\ibd\in\brnbd$. We let $\brnbdsep:=\{\{\ibd,\jbd\}\subset\brnbd\mid j\not\equiv i,i\pm1\ \mod\ n\}$ and
\begin{equation}\label{eq:brnLbdsep_dfn}
 \brnLbdsep:=\brnbdsep\sqcup(\brnbd\times\brnL)\sqcup {\brnL\choose 2}.
\end{equation}
We say that an \Lfunc graph $\Gfunc$ is \emph{\fullysep} if for any $\sepst\in\brnLbdsep$, the faces $\ploc_\seps$ and $\ploc_\sept$ are \twosep.
\end{definition}

Recall from \cref{dfn:LPUNC:Hknk} that $\Hknk:\Faces\to\Rknk$ is defined up to an additive constant. From now on, we assume that it is \emph{in normal form} meaning $\Hknk(\bdf_1)=\bzero_{\knk}$. 

\begin{definition}[\Lfunc boundary measurement map]\label{dfn:LPUNC:Meas_func}
We denote $\Hknkloc_\sepr:=\Hknk(\ploc_\sepr)$ for $\sepr\in\brnLbd$, 
$\Hknklocsbd=(\Hknkloc_{\xbd1},\dots,\Hknkloc_{\xbd{n}})$, 
$\Hknklocs:=(\Hknkloc_1,\dots,\Hknkloc_\nL)$, and 
\begin{align}\label{eq:Measknk_dfn}
 \Measknk(\Gfunc,\wt)&:=\CHL\in\TLbundle\Gr(k,n), \quad\text{where}\\
 \TLbundle\Gr(k,n)&:=\{(C;\Hloc_1,\dots,\Hloc_\nL)\mid C\in\Gr(k,n),
\ 
\Hloc_1,\dots,\Hloc_{\nL}\in\TC\Gr(k,n)\}
\end{align}
is the \emph{$\nL$-fold tangent bundle}; cf. \cref{rmk:LPUNC:tangent}. 
We view the boundary measurement $\Measknk(\Gfunc,\wt)$ as defined modulo the simultaneous action of $\GL_{n-k}(\R)\times\GL_k(\R)$ on $C$, $C^\perp$, and $\Hknklocs$; cf. \cref{rmk:LPUNC:GL_x_GL}.
\end{definition}

\begin{remark}\label{rmk:H_bdry}
By~\eqref{eq:TE:t_imm_bdry_vs_pFw_pFb}, the boundary values 
$\Hknkloc_{\jbd} = \Hknkloc_{\xbd1} + \sum_{i=2}^j (C^\perp)_i\cdot (C_i)^T$ are fully determined by $C$ and $C^\perp$. 
\end{remark}

\begin{definition}[\Lfunc totally nonnegative Grassmannian]\label{dfn:Lfunc_tnn_Gr}
Let
\begin{align*}
 \GrtnnLfunc(k,n)&:=\left\{\Measknk(\Gfunc,\wt)\in\TLbundle\Gr(k,n)\middle|\text{
\begin{tabular}{c}
$(\Gfunc,\wt)$ is a weighted \Lfunc \\
planar bipartite graph of type $(k,n)$\\
admitting an \APM
\end{tabular}
}\right\},\\
 \GrtnnLfuncx_2(k,n)&:=\left\{\Measknk(\Gfunc,\wt)\in\GrtnnLfunc(k,n)\middle| \text{$\Gfunc$ is \fullysep}\right\}.
\end{align*}
\end{definition}

We leave the following important problem for future work; see also \cref{que:LPUNC:GrtnnLvuncproj=GrtnnLvuncamb} below.
\begin{problem}\label{problem:GrtnnLfunc_by_ineq}
Describe $\GrtnnLfunc(k,n)\subset\TLbundle\Gr(k,n)$ by algebraic inequalities.
\end{problem}

\begin{remark}
Following~\Mref{ssec:annular_immanants}, we consider a natural collection of functions on a weighted \Lfunc graph $(\Gfunc,\wt)$ called \emph{\Lfunc immanants}. Every double-dimer configuration $\Om\in\OmsG$ gives rise to a \emph{lamination} $\lamin_\Om$ of an \Lpunc disk, i.e., a collection of disjoint closed curves and \bdbd paths avoiding the \Lfunctures, considered up to isotopy. We also let $\TOm\subset\brn$ be such that $\TOm=\partial\apm\cap\partial\apm'$ for any splitting $\Om=\apm\cup\apm'$ of $\Om$ as a union of two \APMs. 
Thus, for each lamination $\lamin$ and each $T\subset\brn$, we get a ``double-dimer immanant'' $\Delta_{\lamin,T}(\Gfunc,\wt):=\sum_{\Om\in\OmsG:\ (\lamin_\Om,\TOm)=(\lamin,T)} \wt(\Om)$. These functions generalize the \emph{Temperley--Lieb immanants} of~\cite{RhSk,Lam_dimers} and are closely related to the constructions of~\cite{FoGo,Kenyon_conf_loops}. This provides a potentially promising approach to \cref{problem:GrtnnLfunc_by_ineq}.
\end{remark}

\begin{definition}\label{dfn:Lfunc_positroid_cell_and_reduced}
We define an \emph{\Lfunc positroid cell} 
$\PtpLfunc_{\Gfunc}:=\{\Measknk(\Gfunc,\wt)\mid \wt\in\Rtpgauge\}$. 
An \Lfunc graph $\Gfunc$ is called \emph{reduced} if the map $\Measknk(\Gfunc,\cdot):\Rtpgauge\to\PtpLfunc_{\Gfunc}$ is a homeomorphism.
\end{definition}

\begin{problem}
Describe reduced \Lfunc graphs and elementary moves relating them.
\end{problem}

\begin{remark}
A natural gauge-invariant coordinate system on $\Rtpgauge$ is given by the \emph{face weights} $\facewtall=(\facewt(\ff))_{\ff\in\Faces}\in\Rtp^{|\Faces|}$ (subject to the relation $\prod_{\ff\in\Faces} \facewt(\ff) = 1$), where $\facewt(\ff)$ is the alternating product of edge weights around the boundary of $\ff$. 
For $\f\in\Faces$, let $\facewtallg(t)$ be obtained from $\facewtall$ by multiplying $\facewt(\f)$ by $t$ and dividing $\facewt(\bdf_1)$ by $t$. 
 It follows from \cref{lemma:LPUNC:H_vs_C'(1)} that the tangent point $\Hknk(\f)\in\TC\Gr(k,n)$ is the image of $\facewtallg'(1)\in T_{\facewtall}\Rtpgauge$ under the differential of the boundary measurement map. 
 The tangent points $(\facewtallg'(1))_{\f\in\Faces}$ form a dual affine linear basis to the $\dlog$ forms $(\dlog\facewt(\ff))_{\ff\in\Faces}$ in the sense that $\<\facewtallg'(1),\dlog\facewt(\ff)\> = \delta_{\ff,\f}$ (Kronecker delta) for all $\ff,\f\in\Faces\setminus\{\bdf_1\}$,
where $\<\cdot,\cdot\>$ denotes the standard pairing between $T_{\facewtall}\Rtpgauge$ and $T^\ast_{\facewtall}\Rtpgauge$. 
\end{remark}

\begin{remark}
Let $\G$ be a reduced (unpunctured) planar bipartite graph as in \cref{dfn:reduced} and let $d:=|\Faces|-1 = \dim\Ptp_{\G}$. 
Then $\Meas(\G,\cdot):\Rtp^d\to\Ptp_{\G}$ is a diffeomorphism,
so the points $(\Hknk(\ff))_{\ff\in\Faces}$
 form an affine linear basis of the tangent space $\TC\Ptp_{\G}$. The dual affine linear basis is given by $(\dlog\facewt(\ff))_{\ff\in\Faces}$, where $\facewt(\ff)$ is now viewed as a function on $\Ptp_{\G}$ (described explicitly in~\cite{MuSp}). 
\end{remark}

\begin{definition}
In the notation of \cref{dfn:LPUNC:Hknk}, let $\la\in\Gr(2,C)$ and let $\laext\in\Hwspace_{\R^2}\HtripK$ be its \wdash holomorphic extension. 
We let $\Hdnk:\Faces\to\Rdnk$ be the \KSprim of $(\laext,\Fbnk)\in\HHspaceRdnk$. 
Given $\lat\in\Gr(2,C^\perp)$, we extend it to $\latext\in\Hbspace_{\R^2}\HtripK$ and let $\Hdnd:\Faces\to\Rdnd$ be the \KSprim of $(\laext,\latext)\in\HHspaceRdnd$. We denote $\Hdnkloc_\rho:=\Hdnk(\ploc_\rho)$ and $\Hdndloc_\rho:=\Hdnd(\ploc_\rho)$ for $\rho\in\brnLbd$ and set 
\begin{equation}\label{eq:LPUNC:Measdnk_def}
 \Measdnk(\Gfunc,\wt):=(C;\Hdnklocs)\quad\text{and}\quad
 \Measdnd(\Gfunc,\wt):=(C;\Hdndlocs).
\end{equation}
\end{definition}

\subsection{\LfuncTITLE graphs of type $(1,n)$}
Let $(\Gfunc,\wt)$ be a weighted \Lfunc graph of type $(1,n)$ and assume that $C:=\Meas(\G,\wt)\in\Grtp(1,n)$. Recall that for each $\sept\in\brnbd$, $\Hknkloc_\sept\in \TC\Grtp(1,n)$ is uniquely determined by $C$ via \cref{rmk:H_bdry}. 

\begin{proposition}\label{prop:ccc}
For each $\sepr\in\brnL$, $\Hknkloc_\sepr\in\TC\Grtp(1,n)$ may be written as a convex combination of the points 
$\Hknklocsbd=(\Hknkloc_{\xbd1},\dots,\Hknkloc_{\xbd{n}})$. That is, there exists a nonempty subset $\Sloc_\rho\subset\brn$ and a tuple 
\begin{equation}\label{eq:bcccloc_dfn_type_1n}
 \bcccloc_\rho=(\cccx(\rho,i))_{i\in\Sloc_\rho}\in\Rtp^{\Sloc_\rho}
 \quad\text{such that}\quad
 \sum_{i\in\Sloc_\rho} \cccx(\rho,i) = 1
 \quad\text{and}\quad
\Hknkloc_\rho = \sum_{i\in\Sloc_\rho} \cccx(\rho,i) \Hknkloc_{\xbd{i}}.
\end{equation}
\end{proposition}
\begin{proof}
Following \cref{rmk:LPUNC:prob_ZST}, 
 for distinct $i,j\in\brn$, we consider the \emph{right passage probability} 
 $\rpp_{i,j}=\rpp_{i,j}(\G,\wt):= \frac{\Delta^{\rho\,\nmid\,[i,j)}_{i,j}(\G,\wt)}{\Delta_i(\G,\wt)\Delta_{j}(\G,\wt)}$,
where $\Delta_i(\G,\wt)\Delta_{j}(\G,\wt)$ is the sum of weights of double-dimer configurations $\Omij\in\Oms_{\G}(i,j)$ on $\G$ with a single \bdbd path $\Pathbdsub_{i,j}(\Omij)$ from $\bdv_i$ to $\bdv_{j}$, and $\Delta^{\rho\,\nmid\,[i,j)}_{i,j}(\G,\wt)$ is the sum of weights of $\Omij\in\Oms_{\G}(i,j)$ such that the path $\Pathbdsub_{i,j}(\Omij)$ \emph{passes to the right of $\ploc_\rho$}, i.e., does not separate $\ploc_\rho$ from the faces $\bdf_i,\bdf_{i+1},\dots,\bdf_{j-1}$. 
Thus, $\rpp_{i,j} + \rpp_{j,i} = 1$. Moreover, %
\begin{equation}\label{eq:rpp_sum=1}
 \rpp_{i,j} + \rpp_{j,l} + \rpp_{l,i} = 1
 \quad\text{for all distinct cyclically ordered $i,j,l\in\brn$}.
\end{equation}
Indeed, choose a cut $\Cut_l$ from $\ploc_\rho$ to $\bdf_l$. Let $\Ebb_i\<\Cut_l,\apm_i\>$ be the expected signed intersection number between $\Cut_l$ and a random \APM $\apm_i\in\APMSGbd(\{i\})$, where we direct all edges of $\apm_i$ from black to white. Then 
$\rpp_{i,j}
 = \Ebb_{i,j} \<\Cut_l,\Pathbdsub_{i,j}(\Omij)\> 
 = \Ebb_i\<\Cut_l,\apm_i\> - \Ebb_j\<\Cut_l,\apm_j\>
$
 is the expected signed intersection number between $\Cut_l$ and a random double-dimer path $\Pathbdsub_{i,j}(\Omij)$ for $\Omij\in\Oms_{\G}(i,j)$. We similarly get 
$\rpp_{j,l}
 = \Ebb_{j,l} \<\Cut_l,\Pathbdsub_{j,l}(\Omxx_{j,l})\>
 = \Ebb_j\<\Cut_l,\apm_j\> - \Ebb_l\<\Cut_l,\apm_l\>
$ 
and
$\rpp_{l,i}
 = 1+\Ebb_{l,i} \<\Cut_l,\Pathbdsub_{l,i}(\Omxx_{l,i})\>
 = 1+\Ebb_l\<\Cut_l,\apm_l\> - \Ebb_i\<\Cut_l,\apm_i\>
$,
which yields~\eqref{eq:rpp_sum=1}.

We claim that 
$\Hknkloc_\rho = \sum_{i=1}^n \rpp_{i,i+1} \Hknkloc_{\xbd{i}}$. Writing $C = \mat[1,c_2,\cdots,c_n]$, we have $\Hknkloc_\rho^T = \mat[\rpp_{2,1}c_2,\cdots,\rpp_{n,1}c_n]$ by~\eqref{eq:LPUNC:minors_dH}. 
 By \cref{rmk:H_bdry}, 
\begin{equation}\label{eq:Hbd_for_k=1} 
\Hknkloc_{\xbd1}^T=\bzero_{1\times(n-1)} 
\quad\text{and}\quad
\Hknkloc_{\ibd}^T = \mat[c_2,\cdots,c_i,0,\cdots,0] 
\quad\text{for each $2\leq i\leq n$}.
\end{equation}
 Thus, 
$\Hknkloc_\rho = (1-\rpp_{2,1})\Hknkloc_{\xbd1} + (\rpp_{2,1}-\rpp_{3,1})\Hknkloc_{\xbd2} + \cdots + 
(\rpp_{n-1,1}-\rpp_{n,1})\Hknkloc_{\xbd{(n-1)}} + \rpp_{n,1}\Hknkloc_{\xbd n}$. Clearly, these coefficients add up to $1$.
Using $\rpp_{i,j}+\rpp_{j,i}=1$ and~\eqref{eq:rpp_sum=1}, we find 
$\rpp_{i,1} - \rpp_{i+1,1} = \rpp_{i,i+1}$, as desired. 
In particular, $\rpp_{i,1} - \rpp_{i+1,1}\geq0$. 
 The set $\Sloc_\rho$ in~\eqref{eq:bcccloc_dfn_type_1n} is given by 
$\Sloc_\rho=\{i\in\brn\mid \rpp_{i,i+1}>0\}$. 
\end{proof}

By~\eqref{eq:Hbd_for_k=1}, the $n$ points in $\Hknklocsbd$ are affinely independent. Consider the \emph{open affine simplex}
\begin{equation}\label{eq:Simplexr_dfn}
 \Simplexr:=\Big\{\bcccloc_\rho=(\cccx(\rho,i))_{i\in\Sloc_\rho}\in\Rtp^{\Sloc_\rho}\Big| \sum_{i\in\Sloc_\rho} \cccx(\rho,i) = 1\Big\},
 \quad\text{and let}\quad
\SimplexbL:=\prod_{\rho\in\brnL}\Simplexr.
\end{equation}
 By~\eqref{eq:bcccloc_dfn_type_1n}, the location of the point $\Hknkloc_\rho$ uniquely determines (and is uniquely determined by) the coefficients $\bcccloc_\rho\in\Simplexr$. Thus, 
 we can think of the \Lfunc positroid cell $\PtpLfunc_{\Gfunc}$ (\cref{dfn:Lfunc_positroid_cell_and_reduced}) as a subset of $\Grtp(1,n)\times\SimplexbL$. Our next goal is to give a sufficient condition for the map $\Meas(\Gfunc,\cdot):\Rtpgauge\to\Grtp(1,n)\times\SimplexbL$ to be a homeomorphism. This will be applied later in \cref{ssec:BCFW_triang_final2} to the BCFW recursion.

\begin{definition}
Recall from~\cite{Pos} that a planar bipartite graph $\G$ of type $(1,n)$ is reduced if and only if $\Fint=\emptyset$, and that any non-reduced graph $\G$ can be transformed into a reduced graph $\CollGW$ using moves \MV1--\MV2 and \RV1 (\cref{fig:moves}). 
We say that $\G$ is \emph{\easyred} if it can be transformed into a reduced graph using only moves \MV1 and \RV1.
\end{definition}
\noindent For example, the first graph in \cref{fig:non-einj} is \easyred while the second one is not. 

Assume that $\G$ is \easyred and let $\CollGW$ be the associated reduced graph. 
Each face of $\G$ either disappears during some reduction move \RV1 or is among the faces of $\CollGW$. 
Consider a digraph $\GflW$ with vertex set $\Faces$ and arrows $\ff_1\leftarrow\ff_2\to\ff_3$ for each face $\ff_2$ that disappears during move \RV1, where $\ff_1$ and $\ff_3$ are the two faces adjacent to $\ff_2$ during that move. Thus, the set of sinks of $\GflW$ is precisely the set $\CollFaces$ of faces of $\CollGW$. 
For $\f\in\Faces$, let $\preKmaxnoloc_{\f}:=\{\ff\in\Faces\mid \GflW\text{ contains a directed path from $\f$ to $\ff$}\}$, and set $\Snoloc_{\f}:=\preKmaxnoloc_{\f}\cap\CollFaces$ and $\Tnoloc_{\f}:=\preKmaxnoloc_{\f}\setminus\CollFaces$. 
For $\rho\in\brnL$, let $\preKmaxloc_\rho:=\preKmaxnoloc_{\ploc_\rho}$, $\Sloc_{\rho}:=\Snoloc_{\ploc_\rho}$, and $\Tloc_\rho:=\Tnoloc_{\ploc_\rho}$. 
 See \figref{fig:S-rho}(b,c) for examples. 
\begin{lemma}\label{lemma:counting_stars_type_1_n}
Let $\Gfunc$ be an \Lfunc graph of type $(1,n)$. Assume that $\G$ is \easyred, \begin{equation}\label{eq:counting_stars_type_1_n}
 \Faces\setminus\CollFaces = \bigsqcup_{\rho\in\brnL} \Tloc_\rho,
 \quad\text{and}\quad
 |\Tloc_\rho| = |\Sloc_\rho|-1 \quad\text{for each $\rho\in\brnL$}. 
\end{equation}
Then $\PtpLfunc_{\Gfunc}=\Grtp(1,n)\times\SimplexbL$ and the map 
$\Measknk(\Gfunc,\cdot):\Rtpgauge\to\PtpLfunc_{\Gfunc}$ is a homeomorphism. 
\end{lemma}
\begin{proof}
Suppose that $\GflW$ contains arrows $\ff_1\leftarrow\ff_2\to\ff_3$ corresponding to some reduction move \RV1 during which the face $\ff_2$ disappeared. Let $a_1,a_3>0$ be the weights of the edges separating $\ff_2$ from $\ff_1$ and $\ff_3$, respectively. Then 
$\Hknk(\ff_2) = \frac{a_3}{a_1+a_3}\Hknk(\ff_1) + \frac{a_1}{a_1+a_3}\Hknk(\ff_3)$.
Thus, the face weight $\facewt(\ff_2)=\frac{a_1}{a_3}$ 
 contains the same information as the coefficients expressing $\Hknk(\ff_2)$ as a convex combination of $\Hknk(\ff_1)$ and $\Hknk(\ff_3)$. 

By construction, $\GflW$ has no directed cycles and satisfies $|\Tnoloc_{\f}|\geq|\Snoloc_{\f}|-1$ for each $\f\in\Faces$. 
Let $\rho\in\brnL$ and suppose that $\GflW$ contains arrows $\ff_1\ot \ploc_\rho\to\ff_3$. 
By~\eqref{eq:counting_stars_type_1_n}, $|\Tloc_\rho| = |\Sloc_\rho|-1$, so we get $\Sloc_\rho=\Snoloc_{\ff_1}\sqcup\Snoloc_{\ff_3}$ with $|\Tnoloc_{\ff_1}|=|\Snoloc_{\ff_1}|-1$, and $|\Tnoloc_{\ff_3}|=|\Snoloc_{\ff_3}|-1$. 
Thus, the coefficients $t,\bcccnoloc_{\ff_1},\bcccnoloc_{\ff_3}$ in the convex combinations 
$\Hknkloc_\rho = (1-t)\Hknk(\ff_1) + t\Hknk(\ff_3)$, 
$\Hknk(\ff_1)=\sum_{i_1\in\Snoloc_{\ff_1}} \cccxnoloc(\ff_1,i_1) \Hknkloc_{\xbd{i_1}}$
and 
$\Hknk(\ff_3)=\sum_{i_3\in\Snoloc_{\ff_3}} \cccxnoloc(\ff_3,i_3) \Hknkloc_{\xbd{i_3}}$
are uniquely determined by the coefficients $\bcccloc_\rho$ in~\eqref{eq:bcccloc_dfn_type_1n}. 
As explained above, the face weight of $\ploc_\rho$ is given by $\facewt(\ploc_\rho)=\frac{t}{1-t}$. 
Continuing in this fashion, we determine the convex combination coefficients $\bcccnoloc_{\ff}$ and the face weight $\facewt(\ff)$ for each $\ff\in\Tloc_\rho$. 
The remaining face weights $(\facewt(\ff))_{\ff\in\CollFaces}$ are determined by $C$. 
Thus, we obtain a continuous inverse
$\Grtp(1,n)\times\SimplexbL\to \Rtpgauge$ of $\Meas(\Gfunc,\cdot)$. 
\end{proof}

\subsection{\LvuncTITLE graphs}\label{ssec:Lvunc}
\begingroup
\def\ddk{k-2}
\def\ddkpd{k}
\def\ddkpdd{k-2+2d}

Let $\ddG$ be a planar bipartite graph of type $(\ddk,n)$ with white boundary. 
Recall that a \emph{black \bivertex} of $\ddG$ is an unordered pair
 $\bivloc_\sepr=\{\ddbloc_\sepr^1,\ddbloc_\sepr^2\}\subset\ddBV$ of vertices sharing a face of $\ddG$. 
In the case when $\ddbloc_\sepr^1$ and $\ddbloc_\sepr^2$ share several faces or when the complement of $\ddG$ contains several non-isotopic chords connecting $\ddbloc_\sepr^1$ to $\ddbloc_\sepr^2$, we assume that $\bivloc_\sepr$ includes a fixed choice $\ddeloc_\sepr$ of such a chord. 
We refer to black \bivertices as \emph{\Lvunctures}. 
We say that two \Lvunctures $\bivloc_\seps = \{\ddbloc_\seps^1,\ddbloc_\seps^2\}$ and 
$\bivloc_\sept = \{\ddbloc_\sept^1,\ddbloc_\sept^2\}$ are \emph{non-crossing} if adding the chords $\ddeloc_\seps$ and $\ddeloc_\sept$ 
 to the edge set of $\ddG$ results in a planar (non-bipartite) graph.
\begin{definition}\label{dfn:Lvunc_ordinary}
An \emph{\Lvunc graph} $\ddGvuncpair$ is a planar bipartite graph $\ddG$ equipped with a distinguished $\nL$-tuple $\bivlocs=(\bivloc_1,\dots,\bivloc_\nL)$ of pairwise non-crossing black \bivertices. 
\end{definition}
\noindent Let $\ddbdbx_1,\ddbdbx_2,\dots,\ddbdbx_n$ be the (black) next-to-boundary vertices of $\ddG$. For $\ibd\in\brnbd$, set $\bdbiv_i:=\{\ddbdbx_i,\ddbdbx_{i+1}\}$.
For $\bivI\subset\brnLbd$, we write $\Bivloc_{\bivI}:=\bigcup_{\sepr\in\bivI} \bivloc_\sepr$. 
\begin{definition}\label{dfn:fullyind}
We say that a subset $\bivI\subset\brnLbd$ is \emph{\Gindependent} if 
the \bivertices in $\{\bivloc_\sepr\mid\sepr\in\bivI\}$ are pairwise disjoint (i.e., $|\Bivloc_{\bivI}|=2|\bivI|$)
and $\ddG\rem\Bivloc_{\bivI}$ admits an \APM. 
We say that $\ddGvunc$ is \emph{\oneind} if any subset $\bivI\subset\brnLbd$ of size at most $1$ is \Gindependent.
We say that $\ddGvunc$ is \emph{\twoind} if it is \oneind and each $\bivI\in\brnLbdsep$ is \Gindependent.
\end{definition}

For the purposes of defining the \Lvunc positive Grassmannian (see \cref{ex:non_ordinary_Meas_Lvunc}), it is necessary to generalize the above definitions following \cref{rmk:add_remove_vs_remove_two}. 

\begin{definition}\label{dfn:gen_Lvunc}
A \emph{generalized \Lvuncture} in $\ddG$ is a nonempty collection $\ddSloc_\rho$ of black \bivertices of $\ddG$ all sharing a common interior black vertex $\ddbloc_\rho$ of $\ddG$. 
Two generalized \Lvunctures $\ddSloc_\seps$ and $\ddSloc_\sept$ are called \emph{non-crossing} if the bivertices in $\ddSloc_\seps\cup\ddSloc_\sept$ are pairwise non-crossing. 
A \emph{generalized \Lvunc graph} $\ddGvunc$ is a planar bipartite graph $\ddG$ equipped with a distinguished $\nL$-tuple $\ddSlocs=(\ddSloc_1,\dots,\ddSloc_\nL)$ of pairwise non-crossing generalized \Lvunctures. 
\end{definition}
\noindent 
A special case of a generalized \Lvuncture is an \emph{ordinary} \Lvuncture when $\ddSloc_\rho=\{\bivloc_\rho\}$ consists of a single \bivertex. 
In this case, the shared vertex $\ddbloc_\rho$ is chosen to be either one of the two vertices in $\bivloc_\rho$ in an arbitrary way. In particular, for each $\ibd\in\brnbd$, we have a generalized \Lvuncture $\ddSloc_{\ibd}=\{\bdbiv_i\}$, with $\ddbloc_{\ibd}:=\ddbdbx_i$. For simplicity, we assume that the vertices $\{\ddbloc_\rho\mid\rho\in\brnLbd\}$ are pairwise distinct.

\begin{definition}\label{dfn:ddGlocs}
For $\sepr\in\brnLbd$, we write $\ddSloc_\sepr=\{\{\ddbloc_\sepr,\ddbloc_\sepr^1\},\dots,\{\ddbloc_\sepr,\ddbloc_\sepr^{\mloc_\sepr}\}\}$ and $\blocout_\sepr:=\{\ddbloc_\sepr^1,\dots,\ddbloc_\sepr^{\mloc_\sepr}\}$. 
Consider a (non-planar) bipartite graph $\ddGaux$ obtained from $\ddG$ by adding a white vertex $\wauxloc_\sepr$ connected to the vertices in $\blocout_\sepr$ for each $\sepr\in\brnLbd$. 
 For $\bivI\subset\brnLbd$, let 
\begin{equation}\label{eq:ddGlocs_dfn}
 \ddGlocs_{\bivI}:=\ddGaux\rem\left(
\{\wauxloc_\rho\mid\rho\in(\brnLbd)\setminus\bivI\}
\sqcup
\{\ddbloc_\rho\mid\rho\in\bivI\}\right).
\end{equation}
In other words, $\ddGlocs_{\bivI}$ is obtained from $\ddG$ by \emph{popping} the black vertex $\ddbloc_\rho$ for each $\rho\in\bivI$, i.e., replacing $\ddbloc_\rho$ with a white vertex $\wauxloc_\rho$ connected to each vertex in $\blocout_\rho$; see \cref{lemma:DIM:remove_add} and \cref{fig:popping}. 
For $\sepr\in\brnLbd$, we denote $\ddGloc_\sepr:=\ddGlocs_{\{\sepr\}}$. 
\end{definition}

\begin{figure}
 \def\inputscale{1.52}
 \setlength{\tabcolsep}{0.5pt}
\begin{tabular}{cccc|cc}
 \includegraphics[scale=\inputscale]{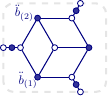}
&
 \includegraphics[scale=\inputscale]{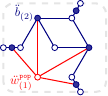}
&
 \includegraphics[scale=\inputscale]{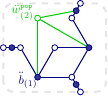}
&
 \includegraphics[scale=\inputscale]{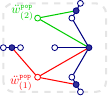} \hspace{2pt}
&
 \includegraphics[scale=\inputscale]{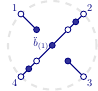}
&
 \includegraphics[scale=\inputscale]{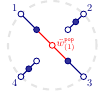}
\\
(a) $\ddG$ & 
(b) $\ddGloc_1$ &
(c) $\ddGloc_2$ &
(d) $\ddGlocs_{\{1,2\}}$ &
(e) $\ddG$ &
(f) $\ddGloc_1$
\end{tabular}
 \caption{\label{fig:popping} Popping black vertices of a generalized \Lvunc graph $\ddGvunc$.}
\end{figure}

Since the \Lvunctures $\{\ddSloc_\rho\mid\rho\in\brnLbd\}$ are pairwise non-crossing, the bipartite graph $\ddGlocs_{\bivI}$ is planar for each $\bivI\subset\brnLbd$. It is of type $(\ddkpdI,n)$ with $\ddkpdI:=k-2+2|\bivI|$. 
\begin{definition}\label{dfn:fullyind_gen}
We say that $\bivI\subset\brnLbd$ is \emph{\Gindependent} if $\ddGlocs_{\bivI}$ admits an \APM.
As before, a generalized \Lvunc graph $\ddGvunc$ is \emph{\oneind} if each $\bivI\subset\brnLbd$ of size at most $1$ is \Gindependent and \emph{\twoind} if additionally each $\bivI\in\brnLbdsep$ is \Gindependent.
\end{definition}
\noindent This generalizes \cref{dfn:fullyind}: if each generalized \Lvuncture of $\ddGvunc$ consists of a single \bivertex then 
each $\wauxloc_\rho$, $\rho\in\bivI$, is a white interior leaf in $\ddGlocs_{\bivI}$, and we have $|\Bivloc_{\bivI}|=2|\bivI|$ if and only if these leaves are connected to pairwise distinct vertices. In this case, deleting these white leaves and their sole neighbors from $\ddGlocs_{\bivI}$ yields the graph $\ddG\rem\Bivloc_{\bivI}$. 

\begin{definition}\label{dfn:ddwtaux}
Let $\Rtpgaux$ be the space of edge weights $\ddwtaux:\ddEaux\to\Rtp$ modulo the group $\Rtp^{|\ddVintaux|}$ of gauge transformations at \emph{interior} (i.e., not belonging to $\ddVbd$) vertices of $\ddGaux$. 
We refer to $(\ddGvunc,\ddwtaux)$ as a \emph{weighted} generalized \Lvunc graph. 
For $\bivI\subset\brnLbd$, let $\ddwtlocs_{\bivI}$ be the restriction of $\ddwtaux$ to the edges of $\ddGlocs_{\bivI}$. Denote $\ddwt:=\ddwtlocs_{\emptyset}$ and $\ddwtloc_{\sepr}:=\ddwtlocs_{\{\sepr\}}$ for $\sepr\in\brnLbd$.
\end{definition}

\begin{definition}[\Lvunc boundary measurement map]\label{dfn:LPUNC:Meas_vunc}
Let $(\ddGvunc,\ddwtaux)$ be a \oneind weighted generalized \Lvunc graph. 
Let $\ddC :=\Meas(\ddG,\ddwt)$ and $\Dbivloc_\sepr:=\Meas(\ddGloc_\sepr,\ddwtloc_\sepr)$ for each $\sepr\in\brnLbd$. 
We let $\DbivlocsL:=(\Dbivloc_1,\dots,\Dbivloc_\nL)$ and 
\begin{equation}\label{eq:LPUNC:Measbiv_dfn}
 \Measbiv(\ddGvunc,\ddwtaux):=\ddCDL\in\Grtnnflamb,
 \quad\text{where}
\end{equation}
\begin{equation*}%
 \Grtnnflamb :=\{\ddCDL\in\Grtnn(\ddk,n)\times\Grtnn(\ddkpd,n)^\nL\mid \ddC\subset\Dbivloc_\rho\text{ for all $\rho\in\brnL$}\}.
\end{equation*}
(By \cref{lemma:DIM:remove_add}, $\ddCDL$ indeed belongs to this set.)
For each $\sepr\in\brnLbd$, 
 choose a $2\times n$ matrix $\Drestloc_\sepr$ such that 
$\Dbivloc_\sepr = \begin{pmatrix}
\Drestloc_\sepr\skipddC \ddC
\end{pmatrix}$. For $\bivI=\{\sepr_1,\dots,\sepr_d\}\subset\brnLbd$, let 
\begin{equation}\label{eq:LPUNC:stack}
 \Dbivlocs_{\bivI}:=\begin{pmatrix}
\Drestloc_{\sepr_1} \\ \vdots \\ \Drestloc_{\sepr_d} \skipddC \ddC
 \end{pmatrix}.
\end{equation}
\end{definition}

\begin{remark}\label{rmk:Drestloc_ibd}
\setlength{\arraycolsep}{2pt}
We have 
$\Drestloc_{\ibd}=\mat[\bzero_{2\times(i-1)}|\Id_2|\bzero_{2\times(n-i-1)}]$ for $i<n$
and
$\Drestloc_{\nbd}=\begin{pmatrix}
0 & 0 & \cdots & 0 & 1\\
(-1)^{k-1}& 0 & \cdots & 0 & 0
\end{pmatrix}$.
\end{remark}

\begin{example}\label{ex:non_ordinary_Meas_Lvunc}
Consider the generalized \Lvunc graph $\ddGvunc$ shown in \figref{fig:popping}(e,f) (with $\knL=(3,4;1)$) and assume that all edge weights $\ddwtaux(\dde)$ are equal to $1$. 
Then $\ddC=\Meas(\ddG,\ddwt)=\mat[0,1,0,1]\in\Grtnn(1,4)$ and $\Dbivloc_1=\Meas(\ddGloc_1,\ddwtloc_1) = \mat[1,0,1,0]^\perp \in\Grtnn(3,4)$. We can choose e.g. $\Drestloc_{1}=\begin{pmatrix}
1 & 0 & -1 & 0\\
0 & 1 & 0 & 0
\end{pmatrix}$. 
We expect that the resulting boundary measurements $\ddCDL$ cannot be obtained from any ordinary weighted \Lvunc graph of type $\knL=(3,4;1)$.
\end{example}

\begin{lemma}\label{lemma:LPUNC:indep<=>full_rank_Dbivloc}
Let $\ddGvunc$ be \oneind. 
Then a subset $\bivI\subset\brnLbd$ is \Gindependent if and only if 
for some (equivalently, any) $\ddwtaux\in\Rtpgaux$, we have 
$\rank\Dbivlocs_{\bivI}=\ddkpdI$. Furthermore, 
\begin{equation}\label{eq:Dbiv=Meas}
 \Dbivlocs_{\bivI} = \Meas(\ddGlocs_{\bivI},\ddwtlocs_{\bivI}) \in\Grtnn(\ddkpdI,n)
 \quad\text{when $\bivI\subset\brnLbd$ is \Gindependent}.
\end{equation}
\end{lemma}
\begin{proof}

 Choose Kasteleyn signs $\ddepsK$ for $\ddG$. 
As explained in the proof of \cref{lemma:DIM:remove_add}, for each $\sepr\in\brnLbd$, there exists a choice $\ddepsKloc_\sepr$ of Kasteleyn signs on $\ddGloc_\sepr$ such that 
the restrictions $\ddepsKloc_\sepr|_{\ddE\cap\ddEloc_\sepr} = \ddepsK|_{\ddE\cap\ddEloc_\sepr}$ coincide. 
Here, $\ddEloc_\sepr$ denotes the edge set of $\ddGloc_\sepr$. 
 Observe that every edge of $\ddGaux$ appears in $\ddGloc_\sepr$ for some $\sepr\in\brnLbd$. Let $\ddepsKaux:\ddEaux\to\{\pm1\}$ be such that
$\ddepsKaux|_{\ddE} = \ddepsK$ and $\ddepsKaux|_{\ddEloc_\sepr} = \ddepsKloc_\sepr$ for all $\sepr\in\brnLbd$. 
For each $\bivI\subset\brnLbd$, let $\ddepsKlocs_{\bivI}:=\ddepsKaux|_{\ddElocs_{\bivI}}$.
 Our first goal is to show that $\ddepsKlocs_{\bivI}$ is a choice of Kasteleyn signs on $\ddGlocs_{\bivI}$. 

Let $\sepr\in\brnLbd$, $\ddbloc_\sepr^i\in\blocout_\sepr$, and let $\ddeloc_\sepr^i$ be the edge of $\ddGloc_\sepr$ connecting $\wauxloc_\sepr$ to $\ddbloc_\sepr^i$. We claim that the sign $\ddepsKloc_\sepr(\ddeloc_\sepr^i)$ coincides with the sign $\ddepsbb_{\ddbloc_\sepr,\ddbloc_\sepr^i}$ given by \cref{dfn:eksbb}. To see that, for any \bivertex $\biv=\{\ddb^1,\ddb^2\}$, let $\psitrip_{\ddb^1,\ddb^2}$ be the \emph{tripod insertion operator}, so that $\psitrip_{\ddb^1,\ddb^2}(\ddG,\ddepsK)=(\ddG',\ddepsK')$ is obtained from $\ddG$ by inserting a black leaf $\ddb^3$ and a white vertex $\ddw$ connected to $\ddb^1,\ddb^2,\ddb^3$ in clockwise order, and extending Kasteleyn signs $\ddepsK$ for $\ddG$ to Kasteleyn signs $\ddepsK'$ for $\ddG'$ as in \cref{dfn:eksbb}. 
By~\Mref{rmk:tripods_commute},
 for any two non-crossing bivertices $\{\ddb^1,\ddb^2\}$ and $\{\ddb^3,\ddb^4\}$, the operators $\psitrip_{\ddb^1,\ddb^2}$ and $\psitrip_{\ddb^3,\ddb^4}$ commute.
By \cref{lemma:OCP:Kast_even}, deleting edges from $\ddG$ does not affect the validity of the Kasteleyn sign condition. 
Let $(\ddGloc_\sepr',\ddepsKloc_\sepr'):=\psitriploc_\sepr(\ddG,\ddepsK)$ be obtained from $(\ddG,\ddepsK)$ by applying $\psitrip_{\ddbloc_\sepr,\ddbloc_\sepr^i}$ for each $\ddbloc_\sepr^i\in\blocout_\sepr$ and then deleting the original edges of $\ddG$ incident to $\ddbloc_\sepr$. Then the faces of $\ddGloc_\sepr$ are in bijection with those of $\ddGloc_\sepr'$, and for every face $\ff$, the difference between the numbers of white vertices of $\ddGloc_\sepr$ and of $\ddGloc_\sepr'$ incident to $\ff$ is even. Thus, setting $\ddepsKloc_\sepr(\ddeloc_\sepr^i):=\ddepsbb_{\ddbloc_\sepr,\ddbloc_\sepr^i}$ gives a choice of Kasteleyn signs for $\ddGloc_\sepr$. 
More generally, given $\bivI\subset\brnLbd$, composing the operators $(\psitriploc_\sepr)_{\sepr\in\bivI}$, we similarly conclude that 
 $\ddepsKlocs_{\bivI}$ is a choice of Kasteleyn signs on $\ddGlocs_{\bivI}$.

Since $\ddGvunc$ is \oneind, $\ddG$ and $\ddGloc_\sepr$ admit \APMs for each $\sepr\in\brnLbd$. Consider the respective Kasteleyn matrices $\ddWKmat$ and $\ddWKmatloc_\sepr$ as in \cref{dfn:MCE:WKmat}. 
By \cref{lemma:MCE:apm_vs_Kast}, the columns of $\ddWKmat$ are linearly independent, and the matrix $\Dbivloc_\sepr$ has rank $\ddkpd$ and extends uniquely to a \wdash holomorphic function $\Dbivlocext_\sepr=\begin{pmatrix}
\Drestlocext_\sepr \\[4pt] \ddCext
\end{pmatrix}\in\Hwspace_{\R^{\ddkpd}}(\ddGloc_\sepr,\ddwtKloc_\sepr)$ whose rows give a basis for the (left) kernel of 
$\ddWKmatloc_\sepr$, i.e., $\Dbivlocext_\sepr = (\ddWKmatloc_\sepr)^\perp$.
For any set $\bivI\subset\brnLbd$ (resp., any set $\bivI$ containing $\sepr$), we extend $\ddCext$ (resp., $\Drestlocext_\sepr$) to a \wdash holomorphic function on $(\ddGlocs_{\bivI},\ddwtKlocs_{\bivI})$ by setting $\ddCext(\wauxloc_{\sept}):=0$ (resp., $\Drestlocext_\sepr(\wauxloc_{\sept}):=0$) for any $\sept\in\bivI$ (resp., $\sept\in\bivI\setminus\{\sepr\}$). We denote by $\ddWKmatlocs_{\bivI}$ the Kasteleyn matrix of $\ddGlocs_{\bivI}$. Since the restriction of $\ddWKmatlocs_{\bivI}$ to the row set $\ddWV\subset\ddWVlocs_{\bivI}$ of $\ddWKmat$ is obtained from $\ddWKmat$ by deleting columns, we see that 
$\rank\ddWKmatlocs_{\bivI} = |\ddBV| - |\bivI|$, and therefore $\dim((\ddWKmatlocs_{\bivI})^\perp) = \ddkpdI$,
 for all $\bivI\subset\brnLbd$. 

Let $\ddWVloc_\sepr:=\ddWV\sqcup\{\wauxloc_\sepr\}$ be the set of white vertices of $\ddGloc_\sepr$. Consider a $|\ddWVloc_\sepr|\times 2$ matrix $\Dveeloc_\sepr$ whose first column has a single $1$ in the row corresponding to $\wauxloc_\sepr$ and whose second column is the column of $\ddWKmat$ corresponding to $\ddbloc_\sepr$ (extended by zero to $\wauxloc_\sepr$). Since $\rank\ddCext=k-2$, $\rank\Dbivlocext_\sepr=\ddkpd$, 
$\ddCext = \mat[\ddWKmatloc_\sepr|\Dveeloc_\sepr]^\perp$, and
$\Dbivlocext_\sepr = (\ddWKmatloc_\sepr)^\perp$, 
 it follows that $\Drestlocext_\sepr\cdot \Dveeloc_\sepr$ is an invertible $2\times 2$ matrix. 
 On the other hand, for distinct $\sepr,\sept\in\brnLbd$, we have
 $\Drestlocext_\sepr\cdot \Dveeloc_\sept=\bzero_{2\times2}$.
Thus, for all $\bivI\subset\brnLbd$, the \wdash holomorphic function $\Dbivlocsext_{\bivI}$ on $(\ddGlocs_{\bivI},\ddwtKlocs_{\bivI})$ obtained by stacking $\Drestlocext_{\sepr_1},\dots,\Drestlocext_{\sepr_d}$ on top of $\ddCext$ as in~\eqref{eq:LPUNC:stack} satisfies $\rank\Dbivlocsext_{\bivI} = \ddkpdI$ and 
$\Dbivlocsext_{\bivI} = (\ddWKmatlocs_{\bivI})^\perp$. 

By \cref{lemma:MCE:Delta_vs_Kast}, $\bivI\subset\brnLbd$ is \Gindependent if and only if 
$\Delta_{\ddWVlocs_{\bivI}\setminus J}(\ddWKmatlocs_{\bivI})\neq0$ 
for some $J\subset\brn$ with $|J| = \ddkpdI$. 
Since $\Dbivlocsext_{\bivI} = (\ddWKmatlocs_{\bivI})^\perp$, we have 
$\Delta_{\ddWVlocs_{\bivI}\setminus J}(\ddWKmatlocs_{\bivI})=\pm \Delta_J(\Dbivlocsext_{\bivI}) = \pm\Delta_J(\Dbivlocs_{\bivI})$. Thus, $\bivI$ is \Gindependent if and only if $\rank\Dbivlocs_{\bivI} = \ddkpdI$.
In this case~\eqref{eq:Dbiv=Meas} follows from~\eqref{eq:MCE:alt(C)_vs_pFw}. 
\end{proof}

We record the following corollary of the above proof for later use. 
\begin{corollary}\label{lemma:LPUNC:D_perp_cap_la=0}
Suppose that $\ddGvunc$ is \oneind. Let $\ddCDL:=\Measbiv(\ddGvunc,\ddwtaux)$ for some $\ddwtaux\in\Rtpgaux$, and let $\la\in\lak$ be such that $\la\subset\ddC^\perp$. Then for each $\sepr\in\brnLbd$, 
\begin{equation}\label{eq:Dbivloc_cap_la=0}
 \Dbivloc_\sepr^\perp\cap\la = \{0\}.
\end{equation}
\end{corollary}
\begin{proof}
Let $\ddlaext\in\Hbspace_{\R^2}(\ddG,\ddwtK)$ be the \bdash holomorphic extension of $\la\subset\ddC^\perp$. 
By~\eqref{eq:Dbiv=Meas} and~\eqref{eq:MCE:alt(Cp)_vs_pFb}, for $\sepr\in\brnLbd$, $\Dbivloc_\sepr^\perp\cap\la$ is the space of (boundary restrictions of) \bdash holomorphic functions $\ddFbloc_\sepr\subset\ddlaext$ satisfying
\begin{equation}\label{eq:GENPUNC:dual_vector}
\ddFbloc_\sepr(\ddbloc_\sepr)=0
\quad\text{and}\quad
 \sum_{i=1}^{\mloc_\sepr} \ddWKmatloc_\sepr(\wauxloc_\sepr,\ddbloc_\sepr^i) \ddFbloc_\sepr(\ddblocx(\sepr,i)) = 0.
\end{equation}
Here, $\ddWKmatloc_\sepr(\wauxloc_\sepr,\ddbloc_\sepr^i)=\ddepsKloc_\sepr(\ddeloc_\sepr^i)\ddwtloc_{\sepr}(\ddeloc_\sepr^i)$ for the edge $\ddeloc_\sepr^i$ connecting $\wauxloc_\sepr$ to $\ddbloc_\sepr^i$ as before.
As explained in the proof of \cref{lemma:LPUNC:indep<=>full_rank_Dbivloc}, the sign $\ddepsKloc_\sepr(\ddeloc_\sepr^i)=\ddepsbb_{\ddbloc_\sepr,\ddbloc_\sepr^i}$ is given by \cref{dfn:eksbb}. 
By \crefi{lemma:TE:brla_nonzero_if_apm}{lak_implies2}, 
$\ddepsbb_{\ddbloc_\sepr,\ddbloc_\sepr^i}[\ddbloc_\sepr\,\ddbloc_\sepr^i]_{\ddla}\geq0$, so the $2\times2$ matrix 
$\ddlaveeloc_\sepr:=\matlr[\ddlaext(\ddbloc_\sepr)\middle|
\sum_{i=1}^{\mloc_\sepr} \ddWKmatloc_\sepr(\wauxloc_\sepr,\ddbloc_\sepr^i) \ddlaext(\ddblocx(\sepr,i))]$ has nonnegative determinant, and moreover, $\det\ddlaveeloc_\sepr=0$ if and only if 
$\ddG\setminus\bivlocx(\sepr,i)$ does not admit an \APM for each $i\in\brx{\mloc_\sepr}$. Since $\ddGvunc$ is \oneind, $\ddGloc_\sepr$ admits an \APM. This \APM contains one edge connecting $\wauxloc_\sepr$ to some vertex $\ddbloc_\sepr^i$, and the remaining edges form an \APM of $\ddG\setminus\bivlocx(\sepr,i)$, a contradiction. Thus, $\det\ddlaveeloc_\sepr>0$. In particular, any \bdash holomorphic function $\ddFbloc_\sepr\subset\ddlaext$ satisfying~\eqref{eq:GENPUNC:dual_vector} must be zero. This implies~\eqref{eq:Dbivloc_cap_la=0}.
\end{proof}

\begin{definition}\label{dfn:Grtnnflamb_CDind}
Given $\ddCDL\in\Grtnnflamb$, 
we say that $\bivI\subset\brnLbd$ is \emph{\CDindependent} if $\rank\Dbivlocs_{\bivI}=\ddkpdI$.
Here, $\Dbivlocs_{\bivI}$ is given by~\eqref{eq:LPUNC:stack} with $(\Dbivloc_1,\dots,\Dbivloc_\nL)=\DbivlocsL$ and
$(\Dbivloc_{\xbd1},\dots,\Dbivloc_{\nbd})$ given by \cref{rmk:Drestloc_ibd}. 
Similarly to \cref{dfn:fullyind}, we call $\ddCDL$ \emph{\oneind} if any subset $\bivI\subset\brnLbd$ of size at most $1$ is \CDindependent, and \emph{\twoind} if in addition each $\bivI\in\brnLbdsep$ is \CDindependent.
\end{definition}

\begin{definition}[\Lvunc totally nonnegative Grassmannian]\label{dfn:LPUNC:Lvunc_Grtnn}
For $d=1,2$, introduce the following subsets of $\Grtnnflamb$. 
\begin{align*}%
 \GrtnnLvuncprojx_d(\ddk,n)&:=\left\{\Measbiv(\ddGvunc,\ddwtaux)\middle|\text{
\begin{tabular}{c}
$(\ddGvunc,\ddwtaux)$ is a \dind weighted \\
generalized \Lvunc graph of type $(\ddk,n)$
\end{tabular}
}\right\};\\
\GrtnnLvuncambx_d(\ddk,n)&:=\left\{\ddCDL\middle|\text{
\begin{tabular}{c}
$\ddCDL$ is \dind and $\Dbivlocs_{\bivI}\in\Grtnn(\ddkpdI,n)$\\
for each \CDindependent subset $\bivI\subset\brnLbd$
\end{tabular}
}\right\}.
\end{align*}
\end{definition}
\noindent By \cref{lemma:LPUNC:indep<=>full_rank_Dbivloc}, $\GrtnnLvuncprojx_d(\ddk,n)\subset\GrtnnLvuncambx_d(\ddk,n)$ for $d=1,2$.
\begin{question}\label{que:LPUNC:GrtnnLvuncproj=GrtnnLvuncamb}
Do we have $\GrtnnLvuncprojx_d(\ddk,n)=\GrtnnLvuncambx_d(\ddk,n)$ for $d=1$ or $d=2$?
\end{question}

We consider the \emph{\Lvunc positroid cell} 
$\PtpLvunc_{\ddGvunc}:=\{\Measbiv(\ddGvunc,\ddwtaux)\mid \ddwtaux\in\Rtpgaux\}$ and say that $\ddGvunc$ is \emph{reduced} if the map $\Measbiv(\ddGvunc,\cdot):\Rtpgaux\to\PtpLvunc_{\ddGvunc}$ is a homeomorphism. 

\endgroup

\subsection{T-duality for \Lpunc graphs}\label{ssec:LPUNC:T_duality_graphs}
Let $\Gfunc$ (resp., $\ddGvunc$) be an \Lfunc (resp., \Lvunc) planar bipartite graph of type $(k,n)$ (resp., $(k-2,n)$). 
In this subsection, we assume that each \Lvuncture of $\ddGvunc$ is ordinary, i.e., consists of a single \bivertex. We extend the below results to generalized \Lpunc graphs in the next subsection.

We say that $\Gfunc$ is \emph{T-dualizable} if so is the underlying planar bipartite graph $\G$ (\cref{dfn:G:T_dualizable}). We assume that $\Gfunc$ is T-dualizable and that $\G$ and $\ddG$ form a T-dual pair (\cref{dfn:SHIFT:ddG}). 
In particular, 
 $\ddG$ satisfies the conclusion of \cref{lemma:SHIFT:ddG_admits_apms}.
 Applying \cref{lemma:DIM:deleting_hel_verts}, we obtain the following. 
\begin{corollary}\label{lemma:ddGvunc_is_oneind}
$\ddGvunc$ is \oneind.
\end{corollary}

Recall that we have a bijection $\WV=\WVint\to\ddBVint=\ddBV$ sending $\w\mapsto\ddbx(\w)$. 

\begin{definition}[T-dual \Lpunc graphs]\label{dfn:LPUNC:T_dual_Lpunc_graphs}
 We say that $\Gfunc$ and $\ddGvunc$ form a \emph{T-dual pair} if the underlying planar bipartite graphs $\G$ and $\ddG$ form a T-dual pair and for each $\rho\in\brnL$, 
the white vertices $\wloc_\rho^1,\wloc_\rho^2$ of $\G$ such that $\bivloc_\rho=\{\ddbx(\wloc_\rho^1),\ddbx(\wloc_\rho^2)\}$
are both incident to the face $\ploc_\rho$.
\end{definition}

\begin{remark}\label{rmk:LPUNC:choices}
Applying T-duality to an \Lfunc graph $\Gfunc$ involves some choices. First, one needs to triangulate the light regions of $\Sig$ as in \cref{dfn:SHIFT:Sig}. 
 Second, in order to construct the T-dual \Lvunc graph $\ddGvunc$, 
 for each $\rho\in\brnL$, 
 one needs to choose a pair $\{\wloc_\rho^1,\wloc_\rho^2\}$ of white vertices incident to $\ploc_\rho$. We claim that these choices are immaterial. First, any two triangulations of an unpunctured light region of $\Sig$ are related by diagonal flips. Such flips correspond to square moves \MV2 on $\ddG$ that preserve the boundary measurements $\Measbiv(\ddG,\ddwtlap)$. Second, similarly to the proof of \cref{lemma:SHIFT:ddepsKla_is_Kast}, the face $\ploc_\rho$ of $\G$ corresponds to either (i) a light region of $\Sig$, (ii) an interior edge of $\Sig$ incident to two dark triangles, or (iii) a boundary edge of $\Sig$ incident to a dark triangle. The only potential choice for $\{\wloc_\rho^1,\wloc_\rho^2\}$ arises in case (i), so suppose that $\ploc_\rho$ contains a light region $\RgSigloc_\rho$ of $\Sig$. 
 Consider a subset $\ddRgloc_\rho\in\BNEI(\ddG)$ consisting of trivalent white vertices of $\ddG$ located inside $\RgSigloc_\rho$ together with black vertices of $\ddG$ located on the boundary of $\RgSigloc_\rho$. %
As explained in the proof of \cref{lemma:SHIFT:ddepsKla_is_Kast}, the triangles in $\RgSigloc_\rho$ form a triangulation of a polygon (containing no vertices of $\Sig$ in the interior), so we have 
$\helBsub_{\ddG}(\ddRgloc_\rho)=2$. Thus, $\helBsub_{\ddG\rem\bivloc_\rho}(\ddRgloc_\rho\rem\bivloc_\rho)=0$, where $\ddRgloc_\rho\rem\bivloc_\rho\in\BNEI(\ddG\rem\bivloc_\rho)$.
 Consequently, any \APM of $\ddG\rem\bivloc_\rho$ restricts to a perfect matching of 
$\ddG\ind[\ddRgloc_\rho\rem\bivloc_\rho]$. 
 It follows that choosing any \bivertex $\bivloc_\rho\subset\ddRBloc_\rho$ gives rise to the same boundary measurements $\Measbiv(\ddGvunc,\ddwtlap)$.
\end{remark}

Let $(\Gfunc,\wt)$ be a weighted \Lfunc graph and let $(C;\Hknklocs):=\Measknk(\Gfunc,\wt)$ be as in~\eqref{eq:Measknk_dfn}. 
Fix $\la\in\lak$ such that $\la\subset C$ and let $(C;\Hdnklocs):=\Measdnk(\Gfunc,\wt)$ be given by~\eqref{eq:LPUNC:Measdnk_def}. 
By \cref{lemma:black_gauge_eq}, the positive edge weights $\wt$ and $\wtlap$ are gauge equivalent.
Let $\ddCDL:=\Measbiv(\ddGvunc,\ddwtlap)$ be as in~\eqref{eq:LPUNC:Measbiv_dfn}.
 By~\eqref{eq:Meas_ddG=Chat_Qla}, $\ddC = C\cdot \Qla$. 
 Our goal is to relate $\Hdnklocs$ to $\DbivlocsL$. 
Choose matrix representatives $C^\perp$ and $\ddC^\perp$ satisfying
\begin{equation}\label{eq:LPUNC:ddCprest_dfn}
\ddC^\perp = \begin{pmatrix}
\la \\ \ddCprest
\end{pmatrix} \quad\text{and}\quad \ddC^\perp\cdot \Qla = \ddCprest \cdot \Qla =C^\perp. 
\end{equation}
\noindent 
The following identity relating $\Measdnk(\Gfunc,\wtlap)$ to $\Measbiv(\ddGvunc,\ddwtlap)$ will play a fundamental role in our study of T-duality for loop amplituhedra in \cref{sec:LOOP}.
Throughout, we denote 
\begin{equation}\label{eq:dual_spinor}
 \CMI:=\begin{pmatrix}
0 & 1 \\ -1 & 0
\end{pmatrix},
\quad\text{so that}\quad 
\la_i^T\cdot \CMI\cdot \la_j = \brla<i,j>
\quad\text{for all $i,j\in\brn$.}%
\end{equation} 

\begin{proposition}[T-duality for \Lpunc boundary measurements]
\label{lemma:LPUNC:T_duality_D_vs_H}
For each $\sepr\in\brnLbd$, 
\begin{equation}\label{eq:LPUNC:T_duality_D_vs_H}
 \Dbivloc_\sepr^\perp = \ddCprest - \Hdnkloc_\sepr\cdot \CMI\cdot \la \quad\text{as elements of $\Gr(n-k,n)$}.
\end{equation}
\end{proposition}
\begin{remark}
The group $\GL_{n-k}(\R)$ acts simultaneously on the columns of $\Dbivloc_\sepr^\perp$, $\ddCprest$, $C^\perp$, and $\Hdnkloc_\sepr$; cf. \cref{rmk:LPUNC:tangent}. Furthermore, for an $(n-k)\times2$ matrix $M$, the transformation $\ddCprest \mapsto \ddCprest + M\cdot \la$ corresponds to shifting each matrix $\Hdnkloc_\sepr$ by $M\CMI^{-1}$. All terms in~\eqref{eq:LPUNC:T_duality_D_vs_H} behave equivariantly under such transformations. We usually assume that $\Hdnk$ and $\ddCprest$ are \emph{in normal form} meaning $\Hdnkloc_{\xbd1}=\mat[\ddCprest_1|\ddCprest_2]=\bzero_{(n-k)\times2}$. 
\end{remark}
\begin{proof}[Proof of \cref{lemma:LPUNC:T_duality_D_vs_H}]
Let $\Fbnk\in\Hbspace_{\R^{n-k}}\HtripK$ be the \bdash holomorphic extension of $C^\perp$. We introduce a function $\munk:\WV\to\R^{n-k}$ defined as follows. 
For $\w\in\WV$ and any face $\ff\in\Faces$ incident to $\w$, we set
\begin{equation}\label{eq:LPUNC:munk_vs_H}
 \munk(\w):= \Hdnk(\ff)\cdot \CMI\cdot \laext(\w).
\end{equation}
First, observe that the right-hand side of~\eqref{eq:LPUNC:munk_vs_H} does not depend on the choice of $\ff$: if $\f$ is another face incident to $\w$ and separated from $\ff$ by a single edge $\e$ then by~\eqref{eq:LPUNC:primitive_knk}, $\Hdnk(\ff) - \Hdnk(\f)=\wtK(\e)\cdot \Fbnk(\b)\cdot \laext(\w)^T$. Multiplying this difference by $\CMI\cdot \laext(\w)$, we get $\bzero_{(n-k)\times1}$ by~\eqref{eq:dual_spinor}. 

Next, let $\b\in\BVint$ be incident to faces $\f_1,\f_2,\f_3$ and vertices $\w_1,\w_2,\w_3$ as in \figref{fig:black3}(a). 
 Denote $\laext_\s:=\laext(\w_\s)$, $\Uext_\s:=\munk(\w_\s)$, and $\Hdnk_\s:=\Hdnk(\f_\s)$ for $\s=1,2,3$.
We claim that 
\begin{equation}\label{eq:ddCext_vs_munk}
\frac{
\Uext_1\brlaw<\wv_2,\wv_3> +
\Uext_2\brlaw<\wv_3,\wv_1> +
\Uext_3\brlaw<\wv_1,\wv_2> 
}{
\brlaw<\wv_2,\wv_3>\cdot \brlaw<\wv_3,\wv_1>\cdot \brlaw<\wv_1,\wv_2>
} = \Fbnk(\b).
\end{equation}
By~\eqref{eq:la_Cramer}, $\laext_3\brlaw<\wv_1,\wv_2> = -\laext_1\brlaw<\wv_2,\wv_3>-\laext_2\brlaw<\wv_3,\wv_1>$, so
\begin{equation}\label{eq:Hdnk_Cramer}
\Hdnk_1\CMI\laext_3\brlaw<\wv_1,\wv_2>
= -\Hdnk_1\CMI\laext_1\brlaw<\wv_2,\wv_3> - \Hdnk_1\CMI\laext_2\brlaw<\wv_3,\wv_1>.
\end{equation}
By~\eqref{eq:LPUNC:munk_vs_H}, $\Hdnk_1\CMI\laext_3= \Uext_3$ 
and $\Hdnk_1\CMI\laext_2=\Uext_2$. Using this and substituting~\eqref{eq:Hdnk_Cramer} into~\eqref{eq:ddCext_vs_munk}, the term $\Uext_2\brlaw<\wv_3,\wv_1>$ cancels out, and the numerator simplifies to
$(\Uext_1-\Hdnk_1\CMI\laext_1)\brlaw<\wv_2,\wv_3>$. 
By~\eqref{eq:LPUNC:munk_vs_H} and~\eqref{eq:LPUNC:primitive_knk}, 
\begin{equation*}%
\Uext_1-\Hdnk_1\CMI\laext_1 
= (\Hdnk_2-\Hdnk_1)\CMI\laext_1 
= \wtlaK(\e_3) \Fbnk(\b) (\laext_3)^T\CMI\laext_1
= \brlaw<\w_1,\w_2>\brlaw<\w_3,\w_1>\Fbnk(\b),
\end{equation*}
since $\wtlaK(\e_3)=\brlaw<\w_1,\w_2>$ by~\eqref{eq:wtlaK_dfn} and $(\laext_3)^T\CMI\laext_1=\brlaw<\w_3,\w_1>$ by~\eqref{eq:dual_spinor}. This shows~\eqref{eq:ddCext_vs_munk}.

Let $\ddCprestb:\ddBV\to\R^{n-k}$ be given by $\ddCprestb(\ddbx(\w)):=\munk(\w)$ for all $\w\in\WVint$. We claim that $\ddCprestb$ is the \bdash holomorphic extension of the matrix $\ddCprest\subset\ddC^\perp$ defined in~\eqref{eq:LPUNC:ddCprest_dfn}. Let $\ddw\in\ddWVint$ be adjacent to $\ddbx(\w_1),\ddbx(\w_2),\ddbx(\w_3)$, with $\w_1,\w_2,\w_3\in\WVint$. 
The vertex $\ddw$ is contained in a unique face $\ff\in\Faces$ of $\G$, and the vertices $\w_1,\w_2,\w_3$ are all incident to $\ff$. By~\eqref{eq:LPUNC:munk_vs_H}, $\ddCprestb_\s:=\ddCprestb(\ddbx(\w_\s))=\munk(\w_\s)=\Hdnk(\ff)\cdot \CMI\cdot \laext(\w_\s)$ for $\s=1,2,3$. 
Therefore, multiplying~\eqref{eq:la_Cramer} by $\Hdnk(\ff)\cdot \CMI$, we obtain 
\begin{equation*}%
 \ddCprestb_1\brlaw<\wv_2,\wv_3> +
\ddCprestb_2\brlaw<\wv_3,\wv_1> +
\ddCprestb_3\brlaw<\wv_1,\wv_2> 
= \bzero_{(n-k)\times1}.
\end{equation*}
By~\eqref{eq:ddwtlaK_dfn}, the coefficients in the above equation are precisely the Kasteleyn edge weights $\ddwtlaK(\dde_\s)$. Thus, $\ddCprestb\in\Hbspace_{\R^{n-k}}(\ddG,\ddwtlaK)$ is indeed \bdash holomorphic and the boundary restriction $\ddCprest$ of $\ddCprestb$ satisfies $\ddCprest \subset \ddC^\perp$. 
Comparing~\eqref{eq:ddCext_vs_munk} with~\eqref{eq:ddCext_dfn}, we see that $\Fbnk$ is obtained from $\ddCprestb$ via inverse T-duality (\cref{rmk:SHIFT:inverse}). Thus, by~\eqref{eq:Meas_ddG=Chat_Qla}, $\ddCprest\cdot \Qla = C^\perp$. In particular, this shows that $\rank\ddCprest = n-k$ and that $\ddCprest$ is linearly independent from the kernel $\la$ of $\Qla$. Thus, $\ddCprest$ satisfies all conditions in~\eqref{eq:LPUNC:ddCprest_dfn}. 

Let $\sepr\in\brnLbd$. Recall that $\Dbivloc_\sepr^\perp$ is the $(n-k)$-dimensional linear subspace of $\ddC^\perp$ consisting of boundary restrictions of \bdash holomorphic functions $\ddFb\in\Hbspace_{\R}(\ddG,\ddwtlaK)$ such that $\ddFb(\ddbloc_\sepr^1)=\ddFb(\ddbloc_\sepr^2)=0$. Let $\wloc_\sepr^1,\wloc_\sepr^2\in\WVint$ be such that $\ddbx(\wloc_\sepr^1)=\ddbloc_\sepr^1$ and $\ddbx(\wloc_\sepr^2)=\ddbloc_\sepr^2$. By \cref{dfn:LPUNC:T_dual_Lpunc_graphs}, $\wloc_\sepr^1$ and $\wloc_\sepr^2$ share the face $\ploc_\sepr$ of $\G$. By~\eqref{eq:LPUNC:munk_vs_H}, $\munk(\wloc_\sepr^\s)-\Hdnkloc_\sepr\cdot \CMI\cdot \laext(\wloc_\sepr^\s)=\bzero_{(n-k)\times1}$ for $\s=1,2$. It follows that the (\bdash holomorphic) function $\ddCprestb - \Hdnkloc_\sepr\cdot \CMI\cdot \ddlaext$, where $\ddlaext\in\Hbspace_{\R^2}(\ddG,\ddwtlaK)$ is the \bdash holomorphic extension of $\la$, vanishes at both $\ddbloc_\sepr^1,\ddbloc_\sepr^2$. Thus, the right-hand side $\ddCprest - \Hdnkloc_\sepr\cdot \CMI\cdot \la$ of~\eqref{eq:LPUNC:T_duality_D_vs_H} is contained in the left-hand side $\Dbivloc_\sepr^\perp$. Since $\ddCprest$ is linearly independent from $\la$, $\rank(\ddCprest - \Hdnkloc_\sepr\cdot \CMI\cdot \la)=n-k$, so the two sides are actually equal as elements of $\Gr(n-k,n)$. 
\end{proof}

\subsection{T-duality for generalized \Lpunc graphs}\label{ssec:T_gen}
\begin{definition}\label{dfn:gen_Lfunc_graph}
Let $\CollGW$ be a planar bipartite graph of type $(k,n)$ with black boundary. A \emph{generalized \Lfuncture} in $\CollGW$ is a nonempty collection $\Sloc_\rho\subset\CollWFaces$ of faces of $\CollGW$, all sharing some (interior) white vertex
$\wloc_\rho$ of $\CollGW$.
A \emph{generalized \Lfunc graph} $\CollGWfunc$ is a planar bipartite graph $\CollGW$ equipped with an $\nL$-tuple $\Slocs=(\Sloc_1,\dots,\Sloc_\nL)$ of generalized \Lfunctures.
\end{definition}
\noindent Thus, a special case of a generalized \Lfuncture is an \emph{ordinary} \Lfuncture (\cref{dfn:LPUNC:Lfunc_graph}) when $\Sloc_\rho=\{\ploc_\rho\}$ consists of a single face. 
This includes boundary \Lfunctures $\Sloc_{\ibd}=\{\bdf_i\}$ with $\wloc_{\ibd}=\bdwx_i$.
We continue to assume that the vertices $\{\wloc_\rho\mid \rho\in\brnLbd\}$ are pairwise distinct. 

\begin{definition}\label{dfn:fullysep_gen}
Two generalized \Lfunctures $\Sloc_\seps$ and $\Sloc_\sept$ are called \emph{\twosep} if there exist faces $\ffx(\seps,i)\in\Sloc_\seps$ and $\ffx(\sept,j)\in\Sloc_\sept$ that are \twosep in $\CollGW$. 
We say that $\CollGWfunc$ is \emph{\fullysep} if $\Sloc_\seps$ and $\Sloc_\sept$ are \twosep for all $\sepst\in\brnLbdsep$.
\end{definition}

\begin{definition}\label{dfn:weighted_gen_Lfunc}
We consider \emph{weighted} generalized \Lfunc graphs $(\CollGWfunc,\wt,\bcccnL)$ with $\wt\in\CollWRtpgauge$ and $\bcccnL=(\bcccloc_1,\dots,\bcccloc_\nL)\in\SimplexbL$ as in~\eqref{eq:Simplexr_dfn}, where each open simplex $\Simplexr$ now consists of coefficient tuples $\bcccloc_\rho=(\cccx(\rho,\ff))_{\ff\in\Sloc_\rho}\in\Rtp^{\Sloc_\rho}$ 
satisfying $\sum_{\ff\in\Sloc_\rho} \cccx(\rho,\ff) = 1$.
Similarly to~\eqref{eq:bcccloc_dfn_type_1n}, we define
\begin{equation}\label{eq:GENPUNC:H_convex}
 \Hknkloc_\rho := \sum_{\ff\in\Sloc_\rho} \cccx(\rho,\ff)\Hknk(\ff),
 \quad
 \Hdnkloc_\rho := \sum_{\ff\in\Sloc_\rho} \cccx(\rho,\ff)\Hdnk(\ff), 
 \quad%
 \Hdndloc_\rho := \sum_{\ff\in\Sloc_\rho} \cccx(\rho,\ff)\Hdnd(\ff).
\end{equation}
We set
$\Measknk(\CollGWfunc,\wt,\bcccnL):=\CHL$, $\Measdnk(\CollGWfunc,\wt,\bcccnL):=(C;\Hdnklocs)$, and 
$\Measdnd(\CollGWfunc,\wt,\bcccnL):=(C;\Hdndlocs)$, where $C := \Meas(\CollGW,\wt)$. Finally, we consider a \emph{generalized \Lfunc positroid cell}
$\PtpLfunc_{\CollGWfunc}:=\{\Measknk(\CollGWfunc,\wt,\bcccnL)\mid(\wt,\bcccnL)\in
\CollWRtpgauge\times\SimplexbL\}$.
\end{definition}

\begin{remark}
By inverting the move-reduction procedure in \cref{lemma:counting_stars_type_1_n} and replacing each white vertex $\wloc_\rho$ with an \easyred graph of type $(1,\nloc_\rho)$, we see that for any weighted generalized \Lfunc graph $(\CollGWfunc,\wt,\bcccnL)$, there exists an ordinary weighted \Lfunc graph 
 with the same \Lfunc boundary measurements. 
As we mentioned in \cref{ex:non_ordinary_Meas_Lvunc}, we do not expect the analogous statement to hold for \Lvunc graphs. 
This is why we used ordinary \Lfunc graphs in \cref{dfn:Lfunc_tnn_Gr} but generalized \Lvunc graphs in \cref{dfn:LPUNC:Lvunc_Grtnn}.
\end{remark}

Consider a weighted generalized \Lvunc graph $(\ddCollGWvunc,\ddwtaux)$; cf. \cref{dfn:ddwtaux}. For $\rho\in\brnL$ and $i\in\brx{\mloc_\sepr}$, denote $\ddcccx(\rho,i):=\ddwtaux(\ddeloc_\rho^i)$, where $\ddeloc_\rho^i$ connects $\wauxloc_\rho$ to $\ddbloc_\rho^i$ as in \cref{ssec:Lvunc}. Thus, the tuple 
$\ddbcccloc_\sepr=(\ddcccx(\rho,1):\dots:\ddcccx(\rho,{\mloc_\sepr}))$
 belongs to the open
\emph{projective} $(\mloc_\sepr-1)$-dimensional simplex 
$\Dimplexr:=\Rtp^{\mloc_\rho} / \Rtp$, 
where the dilation $\Rtp$-action corresponds to gauge transformations at $\wauxloc_\rho$. 
We denote $\ddbcccnL = (\ddbcccloc_1,\dots,\ddbcccloc_\nL)$ and write $(\ddCollGWvunc,\ddwt,\ddbcccnL)$ instead of $(\ddCollGWvunc,\ddwtaux)$ from now on.

\begin{definition}[Generalized T-duality]\label{dfn:generalized_T-duality}
We say that generalized \Lpunc graphs $(\CollGWfunc,\ddCollGWvunc)$ form a \emph{T-dual pair} if the planar bipartite graphs $(\CollGW,\ddCollGW)$ form a T-dual pair and in addition, the 
generalized \Lfunctures of $\CollGWfunc$ correspond to the generalized \Lvunctures of $\ddCollGWvunc$ via \cref{dfn:LPUNC:T_dual_Lpunc_graphs}. 
The edge weight correspondence $\wtlap\mapsto\ddwtlap$ is extended so that $\bcccnL$ and $\ddbcccnL$ are related by
\begin{equation}\label{eq:GENPUNC:Simplex_vs_Dimplex}
 \ddcccx(\rho,i) = \frac{\cccx(\rho,i)}{|\wrlai|} \quad\text{and}\quad
 \cccx(\rho,i) = \frac{\ddcccx(\rho,i)\cdot |\wrlai|}{
\sum_{j=1}^{\mloc_\rho} \ddcccx(\rho,j)\cdot |\wrlaj|
}
\quad\text{for all $\rho\in\brnL$,}
\end{equation}
where we denote $\wrlai:=\brlaw<\wloc_\rho,\wlocx(\rho,i)>$ and $\wrlaj:=\brlaw<\wloc_\rho,\wlocx(\rho,j)>$.
(These coefficients are nonzero by~\eqref{eq:la_generic}.)
\end{definition}
\noindent 
When applying generalized T-duality to a given (\bdash trivalent) generalized \Lfunc graph $\Gfunc$, we first triangulate its faces obtaining a complex $\Sig$ as in \cref{dfn:SHIFT:Sig}. 
For each $\rho\in\brnL$ and $\ffx(\rho,i)\in\Sloc_\rho$, we choose an edge $\{\wloc_\rho,\wloc_\rho^i\}$ of $\Sig$ contained inside the face $\ffx(\rho,i)$ and let $\{\ddbloc_\rho:=\ddbx(\wloc_\rho),\ddbloc_\rho^i:=\ddbx(\wloc_\rho^i)\}$ be the corresponding \bivertex in $\ddSloc_\rho$. 
This ensures that the resulting generalized \Lvunctures $\ddSloc_\rho$ of $\ddCollGWvunc$ are pairwise non-crossing.

\begin{remark}\label{rmk:oneind_gen}
By \cref{lemma:SHIFT:ddG_admits_apms}, $\ddCollGW$ admits an \APM. 
Furthermore, 
since $\helWmin(\ddCollGW)\geq0$ and $\helBmin(\ddCollGW)\geq2$,
by \cref{lemma:DIM:deleting_hel_verts}, 
$\ddCollGW\rem\{\ddbloc_\sepr,\ddbloc_\sepr^i\}$ admits an \APM for all $\sepr\in\brnLbd$ and $i\in\brx{\mloc_\sepr}$. Thus, 
 $\ddCollGWvunc$ is \oneind.
\end{remark} 

\begin{figure}
 \def\inputscale{1.3}
 \setlength{\tabcolsep}{1pt}
\begin{tabular}{cccc}
 \includegraphics[scale=\inputscale]{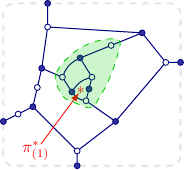}
&
 \includegraphics[scale=\inputscale]{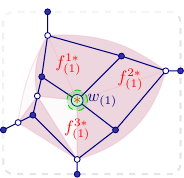}
&
 \includegraphics[scale=\inputscale]{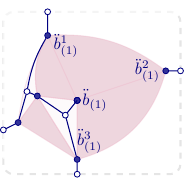}
&
 \includegraphics[scale=\inputscale]{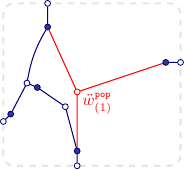}
\\
(a) $\Gfunc$ &
(b) $\CollGWfunc$ &
(c) $\ddCollGW$ &
(d) $\ddCollGWloc_1$ 
\end{tabular}
 \caption{\label{fig:BCFW-T-dual} Applying generalized T-duality to the graph $\Gfunc$ in \protect\figref{fig:BCFW-full}(e).}
\end{figure}

\begin{example}\label{ex:BCFW-T-dual}
Consider the \Lfunc graph $\Gfunc$ obtained as the output of the \ora in \figref{fig:BCFW-full}(e).
 We have $\knL=(2,4;1)$ and $\helWmin(\G) = \helBmin(\G) = 1$, with a maximal \wdash collapsible subset containing $\ploc_1$ circled in green in \figref{fig:BCFW-T-dual}(a). 
Collapsing this subset into a single white vertex $\wloc_1$ via \cref{prop:ccc}, we obtain a generalized \Lfunc graph $\CollGWfunc$ shown in \figref{fig:BCFW-T-dual}(b), with $\Sloc_1=\{\ffx(1,1),\ffx(1,2),\ffx(1,3)\}$.
It satisfies $\helWmin(\CollGW) = 2$ and $\helBmin(\CollGW) = 1$, so it is T-dualizable. 
 Applying generalized T-duality, we obtain a generalized \Lvunc graph $\ddCollGWvunc$ with the graphs $\ddCollGW$ and $\ddCollGWloc_1$ shown in \figref{fig:BCFW-T-dual}(c--d). 

We have $\helWmin(\ddCollGW)=0$ and $\helBmin(\ddCollGW)=2$, in agreement with \cref{lemma:SHIFT:ddG_admits_apms}. 
Both $\ddCollGW$ and $\ddCollGWloc_1$ admit \APMs, and for each $\ibd\in\brFOURbd$, the graph $\ddCollGWloc_{\ibd}$ obtained from $\ddCollGW$ by deleting two consecutive next-to-boundary black vertices $\ddbdbx_i,\ddbdbx_{i+1}$ also admits an \APM. Thus, $\ddCollGWvunc$ is \oneind, in agreement with \cref{lemma:ddGvunc_is_oneind,rmk:oneind_gen}. 
For each $\ibd\in\brFOURbd$, the graph $\ddCollGWlocs_{\{1,\ibd\}}=\ddCollGWloc_{1}\rem\{\ddbdbx_i,\ddbdbx_{i+1}\}$ also admits an \APM, and so do the graphs $\ddCollGWlocs_{\{\xbd1,\xbd3\}}=\ddCollGWlocs_{\{\xbd2,\xbd4\}}$. Thus, $\ddCollGWvunc$ is \twoind. 
On the other hand, one can check directly that any two faces $\ploc_\seps,\ploc_\sept$ for $\sepst\in\brnLbdsep$ can be separated by a pair of \bdbd paths in some double-dimer configuration on $\G$, so $\Gfunc$ is \fullysep.
 This agrees with \cref{lemma:LPUNC:twosep_vs_twoind,ex:fullysep_BCFW_check} below. 
\end{example}

\begin{remark}[Inverse generalized T-duality]\label{rmk:SHIFT:inverse_gen}
Generalized T-duality is invertible. First, since the generalized \Lvunctures in \cref{dfn:gen_Lvunc} are pairwise non-crossing, we can triangulate the faces of the \wdash trivalent graph $\ddCollGW$ by dark triangles of $\Sig$ so that for each $\rho\in\brnLbd$ and each \bivertex $\bivloc_\rho^i\in\ddSloc_\rho$, the two vertices in $\bivloc_\rho^i$ are connected by an edge of $\Sig$. Under inverse T-duality, the 
remaining edges of $\Sig$ incident to two dark triangles 
 correspond to square faces of $\CollGW$ that do not appear in $\Sloc_\rho$ for any $\rho\in\brnLbd$. Similarly to \cref{rmk:LPUNC:choices}, flipping such edges of $\Sig$ and applying square moves~\MV2 at the corresponding faces does not affect the \Lfunc boundary measurements. As explained in \cref{rmk:SHIFT:inverse}, the edge weight transformation $\wtlap\mapsto\ddwtlap$ is invertible. It is clear that the map $\bcccloc_\rho\mapsto\ddbcccloc_\rho$ given by~\eqref{eq:GENPUNC:Simplex_vs_Dimplex} yields a homeomorphism $\Simplexr\xrasim\Dimplexr$ for each $\rho\in\brnL$.
\end{remark}

\begin{proposition}%
\label{lemma:GENPUNC:T_duality_D_vs_H}
Let $(\CollGWfunc,\wtlap,\bcccnL)$ and $(\ddCollGWvunc,\ddwtlap,\ddbcccnL)$ be a T-dual pair of weighted generalized \Lpunc graphs.
 Then~\eqref{eq:LPUNC:T_duality_D_vs_H} holds for each $\rho\in\brnLbd$.
\end{proposition}
\begin{proof}

Recall from the proof of \cref{lemma:LPUNC:T_duality_D_vs_H} that $\la\subset\ddC^\perp$ extends to 
a \bdash holomorphic function
 $\ddlaext\in\Hbspace_{\R^2}(\ddCollGW,\ddCollwtlaK)$
given by $\ddlaext(\ddbx(\wv)) = \laext(\wv)$ for all $\wv\in\WVint$. 
As explained in the proof of \cref{lemma:LPUNC:D_perp_cap_la=0},
 $\ddepsKloc_\sepr(\ddeloc_\sepr^i)=\ddepsbb_{\ddbloc_\sepr,\ddbloc_\sepr^i}$
is the sign of 
$[\ddbloc_\rho\,\ddblocx(\rho,i)]_{\ddot\la} = \brlaw<\wloc_\rho,\wlocx(\rho,i)> = \wrlai$.
Thus, by~\eqref{eq:GENPUNC:Simplex_vs_Dimplex}, the coefficients in~\eqref{eq:GENPUNC:dual_vector} are 
$\ddWKmatloc_\sepr(\wauxloc_\sepr,\ddbloc_\sepr^i)
=\ddepsKloc_\sepr(\ddeloc_\sepr^i)\ddcccx(\rho,i)
=\frac{\cccx(\rho,i)}{\wrlai}$.
 Let 
\begin{equation}\label{eq:laextloc_munkloc_dfn}
 \laextloc_\rho := \sum_{i=1}^m \frac{\cccx(\rho,i)}{\wrlai} \laext(\wlocx(\rho,i))
 \quad\text{and}\quad
 \munkloc_\rho := \sum_{i=1}^m \frac{\cccx(\rho,i)}{\wrlai} \munk(\wlocx(\rho,i)),
\end{equation}
where $m:=\mloc_\rho$ and $\munk$ was introduced in~\eqref{eq:LPUNC:munk_vs_H}.
For each $i\in\brm$, the face $\ffx(\rho,i)\in\Sloc_\rho$ of $\CollGW$ is incident to both $\wloc_\rho$ and $\wlocx(\rho,i)$. Set $\Hx(\rho,i):=\Hdnk(\ffx(\rho,i))$. By~\eqref{eq:LPUNC:munk_vs_H}, 
\begin{equation}\label{eq:Hx_CMI_laext}
 \Hx(\rho,i)\cdot \CMI\cdot \laext(\wloc_\rho) = \munk(\wloc_\rho)
 \quad\text{and}\quad
\Hx(\rho,i)\cdot \CMI\cdot \laext(\wlocx(\rho,i)) = \munk(\wlocx(\rho,i)) 
\quad\text{for all $i\in\brm$}. 
\end{equation}
Applying~\eqref{eq:la_Cramer} to $\laext(\wloc_\rho),\laext(\wlocx(\rho,i)),\laext(\wlocx(\rho,j))$ and multiplying both sides by $\Hx(\rho,i)\cdot \CMI$, we get
\begin{equation}\label{eq:GENPUNC:Hx_rho_i}
 \Hx(\rho,i)\cdot \CMI\cdot\wrlai\cdot \laext(\wlocx(\rho,j)) = \wrlaj \munk(\wlocx(\rho,i)) + 
\wrlaji \munk(\wloc_\rho)
\quad\text{for $i,j\in\brm$,}
\end{equation}
where $\wrlaji:=\<\wlocx(\rho,j)\,\wlocx(\rho,i)\>_{\la}$.
By~\eqref{eq:GENPUNC:H_convex}, \eqref{eq:laextloc_munkloc_dfn}, and~\eqref{eq:GENPUNC:Hx_rho_i}, we find
\begin{equation*}%
 \Hdnkloc_\rho \cdot \CMI\cdot \laextloc_\rho 
= \sum_{i,j=1}^m \cccx(\rho,i) \cccx(\rho,j) \frac{\Hx(\rho,i)\cdot \CMI\cdot \laext(\wlocx(\rho,j))}{\wrlaj}
= \sum_{i,j=1}^m \cccx(\rho,i) \cccx(\rho,j) \left( 
\frac{\munk(\wlocx(\rho,i))}{\wrlai} 
+ \frac{\wrlaji\munk(\wloc_\rho)}
{\wrlaj\wrlai} 
\right).
\end{equation*}
The terms $a_{j,i}:=\cccx(\rho,i) \cccx(\rho,j)\frac{\wrlaji\munk(\wloc_\rho)}
{\wrlaj\wrlai}$ cancel out since $a_{j,i}=-a_{i,j}$ for all $i,j\in\brm$. Since $\sum_{j=1}^m \cccx(\rho,j)=1$, the remaining terms add up to $\munkloc_\rho$ by~\eqref{eq:laextloc_munkloc_dfn}. Thus,
$\Hdnkloc_\rho \cdot \CMI\cdot \laextloc_\rho = \munkloc_\rho$. 

Recall that $\ddCprestb:\ddBV\to\R^{n-k}$ is given by $\ddCprestb(\ddbx(\w)):=\munk(\w)$ for all $\w\in\WVint$.
Let 
$\ddFbloc_\rho:=(\ddCprestb - \Hdnkloc_\sepr\cdot \CMI\cdot\ddlaext) \in\Hbspace_{\R^{n-k}}(\ddCollGW,\ddCollwtlaK)$. 
Applying~\eqref{eq:GENPUNC:H_convex} and using the first identity in~\eqref{eq:Hx_CMI_laext}, we see that $\ddFbloc_\rho$ satisfies the first identity in~\eqref{eq:GENPUNC:dual_vector}. Combining the second identity in~\eqref{eq:Hx_CMI_laext} with $\Hdnkloc_\rho \cdot \CMI\cdot \laextloc_\rho = \munkloc_\rho$, we obtain the second identity in~\eqref{eq:GENPUNC:dual_vector}. As explained in the last paragraph of the proof of \cref{lemma:LPUNC:T_duality_D_vs_H}, this implies~\eqref{eq:LPUNC:T_duality_D_vs_H}.
\end{proof}

\section{T-duality for loop amplituhedra}\label{sec:LOOP}
The goal of this section is to relate tilings of loop amplituhedra in momentum space to those in momentum-twistor space. We extend the results of~\Mref{sec:TREE} from tree level to loop level, relying on T-duality for \Lpunc graphs developed in \crefrange{sec:shift}{sec:LPUNC}.

\subsection{Ambient loop amplituhedra}\label{ssec:LOOP:ambient_ampl_dfn}

Fix integers $2\leq k\leq n-2$. We extend the notion of the \emph{ambient tree momentum amplituhedron} $\MPkntree$ (\cref{dfn:TREE:MPkn}) to loop level.

\begin{definition}[Ambient loop momentum amplituhedron]\label{dfn:LPUNC:amb_loop_ampl}
Fix an integer $\nL\geq0$. The \emph{ambient $\nL$-loop momentum amplituhedron} $\MPknL$ is the space of triples $\llPllL$ with $\byL=(\yloc_1,\dots,\yloc_\nL)\in(\Rdd)^{\nL}$, 
such that 
\begin{enumerate}[label=(\alph*)]
\item\label{MPknL1} $\llPll\in\MPkntree$, 
\item\label{MPknL2} for each $\sepst\in\brnLbdsep$, we have $(\yloc_\seps - \yloc_\sept)^2>0$, and
\item\label{MPknL4} for each $\rho\in\brnL$, $\yTloc_\rho$ is located strictly inside the (simple by \cref{lemma:TOP:Mpos=>simple}) polygon $\PllT$.
\end{enumerate}
Here, we denote $\yloc_{\ibd}:=\bdx_i$ for $\ibd\in\brnbd$, with $\Pll=(\bdx_1=0,\bdx_2,\dots,\bdx_n)$ as in \cref{dfn:TREE:lalat_vs_xbd}. 
\end{definition}

\begin{remark}\label{rmk:GGsh_action_preserves_MPknL}
We view the triple $\llPllL$ as defined up to the natural $\GGsh$-action (\cref{dfn:BACKGR:GGsh}). %
Since $\GGsh$ is generated by scaled Lorentz transformations and translations, conditions~\itemref{MPknL1}--\itemref{MPknL2} are preserved by $\GGsh$-action. By \cref{lemma:TOP:Mpos=>simple}, when conditions~\itemref{MPknL1}--\itemref{MPknL2} are satisfied, each point $\yTloc_\rho$, $\rho\in\brnL$, lies either strictly inside or strictly outside $\PllT$. Since $\GGsh$ is connected, it follows that condition~\itemref{MPknL4} is also invariant under $\GGsh$-action. Alternatively, this may be deduced by noting that $\GLp$-winding numbers are invariant under the left and right $\GLp$-action; see~\Mref{ssec:GLp_winding}.
\end{remark}

\begin{definition}
For an \Lfunc graph $\Gfunc$, we set 
\begin{equation}\label{eq:LPUNC:MPG_dfn}
 \MPG:=\left\{\llPllL\in\MPknL\middle| 
\text{
\begin{tabular}{c}
there exists $\wt\in\Rtpgauge$ such that for\\
 $(C;\Hdndlocs):=\Measdnd(\Gfunc,\wt)$, we have\\
$\la\subset C\subset\latp$ and 
$\yMloc_\rho=\Hdndloc_\rho$ for all $\rho\in\brnL$.
\end{tabular}
}
\right\}.
\end{equation}
\noindent Here, $\yMloc_\rho$ was defined in~\eqref{eq:x_vs_ap_vs_am} and $\Hdndloc_\rho$ was defined in~\eqref{eq:LPUNC:Measdnk_def}.
\end{definition}

\begin{definition}
For a nonzero vector $v=(v_1,v_2,\dots,v_d)\in\R^d\setminus\{0\}$, let $\var(v)$ be the number of times the sequence $v_1,v_2,\dots,v_d$ changes sign (omitting all zero entries in $v$). 
\end{definition}
\noindent It is easy to see that if $\la\in\Gror(2,n)$ satisfies $\brla<i,i+1> >0$ for all $i\in\brn$ then 
\begin{equation}\label{eq:TREE:wind_vs_varx}
 \wind(\la) = (\varx(\la) +1)\pi, \quad\text{where}\quad \varx(\la):=\var\left(\brla<1,2>,\brla<1,3>,\dots,\brla<1,n>\right).
\end{equation}
\noindent For $V=\mat[V_1|V_2|\cdots|V_n]\in\Gror(4,n)$, we set $V_{i+n}=(-1)^{k-1}V_i$ for all $i\in\Z$ as in \cref{notn:BACKGR:cs}. For $a,b,c,d\in\Z$, we denote $\brV[a,b,c,d]:=\det\mat[V_a|V_b|V_c|V_d]$. We set
\begin{equation*}%
 \varxxx(V):=\var\left(\brV[1,2,3,4],\brV[1,2,3,5],\dots,\brV[1,2,3,n]\right).
\end{equation*}

\begin{definition}\label{dfn:TREE:AAkn}
The \emph{ambient tree \mta} is given by
\begin{equation}\label{eq:TREE:AAkn_dfn}
 \AAkntree := \left\{V\in\Gror(4,n)\middle| 
\text{
\begin{tabular}{cc}
$\brV[i,i+1,j,j+1] >0$ for all $i+2\leq j\leq i+n-2$, and\\
$\varxxx(V)=k-2$
\end{tabular}
} \right\}.
\end{equation}
\end{definition}

\begin{definition}
The \emph{$\nL$-loop Grassmannian} $\GrL(4,n)$ is the quotient of the space of 
$4\times (n+2L)$ matrices $\VLpunc:=\mat[V|\Line_1|\Line_2|\cdots|\Line_\nL]$ %
(with $V\in\Mator_{4,n}$ and $\Liner\in\Mator_{4,2}$ for $\rho\in\brnL$)
 modulo the $\GL^+_4(\R)\times\GLp^\nL$-action, where $\GL^+_4(\R):=\{g\in\GL_4(\R)\mid \det g>0\}$ acts on $\mat[V|\Line_1|\Line_2|\cdots|\Line_\nL]$ by left multiplication and $\GLp^\nL$ acts on $\mat[\Line_1|\Line_2|\cdots|\Line_\nL]$ by right multiplication. 
\end{definition}
\noindent %
For $\rho\in\brnL$, we let $\pLr\in\Gror(2,4)$ be the $2$-plane orthogonal to $\Liner$, oriented so that $\det\mat[(\pLr)^T|\Liner]>0$, and we set $\Vloc_\rho:=\pLr\cdot V\in\Gror(2,V)$. Thus, 
 $\Vloc_\rho$ and $\Liner$ determine each other.
For $\rho\in\brnL$ and $i,j\in\Z$, we set $\brVL[i,j,\rho]:=\det\mat[V_i|V_j|\Liner]$. For $\rho,\gamma\in\brnL$, we set $\brLL[\rho,\gamma]:=\det\mat[\Liner|\Lineg]$. For $\rho\in\brnL$, we set
\begin{equation}\label{eq:varVL_dfn}
 \varVL_\rho:=
\var\left(\brVL[1,2,\rho],\brVL[1,3,\rho],\dots,\brVL[1,n,\rho]\right).
\end{equation}
Finally, we set $\brLL[\ibd,\rho]:=\brVL[i,i+1,\rho]$ and $\brLL[\ibd,\jbd]:=\brV[i,i+1,j,j+1]$ for $\ibd,\jbd\in\brnbd$ and $\rho\in\brnL$. 

\begin{definition}[Ambient loop \mta]\label{dfn:LSH:AAknL}
The \emph{ambient $\nL$-loop \mta} $\AAknL$ is the space of points $\VLpunc\in\GrL(4,n)$ such that 
\begin{enumerate}[label=(\"{\alph*})]
\item\label{AAknL1} $V\in\AAkntree$,
\item\label{AAknL2} for each $\sepst\in\brnLbdsep$, we have $\brLL[\seps,\sept]>0$, and
\item\label{AAknL4} for each $\rho\in\brnL$, we have $\varVL_\rho = k$.
\end{enumerate}
\end{definition}

\begin{definition}\label{dfn:LSHIFT:AAddG}
For an \Lvunc graph $\ddGvunc$, we set 
\begin{equation}\label{eq:LPUNC:AAddG_dfn}
 \AAddG:=\left\{\VLpunc\in\AAknL\middle| 
\text{
\begin{tabular}{c}
there exists $\ddwt\in\ddRtpgauge$ such that for\\
$\ddCDL:=\Measbiv(\ddGvunc,\ddwt)$, we have\\
$V\subset\ddC^\perp$ and $\VLr\subset \Dbivloc_\rho^\perp$ for all $\rho\in\brnL$
\end{tabular}
}
\right\}.
\end{equation}
\end{definition}
\begin{remark} \label{rmk:LSHIFT:D_perp_cap_V}
The condition $\VLr\subset \Dbivloc_\rho^\perp$
 in~\eqref{eq:LPUNC:AAddG_dfn} may be replaced with $\VLr = V\cap \Dbivloc_\rho^\perp$ as elements of $\Gr(2,n)$. Indeed, since $V\subset \ddC^\perp$ and $\ddC$ is codimension-$2$ inside $\Dbivloc_\rho$, we have ${\dim(V\cap \Dbivloc_\rho^\perp)\geq2}$. On the other hand, by \cref{thm:AAshift_tree} below, for each $\VLpunc\in\AAknL$, there exists $\la\in\lak$ such that $\la\subset V$. Since $V\subset\ddC^\perp$, by \cref{lemma:LPUNC:D_perp_cap_la=0}, we get $\la\cap\Dbivloc_\rho^\perp=\{0\}$. Thus, $\dim(V\cap \Dbivloc_\rho^\perp)\leq2$. 
\end{remark}

We set $\FlnL(2,4):=\{\laVLpunc\in\Gror(2,n)\times\GrL(4,n)\mid \la\subset V\}$,
\begin{align}
\label{eq:LSHIFT:MAknL_dfn}
 \MAknL&:=\{\laVLpunc\in\lak\times\AAknL\mid \la\subset V\},\quad\text{and}\\
\label{eq:LSHIFT:MAddG_dfn}
 \MAddG&:=\{\laVLpunc\in\lak\times\AAddG\mid \la\subset V\}.
\end{align}

\subsection{T-duality for ambient loop amplituhedra}\label{ssec:LOOP:ambient_ampl_T_dual}
Our next goal is to relate the ambient loop amplituhedra $\MPknL$, $\MAknL$, and $\AAknL$ by explicit maps. 
Observe that 
\begin{equation*}%
 \dim\MPknL = \dim\MAknL = 4(n+L-3) \quad\text{and}\quad \dim\AAknL = 4(n+L-4).
\end{equation*}

In what follows, we continue to denote $\Pll=(\bdx_1=0,\bdx_2,\dots,\bdx_n)$. %
\begin{lemma}
Let $\llPllL\in\MPknL$. Then $V:=\Qlapp(\lat)$ can be represented by a $4\times n$ matrix 
\begin{equation}\label{eq:mu_vs_bdxM_vs_Qlapp}
V=\begin{pmatrix}
\la \\ \mu
\end{pmatrix}
\quad\text{with}\quad
 \mu_i = \bdxM_i\cdot \CMI\cdot \la_i = \bdxM_{i-1}\cdot \CMI\cdot \la_i
 \quad\text{for all $i\in\brn$}.
\end{equation}
\end{lemma}
\begin{proof}
Recall from~\eqref{eq:TE:decor_dfn} that $\bdxM_i - \bdxM_{i-1} = \lat_i\cdot \la_i^T$ for $i\in\brn$. By~\eqref{eq:dual_spinor}, $(\bdxM_i - \bdxM_{i-1})\cdot \CMI\cdot \la_i = \bzero_{2\times1}$. For a computation showing that the $2\times n$ matrix $\mu$ defined by~\eqref{eq:mu_vs_bdxM_vs_Qlapp} satisfies $\lat = \Qla(\mu)$, see the proof of~\Mref{prop:MPnla_AAnla_homeo}.
\end{proof}

The \emph{ambient loop amplituhedron T-duality map} is defined by
\begin{align}
\label{eq:AAshift_dfn}
\AAshift:\MPknL &\to \GrL(4,n),%
& \llPllL&\mapsto \matlr[V=\Qlapp(\lat)\,\middle|\,\Liner=\begin{pmatrix}
\Id_2\\ \xM_{\yloc_\rho}\cdot \CMI
\end{pmatrix},\ \rho\in\brnL];\\%\text{ for $\rho\in\brnL$};\\
\label{eq:AAiso_dfn}
\AAiso:\MPknL&\to\FlnL(2,4), & \llPllL
&\mapsto (\la,\AAshift\llPllL).
\end{align}
Here, we represent $V=\Qlapp(\lat)$ by a specific matrix given by~\eqref{eq:mu_vs_bdxM_vs_Qlapp}. 
It follows that $\AAshift = \AAforg\circ\AAiso$, where $\AAforg:\FlnL(2,4)\to\GrL(4,n)$ is the projection map. For $\ibd\in\brnbd$, we denote $\Line_{\ibd}:=\begin{pmatrix}
\Id_2\\ \bdxM_i\cdot \CMI
\end{pmatrix}$. By~\eqref{eq:mu_vs_bdxM_vs_Qlapp}, $\Line_{\ibd}=\begin{pmatrix}
\la_i&\la_{i+1}\\ \mu_i & \mu_{i+1}
\end{pmatrix}\cdot \mat[\la_i|\la_{i+1}]^{-1}$ with $\mat[\la_i|\la_{i+1}]\in\GLp$ by \crefi{dfn:LPUNC:amb_loop_ampl}{MPknL1}.

\begin{theorem}[T-duality for ambient loop amplituhedra] \label{thm:AAshift}
Let $2\leq k\leq n-2$ and $\nL\geq0$. 
\begin{enumerate}[label=(\arabic*)]
\item\label{AAshift_iso} The map $\AAiso:\MPknL\xrasim\MAknL$ is a homeomorphism. It restricts to a homeomorphism $\MPG\xrasim \MAddG$ for all T-dual pairs $(\Gfunc,\ddGvunc)$ of \Lpunc graphs.
\item\label{AAshift_forg} The map $\AAforg:\MAknL\to\AAknL$ is surjective and open. It restricts to a surjective open map $\MAddG\to\AAddG$ for all T-dual pairs $(\Gfunc,\ddGvunc)$ of \Lpunc graphs.%
\end{enumerate}
\end{theorem}
\begin{theorem}[{\Mref{thm:AAshift}}]\label{thm:AAshift_tree}
 \Cref{thm:AAshift} holds at tree level ($\nL=0$).
\end{theorem}

\begin{definition}[Normal form]\label{dfn:LSHIFT:normal_form}
We say that $V\in\AAkntree$ is in \emph{normal form} if $V=\begin{pmatrix}
\labase \\ \mu
\end{pmatrix}$ for some $\labase\in\lakMAT$ and $\mu\in\Mator_{2,n}$ satisfying 
 $\mat[\mu_1|\mu_2]=\bzerodd$. Next, $\VLpunc\in\AAknL$ is in \emph{normal form} if $V$ is in normal form and for each $\rho\in\brnL$, $\Liner=\begin{pmatrix}
\Id_2\\ \MLr\CMI
\end{pmatrix}$ for some $\MLr\in\Matddr$. Finally, $\llPllL\in\MPknL$ is in \emph{normal form} if $\Pll=(\bdx_1=0,\bdx_2,\dots,\bdx_n)$. 
\end{definition}
\noindent It is clear that $\llPllL\in\MPknL$ can be written (non-uniquely) in normal form using $\GGsh$-action. 
The analog of this result for $\AAknL$ is surprisingly non-trivial to prove. 

\begin{lemma}\label{lemma:LSHIFT:VLpunc_normal_form}
Any $\VLpunc\in\AAknL$ can be written (non-uniquely) in normal form using left $\GL^+_4(\R)$-action and right $\GLp^\nL$-action.
\end{lemma}
\begin{proof}
Let $\VLpunc\in\AAknL$. By \cref{thm:AAshift_tree}, $\AAforg:\MAkntree\to\AAkntree$ is surjective. Since $V\in\AAkntree$, there exists $\la\in\lak$ satisfying $\la\subset V$. 
Applying left $\GL^+_4(\R)$-action, we may assume that $V=\begin{pmatrix}
\labase \\ \mu
\end{pmatrix}$ is in normal form. 
In particular, $\mat[\mu_1|\mu_2]=\bzerodd$ and $\brla<1,2> >0$. Applying left $\GLp$-action to $\la$, we assume that $\mat[\labase_1|\labase_2]=\Id_2$. 
Fix $\rho\in\brnL$. We have $\Delta_{34}(\Liner)=\brVL[1,2,\rho]>0$, so after applying right $\GLp$-action to $\Liner$, we may assume that $\Liner = \begin{pmatrix}
\Wloc_\rho \\ \Id_2
\end{pmatrix}$ for some $\Wloc_\rho\in\Matddr$. It suffices to show that $\det\Wloc_\rho>0$, since in this case, $\Liner\cdot \Wloc_\rho^{-1}$ will be in normal form. 

Suppose first that $\det\Wloc_\rho = 0$. 
Projectivizing the oriented $2$-planes $\Liner$ and $\Kinebase:=\begin{pmatrix}
\bzerodd\\ \Id_2
\end{pmatrix}$, we obtain two intersecting oriented lines in $\RP^3$ (cf. \cref{ssec:APP:RP3}).
Suppose first that $\Wloc_\rho\neq\bzero_{2\times2}$, i.e., $\Liner\neq\Kinebase$. 
Let $\Piner$ be the plane in $\RP^3$ spanned by these two lines. We identify $\Piner$ with the standard Cartesian plane and $\Liner$ and $\Kinebase$ with the $x$- and the $y$-axis, respectively. 
For $i\in\brn$, we have $\det\mat[V_i|V_{i+1}|\Kinebase] = \brlabase<i,i+1> >0$ and $\brVL[i,i+1,\rho]>0$. 
Thus, 
for each $i\in\brn$, the intersection point of the line $\Kine_i=\mat[V_i|V_{i+1}]$ with $\Piner$ lies in either the second or the fourth quadrant. In particular, we may rotate $\Liner$ towards $\Kinebase$ around the origin through the first and the third quadrants until the two oriented lines coincide without violating the inequalities $\brVL[i,i+1,\rho]>0$ throughout the deformation. On the other hand, since $\VLpunc\in\AAknL$, we have $\varVL_\rho = k$ and by~\eqref{eq:TREE:wind_vs_varx}, $\varx[V|\Kinebase] = \varx(\la) = k-2$ since $\wind(\la) = (k-1)\pi$. This leads to a contradiction since by~\eqref{eq:TREE:wind_vs_varx}, $\varx[V|\Liner]$ is locally constant on the set of oriented lines $\Liner$ satisfying $\brVL[i,i+1,\rho]>0$ for all $i\in\brn$.
Similarly, we cannot have $\Liner = \Kinebase$ because $\varVL_\rho = k$ but $\varx[V|\Kinebase] = k-2$.

Suppose now that $\Wloc_\rho\in\GLm$, i.e., $\det\Wloc_\rho<0$. Let $\MLr\in\GLm$ be such that $\MLr\CMI = \Wloc_\rho^{-1}$, and let $\yloc_\rho\in\Rdd$ be the point corresponding to $\MLr$ via~\eqref{eq:x_vs_ap_vs_am}. Using~\eqref{eq:det_xM_vs_Pmom^2}, we calculate $\frac14(\yloc_\rho - \bdx_i)^2 = \brVL[i,i+1,\rho]\brla<i,i+1>^{-1}\det\Wloc_\rho^{-1} < 0$ for all $i\in\brn$; cf.~\eqref{eq:LOOP:br=xdiff_combined} below. Thus, $(\yloc_\rho-\bdx_i)^2<0$ for all $i\in\brn$.
On the other hand, by \crefi{dfn:LPUNC:amb_loop_ampl}{MPknL1}, $(\bdx_i-\bdx_j)^2>0$ for all $i+2\leq j\leq i+n-2$. 

 We claim that $\yOloc_\rho$ lies outside the convex hull $\Conv\{\bdxO_1,\dots,\bdxO_n\}$. 
Suppose otherwise that there exists a convex combination $\yOloc_\rho = \sum_{i=1}^n c_i \bdxO_i$ for some coefficients $c_i\geq0$ such that $\sum_{i=1}^n c_i = 1$. Let $\yloc_\rho'\in\Rdd$ be given by $\yloc_\rho' = \sum_{i=1}^n c_i \bdx_i$ so that $\yOloc_\rho'=\yOloc_\rho$. 
Consider a discrete probability measure 
on $\brn$ 
 defined by $(c_1,c_2,\dots,c_n)$. For a random variable $X:\brn\to\Rdd$ on this probability space, 
 denote $\Vardd(X):=\Ebb(X-\Ebb X)^2$. Thus, for an i.i.d.\ copy $X'$ of $X$, we have $\frac12\Ebb(X-X')^2 = \frac12(\Vardd(X)+\Vardd(X')) = \Vardd(X)$. 
Suppose that $X$ takes value $\bdx_i$ with probability $c_i$.
Using $\Ebb(X-\yloc_\rho)^2 = \Vardd(X) + (\Ebb X - \yloc_\rho)^2$ and $\Ebb X = \yloc_\rho'$, we find
\begin{equation*}
 \Ebb(X-\yloc_\rho)^2 = \frac12\Ebb(X-X')^2 + (\yloc_\rho'-\yloc_\rho)^2
 = \sum_{1\leq i<j\leq n} c_i c_j (\bdx_i - \bdx_j)^2 + |\yTloc_\rho'-\yTloc_\rho|^2 \geq 0.
\end{equation*}
On the other hand, $\Ebb(X-\yloc_\rho)^2 = \sum_{i=1}^n c_i (\bdx_i - \yloc_\rho)^2 < 0$, a contradiction.

Similarly to~\Mref{lemma:turnGL_trT}, the map $\trO:\GLm\to\R^2\setminus\{0\}$ sending $\begin{pmatrix}
a & b\\ 
c & d
\end{pmatrix}\mapsto \begin{pmatrix}
a-d \\
-b-c
\end{pmatrix}$ induces an isomorphism on the respective fundamental groups (both of which are isomorphic to $\Z$). 
Since $\yOloc_\rho$ lies outside $\Conv\{\bdxO_1,\dots,\bdxO_n\}$, we conclude that the loop in $\GLm$ corresponding to the null polygon $(\bdx_1-\yloc_\rho,\dots,\bdx_n-\yloc_\rho)$ is contractible. 
 Similarly to \Meqref{eq:windaround_muof}, 
 we again get $\varx[V|\Liner]=\varx[V|\Kinebase] = k-2$, contradicting $\VLpunc\in\AAknL$.
\end{proof}

\begin{proof}[Proof of \cref{thm:AAshift}]
We prove part~\itemref{AAshift_iso}. Let $\llPllL\in\MPknL$ and $\AAiso\llPllL=\laVLpunc$. This map is invertible: 
we have $\lat=V\cdot \Qla = \mu\cdot \Qla$, and by \cref{lemma:LSHIFT:VLpunc_normal_form}, $\laVLpunc$ can be written in normal form, in which case each $\yloc_\rho$ is recovered uniquely from $\Liner=\begin{pmatrix}
\Id_2\\ \xM_{\yloc_\rho}\cdot \CMI
\end{pmatrix}$.

We claim that the map $\AAiso$ translates conditions~\itemref{MPknL1}--\itemref{MPknL4} in \cref{dfn:LPUNC:amb_loop_ampl} into the corresponding conditions~\itemref{AAknL1}--\itemref{AAknL4} in \cref{dfn:LSH:AAknL}. 
 By \cref{thm:AAshift_tree},~\itemref{MPknL1} is equivalent to~\itemref{AAknL1}. Assume now that~\itemref{MPknL1} holds for $\llPllL$ and~\itemref{AAknL1} holds for $\VLpunc$. 
For $\sepr\in\brnLbd$, write $\brlasepr:=\brla<i,i+1>$ if $\sepr=\ibd\in\brnbd$ and $\brlasepr:=1$ if $\sepr\in\brnL$. 
Similarly to~\Mref{eq:brV_vs_Mand}, for $\sepst\in\brnLbdsep$, we get 
\begin{equation}\label{eq:LOOP:br=xdiff_combined}
 \frac14(\yloc_\seps - \yloc_\sept)^2 = \brLL[\seps,\sept] \brlaseps^{-1}\brlasept^{-1}.
\end{equation}
Thus,~\itemref{MPknL2} is equivalent to~\itemref{AAknL2}, so assume that~\itemref{MPknL1}--\itemref{MPknL2} hold for $\llPllL$ and~\itemref{AAknL1}--\itemref{AAknL2} hold for $\VLpunc$. 
By~\Mref{cor:var_vs_inside},~\itemref{MPknL4} is equivalent to~\itemref{AAknL4}. Thus, $\AAiso:\MPknL\xrasim\MAknL$ is a homeomorphism. 

Let $(\Gfunc,\ddGvunc)$ be a T-dual pair of \Lpunc graphs. Let $\llPllL\in\MPG$ and $\laVLpunc:=\AAiso\llPllL\in\MAknL$. We check that $\VLpunc\in\AAddG$ satisfies all conditions in~\eqref{eq:LPUNC:AAddG_dfn}. 
 Let $\wt\in\Rtpgauge$ and $(C;\Hdndlocs):=\Measdnd(\Gfunc,\wt)$ be as in~\eqref{eq:LPUNC:MPG_dfn}. By \cref{lemma:black_gauge_eq}, we may assume that $\wt=\wtlap$. 
Let $\ddwt:=\ddwtlap$ be the T-dual edge weights and set $\ddCDL:=\Measbiv(\ddGvunc,\ddwt)$. 
Since $V=\Qlapp(\lat)$ by~\eqref{eq:AAshift_dfn} and $\ddC=C\cdot \Qla$ by~\eqref{eq:Meas_ddG=Chat_Qla}, we get $V\subset \ddC^\perp$ because $V\cdot \ddC^T = V\cdot \Qla\cdot C^T = \lat\cdot C^T = \bzero_{2\times k}$, where we used that $\Qla$ is self-adjoint and $C\subset\latp$.

We show $\VLr\subset\Dbivloc_\rho^\perp$. 
Let $\Atmat\in\Mator_{2,n-k}$ be such that $\lat = \Atmat \cdot C^\perp$. Then $\Hdndloc_\sepr=\Atmat\cdot \Hdnkloc_\sepr$ for each $\sepr\in\brnLbd$. 
 In the notation of~\eqref{eq:LPUNC:ddCprest_dfn}, we set $\mu:=\Atmat\cdot \ddCprest$ so that
 $\mu\cdot \Qla = \lat$ (as matrices). We set $V':=\begin{pmatrix}
\la \\ \mu
\end{pmatrix}$. Since $\la\subset V'$ and $V'\cdot \Qla=\lat$ are both $2$-dimensional, we get $\dim V' = 4$, and thus $V'=\Qlapp(\lat)=V$. Letting 
\begin{equation}\label{eq:LPUNC:V_cap_Drho=pLr*V}
 \pLr:=\mat[-\xM_{\yloc_\sepr} \cdot \CMI|\Id_2], \quad\text{we find}\quad
 \Vloc_\sepr = \pLr\cdot V = \mu - \xM_{\yloc_\sepr}\cdot \CMI\cdot \la.
\end{equation}
We have $\xM_{\yloc_\sepr}=\Hdndloc_\sepr$ by~\eqref{eq:LPUNC:MPG_dfn}. Since $\Hdndloc_\sepr=\Atmat\cdot \Hdnkloc_\sepr$ and $\mu=\Atmat\cdot \ddCprest$, by~\eqref{eq:LPUNC:T_duality_D_vs_H} and~\eqref{eq:LPUNC:V_cap_Drho=pLr*V}, we get
$\Vloc_\sepr = \Atmat\cdot (\ddCprest - \Hdnkloc_\sepr\cdot \CMI\cdot \la)=\Atmat\cdot \Dbivloc_\sepr^\perp$. 
 Thus, $\VLpunc\in\AAddG$, and therefore $\laVLpunc\in\MAddG$. 

Conversely, let $\laVLpunc\in\MAddG$ be in normal form and let $\llPllL:=\AAiso^{-1}\laVLpunc\in\MPknL$. 
Let $\ddCDL=\Measbiv(\ddGvunc,\ddwt)$ be as in~\eqref{eq:LPUNC:AAddG_dfn}. 
Set $C:=\Qlapp(\ddC)$. 
We have $\lat=V\cdot \Qla = \mu\cdot \Qla$ for $\mu$ as in \cref{dfn:LSHIFT:normal_form}. 
Applying inverse T-duality (\cref{rmk:SHIFT:inverse}), we obtain a weighted \Lfunc graph $(\Gfunc,\wtlap)$ with boundary measurements $(C;\Hdndlocs)=\Measdnd(\Gfunc,\wtlap)$. To show that $\llPllL\in\MPG$, we need to check the conditions in~\eqref{eq:LPUNC:MPG_dfn}. By construction, $\la\subset C$. We have $C\subset\latp$ since $C=\Qlapp(\ddC)$ and $\lat = V\cdot \Qla$ with $V\subset\ddC^\perp$. 
It remains to check that $\yMloc_\sepr=\Hdndloc_\sepr$ for all $\sepr\in\brnL$. We check it more generally for $\sepr\in\brnLbd$.

By \cref{rmk:LSHIFT:D_perp_cap_V}, $\Vloc_\sepr= V\cap \Dbivloc_\sepr^\perp$ as elements of $\Gr(2,n)$. 
Since $V\subset\ddC^\perp$, in the notation of~\eqref{eq:LPUNC:ddCprest_dfn}, there exists $\Atmat\in\Mator_{2,n-k}$ such that $\mu=\Atmat\cdot \ddCprest$. Thus, $\lat = \Atmat\cdot C^\perp$, and so $\Hdndloc_\sepr = \Atmat\cdot \Hdnkloc_\sepr$. Multiplying both sides of~\eqref{eq:LPUNC:T_duality_D_vs_H} by $\Atmat$, we get $\Atmat\cdot \Dbivloc_\sepr^\perp = \mu - \Hdndloc_\sepr\cdot \CMI\cdot \la$. The left-hand side of this equation defines a $2$-plane $\Atmat\cdot \Dbivloc_\sepr^\perp$ contained in $\Dbivloc_\sepr^\perp$, while the right-hand side defines a $2$-plane contained in $V=\begin{pmatrix}
\la \\ \mu
\end{pmatrix}$. 
Since $V\cap \Dbivloc_\sepr^\perp=\Vloc_\sepr$, we get $\Atmat\cdot \Dbivloc_\sepr^\perp=\Vloc_\sepr$ 
 as elements of $\Gr(2,n)$. Since the rows of $V$ are linearly independent, it follows that the matrix representatives $\mu - \Hdndloc_\sepr\cdot \CMI\cdot \la$ and $\mu - \xM_{\yloc_\sepr}\cdot \CMI\cdot \la$ must agree, so $\yMloc_\sepr=\Hdndloc_\sepr$. This completes the proof of part~\itemref{AAshift_iso} of the theorem.

We show part~\itemref{AAshift_forg}. By \cref{thm:AAshift_tree}, the result holds for $\nL=0$. Since the only condition involving $\la$ in~\eqref{eq:LSHIFT:MAknL_dfn}--\eqref{eq:LSHIFT:MAddG_dfn} is that $\la\subset V$, it follows immediately that the maps $\AAforg:\MAknL\to\AAknL$ and $\AAforg:\MAddG\to\AAddG$ are both surjective. The proof that both maps are open is identical to the proof in the $\nL=0$ case; see~\Mref{thm:AAshift}. %
\end{proof}

\subsection{\Fullysep and \twoind graphs}
We use T-duality for amplituhedra (\cref{thm:AAshift}) to deduce the following combinatorial result; cf. \cref{ex:BCFW-T-dual}.
\begin{proposition}\label{lemma:LPUNC:twosep_vs_twoind}
Let $\Gfunc$ and $\ddGvunc$ be T-dual generalized \Lpunc planar bipartite graphs. Then $\Gfunc$ is \fullysep (\cref{dfn:fullysep_gen}) if and only if $\ddGvunc$ is \twoind (\cref{dfn:fullyind_gen}).
\end{proposition}
\noindent 
First, we give a criterion for \fullysepon in terms of \wtimms.
Denote by $\xll:\Faces\to\Rdd$ the \KSprim of $(\Fw,\Fb)\in\HHspaceC$ obtained from $\lalat$ 
via~\eqref{eq:TE:y_to_lalat}--\eqref{eq:TE:lalat_vs_pFw_pFb}.

\begin{proposition}[\Mref{lemma:PROP:Mpos}]\label{lemma:PROP:Mpos}
Assume that $\G$ admits an \APM. Let $(\La,\Lat)\in\LaLaimmnn$, $C:=\Meas(\G,\wt)$, and $\lalat = \PhiLL(C)$. 
Then for any $\ff,\f\in\Faces$, we have $(\xll(\ff)-\xll(\f))^2>0$ if $\ff,\f$ are \twosep in $\G$ and $(\xll(\ff)-\xll(\f))^2=0$ otherwise. 
\end{proposition}

\begin{remark}%
\label{rmk:TE:Mpos_existence}
Similarly to the proof of \cref{thm:TE:existence}, we see from \cref{lemma:TOP:Mpos=>simple,lemma:PROP:Mpos} that if $\bdf_i,\bdf_j$ are \twosep in $\G$ for all $i+2\leq j\leq i+n-2$ then for all $\wt\in\Rtpgauge$, $(\G,\wt)$ admits a \wtemb with \Mdash positive boundary polygon.%
\end{remark}

\begin{proof}[Proof of \cref{lemma:LPUNC:twosep_vs_twoind}]
We first show the result for a pair of T-dual ordinary \Lpunc graphs $\Gfunc$ and $\ddGvunc$ as in \cref{ssec:LPUNC:T_duality_graphs}. 
 Let $\sepst\in\brnLbdsep$, $\wt\in\Rtpgauge$, $\LaLat\in\LaLaimmnn$, and $(C;\Hdndlocs)=\Measdnd(\Gfunc,\wt)$, where $\lalat:=\PhiLL(C)$. By \cref{lemma:PROP:Mpos}, we get $\det(\Hdndloc_\seps-\Hdndloc_\sept)=\frac14(\xll(\ploc_\seps)-\xll(\ploc_\sept))^2>0$ 
if $\ploc_\seps$ and $\ploc_\sept$ are \twosep and $\det(\Hdndloc_\seps-\Hdndloc_\sept)=0$ otherwise.
Since $\rank(\Hdndloc_\seps-\Hdndloc_\sept) \leq \rank(\Hdnkloc_\seps-\Hdnkloc_\sept)\leq \rank(\Hknkloc_\seps-\Hknkloc_\sept)$, by~\eqref{eq:distrk<distsep}, we get
\begin{equation}\label{eq:TE:twosep_vs_rk_Hdnk}
 \text{$\ploc_\seps$ and $\ploc_\sept$ are \twosep}
 \quad\Longleftrightarrow\quad
 \rank(\Hdnkloc_\seps-\Hdnkloc_\sept) = 2. 
\end{equation}

Applying T-duality, we obtain edge weights $\ddwt:=\ddwtlap\in\ddRtpgauge$ such that $\Measbiv(\ddGvunc,\ddwt)=\ddCDL$ satisfies $\ddC = C\cdot \Qla$.
By~\eqref{eq:LPUNC:T_duality_D_vs_H}, we find
$\dim(\Dbivloc_\seps^\perp+\Dbivloc_\sept^\perp) = (n-k) + \rank(\Hdnkloc_\seps-\Hdnkloc_\sept)$. 
Consequently, we have $\rank(\Hdnkloc_\seps-\Hdnkloc_\sept)=2$ if and only if $\rank\Dbivlocs_{\sepst} = k+2$. 
By \cref{lemma:LPUNC:indep<=>full_rank_Dbivloc}, the latter condition is equivalent to $\sepst$ being \Gindependent.

Suppose now that $\CollGWfunc$ and $\ddCollGWvunc$ are T-dual generalized \Lpunc graphs as in \cref{ssec:T_gen}. Let $\sepst\in\brnLbdsep$. Since the faces in $\Sloc_\seps$ (resp., $\Sloc_\sept$) share a white vertex,
by~\eqref{eq:GENPUNC:H_convex}, the function 
$\afflin_{\bcccloc_\seps}(\bcccloc_\sept) = \det(\Hdndloc_\seps-\Hdndloc_\sept)$ of 
$(\bcccloc_\seps,\bcccloc_\sept)\in\Simplexloc_\seps\times\Simplexloc_\sept$ 
 is affine linear in each argument; see \cref{lemma:MCE:clique_affine_linear} below. 
It takes nonnegative values on the vertices of the closed simplices $\clSimplexloc_\seps$ and $\clSimplexloc_\sept$
since $\det(\Hdnd(\ff)-\Hdnd(\f))\geq0$ for all $\ff,\f\in\CollFaces$ as above. 
By \cref{dfn:weighted_gen_Lfunc}, both points $\bcccloc_\seps\in\Simplexloc_\seps$ and $\bcccloc_\sept\in\Simplexloc_\sept$ belong to the interiors of the respective simplices. 
Thus, $\det(\Hdndloc_\seps-\Hdndloc_\sept)>0$ 
 if and only if there exist $\ffx(\seps,i)\in\Sloc_\seps$ 
and $\ffx(\sept,j)\in\Sloc_\sept$ such that $\det(\Hdnd(\ffx(\seps,i))-\Hdnd(\ffx(\sept,j)))>0$.
As we showed above, this is equivalent to $\ffx(\seps,i)$ and $\ffx(\sept,j)$ being \twosep in $\CollGW$. 
Thus, $\det(\Hdndloc_\seps-\Hdndloc_\sept)>0$ is equivalent to $\Sloc_\seps$ and $\Sloc_\sept$ being \twosep in the sense of \cref{dfn:fullysep_gen}. 

Similarly, we see that $\sepst$ is \CollGindependent in the sense of \cref{dfn:fullyind_gen} if and only if there exist \bivertices $\bivloc_\seps^i\in\ddSloc_\seps$ and $\bivloc_\sept^j\in\ddSloc_\sept$ that are \CollGindependent in the sense of \cref{dfn:fullyind}, i.e., such that $\bivloc_\seps^i\cap\bivloc_\sept^j=\emptyset$ and $\ddCollGW\rem(\bivloc_\seps^i\sqcup\bivloc_\sept^j)$ admits an \APM. 
By \cref{dfn:generalized_T-duality}, the individual faces $\ffx(\seps,i)\in\Sloc_\seps$ and $\ffx(\sept,j)\in\Sloc_\sept$ correspond under generalized T-duality to the individual \bivertices $\bivloc_\seps^i\in\ddSloc_\seps$ and $\bivloc_\sept^j\in\ddSloc_\sept$. As we showed above, $\ffx(\seps,i)$ and $\ffx(\sept,j)$ are \twosep in $\CollGW$ if and only if $\bivloc_\seps^i$ and $\bivloc_\sept^j$ are \CollGindependent.
\end{proof}

We record the following consequence of \cref{lemma:PROP:Mpos} for future use.
\begin{corollary}\label{lemma:PROP:if_not_twosep_then_M=0}
Assume that $\G$ admits an \APM. 
If $\ff,\f\in\Faces$ are not \twosep then $(\xd(\ff)-\xd(\f))^2=0$ for all \wtimms $\datrQL=\datrQ\in\Mdti(\G)$.
\end{corollary}
\begin{proof}
Indeed, by \cref{thm:TE:OAC,prop:from_la_to_La}, any \wtimm $\datrQL$ of $(\G,\wt)$ is of the form $\datrQLll$ for some $\LaLat\in\LaLak$, $C:=\Meas(\G,\wt)$, and $\lalat=\PhiLL(C)$. Since $\ff,\f$ are not \twosep, by \cref{lemma:PROP:Mpos}, we have $(\xd(\ff)-\xd(\f))^2=0$ when $\LaLat\in\LaLaimmnn$. By \cref{lemma:LaLaimmp_nonempty_Z_dense}, the same result holds for $\LaLat\in\LaLak$. 
Thus, $(\xd(\ff)-\xd(\f))^2$, viewed as a function of $\LaLat\in\LaLak$, is identically zero, so it vanishes for all $\datrQL\in\Mdti(\G)$.
\end{proof}

\subsection{T-duality for loop amplituhedron tilings} \label{ssec:T_duality_loop_tilings}
Following~\Mref{dfn:BCFW:tiling_amb}, we consider a ``multivalued'' generalization of the notion of a tiling. %

\begin{definition}[\Mtiling]\label{dfn:BCFW:tiling_amb}
Let $\RX,\RY$ be topological spaces equipped with a relation $\Rel\subset \RX\times \RY$. 
Let $\RGbf$ be a finite set and let $\{\RXo_\RG\mid\RG\in\RGbf\}$ be a collection of subsets of $\RX$. 
For $\RG\in\RGbf$, define a \emph{tile}
\begin{equation}\label{eq:BCFW:tile_amb_dfn}
 \RYo_\RG:=\{y\in \RY\mid (x,y)\in\Rel\text{ for some $x\in \RXo_\RG$}\}.
\end{equation}
In other words, the tile $\RYo_\RG=\Rproj_\RY(\Relo_\RG)$ is the image of $\Relo_\RG:=\Rel\cap(\RXo_\RG\times \RY)$ under the projection map $\Rproj_\RY:\RX\times \RY\to \RY$. 
We say that the tiles $\{\RYo_\RG\mid \RG\in\RGbf\}$ form an \emph{\mtiling} of $\RY$ if the following conditions are satisfied.
\begin{enumerate}[label=(\alph*)]
\item\label{Rtiling1} \emph{Injectivity:} For each $\RG\in\RGbf$, $\Rproj_\RY$ restricts to a homeomorphism $\Relo_\RG\xrasim\RYo_\RG$. 
\item\label{Rtiling2} \emph{Disjointness:} The tiles $\{\RYo_\RG\mid \RG\in\RGbf\}$ are pairwise disjoint.
\item\label{Rtiling3} \emph{Surjectivity:} The union $\bigsqcup_{\RG\in\RGbf}\RYo_\RG$ is dense in $\RY$.
\end{enumerate}
\end{definition}

Unlike in~\justpapone, where all graphs $\RG\in\RGbf$ were automatically reduced (cf. \cref{dfn:reduced}), we will need to work with the lift of the notion of an \Mtiling under the boundary measurement map.

\begin{proposition}[\tMtiling]\label{lemma:tMtiling}
In the notation of \cref{dfn:BCFW:tiling_amb}, suppose that for each $\RG\in\RGbf$, we are given a continuous surjective map
$\tMeas_\RG:\tRXo_\RG\to\RXo_\RG$. Denote
\begin{equation}\label{eq:tRel_dfn}
	\tRelo_\RG:=\{(\trx,y)\in\tRXo_\RG\times\RY\mid (\tMeas_\RG(\trx),y)\in\Rel\}
 \quad\text{and}\quad
 \tRel:=\bigsqcup_{\RG\in\RGbf}\tRelo_\RG.
\end{equation}
If the tiles $\{\RYo_\RG\mid \RG\in\RGbf\}$ form \atmtiling of $\RY$ then the tiles $\{\RYo_\RG\mid \RG\in\RGbf\}$ form an \mtiling of $\RY$.
Furthermore, if the projection
$\Rproj_{\RXo_{\RG}}:\Relo_\RG\to\RXo_{\RG}$
admits a continuous section 
then the map $\tMeas_\RG:\tRXo_\RG\to\RXo_\RG$ is a homeomorphism.
\end{proposition}
\begin{proof}
By part~\itemref{Rtiling1} of \cref{dfn:BCFW:tiling_amb}, the composition
$\tRelo_\RG\xrightarrow{(\tMeas_\RG,\id_{\RY})} \Relo_\RG\xrightarrow{\Rproj_\RY} \RYo_\RG$
is a homeomorphism. Since $\tMeas_\RG$ is surjective, both maps in the composition are homeomorphisms. This shows part~\itemref{Rtiling1} for $\Rel$.
Furthermore, by the surjectivity of $\tMeas_\RG$, the set $\RYo_\RG$ given by~\eqref{eq:BCFW:tile_amb_dfn} coincides with the projection of $\tRelo_\RG$ under $\tRproj_\RY:\tRel\to\RY$. 
Thus, parts~\itemref{Rtiling2}--\itemref{Rtiling3} are identical for $\Rel$ and $\tRel$.

Finally, a section of $\Rproj_{\RXo_{\RG}}:\Relo_\RG\to\RXo_{\RG}$ is a map of the form
$(\id_{\RXo_{\RG}},\RPhi)$ where 
$\RPhi:\RXo_\RG\to\RYo_\RG$ satisfies $(x,\RPhi(x))\in\Rel$ for all $x\in\RXo_\RG$. If such a continuous map exists
 then $\tMeas_\RG:\tRXo_\RG\to\RXo_\RG$ admits a continuous inverse
$\RXo_{\RG}\xrightarrow{(\id_{\RXo_{\RG}},\RPhi)}\Relo_\RG\xrightarrow{(\tMeas_\RG,\id_{\RY})^{-1}} \tRelo_\RG \xrightarrow{\tRproj_{\tRXo_{\RG}}} \tRXo_\RG$ and is thus a homeomorphism.
\end{proof}

In what follows, we will mostly work with \tmtilings; the proofs of the corresponding \mtiling results will follow automatically by \cref{lemma:tMtiling}. 

\begin{proposition}[{\Mref{lemma:RX_triple}}]\label{lemma:RX_triple}
For $\s=2,3$ and $\RG\in\RGbf$, let $\RelS\subset\RXS\times\RYS$, $\RXoS_\RG\subset\RXS$, $\ReloS_\RG=\RelS\cap(\RXoS_\RG\times\RYS)$, and $\RYoS_\RG=\RprojS(\ReloS_\RG)$ be as in \cref{dfn:BCFW:tiling_amb}.
Assume that $\RXB=\RXC$ and $\RXoB_\RG=\RXoC_\RG$ for all $\RG\in\RGbf$.
 Suppose that we have a commutative diagram
\begin{equation}\label{eq:TREE:comm_diag}
\begin{tikzcd}[column sep=3em]
\RelB \arrow[r,"{(\id_{\RXB},\Rpi)}"] \arrow[d,"\RprojB"] 
\drar[phantom, "\ \square",pos=0.47]
& \RelC \arrow[d,"\RprojC"] 
\\
\RYB\arrow[r,"\Rpi",twoheadrightarrow] 
& \RYC.
\end{tikzcd} 
\end{equation}
Suppose in addition that the map $\Rpi:\RYB\to\RYC$ is continuous, surjective, and open, and that the diagram is \emph{Cartesian}, i.e., 
$\RelB = \{(x,y)\in \RXB\times\RYB\mid (x,\Rpi(y))\in\RelC\}$. 
Then 
\begin{equation}\label{eq:RtilingS}
 \text{$\{\RYoB_\RG\mid \RG\in\RGbf\}$ is \amtilingB of $\RYB$}
 \quad\Longleftrightarrow\quad
 \text{$\{\RYoC_\RG\mid \RG\in\RGbf\}$ is \amtilingC of $\RYC$}.
\end{equation}
\end{proposition}

\begin{definition}\label{dfn:RX_triple}
In the notation of \cref{lemma:RX_triple}, we set 
\begin{equation*}%
 \RXA:=\GrtnnLfuncx_2(k,n), \ \ 
\RXB = \RXC := \GrtnnLvuncprojx_2(k-2,n),\ \ 
\RYA:=\MPknL, \ \ 
\RYB:=\MAknL, \ \ 
\RYC:=\AAknL.
\end{equation*}

Let $\RelA\subset \RXA\times\RYA$ be the set of all pairs $(\CHL,\llPllL)\in\RXA\times \RYA$ such that 
\begin{equation}\label{eq:LOOP:Rel_dfn}
 \la\subset C\subset\latp, \quad
 \la = \Amat\cdot C,\quad
 \lat = \Atmat\cdot C^\perp, \quad\text{and}\quad
 \yMloc_\rho = \Atmat\cdot \Hknkloc_\rho\cdot \Amat^T \quad\text{for all $\rho\in\brnL$},
\end{equation}
where $\Amat\in\Mator_{2,k},\Atmat\in\Mator_{2,n-k}$ are uniquely determined by matrix representatives of $C,C^\perp,\la,\lat$. 
Let $\RelB\subset \RXB\times\RYB$ consist of all pairs $\big(\ddCDL,\laVLpunc\big)\in\RXB\times\RYB$ such that 
\begin{equation}\label{eq:TREE:ddRel_dfn}
 V\subset\ddC^\perp \quad\text{and}\quad \VLr:=\pLr\cdot V\subset \Dbivloc_\rho^\perp \quad\text{for all $\rho\in\brnL$},
\end{equation}
and let $\RelC\subset\RXC\times\RYC$ consist of all pairs $\big(\ddCDL,\VLpunc\big)\in\RXC\times\RYC$ satisfying~\eqref{eq:TREE:ddRel_dfn}.

For each T-dual pair $(\Gfunc,\ddGvunc)$ of \Lpunc graphs, we let 
\begin{equation*}%
 \RXoA_{\Gfunc}:=\PtpLfunc_{\Gfunc}, \ \ 
 \RXoB_{\ddGvunc}=\RXoC_{\ddGvunc} := \PtpLvunc_{\ddGvunc}; \ \ 
 \RYoA_{\Gfunc}:=\MPG,\ \ 
 \RYoB_{\ddGvunc}:=\MAddG,\ \ 
 \RYoC_{\ddGvunc}:=\AAddG.
\end{equation*}
We let $\tRXoA_{\Gfunc}:=\Rtpgauge$ and $\tRXoB_{\ddGvunc}=\tRXoC_{\ddGvunc}:=\ddRtpgauge$, and let 
$\tReloA_{\Gfunc}$, $\tReloB_{\ddGvunc}$, $\tReloC_{\ddGvunc}$ denote the respective lifts under
$\Measknk_{\Gfunc}:=\Measknk(\Gfunc,\cdot)$ and
$\Measbiv_{\ddGvunc}:=\Measbiv(\ddGvunc,\cdot)$ as in~\eqref{eq:tRel_dfn}.
\end{definition}

\begin{corollary}[T-duality for \tmtilings of ambient loop amplituhedra]\label{lemma:TREE:amb_tiling_equivalence}
Let $\GGknL$ be a collection of \fullysep T-dualizable \Lfunc graphs and let $\ddGGknL$ be the collection of the corresponding T-dual \Lvunc graphs. 
The tiles $\{\MPG\mid \Gfunc\in\GGknL\}$ form \atmtilingA of $\MPknL$ if and only if the tiles $\{\AAddG\mid \ddGvunc\in\ddGGknL\}$ form \atmtilingC of $\AAknL$. 
\end{corollary}
\begin{proof}
By \cref{lemma:LPUNC:twosep_vs_twoind}, the graphs in $\ddGGknL$ are \twoind. 
Thus, $\RXoA_{\Gfunc}\subset\RXA$ and $\RXoB_{\ddGvunc}=\RXoC_{\ddGvunc}\subset\RXB=\RXC$ for all $\Gfunc\in\GGknL$
and $\ddGvunc\in\ddGGknL$. 
Let $\Rphibot$ and $\Rpi$ be as in \cref{thm:AAshift}. 

Let $\tRphitop:\tRelA\to\tRelB$ be given by
$\tRphitop\big(\wt,\llPllL\big):=\big(\ddwtlap,\laVLpunc\big)$,
where $\ddwtlap$ is given by~\eqref{eq:ddwtlap_dfn} and $\laVLpunc:=\AAiso\llPllL$ is given by~\eqref{eq:AAiso_dfn}.
By \cref{rmk:SHIFT:inverse} and the proof of \cref{thm:AAshift}, $\tRphitop:\tRelA\xrasim\tRelB$ is a homeomorphism that restricts to a homeomorphism $\tReloA_{\Gfunc}\xrasim\tReloB_{\ddGvunc}$ for each $\Gfunc\in\GGknL$. 
Thus, the tiles $\{\MPG\mid \Gfunc\in\GGknL\}$ form \atmtilingA of $\RYA=\MPknL$ if and only if 
the tiles $\{\MAddG\mid \ddGvunc\in\ddGGknL\}$ form \atmtilingB of $\RYB=\MAknL$. 

To relate \tmtilingsB of $\RYB=\MAknL$ to \tmtilingsC of $\RYC=\AAknL$, we apply \cref{lemma:RX_triple}. 
 By construction, the diagram~\eqref{eq:TREE:comm_diag} commutes.
By \cref{thm:AAshift}, the continuous map $\Rpi$ is open and surjective. 
Finally, the Cartesian square condition
$\tRelB = \{(\trx,y)\in \tRXB\times\RYB\mid (\trx,\Rpi(y))\in\tRelC\}$
(where $\tRXB:=\bigsqcup_{\ddGvunc\in\ddGGknL} \tRXoB_{\ddGvunc}$ similarly to~\eqref{eq:tRel_dfn})
follows from the fact that the relation~\eqref{eq:TREE:ddRel_dfn} does not involve $\la$. 
\end{proof}

\begin{definition}[Tiling]\label{dfn:BCFW:tiling}
Let $\RPhi: \RX\to \W$ be a continuous map. %
Let $\RGbf$ be a finite set and let $\{\RXo_\RG\mid\RG\in\RGbf\}$ be a collection of subsets of $\RX$. 
For $\RG\in\RGbf$, define a \emph{tile} $\Wo_\RG:=\RPhi(\RXo_\RG)$. 
We say that the tiles $\{\Wo_\RG\mid \RG\in\RGbf\}$ form a \emph{tiling} of $\W$ if the following conditions are satisfied.
\begin{enumerate}[label=(\alph*)]
\item\label{tiling1} \emph{Injectivity:} For each $\RG\in\RGbf$, the map $\RPhi$ restricts to a homeomorphism 
$\RXo_\RG\xrasim\Wo_\RG$.%
\item\label{tiling2} \emph{Disjointness:} The tiles $\{\Wo_\RG\mid \RG\in\RGbf\}$ are pairwise disjoint.
\item\label{tiling3} \emph{Surjectivity:} The union $\bigsqcup_{\RG\in\RGbf} \Wo_\RG$ is dense in $\W$. 
\end{enumerate}
\end{definition}

The following result will be used in \cref{ssec:flip_proj} to deduce BCFW tiling results for ordinary (non-ambient) loop amplituhedra.
\begin{proposition}[{\Mref{lemma:Rtiling_to_tiling}}]\label{lemma:Rtiling_to_tiling}
Let $\RX\subset\RXcl$, $\RY\subset\RYcl$, $\W=\Wcl\cap\RY$, and $\Rel=\Relcl\cap(\RX\times\RY)$ for some $\Wcl\subset\RYcl$ and $\Relcl\subset\RXcl\times\RYcl$. 
Suppose that the tiles $\{\RYo_\RG\mid \RG\in\RGbf\}$ form an \emph{\mtiling} of $\RY$.
Suppose in addition that $\RPhi: \RX\to \W$ extends to a continuous map $\RPhicl:\RXcl\to\Wcl$ 
 whose graph 
\begin{equation}\label{eq:RGraph}
\RGraph_{\RPhicl}:=\{(x,\RPhicl(x))\mid x\in\RXcl\} 
\quad\text{satisfies}\quad
 \RGraph_{\RPhicl} = \Relcl\cap (\RXcl\times \Wcl). 
\end{equation}
Assume that $\Relcl\subset\RXcl\times\RYcl$ is a closed subset and that the closure $\RXocl_\RG$ of $\RXo_\RG$ in $\RXcl$ is compact for each $\RG\in\RGbf$. 
Then the tiles $\{\Wo_\RG\mid \RG\in\RGbf\}$ form a tiling of $\W$.
\end{proposition}

\part{\ORATITLE}\label{part2}

\section{\MNEsTITLE}\label{sec:MNE}

The goal of this section is to introduce \emph{\Mdash nonnegative point configurations} and \emph{\MNEs} of planar graphs.

\subsection{Cliques in \Mdash nonnegative point configurations}\label{ssec:MCE:cliques}
Let $\Faces$ be a finite set.

\begin{definition}
A map $\xd:\Faces\to\Rdd$ is called \emph{\Mdash nonnegative} if 
$(\xd(\ff)-\xd(\f))^2\geq0$ for all $\ff,\f\in\Faces$.
\end{definition}
\noindent We fix an \Mdash nonnegative map $\xd:\Faces\to\Rdd$ for the rest of this subsection. 

Consider a simple undirected graph 
\begin{equation}\label{eq:MCE:NullG_dfn}
 \NullG(\GD,\xd) = (\Faces,\NullE(\GD,\xd)), \quad\text{where}\quad
\NullE(\GD,\xd):=\{\{\ff,\f\}\subset\Faces\mid (\xd(\ff) - \xd(\f))^2 = 0\}.
\end{equation}

\begin{definition}\label{dfn:MCE:white_black_deg}
A subset $\preK\subset\Faces$ is called a \emph{clique} if any two vertices $\ff,\f\in\preK$ are connected by an edge in $\NullG(\GD,\xd)$. In this case, we denote $\clique:=\xd(\preK)$ and also refer to $\clique$ as a \emph{clique}. 
The point configurations $\KT := \{\xT(\f)\mid\f\in\preK\}$ and $\KO := \{\xO(\f)\mid\f\in\preK\}$ in the plane are related by an isometry $\phi$. We say that the clique $\clique$ is \emph{white} (resp., \emph{black}) if $\phi$ is orientation-preserving (resp., orientation-reversing). We say that $\clique$ is \emph{degenerate} if all points in $\KT$ lie on a line; in this case, $\clique$ is considered both white and black.
\end{definition}

\begin{definition}\label{dfn:MCE:col}
When $\clique$ is a nondegenerate clique, we denote its color by $\colop(\clique)\in\{\colW,\colB\}$, and we denote the opposite color by $\colbarop(\clique)$. (E.g., when $\clique$ is white, we have $\colop(\clique)=\colW$ and $\colbarop(\clique) = \colB$.) 
\end{definition}

\begin{lemma}\label{lemma:MCE:clique_convex}
Let $\clique=\xd(\preK)$ be a clique, and let $z=(\Tcomp{z},\Ocomp{z})\in\Rdd$ be such that $(z-\xd(\f))^2\geq0$ for all $\f\in\preK$. If $\Tcomp{z}\in\Conv\KT$ then $(z-\xd(\f))^2=0$ for all $\f\in\preK$.
\end{lemma}
\begin{proof}
After rotating and possibly reflecting the $\O$-plane, we may assume that $\xT(\f) = \xO(\f)$ for all $\f\in\preK$. We would like to show that $\Tcomp{z}=\Ocomp{z}$. Otherwise, let $\elline$ be the perpendicular bisector to $[\Tcomp{z},\Ocomp{z}]$. Let $H\subset\C$ be the closed half-plane bounded by $\elline$ and containing $\Ocomp{z}$. 
For $\f\in\preK$, since $(z-\xd(\f))^2\geq0$, the point $\xT(\f)=\xO(\f)$ belongs to $H$. Thus, $\Tcomp{z}$ is separated by $\elline$ from $\Conv\KT$, a contradiction.
\end{proof}

\begin{lemma}\label{lemma:MCE:clique_affine_linear}
Let $\clique=\xd(\preK)$ be a clique. Then for each fixed $y\in\Rdd$, the function $\afflin_y:\Conv\clique\to\R$ given by $\afflin_y(z) :=(y-z)^2$ is affine linear on $\Conv\clique$. 
\end{lemma}
\begin{proof}
After rotating and possibly reflecting the $\O$-plane, we may assume that $\xT(\f) = \xO(\f)$ for all $\f\in\preK$, and thus $\Tcomp{z}=\Ocomp{z}$ for all $z\in\Conv\clique$. 
Using $|a-b|^2 = |a|^2 + |b|^2 - 2\Re(a\bar b)$ for $a,b\in\C$, we find
\begin{equation}\label{eq:*}
 \afflin_y(z) = (y-z)^2 
= |\Tcomp{y}|^2+|\Tcomp{z}|^2 - 2\Re(\Tcomp{y}\overline{\Tcomp{z}}) - |\Ocomp{y}|^2 - |\Ocomp{z}|^2 + 2\Re(\Ocomp{y}\overline{\Ocomp{z}})
= y^2 - 2\Re((\Tcomp{y}-\Ocomp{y})\overline{\Tcomp{z}}).
\qedhere
\end{equation}
\end{proof}

\begin{notation}\label{dfn:MCE:interior_edge_face}
For $\east = \{\ff,\f\}\subset\Faces$, $\Tarb(\east)=[\Tarb(\ff),\Tarb(\f)]$ denotes a closed line segment. The open line segment is denoted $\xTint(\east)=\open[\xT(\ff),\xT(\f)]$. 
For a subset $\KT\subset\C$, we let $\Convint\KT$ (resp., $\Convrelint\KT$) denote the interior (resp., the relative interior) of $\Conv\KT$. 
\end{notation}

\begin{corollary}\label{lemma:MCE:clique_union}
Let $\clique_1=\xd(\preK_1)$ and $\clique_2=\xd(\preK_2)$ be two cliques such that %
 $\Convrelint\KT_2$ intersects $\Conv\KT_1$. Then $\preK_1\cup \preK_2$ is a clique.
\end{corollary}
\begin{proof}
Let $y:=\xd(\f)\in\clique_1$ with $\f\in\preK_1$ and consider the affine linear function $\afflin_y(z)=(y-z)^2$ on $\Conv\clique_2$ as in \cref{lemma:MCE:clique_affine_linear}.
 Because $\xd$ is \Mdash nonnegative, $\afflin_y$ takes nonnegative values on $\Conv\clique_2$. 
By assumption, there exists $z\in\Convrelint\clique_2$ such that $\Tcomp{z}\in\Convrelint\KT_2\cap\Conv\KT_1$, in which case we have $\afflin_y(z)=0$ by \cref{lemma:MCE:clique_convex}. Thus, $\afflin_y$ must be identically zero on $\Conv\clique_2$. 
\end{proof}

In the special case when $\clique_2$ consists of a single point, \cref{lemma:MCE:clique_union} reduces to \cref{lemma:MCE:clique_convex}. When $\clique_2$ consists of two distinct points, \cref{lemma:MCE:clique_union} yields the following observation which will be used later.

\begin{corollary}\label{lemma:MCE:clique_edge}
Let $\clique_1=\xd(\preK_1)$ be a clique and let $\ff,\f\in\Faces$ be such that $(\xd(\ff)-\xd(\f))^2=0$ and such that the open line segment $\open[\xT(\ff),\xT(\f)]$ intersects $\Conv\KT_1$. 
Then $\preK_1\cup\{\ff,\f\}$ is a clique.
\end{corollary}

\subsection{\Wembs of planar graphs}\label{ssec:MCE:near-embedd-curvy}
For the rest of this section, we assume the following. 
\begin{assumption}\label{ass:lgg_dual}
Assume that $\GDarb=(\FDarb,\EDarb)$ is a graph embedded in a disk $\Disk$, with parallel edges allowed but without any loop edges. 
We assume that the boundary of the disk $\Disk$ is a simple cycle of $\GDarb$ with vertices $(\bdf_1,\bdf_2,\dots,\bdf_n=\bdf_0)$. For all \wembs $\xT$ of $\GDarb$ considered below, we assume that the boundary polygon $(\bdxT_1,\bdxT_2,\dots,\bdxT_n=\bdxT_0)$ (with $\bdxT_i = \xT(\bdf_i)$ as before) is oriented clockwise. 
\end{assumption}
\noindent We do not assume that $\GDarb$ \hasnofloat or that its planar dual graph is bipartite.

For an edge $\east\in\EDarb$ with endpoints $\ff,\f\in\Faces$, we denote $\ebarast:=\{\ff,\f\}$ and set $\Ebarast:=\{\ebarast\mid \east\in\EDarb\}$. 
We denote by 
$\Evecast$
 the set of oriented edges of $\GDarb$. Thus, every edge $\east\in\East$ appears in $\Evecast$ twice, with opposite orientations. For an oriented edge 
$\evecast$, we denote the oppositely oriented edge by $-\evecast$. 

\begin{definition}\label{dfn:partF}
The set of faces of $\GD$ (excluding the outer face which is not considered a face) is denoted by $\Vint$. 
Recall that for $\v\in\Vint$, we denote by $\partF\v\subset\Faces$ the set of vertices of $\GD$ incident to $\v$. This includes the (incident to $\v$) vertices of floating connected components of $\GD$ located inside~$\v$, and in particular, isolated vertices inside $\v$. We also denote by $\partEast\v\subset\East$ the set of edges $\east$ of $\GD$ incident to $\v$. We let $\partEvecast\v\subset\Evecast$ be the \emph{clockwise boundary} of $\v$, i.e., the set of oriented edges $\evecast\in\Evecast$ of $\GD$ such that $\east\in\partEast\v$ and $\evecast$ is directed clockwise around the boundary of $\v$.
\end{definition}

\begin{definition}\label{dfn:MCE:bent_line_seg}
Let $a,b\in\C$ with $a\neq b$. 
A \emph{\bending of $[a,b]$} is a family $(\bendh)_{\eps>0}$ of embedded piecewise-linear curves $\bendh:[a,b]\to\C$ depending continuously on $\eps>0$ and converging uniformly to 
the identity map on $[a,b]$ 
 as $\eps\to0$. Such a triple $(a,b,\bendh)$ is called a \emph{\bent line segment}.
 For $\eps>0$, we set $[a,b]_{\eps}:=\bendh([a,b])$.%
\end{definition}

\begin{definition}\label{dfn:MCE:face_inj}
A map $\Tarb:\Faces\to\C$ is called \emph{\finj} if its restriction to $\partF\v$ is injective for each face $\v\in\Vint$ of $\GD$ and if its restriction to the boundary vertex set $\{\bdf_1,\bdf_2,\dots,\bdf_n\}$ is also injective.
\end{definition}
\noindent We extend every map $\Tarb:\Faces\to\C$ to $\SkelGD$ by linearity and consider \wembs of $\GD$ as in \cref{dfn:TOP:wtimm}.
We start by discussing elementary topological properties of \wembs. %

\begin{figure}
 \def\inputscale{1.5}
 \setlength{\tabcolsep}{4pt}
\begin{tabular}{cccc}
\includegraphics[scale=\inputscale]{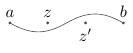}
&
\includegraphics[scale=\inputscale]{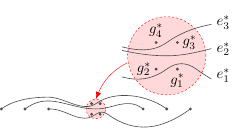}
&
\hspace{-0.2in}
\includegraphics[scale=\inputscale]{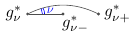}
&
\hspace{-0.2in}
\includegraphics[scale=\inputscale]{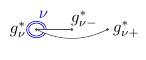}
\\
 (a) Above/below
& (b) \Cref{lemma:MCE:non_inj}
& (c) $\sumTx(\cor)=0$
& (d) $\sumTx(\cor)=2\pi$ %
\end{tabular}
 \caption{\label{fig:bend} Examples of \bent line segments.}
\end{figure}

\begin{definition}
Consider a \bent line segment $(a,b,\bendh)$ and a family $z_\eps\in\C$ of points depending continuously on $\eps$ and converging to some $z\in\open[a,b]$. Suppose that $z_\eps\notin[a,b]_\eps$ for all small $\eps>0$. We say that 
$[a,b]_{\eps}$
 \emph{passes above} (resp., \emph{below}) $z_\eps$ if for small $\eps>0$, a generic bent line segment connecting $z_\eps$ to $z_\eps + \I(b-a)$ 
 intersects $[a,b]_\eps$ an odd (resp., even) number of times. %
Here and below, $\I:=\sqrt{-1}$.
\end{definition}
\noindent The above definition applies to \emph{oriented} \bent line segments $(a,b,\bendh)$ as it depends on the ordering of the endpoints $a,b$. If 
$[a,b]_{\eps}$ 
 passes above $z_\eps$ and below $z'_\eps$ or vice versa then we say that $[a,b]_{\eps}$ \emph{separates} $z_\eps$ from $z'_\eps$. This notion is well defined for unoriented \bent line segments. For example, the \bent line segment $(a,b,\bendh)$ in \figref{fig:bend}(a) 
passes below $z$ and above $z'$ so it
separates $z$ from $z'$. 

\begin{lemma}\label{lemma:MCE:non_inj}
Let $\Tarb$ be a face-injective \wemb of $\GD$ and let $z\in\Tarb(\Faces)$. Let $\preK:=\{\f\in\Faces\mid\Tarb(\f)=z\}$ and $d:=|\preK|\geq1$.
\begin{enumerate}[label=(\arabic*)]
\item\label{non_inj1} There exists a line $\elline\subset \C$ containing all edges $\Tarb(\east)$ of $\Tarb(\GD)$ that contain $z$ in their relative interiors.
\item\label{non_inj2} There exist $d-1$ such edges $\east_1,\east_2,\dots,\east_{d-1}$ and an ordering $\preK=\{\f_1,\f_2,\dots,\f_d\}$ such that for each $1\leq i\leq d-1$, the edge $\Tcur(\east_i)$ 
separates $\Tcur(\{\f_1,\dots,\f_i\})$ from $\Tcur(\{\f_{i+1},\dots,\f_d\})$. See \figref{fig:bend}(b).
\end{enumerate}
\end{lemma}
\begin{proof}
Let us say that two line segments $[a,b]$ and $[c,d]$ form an \emph{essential crossing} if $\open[a,b]\cap\open[c,d]\neq\emptyset$ and $[a,b]\cup[c,d]$ is not contained in a line. Thus, if two edges of $\Tarb(\GD)$ form an essential crossing then $\Tarb$ is not \awemb. 
In particular, if two edges $\Tarb(\east_1),\Tarb(\east_2)$ of $\Tarb(\GD)$ contain $z$ in their relative interiors and do not lie on the same line then $\Tarb$ is not \awemb of $\GD$. This shows part~\itemref{non_inj1}.

To show part~\itemref{non_inj2}, let $\ff,\f\in\preK$ be two distinct vertices. Since $\Tarb(\ff)=\Tarb(\f)$, $\ff$ and $\f$ cannot share a face of $\GD$. 
Thus, there exists an edge $\east\in\East$ sharing a face with $\ff$ such that $\Tcur(\east)$ intersects a generic \bent line segment connecting $\Tcur(\ff)$ to $\Tcur(\f)$ an odd number of times. 
We have $\ebarast\cap\preK=\emptyset$ since $\xT$ is \finj. 
Thus, $\Tarb(\east)$ contains 
 $z$ in its relative interior and $\Tcur(\east)$ separates $\Tcur(\ff)$ from $\Tcur(\f)$. The edge $\Tcur(\east)$ separates $\Tcur(\preK)$ into two nonempty subsets. Iterating the process for each of the two subsets, we obtain a proof of part~\itemref{non_inj2}.
\end{proof}

\begin{corollary}\label{lemma:MCE:non_inj_edge}
Let $\Tarb$ be a face-injective \wemb of $\GD$.
If $\ff\in\Faces$ is such that $\Tarb(\ff)$ is not contained in the relative interior of any edge $\Tarb(\east)$ of $\Tarb(\GD)$ then $\Tarb(\ff)\neq\Tarb(\f)$ for any $\f\in\Faces\setminus\{\ff\}$.
\end{corollary}

\begin{remark}\label{rmk:MCE:near_emb_combin}
The \bending $\hall$ of $\Tarb$ encodes the following combinatorial information. 
 Let $z\in\Tarb(\SkelGD)$, and consider all edges $\Tarb(\east_1),\Tarb(\east_2),\dots,\Tarb(\east_k)$ that contain $z$ in their relative interiors. Let $\elline$ be the line containing $\Tarb(\east_1)\cup\Tarb(\east_2)\cup\dots\cup\Tarb(\east_k)$, and let $\elline^\perp$ be the line perpendicular to $\elline$ satisfying $\elline\cap \elline^\perp = \{z\}$. Without loss of generality, we may assume that $z=0$, $\elline=\R$, and $\elline^\perp=\I\R$. 
Suppose that all \bent line segments $\Tarb(\evecast_i)$, $i\in\brk$, are oriented from left to right along $\elline$.
Let $\preK=\{\f_1,\dots,\f_d\}:=\Tarb^{-1}(z)$. For each $i\in\brd$ and $j\in\brk$, $\hall$ encodes whether $\Tcur(\evecast_j)$ passes above or below $\Tcur(\f_i)$. Thus, in view of \cref{lemma:MCE:non_inj}, $\hall$ induces a total ordering $\preceq_z$ ``by the vertical coordinate'' on the set $\{\east_1,\dots,\east_k,\f_1,\dots,\f_d\}$, with at least one edge separating any two vertices. 
These relative orderings depend on $z\in \elline$ in a locally constant fashion, changing only when $z$ passes through a vertex of $\Tarb(\GD)$. 
Furthermore, suppose that an edge $\east$ of $\GD$ is incident to some $\f_i\in\preK$, and suppose that its other endpoint $\ff$ satisfies $\Tarb(\ff)\notin \elline$. Then if $\Tarb(\ff)$ lies strictly above (resp., below) $\elline$, $\f_i$ is the $\preceq_z$-maximal (resp., $\preceq_z$-minimal) element of $\{\east_1,\dots,\east_k,\f_1,\dots,\f_d\}$.
\end{remark}

\subsection{\MNEs}%
We continue to fix a graph $\GD$ satisfying \cref{ass:lgg_dual}. 

\begin{definition}%
\label{dfn:MCE:MNE}
An \emph{\MNE} is a pair $(\GD,\xd)$ where $\GD$ is a graph satisfying \cref{ass:lgg_dual} and $\xd:\Faces\to\Rdd$ is a map satisfying the following conditions.
{\setlength{\leftmargini}{50pt}
\begin{enumerate}[label=(MCE\arabic*)]
\item\label{MCE1:emb} \emph{\Wemb}: $\xdT$ is a face-injective \wemb of $\GD$.%
\item\label{MCE2:null_edges} \emph{Null edges}: for any edge $\{\ff,\f\}$ of $\GD$, $(\xdf(\ff)-\xdf(\f))^2 = 0$ 
(i.e., $\Ebarast\subset\NullE(\GD,\xd)$).
\item\label{MCE3:M-tnn} \emph{\Mdash nonnegativity}: we have $(\xdf(\ff)-\xdf(\f))^2\geq0$ for all $\ff,\f\in\Faces$. 
\end{enumerate}
}
\end{definition}

\begin{remark}
Throughout, we additionally assume that $(\GD,\xd)$ has \emph{\Mbd}, i.e., 
\begin{equation}\label{eq:MCE3':Mbd}
 (\bdx_i - \bdx_j)^2 >0 \quad\text{for all $i+2\leq j\leq i+n-2$.}%
\end{equation}
\noindent 
 Strictly speaking, this assumption is not necessary for the main results in \crefrange{sec:MNE}{sec:ORA} to hold. However, it is sufficient (and more natural) for our applications, and simplifies some of the arguments. 
\end{remark}

We denote by $\Fint:=\Faces \setminus\{\bdf_1,\bdf_2,\dots,\bdf_n\}$ the set of interior vertices of $\GD$. We denote by $\Facesgr\subset\Faces$ the set of non-isolated vertices of $\GD$, and set $\Fintgr:=\Fint\cap\Facesgr$. 
Similarly to \cref{dfn:corners}, 
for $\f\in\Facesgr$, let $\corners(\f)$ denote the set of corners of $\GD$ incident to $\f$. 
We set $\corners(\GD)=\bigsqcup_{\f\in\Facesgr}\corners(\f)$. 
 For a face $\v\in\Vint$, let $\corners(\v):=\{\corner\in\corners(\GD)\mid \corv=\v\}$ be the set of corners contained in $\v$. We refer to the images of edges, faces, and corners of $\GDarb$ under $\Tarb$ as edges, faces, and corners of $(\GDarb,\xd)$, respectively.

For the rest of this subsection, we assume that $(\GDarb,\xd)$ is an \MNE. 

\begin{definition}\label{dfn:MCE:deg_emb_triang_big_faces}
A face $\v\in\Vint$ of $(\GDarb,\xd)$ is called \emph{degenerate} if $\dim\Conv\Tarb(\partF\v) <2$.
 We say that $\v$ is \emph{embedded} if $\Tarb$ restricts to a (straight-edge) embedding of the subgraph $\partGD\v:=(\partF\v,\partEast\v)$ of $\GD$. 
We say that $\v$ is a \emph{triangle} if $\partGD\v$ is a $3$-cycle and a \emph{bigon} if $\partGD\v$ is a $2$-cycle.
\end{definition}
\noindent The graph $\partGD\v$ need not be connected when $\GD$ has floating connected components. 

\begin{lemma}
If $\v\in\Vint$ is a degenerate face of $(\GDarb,\xd)$ then $\dim\Conv\Tarb(\partF\v)=1$ and 
 $\partF\v$ is a degenerate clique.
\end{lemma}
\begin{proof}
Since $\xT$ is \finj, we have $\dim\Conv\Tarb(\partF\v)=1$. If $\ff\in\partF\v$ and $\east\in\partEast\v$ are such that $\xT(\ff)\in\xTint(\east)$ then $\ff\sqcup\ebarast$ is a clique by \cref{lemma:MCE:clique_convex}. It follows by induction that $\partF\v$ is a clique. 
\end{proof}

\begin{definition}
For a corner $\corner$ of $(\GD,\xd)$, we define an angle $\sumTx(\corner)\in[0,2\pi]$ such that 
\begin{equation}\label{eq:MCE:angle_dfn}
 \sumTx(\corner)\equiv\arg \frac{\Tarb(\corfp) - \Tarb(\corf)}{\Tarb(\corfm) - \Tarb(\corf)}\quad\mod\ {2\pi}.
\end{equation}
Here, in the case when the argument is equal to $0$ modulo $2\pi$, we choose the angle $\sumTx(\corner)\in\{0,2\pi\}$ as follows. 
If $\coream=\coreap$ then we set $\sumTx(\corner):=2\pi$. 
Otherwise, pick a point $z_\eps\in\xtehint(\coreap)$ close to $\Tcur(\corf)$. We set $\sumTx(\corner):=0$ (resp., $\sumTx(\corner):=2\pi$) if $\xtehint(\corevecam)$ passes below (resp., above) $z_\eps$, where $\corevecam$ is oriented away from $\corf$. See \figref{fig:bend}(c,d).
We set
\begin{equation*}
 \sumOx(\corner):=\arg_{[0,2\pi)}\frac{\xO(\corfp) - \xO(\corf)}{\xO(\corfm) - \xO(\corf)} \quad\in[0,2\pi).
\end{equation*}
Unlike in~\eqref{eq:MCE:angle_dfn}, here we do not break ties between $\sumOx(\corner) = 0$ and $\sumOx(\corner) = 2\pi$. When the map $\xd$ is clear from the context, we omit it from the notation and denote $\sumT(\corner):=\sumTx(\corner)$ and $\sumO(\corner):=\sumOx(\corner)$. 
\end{definition}

Our next goal is to define angles 
\begin{equation}\label{eq:nearGr:angles_exist}
 \sumwT(\corner),\sumbT(\corner)\in[0,\pi] \quad\text{such that}\quad 
\sumwT(\corner) + \sumbT(\corner) = \sumT(\corner)
\quad\text{and}\quad
\sumwT(\corner) - \sumbT(\corner) \equiv \sumO(\corner)\ \ \mod\ 2\pi.
\end{equation}

The following trivial observation will be used several times in what follows.
\begin{lemma}%
Given nonzero null vectors $\Pmom,\Qmom\in\Rdd$ with $\sumT:=\arg_{(-\pi,\pi]}(\PmomT/\QmomT)$, $\sumO:=\arg_{(-\pi,\pi]}(\PmomO/\QmomO)$, 
\begin{equation}\label{eq:M_tnn_vs_angles}
 (\Pmom - \Qmom)^2 \geq 0 \quad\Longleftrightarrow\quad
 \cos(\sumT)\leq \cos(\sumO) \quad\Longleftrightarrow\quad
 |\sumT| \geq |\sumO|;
\end{equation}
\begin{equation}
\label{eq:M_tnn_vs_angles_eq}
 (\Pmom - \Qmom)^2 = 0 \quad\Longleftrightarrow\quad
 \cos(\sumT) = \cos(\sumO) \quad\Longleftrightarrow\quad
 |\sumT| = |\sumO|.
\end{equation}

\end{lemma}

Our next result classifies solutions to~\eqref{eq:nearGr:angles_exist}.

\begin{lemma}\label{lemma:MCE:corner_angle}
Let $(\GD,\xd)$ be an \MNE, and let $\corner$ be a corner of $\GD$.
\begin{enumerate}[label=(\arabic*)]
\item\label{corner_angle1}
 If $(\sumT(\corner), \sumO(\corner)) = (\pi,\pi)$ then~\eqref{eq:nearGr:angles_exist} has two solutions: 
$(\sumwT(\corner),\sumbT(\corner))\in\{(\pi,0),(0,\pi)\}$.
\item\label{corner_angle2}
 Otherwise, $\sumO(\corner)\neq\pi$, and the unique solution to~\eqref{eq:nearGr:angles_exist} is given by
\begin{equation}\label{eq:sumwT_sumbT_vs_sumT_sumO}
 (\sumwT(\corner),\sumbT(\corner)) = 
 \begin{cases}
 \left(\frac{\sumT(\corner) + \sumO(\corner)}2,\frac{\sumT(\corner) - \sumO(\corner)}2\right), &\text{if $\sumO(\corner)\in[0,\pi)$,}\\
 \left(\frac{\sumT(\corner) + \sumO(\corner)}2-\pi,\frac{\sumT(\corner) - \sumO(\corner)}2+\pi\right), &\text{if $\sumO(\corner)\in(\pi,2\pi)$.}\\
 \end{cases}
\end{equation}
\end{enumerate}
\end{lemma}
\begin{proof}
Part~\itemref{corner_angle1} is clear. 
 Suppose that $(\sumT(\corner), \sumO(\corner)) \neq (\pi,\pi)$. 
By~\itemref{MCE3:M-tnn},
 $(\xd(\corfp) - \xd(\corfm))^2 \geq0$, which by~\eqref{eq:M_tnn_vs_angles} implies that 
\begin{equation}\label{eq:cos_M_tnn}
 \cos(\sumT(\corner)) \leq \cos(\sumO(\corner)).
\end{equation}
In particular, $\sumO(\corner)=\pi$ would imply $\sumT(\corner)=\pi$, a contradiction. Thus, $\sumO(\corner)\neq\pi$. Taking both equations in~\eqref{eq:nearGr:angles_exist} modulo $2\pi$, we get $(\sumwT(\corner),\sumbT(\corner)) \equiv \left(\frac{\sumT(\corner) + \sumO(\corner)}2,\frac{\sumT(\corner) - \sumO(\corner)}2\right)\ \mod\ \pi$. Since $\sumwT(\corner) + \sumbT(\corner) = \sumT(\corner)$, we must have $(\sumwT(\corner),\sumbT(\corner)) = \left(\frac{\sumT(\corner) + \sumO(\corner)}2 - d\pi,\frac{\sumT(\corner) - \sumO(\corner)}2 + d\pi\right)$ for some $d\in\Z$. Furthermore, both angles must belong to $[0,\pi]$. 
If $\sumO(\corner)\in[0,\pi)$ then by~\eqref{eq:cos_M_tnn}, $\sumT(\corner)\in[\sumO(\corner),2\pi-\sumO(\corner)]$, so we must have $d=0$. If $\sumO(\corner)\in(\pi,2\pi)$ then by~\eqref{eq:cos_M_tnn}, $\sumT(\corner)\in[2\pi-\sumO(\corner),\sumO(\corner)]$, so we must have $d=1$. 
\end{proof}

\begin{definition}\label{dfn:corner_active}
A corner $\corner$ of $(\GD,\xd)$ is called \emph{ambiguous} if $(\sumT(\corner), \sumO(\corner)) = (\pi,\pi)$. 
We say that $\corner$ is \emph{\bic} if it is unambiguous and the unique solution to~\eqref{eq:nearGr:angles_exist} satisfies $\sumwT(\corner),\sumbT(\corner)>0$.
\end{definition}

\begin{corollary}\label{lemma:MCE:corner_M_pos} Let $\cor$ be a corner of $(\GD,\xd)$.
\begin{enumerate}[label=(\arabic*)]
\item\label{corner_M_pos1} If $(\xdf(\corfm) - \xdf(\corfp))^2>0$ then $\corner$ is \bic.
\item\label{corner_M_pos2} If $(\xdf(\corfm) - \xdf(\corfp))^2=0$ then $\corner$ is \bic if and only if $\pi<\sumT(\cor)\leq 2\pi$.
\end{enumerate}
\end{corollary}
\begin{proof}
We prove~\itemref{corner_M_pos1}. We have $(\xdf(\corfm) - \xdf(\corfp))^2>0$ if and only if the inequality~\eqref{eq:cos_M_tnn} is strict. In this case, $0<\sumT(\corner)<2\pi$ and $\sumO(\corner)\neq\pi$, and thus $\sumwT(\corner),\sumbT(\corner)$ are given by~\eqref{eq:sumwT_sumbT_vs_sumT_sumO}. 
 In the first case in~\eqref{eq:sumwT_sumbT_vs_sumT_sumO}, we have $0\leq\sumO(\corner)<\sumT(\corner)<2\pi-\sumO(\corner)$, so $\sumwT(\corner),\sumbT(\corner)>0$. In the second case in~\eqref{eq:sumwT_sumbT_vs_sumT_sumO}, we have $0<2\pi-\sumO(\corner)<\sumT(\corner)<\sumO(\corner)<2\pi$ which again implies $\sumwT(\corner),\sumbT(\corner)>0$.

The proof of~\itemref{corner_M_pos2} is similar, where we now use that~\eqref{eq:cos_M_tnn} becomes an equality.
\end{proof}

\begin{lemma}\label{lemma:MCE:ambig_corner_clique}
Let $\cor$ be an ambiguous corner of $(\GD,\xd)$. Then the neighborhood $\Neigh_{\GD}(\corf)$ of $\corf$ in $\GD$ is a clique (in the sense of \cref{dfn:MCE:white_black_deg}). 
\end{lemma}
\begin{proof}
For each $\ff\in\Neigh_{\GD}(\corf)$, we have $\xT(\corf)\in[\xT(\ff),\xT(\corf)]\cap\open[\xT(\corfm),\xT(\corfp)]$, so $\{\ff,\corf,\corfm,\corfp\}$ is a clique by \cref{lemma:MCE:clique_edge}. Furthermore, for any $\ff,\fff\in\Neigh_{\GD}(\corf)$, the line segments $[\xT(\ff),\xT(\corf)]$ and $[\xT(\fff),\xT(\corf)]$ lie weakly on the same side of the line containing $[\xT(\corfm),\xT(\corfp)]$. 
Thus, by \cref{lemma:MCE:clique_union}, $\{\ff,\fff,\corf,\corfm,\corfp\}$ is a clique. 
\end{proof}

Combining~\eqref{eq:MCE3':Mbd} and \cref{lemma:MCE:ambig_corner_clique}, we obtain the following.
\begin{corollary}\label{lemma:MCE:bdry_unambig}
For each $i\in\brn$, $\bdf_i$ is not incident to any ambiguous corners of $(\GD,\xd)$.
\end{corollary}

\subsection{\KawAngle condition} \label{ssec:PMNEs}
We relate the \Kawangle condition (\cref{ssec:Kawangle}) to the geometry of \MNEs.

\begin{definition}
A \emph{colored \MNE} is an \MNE $(\GD,\xd)$ equipped with a black and white coloring of 
ambiguous corners of $(\GD,\xd)$.
For an ambiguous corner $\cor$ of $(\GD,\xd)$, we set $(\sumwTcr(\cor),\sumbTcr(\cor)):=(\pi,0)$ if $\cor$ is colored white and $(\sumwTcr(\cor),\sumbTcr(\cor)):=(0,\pi)$ if $\cor$ is colored black. For unambiguous $\cor$, we let $(\sumwTcr(\cor),\sumbTcr(\cor))$ be the unique solution to~\eqref{eq:nearGr:angles_exist}.
For $\f\in\Facesgr$, let
\begin{equation}\label{eq:sumwT_sumbT_of_f_dfn}
 \sumwTcr(\f):=\sum_{\cor\in\corners(\f)}\sumwTcr(\cor) \quad\text{and}\quad
 \sumbTcr(\f):=\sum_{\cor\in\corners(\f)}\sumbTcr(\cor).
\end{equation}
\end{definition}

\begin{definition}[\Kawangle condition]%
A coloring of $(\GD,\xd)$ is called \emph{proper} if 
\begin{equation}\label{eq:MCE:sumwT_sumbT=pi}
 \sumwTcr(\f) = \sumbTcr(\f) = \pi \quad\text{for all $\f\in\Fintgr$}.
\end{equation}
\end{definition}

\begin{lemma}\label{lemma:MCE:sumwT_sumbT_options}
Let $(\GD,\xd)$ be a colored \MNE. 
For all $\f\in\Fintgr$, 
\begin{equation}\label{eq:sumwT_sumbT_options}
 (\sumwTcr(\f), \sumbTcr(\f)) \in \{(\pi,\pi),(0,2\pi),(2\pi,0)\}.
\end{equation}
In particular, if there exist $\corner_1,\corner_2\in\corners(\f)$ such that $\sumwTcr(\corner_1),\sumbTcr(\corner_2)>0$ then~\eqref{eq:MCE:sumwT_sumbT=pi} is satisfied for $\f$.
\end{lemma}
\begin{proof}
Since $\xT$ is a face-injective \wemb of $\GD$,
\begin{equation}\label{eq:near:emb_vertex_angle_sum}
 \sum_{\corner\in\corners(\f)} \sumT(\corner) = 2\pi \quad\text{for all $\f\in\Fintgr$.}
\end{equation}
By~\eqref{eq:nearGr:angles_exist} and~\eqref{eq:near:emb_vertex_angle_sum}, we get $\sumwTcr(\f)+\sumbTcr(\f)=2\pi$ and $\sumwTcr(\f)-\sumbTcr(\f)\equiv0\pmod{2\pi}$. Because $0\leq \sumwTcr(\corner),\sumbTcr(\corner)\leq\pi$ for all corners $\corner$, we have $\sumwTcr(\f),\sumbTcr(\f)\geq0$ which implies~\eqref{eq:sumwT_sumbT_options}. If moreover $\sumwTcr(\f),\sumbTcr(\f)>0$ then indeed~\eqref{eq:MCE:sumwT_sumbT=pi} is the only possibility.
\end{proof}

Let $(\GD,\xd)$ be an \MNE. A natural obstruction to the existence of a proper coloring of $(\GD,\xd)$ is a non-isolated vertex $\f\in\Fintgr$ such that every corner in $\corners(\f)$ is unambiguous but~\eqref{eq:MCE:sumwT_sumbT=pi} is not satisfied for $\f$. In this case, $ (\sumwTcr(\f), \sumbTcr(\f)) \in \{(0,2\pi),(2\pi,0)\}$ by~\eqref{eq:sumwT_sumbT_options}. 

\begin{definition}\label{dfn:improper}
We say that $\f\in\Facesgr$ is 
\emph{improper} if there exists a clique $\clique\subset\xd(\Faces)$ such that $\xT(\f)\in\Convint\KT$, and for all $\corner\in\corners(\f)$, we have $0\leq \sumT(\corner)<\pi$. Otherwise, $\f$ is called \emph{proper}. 
\end{definition}

\begin{lemma}\label{lemma:MCE:improper_char}
A non-isolated vertex $\f\in\Fintgr$ is improper if and only if every corner in $\corners(\f)$ is unambiguous but~\eqref{eq:MCE:sumwT_sumbT=pi} is not satisfied for $\f$.
\end{lemma}
\begin{proof}
 $(\Longleftarrow)$: By \cref{lemma:MCE:sumwT_sumbT_options}, if $\cor\in\corners(\f)$ is \bic then~\eqref{eq:MCE:sumwT_sumbT=pi} is satisfied for $\f$, a contradiction. Otherwise, by~\eqref{eq:M_tnn_vs_angles_eq} and \cref{lemma:MCE:corner_M_pos}, $\cos(\sumT(\corner)) = \cos(\sumO(\corner))$
and $0\leq \sumT(\cor)\leq\pi$. By assumption, $\cor$ is unambiguous, so $\sumO(\corner)\neq\pi$ and thus $\sumT(\cor)\neq\pi$. 
 Thus, $0\leq \sumT(\corner)<\pi$ for all $\corner\in\corners(\f)$. By~\eqref{eq:sumwT_sumbT_options}, $(\sumwTcr(\f), \sumbTcr(\f)) \in \{(0,2\pi),(2\pi,0)\}$. Suppose that, say, $(\sumwTcr(\f), \sumbTcr(\f)) = (2\pi,0)$. For every corner $\corner\in\corners(\f)$ such that $\sumT(\corner)\in(0,\pi)$, the triangle $\Conv\xT(\{\corf,\corfm,\corfp\})$ is nondegenerate and isometric to the triangle $\Conv\xO(\{\corf,\corfm,\corfp\})$, and since $\sumbT(\cor)=0$, this isometry is orientation-preserving. Thus, $\xd(\{\f,\corfm,\corfp\})$ is a white clique. When two such triangles have $1$-dimensional intersection, their union is also a white clique. Thus, the neighbors of $\f$ form a white clique $\clique := \xd(\Neigh_{\GD}(\f))$, and since each corner $\cor\in\corners(\f)$ satisfies $0\leq \sumT(\corner)<\pi$, the vertex $\xT(\f)$ belongs to $\Convint\KT$. Thus, $\f$ is improper.

 $(\Longrightarrow)$: Suppose that $\f$ is improper, and let $\clique=\xd(\preK)$ be as in \cref{dfn:improper}. 
Let $\ff\in\Neigh_{\GD}(\f)$.
The line segment $[\xT(\ff),\xT(\f)]$ intersects $\Convint\KT$, so by \cref{lemma:MCE:clique_union}, $\{\ff\}\cup\preK$ is a clique. Thus, we may assume that $\Neigh_{\GD}(\f)\subset\preK$.
 Since $\Convint\KT$ is nonempty, we have $\dim\Conv\KT = 2$, so $\clique$ is nondegenerate. Thus, it must be either black or white; suppose that it is white. Let $\cor\in\corners(\f)$ be a corner such that $\sumT(\corner)\in(0,\pi)$. Since $\{\f,\corfm,\corfp\}\subset\preK$ and since $\dim\Conv\xT(\{\f,\corfm,\corfp\})=2$, the clique $\{\f,\corfm,\corfp\}$ is also nondegenerate and white. It follows that $\sumwT(\cor)=\sumT(\cor)$ and $\sumbT(\cor) = 0$. Thus, $(\sumwTcr(\f), \sumbTcr(\f)) = (2\pi,0)$, so~\eqref{eq:MCE:sumwT_sumbT=pi} is not satisfied for $\f$.
\end{proof}

\begin{remark}\label{rmk:MCE:bdry_proper}
For each $i\in\brn$, $\xT(\bdf_i)=\bdxT_i$ is not contained in $\Convint\KT$ for any clique $\clique=\xd(\preKmax)$. Otherwise, by \cref{lemma:MCE:clique_union}, $\preKmax\cup\{\bdf_{i-1},\bdf_{i+1}\}$ would also be a clique, contradicting $(\bdx_{i-1} - \bdx_{i+1})^2>0$; cf.~\eqref{eq:MCE3':Mbd}. Thus, the boundary vertices of $\GD$ are always proper. Similarly to \cref{lemma:MCE:sumwT_sumbT_options}, we get
\begin{equation}\label{eq:MCE:sumwT_sumbT_bdry_0_pi}
 0<\sumwT(\bdf_i),\sumbT(\bdf_i)<\pi \quad\text{for all $i\in\brn$}. 
\end{equation}
\end{remark}

\begin{definition}\label{dfn:MCE:pMNE}
An \MNE is called \emph{proper} if it has no improper vertices. 
\end{definition}

We say that $\f\in\Fintgr$ is \emph{doubly ambiguous} if it is incident to two ambiguous corners.
\begin{lemma}\label{lemma:MCE:coloring_improper_verts}
Suppose that $(\GD,\xd)$ is an \MNE. Then $(\GD,\xd)$ admits a proper coloring if and only if it is proper. 
 In this case, the number of such proper colorings is $2^d$, where $d$ is the number of doubly ambiguous vertices of $\GD$.
\end{lemma} 
\begin{proof}
By \cref{lemma:MCE:improper_char}, if $\GD$ has an improper vertex then it admits no proper colorings. Conversely, suppose that $\GD$ has no improper vertices. 
By \cref{lemma:MCE:bdry_unambig,rmk:MCE:bdry_proper}, each boundary vertex of $\GD$ is proper and incident to no ambiguous corners. 
Let $\f\in\Fintgr$. By~\eqref{eq:near:emb_vertex_angle_sum}, $\f$ can only be incident to $l=0$, $l=1$, or $l=2$ ambiguous corners. If $l=0$ then $\f$ satisfies~\eqref{eq:MCE:sumwT_sumbT=pi} for any coloring since it would be improper otherwise. 
If $l=1$ then exactly one choice of the color of the ambiguous corner $\cor$ incident to $\f$ satisfies~\eqref{eq:MCE:sumwT_sumbT=pi}, and if $l=2$ then there are two choices, where we have to use opposite colors for the two ambiguous corners incident to $\f$.
\end{proof}

Next, we discuss the behavior of \pcMNEs under taking subgraphs and limits. 
\begin{definition}%
Let $\GD$ be a graph satisfying \cref{ass:lgg_dual}. Define 
\begin{equation}\label{eq:MCE:MCMalg_dfn}
 \MCMalg(\GD):=\{(\GD,\xd)\mid \xd:\Faces\to\Rdd\text{ satisfies }(\xd(\ff) - \xd(\f))^2 = 0\text{\ \ for all $\{\ff,\f\}\in\Ebarast$}\}.
\end{equation}
\end{definition}

\begin{proposition}\label{lemma:MCE:Res}
Suppose that $\GDout$ satisfies \cref{ass:lgg_dual}, and let $\GD$ be a subgraph of $\GDout$ with the same outer boundary cycle $(\bdf_1,\bdf_2,\dots,\bdf_n=\bdf_0)$. 
Let $\Faces\subset\Fout$ denote their respective vertex sets, and consider the \emph{restriction operator}
\begin{equation}\label{eq:MCE:Res_dfn}
 \ResG:\MCMalg(\GDout)\to\MCMalg(\GD),\quad (\GDout,\xdout)\mapsto (\GD,\xdout|_{\Faces}).
\end{equation}
If $(\GDout,\xdout)\in\MCMalg(\GDout)$ is a \pMNE then so is $(\GD,\xd) := \ResG(\GDout,\xdout)$, provided that $\xT$ is \finj. 
Furthermore, any proper coloring of $(\GDout,\xdout)$ induces a unique proper coloring of $(\GD,\xd)$ satisfying
\begin{equation}\label{eq:induced_angle_sums}
\sumwTx(\cor)= \sum_{\corout\in\corsout(\cor)} \sumwTxout(\corout) \quad\text{and}\quad 
\sumbTx(\cor)= \sum_{\corout\in\corsout(\cor)} \sumbTxout(\corout)
\quad\text{for all $\cor\in\corners(\GD)$}, 
\end{equation}
where
 $\corsout(\cor)$ is the set of corners of $\GDout$ incident to $\corf$ and contained inside $\cor$.
\end{proposition}
\begin{proof}
First, observe that $(\GD,\xd)$ satisfies \cref{ass:lgg_dual} and~\eqref{eq:MCE3':Mbd} since its boundary cycle is the same as that of $(\GDout,\xdout)$. Second, if $(\GDout,\xdout)$ is an \MNE and $\xT$ is \finj then $(\GD,\xd)$ is also an \MNE.
 Third, we claim that if $(\GDout,\xdout)$ is proper then so is $(\GD,\xd)$. Suppose first that $(\GD,\xd)$ is obtained from $(\GDout,\xdout)$ by removing a single edge $\east$ with endpoints $\ff,\f\in\Fout$, and suppose for contradiction that $\f$ is improper in $(\GD,\xd)$. 
By \cref{rmk:MCE:bdry_proper}, $\f\in\Fint$. 
By \cref{dfn:improper}, $\f$ is not isolated in $\GD$, there exists a clique $\clique\subset\xd(\Faces)$ such that $\xT(\f)\in\Convint\KT$, and for all $\corner\in\cornersGD(\f)$, we have $0\leq \sumT(\corner)<\pi$. Therefore, $\f$ is not isolated in $\GDout$ and $\clique\subset\xd(\Faces)\subset\xdout(\Fout)$ is a clique in $(\GDout,\xdout)$. Since $\east$ is not an outer boundary edge, it subdivides one of the corners $\cor$ of $(\GD,\xd)$ incident to $\f$ into corners $\cor',\cor''$ and we have $\sumT(\cor) = \sumT(\cor') + \sumT(\cor'')$. Thus, $0\leq \sumT(\cor'),\sumT(\cor'')<\pi$, so $\f$ is an improper vertex of $(\GDout,\xdout)$, a contradiction. 

Any subgraph $\GD$ of $\GDout$ can be obtained from $\GDout$ via a sequence of edge deletions and isolated vertex deletions. By \cref{dfn:improper}, isolated vertex deletions also preserve properness.

Fix a proper coloring of $(\GDout,\xdout)$ and let $\cor\in\corners(\GD)$.
By~\eqref{eq:MCE:sumwT_sumbT=pi} (resp.,~\eqref{eq:MCE:sumwT_sumbT_bdry_0_pi}) applied to 
$(\GDout,\xdout)$, the angles $\sumwTx(\cor),\sumbTx(\cor)$ defined in~\eqref{eq:induced_angle_sums} belong to $[0,\pi]$. Since $\sumTx(\cor) = \sum_{\corout\in\corsout(\cor)} \sumTxout(\corout)$ and $\sumOx(\cor) \equiv \sum_{\corout\in\corsout(\cor)} \sumOxout(\corout)$ $\mod$ $2\pi$, $(\sumwTx(\cor),\sumbTx(\cor))$ is a solution to~\eqref{eq:nearGr:angles_exist}. It automatically satisfies~\eqref{eq:MCE:sumwT_sumbT=pi} (resp.,~\eqref{eq:MCE:sumwT_sumbT_bdry_0_pi}) for $\corf$. %
\end{proof}

\begin{proposition}\label{lemma:limit_of_pcMNE_is_pMNE}
Assume that $(\GD,\xdtls)$ is a \pcMNE for all small $\tlim>0$ and assume that the limit $\xd=\lim_{\tlim\to0} \xdtls$ exists, satisfies~\eqref{eq:MCE3':Mbd}, and $\xT$ is \finj. Then $(\GD,\xd)$ is a \pMNE. 
\end{proposition}
\begin{proof}
Since $\xT$ is a limit of \wembs of $\GD$, it is itself a \wemb by \cref{dfn:TOP:wtimm}. Thus, $(\GD,\xd)$ satisfies~\itemref{MCE1:emb}. Since~\itemref{MCE2:null_edges}--\itemref{MCE3:M-tnn} are given by closed conditions, they are satisfied for $(\GD,\xd)$. Let $\ff\in\Fintgr$. Suppose that $\ff$ is incident to $l\in\{0,1,2\}$ ambiguous corners of $(\GD,\xd)$ and let $\cor\in\corners(\ff)$. 
After possibly passing to a subsequence, we may assume that 
 $\lim_{\tlim\to0}\sumT_{\xdtls}(\cor) = \sumT(\cor)$ in $[0,2\pi]$ 
and $\lim_{\tlim\to0}\sumO_{\xdtls}(\cor) = \sumO(\cor)$ in $\R/2\pi\Z$.
 By \cref{lemma:MCE:corner_angle}, if $\cor$ is unambiguous in $(\GD,\xd)$ then 
$\lim_{\tlim\to0}\sumwT_{\xdtls}(\cor) = \sumwT_{\xd}(\cor)$ and $\lim_{\tlim\to0}\sumbT_{\xdtls}(\cor) = \sumbT_{\xd}(\cor)$. In particular, if $l=0$ then~\eqref{eq:MCE:sumwT_sumbT=pi} is satisfied for $\ff$ in the $\tlim\to0$ limit, so $\ff$ is proper in $(\GD,\xd)$ by \cref{lemma:MCE:improper_char}. If $l\in\{1,2\}$ then $\ff$ is automatically proper in $(\GD,\xd)$. %
\end{proof}

\section{\MCEsnoacrTITLE}\label{sec:MCE}
We introduce \emph{\MCEsnoacr} which will play a central role in our proof. 

\subsection{Definition and basic properties}

\begin{definition}\label{dfn:pchord}
Let $(\GD,\xd)$ be an \MNE. 
For $\{\f_a,\f_b\}\in\NullE(\GD,\xd)$, 
 a \bent line segment $(a,b,\bendh)$ connecting $a:=\xT(\f_a)$ to $b:=\xT(\f_b)$ is called a \emph{\pchord} of $(\GD,\xd)$ if %
\begin{enumerate}[label=(\roman*)]
\item%
 $(a,b,\bendh)$ is \emph{\addable}: for all small $\eps>0$, we have 
$[a,b]_{\ebendh} \cap \Skelteh = \xteh(\{\f_a,\f_b\})$, and
\item\label{pchord2} $(a,b,\bendh)$ is \emph{proper}: adding the edge $(a,b,\bendh)$ to $(\GD,\xd)$ does not create any improper vertices.
\end{enumerate}
\end{definition}

\noindent Thus, in the case when $(\GD,\xd)$ is itself proper, $(a,b,\bendh)$ is a \pchord if adding it to the set of edges of $(\GD,\xd)$ still results in a \pMNE; cf. \cref{lemma:MCE:add_chords} below. 

\begin{definition}
For a fixed \MNE $(\GD,\xd)$, two \bent line segments $(a,b,\bendhx0)$, $(a,b,\bendhx1)$ are called \emph{isotopic} if there exists a homotopy $\bendhx t$, $t\in[0,1]$ that coincides with $\bendhx0$ (resp., $\bendhx1$) for $t=0$ (resp., $t=1$), such that $(a,b,\bendhx t)$ is addable for all $0<t<1$. 
\end{definition}

\begin{remark}\label{rmk:MCE:pchord_outer_face}
By~\eqref{eq:MCE3':Mbd}, any \pchord contained in the outer face of $(\GD,\xd)$ is isotopic to a boundary edge. 
\end{remark}

\begin{definition}[\MCEnoacr]\label{dfn:MCE:MCE}
An \emph{\MCEnoacr} (\emph{\MCE})
is an \MNE $(\GD,\xd)$ satisfying~\itemref{MCE1:emb}--\itemref{MCE3:M-tnn} together with the following.
{\setlength{\leftmargini}{50pt}
\begin{enumerate}[label=(MCE\arabic*)] 
\setcounter{enumi}{3}
\item\label{MCE4:properly_colored}
 \emph{Properly colored}: 
The \MNE $(\GD,\xd)$ is properly colored.
\setcounter{enumi}{4}
\item\label{MCE5:Mpos_chords} 
\emph{No \pchords:}
Every \pchord in $(\GD,\xd)$ is isotopic to some edge of $(\GD,\xd)$.
\end{enumerate}
}
\end{definition}

We discuss some basic properties of \MCEs.

\begin{lemma}\label{lemma:ambig=>triangular}
If $\cor$ is an ambiguous corner of an \MCE $(\GD,\xd)$ then the face $\corv$ is a degenerate triangle.
\end{lemma}
\begin{proof}
Since $\cor$ is ambiguous, $(\xd(\corfp)-\xd(\corfm))^2=0$ and $\GD$ admits a \pchord connecting $\xT(\corfp)$ to $\xT(\corfm)$. By~\itemref{MCE5:Mpos_chords}, this \pchord is isotopic to an edge of $(\GD,\xd)$.
\end{proof}

\begin{lemma}
\label{lemma:MCE:add_chords}
Suppose that $(\GD,\xd)$ is an \MNE. Then one can add some \pchords of $(\GD,\xd)$ to the set of edges so that the resulting pair $(\GDout,\xdout)$ satisfies~\axrangeMCEchord. 
Furthermore, if $(\GD,\xd)$ is proper then so is $(\GDout,\xdout)$. 
In this case, with an arbitrary choice of a proper coloring (cf. \cref{lemma:MCE:coloring_improper_verts}), $(\GDout,\xdout)$ becomes an \MCE. 
\end{lemma}
\begin{proof}

Let $(\GD_0,\xd_0):=(\GD,\xd)$. Suppose that a \pchord $(a,b,\bendh)$ with $a=\xT(\f_a)$ and $b=\xT(\f_b)$ violates~\itemref{MCE5:Mpos_chords}. Let $(\GD_1,\xd_1)$ be obtained from $(\GD_0,\xd_0)$ by adding $(a,b,\bendh)$ to the set of edges. Since $(a,b,\bendh)$ is addable, $(\GD_1,\xd_1)$ satisfies~\itemref{MCE1:emb}. 
Since $\{\f_a,\f_b\}\in\NullE(\GD,\xd)$, 
 $(\GD_1,\xd_1)$ satisfies~\itemref{MCE2:null_edges}. Since $\xd_1(\Faces) = \xd_0(\Faces)$, $(\GD_1,\xd_1)$ satisfies~\itemref{MCE3:M-tnn}. 
Iterating this process, we eventually arrive at a pair $(\GDout,\xdout) = (\GD_{\Tmax},\xd_{\Tmax})$, $\Tmax\geq0$, such that $(\GDout,\xdout)$ satisfies~\axrangeMCEchord. 
Furthermore, if $(\GD_0,\xd_0)$ was proper then by~\crefi{dfn:pchord}{pchord2}, so is $(\GDout,\xdout)$. After choosing a proper coloring of $(\GDout,\xdout)$ (cf. \cref{lemma:MCE:coloring_improper_verts}), we indeed obtain an \MCE. 
\end{proof}

\begin{lemma}\label{lemma:MCE:sticky}
Suppose that $(\GD,\xd)$ is an \MCE and let $z\in\xT(\GD)$. 
Let $\elline$ and $\{\east_1,\dots,\east_k,\f_1,\dots,\f_d\}$ be as in \cref{rmk:MCE:near_emb_combin}, and 
 let $\bmv_+$ (resp., $\bmv_-$) be the set of faces $\v\in\Vint$ of $\GD$ such that 
$\xT(\v)$ contains points arbitrarily close to $z$ located strictly above (resp., below) $\elline$.
 Then an element of $\{\east_1,\dots,\east_k,\f_1,\dots,\f_d\}$ is only incident to degenerate \btrar faces contained in $\elline$ unless it is $\preceq_z$-maximal (resp., $\preceq_z$-minimal), in which case 
it is also incident to the faces in $\bmv_+$ (resp., $\bmv_-$). 
\end{lemma}
\begin{proof}
Consider a covering relation $\past\precdot_z\qast$ with respect to the total ordering $\preceq_z$, where $\past,\qast\in\{\east_1,\dots,\east_k,\f_1,\dots,\f_d\}$. By \cref{lemma:MCE:non_inj}, at least one of $\past,\qast$ is an edge. If one of them is a vertex, say, $\past = \f_i$ and $\qast=\east_j$ then $(\GD,\xd)$ admits \pchords connecting $\f_i$ to each endpoint of $\east_j$. By~\itemref{MCE5:Mpos_chords}, both of these \pchords must be isotopic to edges of $(\GD,\xd)$. Next, consider the case where $\past=\east_i$ and $\qast=\east_j$ are both edges. 
Following \cref{rmk:MCE:near_emb_combin}, assume that $\elline=\R$ is horizontal. 
Let $z_-\in\xT(\Faces)$ (resp., $z_+\in\xT(\Faces)$) be the point on $\elline$ closest to $z$ located to the left (resp., to the right) of $z$ such that we do not have $\past\precdot_{z_-}\qast$ (resp., $\past\precdot_{z_+}\qast$). 
 By \cref{rmk:MCE:near_emb_combin}, there exists a vertex $\f_\pm\in\Faces$ satisfying $z_\pm=\xT(\f_\pm)$ such that $\f_\pm$ is either a vertex
of $\past$ or $\past\precdot_{z_\pm} \f_{\pm}$, and similarly $\f_\pm$ is either a vertex of $\qast$ or $\f_{\pm} \precdot_{z_\pm} \qast$. 
If $\f_\pm$ is not a vertex of $\past$ then as we showed above, $\f_\pm$ and $\past$ share a degenerate triangular face of $(\GD,\xd)$. This face contains an edge $\east_\pm$ incident to $\f_\pm$ such that $z\in\xTint(\east_\pm)$
 and $\past\precdot_z \east_\pm\preceq_z\qast$. Since $\past\precdot_z\qast$, this implies $\east_\pm=\qast$, so $\past$ and $\qast$ share a triangular face, as desired. Otherwise, $\f_\pm$ must be a vertex of $\past$. Similarly, $\f_\pm$ must be a vertex of $\qast$. Thus, $\past$ and $\qast$ share a bigonal face, as desired. 

We have shown that any edge $\east_j$, $j\in\brk$, that is not $\preceq_z$-maximal (resp., $\preceq_z$-minimal) must be incident to a \btrar face located immediately above (resp., below) $\east_j$. Suppose that a vertex $\f_i$, $i\in\brd$, is not, say, $\preceq_z$-maximal, and let $\east_j$ be the edge immediately above it so that $\f_i\precdot_z \east_j$. As we showed above, $\f_i$ and $\east_j$ share a triangular face. For any edge $\east$ incident to $\f_i$ such that $\xT(\east)\subset \elline$, we have $\east\prec_y \east_j$ for any $y\in\xTint(\east)\cap\xTint(\east_j)$. Thus, the face immediately above $\xT(\east)$ is also \btrar. If $\f_i$ is $\preceq_z$-minimal then it follows that all faces incident to $\f_i$ that are neither triangles nor bigons must belong to $\bmv_-$; otherwise, all faces incident to $\f_i$ are \btrgles.
\end{proof}

\subsection{Maximal cliques}%
Suppose that $(\GD,\xd)$ is an \MNE.
We denote by $\preKmaxs$ the set of maximal by inclusion cliques $\preKmax\subset\Faces$ consisting of at least two vertices, and let $\Kmaxs:=\{\xd(\preKmax)\mid\preKmax\in\preKmaxs\}$. We start with the following trivial observation.

\begin{lemma}\label{lemma:MCE:trivial_\Mdash zero<=>clique}
Let $\ff,\f\in\Faces$ be distinct vertices. Then we have $(\xd(\ff) - \xd(\f))^2 = 0$ if and only if there exists a maximal clique $\preKmax\in\preKmaxs$ containing $\ff$ and $\f$. %
\end{lemma}

\begin{lemma}\label{lemma:MCE:Conv_vertices_injective}
For each $\Kmax=\xd(\preKmax)\in\Kmaxs$, we have $\dim\Conv\KTmax\in\{1,2\}$, and for each vertex $z$ of $\Conv\KTmax$, the preimage $\{\f\in\Faces\mid\xT(\f)=z\}$ has size $1$. %
\end{lemma}
\begin{proof}
Let $z$ be a vertex of $\Conv\KTmax$. 
By \cref{lemma:MCE:clique_convex}, for any edge $\xT(\east)$ containing $z$ in its relative interior, $\preKmax\cup\ebarast$ must be a clique, so by the maximality of $\preKmax$, we have $\ebarast\subset\preKmax$. Thus, $\xT(\east)\subset\Conv\KTmax$, so $z$ is not a vertex of $\Conv\KTmax$, a contradiction. 
 By \cref{lemma:MCE:non_inj_edge}, we get $|\{\f\in\Faces\mid\xT(\f)=z\}|=1$. If $\dim\Conv\KTmax=0$ then $\Conv\KTmax=\{z\}$, so $|\preKmax|=1$, a contradiction.
\end{proof}

\begin{lemma}\label{lemma:MCE:two_cliques_intersection}
Let $\Kmax_1,\Kmax_2\in\Kmaxs$ be distinct maximal cliques. Then the intersection $\Conv\KTmax_1\cap\Conv\KTmax_2$ is either empty, a common vertex, or a common edge of $\Conv\KTmax_1$ and $\Conv\KTmax_2$. In the case of a common edge, we have $\dim\Conv\KTmax_1 = \dim\Conv\KTmax_2 = 2$ and the cliques $\Kmax_1,\Kmax_2$ are of different color.
\end{lemma}
\begin{proof}
Assuming the intersection $\PolygT:=\Conv\KTmax_1\cap\Conv\KTmax_2$ is nonempty, it is a convex subset of the plane. Let $z\in\PolygTrelint$ be a point in its relative interior. 
 Let $\Conv\dotKTmax_1$ and $\Conv\dotKTmax_2$ be the faces of $\Conv\KTmax_1$ and $\Conv\KTmax_2$ that contain $z$ in their relative interiors so that $\PolygT=\Conv\dotKTmax_1\cap\Conv\dotKTmax_2$. 
Observe that $\Conv\KTmax_1$ intersects $\Convrelint\dotKTmax_2$ and $\Conv\KTmax_2$ intersects $\Convrelint\dotKTmax_1$. By \cref{lemma:MCE:clique_union}, since $\Kmax_1,\Kmax_2\in\Kmaxs$, we have $\dotKmax_1\cup\dotKmax_2 \subset \Kmax_1\cap\Kmax_2$. Thus, $\Conv\dotKTmax_1=\Conv\dotKTmax_2=\PolygT$, so $\PolygT$ is a common face of $\Conv\KTmax_1$ and $\Conv\KTmax_2$. Since the cliques $\Kmax_1,\Kmax_2$ are maximal and distinct, 
 $\PolygT$ is a proper face of each of $\Conv\KTmax_1$ and $\Conv\KTmax_2$. Thus, if $\dim\PolygT=1$ then $\dim\Conv\KTmax_1 = \dim\Conv\KTmax_2 = 2$. In this case, if the cliques $\Kmax_1,\Kmax_2$ were of the same color then $\KTmax_1\cup\KTmax_2$ would be isometric to $\KOmax_1\cup\KOmax_2$, so $\Kmax_1\cup\Kmax_2$ would be a clique, contradicting the maximality of $\Kmax_1,\Kmax_2$. 
\end{proof}

\begin{definition}\label{dfn:MCE:ext_on_edge}
Let $\Kmax=\xd(\preKmax)\in\Kmaxs$ and let $(b_1,\dots,b_k=b_0)$ be the sequence of vertices of $\Conv\KTmax$ listed in clockwise order. 
 For $j\in\brk$, let $\ff_j\in\Faces$ be the unique (by \cref{lemma:MCE:Conv_vertices_injective}) vertex such that $\xT(\ff_j)=b_j$, and 
 let $\normal_j:=\I(b_j-b_{j-1})$ be the outward normal vector to the edge $[b_{j-1},b_j]$. Thus, the maximum of the supporting linear function\footnote{Here, $\<\cdot,\cdot\>$ denotes the standard dot product on $\R^2\cong\C$.} $\<\cdot,\normal_j\>$ on $\Conv\KTmax$ is achieved on the edge $[b_{j-1},b_j]$. We say that a vertex $\f\in\preKmax$ is \emph{$(b_{j-1},b_j)$-external} if $\xT(\f)\in\open[b_{j-1},b_j]$ and there exists an edge $\{\f,\ff\}\in\Ebarast$ such that $\<\xT(\f),\normal_j\> < \<\xT(\ff),\normal_j\>$. 
\end{definition}
\begin{remark}\label{rmk:MCE:extpt_ordering}
For any $z\in\open[b_{j-1},b_j]$, there exists at most one $(b_{j-1},b_j)$-external vertex $\f\in\preKmax$ satisfying $\xT(\f)=z$, because by \cref{rmk:MCE:near_emb_combin}, in order to be $(b_{j-1},b_j)$-external, $\f$ has to be $\preceq_z$-maximal. Thus, the $(b_{j-1},b_j)$-external vertices are linearly ordered along the line segment $\open[b_{j-1},b_j]$.
\end{remark}

\begin{figure}
 \def\inputscale{2}
 \setlength{\tabcolsep}{7pt}
\begin{tabular}{cc}
\includegraphics[scale=\inputscale]{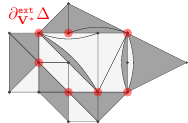}
&
\includegraphics[scale=\inputscale]{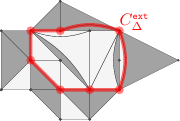}
\end{tabular}
 \caption{\label{fig:extcycle} \Extpts (left) and \extpchords (right) are marked in red.}
\end{figure}

\begin{definition}[\Extpt]\label{dfn:MCE:extpt}
In the notation of \cref{dfn:MCE:ext_on_edge}, we say 
 that $\f\in\preKmax$ is an \emph{\extpt} of $\Kmax$ if either $\xT(\f)$ is a vertex of $\Conv\KTmax$ or 
$\f$ is $(b_{j-1},b_j)$-external for some $j\in\brk$. We denote by $\partFext\Kmax = (\f_1,\dots,\f_m=\f_0)$
 the sequence of \extpts of $\Kmax$ listed in clockwise order around the boundary of $\Conv\KTmax$.\footnote{More precisely, $(\f_1,\dots,\f_m=\f_0)$ contains the vertex sequence $(\ff_1,\dots,\ff_k=\ff_0)$ of $\Conv\KTmax$ as a subsequence, and for each $j\in\brk$, the $(b_{j-1},b_j)$-external vertices appear between $\ff_{j-1}$ and $\ff_j$ in the order specified in \cref{rmk:MCE:extpt_ordering}.}
 See \figref{fig:extcycle}(left) for an example.
\end{definition}

\begin{lemma}\label{lemma:MCE:extpt_vs_vertex}
Let $\Kmax=\xd(\preKmax)\in\Kmaxs$ and suppose that $\Kmax$ admits an \extpt $\f\in\preKmax$ that is not a vertex of $\Conv\KTmax$. Then $\dim\Conv\KTmax = 2$ and there exists another maximal clique $\Kmaxp\in\Kmaxs$ such that $\dim\Conv\KTmaxp = 2$, the cliques $\Kmax$ and $\Kmaxp$ are of different color, and $\xT(\f)$ belongs to a common edge of the polygons $\Conv\KTmax$ and $\Conv\KTmaxp$.
\end{lemma}
\begin{proof}
Let $b_-=\xT(\f_-)$, $b_+ = \xT(\f_+)$ be vertices of $\Conv\KTmax$ such that $\f$ is $(b_-,b_+)$-external. Let $\normal:=\I(b_+-b_-)$ and consider an edge $\{\f,\ff\}\in\Ebarast$ such that $\<\xT(\f),\normal\> < \<\xT(\ff),\normal\>$ as in \cref{dfn:MCE:ext_on_edge}. 
By \cref{lemma:MCE:clique_union}, $\xd(\{\ff,\f,\f_-,\f_+\})$ is a clique, and since
 $\<\xT(\f),\normal\> < \<\xT(\ff),\normal\>$,
we have $\dim\Conv\xT(\{\ff,\f,\f_-,\f_+\})=2$. 
Let $\Kmaxp=\xd(\preKmaxp)\in\Kmaxs$ be a maximal clique containing $\xd(\{\ff,\f,\f_-,\f_+\})$. By \cref{lemma:MCE:two_cliques_intersection}, $\dim\Conv\KTmax = \dim\Conv\KTmaxp = 2$, the cliques $\Kmax$ and $\Kmaxp$ are of different color, and $[b_-,b_+]$ is a common edge of $\Conv\KTmax$ and $\Conv\KTmaxp$. 
\end{proof}

\begin{remark}\label{rmk:MCE:clique_boundary_outward_bent}
In particular, when $\Kmax\in\Kmaxs$ satisfies $\dim\Conv\KTmax = 1$, every \extpt of $\Kmax$ is a vertex of $\Conv\KTmax$. Thus, when $\dim\Conv\KTmax = 1$ with $\Conv\KTmax = [\xT(\ff),\xT(\f)]$, we have $\partFext\Kmax = (\ff,\f)$, and the open line segment $\open[\xT(\ff),\xT(\f)]$ does not intersect $\Conv\KTp$ for any other clique $\cliquep\in\Kmaxs$.
\end{remark}

\begin{lemma}[\Extpchord]\label{lemma:MCE:extpchord}
Let $\Kmax\in\Kmaxs$, and let $\partFext\Kmax = (\f_1,\dots,\f_m = \f_0)$ and $a_i:=\xT(\f_i)$ for $i\in\brm$. For each $i\in\brm$, choose a \bent line segment $(a_{i-1},a_i,\bendhx i)$ that is $\preceq_z$-maximal in the direction of $\normal_i:=\I(a_i-a_{i-1})$ for all $z\in\open[a_{i-1},a_i]$ (cf. \cref{rmk:MCE:extpt_ordering}). Then $(a_{i-1},a_i,\bendhx i)$ is a \pchord of $(\GD,\xd)$.
\end{lemma}
\noindent We refer to $(a_{i-1},a_i,\bendhx i)$ as an \emph{\extpchord} of $\Kmax$. See \figref{fig:extcycle}(right) for an example.
\begin{proof}
Since $\open[a_{i-1},a_i]$ contains no \extpts of $\Kmax$, we see that $(a_{i-1},a_i,\bendhx i)$ is addable. 
It is proper because neither $a_{i-1}$ nor $a_i$ can be contained in the interior $\Convint\KTmaxp$ of some clique $\Kmaxp$ by \cref{lemma:MCE:clique_union}.
\end{proof}

\begin{proposition}[\Extcycle]\label{lemma:MCE:outward_simple}
Suppose that $(\GD,\xd)$ is an \MCE.
Let $\Kmax=\xd(\preKmax)\in\Kmaxs$ and assume that 
$\GD\ind[\preKmax]$ is not isomorphic to a graph consisting of two vertices connected by a single edge.
 Let 
 $\partFext\Kmax = (\f_1,\dots,\f_m = \f_0)$. Then 
 $\GD$ contains a simple cycle $\CycKmax$, called the \emph{\extcycle} of $\Kmax$, with edges $\east_1,\east_2,\dots,\east_m$ such that $\east_i$ connects $\f_{i-1}$ to $\f_i$ for each $i\in\brm$, and such that every face $\v\in\Vint$ satisfying $\partF\v\subset\preKmax$ (resp., every vertex $\f\in\preKmax\setminus\partFext\Kmax$) lies strictly inside the cycle $\CycKmax$. 
\end{proposition}
\begin{proof}
The union of \extpchords $[a_{i-1},a_i]_{\ebendhx i}$ of $\Kmax$ for $i\in\brm$ is an embedded cycle which contains all points in 
$\xteh(\preKmax\setminus\partFext\Kmax)$ 
 in its interior. By~\itemref{MCE5:Mpos_chords}, each edge of this cycle must be isotopic to some edge of $\xT(\GD)$. 
If $|\preKmax|=2$ and $\GD\ind[\preKmax]$ is a single edge $\east$ then we define $\CycKmax$ to be the closed walk traversing this edge twice in opposite directions. Otherwise, $\CycKmax$ is indeed a simple cycle.
\end{proof}

\subsection{\Flex faces}%
Let $(\GD,\xd)$ be an \MCE. 

\begin{definition}
A face $\v\in\Vint$ of $\GD$ 
 is called \emph{rigid} if $\xd(\partF\v)$ is a clique; 
 otherwise, $\v$ is called \emph{\flex}. The set of \flex (resp., rigid) faces of $\GD$ is denoted $\Vact$ (resp., $\Vrig$) so that $\Vint = \Vact\sqcup\Vrig$.
\end{definition}

\begin{corollary}\label{lemma:MCE:inside_CycKmax=>rigid}
A face $\v\in\Vint$ is \flex if and only if it lies outside the cycle $\CycKmax$ for each $\Kmax\in\Kmaxs$.
\end{corollary}
\begin{proof}
If $\v$ lies inside $\CycKmax$ then $\xd(\partF\v)\subset\Kmax$ is a clique so $\v$ is rigid. Conversely, if $\v$ is rigid then by \cref{lemma:MCE:outward_simple}, $\v$ is located inside $\CycKmax$ for some $\Kmax\in\Kmaxs$ containing $\xd(\partF\v)$.
\end{proof}

\begin{lemma}\label{lemma:MCE:non_vertex_is_not_incident_to_active_face}
Let $\Kmax=\xd(\preKmax)\in\Kmaxs$ and $\f\in\preKmax$. 
If $\xT(\f)$ is not a vertex of $\Conv\KTmax$ then $\f$ is not incident to any \flex face of $(\GD,\xd)$.
\end{lemma}
\begin{proof}
Suppose first that $\xT(\f)$ is an \extpt of $\Kmax$. By \cref{lemma:MCE:extpt_vs_vertex}, an open neighborhood of $\xT(\f)$ is contained in the union of $\Conv\KTmax\cup\Conv\KTmaxp$ for some other clique $\Kmaxp\in\Kmaxs$. By \cref{lemma:MCE:outward_simple}, any face $\v$ incident to $\f$ is contained inside the \extcycle of $\Kmax$ or of $\Kmaxp$. Similarly, if $\xT(\f)$ is not an \extpt of $\Kmax$ then any face $\v$ incident to $\f$ lies inside the \extcycle of $\Kmax$. In either case, $\partF\v$ is contained in a clique so $\v$ is rigid. 
\end{proof}

\begin{corollary}\label{cor:MCE:active_vertex_not_mid_edge}
Let $\v$ be \aflex face and $\f\in\partF\v$. Then $\xT(\f)$ is not contained in 
$\xTint(\east)$ for any edge $\east\in\East$ of $\GD$. More generally, $\xT(\f)\notin\open[\xT(\ff_1),\xT(\ff_2)]$ for any $\{\ff_1,\ff_2\}\in\NullE(\GD,\xd)$.
\end{corollary}
\begin{proof}
Suppose otherwise that $\xT(\f)\in\open[\xT(\ff_1),\xT(\ff_2)]$ for some $\{\ff_1,\ff_2\}\in\NullE(\GD,\xd)$.
By \cref{lemma:MCE:clique_convex}, $\xd(\{\ff_1,\f,\ff_2\})$ is a degenerate clique.
Let $\Kmax\in\Kmaxs$ be a maximal clique containing $\xd(\{\ff_1,\f,\ff_2\})$. Then $\xT(\f)$ is not a vertex of $\Conv\KTmax$, so by \cref{lemma:MCE:non_vertex_is_not_incident_to_active_face}, it cannot be incident to \aflex face. %
\end{proof}

\begin{corollary}\label{lemma:MCE:active_face_embedded}
Let $\v$ be \aflex face. 
Then the restriction of $\xT$ to $\partF\v$ gives rise to a straight-edge embedding of the simple graph 
$\overline{\partGD\v}:=(\partF\v,\{\ebarast\mid\east\in\partEast\v\})$ obtained from the induced graph $\partGD\v$ (cf. \cref{dfn:MCE:deg_emb_triang_big_faces}) by identifying parallel edges.
\end{corollary}
\begin{proof}
By~\itemref{MCE1:emb}, $\xT$ is injective on $\partF\v$. 
By \cref{cor:MCE:active_vertex_not_mid_edge}, no vertex of $\xT(\partGD\v)$ is contained in the interior of an edge of $\xT(\partGD\v)$. Since $\xteh\ind[\partF\v]$ is an embedding of $\partGD\v$, in the $\eps\to0$ limit, no two edges can form an essential crossing. Thus, $\xT|_{\partF\v}$ is a straight-edge embedding of $\overline{\partGD\v}$.
\end{proof}

For $\v\in\Vact$, we denote $\partgeom\xT(\v):=\xT(\v)\cap\SkelT$ and $\xTint(\v):=\xT(\v)\setminus \partgeom\xT(\v)$. 
\begin{corollary}
\label{lemma:MCE:xTint_connected_open_avoids_cliques}
For each \flex face $\v\in\Vact$, $\xTint(\v)$ is a nonempty connected open subset of the plane. It is disjoint from $\Conv\KT$ for any clique $\clique\subset\xd(\Faces)$.
\end{corollary}
\begin{proof}
By \cref{lemma:MCE:active_face_embedded}, $\xTint(\v)$ is the interior of a face of an embedded graph $\xT(\overline{\partGD\v})$. Thus, $\xTint(\v)$ is nonempty, connected, and open. By \cref{lemma:MCE:inside_CycKmax=>rigid}, $\xTint(\v)$ is disjoint from $\Conv\KTmax$ for any maximal clique $\Kmax\in\Kmaxs$ (and therefore for any clique of size at least $2$). Since $\xTint(\v)\cap\xT(\Faces)=\emptyset$, $\xTint(\v)$ is disjoint from any clique of size $1$.
\end{proof}

For a corner $\corner$ of $\GD$ and an angle $0<\alpha<\sumT(\corner)$, let $\RmomT_{\corner,\alpha}\in\C$ be the unit complex number such that 
\begin{equation}\label{eq:MCE:Rcc_dfn}
 \arg_{[0,2\pi)}\left(\RmomT_{\corner,\alpha} / (\xT(\corfm) - \xT(\corf)) \right) = \alpha.
\end{equation}

\begin{corollary}%
\label{lemma:MCE:collared_corner_active}
Let $\v\in\Vact$ be \aflex face. For any corner $\corner$ of $\v$ and any angle $0<\alpha<\sumT(\corner)$, there exists a constant $c=c_{\corner,\alpha}>0$ such that $\open[\xT(\corf),\xT(\corf) + c\RmomT_{\corner,\alpha}]\subset\xTint(\v)$.
\end{corollary}
\begin{proof}
Follows immediately from \cref{lemma:MCE:active_face_embedded}.
\end{proof}

\begin{corollary}\label{lemma:MCE:active_face=>active}
Let $\v\in\Vact$ be \aflex face and let $\cor$ be a corner of $\v$. Then $\cor$ is \bic.
\end{corollary}
\begin{proof}
 Suppose otherwise that $\sumwT(\cor)=0$ or $\sumbT(\cor) = 0$; thus, $0\leq \sumT(\cor)\leq\pi$. We have ${(\xd(\corfm) - \xd(\corfp))^2 = 0}$ by~\eqref{eq:M_tnn_vs_angles_eq}, so $\clique:=\xd(\{\corfm,\corf,\corfp\})$ is a clique. By \cref{lemma:MCE:xTint_connected_open_avoids_cliques}, $\xTint(\v)$ is disjoint from $\Conv\KT$. By \cref{lemma:MCE:collared_corner_active}, $\xTint(\v)$ contains points of the form $\xT(\corf) + c\RmomT_{\corner,\alpha}$ for $0<\alpha<\sumT(\cor)$ and small $c>0$. Thus, $\Conv\KT$ cannot contain such points, so $\sumT(\cor)\in\{0,\pi\}$, contradicting \cref{cor:MCE:active_vertex_not_mid_edge}.
\end{proof}

\subsection{Pointed \ptrons}\label{ssec:pointed}
The goal of this section is to show that an \MCE $(\GD,\xd)$ restricts to a (weakly embedded) \emph{pointed \ptron}~\cite{RSS} of $\Conv\KTmax$ for each $\Kmax\in\Kmaxs$.

\begin{definition}\label{dfn:MCE:ptrar}
A rigid face $\v\in\Vrig$ is called \emph{\ptrar} if 
 $\partGD\v$ is a simple $d$-cycle ($d\geq3$) such that $\xT(\partGD\v)$ is a simple $d$-gon with 
exactly $3$ \emph{strictly convex} angles in $(0,\pi)$ and $d-3$ \emph{strictly reflex} angles in $(\pi,2\pi)$.
See \cref{fig:ORA-rigid} for examples. 
\end{definition}
\begin{remark}\label{rmk:MCE:angles_imply_ptrgle}
In the notation of \cref{dfn:DIM:degenerate_convex_polygon}, consider an \einj closed polygonal chain $\PcurveT$. Suppose that $\PcurveT$ has $3$ strictly convex angles ($-\pi<\turnanglex_i(\PcurveT)<0$) and $d-3$ strictly reflex angles ($0<\turnanglex_i(\PcurveT)<\pi$), and that $\TURN(\PcurveT) = -2\pi$. Then it is straightforward to check that $\PcurveT$ is an (embedded) \ptrgle.
\end{remark}

 The following basic observation, proved in~\cite[Section~2.2]{RSS}, provides a connection between \ptrgles and \MCEs. 
\begin{lemma}[\cite{RSS}]\label{lemma:MCE:ptrar_no_chords}
Suppose that $\xT(\v)$ is a rigid embedded face, with every corner either strictly convex or strictly reflex. Then $\xT(\v)$ satisfies~\itemref{MCE5:Mpos_chords} (i.e., every \pchord contained inside $\xT(\v)$ is isotopic to a boundary edge of $\xT(\v)$) if and only if $\xT(\v)$ is \ptrar.
\end{lemma}
\begin{proof}
Our notion of a \pchord inside a rigid face is equivalent to the notion of a \emph{bitangent} in~\cite{RSS}.
 Since $\xT(\v)$ is embedded, the result follows from~\cite[Theorem~2.6]{RSS}.
\end{proof}

\begin{figure}
 \def\inputscale{1.5}
 \def\inputscaledd{1.6}
 \setlength{\tabcolsep}{4pt}
\begin{tabular}{cccccc}
\includegraphics[scale=\inputscale]{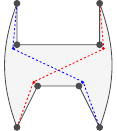}
&
\includegraphics[scale=\inputscale]{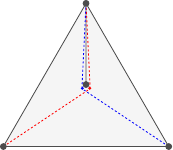}
&
\includegraphics[scale=\inputscale]{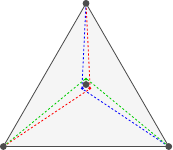}
&
\includegraphics[scale=\inputscaledd]{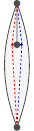}
&
\includegraphics[scale=\inputscaledd]{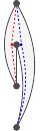}
&
\includegraphics[scale=\inputscaledd]{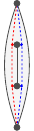}
\\
 $m=4$
& $m=4$
& $m=3$
& $m=3$
& $m=3$
& $m=2$
\end{tabular}
 \caption{\label{fig:geodesics}Geodesics inside a rigid non-\ptrar face $\v$; see the proof of \cref{lemma:MCE:ptrar_faces}.
Here, $m$ is the number of corners $\cor$ of $\v$ such that $\sumT(\cor)\in[0,\pi)$.
}
\end{figure}

We generalize \cref{lemma:MCE:ptrar_no_chords} to the case where $\xT(\v)$ is weakly embedded.
\begin{proposition}\label{lemma:MCE:ptrar_faces}
Let $(\GD,\xd)$ be an \MCE and let $\v\in\Vrig$ be a rigid face of $(\GD,\xd)$ that is not a degenerate \btrgle. Then $\v$ is \ptrar.
\end{proposition}
\begin{proof}
The argument is entirely analogous to the one presented in~\cite[Section~2.2]{RSS}. Given $\ff,\f\in\partF\v$ and an isotopy class of curves $\geps:[0,1]\to\xteh(\v)$ connecting $\geps(0)=\xteh(\ff)$ to $\geps(1)=\xteh(\f)$, a \emph{geodesic} $\gzero:[0,1]\to\xT(\v)$ 
is an $\eps\to0$ limit of curves in this isotopy class that has minimal possible length. Thus, each geodesic is a concatenation of several \addable bent line segments $(a_0=\xT(\ff),a_1,\bendhx 1),\dots,(a_{d-1},a_d=\xT(\f),\bendhx d)$. 
As explained in~\cite[Section~2.2]{RSS}, $(a_{i-1},a_i,\bendhx i)$ is a \pchord for each $1<i<d$. For $i=1$ (resp., $i=d$), $(a_{i-1},a_i,\bendhx i)$ is a \pchord if and only if it is proper at $a_0$ (resp., $a_d$). We refer to $(a_{i-1},a_i,\bendhx i)$ as the \emph{segments} of the geodesic $\gzero$.

Let $\cor_1,\cor_2,\dots,\cor_m$ be the corners of $\xT(\v)$ such that $\sumT(\cor_i)\in[0,\pi)$. Thus, every segment of a geodesic that starts and ends at one of these corners is automatically a \pchord. 
We have $m\geq2$ since the clockwise boundary $\partEvecast\v$ contains at least one cycle $\vec\Cyc$ directed clockwise, so the clockwise weakly embedded polygon $\xteh(\vec\Cyc)$ must have at least two corners $\cor_i$ satisfying $\sumT(\cor_i)\in[0,\pi)$. 
 If $m=2$ then drawing a geodesic in each isotopy class and analyzing their segments, we conclude that $\xT(\v)$ satisfies~\itemref{MCE5:Mpos_chords} if and only if $\xT(\v)$ is a degenerate \btrgle. 
Similarly to~\cite[Lemma~2.3]{RSS}, we see that if $m\geq4$ then $\xT(\v)$ admits at least two \pchords and thus violates~\itemref{MCE5:Mpos_chords}. The only remaining possibility is $m=3$. In this case,~\itemref{MCE5:Mpos_chords} is satisfied if and only if each segment of each geodesic connecting $\corfx_i$ to $\corfx_j$ (for $1\leq i,j\leq 3$) is an edge of $\xT(\GD)$. This implies that $\xT(\v)$ is a \ptrgle. See \cref{fig:geodesics}.
\end{proof}

\begin{corollary}\label{lemma:MCE:face_classification}
Let $(\GD,\xd)$ be an \MCE. Then every face of $(\GD,\xd)$ is either 
\flex, a rigid \ptrgle, or a degenerate \btrgle.
\end{corollary}
\begin{definition}\label{dfn:MCE:Vsact_Vbigons_Vtrs}
We denote by $\Vbigons$ (resp., $\Vtrs$) the set of bigonal (resp., triangular) faces of $\GD$, and we let $\Vsact:=\Vint\setminus(\Vbigons\sqcup\Vtrs)$. 
\end{definition}
\noindent By \cref{lemma:MCE:face_classification}, $\Vsact=\Vact\sqcup\Vrigsact$ consists of all \flex faces of $(\GD,\xd)$ and of rigid \ptrar faces with at least $4$ edges.
By \cref{lemma:MCE:active_face=>active,lemma:MCE:face_classification}, we obtain the following. 
\begin{corollary}\label{lemma:zero_corner=>rigid}
Let $(\GD,\xd)$ be an \MCE and let $\cor\in\corners(\GD)$. If $\sumbT(\cor)=0$ then $\corv$ is a (possibly degenerate) rigid white face, and if $\sumwT(\cor)=0$ then $\corv$ is a (possibly degenerate) rigid black face.
\end{corollary}

\section{\ORATITLE}\label{sec:ORA}
We describe the \emph{\ORA} in \cref{sec:ORA_overview} and prove its correctness in the remaining subsections. 
We will relate it to the loop BCFW recursion in \cref{sec:BCFW}. 
An interactive implementation of the algorithm can be found at~\cite{applets}.

\subsection{Overview}\label{sec:ORA_overview}

Let $(\GD,\xd)$ be an \MCE. 

\begin{definition}\label{dfn:OR_move}
An \emph{input datum} is a pair $\BCin=(\corner,\conven)$, where $\corner$ is a corner satisfying $\corv\in\Vsact$ and $\conven\in\{\leftcon,\rightcon\}$ is one of two \emph{coloring conventions}. 
The input datum $\BCin$ is called \emph{\flex} if the face $\corv$ is \flex, and \emph{rigid} if $\corv$ is rigid.
\end{definition}
\noindent 
For example, in \cref{fig:BCFW-full}, the input datum is \flex for the first three steps and rigid for the last step. 

In \cref{dfn:ORA:rigid_ORst_descr}, we introduce the notion of a \emph{valid} rigid input datum. By convention, any \flex input datum is considered valid. 
Given a valid input datum $\BCin=(\corner,\conven)$ for $(\GD,\xd)$, the associated \emph{\ORst} produces another \MCE denoted $\ORmv(\GD,\xd) = (\GDout,\xdout)$. 
We show in \cref{lemma:ORA:ORA_termination} that every face $\v\in\Vsact$ admits at least one valid input datum, and that any sequence of valid (\flex and rigid) \ORsts eventually produces a \emph{\terminal} \MCE, i.e., an \MCE with all faces \btrar. In the case of the loop BCFW recursion discussed in \cref{sec:BCFW}, every possible input datum will be automatically valid; see \cref{lemma:ORA:iccar_valid}.

We describe \flex and rigid \ORsts in more detail. 
For each (valid) input datum $\BCin$, the graph $\GDout$ is obtained from $\GD$ by adding a single vertex $\ffout$ inside $\corv$ so that $\Fout=\Faces\sqcup\{\ffout\}$. 
The new vertex $\ffout$ is connected by a single edge to each of $\corfm,\corf,\corfp$. In addition, it is connected by one or several \emph{outgoing edges} to some of the vertices in 
\begin{equation}\label{eq:Pivots_dfn}
 \Pivots(\corner) := \partF\corv \setminus \{\corfm,\corf,\corfp\}.
\end{equation}
Recall from \cref{dfn:partF} that $\Pivots(\corner)$ includes isolated vertices located inside $\corv$.

\begin{definition}[Folding ray]\label{dfn:folding_ray}
Set $(\betaT,\betaO):= (\sumwT(\corner),\sumwT(\corner))$ if $\conven=\leftcon$ and $(\betaT,\betaO):=(\sumbT(\corner),2\pi-\sumbT(\corner))$ if $\conven=\rightcon$. 
Let $\Rcc\in\Rdd$ be a null vector with $|\RccT| = |\RccO| = 1$ such that 
\begin{equation*}%
 \arg_{[0,2\pi)} \left(\RccT/(\xT(\corfm)-\xT(\corf))\right)=\betaT
 \quad\text{and}\quad
\arg_{[0,2\pi)} \left(\RccO/(\xO(\corfm)-\xO(\corf))\right)=\betaO;
\end{equation*}
cf.~\eqref{eq:MCE:Rcc_dfn}. 
 We let $\xrout:=\xdf(\corf) + \r\Rcc$ for all $\r\geq0$, and refer to the ray 
$\FRay:=\{\xrout\mid \r\geq0\}$ as 
the \emph{folding ray} associated to $\BCin=(\corner,\conven)$. 
\end{definition}
\noindent 
For $\conven=\leftcon$, we write $(\conp,\conm):=(\colB,\colW)$ and for $\conven = \rightcon$, we write $(\conp,\conm):=(\colW,\colB)$. 
Thus, the angle between $\RccT$ and $\xT(\corfpm)-\xT(\corf)$ 
(resp., between $\RccO$ and $\xO(\corfpm)-\xO(\corf)$)
is equal to $\sumcpmT(\corner)$ in absolute value.
By~\eqref{eq:M_tnn_vs_angles_eq},
\begin{equation}\label{eq:zero_type_corfpm}
 (\xrout - \xdf(\corfm))^2 = 
 (\xrout - \xdf(\corf))^2 = 
 (\xrout - \xdf(\corfp))^2 = 0
 \quad\text{for all $r\geq0$.}
\end{equation}

The location $\xdout(\ffout)$ of the new vertex $\ffout$ is determined as follows. Consider an intermediate graph $\GDr$ in which the vertex $\ffout$ is only connected to $\{\corfm,\corf,\corfp\}$. For each $r>0$, we extend $\xd:\Faces\to\Rdd$ %
 to a map
$\xdr:\Fout\to\Rdd$ with $\xdr(\ffout):=\xd(\corf) + r\Rcc$.
 We define 
\begin{equation}\label{eq:rcrit_dfn}
 \rcrit:=\inf\left\{r>0\mid \xdr\text{ is \finj but $(\GDr,\xdr)$ is not an \MCE}\right\}.
\end{equation}
Here, the \finj requirement only affects the result when $\BCin$ is rigid. 
We show in \cref{lemma:ORA:correctness_vertex} that $0<\rcrit<\infty$. We set $\xdout:=\xdrcrit$ and let $(\GDout,\xdout)$ be obtained from $\GDr$ by adding all \pchords of $(\GDr,\xdout)$ incident to $\ffout$ to the set of edges as we did in \cref{lemma:MCE:add_chords}.

For valid $\BCin$, we show in \cref{thm:ORA_output_is_MCE,lemma:ORA:rigid_MCNE=>MCNE} that $(\GDout,\xdout)=\ORmv(\GD,\xd)$ is an \MCE.
The \emph{\ORA} consists of applying an arbitrary sequence of valid \ORsts until the output \MCNE $(\GDout,\xdout)$ is \terminal.

\subsection{Bounds on \texorpdfstring{$\rcrit$}{\~{$r$}}}\label{sec:ORA_new_vert}
We start by analyzing the structure of $(\GDr,\xdr)$ for small and large $r>0$. Suppose that $\BCin$ is a (\flex or rigid) input datum such that the corner $\corner$ is \bic.
(The case of \unic $\cor$ is treated in \cref{sec:ORA_TORA}.)

\begin{lemma}\label{lemma:ORA:r_small}
 For all sufficiently small $\r>0$, the point $\xtrout$ belongs to $\xTint(\corv)$ and satisfies 
\begin{equation}\label{eq:xTrout_geq_0}
(\xrout - \xdf(\f))^2\geq0 \quad\text{for all $\f\in\Faces$.}
\end{equation}
\end{lemma}
\begin{proof}
If $\BCin$ is rigid then $\xT(\corv)$ is an (embedded) \ptrar face by \cref{lemma:MCE:ptrar_faces}. 
Since $\cor$ is \bic, the corner $\cor$ is strictly reflex by \cref{lemma:MCE:corner_M_pos}. 
Thus, $\xtrout\in\xTint(\corv)$ for small $\r>0$. 
Let $\clique:=\xd(\partF\corv)$. 
By~\eqref{eq:sumwT_sumbT_vs_sumT_sumO}, $\sumxT^{\colbarop(\clique)}(\cor) = \pi$, so 
$\{\xrout\} \cup \xd(\partF\corv)$ is a clique. 
 Thus,~\eqref{eq:xTrout_geq_0} follows from~\itemref{MCE3:M-tnn} and \cref{lemma:MCE:clique_affine_linear}.

Suppose now that $\BCin$ is \flex. 
In this case, $\xtrout$ belongs to $\xTint(\corv)$ for small $r>0$ by \cref{lemma:MCE:collared_corner_active}. 

Let $\f\in\Faces$, $\Qmom(\r):=\xdf(\f) - \xrout$, and $\Qmom:=\Qmom(0)=\xdf(\f)-\xdf(\corf)$. We have $\Qmom(\r)^2 = \Qmom^2 - 2\r\Qmom\cdot \Rcc$ since $\Rcc$ is null. Thus, $\Qmom(\r)^2$ is an affine linear function of $\r$ with $\Qmom(0)^2\geq0$ by~\itemref{MCE3:M-tnn}. If $\Qmom^2>0$ or $\Qmom\cdot \Rcc\leq0$ then we are done. Assume that we are in the remaining case $\Qmom^2=0$ and $\Qmom\cdot \Rcc>0$. 

Set $\Pmom_\pm:=\xdf(\corfpm)-\xdf(\corf)$. 
Consider the vectors $\QmomT,\PmT,\PpT,\RccT$ emanating from $\xT(\corf)$. The line segment $[\xTf(\f),\xTf(\corf)]$ cannot intersect $\xTint(\corv)$ by \cref{lemma:MCE:xTint_connected_open_avoids_cliques}. 
Thus, $\QmomT$ is not located in the sector between $\PmT$ and $\PpT$ containing $\RccT$. That is, for $\beta:=\arg_{[0,2\pi)}(\QmomT/\RccT)$, 
we have $\beta \in [\alphaT^{\conp}(\corner),2\pi-\alphaT^{\conm}(\corner)]$.
Assume for example that $\conven = \leftcon$. Since $(\xdf(\f)-\xdf(\corfp))^2\geq0$ by~\itemref{MCE3:M-tnn}, we have $|\arg(\QmomO/\PpO)| \leq |\arg(\QmomT/\PpT)|$ by~\eqref{eq:M_tnn_vs_angles}, where we denote $\arg:=\arg_{(-\pi,\pi]}$. 
Assume first that $\beta \leq \pi$, i.e., $\beta \in [\sumbT(\corner),\pi]$. Then, $|\arg(\QmomT/\PpT)| = \beta-\sumbT(\corner)$. Using $|\arg(\PpO/\RccO)| = |\arg(\PpT/\RccT)| = \sumbT(\corner)$ and the triangle inequality, we find
\begin{equation*}%
 |\arg(\QmomO/\RccO)| 
\leq |\arg(\QmomO/\PpO)| + |\arg(\PpO/\RccO)|
\leq (\beta - \sumbT(\corner)) + \sumbT(\corner) = \beta = |\arg(\QmomT/\RccT)|.
\end{equation*}
By~\eqref{eq:M_tnn_vs_angles}, $\Qmom^2 - 2\Qmom\cdot \Rcc = (\Qmom-\Rcc)^2\geq0$, contradicting the assumptions $\Qmom^2=0$, $\Qmom\cdot \Rcc >0$. 
In the case $\beta\geq\pi$, we use an analogous triangle inequality argument involving $\Pm$ instead of $\Pp$. 
\end{proof}

\begin{figure}
 \def\inputscale{1.3}
 \begin{tabular}{cccc}
\includegraphics[scale=\inputscale]{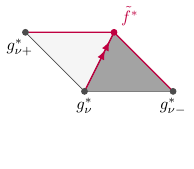}
&
\includegraphics[scale=\inputscale]{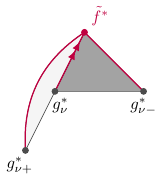}
&
\includegraphics[scale=\inputscale]{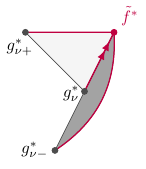}
&
\includegraphics[scale=\inputscale]{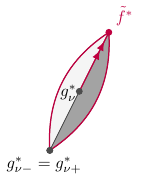}
\\
 (a) $\sumwT(\cor),\sumbT(\cor)\in(0,\pi)$ 
& (b) $\sumwT(\cor)=\pi$ 
& (c) $\sumbT(\cor)=\pi$
& (d) $\sumwT(\cor)=\sumbT(\cor)=\pi$ 
 \end{tabular} 

\vspace{-0.15in}
 \caption{\label{fig:fout-bend} Bending the edges of $\protect\GDr$.}
\end{figure}

\begin{definition}\label{dfn:GDr}
We connect each of the vertices $\corfm,\corf,\corfp$ to $\ffout$ by an edge, and denote the resulting graph $\GDr$. The \bending of each edge of $(\GDr,\xdr)$ is given as follows. The edge connecting $\xTr(\ffout)$ to $\xTr(\corf)$ is always straight. The edge connecting $\xTr(\ffout)$ to $\xTr(\corfpm)$ is straight if the angle $\sumcpmT(\corner)$ 
between $\RccT$ and $\PmomT_\pm$
 satisfies $0<\sumcpmT(\corner)<\pi$. If $\sumcpT(\cor) = \pi$ (resp., $\sumcmT(\cor) = \pi$) then we slightly bend the edge from $\xTr(\ffout)$ to $\xTr(\corfpm)$ 
 so that it avoids $\xTr(\corf)$ and the edges incident to it. See \cref{fig:fout-bend}.
\end{definition}

\begin{proposition}\label{lemma:ORA:r_small_MCE}
For all sufficiently small $r>0$, $(\GDr,\xdr)$ is an \MCE. 
\end{proposition}
\begin{proof}
We check that $(\GDr,\xdr)$ satisfies the conditions in \cref{dfn:MCE:MCE}. \itemref{MCE1:emb} follows from \cref{lemma:MCE:active_face_embedded,lemma:MCE:collared_corner_active}.
\itemref{MCE2:null_edges} follows from~\eqref{eq:zero_type_corfpm} and \itemref{MCE3:M-tnn} follows from~\eqref{eq:xTrout_geq_0}. 

To check~\itemref{MCE4:properly_colored}, we need to check~\eqref{eq:MCE:sumwT_sumbT=pi} for the vertices $\ffout,\corf,\corfp,\corfm$. 
(When some of these vertices are on the boundary of $\GD$, they are automatically proper by \cref{rmk:MCE:bdry_proper}.) 
By \cref{lemma:MCE:sumwT_sumbT_options}, for each of these vertices, it is enough to find two corners $\cor_1,\cor_2$ (with possibly $\cor_1=\cor_2$) such that $\sumwT(\cor_1),\sumbT(\cor_2)>0$. 
The two corners $\corout_+,\corout_-$ incident to $\ffout$ (contained in $\Conv\xTr(\{\ffout,\corf,\corfpm\})$) satisfy $0\leq \sumT(\corout_+),\sumT(\corout_-)<\pi$. If $\sumT(\corout_+),\sumT(\corout_-)>0$ then we are done since 
$\sumcpmT(\corout_\pm)=\sumT(\corout_\pm)$. 
 Otherwise, $\sumT(\corout_+)+\sumT(\corout_-)<\pi$, so the third corner $\corout'$ incident to $\ffout$ satisfies $\sumT(\corout')>\pi$. Therefore, $\sumwT(\corout'),\sumbT(\corout')>0$
by \cref{lemma:MCE:corner_M_pos}. 
Next, for $\corf$, we have $\sumwT(\cor),\sumbT(\cor)>0$ since $\cor$ is \bic. The corners of $\corv$ at $\corfp,\corfm$ are unambiguous 
 since $\corv$ is not triangular; cf. \cref{lemma:ambig=>triangular}. 
Thus, for small $\r>0$, the (locally constant by~\eqref{eq:sumwT_sumbT_options}) functions $\sumwT(\corfpm),\sumbT(\corfpm)$ depend continuously on $\r$.
Since~\eqref{eq:MCE:sumwT_sumbT=pi} was satisfied for $\corfp,\corfm$ in $(\GD,\xd)$, it is satisfied for $\corfp,\corfm$ in $(\GDr,\xdr)$ for small $r>0$.

We check~\itemref{MCE5:Mpos_chords}. Suppose that for some $\f\in\Faces$, $(a_r,b,\bendhr) :=(\xTr(\ffout),\xT(\f),\bendhr)$ is a \pchord in $(\GDr,\xdr)$ for all small $r>0$. Set $a:=\xT(\corf)$. 
Consider a \bent line segment $(a,b,\bendh)$ such that $[a,b]_\eps$ is a concatenation of $[a,a_{r(\eps)}]_{\eps}$ and $[a_{r(\eps)},b]_\eps$ for some function $\r(\eps)>0$ with $\r(\eps)\to0$ as $\eps\to0$.
Since $(a_r,b,\bendhr)$ is addable in $(\GDr,\xdr)$, 
 $(a,b,\bendh)$ is addable in $(\GD,\xd)$. Since $(a_r,b,\bendhr)$ was proper in $(\GDr,\xdr)$, its $r\to0$ limit $(a,b,\bendh)$ is proper in $(\GD,\xd)$. Thus, $(a,b,\bendh)$ is a \pchord in $(\GD,\xd)$. 
 Since $(\GD,\xd)$ satisfies~\itemref{MCE5:Mpos_chords}, $(a,b,\bendh)$ must be isotopic to an edge of $(\GD,\xd)$. Since $\xTr(\ffout)$ lies inside $\xTint(\corv)$ for small $\r>0$, we must have $\f\in\{\corfp,\corfm\}$. But then $(a_r,b,\bendhr)$ is isotopic to an edge of $\xTr(\GDr)$. Thus,~\itemref{MCE5:Mpos_chords} holds for $(\GDr,\xdr)$ for small $r>0$. 
\end{proof}

By \cref{lemma:ORA:r_small}, $\xTr(\ffout)\in\xTint(\corv)$ for small $\r>0$. 
Let $\rmax := \min\{\r>0 \mid \xTr(\ffout)\in\partial\xT(\corv)\}$. Thus, 
\begin{equation}\label{eq:ORA:r_large}
 \rmax>0 \quad\text{and}\quad \xtrout\in\xTint(\corv) \quad\text{for all $0<\r<\rmax$}.
\end{equation}

\begin{lemma}\label{lemma:ORA:r_large_not_MCE}
$(\GDr,\xdr)$ is not an \MCE for all $r>\rmax$.
\end{lemma}
\begin{proof}
Let $r>\rmax$ and suppose for contradiction that $(\GDr,\xdr)$ is an \MCE. 
By~\itemref{MCE1:emb}, $\open[\xTr(\corf),\xTr(\ffout)]$ cannot form an essential crossing with any edge of $\xT(\GD)$. 
Therefore, $\xTrmax(\ffout)=\xT(\f)$ for some $\f\in\Faces$. 
By~\itemref{MCE3:M-tnn} and \cref{lemma:MCE:clique_convex}, $\xdr(\{\corf,\f,\ffout\})$ is a clique. Since $[\xT(\corf),\xT(\f)]$ intersects $\xTint(\corv)$, we obtain a contradiction by \cref{lemma:MCE:xTint_connected_open_avoids_cliques} when $\BCin$ is \flex. When $\BCin$ is rigid \ptrar, $\xTr(\ffout)$ is located outside the simple polygon $\xT(\corv)$ for all $r>\rmax$, contradicting~\itemref{MCE1:emb}.
\end{proof}

\begin{corollary}\label{lemma:ORA:correctness_vertex}
The set 
$\left\{r>0\mid \xdr\text{ is \finj but $(\GDr,\xdr)$ is not an \MCE}\right\}$
 on the right-hand side of~\eqref{eq:rcrit_dfn} is nonempty and bounded away from $0$. In particular, $\rcrit\in(0,\infty)$. 
\end{corollary}
\begin{proof}
This set is nonempty (and thus $\rcrit<\infty$) by \cref{lemma:ORA:r_large_not_MCE}. It is bounded away from $0$ (and thus $\rcrit>0$) by \cref{lemma:ORA:r_small_MCE}. 
\end{proof}

\subsection{New vertex location (\flex \texorpdfstring{$\BCin$}{δ})} 
From now on, we assume that $\BCin$ is \aflex input datum. 
 
\begin{lemma}\label{lemma:ORA:r_large}
 There exists $\f\in\Pivots(\corner)$ such that $(\xdrmax(\ffout) - \xdf(\f))^2<0$. In particular, $(\GDr,\xdr)$ is not an \MCE for small $|r-\rmax|$.
\end{lemma}
\begin{proof}

Set $r:=\rmax$. Let $[\xt(\f_1),\xt(\f_2)]$ be a boundary edge of $\xT(\corv)$ containing the point $\xtrout$, and let $t\in[0,1]$ be such that $\ys:=(1-t)\xd(\f_1)+t\xd(\f_2)\in\Skel$ satisfies $\ysT = \xtrout$. Let $\Pmom:=\ys-\xdf(\corf)$; thus, $\PmomT = \rmax\RccT$. By \cref{lemma:MCE:clique_affine_linear} and~\itemref{MCE3:M-tnn}, $\Pmom^2\geq0$. 
Suppose that $\Pmom^2 = 0$. 
If $0<t<1$ (resp., $t=0$ or $t=1$) then \cref{lemma:MCE:clique_affine_linear} implies that $\{\f_1,\f_2,\corf\}$ (resp., $\{\f_1,\corf\}$ or $\{\f_2,\corf\}$) is a clique. Denote this clique by $\preK$. The line segment $[\xTf(\corf),\ysT]\subset\Conv\xT(\preK)$ intersects $\xTint(\corv)$ by \cref{lemma:ORA:r_small}, contradicting \cref{lemma:MCE:xTint_connected_open_avoids_cliques}. 
Thus, $\Pmom^2>0$, i.e., 
 $|\PmomT|>|\PmomO|$. On the other hand, since $\Rcc$ is null, $|\rmax\RccO|=|\rmax\RccT|=|\PmomT|$. It follows that $\rmax\RccO\neq\PmomO$, and since $\PmomT = \rmax\RccT$, we have $(\Pmom-\rmax\Rcc)^2<0$.

By \cref{lemma:MCE:clique_affine_linear}, the function $\afflin(t'):= ((1-t')\xd(\f_1)+t'\xd(\f_2) - \xdrmax(\ffout))^2$ is affine linear in $t'$ and is negative for $t'=t\in[0,1]$. Therefore,
 $(\xdf(\f)- \xdrmax(\ffout))^2<0$ for some $\f\in\{\f_1,\f_2\}$. We have $\f\in\partF\corv$. 
By~\eqref{eq:zero_type_corfpm}, $\f\in\partF\corv\setminus\{\corfm,\corf,\corfp\} = \Pivots(\corner)$. 
\end{proof}

\begin{corollary}\label{lemma:ORA:ffout_in_xTint_corv}
The vertex $\xTout(\ffout) = \xTrcrit(\ffout)$ is located inside $\xTint(\corv)$.
\end{corollary}

\begin{proposition}\label{lemma:ORA:GDr_MCE1235}
$(\GDr,\xdout)$ is a \pMNNE. 
\end{proposition}
\begin{proof}
By \cref{lemma:ORA:ffout_in_xTint_corv}, $\xTout(\ffout)\in\xTint(\corv)$. 
By \cref{lemma:ORA:r_large}, $\rcrit<\rmax$, so $\xTr$ is \finj for all $0<r\leq \rcrit$. 
Thus, $(\GDr,\xdr)$ is an \MCE for $0<r<\rcrit$ by~\eqref{eq:rcrit_dfn}.
In particular, no two edges of $\xTout(\GDr)$ form an essential crossing. 
 Furthermore, we claim that 
\begin{equation}\label{eq:vertices_on_two_triangles}
 \text{if $\f\in\Faces$ satisfies $\xT(\f)\in\Conv\xTout(\{\ffout,\corf,\corfpm\})$ then $\xT(\f)\in[\xT(\corf),\xT(\corfpm)]$.}
\end{equation}
 Indeed, if $\xT(\f)\in\Conv\xTout(\{\ffout,\corf,\corfpm\})$ then by \cref{lemma:MCE:clique_convex}, $\xdout(\{\f,\ffout,\corf,\corfpm\})$ is a clique. 
 If $\xT(\f)\notin [\xT(\corf),\xT(\corfpm)]$ then the line segment $[\xT(\f),\xT(\corf)]$ intersects $\xTint(\corv)$, contradicting \cref{lemma:MCE:xTint_connected_open_avoids_cliques}. The \bendings in \cref{dfn:GDr} are chosen so that the edges incident to $\xTout(\ffout)$ avoid the line segment $[\xT(\corf),\xT(\corfpm)]$. 
 Thus, $(\GDr,\xdout)$ satisfies~\itemref{MCE1:emb}. \itemref{MCE2:null_edges} is satisfied by~\eqref{eq:zero_type_corfpm}. \itemref{MCE3:M-tnn} is satisfied for $r = \rcrit$ because it is given by a closed condition satisfied for $0<r<\rcrit$. 

Since $(\GD,\xd)$ is an \MCE,~\eqref{eq:MCE:sumwT_sumbT=pi} holds for all $\f\in\Fintgr\setminus\{\corf,\corfp,\corfm\}$. 
We check that each vertex $\f\in\{\ffout,\corf,\corfp,\corfm\}$ is proper. 
By \cref{dfn:improper}, if a vertex $\f\in\{\ffout,\corf,\corfm,\corfp\}$ is incident to an ambiguous corner of $(\GDr,\xdout)$ then $\f$ is proper. Otherwise, the functions $\sumwT(\f),\sumbT(\f)$ defined in~\eqref{eq:sumwT_sumbT_of_f_dfn} depend continuously on the geometry of $\xdr(\Fout)$ for $\r$ in a neighborhood of $\rcrit$, 
 so the vertex $\f$ is proper in $(\GDr,\xdout)$ since it was proper in $(\GDr,\xdr)$ for $0<r<\rcrit$. 
\end{proof}

\subsection{Creating new edges (\flex \texorpdfstring{$\BCin$}{δ})}\label{ssec:ORA:new_edges}
We continue to assume that $\BCin$ is \aflex input datum. 
In view of \cref{lemma:MCE:add_chords,lemma:ORA:GDr_MCE1235}, our next goal is to add some \pchords 
to $(\GDr,\xdout)$ and choose a coloring of ambiguous corners so that~\axrangeMCElasttwo would be satisfied for the resulting graph. 

Let $\GDo$ be the subgraph of $\GDr$ obtained by deleting the three edges incident to $\ffout$. Thus, $\ffout$ is an isolated vertex of $\GDo$, and $(\GDo,\xdout)$ 
is a \pMNNE
 by \cref{lemma:MCE:Res,lemma:ORA:GDr_MCE1235}. 
For the rest of this subsection, the term \emph{\pchord} refers to \pchords of $(\GDo,\xdout)$ incident to $\xTout(\ffout)$. We denote the set of isotopy classes of such \pchords by $\Chordsiso$. We aim to choose a collection $\Chords$ of pairwise non-crossing representative chords, one per each isotopy class in $\Chordsiso$. The output graph $\GDout$ will be obtained from $\GDo$ by adding all chords in $\Chords$ to the set of edges.

\begin{definition}\label{dfn:ORA:fmin_fmax}
For a ray $\Ray\subset\C$ originating at $\xTout(\ffout)$, denote $\Rayopen:=\Ray\setminus\{\xTout(\ffout)\}$ and
\begin{equation*}%
\fall(\Ray):=\{\f\in\Faces\mid \text{$\xTout(\f)\in\Rayopen$ and $(\xdout(\ffout) - \xdout(\f))^2 = 0$}\}.
\end{equation*}
 Let $\Raysffout$ be the set of rays $\Ray$ originating at $\xTout(\ffout)$ and satisfying $\fall(\Ray)\neq\emptyset$. 
For $\Ray\in\Raysffout$, we denote by $\bmin(\Ray)$ (resp., $\bmax(\Ray)$) the point in $\xTout(\fall(\Ray))$ that is closest to (resp., farthest from) $\xTout(\ffout)$.
\end{definition}

\begin{remark}\label{rmk:Ray_fall}
 By \cref{lemma:MCE:clique_convex}, the vertices in 
$\preKray:=\fall(\Ray)\sqcup\{\ffout\}$
 form a clique. Let $\Kray:=\xdout(\preKray)$. Then for any $\f\in\Faces$ satisfying 
$\xTout(\f)\in\Conv\KTray=[\xTout(\ffout),\bmax(\Ray)]$,
 we have $\f\in\fall(\Ray)$.
\end{remark}

\begin{definition}\label{dfn:Kbigs}
Let $\Kbigs$ be the set of maximal by inclusion cliques $\Kbig\subset\xdout(\Fout)$ containing $\xdout(\ffout)$. For $\Kbig\in\Kbigs$, we denote $\Ksm:=\Kbig\setminus\{\xdout(\ffout)\}$. 
\end{definition}

\begin{lemma}\label{rmk:ffout_is_a_vertex_of_KTbig}
For each $\Kbig\in\Kbigs$, $\xTout(\ffout)$ is a vertex of $\Conv\KTbig$.
\end{lemma}
\begin{proof}
Indeed, otherwise we would have $\Conv\KTbig = \Conv\KTsm$, and since $\Ksm\subset\xd(\Faces)$ satisfies $\xTout(\ffout)\in\Conv\KTsm\cap\xTint(\corv)$ by \cref{lemma:ORA:ffout_in_xTint_corv}, we get a contradiction with \cref{lemma:MCE:xTint_connected_open_avoids_cliques}.
\end{proof} 

We let $\Kbig_\pm\in\Kbigs$ be a maximal by inclusion clique containing 
the clique $\xdout(\{\ffout,\corf,\corfpm\})$; 
 cf.~\eqref{eq:zero_type_corfpm}. (We may have $\Kbig_- = \Kbig_+$; see \figref{fig:fout-bend}(b--d).) 
In particular, the clique $\{\xdout(\ffout)\}$ is not maximal, so any maximal clique $\Kbig\in\Kbigs$ contains at least two vertices and satisfies $\dim\Conv\KTbig \in\{1,2\}$. 

\begin{lemma}\label{lemma:ORA:Ksm_contained_in_clique}
Let $\Kbig\in\Kbigs$ be such that $\dim\Conv\KTbig = 2$. Then $\Ksm$ is contained in a unique maximal clique $\Ksm_2\in\Kmaxs$. 
\end{lemma}
\begin{proof}
We have $\dim\Conv\KTsm\geq1$. If $\Ksm$ is contained in two different maximal cliques $\Ksm_2,\Ksm_3\in\Kmaxs$ then $\dim\Conv\KTsm = 1$ and the union $\Conv\KTsm_2\cup\Conv\KTsm_3$ contains an open neighborhood of $\Convrelint\KTsm$. Since $\dim\Conv\KTbig = 2$, 
 $\Conv\KTbig$ must intersect this open neighborhood, so it intersects either $\Convint\KTsm_2$ or $\Convint\KTsm_3$. By \cref{lemma:MCE:clique_union}, the clique $\Kbig$ is therefore not maximal, a contradiction. 
\end{proof}

\begin{definition}
For $\Kbig\in\Kbigs$, we denote by $\Rext_+(\Kbig)$ (resp., $\Rext_-(\Kbig)$) the \emph{left} (resp., \emph{right}) \emph{$\Kbig$-external ray}, i.e., the unique ray $\Ray\in\Raysffout$ such that $\Rayopen$ intersects $\Conv\KTbig$ and lies weakly to the left (resp., to the right) of $\Conv\KTbig$ when viewed from $\xTout(\ffout)$. A ray $\Ray\in\Raysffout$ is called \emph{external} if it is left or right $\Kbig$-external for some $\Kbig\in\Kbigs$; otherwise, it is called \emph{internal}.
\end{definition}

\begin{lemma}\label{lemma:MCE:fmin_fmax_unique}
Let $\Ray\in\Raysffout$ be a $\Kbig$-external ray for some $\Kbig\in\Kbigs$. Then the preimage of $\bmin(\Ray)$ (resp., $\bmax(\Ray)$) under $\xTout$ consists of a single vertex of $\GD$ denoted $\fmin(\Ray)$ (resp., $\fmax(\Ray)$).
\end{lemma}
\begin{proof}
Since $\bmax(\Ray)$ is a vertex of $\Conv\KTbig$, it has a unique preimage by \cref{lemma:MCE:Conv_vertices_injective}. Let $\bmin'(\Ray)$ be the point in $\Ray\cap\partgeom\xT(\corv)$ closest to $\xTout(\ffout)$. If $\bmin'(\Ray)\neq\bmin(\Ray)$ then $\bmin'(\Ray)$ belongs to the relative interior of some edge $\xT(\east)$ of $\xT(\GD)$ not contained in $\Ray$. By \cref{lemma:MCE:clique_union}, $\preKray\cup\ebarast$ is a clique, so $\ebarast\subset\Kbig$ by maximality of $\Kbig$. Since $\xT(\east)\not\subset\Ray$, $\Ray$ is not $\Kbig$-external, a contradiction. Thus, $\bmin'(\Ray)=\bmin(\Ray)$. Since the vertex $\bmin(\Ray)$ is incident to $\xT(\corv)$, the preimage of $\bmin(\Ray)$ is also unique by \cref{lemma:MCE:non_inj_edge,cor:MCE:active_vertex_not_mid_edge}.
\end{proof}

\def\minusone{-1}
\begin{definition}\label{dfn:ORA:pmspec}
Let $\Kbig\in\Kbigs$ and let $\partFextout\Kbig = (\f_1,\f_2,\dots,\f_m=\f_0)$ with $\f_0=\f_m=\ffout$. Set $\f_{-1}:=\f_{m-1}$. 
For $\opm\in\{\oplus,\ominus\}$, we say that $\Kbig$ is \emph{\pmspec} if $\f_{\pm1} \neq \fmin(\Rext_\pm(\Kbig))$. In this case, the \extpchord (cf. \cref{lemma:MCE:extpchord}) of $\Kbig$ connecting $\xTout(\ffout)$ to $\xTout(\f_{\pm1})$ is also called \emph{\pmspec}.
\end{definition}

We are now ready to describe the collection $\Chords$ of \pchords.
\begin{definition}\label{dfn:ORA:Chords}
The set $\Chords$ consists of straight line segments $[\xTout(\ffout),\bmin(\Ray)]$ for all external rays $\Ray$, together with all \pandmspec chords.
\end{definition}

\begin{figure}
 \def\inputscale{1.2}
 \setlength{\tabcolsep}{2pt}
\begin{tabular}{ccc} \includegraphics[scale=\inputscale]{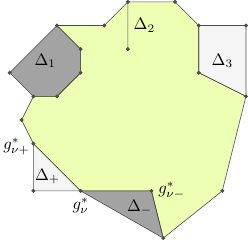}
&
\includegraphics[scale=\inputscale]{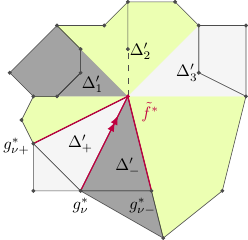}
&
\includegraphics[scale=\inputscale]{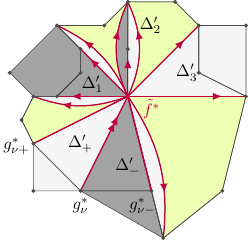}
\\
$(\GD,\xd)$ & $(\GDr,\xdout)$ & $(\GDout,\xdout)$
\end{tabular}
 \caption{\label{fig:fout-cliques} New edges created during an \ORst (\cref{dfn:ORA:Chords}).}
\end{figure}

For example, in \figref{fig:fout-cliques}(middle), the cliques $\Kbig_1,\Kbig_2$ are both \pspec and \mspec, $\Kbig_3,\Kbig_+$ are neither \pspec nor \mspec, and $\Kbig_-$ is \pspec but not \mspec. 

\begin{lemma}\label{lemma:ORA:spec}
Let $\opm\in\{\oplus,\ominus\}$ and suppose that $\Kbig\in\Kbigs$ is \pmspec. 
Denote $\Ray:=\Rext_\pm(\Kbig)$, $\fmax := \fmax(\Ray)$, and $\fmin:=\fmin(\Ray)$. Then, in the notation of \cref{dfn:ORA:pmspec}, 
\begin{enumerate}[label=(\arabic*)]
\item\label{algo:spec2} $\Ksm\in\Kmaxs$, where $\Ksm:=\Kbig\setminus\{\xdout(\ffout)\}$ as in \cref{dfn:Kbigs},
\item\label{algo:spec3} $\xT(\fmax),\xT(\fmin)$ are adjacent vertices of the polygon $\Conv\KTsm$, 
\item\label{algo:spec1} $\f_{\pm1} = \fmax \neq\fmin$,
\item\label{algo:spec4} $\CycKsm$ contains a clockwise edge connecting $\fmin$ to $\fmax$ (resp., $\fmax$ to $\fmin$) if $\opm=\oplus$ (resp., $\opm=\ominus$),
\item\label{algo:spec5} if $\dim\Conv\KTbig=1$ then $\Kbig$ is both \pspec and \mspec. 
\end{enumerate} 
\end{lemma}
\begin{proof}
Denote $\bmax:=\xT(\fmax)$ and $\bmin:=\xT(\fmin)$.
By construction, $\f_{\pm1}\in\fall(\Ray)$. Since $\f_{\pm1}\neq\fmin$, we must have $\fmax\neq\fmin$ by \cref{rmk:Ray_fall}. Since $\Kbig$ contains $\xdout(\ffout)$ and $\xd(\fmin)$ and is maximal, it must contain the entire clique $\Kray$ by \cref{lemma:MCE:clique_union}. Thus, $\xd(\fmin),\xd(\fmax)\in\Ksm$. 
Suppose that $\Ksm\notin\Kmaxs$, and let $\Ksm_2\in\Kmaxs$ be such that $\Ksm\subsetneq\Ksm_2$.
 Then $\Conv\KT_2$ intersects $\Convrelint\KTray$, so by \cref{lemma:MCE:clique_union}, $\Kray\cup\Ksm_2$ is a clique of $(\GDo,\xdout)$ containing $\Kbig$, contradicting the maximality of $\Kbig$. This shows part~\itemref{algo:spec2}. Part~\itemref{algo:spec3} follows from \cref{rmk:Ray_fall}.

Suppose that $\dim\Conv\KTbig=1$. Then $\Conv\KTbig = [\xTout(\ffout),\bmax]$ and $\Conv\KTsm = [\bmin,\bmax]$, so parts~\itemref{algo:spec1}, \itemref{algo:spec4}, \itemref{algo:spec5} follow from \cref{rmk:MCE:clique_boundary_outward_bent}. 

Suppose now that $\dim\Conv\KTbig=2$. 
 If $\Conv\KTsm$ contains an \extpt $\xT(\f)\in\open[\bmin,\bmax]$ then by \cref{lemma:MCE:extpt_vs_vertex}, there exists $\clique_2\in\Kmaxs$ such that $\clique_2\neq\Ksm$, $\dim\Conv\KT_2=2$, and such that $[\bmin,\bmax]$ is a common edge of $\Conv\KTsm$ and $\Conv\KT_2$. The \pchord connecting $\xT(\f_{\pm1})$ to $\xTout(\ffout)$ cannot intersect the \extcycles of $\Conv\KTsm$ and $\Conv\KT_2$, so we must have $\f_{\pm1} = \fmin$, a contradiction. Thus, none of the \extpts of $\Conv\KTsm$ belong to $\open[\bmin,\bmax]$. In particular, $\CycKsm$ contains a clockwise edge connecting $\fmin$ to $\fmax$ (resp., $\fmax$ to $\fmin$), which shows part~\itemref{algo:spec4}. The \pchord connecting $\xT(\f_{\pm1})$ to $\xTout(\ffout)$ cannot intersect $\CycKsm$, so $\f_{\pm1} \in\{\fmin, \fmax\}$, and since $\f_{\pm1}\neq\fmin$, we have $\f_{\pm1} = \fmax$, finishing the proof of part~\itemref{algo:spec1}. 
\end{proof}

\begin{proposition}\label{lemma:ORA:ORA_Cext_chords}
$\Chords$ consists of \pchords of $(\GDo,\xdout)$. They are pairwise non-crossing in their relative interiors, and
each isotopy class in $\Chordsiso$ contains a unique representative \pchord in $\Chords$.
\end{proposition}
\begin{proof}
 Let $(a,b,\bendh)\in\Chords$ with $a=\xTout(\ffout)$ and $b=\xT(\f)$ for some $\f\in\Faces$, and let $\Ray\in\Raysffout$ be the ray containing $[a,b]$. Since $(a,b,\bendh)\in\Chords$, the ray $\Ray$ is $\Kbig$-external for some $\Kbig\in\Kbigs$. Assume first that $(a,b,\bendh)=[a,b]$ is a straight line segment (with $\f=\fmin(\Ray)$). By \cref{dfn:ORA:fmin_fmax,rmk:Ray_fall}, $\xT(\Faces)\cap\open[a,b]=\emptyset$. 
By \cref{lemma:MCE:xTint_connected_open_avoids_cliques,lemma:ORA:ffout_in_xTint_corv}, 
 $\xTout(\ffout)\notin \xT(\east)$ for any $\east\in\East$. 
Furthermore, if $\xTint(\east)$ and $\open[a,b]$ form an essential crossing then similarly to \cref{lemma:MCE:fmin_fmax_unique}, we see that 
 $\Ray$ is not $\Kbig$-external, a contradiction. Thus, $(a,b,\bendh)$ is addable. 
Since $\ffout$ is isolated in $\GDo$, $(a,b,\bendh)$ is proper at $a$. 
Since $b$ lies on the boundary of $\Conv\KTbig$, it cannot belong to the interior of $\Conv\KTbig_2$ for any clique $\Kbig_2\subset\xdout(\Fout)$, so by \cref{lemma:MCE:improper_char}, 
$(a,b,\bendh)$ is proper at $b$. 
 Thus, $(a,b,\bendh)$ is a \pchord of $(\GDo,\xdout)$. Next, assume that $(a,b,\bendh)$ is \pmspec. Then it is a \pchord of $(\GDo,\xdout)$ by \cref{lemma:MCE:extpchord} (which applies to $(\GDo,\xdout)$ by \cref{lemma:ORA:GDr_MCE1235}). Thus, $\Chords$ consists of \pchords of $(\GDo,\xdout)$. 

We show that the \pchords in $\Chords$ are pairwise non-crossing. Indeed, two \pchords $(a,b,\bendh)$ and $(a,b',\bendh')$ in $\Chords$ can possibly intersect only when they are contained in the same ray $\Ray$. But the only \pchords contained in $\Ray$ are the straight line segment $[a,\bmin(\Ray)]$ and one or two \pchords connecting $a$ to $\bmax(\Ray)$. By construction, these \pchords do not intersect, i.e., $[a,b]_{\eps}\cap[a,b']_{\eps}=\{a\}$ for small $\eps>0$. 

Finally, let $(a,b,\bendh)$ be any \pchord of $(\GDo,\xdout)$ with $a=\xTout(\ffout)$ and $b=\xT(\f)$ for some $\f\in\Faces$, and let $\Ray\in\Raysffout$ be the ray containing $[a,b]$. 
 Let $\Kbig\in\Kbigs$ be a maximal clique containing $\{\xdout(\ffout),\xd(\f)\}$. 

Suppose first that $\Ray$ is internal. Since $\Ray$ is not $\Kbig$-external, it intersects $\Convint\KTbig$, i.e., it is located strictly between the rays $\Rext_{\pm}(\Kbig)$ when viewed from $\xTout(\ffout)$. 
Thus, $\dim\Conv\KTbig = 2$. Let $\Ksm_2\in\Kmaxs$ be the unique (by \cref{lemma:ORA:Ksm_contained_in_clique}) maximal clique containing $\Ksm$. 
By \cref{lemma:MCE:non_vertex_is_not_incident_to_active_face}, $b$ is a vertex of $\Conv\KTsm_2$. Since the ray $\Ray$ containing it is located strictly between the rays $\Rext_{\pm}(\Kbig)$, it follows that the corner $\cor_b$ of $\GDo$ at $b$ containing the \pchord $(a,b,\bendh)$ satisfies $\pi< \sumT(\cor_b)< 2\pi$, 
and the \pchord $(a,b,\bendh)$ splits $\cor_b$ into two new corners $\cor'_b,\cor''_b$ 
satisfying $0<\sumT(\cor'_b),\sumT(\cor''_b)<\pi$. 
 Thus, $(a,b,\bendh)$ is not proper at $b$, a contradiction.

Suppose now that $\Ray$ is external (and therefore $\Kbig$-external). 
We claim that $\Chords$ contains a \pchord $(a,b,\bendh')$ isotopic to $(a,b,\bendh)$. If $b = \bmin(\Ray)$ then clearly $(a,b,\bendh)$ is isotopic to the straight line segment $[a,b]$. Otherwise, $b\in(\bmin(\Ray),\bmax(\Ray)]$ by \cref{rmk:Ray_fall}, so $\fmax(\Ray) \neq \fmin(\Ray)$. It follows that $\Kbig$ is \pormspec and $\f$ is an \extpt of $\Conv\KTbig$. 
Since $(a,b,\bendh)$ cannot intersect the edges of $\CycKsm$, by parts~\itemref{algo:spec1}--\itemref{algo:spec4} of \cref{lemma:ORA:spec}, we have 
 $\f=\fmax(\Ray)$ and the \pchord $(a,b,\bendh)$ is isotopic to the corresponding \extpchord of $\Kbig$ connecting $a$ to $b$.
\end{proof}

\begin{definition}%
\label{dfn:GDout}
Let $(\GDout,\xdout)$ be obtained from $(\GDo,\xdout)$ by adding all \pchords in $\Chords$ to the set of edges.
The \bending of the edges of $\xTout(\GDout)$ is described in \cref{dfn:ORA:pmspec} and \cref{lemma:MCE:extpchord}.
\end{definition}

\begin{corollary}\label{lemma:ORA_output_is_MNE}
$(\GDout,\xdout)$ satisfies~\axrangeMCEchord.
\end{corollary}
\begin{proof}
 By \cref{lemma:ORA:ffout_in_xTint_corv}, the point $a:=\xTout(\ffout)$ belongs to $\xTint(\corv)$. By \cref{lemma:ORA:ORA_Cext_chords}, $\Chords$ consists of \pchords that are pairwise non-crossing. 
 Thus, by \cref{lemma:ORA:GDr_MCE1235}, conditions~\itemref{MCE1:emb}--\itemref{MCE3:M-tnn} are satisfied. By \cref{lemma:ORA:ORA_Cext_chords}, each \pchord in $(\GDo,\xdout)$ incident to $\xTout(\ffout)$ is isotopic to an edge of $\xTout(\GDout)$; thus, $(\GDout,\xdout)$ satisfies~\itemref{MCE5:Mpos_chords}. 
\end{proof}

To construct a proper coloring of $(\GDout,\xdout)$ and verify~\itemref{MCE4:properly_colored}, we describe the rigid faces of $(\GDout,\xdout)$.

\begin{corollary} \label{lemma:ORA:ORA_rigid_faces_descr} 
The set $\Vrigout(\ffout)$ of rigid faces of $(\GDout,\xdout)$ incident to $\ffout$ is described as follows.
\begin{enumerate}[label=(\arabic*)]
\item\label{algo_face1} For each \pspec (resp., \mspec) clique $\Kbig\in\Kbigs$, $\Vrigout(\ffout)$ contains a degenerate triangular face with vertices $\ffout,\fmax(\Ray),\fmin(\Ray)$ (resp., $\ffout,\fmin(\Ray),\fmax(\Ray)$) listed in clockwise order, where $\Ray = \Rext_{+}(\Kbig)$ (resp., $\Ray = \Rext_{-}(\Kbig)$).
\item\label{algo_face2} For each $\Kbig\in\Kbigs$ with $\dim\Conv\KTbig = 2$, let $\f_\pm:=\fmin(\Rext_\pm(\Kbig))$ and let $\Ksm_2\in\Kmaxs$ be such that $\Ksm\subset\Ksm_2$; cf. \cref{lemma:ORA:Ksm_contained_in_clique}. 
Then $\Vrigout(\ffout)$ contains a rigid \ptrar face whose clockwise boundary consists of the straight line segments $[\xT(\f_-),\xTout(\ffout)]$, $[\xTout(\ffout),\xT(\f_+)]$ followed by the part of the counterclockwise \extcycle of $\Conv\Ksm_2$ connecting $\f_+$ to $\f_-$.
\end{enumerate}
Rigid faces of type~\itemref{algo_face1} are called \emph{external}, and rigid faces of type~\itemref{algo_face2} are called \emph{internal}.
\end{corollary}
\noindent See \figref{fig:fout-cliques}(right) for an example. 
\begin{proof}
Clearly, $\Vrigout(\ffout)$ contains all rigid faces of type~\itemref{algo_face1} and~\itemref{algo_face2}. Conversely, let $\vout\in\Vrigout(\ffout)$, and let $\Kbig\in\Kbigs$ be a maximal clique containing $\xdout(\partFout\vout)$. If the boundary of $\vout$ contains a \pmspec \pchord contained in $\Rext_\pm(\Kbig)$ then $\vout$ is of type~\itemref{algo_face1}. Otherwise, the only edges of $\vout$ incident to $\ffout$ are straight line segments contained in $\Conv\KTbig$, so $\vout$ is of type~\itemref{algo_face2}.
\end{proof}

\begin{definition}\label{dfn:ORA:coloring}
The coloring of $(\GDout,\xdout)$ is described as follows. Let $\Kbig\in\Kbigs$ be \pmspec. If $\dim\Conv\KTbig = 2$ then we color the ambiguous corner of the corresponding degenerate external triangle 
 in the color opposite to that of $\Kbig$. If $\dim\Conv\KTbig = 1$ then it is both \pspec and \mspec by \crefi{lemma:ORA:spec}{algo:spec5}. In this case, the vertex $\fmin(\Ray)$ (where $\Ray\in\Raysffout$ is the ray containing $\Conv\KTbig$) is doubly ambiguous, and we choose an arbitrary (black-white or white-black) coloring of the two ambiguous corners incident to $\fmin(\Ray)$. For example, we made such a choice for the clique denoted $\Kbig_2$ in \cref{fig:fout-cliques}.
\end{definition}

\begin{proposition}\label{thm:ORA_output_is_MCE}
$(\GDout,\xdout)$ is an \MCE.
\end{proposition}
\begin{proof}

 By \cref{lemma:ORA_output_is_MNE}, $(\GDout,\xdout)$ satisfies~\axrangeMCEchord. 
It remains to check that~\eqref{eq:MCE:sumwT_sumbT=pi} holds for each non-isolated $\f\in\Fintout$; it suffices to check it for $\f\in\{\ffout\}\sqcup\Neigh_{\GDout}(\ffout)$.

We start with $\f = \ffout$. 
By \cref{lemma:MCE:xTint_connected_open_avoids_cliques,lemma:ORA:ffout_in_xTint_corv}, $\xTout(\ffout)\notin\Conv\KTmax_2$ for any clique $\Kmax_2\subset\xd(\Faces)$. 
 Thus, $\xTout(\ffout)$ is not incident to any ambiguous corners and is not contained in the interior of any clique of $(\GDout,\xdout)$. By \cref{lemma:MCE:improper_char},~\eqref{eq:MCE:sumwT_sumbT=pi} is satisfied for $\ffout$.

Suppose now that $\f\in\Neigh_{\GDout}(\ffout)$. Set $a:=\xTout(\ffout)$, $b:=\xT(\f)$, and let $\Ray\in\Raysffout$ be the (necessarily external) ray containing $[a,b]$. If $\f\neq\fmin(\Ray)$ then any \pchord in $\Chords$ connecting $a$ to $b$ splits a corner $\cor_b$ at $b$ into two corners $\cor'_b,\cor''_b$ such that one of $\sumT(\cor'_b),\sumT(\cor''_b)$ is zero. Thus,~\eqref{eq:MCE:sumwT_sumbT=pi} holds for $\f$. Suppose now that $\f=\fmin(\Ray)$. 
Then $a$ and $b$ are connected by a single (straight) edge in $\xTout(\GDout)$, and by \cref{lemma:ORA:ORA_Cext_chords},
the \pchord $[a,b]$ is proper at $b$. 
 In the case where $b$ is a vertex of one or two external degenerate triangles of $(\GDout,\xdout)$, the coloring choice in \cref{dfn:ORA:coloring} ensures that~\eqref{eq:MCE:sumwT_sumbT=pi} is satisfied for $\f$.
\end{proof}

\subsection{\texorpdfstring{Comparing $\GD$ and $\GDout$ (\flex $\BCin$)}{Comparing Γ\textsuperscript{*} and \~{Γ}\textsuperscript{*} (\flex δ)}}

We continue to assume that $\BCin$ is \aflex input datum. 
Recall that $\GDr$ contains edges connecting $\ffout$ to $\corfm,\corf,\corfp$.
 Let $\Raym,\Rayo,\Rayp\in\Raysffout$ be the rays 
originating at $\xTout(\ffout)$ and
 passing through $\xT(\corfm),\xT(\corf),\xT(\corfp)$, respectively. 

\begin{corollary}\label{lemma:ORA:GDout_contains_GDr}
The graph $\GDout$ contains all edges of $\GDr$, and each ray $\Raym,\Rayo,\Rayp$ is external.
\end{corollary}
\begin{proof}
Let $a:=\xTout(\ffout)$, $b:=\xT(\corf)$, and $b_\pm:=\xT(\corfpm)$. Let $(a,b,\bendh)$ and $(a,b_\pm,\bendhpm)$ be the three edges of $\xTout(\GDr)$ incident to $\ffout$ (cf. \cref{dfn:GDr}). By \cref{lemma:ORA:GDr_MCE1235}, each of them is a \pchord in $(\GDo,\xdout)$. 
 By \cref{lemma:ORA:ORA_Cext_chords}, each of them is isotopic to some edge of $(\GDout,\xdout)$. By \cref{dfn:ORA:Chords}, the ray in $\Raysffout$ containing each such edge is external.
\end{proof}

\begin{lemma}\label{lemma:ORA:vr_active}
Let $\vr$ be the face of $\GDr$ whose boundary $\partFout\vr$ contains $(\{\ffout\}\sqcup\partF\corv)\setminus\{\corf\}$. Then $\vr$ is \aflex face of $(\GDr,\xdout)$.
\end{lemma}
\begin{proof}
Suppose otherwise that $\vr$ is rigid; thus, 
$\Kbig_1:=\xdout(\partFout\vr)$
 is a clique. Since $\xdout(\{\ffout,\corfm,\corfp\})\subset\Kbig_1$ and since $(\xd(\corf)-\xdout(\f))^2=0$ for $\f\in\{\ffout,\corfm,\corfp\}$ by construction, 
 $\Kbig_2:=\xdout(\{\ffout,\corfm,\corf,\corfp\})$ is a clique. 
However, the face $\corv$ is \flex in $(\GD,\xd)$, so $\Kbig_1\cup\Kbig_2$ cannot be a clique as it contains $\xd(\partF\corv)$. 
By \cref{rmk:ffout_is_a_vertex_of_KTbig}, $\xTout(\ffout)$ is a vertex of $\Conv\KTbig_1$ and of $\Conv\KTbig_2$. But the sum of angles of $\Conv\KTbig_1$ and of $\Conv\KTbig_2$ at $\xTout(\ffout)$ is $2\pi$, a contradiction. 
\end{proof}

\begin{proposition}\label{lemma:ORA:new_edge_at_least_one}
The graph $\GDout$ contains at least one \emph{outgoing edge}, i.e., an edge not present in $\GDr$. All outgoing edges connect $\ffout$ to vertices in $\Pivots(\cor)$. 
\end{proposition}
\begin{proof}
 Let $\Chordsr$ consist of the three \pchords in $\Chords$ which are isotopic to the edges of $\xTout(\GDr)$, and let $\Chordsout:=\Chords\setminus\Chordsr$ be the set of outgoing edges. 
By~\eqref{eq:vertices_on_two_triangles}, 
 all \pchords in $\Chordsout$ are located inside the face $\xTout(\vr)$ of $\xTout(\GDr)$ which is \flex by \cref{lemma:ORA:vr_active}.

Suppose for contradiction that $\Chordsout=\emptyset$. 
Then by \cref{thm:ORA_output_is_MCE}, $(\GDr,\xdout)=(\GDout,\xdout)$ is an \MCE.
By \cref{lemma:MCE:active_face=>active}, all corners of $\vr$ are unambiguous, so for $\f\in\{\ffout,\corfp,\corfm\}$, the angle sums $\sumwT(\f),\sumbT(\f)$ depend continuously on $r$ in an open neighborhood of $\rcrit$. It follows that $(\GDr,\xdr)$ satisfies~\itemref{MCE4:properly_colored} for small $r>\rcrit$.
\itemref{MCE2:null_edges} is satisfied for $(\GDr,\xdr)$ for all $r>0$ by~\eqref{eq:zero_type_corfpm}. 
By~\eqref{eq:vertices_on_two_triangles}, 
 the edges of $\xTout(\GDr)$ incident to $\xTout(\ffout)$ contain no vertices in $\xTout(\Faces)$ in their relative interiors. 
 Thus, $(\GDr,\xdr)$ satisfies~\itemref{MCE1:emb} for small $r-\rcrit>0$.
If~\itemref{MCE3:M-tnn} holds but~\itemref{MCE5:Mpos_chords} is violated for $(\GDr,\xdr)$ for small $\r-\rcrit>0$ then we can add several \pchords (at least one) to the edge set of $(\GDr,\xdr)$ (cf. \cref{lemma:MCE:add_chords}) to obtain an \MCE. 
By \cref{lemma:limit_of_pcMNE_is_pMNE}, in the $\r\to\rcrit$ limit, the added \pchords become \pchords of $(\GDr,\xdout)$, violating~\itemref{MCE5:Mpos_chords} for $(\GDr,\xdout)$, a contradiction.
Thus, for small $r-\rcrit>0$, either $(\GDr,\xdr)$ is an \MCE or it violates~\itemref{MCE3:M-tnn}.

By~\eqref{eq:rcrit_dfn}, for any $\eps>0$, there exists $r$ satisfying $\rcrit<r<\rcrit+\eps$ such that $(\GDr,\xdr)$ is not an \MCE. As we showed above, this implies that~\itemref{MCE3:M-tnn} is violated for $(\GDr,\xdr)$. 
 Thus, there is a vertex $\f\in\Faces$ such that $(\xdr(\ffout) - \xdr(\f))^2<0$ for $\rcrit<r<\rcrit+\eps$. Recall that $(\xdr(\ffout) - \xdr(\f))^2\geq0$ for $0<r\leq \rcrit$. Since the function $(\xdr(\ffout) - \xdr(\f))^2$ is affine linear in $\r$
by \cref{lemma:MCE:clique_affine_linear}, 
 it follows that 
$(\xdr(\ffout) - \xdr(\f))^2 = c_{\f} (\rcrit-\r)$ for some $c_{\f}>0$ and all $\r>0$.
 In particular, $(\xd(\corf) - \xd(\f))^2>0$, so by \cref{lemma:MCE:clique_union}, the open line segment $\open[\xTout(\f),\xTout(\ffout)]$ cannot intersect $\Conv\xTout(\{\ffout,\corf,\corfpm\})$. By \cref{lemma:MCE:xTint_connected_open_avoids_cliques,lemma:ORA:vr_active}, $\open[\xTout(\f),\xTout(\ffout)]$ cannot intersect $\xToutint(\vr)$. Thus, $\open[\xTout(\f),\xTout(\ffout)]$ cannot intersect an open neighborhood of $\xTout(\ffout)$, a contradiction. 

We have shown that $\Chordsout\neq\emptyset$. Let $(a,b,\bendh)\in\Chordsout$ with $a=\xTout(\ffout)$ and $b=\xTout(\f)$ for some $\f\in\Faces$. We need to show that $\f\in\Pivots(\cor)$. 
Suppose otherwise that $\f\in\{\corfm,\corf,\corfp\}$. 
 In order for $(a,b,\bendh)$ to not be isotopic to an edge of $(\GDr,\xdout)$ connecting $a$ to $b$, there must be a vertex $\xT(\ff)\in[a,b]$ such that $\xteh(\ff)$ is 
 located between these two \pchords. By~\eqref{eq:vertices_on_two_triangles}, 
 this is impossible. 
\end{proof}

\begin{remark}\label{rmk:one_edge}
We will be particularly interested in the case where $\GDout$ contains exactly one outgoing edge $\eastout$. 
Let $\fout\in\Faces$ be such that $\eastout$ connects $\ffout$ to $\fout$. 
 We have the following options.
\begin{enumerate}[label=(\arabic*)]
\item\label{ORA-single1} $\fout$ is an isolated vertex inside $\corv$, or, more generally, belongs to a different connected component of $\GDr$ than $\ffout$. We denote the sole face of $\GDout$ incident to $\eastout$ by $\vout$; see \figref{fig:BCFW-vertices}(left).
\item\label{ORA-single2} $\fout$ and $\ffout$ belong to the same connected component of $\GDr$. We denote the two faces of $\GDout$ incident to $\eastout$ by $\vL$ and $\vR$; see \figref{fig:BCFW-vertices}(right). 
Each of $\vLR$ is either
\begin{enumerate}[label=(2.\arabic*)]
\item\label{ORA-single2.1p}\label{ORA-single2.1m} \aflex face, or
\item\label{ORA-single2.2p}\label{ORA-single2.2m} a nondegenerate rigid \ptrar face of color $\conmp$, or
\item\label{ORA-single2.3p}\label{ORA-single2.3m} a degenerate triangular face of color $\conmp$ contained in $\Raypm$.
\end{enumerate}
\end{enumerate}
\noindent For example, the first step in \cref{fig:BCFW-full} was of type~\itemref{ORA-single1}. In the second step, $\vL$ was \flex (type~\itemref{ORA-single2.1m}) while $\vR$ was nondegenerate rigid triangular (type~\itemref{ORA-single2.2m}). In the third step, $\vL$ was \ptrar and $\vR$ was triangular (both of type~\itemref{ORA-single2.2m}). The fourth step was rigid; see the next subsection.
A degenerate triangular white face (type~\itemref{ORA-single2.3m}) is adjacent to $\KTbig_-$ in \cref{fig:fout-cliques} 
(though in that example, there is more than one outgoing edge).
\end{remark}

\subsection{Rigid \texorpdfstring{$\BCin$}{δ}}\label{sec:ORA_TORA}

Assume now that $\BCin=(\cc)$ is rigid (and that $\cor$ is not necessarily \bic). 
Thus, $\xT(\corv)$ is a \ptrar face with $m\geq4$ vertices; cf. \cref{lemma:MCE:face_classification}. In particular, $\xT(\corv)$ is an embedded $m$-gon. Let us denote its vertices by $(a_1,a_2,\dots,a_m=a_0)$ in clockwise order, with $a_i = \xT(\f_i)$ for $\f_i\in\partF\corv$ for $i\in\brm$. The indices $i\in\brm$ in the subscripts are always taken modulo $m$. By \cref{dfn:MCE:ptrar}, every corner $\cor_i$ of $\corv$ at $\f_i$ is either strictly convex or strictly reflex, and exactly three of the corners are strictly convex. 
By \crefi{lemma:MCE:corner_M_pos}{corner_M_pos2}, a corner $\cor_i$ is \bic if and only if it is strictly reflex. 
Our first goal is to give an alternative description of the location of $\xTout(\ffout)$. We will compare it to~\eqref{eq:rcrit_dfn} in \cref{lemma:ORA:rigid_rcrit_vs_GDr}. 

\begin{figure}
 \def\inputscale{1.35}
 \setlength{\tabcolsep}{1pt}
\def\textscl{0.98}
\begin{tabular}{cccc} \includegraphics[scale=\inputscale]{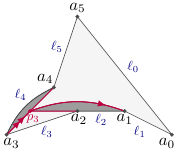}
&
\includegraphics[scale=\inputscale]{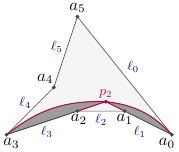}
&
\includegraphics[scale=\inputscale]{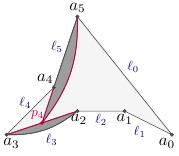}
&
\includegraphics[scale=\inputscale]{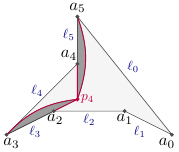}
\\
 (a) \scalebox{\textscl}{$\BCin=(\cor_3,\BWcon)$}
& (b) \scalebox{\textscl}{$\BCin\in\{(\cor_2,\BWcon),(\cor_1,\WBcon)\}$}
& (c) \scalebox{\textscl}{$\BCin\in\{(\cor_4,\BWcon),(\cor_3,\WBcon)\}$}
& (d) \scalebox{\textscl}{$\BCin\in\{(\cor_4,\BWcon),(\cor_3,\WBcon)\}$}
\end{tabular}
 \caption{\label{fig:ORA-rigid} Examples of (valid) rigid \ORsts.}
\end{figure}

\begin{lemma}\label{lemma:ORA:rigid_lines}
For $i\in\brm$, let $\elline_i$ be the line containing $[a_{i-1},a_i]$. Then for all $i\in\brm$, the lines $\elline_{i-1}$ and $\elline_{i+1}$ intersect at a single point denoted $p_i$, and we have $p_i\in\xT(\corv)$. Furthermore, if $p_i$ is not a vertex of $\xT(\corv)$ then the polygon with vertices 
\begin{equation}\label{eq:ORA:new_ptrgle}
 (a_1,\dots,a_{i-2},p_i,a_{i+1},\dots,a_m=a_0)
\end{equation}
 is again a \ptrgle. See \cref{fig:ORA-rigid}.
\end{lemma}
\begin{proof}
Suppose we are given any \einj closed polygonal chain $\PcurveT$ with vertices $(a_1,a_2,\dots,a_m=a_0)$.
 For $j\in\brm$, let $\normal_j:=\I(a_j - a_{j-1})$ be the corresponding outward normal vector. 
 It is not hard to see that the following two conditions are equivalent:
\begin{enumerate}[label=(\arabic*)]
\item\label{ptrgle_char1} $\PcurveT$ is a \ptrgle with clockwise boundary vertices $(a_1,a_2,\dots,a_m=a_0)$, with strictly convex angles at $a_0,a_s,a_t$ for some $0<s<t<m$;
\item\label{ptrgle_char2} the vectors $(z_1,\dots,z_m) := (\normal_1,\dots,\normal_s,-\normal_{s+1},\dots,-\normal_t,\normal_{t+1},\dots,\normal_m)$
satisfy 
\begin{equation}\label{eq:ptrgle_char2}
 0 < \arg(z_2/z_1) < \arg(z_3/z_1) < \cdots < \arg(z_m/z_1)<\pi,\quad\text{where}\quad \arg:=\arg_{[0,2\pi)}.
\end{equation}
\end{enumerate}
Next, suppose that the \ptrgle $\PcurveT$ is the clockwise boundary of $\xT(\corv)$, with $a_j = \xT(\f_j)$ as above. After cyclically relabeling the vertices of $\PcurveT$, we may assume that its convex vertices are $a_0,a_s,a_t$ for $0<s<t<m$, and that the index $i$ from the statement of the lemma satisfies $s<i\leq t$. Let $(z_1,\dots,z_m)$ be as in~\itemref{ptrgle_char2} above. Let $\PcurveT'$ be the closed polygonal chain with vertices given by~\eqref{eq:ORA:new_ptrgle}. The outward normal vectors of $\PcurveT'$ are given by $(\normal'_1,\dots,\normal'_{m-1})=(\normal_1,\dots,\normal_{i-2},\epsilon_{i-1}\normal_{i-1},\epsilon_{i+1}\normal_{i+1},\normal_{i+2},\dots,\normal_m)$ for some signs $\epsilon_{i-1},\epsilon_{i+1}\in\{+,-\}$. 
Furthermore, the sign $\epsilon_{i-1}$ (resp., $\epsilon_{i+1}$) is $+$ unless $i=s+1$ (resp., $i=t$), in which case it can be either $+$ or $-$, depending on whether $p_i\in\open[a_{i-2},a_{i-1}]$ (resp., $p_i\in\open[a_i,a_{i+1}]$); see \figref{fig:ORA-rigid}(c,d). 
In the case $s+1=i=t$, at least one of the two signs has to be $-$. It follows that in each case, one can choose $0<s'<t'<m-1$ such that the vectors $(z'_1,\dots,z'_{m-1}) := (\normal'_1,\dots,\normal'_{s'},-\normal'_{s'+1},\dots,-\normal'_{t'},\normal'_{t'+1},\dots,\normal'_{m-1})$ satisfy~\eqref{eq:ptrgle_char2}. Thus, $\PcurveT'$ is a \ptrgle with clockwise boundary vertices given by~\eqref{eq:ORA:new_ptrgle}. 
It is straightforward to check that $p_i$ (or more generally, the intersection point of $\elline_j$ and $\elline_k$ for any $j\neq k$) is located either in the interior or on the boundary of $\PcurveT$.
\end{proof}

\begin{definition}\label{dfn:ORA:col_of_v}
For $\v\in\Vrig$ such that $\xT(\v)$ is nondegenerate, following \cref{dfn:MCE:col}, we denote $\colop(\v):=\colop(\xd(\partF\v))$. 
If $\xT(\v)$ is a degenerate triangle then we set $\colop(\v)$ to be the color of its sole ambiguous corner.
\end{definition}

\begin{definition}\label{dfn:ORA:rigid_ORst_descr}
Let $\BCin = (\corner,\conven)$ be a rigid input datum, with $\corner=\cor_j$ for some $j\in\brm$. 
For $i\in\brm$, let 
$p_i$ be the intersection point of $\elline_{i-1}$ and $\elline_{i+1}$ as in \cref{lemma:ORA:rigid_lines}. We set 
\begin{equation}\label{eq:ORA:rigid_ORst_descr}
 \xTout(\ffout):=
 \begin{cases}
 p_{j+1}, &\text{if $\conven = (\colop(\corv),\colbarop(\corv))$;}\\
 p_{j}, &\text{if $\conven = (\colbarop(\corv),\colop(\corv))$;}\\
 \end{cases} 
\quad
 \fout:=
 \begin{cases}
 \f_{j+2}, &\text{if $\conven = (\colop(\corv),\colbarop(\corv))$;}\\
 \f_{j-2}, &\text{if $\conven = (\colbarop(\corv),\colop(\corv))$.}\\
 \end{cases} 
\end{equation}
We let $\rcrit\in\Rtp$ be such that $\xTout(\ffout) = \xTrcrit(\ffout)$. 
We say that $\BCin$ is \emph{valid} if $\xTout(\ffout)\notin\xT(\partF\corv)$.
\end{definition}
\noindent See \cref{fig:ORA-rigid}. 
Note that if we continuously deform \figref{fig:ORA-rigid}(c) into \figref{fig:ORA-rigid}(d), for some intermediate \ptrgle, the point $p_4$ will coincide with $a_2$, in which case the input data $\BCin=(\cor_4,\BWcon)$ and $\BCin = (\cor_3,\WBcon)$ become invalid.

 From now on, we assume that $\BCin=(\cc)$ is a valid rigid input datum. 
We choose the \bending of $(\GDr,\xdout)$ so that $\xtehout(\ffout)$ lies inside $\xteh(\corv)$ for all small $\eps>0$. 
As before, let $\GDo$ be obtained from $\GDr$ by deleting the edges incident to $\ffout$, making it isolated. 
For the following result, assume that $\corf = \f_j$ as in \cref{dfn:ORA:rigid_ORst_descr}. Thus, $\corfpm = \f_{j\pm1}$ and
 $\fout\in\{\f_{j-2},\f_{j+2}\}$.

\begin{lemma}\label{lemma:ORA:rigid_Chords}
Let $\Chordsiso$ be the set of isotopy classes of \pchords in $(\GDo,\xdout)$ incident to $\xTout(\ffout)$. 
Then $|\Chordsiso|=4$, and the four representative \pchords connect $\xTout(\ffout)$ to each of $\xT(\corfp)$, $\xT(\corf)$, $\xT(\corfm)$, and $\xT(\fout)$. 
\end{lemma}
\begin{proof}
Suppose that we are in the case $\conven = (\colbarop(\corv),\colop(\corv))$. The five vertices $\xTout(\{\ffout,\corfp,\corf,\corfm,\fout\})$ are contained in $\elline_{j-1}\cup \elline_{j+1}$. The ``deformed'' intersection point $\xtehout(\ffout)$ of $\elline_{j-1}$ and $\elline_{j+1}$ lies inside the face $\xteh(\corv)$ for small $\eps>0$. Since $\xT(\corv)$ is an (embedded) \ptrgle, one can indeed connect $\xTout(\ffout)$ to each of $\xT(\corfp)$, $\xT(\corf)$, $\xT(\corfm)$, $\xTout(\fout)$ by a \pchord, unique up to isotopy, so that these four \pchords are pairwise non-crossing. Let $\Chords$ consist of these four \pchords. 
 Let us add the \pchords in $\Chords$ to the set of edges of $(\GDo,\xdout)$, obtaining the graph $(\GDout,\xdout)$ (cf. \cref{dfn:ORA:rigid_GDout_descr} below). By \cref{lemma:ORA:rigid_lines}, the face $\xTout(\vL)$ of $(\GDout,\xdout)$ (where $\vL$ is as in \cref{rmk:one_edge} and \figref{fig:BCFW-vertices}(right)) with vertices given by~\eqref{eq:ORA:new_ptrgle} is a \ptrgle, and the remaining three faces incident to $\xTout(\ffout)$ are triangular. By \cref{lemma:MCE:ptrar_no_chords}, any \pchord contained inside one of these faces is isotopic to an edge of $\xTout(\GDout)$. Thus, every isotopy class in $\Chordsiso$ contains a \pchord in $\Chords$. 
\end{proof}

\begin{definition}\label{dfn:ORA:rigid_GDout_descr}
Let $\ORmv(\GD,\xd)=(\GDout,\xdout)$ be obtained from $(\GDo,\xdout)$ by adding all four \pchords in $\Chords$ to the set of edges. The colors of the two newly created ambiguous corners (and the degenerate faces containing them) are chosen so that the colors of the four faces of $\GDout$ incident to $\ffout$ alternate between black and white with $\colop(\uLR) = \conpm$, where $\uLR$ is as in \cref{fig:BCFW-vertices}. 
\end{definition}

\begin{corollary}\label{lemma:ORA:rigid_MCNE=>MCNE}
$(\GDout,\xdout)$ is an \MCNE.
\end{corollary}
\begin{proof}
 As explained in \cref{lemma:ORA:rigid_Chords}, $(\GDout,\xdout)$ satisfies~\itemref{MCE1:emb} and~\itemref{MCE5:Mpos_chords}. Since $\xdout(\ffout)\in\Conv\xd(\partF\corv)$ with $\corv\in\Vrig$, $(\GDout,\xdout)$ satisfies~\itemref{MCE2:null_edges}--\itemref{MCE3:M-tnn} by \cref{lemma:MCE:clique_affine_linear}. 
We check~\itemref{MCE4:properly_colored}. Assume without loss of generality that $\colop(\corv)=\colW$. The vertex $\xTout(\ffout)$ is incident to two degenerate black triangles and two nondegenerate white faces, one of which (say, $\xTout(\uR)$) is a triangle and the other one ($\xTout(\vL)$) is a \ptrgle. If both black angles at $\xTout(\ffout)$ are zero then $\xTout(\ffout)$ is incident to a strictly reflex corner $\corout$ of $\xTout(\vL)$ with $\sumbT(\corout)=\pi$. Otherwise, one of the black angles at $\xTout(\ffout)$ equals $\pi$ and the other one equals $0$. Thus,~\eqref{eq:MCE:sumwT_sumbT=pi} holds for $\xTout(\ffout)$. Similarly, let $\f\in\{\corfp,\corf,\corfm,\fout\}$. Then $\f$ is incident to one newly created degenerate black triangle. 
Let $\cor_{\f}$ be the corner of that triangle at $\f$. 
Let $\sumT(\cor_{\f})=\sumbT(\cor_{\f})$ be the angle of that triangle at $\f$. A simple case analysis (see \cref{fig:ORA-rigid}) shows that $\xT(\f)$ is incident to a strictly reflex corner of $\corv$ if and only if $\sumbT(\cor_{\f})=\pi$; otherwise, $\sumbT(\cor_{\f})=0$. Thus, $(\GDout,\xdout)$ satisfies~\itemref{MCE4:properly_colored}.
\end{proof}

\begin{lemma}\label{lemma:ORA:rigid_rcrit_vs_GDr}
The descriptions of $\xTout(\ffout)$ given in~\eqref{eq:rcrit_dfn} and in \cref{dfn:ORA:rigid_ORst_descr} agree. 
In other words, for $\rcrit$ as in \cref{dfn:ORA:rigid_ORst_descr} and all $0<r<\rcrit$, either $\xTr(\ffout)$ is a vertex of $\xT(\corv)$ or $(\GDr,\xdr)$ is an \MCNE.
\end{lemma}
\begin{proof}
Suppose that, say, $\conven = (\colbarop(\corv),\colop(\corv))$. Thus, the clique $\xdr(\{\ffout,\corf,\corfp\})$ is of color $\colbarop(\corv)$, and since it is contained inside $\Conv\xT(\partF\corv)$, it must be degenerate. Therefore, $p_j(\r):=\xTr(\ffout)\in \elline_{j+1}$ for all $r>0$, where $\cor=\cor_j$ as in \cref{dfn:ORA:rigid_ORst_descr}. Recall from \cref{dfn:ORA:rigid_ORst_descr} that $\rcrit>0$ satisfies $p_j(\rcrit) = p_j$. It is clear that $(\GDr,\xdr)$ is a \pcMNNE 
 for all $0<r<\rcrit$ such that $\xTr$ is \finj (i.e., such that $\xTr(\ffout)$ is not a vertex of $\xT(\corv)$). 
 Similarly to \cref{lemma:ORA:rigid_lines}, we see that for all such $0<r<\rcrit$, the polygon $\xTr(\vL)$ with vertices $(a_1,\dots,a_{j-1},p_j(\r),a_{j+1},\dots,a_m=a_0)$ is a \ptrgle
 with the same set of indices of strictly convex corners as $\xT(\corv)$.
 By \cref{lemma:MCE:ptrar_no_chords}, $(\GDr,\xdr)$ satisfies~\itemref{MCE5:Mpos_chords}, and thus is an \MCE. 

Note that $(\GDr,\xdrcrit)$ is \finj but is not an \MCE since it violates~\itemref{MCE5:Mpos_chords}
because $(\GDout,\xdrcrit)$ is an \MCE by \cref{lemma:ORA:rigid_MCNE=>MCNE}. 
 Thus, the values of $\rcrit$ in~\eqref{eq:rcrit_dfn} and in \cref{dfn:ORA:rigid_ORst_descr} indeed coincide. 
\end{proof}

Next, we show that it is always possible to find a valid \ORst.
\begin{lemma}\label{lemma:ORA:rigid_admits_valid_ORst}
Every rigid face $\v\in\Vrigsact$ admits a valid input datum $\BCin=(\cc)$ with $\corv = \v$.
\end{lemma}
\begin{proof}
By \cref{lemma:MCE:face_classification}, $\xT(\v)$ is an embedded \ptrgle with $m\geq4$ sides. 
By the Two Ears Theorem, $\xT(\v)$ admits an \emph{ear}, i.e., a strictly convex corner $\cor = \cor_s$, $s\in\brm$, such that $\open[\xT(\corfm),\xT(\corfp)]\subset\xTint(\v)$. 
We have $\xT(\corf)=a_s$ and $\xT(\corfpm)=a_{s\pm1}$.
 Let $\alpha,\beta,\gamma$ be the angles of the triangle $\Conv\{a_{s-1},a_s,a_{s+1}\}$ at the respective vertices. Let $z:=\I(a_{s+1} - a_{s-1})$. 
Without loss of generality, assume that the strictly convex corners of $\xT(\v)$ are $\cor_0,\cor_s,\cor_t$ with $0<s<t<m$. 
Let $(z_1,\dots,z_m)$ be as in the proof of \cref{lemma:ORA:rigid_lines}, extended modulo $m$ by $z_{i\pm m} = -z_i$ for $i\in\brm$. We have $z_s:=\normal_s$ and $z_{s+1}:=-\normal_{s+1}$. Denote $\arg:=\arg_{[0,2\pi)}$. We have $0<\arg(z_s/z) = \alpha<\arg(z_{s+1}/z) = \alpha+\beta < \pi$. 
 Since $(z_1,\dots,z_m)$ satisfies~\eqref{eq:ptrgle_char2}, we have either $0\leq\arg(z_{s-1}/z)<\alpha$ or $\alpha+\beta<\arg(z_{s+2}/z)\leq\pi$ or both. 
By~\eqref{eq:ptrgle_char2}, $0<\arg(z_{s+2}/z_{s-1})<\pi$, so 
 we cannot simultaneously have $0=\arg(z_{s-1}/z)$ and $\arg(z_{s+2}/z)=\pi$. Thus, we have either $0<\arg(z_{s-1}/z)<\alpha$ or $\alpha+\beta<\arg(z_{s+2}/z)<\pi$ or both. 
We choose the input datum $\BCin=(\cor_j,\conven)$ as follows. 
If $0<\arg(z_{s-1}/z)<\alpha$ then we set $j:=s-1$ and $(\conp,\conm) := (\colop(\v),\colbarop(\v))$. If $\alpha+\beta<\arg(z_{s+2}/z)<\pi$, we set $j:=s+1$ and $(\conp,\conm) := (\colbarop(\v),\colop(\v))$. In each case, 
 $\xTout(\ffout)\in\open[a_s,a_{s+1}]$ (resp., $\xTout(\ffout)\in\open[a_{s-1},a_s]$) is not a vertex of $\xT(\v)$, so $\BCin$ is a valid input.
\end{proof}

\begin{definition}\label{dfn:convex_rigid_face}
A face $\xT(\v)$, $\v\in\Vrigsact$, is called 
 \emph{\iccar} if its three strictly convex corners $\cor_{\s-1},\cor_\s,\cor_{\s+1}$ (for some $\s\in\brm$) are cyclically adjacent along the boundary of $\xT(\v)$. In this case, the vertex $a_\s$ is called the \emph{apex} of $\xT(\v)$.
\end{definition}
\noindent 
For example, all nondegenerate rigid faces incident to $\xTout(\ffout)$ in \figref{fig:fout-cliques}(right) are \iccar. 
The following result will later be applied to the BCFW recursion.
\begin{lemma}\label{lemma:ORA:iccar_valid}
Suppose that $\BCin=(\cc)$ is a rigid input datum. If $\xT(\corv)$ is \iccar then $\BCin$ is valid. In particular, if
 $(\GD,\xd)$ satisfies
\begin{equation}\label{eq:ass_ptrar}
 \text{each face $\v\in\Vrigsact$ is \iccar}
\end{equation} 
 then any input datum $\BCin$ for $(\GD,\xd)$ is valid, and for any such $\BCin$, $(\GDout,\xdout):=\ORmv(\GD,\xd)$ also satisfies~\eqref{eq:ass_ptrar}.
\end{lemma}
\begin{proof}
For \iccar faces, no point $p_i$ defined in \cref{lemma:ORA:rigid_lines} can coincide with another vertex of $\xT(\v)$. Thus, $\BCin$ is automatically valid. 
Suppose that $(\GD,\xd)$ satisfies~\eqref{eq:ass_ptrar}. 
If $\BCin$ is \flex, $(\GDout,\xdout)$ satisfies~\eqref{eq:ass_ptrar} by \cref{lemma:ORA:ORA_rigid_faces_descr}. It is straightforward to see that if $\xT(\v)$ is \iccar then in the notation of \cref{lemma:ORA:rigid_lines}, the new \ptrar face with vertices given by~\eqref{eq:ORA:new_ptrgle} is again \iccar.
\end{proof}

\subsection{Uniqueness and termination}\label{sec:ORA_terminates} 
Our next goal is to show that $\xTout(\ffout)$ is the only possible location for a new vertex on the folding ray $\RccT$.
\begin{proposition}\label{lemma:ORA:uniqueness}
Let $(\GD,\xd)$ be an \MCNE and let $\BCin=(\cc)$ be an input datum. Let $\rvalid$ be the set of points $r\in\Rtp$ such that $(\GDr,\xdr)$ is a \pMNE that is not an \MCNE.
Then $\rvalid = \{\rcrit\}$ if $\BCin$ is valid and $\rvalid = \emptyset$ otherwise. 
\end{proposition}
\begin{proof}
Let $r>0$ be such that $\xdr$ is \finj.
When $0<r<\rcrit$, $(\GDr,\xdr)$ is an \MCNE by~\eqref{eq:rcrit_dfn}; cf. \cref{lemma:ORA:rigid_rcrit_vs_GDr}.
 Thus, $r\notin\rvalid$ for all $0<r<\rcrit$. 
 We have $\rcrit\in\rvalid$ when $\BCin$ is \flex by \cref{thm:ORA_output_is_MCE,lemma:ORA:new_edge_at_least_one}. When $\BCin$ is rigid, $\rcrit\in\rvalid$ by \cref{lemma:ORA:rigid_MCNE=>MCNE} if $\BCin$ is valid and $\rcrit\notin\rvalid$ otherwise. 

It remains to show that $r\notin\rvalid$ for all $r>\rcrit$. Fix $r>\rcrit$ and assume that $(\GDr,\xdr)$ is a \pMNE. Let $\fout\in\Pivots(\corner)$ be such that $(\GDout,\xdout)$ contains an outgoing edge connecting $\ffout$ to $\fout$. Thus, $(\xdout(\ffout) - \xdout(\fout))^2 = 0$. Since $\xdout = \xdrcrit$ with $\rcrit>0$, and since $(\xd(\corf) - \xd(\fout))^2 \geq0$, we see by \cref{lemma:MCE:clique_affine_linear} that $(\xdr(\ffout) - \xd(\fout))^2 \leq0$ for $r>\rcrit$. Thus, in order for $(\GDr,\xdr)$ to satisfy~\itemref{MCE3:M-tnn}, 
 $(\xdr(\ffout) - \xd(\fout))^2$ must be identically zero as a function of $\r$. 
In particular, $\{\xd(\corf),\xd(\fout),\xdout(\ffout),\xdr(\ffout)\}$ is a clique. 

Assume first that $\BCin$ is \flex. By \cref{lemma:MCE:xTint_connected_open_avoids_cliques}, $\xTint(\corv)$ is disjoint from $[\xT(\fout),\xT(\corf)]$. Since $\xTrp(\ffout)\in\xTint(\corv)$ for small $\r'>0$, $\open[\xT(\fout),\xTrp(\ffout)]$ must intersect either $[\xT(\corf),\xT(\corfm)]$ or $[\xT(\corf),\xT(\corfp)]$ or both. Suppose that it intersects $[\xT(\corf),\xT(\corfm)]$. 
By \cref{lemma:MCE:clique_union}, $\xdrp(\{\corf,\fout,\ffout,\corfm\})$ is a clique,
and therefore so is $\xdout(\{\corf,\fout,\ffout,\corfm\})$. 
 This clique is contained in $\Kbig_-\in\Kbigs$. 
If $\xT(\fout)\in\Rayo$ then we must have either $\xT(\corfm)\in\Rayo$ or $\xT(\corfp)\in\Rayo$ or both. 
 By \cref{dfn:ORA:Chords}, each ray in $\Raysffout$ contains at most two neighbors of $\ffout$ in $\GDout$, 
but we have $\xT(\fout),\xT(\corfpm),\xT(\corf)\in\Rayo$ with $\fout\in\Pivots(\cor)$, a contradiction. Thus, $\xT(\fout)\notin\Rayo$, so we must have $\xT(\fout)\in\Raym\neq\Rayo$.
See \figref{fig:fout-cliques}(right) for an example. 
 Thus, for $r>\rcrit$, $\xT(\corfm)\in\Convint\{\xT(\fout),\xT(\corf),\xTr(\ffout)\}$, so the vertex $\corfm$ is improper in $(\GDr,\xdr)$, a contradiction. 

Assume now that $\BCin$ is rigid. Suppose that, say, $\conven = (\colbarop(\corv),\colop(\corv))$. Then $\xTout(\ffout) = p_j$ and $\fout=\f_{j-2}$ in the notation of \cref{dfn:ORA:rigid_ORst_descr}. The angle $\alphaT_{j-1}(r)$ of $\xTr(\vr)$ at $\corfm=\f_{j-1}$ is a strictly decreasing function of $r$. We have $\alphaT_{j-1}(\rcrit)=\pi$ if $\cor_{j-1}$ was strictly reflex and $\alphaT_{j-1}(\rcrit)=0$ if $\cor_{j-1}$ was strictly convex. 
 If $\cor_{j-1}$ was strictly reflex then $\f_{j-1}$ is improper in $(\GDr,\xdr)$ for all $r>\rcrit$, and if $\cor_{j-1}$ was strictly convex then 
 $(\GDr,\xTr)$ is not \awemb for $r>\rcrit$. In either case, we see that $r\notin\rvalid$ for $r>\rcrit$.
\end{proof}

Recall from \cref{sec:ORA_overview} that $(\GD,\xd)$ is called \emph{terminal} if every face of $(\GD,\xd)$ is a \btrgle.

\begin{theorem}\label{lemma:ORA:ORA_termination}
The \ORA terminates: any non-\terminal \MCNE $(\GD,\xd)$ admits at least one valid \ORst, and applying any sequence of valid \ORsts eventually results in a terminal \MCNE $(\GDout,\xdout)$. 
\end{theorem}
\begin{proof}
By \cref{lemma:ORA:rigid_admits_valid_ORst}, any non-\terminal \MCNE admits at least one valid \ORst. 
Suppose that the algorithm does not terminate for some $(\GD,\xd)$. Out of all such graphs $\GD$, choose the one with the smallest number of connected components. Out of those, choose the one with the lexicographically smallest \emph{perimeter sequence} 
$\bperim(\GD):=(|\partEvecast\v|)_{\v\in\Vint}$, 
 sorted in the weakly decreasing order. 
If $\max(\bperim(\GD))\leq 3$ then $(\GD,\xd)$ is \terminal by \cref{lemma:MCE:face_classification}. 
Otherwise, by \cref{lemma:ORA:new_edge_at_least_one,lemma:ORA:rigid_Chords}, the graph $\GDout$ obtained during a valid \ORst contains at least one outgoing edge $\eout$
 (connecting $\ffout$ to some vertex $\fout\in\Pivots(\corner)$). If $\fout$ and $\ffout$ belong to different connected components of $\GDr$ then this step decreases the number of connected components of $\GD$, and thus the algorithm terminates by the induction hypothesis. Otherwise, this step replaces the face $\corv$ with two triangles 
$\xTout(\uLR)$ 
 and at least two other faces, each with 
smaller perimeter than $|\partEvecast\corv|$.
 Thus, $\bperim(\GDout)<\bperim(\GD)$ in the lexicographic order, 
and the algorithm again terminates by induction.
\end{proof}

\part{BCFW tilings}\label{part3}

\section{\MCMSsTITLE}\label{sec:MCMS}
The goal of this section is to introduce \emph{\lggs} $\Gloop$ and study the moduli space $\MCM(\Gloop)$ of \MCEs planar dual to $\Gloop$, satisfying a natural compatibility condition~\eqref{eq:MCMS_angles_vs_k}. 
For \terminal $\Gloop$ (\cref{dfn:BCFW:terminal}), we relate such compatible \MCEs to \wtembs in \cref{lemma:TE:TE_to_MCE_and_back}.

\subsection{\LggsTITLE}\label{ssec:MCMS:lggs}
The following is a variation of~\cite[Definition~4.1]{Pos_ICM}.
\begin{definition}%
A \emph{\ggg} is an undirected graph $\Ggen=(\Vgen,\Egen)$ (with loops and parallel edges allowed) equipped with functions $\gelWgen,\gelBgen:\Vgen\to\Ztnn$ satisfying 
\begin{align}
\label{eq:BCFW:helW+helB=deg}
 \gelWgen(\v)+\gelBgen(\v) &= \degGgen(\v)\quad\text{for all $\v\in\Vgen$,\quad and}\\
\label{eq:BCFW:sum_helW_helB=Egen}
 \sum_{\v\in\Vgen}\gelWgen(\v) &= \sum_{\v\in\Vgen}\gelBgen(\v) = |\Egen|.
\end{align}
\end{definition}

We continue to use notation ($\RgEbd$, $\RgEint$, $\RgE$, $n(\Rg)$, $\gelWgen(\Rg)$, $\gelBgen(\Rg)$, etc.) from \cref{ssec:BACKGR:surplus} for subsets $\Rg\subset\Vgen$. 
By~\eqref{eq:BACKGR:hel=sum_hel_minus_E}, $\gelWgen(\Rg) + \gelBgen(\Rg) = n(\Rg)$. 
In general, we need not have $0\leq \gelWgen(\Rg),\gelBgen(\Rg)\leq n(\Rg)$ for $\Rg\subset\Vgen$; however, see \cref{lemma:BCFW:MCM_nonempty=>hel>=1} below. By~\eqref{eq:BCFW:sum_helW_helB=Egen},
\begin{equation}\label{eq:BCFW:helWgen=helBgen=0}
 \gelWgen(\Vgen) = \gelBgen(\Vgen) = 0.
\end{equation}

\begin{lemma}\label{lemma:BCFW:Rg_vs_Rg'_helW_vs_helB}
Let $\Rg,\Rg'\subset\Vgen$ be disjoint nonempty subsets such that $\RgpEbd\subset\RgEbd$. Then 
\begin{equation}\label{eq:BCFW:Rg_vs_Rg'_helW_vs_helB}
 \gelWgen(\Rg) = \gelWgen(\Rg\sqcup\Rg') + \gelBgen(\Rg') 
 \quad\text{and}\quad
 \gelBgen(\Rg) = \gelBgen(\Rg\sqcup\Rg') + \gelWgen(\Rg').
\end{equation}
\end{lemma}
\begin{proof}
Indeed, by~\eqref{eq:BACKGR:hel=sum_hel_minus_E}, we find 
$\gelWgen(\Rg\sqcup\Rg') = \gelWgen(\Rg) + \gelWgen(\Rg') - n(\Rg') = \gelWgen(\Rg) - \gelBgen(\Rg')$.
\end{proof}
\begin{corollary}
Let $\Rg\subsetneq\Vgen$ be a proper nonempty subset of $\Vgen$ and let $\Rgc:=\Vgen\setminus\Rg$. Then
\begin{equation}\label{eq:BCFW:helWgen_Rc=helBgen_R}
 \gelWgen(\Rg) = \gelBgen(\Rgc) \quad\text{and}\quad \gelBgen(\Rg) = \gelWgen(\Rgc).
\end{equation}
\end{corollary}

Similarly to~\eqref{eq:BACKGR:gelWmin_gelBmin_dfn}, we set $\gelmin(\Ggen):=\min(\gelWmin(\Ggen),\gelBmin(\Ggen))$ with 
\begin{equation}\label{eq:BCFW:helWmin_dfn}
 \gelWmin(\Ggen):=\min\{\gelWgen(\Rg)\mid \emptyset\neq\Rg\subsetneq\Vgen\} \quad\text{and}\quad
 \gelBmin(\Ggen):=\min\{\gelBgen(\Rg)\mid \emptyset\neq\Rg\subsetneq\Vgen\}.
\end{equation}

A \emph{perfect orientation} of $\Ggen$ is an orientation of the edges of $\Ggen$ such that every vertex $\v\in\Vgen$ has exactly $\gelWgen(\v)$ incoming and $\gelBgen(\v)$ outgoing arrows. 
 By~\eqref{eq:BACKGR:hel=sum_hel_minus_E}, given a perfect orientation of $\Ggen$, for any nonempty $\Rg\subset\Vgen$, the number of edges in $\RgEbd$ oriented towards (resp., away from) $\Rg$ is $\gelWgen(\Rg)$ (resp., $\gelBgen(\Rg)$). In particular, if $\Ggen$ admits a perfect orientation then we must have $\gelmin(\Ggen)\geq0$. 

\begin{proposition}[{\cite[Theorem~4]{Hakimi}}]\label{lemma:OCP:Hakimi}
$\Ggen$ admits a perfect orientation if and only if ${\gelmin(\Ggen)\geq0}$.
\end{proposition}
\begin{proof}
First, observe that deleting loop edges from $\Ggen$ (and decreasing the values of $\gelWgen$ and $\gelBgen$ at the corresponding vertex by $1$) does not affect $\gelmin(\Ggen)$ and the existence of a perfect orientation. Thus, we may assume that $\Ggen$ has no loop edges. By~\cite[Theorem~4]{Hakimi}, $\Ggen$ admits a perfect orientation if and only if $\gelWgen(\Vgen)=0$ and $\gelWgen(\Rg)\geq0$ for all $\emptyset\neq\Rg\subsetneq\Vgen$.\footnote{In~\cite[Theorem~4]{Hakimi}, this inequality was imposed more generally on arbitrary subgraphs of $\Ggen$ but it is clear that one can restrict to only induced subgraphs $\Ggen\ind[\Rg]$.} By~\eqref{eq:BCFW:helWgen=helBgen=0}, the former condition is automatically satisfied. By~\eqref{eq:BCFW:helWgen_Rc=helBgen_R}, the latter condition is equivalent to $\gelmin(\Ggen)\geq0$.
\end{proof}

\begin{definition}[{\cite[Definition~4.1]{Pos_ICM}}]
A \emph{Grassmannian graph} is a graph $\G=(\Verts,\E)$ embedded in a disk $\Disk$ with $n$ boundary vertices $\bdv_1,\bdv_2,\dots,\bdv_n\in\Verts$ of degree $1$, equipped with functions $\gelW,\gelB:\Vint\to\Ztnn$ on interior vertices of $\G$ satisfying~\eqref{eq:BCFW:helW+helB=deg} for all $\v\in\Vint$.
\end{definition}

For \ggs, we restrict our attention to subsets $\emptyset\neq\Rg\subset\Vint$. In particular, we define $\gelWmin(\G),\gelBmin(\G)$ by~\eqref{eq:BACKGR:gelWmin_gelBmin_dfn} and set $\gelmin(\G):=\min(\gelWmin(\G),\gelBmin(\G))$. 
We let 
\begin{equation}\label{eq:BCFW:k(G)_dfn}
 k=\gelW(\Vint) \quad\text{and}\quad n-k=\gelB(\Vint).
\end{equation}
 We assume that $0\leq k\leq n$, and we say that $\G$ is a Grassmannian graph \emph{of type $(k,n)$}. 

\begin{remark}\label{rmk:BCFW:gluing_bdv_into_bdvall}
We do not specify the values of $\gelW(\bdv_i),\gelB(\bdv_i)$ at the boundary vertices. Instead, we introduce a \ggg $\Ggen$ obtained from $\G$ by identifying the boundary vertices $\bdv_1,\bdv_2,\dots,\bdv_n$ into a single vertex $\bdvall$ with $\gelWgen(\bdvall):=n-k$ and $\gelBgen(\bdvall):=k$. This choice ensures that~\eqref{eq:BCFW:sum_helW_helB=Egen} is satisfied. By~\eqref{eq:BCFW:helWgen_Rc=helBgen_R},
$\gelmin(\Ggen)\geq0$ if and only if $\gelmin(\G)\geq0$.
\end{remark}

Following~\cite[Section~4]{Pos_ICM}, a \emph{perfect orientation} of $\G$ is an orientation of $\G$ such that every interior vertex $\v\in\Vint$ is incident to exactly $\gelW(\v)$ incoming and $\gelB(\v)$ outgoing edges (with no conditions imposed on the boundary vertices). Thus, perfect orientations of $\G$ are in bijection with those of $\Ggen$.

\begin{corollary}\label{lemma:OCP:Hakimi_cor}
A \gg $\G$ admits a perfect orientation if and only if $\gelmin(\G)\geq0$.
\end{corollary}

We define \emph{\holess} and \emph{\sconn} subsets $\Rg\subset\Vint$ for \ggs and use notation $\Rghol$ and $\Rgcl:=\Rg\sqcup\Rghol$ similarly to \cref{dfn:BACKGR:holess,dfn:BACKGR:sconn}.

\begin{definition}\label{dfn:ORA:black_white_verts}
Let $\G$ be a \gg. 
A vertex $\v\in\Vint$ is called \emph{white} if $\gelW(\v) = 1$ and \emph{black} if $\gelB(\v) = 1$. In the case when $\degG(\v)=2$ and $\gelW(\v)=\gelB(\v)=1$, we declare $\v$ to be either white or black, making an arbitrary choice. 
 The set of white (resp., black) interior vertices of $\G$ is denoted by $\WVint$ (resp., $\BVint$).
\end{definition}

\begin{definition}\label{dfn:BCFW:terminal}
A \gg $\G$ is called \emph{\terminal} if each interior vertex of $\G$ is either black or white, i.e., $\Vint = \WVint\sqcup\BVint$. 
\end{definition}

\begin{remark}%
\label{rmk:BCFW:terminal_vs_plabic_vs_plabip}
\Terminal \ggs are precisely the \emph{plabic (planar bicolored) graphs} of~\cite{Pos}. 
 We make each \terminal \gg bipartite by inserting a degree-$2$ vertex of opposite color in the middle of each interior unicolored edge; cf. \cref{dfn:DIM:MV1}. 
 {\bf From now on, we assume that all \terminal \ggs are planar bipartite}. 
In particular, when $\G$ is terminal, the statistics 
$\gelWmin(\G),\gelBmin(\G)$ and $\helWmin(\G),\helBmin(\G)$ are related by \cref{lemma:BCFW:gelWmin_vs_helWmin}.
\end{remark}

Following \cref{dfn:MCE:Vsact_Vbigons_Vtrs}, we denote $\Vsact:=\{\v\in\Vint\mid \degG(\v)\geq4\}$.
\begin{definition}\label{dfn:BCFW:lgg}
An \emph{\lgg $\Gloop$ of type $\knl$} is 
a \gg $\Gloopgr$ of type $(k,n)$ equipped with the following additional \emph{loop data}:
\begin{enumerate}[label=(\arabic*)]
\item\label{lgg1} 
a partition $\brnL=\Pfl\sqcup\Pfix$ of the set $\brnL$ of \emph{\Lfunctures} into subsets of \emph{floating} and \emph{fixed} \Lfunctures, respectively. 
\item\label{lgg2} a subset $\brnLv\subset\Pfl$ for each $\v\in\Vsact$ such that $\Pfl = \bigsqcup_{\v\in\Vsact} \brnLv$;
\item\label{lgg3} a face $\ploc_\rho$ of $\G$ for each fixed \Lfuncture $\rho\in\Pfix$.
\end{enumerate}
\end{definition}

\begin{definition}\label{dfn:BCFW:lgg_dual}
 When the underlying \gg $\Gloopgr$ \hasnofloat, we denote its planar dual by $\GDgr$ and denote the set of faces of $\Gloopgr$ by $\Facesgr$.
The planar dual $\GD$ of \algg $\Gloop$ is obtained from $\GDgr$ by placing a set $\{\ploc_\rho\mid\rho\in\brnLv\}$ of isolated vertices inside the face $\v$ for each $\v\in\Vsact$. 
 Thus, the vertex set $\Faces:=\Facesgr\sqcup\{\ploc_\rho\mid\rho\in\Pfl\}$ of $\GD$ contains 
$\nL$ distinguished vertices $\Puncs:=(\ploc_1,\ploc_2,\dots,\ploc_\nL)$.
\end{definition}

\begin{assumption}\label{ass:lgg}
For the rest of the paper, we assume that each \lgg $\Gloop$ is of type $\knL$ and satisfies the following.%
\begin{itemize}
\item the underlying \gg $\Gloopgr$ is connected, %
\item the planar dual $\GD$ satisfies \cref{ass:lgg_dual}, 
\item $\gelW(\v),\gelB(\v)\geq1$ for all $\v\in\Vint$, and 
\begin{equation}\label{eq:ass_lgg}
 \text{for each $\v\in\WVint\sqcup\BVint$, we have $\degG(\v)\in\{2,3\}$.}
\end{equation}
\end{itemize}
\end{assumption}

\begin{remark}\label{rmk:Vsact=Vfat}
When $\Gloop$ satisfies \cref{ass:lgg}, it follows that $\Vsact = \Vint\setminus(\WVint\sqcup\BVint)$.
\end{remark}

We record the following trivial consequences of \cref{ass:lgg} for later use.
\begin{corollary}\label{lemma:BCFW:ass_on_GD}\ Let $\Gloop$ be \algg satisfying \cref{ass:lgg}. 
\begin{enumerate}[label=(\arabic*)]
\item $\GD$ is a union of a single connected component and several isolated vertices, with each isolated vertex located inside some face $\v\in\Vsact$;
\item $\GD$ is equipped with a distinguished collection $\Puncs=(\ploc_1,\ploc_2,\dots,\ploc_\nL)$ of vertices that includes all isolated vertices.%
\end{enumerate}
\end{corollary}

\begin{remark}\label{rmk:BCFW:ass_propagates}
For an arbitrary graph $\GD$ satisfying \cref{ass:lgg_dual}, both properties listed in \cref{lemma:BCFW:ass_on_GD} propagate under \ORsts $(\GD,\xd)\mapsto(\GDout,\xdout)=\ORmv(\GD,\xd)$ for any valid input datum $\BCin$ for an \MCE $(\GD,\xd)$. Here, the distinguished vertices of $\GDout$ are the same as those of $\GD$.
\end{remark}

\begin{remark}\label{rmk:BCFW:dual_loop_data_extraction}
The loop data for $\Gloop$ is fully determined by the locations of the distinguished vertices $\Puncs=(\ploc_1,\ploc_2,\dots,\ploc_\nL)$ of $\GD$: we have $\Pfl=\{\rho\in\brnL\mid\ploc_\rho\text{ is isolated}\}$, and for a face $\v\in\Vsact$ of $\GD$, we have $\brnLv = \{\rho\in\Pfl\mid\ploc_\rho\text{ is located inside $\v$}\}$. 
\end{remark}

\subsection{\MCMSs}
For any graph $\GD$ satisfying \cref{ass:lgg_dual}, we set
\begin{equation*}%
 \Mmce(\GD) :=\{(\GD,\xd)\mid (\GD,\xd)\text{ is an \MCNE}\}.
\end{equation*}
Note that when $\GD$ is the planar dual of \algg $\Gloop$ satisfying \cref{ass:lgg}, 
$\Mmce(\GD)$ does not take into account any information about the functions $\gelW,\gelB:\Vint\to\Z_{\geq1}$. 
We introduce a subset $\MCM(\Gloop)\subset\Mmce(\GD)$ that depends on $\gelW,\gelB$ in a natural way.

\begin{definition}[\MCMS]\label{dfn:BCFW:MCE_compatible}
Let $\Gloop$ be \algg satisfying \cref{ass:lgg}. We say that $(\GD,\xd)\in\Mmce(\GD)$
 is \emph{compatible with $\Gloop$} if for each $\v\in\Vint$, 
\begin{equation}\label{eq:MCMS_angles_vs_k}
 \sumwT(\v):=\sum_{\cor\in\corners(\v)} \sumwT(\cor) = \pi(\gelB(\v)-1) \quad\text{and}\quad
 \sumbT(\v):=\sum_{\cor\in\corners(\v)} \sumbT(\cor) = \pi(\gelW(\v)-1).
\end{equation}
 We let $\MCM(\Gloop):=\{(\GD,\xd)\in\Mmce(\GD)\mid(\GD,\xd)\text{ is compatible with }\Gloop\}$.
\end{definition}

\begin{lemma}\label{lemma:BCFW:from_MCM_to_Glout}
Suppose that $\GD$ is a graph satisfying \cref{ass:lgg_dual} and both conditions in \cref{lemma:BCFW:ass_on_GD}, and let $(\GD,\xd)\in\Mmce(\GD)$. 
 Then there exists a unique \lgg $\Gloop$ such that $(\GD,\xd)$ is compatible with $\Gloop$. 
Furthermore, this graph $\Gloop$ satisfies \cref{ass:lgg}.
\end{lemma}
\begin{proof}
The underlying \gg $\Gloopgr$ of $\Gloop$ is the planar dual of the graph $\GDgr$ obtained from $\GD$ by removing isolated vertices. The loop data for $\Gloop$ is fully determined by $\GD$ and $\Puncs$ by \cref{rmk:BCFW:dual_loop_data_extraction}. 
It remains to determine the functions $\gelW,\gelB:\Vint\to\Z_{\geq1}$.

For a face $\v\in\Vint$ of $\GD$, let $\gelW(\v),\gelB(\v)\in\R$ be such that~\eqref{eq:MCMS_angles_vs_k} holds. 
First, since the left-hand sides of both equations in~\eqref{eq:MCMS_angles_vs_k} are nonnegative by~\eqref{eq:nearGr:angles_exist}, $\gelB(\v),\gelW(\v)\geq1$. Next, observe that $\xT(\partEvecast\v)$ and $\xO(\partEvecast\v)$ are closed polygonal chains, and the face $\xT(\partEvecast\v)=\lim_{\eps\to0}\xteh(\partEvecast\v)$ is weakly embedded. 
 The boundary turning angle (\cref{dfn:DIM:degenerate_convex_polygon}) of $\xT(\partEvecast\v)$ (resp., $\xO(\partEvecast\v)$) at a corner $\cor\in\corners(\v)$ is given by 
$\sumT(\cor) - \pi$ 
(resp., $\sumO(\cor)-\pi$ modulo $2\pi$).
 The sum of these turning angles is equal to $-2\pi$ (resp., $0$ modulo $2\pi$). Thus, by~\eqref{eq:nearGr:angles_exist} and~\eqref{eq:MCMS_angles_vs_k}, 
\begin{equation*}%
 \gelB(\v) + \gelW(\v) = \degG(\v) \quad\text{and}\quad
 \gelB(\v) - \gelW(\v) \equiv \degG(\v)\ \ \mod\ 2.
\end{equation*}
Therefore, $\gelB(\v),\gelW(\v)$ are integers, and thus $\Gloop$ is \algg. 

Let $\wv\in\WVint$ be a white vertex of $\Gloop$, so $\gelW(\wv)=1$. By~\eqref{eq:MCMS_angles_vs_k}, $\sumbT(\cor) = 0$ for all $\cor\in\corners(\wv)$. 
By \cref{lemma:MCE:face_classification,lemma:zero_corner=>rigid}, 
 $\wv$ must be \btrar, so $\degG(\wv) \in\{2,3\}$. Similarly, for $\bv\in\BVint$, we have $\degG(\bv) \in\{2,3\}$. 
The graph $\Gloopgr$ is connected since $\GD$ satisfies \cref{ass:lgg_dual}. 
 Thus, $\Gloop$ satisfies \cref{ass:lgg}.
\end{proof}

\begin{lemma}\label{lemma:BCFW:ptrar_face_k=2}
Let $(\GD,\xd)\in\MCM(\Gloop)$ and let $\v\in\Vrig\setminus\Vbigons$ be a rigid non-bigonal face of $(\GD,\xd)$. Then $\gelG^{\colop(\v)}(\v)=\degG(\v) - 2$ and $\gelG^{\colbarop(\v)}(\v)=2$.
\end{lemma}
\begin{proof}
 The result clearly holds when $\v\in\Vtrs$ is a triangle. Suppose now that $\v\in\Vrigsact$. By~\cref{lemma:MCE:face_classification}, $\xT(\v)$ is \ptrar, so it contains $3$ strictly convex corners $\cor_0,\cor_s,\cor_t$ and $m-3$ strictly reflex corners, where $m=\degG(\v)\geq4$. 
If e.g. $\colop(\v)=\colB$ then by \cref{lemma:MCE:corner_M_pos}, $\sumwT(\cor_0)=\sumwT(\cor_s)=\sumwT(\cor_t)=0$ and $\sumwT(\cor_i) = \pi$ for all $i\neq0,s,t$. Thus, by~\eqref{eq:MCMS_angles_vs_k}, $\gelB(\v) = m-2$, and therefore $\gelW(\v) = 2$.
\end{proof}

\begin{remark}
We warn that the white/black 
 vertices of $\Gloop$
 do \emph{not} correspond to the white/black
rigid faces of $(\GD,\xd)\in\MCM(\Gloop)$.
 Rather, they correspond only to bigonal and triangular white/black faces; 
the remaining vertices $\v\in\Vsact=\Vint\setminus(\WVint\sqcup\BVint)$ of $\Gloop$ (cf. \cref{rmk:Vsact=Vfat}) correspond to 
faces of $(\GD,\xd)$ that are either \flex or rigid \ptrar with $m\geq4$ sides.
For example, the \lgg in \figref{fig:BCFW-full}(d) contains a vertex $\v\in\Vsact$ labeled $(2,4)$ (i.e., satisfying $\gelW(\v)=\gelB(\v)=2$), but the corresponding face of $(\GD,\xd)$ is a rigid white quadrilateral. 
\end{remark}

Similarly to \cref{notn:TE:Rg_corners_E_WV_BV_sumTcond}, for $\Rg\subset\Vint$ and $\f\in\Facesgr$, let 
$\corners(\Rg):=\bigsqcup_{\v\in\Rg}\corners(\v)$, 
$\corners\cond(\f|\Rg):=\corners(\f)\cap\corners(\Rg)$, 
$\sumbTcond(\f|\Rg):=\sum_{\cor\in \corners\cond(\f|\Rg)} \sumbT(\cor)$, 
$\sumwTcond(\f|\Rg):=\sum_{\cor\in \corners\cond(\f|\Rg)} \sumwT(\cor)$, and 
$\sumTcond(\f|\Rg):=\sumbTcond(\f|\Rg) + \sumwTcond(\f|\Rg)$. 

\begin{notation}\label{notn:RgFbd}
For $\Rg\subset\Vint$, we let $\RgFac$ be the set of faces of $\Gloopgr$ incident to some vertex in $\Rg$. Let $\RgFint\subset\RgFac$ be the set of interior faces $\ff$ of $\G$ such that all vertices of $\G$ incident to $\ff$ belong to $\Rg$. We let 
 $\RgFbd:=\RgFac\setminus\RgFint$. 
\end{notation}

\begin{lemma}\label{lemma:Rg_sum_bd_vs_chi}
Let $(\GD,\xd)\in\MCM(\Gloop)$ and $\emptyset\neq\Rg\subset\Vint$. Then
\begin{equation}\label{eq:Rg_sum_bd_vs_chi}
 \sum_{\f\in\RgFbd} \sumbTcond(\f|\Rg) = \pi(\gelW(\Rg) - \chind[\Rg])
 \quad\text{and}\quad
 \sum_{\f\in\RgFbd} \sumwTcond(\f|\Rg) = \pi(\gelB(\Rg) - \chind[\Rg]),
 \quad\text{where}
\end{equation}
\begin{equation*}%
 \chind[\Rg] := |\Rg| - |\RgEint| + |\RgFint|.
\end{equation*}
\end{lemma}
\begin{proof}
For $\emptyset\neq\Rg\subset\Vint$, by~\eqref{eq:MCMS_angles_vs_k} and~\eqref{eq:BACKGR:hel=sum_hel_minus_E},
\begin{equation}\label{eq:BCFW:sumbT_vs_chi}
\sum_{\f\in\RgFac} \sumbTcond(\f|\Rg)
 = \sum_{\v\in\Rg} \sumbT(\v)
 = \sum_{\v\in\Rg} (\gelW(\v) - 1)\pi
 = (\gelW(\Rg) + |\RgEint| - |\Rg|)\pi.
\end{equation}
On the other hand, by~\eqref{eq:MCE:sumwT_sumbT=pi}, for each $\f\in\RgFint$, we have $\sumbTcond(\f|\Rg)=\sumbT(\f)=\pi$, so
\begin{equation}\label{eq:BCFW:sumbT_vs_RgFint}
 \sum_{\f\in\RgFac} \sumbTcond(\f|\Rg) = |\RgFint|\pi + \sum_{\f\in\RgFbd} \sumbTcond(\f|\Rg).
\end{equation}
Equating the right-hand sides of~\eqref{eq:BCFW:sumbT_vs_chi} and~\eqref{eq:BCFW:sumbT_vs_RgFint}, we obtain~\eqref{eq:Rg_sum_bd_vs_chi}.
\end{proof}

\begin{corollary}\label{lemma:BCFW:MCM_nonempty=>hel>=1}
Suppose that $\MCM(\Gloop)\neq\emptyset$. Then $\gelmin(\G)\geq1$.
\end{corollary}
\begin{proof}
By \crefi{lemma:BCFW:Rg_simply_conn_lower_bound}{sconn_lower2} (which applies to \ggs with the same proof), 
 it suffices to show that for each \sconn $\emptyset\neq\Rg\subset\Vint$, we have $\gelW(\Rg),\gelB(\Rg)\geq1$. 
Indeed, for \sconn $\Rg$, we have $\chind[\Rg]=1$ by~\eqref{eq:DIM:Euler} and \cref{lemma:BCFW:dualizable_holess_sconn}. 
Since the left-hand side of each equation in~\eqref{eq:Rg_sum_bd_vs_chi} is manifestly nonnegative by~\eqref{eq:nearGr:angles_exist}, we find $\gelW(\Rg)\geq\chind[\Rg]$ and $\gelB(\Rg)\geq\chind[\Rg]$. 
\end{proof}

\subsection{Decorated \MCMSs and \wtembs}\label{ssec:BCFW:MCM_vs_wtemb}
Recall from \cref{rmk:BCFW:terminal_vs_plabic_vs_plabip} that when \algg $\Gloop$ is \terminal, its underlying graph $\G$ is planar bipartite. In this case, our goal is to relate the spaces $\MCM(\Gloop)$ and $\Mdte(\G)$ introduced in \cref{dfn:BCFW:MCE_compatible,dfn:TE:tembending}, respectively. 
 Note that by~\itemref{MCE1:emb} and~\eqref{eq:MCE3':Mbd}, each \MCE of $\Gloop$ is \einj and has \Mbd, 
while neither of these assumptions is imposed on \wtembs of $\G$. On the other hand, each \wtemb $\datrQ\in\Mdte(\G)$ comes equipped with a decoration $\lalat\in\lalakMAT$ of the boundary polygon $\Pbd_{\xd}$; cf. \cref{dfn:BACKGR:decoration}. %
We show that once these discrepancies are taken into account, the two moduli spaces become homeomorphic; see \cref{lemma:TE:TE_to_MCE_and_back}.

\begin{definition}\label{dfn:MCMp_MCMpdec}
 For \algg $\Gloop$ satisfying \cref{ass:lgg}, let 
\begin{equation}\label{eq:***}
 \MCMpdec(\Gloop):=\{(\GD,\xd,\la,\lat)\in\MCM(\Gloop)\times\MPkntreeMAT\mid \lalat\text{ is a decoration of $\Pbd_{\xd}$}\}.
\end{equation}
 For a planar bipartite graph $\Gbip$ that admits an \APM, we denote 
\begin{equation}\label{eq:MdteoMP_dfn}
 \MdteoMP(\Gbip):=\{\datrQL=\datrQ\in\Mdteo(\Gbip)\mid (\GD,\xd)\text{ has \Mdash positive boundary}\}.
\end{equation}
\end{definition}

\begin{notation}\label{notn:BCFW:moduli_of_Gfin_vs_Gbip}
When $\Gbip$ is planar bipartite, we denote by $\MCMdec(\Gbip)$ the space of decorated \MCEs of the corresponding \gg (cf. \cref{dfn:gelWB_for_bip}). Similarly, when $\Gfin$ is \terminal with underlying planar bipartite graph $\Gbip$, we denote $\Mdatr(\Gfin):=\Mdatr(\Gfingr)$, $\Mdte(\Gfin):=\Mdte(\Gfingr)$, etc.
\end{notation}

\begin{remark}\label{rmk:MCMp->MCM_surj}
 Given $(\GD,\xd)\in\MCMp(\Gloop)$, by~\eqref{eq:MCE3':Mbd} and \cref{lemma:TOP:Mpos=>simple}, $\PbdxT$ is a simple polygon in the plane. 
By \cref{ass:lgg_dual}, $\TURN(\PbdxT)=-2\pi$, so $\Pbdx$ admits a (unique up to little group action) positive decoration $\lalat\in\lalakMPMAT$ by \cref{lemma:BCFW:null}.
Thus, the forgetful map $\MCMpdec(\Gloop)\to\MCMp(\Gloop)$ is surjective.
\end{remark}

\begin{proposition}\label{lemma:TE:TE_to_MCE_and_back}\label{lemma:TE:MCE_to_TE}\label{lemma:TE:TE_to_MCE}
\label{lemma:ORA:terminal_vs_atr_no_dim}
Let $\Gbip$ be a connected planar bipartite graph satisfying $\helmin(\Gbip)\geq1$ and~\eqref{eq:ass_lgg}.
Then we have a homeomorphism
\begin{equation}\label{eq:ORA::mceTOwte_dfn_iso} 
 \mceTOwte:\MCMpdec(\Gbip) \xrasim \MdteoMP(\Gbip).
\end{equation}
\end{proposition}
\begin{proof}
Let $(\GD,\xd,\la,\lat)\in\MCMdec(\Gbip)$. 
Since $\xT$ is \finj, it is \einj. 
 Let $\wt\in\Rtpgauge$ be given by $\wt(\e) := |\xT(\ff) - \xT(\f)|$ for all $\e\in\Edges$, where $\{\ff,\f\}:=\ebarast$ are the endpoints of the dual edge. Let $\epsK$ be any choice of Kasteleyn signs for $\Gbip$; this choice is immaterial in view of \cref{rmk:DIM:Kast_gauge_eq}. 
 By~\Mref{prop:t_imm=>holom}, $\xd$ may be obtained as the \KSprim of some pair $(\Fw,\Fb)\in\HHspaceC$ of discrete holomorphic functions, well defined up to gauge equivalence. 
Thus, we obtain a map $\mceTOwte:\MCMdec(\Gbip)\to \Mdatro(\Gbip)$
sending $(\GD,\xd,\la,\lat)\mapsto \datrQL=\datrQ$. 
To see that this map is injective, observe that $\lalat$ is uniquely determined by $(\Fw,\Fb)$ via~\eqref{eq:TE:y_to_lalat}--\eqref{eq:TE:lalat_vs_pFw_pFb}.

We check that the image $\mceTOatr(\MCMpdec(\Gbip))$ is contained in $\MdteoMP(\Gbip)$. 
Let $(\GD,\xd,\la,\lat)\in\MCMpdec(\Gbip)$. 
By \cref{rmk:MCMp->MCM_surj}, it suffices to check that 
its image $\datrQL\in\Mdatro(\Gbip)$
satisfies conditions~\itemref{TE:bdry_angle}--\itemref{TE:t_imm} of \cref{dfn:TE:tembending}. 
Since $\xT$ is an \einj \wemb of $\GD$, by~\eqref{eq:OCP:primitive}, we have 
\begin{equation*}%
 \sumwT(\cor_\s) = \arg_{[0,2\pi)} \frac{-\epsK(\e_{\s+1})\Fb(\b_{\s+1})}{\epsK(\e_\s)\Fb(\b_\s)},%
\quad\text{resp.,}\quad
 \sumbT(\cor_\s) = \arg_{[0,2\pi)} \frac{-\epsK(\e_{\s+1})\Fw(\w_{\s+1})}{\epsK(\e_\s)\Fw(\w_\s)} %
\end{equation*}
in the notation of~\eqref{eq:TE:weak_t_imm_dfn}, where $\cor_\s$ is the corner of $\GD$ with $\corvx_{\cor_\s}=\w$ 
(resp., $\corvx_{\cor_\s}=\b$) 
located between the edges $\east_\s$ and $\east_{\s+1}$.
By~\eqref{eq:nearGr:angles_exist}, $\sumwT(\cor_\s),\sumbT(\cor_\s)\in[0,\pi]$, so~\itemref{TE:t_imm} is satisfied. 
Since $(\GD,\xd)$ \hasMbd, by~\eqref{eq:TE:t_imm_bdry_vs_pFw_pFb}, we have $\pFw_i,\pFb_i\neq0$ for all $i\in\brn$. Similarly to 
\Mref{lemma:angle_sum_arg_rat}, 
we see that the angles $\sumbT_i,\sumwT_i$ introduced in~\eqref{eq:TE:dfn} satisfy $\sumbT_i=\sumbT(\bdf_i)$ and $\sumwT_i=\sumwT(\bdf_i)$, so~\itemref{TE:bdry_angle} holds by~\eqref{eq:MCE:sumwT_sumbT_bdry_0_pi}. 
Finally,~\itemref{TE:bdry_winding} follows by applying~\eqref{eq:Rg_sum_bd_vs_chi} to $\Rg=\Vint$; cf.~\eqref{eq:BCFW:k(G)_dfn}.

Conversely, let $\datrQL=\datrQ\in\MdteoMP(\Gbip)$. Let $\lalat$ be obtained from $(\Fw,\Fb)$ via~\eqref{eq:TE:y_to_lalat}--\eqref{eq:TE:lalat_vs_pFw_pFb}. We claim that $(\GD,\xd,\la,\lat)\in\MCMpdec(\Gbip)$. 
 By \cref{thm:TE:OAC}, we have $\lalat\in\lalakMAT$, and since $\datrQL$ \hasMbd, we get $\lalat\in\lalakMPMAT$. It remains to check \axrangeMCEall and~\eqref{eq:MCMS_angles_vs_k} for $(\GD,\xd)$. 
The map $\xT$ is an (\einj) \wemb of $\GD$ by \cref{thm:TOP:weak_imm=>imm}. 
By~\eqref{eq:ass_lgg}, every face of $(\GD,\xd)$ is \btrar. Thus, $\xT$ is also \finj, so \itemref{MCE1:emb} is satisfied for $(\GD,\xd)$. 
\itemref{MCE2:null_edges} is satisfied by~\eqref{eq:OCP:primitive}. 
\itemref{MCE3:M-tnn} is satisfied by \cref{lemma:TE:hasMbd=>simple_and_MCE3}. 
\itemref{MCE4:properly_colored} is satisfied by \cref{lemma:TE:angle_cond}. 
For~\itemref{MCE5:Mpos_chords}, any \pchord in $(\GD,\xd)$ is contained either in a \btrar face (in which case it is necessarily isotopic to an edge of $(\GD,\xd)$) or in the outer face of $(\GD,\xd)$ (in which case we are done by \cref{rmk:MCE:pchord_outer_face}). 
Finally,~\eqref{eq:MCMS_angles_vs_k} follows from~\eqref{eq:OCP:primitive} and~\eqref{eq:TE:weak_t_imm_dfn}.
\end{proof}

\subsection{Dimension counting}
We give a bound on the dimension of $\MCM(\Gloop)$ for an \lgg $\Gloop$. 
We start with the case where $\Gloop$ is \terminal, i.e., planar bipartite; cf. \cref{rmk:BCFW:terminal_vs_plabic_vs_plabip}. 
For a real semialgebraic set $X$, we denote by $\dimZ X$ the dimension of its Zariski closure. 

\begin{lemma}\label{lemma:OCP:dim_atr}
Assume that $\Gbip$ is a planar bipartite graph that admits an \APM. 
Then 
\begin{equation}\label{eq:OCP:dim_atr}
 \dimZ\Mdatr(\Gbip) = n+\gdimatr(\Gbip), \quad\text{where}\quad \gdimatr(\Gbip):=|\Faces| + n + 3.
\end{equation}
\end{lemma}
\begin{proof}
By \cref{lemma:MCE:apm_vs_Kast}, for any $\wt\in\Rtpgauge$, we have $\dim_{\R}\Hwspace_{\C}\HtripK = 2k$ and $\dim_{\R}\Hbspace_{\C}\HtripK = 2(n-k)$. 
 Once we have chosen $\datrQnox$, the \KSprim $\xd$ is defined up to a global shift in $\Rdd$. Taking these four extra degrees of freedom into account, we find
\begin{equation*}%
 \dimZ\Mdatr(\Gbip) = |\Faces|-1 + 2k + 2(n-k) + 4 = |\Faces| + 2n + 3. \qedhere
\end{equation*} 
\end{proof}

 Recall that 
 $\Vbigons$ denotes the set of bigonal faces of $\GD$. For any graph $\GD$ satisfying \cref{ass:lgg_dual}, set
\begin{equation}\label{eq:gdim_dfn}
 \gdim(\GD):=4|\Faces| - |\East| + |\Vbigons|.
\end{equation}
\begin{lemma}\label{lemma:ORA:terminal_vs_atr_dim_only}
If $\Gbip$ is a connected planar bipartite graph that admits an \APM and satisfies~\eqref{eq:ass_lgg} then %
 \begin{equation}\label{eq:ORA:terminal_vs_atr_dim_only}
 \dimZ\MCM(\Gbip) \leq \gdimatr(\Gbip) = \gdim(\GD).
 \end{equation}
\end{lemma}
\begin{proof}

The little group $\LGpm$ (\cref{dfn:little_group}) acts freely on $\MPkntreeMAT$. 
 Thus, $\LGpm$ acts freely on $\MCMdec(\Gbip)$. 
By \cref{rmk:MCMp->MCM_surj}, 
 the quotient map $\MCMdec(\Gbip)/\LGpm \cong \MCM(\Gbip)$ is a homeomorphism. It follows that $\dimZ\MCM(\Gbip) = \dimZ\MCMdec(\Gbip) - n$. Next, by \cref{lemma:ORA:terminal_vs_atr_no_dim,lemma:OCP:dim_atr}, $\dimZ\MCMdec(\Gbip)\leq \dimZ\Mdatr(\Gbip)=n+\gdimatr(\Gbip)$. This proves the inequality
$\dimZ\MCM(\Gbip) \leq \gdimatr(\Gbip)$ in~\eqref{eq:ORA:terminal_vs_atr_dim_only}.

It remains to show that $\gdimatr(\Gbip) = \gdim(\GD)$. 
By~\eqref{eq:ass_lgg}, $\Vint=\Vbigons\sqcup\Vtrs$. 
 Thus, $|\Edges| = \frac32|\Vtrs| + |\Vbigons| + \frac n2$. 
Since $\Gbip$ is connected, by~\eqref{eq:DIM:Euler_no_float}, $|\Faces| = 1 + |\Edges| - |\Vbigons| - |\Vtrs|$. 
 Using these two identities, 
we get $3|\Faces| - |\Edges| + |\Vbigons| = n+3$, so 
 $\gdim(\GD) = 4|\Faces| - |\Edges| + |\Vbigons| = |\Faces| + n + 3 = \gdimatr(\Gbip)$.
\end{proof}

Next, we give an upper bound on the dimension of $\MCM(\Gloop)$ when $\Gloop$ is not necessarily \terminal. 
\begin{proposition}\label{lemma:OCP:gdim_Mmce}
For any graph $\GD$ satisfying \cref{ass:lgg_dual}, 
\begin{equation}\label{eq:OCP:gdim_Mmce}
 \dimZ\Mmce(\GD)\leq \gdim(\GD).%
\end{equation}
\end{proposition}
\begin{proof}
We apply the \ORA. Let $(\GD,\xd)\in\Mmce(\GD)$ and $(\GDout,\xdout) := \ORmv(\GD,\xd)$, so that $\GDout$ is obtained from $\GD$ by adding $|\Eastout| - |\East|$ edges. Since \ORsts do not create any bigonal faces, $\gdim(\GDout) = \gdim(\GD) - |\Eastout| + |\East| + 4$. 
By \cref{lemma:ORA:new_edge_at_least_one,lemma:ORA:rigid_Chords}, 
\begin{equation}\label{eq:gdim_diff}
 |\Eastout|\geq|\East|+4,\quad\text{and thus}\quad
 \begin{cases}
 \gdim(\GDout) = \gdim(\GD), &\text{if $|\Eastout| = |\East| + 4$;}\\
 \gdim(\GDout) < \gdim(\GD), &\text{if $|\Eastout| > |\East| + 4$.}
 \end{cases}
\end{equation} 

We prove~\eqref{eq:OCP:gdim_Mmce} by induction.
For the induction base, if $(\GD,\xd)$ is terminal, its planar dual $\G$ (cf. \cref{rmk:BCFW:terminal_vs_plabic_vs_plabip}) admits an \APM by \cref{lemma:OCP:Hakimi_cor,lemma:BCFW:MCM_nonempty=>hel>=1}. Thus,~\eqref{eq:OCP:gdim_Mmce} follows from~\eqref{eq:OCP:dim_atr}--\eqref{eq:ORA:terminal_vs_atr_dim_only} for \terminal $(\GD,\xd)$. For the induction step, if $(\GD,\xd)$ is a non-terminal \MCNE then by \cref{lemma:ORA:ORA_termination}, $(\GD,\xd)$ admits a valid input datum $\BCin$. Let $(\GDout,\xdout) = \ORmv(\GD,\xd)$. By the induction hypothesis and~\eqref{eq:gdim_diff}, $\dimZ\Mmce(\GDout)\leq \gdim(\GDout) \leq \gdim(\GD)$. 
Note that $(\GD,\xd)=\ResG(\GDout,\xdout)$ 
belongs to the image of the restriction map $\ResG:\Mmce(\GDout)\to\MCMalg(\GD)$ introduced in~\eqref{eq:MCE:Res_dfn}. Thus, $\Mmce(\GD)$ is contained in a finite union of such images $\ResG\Mmce(\GDout)$, each of which has dimension at most $\gdim(\GD)$.
\end{proof}

\subsection{Combinatorial invariance of \MCEsnoacr}\label{ssec:combin_invar}
Let $\Gloop$ be \algg satisfying \cref{ass:lgg} and let $(\GD,\xd)\in\MCM(\Gloop)$. Our goal is to show that the following geometric properties of $(\GD,\xd)$ are fully determined by the combinatorics of $\Gloop$:
\begin{itemize}
\item whether a given face of $(\GD,\xd)$ is rigid/\flex/degenerate or contained in a white or a black clique;
\item whether a given corner $\cor$ of $(\GD,\xd)$ satisfies $\sumwT(\cor)\in\{0,\pi\}$ or $0<\sumwT(\cor)<\pi$, and similarly for $\sumbT(\cor)$;
\item whether two given vertices $\ff,\f\in\Faces$ satisfy $(\xd(\ff)-\xd(\f))^2>0$ or $(\xd(\ff)-\xd(\f))^2=0$;
\item whether the restriction~\eqref{eq:MCE:Res_dfn} of $(\GD,\xd)$ to a subgraph of $\GD$ is an \MCE.%
\end{itemize}

Following \cref{notn:RgFbd}, for $\Rg\subset\Vint$, let 
$\RgFacl:=\RgFac\sqcup\left(\bigsqcup_{\v\in\Rg} \{\ploc_\rho\mid\rho\in\brnLv\}\right)$
 be the set of vertices of $\GD$ 
 incident to some $\v\in\Rg$. 
\begin{proposition}\label{lemma:BCFW:clique_vs_Rg}
Let $(\GD,\xd)\in\MCM(\Gloop)$ 
 and let $\emptyset\neq\Rg_0\subset\Vint$ be such that $\Gloopgr\ind[\Rg_0]$ is connected. Then $\xd(\RgFaclx_0)$ is a white (resp., black) possibly degenerate clique 
if and only if $\Rg_0$ is contained in a \sconn subset $\Rg\subset\Vint$ satisfying $\gelW(\Rg) = 1$ (resp., $\gelB(\Rg) = 1$).
\end{proposition}
\begin{proof}%
 $(\Longleftarrow)$: Let $\Rg\supset\Rg_0$ be a \sconn set such that, say, $\gelW(\Rg) = 1$. By \cref{lemma:ORA:ORA_termination}, one can apply valid \ORsts to $(\GD,\xd)$ to obtain a terminal \MCNE $(\GDout,\xdout)$. 
 Each face of $\GD$ is a union of faces of $\GDout$, so we have naturally defined subsets $\Rgout$ and $\Rgout_0$ of $\Vintout$ consisting of all faces of $\GDout$ contained in $\Rg$ and $\Rg_0$, respectively.
By \cref{lemma:BCFW:dualizable_holess_sconn}, $\Rgout$ is \sconn. Thus, $\chind[\Rg]=\chind[\Rgout]=1$ by~\eqref{eq:DIM:Euler}. 
By~\eqref{eq:induced_angle_sums} and~\eqref{eq:Rg_sum_bd_vs_chi} applied to $(\GD,\xd)$ and $(\GDout,\xdout)$, we get $\gelWout(\Rgout) = \gelW(\Rg)$, so $\gelWout(\Rgout) = 1$. 
By \cref{rmk:BCFW:terminal_vs_plabic_vs_plabip}, the planar dual $\Glout$ of $(\GDout,\xdout)$ is planar bipartite. 
By \cref{lemma:OCP:Hakimi_cor,lemma:BCFW:MCM_nonempty=>hel>=1}, $\gelmin(\Gbipout)\geq1$ and $\Gbipout$ admits an \APM. By \cref{lemma:ORA:terminal_vs_atr_no_dim}, $\xdout$ is the \KSprim of some pair $(\Fw,\Fb)$ of discrete holomorphic functions on $\Gbipout$. 
Similarly to the proof of \cref{lemma:TOP:restr}, we see that the values of $\Fw|_{\RWbipout}$ are all real multiples of each other. Thus, each black face of $(\GDout,\xdout)$ contained in $\Rgbipout$ must be degenerate. 
It follows that $\xdout(\Faclxout[\Rgbipout])$ is a white clique, and therefore so is $\xd(\RgFacl)\subset\xdout(\Faclxout[\Rgbipout])$.

$(\Longrightarrow)$: Suppose that $\xd(\RgFaclx_0)$ is, say, a white (possibly degenerate) clique. Let $\Rg\subset\Vint$ be a maximal by inclusion subset containing $\Rg_0$ such that $\GR$ is connected and $\clique:=\xd(\RgFacl)$ is a white clique. 
If $\Rg$ is not \holess then $\Rgcl=\Rg\sqcup\Rghol$ is still connected, and since $\xT(\Faces\ind[\Rghol])\subset\Conv\KT$, $\xd(\Faces\ind[\Rgcl])$ is a white clique by \cref{lemma:MCE:clique_convex}, contradicting the maximality of $\Rg$. Thus, $\Rg$ is \holess and therefore \sconn. 
Consider the boundary vertices $(\f_1,\f_2,\dots,\f_m)$ of the simple cycle $\Cyc$ bounding $\Rg$ as in \cref{lemma:BCFW:dualizable_holess_sconn}. 
We claim that 
\begin{equation}\label{eq:BCFW:sumbTcond=0}
 \sumbTcond(\f_i|\Rg)=0 \quad\text{for each $i\in\brm$}.
\end{equation}
Indeed, 
every face $\xT(\v)$, $\v\in\Rg$, is either a bigon, a degenerate black triangle, or a white rigid face. Thus, for each $\cor\in\corners\cond(\f_i|\Rg)$, we have $\sumbT(\cor)\in\{0,\pi\}$. 
If $\sumbT(\cor)=\pi$ then we have 
$\f_i\in\Fint$ by~\eqref{eq:MCE:sumwT_sumbT_bdry_0_pi} and
$\sumbTcond(\f_i|\Rg)=\sumbT(\f_i)=\pi$ by~\eqref{eq:MCE:sumwT_sumbT=pi}. 
Thus, 
 for each $\cor'\in\corners(\f_i)\setminus \{\cor\}$, we have $\sumbT(\cor')=0$. 
Given such $\cor'$, by \cref{lemma:zero_corner=>rigid}, $\xd(\partF\corvx_{\cor'})$ is a white (possibly degenerate) clique. Let $\Rg'\supsetneq\Rg$ be obtained from $\Rg$ by adding all vertices of $\G$ incident to the face $\f_i$. Since $\f_i\in\Fint$, $\G\ind[\Rg']$ is connected. Since $\xd(\partF\v')$ is a white clique for each vertex $\v'\neq\corv$ incident to $\f_i$, $\xd(\Faces\ind[\Rg'])$ is a white clique, contradicting the maximality of $\Rg$. 
 This shows~\eqref{eq:BCFW:sumbTcond=0}. By~\eqref{eq:Rg_sum_bd_vs_chi}, we get $\gelW(\Rg) = \chind[\Rg] = 1$
 since $\Rg$ is \sconn. 
\end{proof}

\begin{corollary}\label{lemma:BCFW:clique_vs_Rg_maximal}
 Let $\Kmax=\xd(\preKmax)\in\Kmaxs$, and let $\Rg$ be the set of faces of $\GD$ contained inside the \extcycle $\CycKmax$ of $\Kmax$.
 Assume that $\GD\ind[\preKmax]$ is not a single edge as in \cref{lemma:MCE:outward_simple}. 
 Then $\Rg$ is nonempty, \sconn, satisfies $\preKmax = \RgFacl$, and if $\Kmax$ is white (resp., black) then $\gelW(\Rg)=1$ (resp., $\gelB(\Rg)=1$). In particular, if $\Kmax$ is degenerate then $\gelW(\Rg)=\gelB(\Rg)=1$.
\end{corollary}
\begin{proof}
It follows from \cref{lemma:MCE:outward_simple} (see also \cref{lemma:MCE:inside_CycKmax=>rigid}) that if $\GD\ind[\preKmax]$ is not a single edge then $\Rg\neq\emptyset$ and $\preKmax = \RgFacl$. By \cref{lemma:BCFW:dualizable_holess_sconn}, $\Rg$ is \sconn. 
The set $\Rg$ is not properly contained in any \sconn subset $\Rg'\subset\Vint$ satisfying $\gelW(\Rg')=1$ or $\gelB(\Rg')=1$ by the maximality of $\preKmax$. By \cref{lemma:BCFW:clique_vs_Rg}, we get $\gelW(\Rg)=1$ (resp., $\gelB(\Rg)=1$) when $\Kmax$ is white (resp., black). 
\end{proof}

\begin{remark}\label{rmk:BCFW:clique_vs_Rg_1_vert}
Recall from \cref{dfn:MCE:white_black_deg} that a clique is degenerate if and only if it is both white and black. Thus, the set of degenerate cliques of the form $\xd(\RgFaclx_0)$ (for $\emptyset\neq\Rg_0\subset\Vint$ such that $\Gloopgr\ind[\Rg_0]$ is connected) is fully determined by the combinatorics of $\Gloop$.
\end{remark}

Recall that an \Mdash nonnegative map $\xd:\Faces\to\Rdd$ gives rise to a graph $\NullG(\GD,\xd)$ introduced in~\eqref{eq:MCE:NullG_dfn}.
\begin{corollary}\label{lemma:BCFW:NullE_vs_NullED}
Given \algg $\Gloop$, let $\NullGD=(\Faces,\NullED)$ be a simple graph with edge set $\NullED$ consisting of all pairs $\{\ff,\f\}\subset\Faces$ such that 
either $\{\ff,\f\}\in\Ebarast$ or 
$\ff,\f\in\RgFacl$ for some \sconn subset $\emptyset\neq\Rg\subset\Vint$ satisfying $\gelW(\Rg)=1$ or $\gelB(\Rg)=1$. 
Then for all $(\GD,\xd)\in\MCM(\Gloop)$, 
\begin{equation}\label{eq:BCFW:NullE_vs_NullED}
\NullG(\GD,\xd) = \NullGD.
\end{equation}
\end{corollary}
\begin{proof}
Let $(\GD,\xd)\in\MCM(\Gloop)$. 
By~\itemref{MCE2:null_edges}, $\Ebarast\subset\NullE(\GD,\xd)$, so it suffices to consider pairs $\{\ff,\f\}\notin\Ebarast$.
 Given such a pair $\{\ff,\f\}\subset\Faces$, by \cref{lemma:MCE:trivial_\Mdash zero<=>clique}, we have $\{\ff,\f\}\in \NullE(\GD,\xd)$ if and only if $\xd(\{\ff,\f\})$ is contained in some maximal clique $\Kmax\in\Kmaxs$. 
Thus, $\{\ff,\f\}\in\NullED$ by \cref{lemma:BCFW:clique_vs_Rg_maximal}. 
 The reverse inclusion $\NullE(\GD,\xd)\supset\NullED$ follows from \cref{lemma:BCFW:clique_vs_Rg}.
\end{proof}

\begin{definition}\label{dfn:twosep_lgg}
Given \algg $\Gloop$, we say that $\ff,\f\in\Faces$ are \emph{\twosepMCE} if $\{\ff,\f\}$ is not an edge in $\NullGD$. We say that $\Gloop$ is \emph{\fullysepMCE} if for any $\sepst\in\brnLbdsep$ (cf. \cref{dfn:fullysep}), the vertices $\ploc_\seps$ and $\ploc_\sept$ are \twosepMCE.
\end{definition}
\noindent In the notation of \cref{conj:dist}, when $\Gloop$ is \terminal, it is \fullysepMCE (resp., \fullysep) if $\distcurve(\ploc_\seps,\ploc_\sept)\geq2$ (resp., $\distsep(\ploc_\seps,\ploc_\sept)\geq2$) for all $\sepst\in\brnLbdsep$. Thus, \cref{conj:dist} implies that the notions of \fullyseponMCE and \fullysepon agree for \terminal graphs. 
We prove this geometrically under the assumption that $\Gloop$ admits \MCEs. 

\begin{corollary}\label{lemma:fullysepMCE_vs_fullysep}
Suppose that $\Gloop$ is \terminal and $\MCM(\Gloop)\neq\emptyset$.
 Then $\Gloop$ is \fullysepMCE if and only if it is \fullysep.
\end{corollary}
\begin{proof}
Assume that $\Gloop$ is \fullysepMCE and let $(\GD,\xd)\in\MCM(\Gloop)$.
By \cref{rmk:MCMp->MCM_surj,lemma:TE:TE_to_MCE_and_back}, 
$(\GD,\xd)$ extends to a \wtemb $\datrQL\in\MdteoMP(\Gbip)$. 
For $\sepst\in\brnLbdsep$, since $(\xd(\ploc_\seps)-\xd(\ploc_\sept))^2>0$ by \cref{lemma:BCFW:NullE_vs_NullED}, 
 $\ploc_\seps$ and $\ploc_\sept$ are \twosep by \cref{lemma:PROP:if_not_twosep_then_M=0}.

Conversely, assume that $\Gloop$ is \fullysep. 
Since $\MCM(\Gloop)\neq\emptyset$, we have $\helmin(\G)\geq1$ by \cref{lemma:BCFW:MCM_nonempty=>hel>=1,lemma:BCFW:gelWmin_vs_helWmin}. 
Pick generic $\wt\in\Rtpgauge$ and let $C:=\Meas(\Gbip,\wt)$, $\LaLat\in\LaLaimmnn$, and $\lalat:=\PhiLL(C)$. 
Let $\datrQLll=\datrQ\in\Mdti(\Gbip)$ be the associated \wtimm (cf. \cref{thm:TE:OAC}). 
By \cref{lemma:PROP:Mpos}, $(\xll(\ploc_\seps)-\xll(\ploc_\sept))^2>0$ for all $\sepst\in\brnLbdsep$. In particular, $(\GD,\xll)$ has \Mbd, so by \cref{lemma:TOP:Mpos=>simple}, $\datrQLll$ is a \wtemb. 
By \crefi{lemma:TE:brla_nonzero_if_apm}{lak_implies4}, $\xll$ is \einj, so $\datrQLll\in\MdteoMP(\Gbip)$. 
By~\eqref{eq:ORA::mceTOwte_dfn_iso}, $(\GD,\xll)\in\MCM(\Gloop)$. Since $(\xll(\ploc_\seps)-\xll(\ploc_\sept))^2>0$ for $\sepst\in\brnLbdsep$, we have $\{\ploc_\seps,\ploc_\sept\}\notin\NullED$ by \cref{lemma:BCFW:NullE_vs_NullED}. Thus, $\Gloop$ is \fullysepMCE.
\end{proof}

\begin{lemma}\label{lemma:BCFW:corner_comparison}
Let $(\GD,\xd),(\GD,\yd)\in\MCM(\Gloop)$.
\begin{enumerate}[label=(\arabic*)]
\item\label{corner_comparison1} For any corner $\cor$ of $\GD$, we have $\sumbTx(\cor)\in\{0,\pi\} \Longleftrightarrow \sumbTy(\cor)\in\{0,\pi\}$.
\item\label{corner_comparison2} For any $\Rg\subset\Vint$ and $\f\in\Faces$, we have 
 $\sumbTx\cond(\f|\Rg)\in\{0,\pi\}\Longleftrightarrow\sumbTy\cond(\f|\Rg)\in\{0,\pi\}$.
\end{enumerate}
Analogous results hold for $\sumwTx$ and $\sumwTy$.
\end{lemma}

\begin{figure}
 \def\inputscale{2}
\begin{tabular}{cc}
\includegraphics[scale=\inputscale]{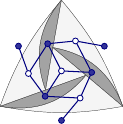}
&
\includegraphics[scale=\inputscale]{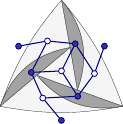}
\end{tabular}
 \caption{\label{fig:corner-0-pi} 
An example of $(\GD,\xd),(\GD,\yd)\in\MCM(\Gloop)$ such that for each interior black corner $\cor$, we have
$\sumbTx(\cor),\sumbTy(\cor)\in\{0,\pi\}$ with $\sumbTx(\cor)\neq\sumbTy(\cor)$; see \cref{ex:BCFW:corner_comparison}.
}
\end{figure}

\begin{example}\label{ex:BCFW:corner_comparison}
It is possible to have $\sumbTx(\cor)=0$ and $\sumbTy(\cor)=\pi$; see \cref{fig:corner-0-pi}.
\end{example}
\begin{proof}
By \cref{lemma:BCFW:clique_vs_Rg,rmk:BCFW:clique_vs_Rg_1_vert}, 
for $\v\in\Vint$, $\xT(\v)$ is \flex (resp., rigid, white, black, or degenerate) if and only if so is $\yT(\v)$. 
We prove~\itemref{corner_comparison1} and~\itemref{corner_comparison2} simultaneously. Let $\f\in\Faces$ and fix a partition $\corners(\f)=\corners_A(\f)\sqcup \corners_B(\f)$. Write $\sumbTx(\f|A):=\sum_{\cor\in \corners_A(\f)} \sumbTx(\cor)$ and similarly for $\sumbTx(\f|B)$, $\sumbTy(\f|A)$, and $\sumbTy(\f|B)$. 
Suppose first that $\sumbTx(\f|A)=0$; thus, $\sumbTx(\cor)=0$ for all $\cor\in\corners_A(\f)$. By \cref{lemma:zero_corner=>rigid}, for each $\cor\in\corners_A(\f)$, $\xT(\corv)$ is 
a (possibly degenerate) rigid white face. 
 Therefore, the same is true for $\yT(\corv)$, which implies $\sumbTy(\f|A)\in\{0,\pi\}$. Next, assume that $\sumbTx(\f|A)=\pi$. 
By~\eqref{eq:MCE:sumwT_sumbT_bdry_0_pi}, 
 $\f\in\Fint$. 
By~\eqref{eq:MCE:sumwT_sumbT=pi}, $\sumbTx(\f|B) = \pi - \sumbTx(\f|A) = 0$ and $\sumbTy(\f|B) = \pi - \sumbTy(\f|A)$. As we showed above, $\sumbTx(\f|B)=0$ implies $\sumbTy(\f|B)\in\{0,\pi\}$, so $\sumbTy(\f|A)\in\{0,\pi\}$. 
\end{proof}

\begin{proposition}\label{lemma:Res_of_MCE_is_MCE}
Let $\Gloop,\Glout$ be \lggs such that $\GD$ is a subgraph of $\GDout$
with the same outer boundary cycle.
 Suppose that $(\GDout,\xdout),(\GDout,\ydout)\in\MCM(\Glout)$ and $(\GD,\xd):=\ResG(\GDout,\xdout)\in\MCM(\Gloop)$. 
Then $(\GD,\yd):=\ResG(\GDout,\ydout)$ also belongs to $\MCM(\Gloop)$.
\end{proposition}

We will deduce the above result from the following lemma.

\begin{lemma}\label{lemma:Res_MCE}
Under the assumptions of \cref{lemma:Res_of_MCE_is_MCE}, we have the following. 
\begin{enumerate}[label=(\arabic*)]
\item\label{Res_MCE1} For each rigid \ptrar face $\xT(\v)$ of $(\GD,\xd)$, $\yT(\v)$ is an (embedded) \ptrgle;
\item\label{Res_MCE2} Fix a maximal clique $\preKmax\in\preKmaxs$ in $(\GD,\xd)$ and let $\Kmaxx:=\xd(\preKmax)$ and $\Kmaxy:=\yd(\preKmax)$. Then $\Conv\KTmaxx$ and $\Conv\KTmaxy$ have the same number of vertices, and 
for all $\ff\in\preKmax$, $\xT(\ff)$ is a vertex (resp., \extpt) of $\Conv\KTmaxx$ if and only if $\yT(\ff)$ is a vertex (resp., \extpt) of $\Conv\KTmaxy$.
\end{enumerate}
\end{lemma}
\begin{proof}
Let $\emptyset\neq\Rg\subset\Vint$ and let $\Rgout\subset\Vintout$ be the set of faces of $\GDout$ contained in some face $\v\in\Rg$ of $\GD$. 
For each $\f\in\Faces$, we claim that
\begin{equation}\label{eq:Rg_Rgout_angle_relation}
 \sumbT_{\xd}\cond(\f|\Rg) = \sumbT_{\xdout}\cond(\f|\Rgout),\quad
\sumbT_{\xdout}\cond(\f|\Rgout)\in\{0,\pi\}\Longleftrightarrow
\sumbT_{\ydout}\cond(\f|\Rgout)\in\{0,\pi\},
\quad
\sumbT_{\ydout}\cond(\f|\Rgout) = \sumbT_{\yd}\cond(\f|\Rg),
\end{equation}
and similarly for $\sumwT$. Indeed, we have 
$\sumbT_{\xd}\cond(\f|\Rg) = \sumbT_{\xdout}\cond(\f|\Rgout)$ 
 by~\eqref{eq:induced_angle_sums} and 
$\sumbT_{\xdout}\cond(\f|\Rgout)\in\{0,\pi\}\Longleftrightarrow
\sumbT_{\ydout}\cond(\f|\Rgout)\in\{0,\pi\}$ by \cref{lemma:BCFW:corner_comparison}. 
Since $\yTout$ is an \einj \wemb of $\GDout$, $\yT$ is an \einj \wemb of $\GD$, and so the angles $\sumwT_{\yd}(\cor),\sumbT_{\yd}(\cor)$ are well defined for all $\cor\in\corners(\GD)$ and satisfy 
$\sumbT_{\ydout}\cond(\f|\Rgout) = \sumbT_{\yd}\cond(\f|\Rg)$ and 
$\sumwT_{\ydout}\cond(\f|\Rgout) = \sumwT_{\yd}\cond(\f|\Rg)$. This shows~\eqref{eq:Rg_Rgout_angle_relation}. 

For part~\itemref{Res_MCE1}, set $\Rg:=\{\v\}$ and $\preKmax:=\RgFacl$. Since $\xT(\v)$ is a \ptrgle, $\Rg$ is \sconn by \cref{lemma:BCFW:dualizable_holess_sconn}. 
For part~\itemref{Res_MCE2}, let $\Rg:=\{\v\in\Vint\mid \xT(\partF\v)\subset\Conv\KTmaxx\}$. 
If $\GD\ind[\preKmax]$ is a single edge, the result is trivial, so we assume that $\GD\ind[\preKmax]$ is not a single edge, in which case $\Rg$ is nonempty and \sconn by \cref{lemma:BCFW:clique_vs_Rg_maximal}. 
For parts~\itemref{Res_MCE1}--\itemref{Res_MCE2}, let $\Cyc=(\f_1,\dots,\f_m)$ be the cycle bounding $\Rg$ in $\GD$.
 By \cref{lemma:BCFW:dualizable_holess_sconn}, $\Rgout$ is also \sconn. 
Since $\chind[\Rg]=\chind[\Rgout]=1$, we get $\gelW(\Rg)=\gelWout(\Rgout)$ by~\eqref{eq:Rg_Rgout_angle_relation} and~\eqref{eq:Rg_sum_bd_vs_chi} applied to $(\GD,\xd)$ and $(\GDout,\xdout)$. By~\eqref{eq:Rg_Rgout_angle_relation}, we see that~\eqref{eq:Rg_sum_bd_vs_chi} also holds for $(\GD,\yd)$. Thus, 
{
\thinmuskip=2mu
\begin{equation}\label{eq:Rg_Rgout_bdry_angle_sum}
 \sum_{i=1}^m \sumbT_{\xd}\cond(\f_i|\Rg) = \sum_{i=1}^m \sumbT_{\yd}\cond(\f_i|\Rg) = (\gelW(\Rg)-1)\pi,
 \ \ \ 
 \sum_{i=1}^m \sumwT_{\xd}\cond(\f_i|\Rg) = \sum_{i=1}^m \sumwT_{\yd}\cond(\f_i|\Rg) = (\gelB(\Rg)-1)\pi.
\end{equation}
}

We show part~\itemref{Res_MCE1}. Suppose that the clique $\Kmaxx=\xd(\partF\v)$ is white.
 Thus, $\sumbT_{\xd}\cond(\f_i|\Rg)\in\{0,\pi\}$ and $\sumwT_{\xd}\cond(\f_i|\Rg)\in(0,\pi)$ for all $i\in\brm$. By~\eqref{eq:Rg_Rgout_angle_relation}, the same is true for $\sumbT_{\yd}\cond(\f_i|\Rg)$ and $\sumwT_{\yd}\cond(\f_i|\Rg)$.
By \cref{lemma:BCFW:ptrar_face_k=2} applied to $(\GD,\xd)$, we get $\gelW(\v)=m-2$. By~\eqref{eq:Rg_Rgout_bdry_angle_sum}, 
 $\sum_{i=1}^m \sumbT_{\yd}\cond(\f_i|\Rg) = (m-3)\pi$. 
 Thus, exactly $3$ corners of $\yT(\v)$ satisfy $\sumbT_{\yd}(\cor_i)=0$ and the rest satisfy $\sumbT_{\yd}(\cor_i)=\pi$. 
Since $\gelW(\v)=m-2$ and $\gelB(\v)=2$, by~\eqref{eq:Rg_Rgout_bdry_angle_sum}, we get $\TURN(\yT(\v))=-2\pi$. 
Since $\sumwT_{\yd}(\cor_i)\in(0,\pi)$ for all $i\in\brm$, by \cref{rmk:MCE:angles_imply_ptrgle}, $\yT(\v)$ is an (embedded) \ptrgle. 

We show part~\itemref{Res_MCE2}. Suppose that the clique $\Kmaxx$ is white. Thus, for each $\f\in\preKmax$, we have $\sumbT_{\xd}\cond(\f|\Rg)\in\{0,\pi\}$. We have $\sumwT_{\xd}\cond(\f|\Rg)\in(0,\pi)$ when $\xT(\f)$ is a vertex of $\Conv\KTmaxx$ and $\sumwT_{\xd}\cond(\f|\Rg)\in\{0,\pi\}$ otherwise. By~\eqref{eq:Rg_Rgout_angle_relation}, the same is true for $\sumbT_{\yd}\cond(\f|\Rg)$ and $\sumwT_{\yd}\cond(\f|\Rg)$. 
 By \cref{lemma:BCFW:clique_vs_Rg,lemma:BCFW:clique_vs_Rg_maximal}, $\gelW(\Rg) = 1$. As we showed above, $\gelWout(\Rgout)=\gelW(\Rg)$, so $\gelWout(\Rgout)=1$. 
If $\Kmaxx$ is degenerate then by \cref{rmk:MCE:clique_boundary_outward_bent}, $\Cyc=\CycKmaxx$ is a $2$-cycle, in which case the result follows trivially. From now on, we assume that $\Kmaxx=\xd(\RgFacl)$ is a nondegenerate white clique. 
By \cref{lemma:BCFW:clique_vs_Rg}, the same is true for $\xdout(\RgoutFacl)$ and $\ydout(\RgoutFacl)$ because $\gelWout(\Rgout)=1$. Therefore, $\Kmaxy=\yd(\RgFacl)$ is a white clique. 
Since $\yTout$ is a \wemb of $\GDout$, $\yTout(\RgoutFacl)$ is contained inside the closed polygonal chain $\yTout(\CycKmaxx)$. 
For $i\in\brm$ such that $\f_i$ is a vertex of $\Conv\KTmaxx$, we have $\sumwT_{\xd}\cond(\f_i|\Rg)\in(0,\pi)$ and thus $\sumwT_{\yd}\cond(\f_i|\Rg)\in(0,\pi)$ by~\eqref{eq:Rg_Rgout_angle_relation}. 
Similarly, $\sumbT_{\yd}\cond(\f_i|\Rg)\in\{0,\pi\}$. 
In particular, $\yTout(\CycKmaxx)$ is not contained in a single line. Thus, the white clique $\Kmaxy$ is also nondegenerate.

Since $\gelW(\Rg)=1$,~\eqref{eq:Rg_Rgout_bdry_angle_sum} yields $\sumbT_{\xd}\cond(\f_i|\Rg)=\sumbT_{\yd}\cond(\f_i|\Rg)=0$ for all $i\in\brm$.
Thus, $\yT(\CycKmaxx)$ is a \wcp (\cref{dfn:DIM:degenerate_convex_polygon}).
Let $I:=\{i\in\brm\mid \xT(\f_i)\text{ is a vertex of }\Conv\KTmaxx\}$ and $J:=\brm\setminus I$.
By~\eqref{eq:Rg_Rgout_angle_relation}, we have $\sumwT_{\yd}\cond(\f_i|\Rg)\in(0,\pi)$ for $i\in I$ and $\sumwT_{\yd}\cond(\f_j|\Rg)\in\{0,\pi\}$ for $j\in J$. 
Since $\Kmaxy$ is nondegenerate, we must have $\sumwT_{\yd}\cond(\f_j|\Rg)=\pi$ for $j\in J$. 
Thus, the set of vertices of $\Conv\KTmaxy$ is given by $\{\yT(\f_i)\mid i\in I\}$. 
Let $j\in J$. Since $\f_j$ is an \extpt of $\Conv\KTmaxx$, $\f_j$ must be connected to some vertex $\ff_j\notin\preKmax$ by an edge of $\GD$. By \cref{lemma:BCFW:NullE_vs_NullED}, 
$\NullG(\GDout,\xdout) = \NullG(\GDout,\ydout)$, and thus the induced subgraphs $\NullG(\GD,\xd)=\NullG(\GD,\yd)$ also coincide. It follows that $\yd(\{\ff_j\}\sqcup\preKmax)$ is not a clique, so $\yT(\ff_j)\notin\Conv\KTmaxy$ by \cref{lemma:MCE:clique_convex}. Thus, $\yT(\f_j)$ is an \extpt of $\Conv\KTmaxy$. Finally, if $\xT(\f)$, $\f\in\preKmax$, is not an \extpt of $\Conv\KTmaxx$ then it is located strictly inside the cycle $\CycKmaxx$, 
 so $\f$ is not an \extpt of $\Conv\KTmaxy$. 
\end{proof}

\begin{proof}[Proof of \cref{lemma:Res_of_MCE_is_MCE}]
In order to apply \cref{lemma:MCE:Res}, we need to check that $\yT$ is \finj. 
Suppose otherwise that for some $\v\in\Vint$ and distinct $\ff,\f\in\partF\v$, we have $\yT(\ff)=\yT(\f)$. 
Since $\ydout$ is \Mdash nonnegative, so is $\yd$. Thus, $\yd(\ff)=\yd(\f)$. 
 We apply \cref{lemma:MCE:face_classification} to the face $\xT(\v)$. If $\xT(\v)$ is a \btrgle then $\ff$ and $\f$ are connected by an edge in $\GDout$, so $\ydout$ is not \finj, a contradiction. 
If $\xT(\v)$ is a \ptrgle then by \crefi{lemma:Res_MCE}{Res_MCE1}, $\yT(\v)$ is also an (embedded) \ptrgle, a contradiction. Finally, suppose that $\v$ is \flex. 
Recall that the graphs $\NullG(\GD,\xd) = \NullG(\GD,\yd)$ coincide because $\NullG(\GDout,\xdout) = \NullG(\GDout,\ydout)$ by \cref{lemma:BCFW:NullE_vs_NullED}.
Thus, $\{\ff,\f\}$ is a clique in $(\GD,\xd)$. 
Let $\preKmax\in\preKmaxs$ be a maximal clique in $(\GD,\xd)$ containing $\{\ff,\f\}$. By \cref{lemma:MCE:non_vertex_is_not_incident_to_active_face}, $\xT(\ff)$ and $\xT(\f)$ are vertices of $\Conv\KTmaxx$. 
These vertices are distinct by \cref{lemma:MCE:Conv_vertices_injective}. 
By \crefi{lemma:Res_MCE}{Res_MCE2}, $\yT(\ff)$ and $\yT(\f)$ are (necessarily distinct) vertices of $\Conv\KTmaxy$, a contradiction. We have shown that $\yT$ is \finj. By \cref{lemma:MCE:Res}, $(\GD,\yd)$ is a \pcMNNE. 

Let $\v\in\Vint$, $\Rg:=\{\v\}$, and let $\Rgout\subset\Vintout$ be the set of faces of $\GDout$ contained inside $\v$. 
Combining~\eqref{eq:Rg_Rgout_angle_relation} with~\eqref{eq:MCMS_angles_vs_k} applied to $(\GD,\xd)$ and~\eqref{eq:Rg_sum_bd_vs_chi} applied to $(\GDout,\xdout)$ and $(\GDout,\ydout)$, we conclude that~\eqref{eq:MCMS_angles_vs_k} is also satisfied for $(\GD,\yd)$. 
 It remains to show that $(\GD,\yd)$ satisfies~\itemref{MCE5:Mpos_chords}. Consider a \pchord 
$(\yT(\ff),\yT(\f),\bendh')$ in $(\GD,\yd)$. 
Since $\GD$ and $\GDout$ have the same boundary cycle, if $(\yT(\ff),\yT(\f),\bendh')$ is contained in the outer face of $\yT(\GD)$ then it is a \pchord of $(\GDout,\ydout)$, so it must be isotopic to a boundary edge by \cref{rmk:MCE:pchord_outer_face}. Suppose now that this \pchord is contained inside some interior face $\yT(\v)$, $\v\in\Vint$. If $\xT(\v)$ is a \btrgle then $(\yT(\ff),\yT(\f),\bendh')$ is isotopic to some edge of $(\GD,\yd)$. If $\xT(\v)$ is a rigid \ptrgle then by \crefi{lemma:Res_MCE}{Res_MCE1}, so is $\yT(\v)$. By \cref{lemma:MCE:ptrar_no_chords}, any \pchord inside $\yT(\v)$ is isotopic to an edge of $\yT(\v)$. Finally, assume that $\v$ is \aflex face of $(\GD,\xd)$. 
 As before, %
 letting $\preKmax\in\preKmaxs$ be a maximal clique in $(\GD,\xd)$ containing $\{\ff,\f\}$, 
we see that $\yT(\ff)$ and $\yT(\f)$ are distinct vertices of $\Conv\KTmaxy$, so $[\yT(\ff),\yT(\f)]\subset\Conv\KTmaxy$. 
Since $\v$ is \flex, it lies outside $\CycKmaxx$. 
Since $\yT(\CycKmaxx)$ is the \extcycle of $\Kmaxy$ by \crefi{lemma:Res_MCE}{Res_MCE2}, we see that $(\yT(\ff),\yT(\f),\bendh')$ must be isotopic to one of the edges of $\yT(\CycKmaxx)$. 
Thus, $(\GD,\yd)$ satisfies~\itemref{MCE5:Mpos_chords}.
\end{proof}

We apply \cref{lemma:Res_of_MCE_is_MCE} to show that generic \MCEs are injective as maps $\Faces\to\Rdd$. 

\begin{lemma}\label{lemma:injective_MCE_exists}
Assume that $\MCM(\Gloop)\neq\emptyset$. 
Then $\MCM(\Gloop)$ contains an open dense subset consisting of $(\GD,\yd)\in\MCM(\Gloop)$ such that $\yT:\Faces\to\C$ is injective.
\end{lemma}
\begin{proof}
Let $(\GD,\xd)\in\MCM(\Gloop)$. We first prove the result assuming $\Gloop$ is \terminal. 
We use convex combination mappings of~\cite{Floater} similarly to \cref{ssec:TOP:proof_of_weak_imm=>imm}. 
By~\itemref{MCE1:emb}, $\xT$ is \finj. Consider the following directed graph $\Gfl$ with vertex set $\Faces$. For each degenerate triangular face $\v$ of $(\GD,\xd)$, let $\partF\v=\{\ff_1,\ff_2,\ff_3\}$, where $\xT(\ff_2)$ is the \emph{middle vertex} of $\xT(\v)$ meaning $\xT(\ff_2)\in\open[\xT(\ff_1),\xT(\ff_3)]$. 
In this case, we add arrows $\ff_1\leftarrow\ff_2\to\ff_3$ to $\Gfl$. 
We denote by $\NeighGfl(\ff)$ the set of outgoing neighbors of $\ff\in\Faces$ in $\Gfl$.
 We let $\Fflsink:=\{\ff\in\Faces\mid \NeighGfl(\ff)=\emptyset\}$ be the set of sinks of $\Gfl$ and let $\Fflmid:=\Faces\setminus\Fflsink$ be the set of middle vertices of degenerate triangles in $(\GD,\xd)$. 
The graph $\Gfl$ is closely related to the fully collapsed graph $\CollGWB$ introduced in \cref{dfn:DIM:CollGBW}; see~\eqref{eq:CollW=FflWsink} below. 

For $\ff\in\Fflmid$, $\xT(\NeighGfl(\ff))$ is a degenerate clique contained in some line $\elline_{\ff}\subset\C$.
 Let $\Gfl\ind[\elline_{\ff}]$ be the induced subgraph of $\Gfl$ with vertex set $\Faces\ind[\elline_{\ff}]:=\{\f\in\Faces\mid \xT(\f)\in\elline_{\ff}\}$. 
Consider the set 
$\clNFL(\ff):=\{\f\in\Faces\ind[\elline_{\ff}]\mid \Gfl\ind[\elline_{\ff}]\text{ contains a directed path from $\ff$ to $\f$}\}$. We claim that $\clNFL(\ff)$ contains exactly two sinks of $\Gfl\ind[\elline_{\ff}]$, denoted $\ff_+,\ff_-$, and that $\xT(\ff)\in\open[\xT(\ff_-),\xT(\ff_+)]$. To see this, suppose that the line $\elline_{\ff}$ is horizontal. 
Since $\ff\in\Fflmid$, $\Gfl\ind[\elline_{\ff}]$ contains arrows $\ff_1\ot \ff \to\ff_3$ with $\ff_1$ and $\ff_3$ located to the left and to the right of $\ff$. Continuing in this fashion, we find directed paths $\Path_\pm$ in $\Gfl\ind[\elline_{\ff}]$ connecting $\ff$ to sinks $\ff_\pm\in\clNFL(\ff)$ located to the left and to the right of $\ff$. 
 Suppose that $\clNFL(\ff)$ contains some other sink $\f$ of $\Gfl\ind[\elline_{\ff}]$, and assume that $\f$ is, say, to the right of $\ff$. Up to swapping $\ff_+$ and $\f$, we may assume that $\ff_+$ is to the right of $\f$. Let $z:=\xT(\f)$. The path $\Path_+$ contains either a vertex $\fff\neq\f$ such that $z=\xT(\fff)$ or an edge $\east$ such that $z\in\xTint(\east)$. In either case, by \cref{lemma:MCE:sticky}, $\xT(\f)$ must be the middle vertex of some degenerate triangle contained in $\elline_{\ff}$, so $\f$ is not a sink in $\Gfl\ind[\elline_{\ff}]$, a contradiction.

Let $\Gflbar$ be the directed graph with vertex set $\Faces$ and arrows $\ff_-\leftarrow\ff\to\ff_+$ for each $\ff\in\Fflmid$. 
Since $\xT(\ff)\in\open[\xT(\ff_-),\xT(\ff_+)]$, let $t_{\ff}\in(0,1)$ be such that $\xT(\ff)=(1-t_{\ff})\xT(\ff_-) + t_{\ff} \xT(\ff_+)$. Since $\clNFL(\ff)$ is a degenerate clique, $\xd(\ff)=(1-t_{\ff})\xd(\ff_-) + t_{\ff} \xd(\ff_+)$. 
Using the discrete maximum principle as in \cref{ssec:TOP:proof_of_weak_imm=>imm}, we see that for each $\ff\in\Fflmid$, 
 $\xd:\Faces\to\Rdd$ is fully determined\footnote{The coefficients expressing $\xd$ in terms of $\xd|_{\Fflsink}$ and $\bt$ are closely related to the natural harmonic measure for a random walk on the associated \emph{T-graph}; see~\cite[Section~3.2]{Kenyon_Sheffield} or~\cite[Section~4]{CLR1}.} by the parameters $\xd|_{\Fflsink}\in(\Rdd)^{|\Fflsink|}$ and $\bt:=(t_{\ff})_{\ff\in\Fflmid}\in(0,1)^{|\Fflmid|}$. For $\bt'\in(0,1)^{|\Fflmid|}$, let $\yd:\Faces\to\Rdd$ be determined similarly by $\yd|_{\Fflsink} := \xd|_{\Fflsink}$ and $\bt'$. We claim that when $\bt'$ belongs to a small open neighborhood of $\bt$, we have $(\GD,\yd)\in\MCM(\Gloop)$, and when $\bt'$ is generic, $\yT:\Faces\to\C$ is injective. 

Since $\xT$ is \finj, so is $\yT$ (for small $|\bt'-\bt|$). By construction, each degenerate triangle of $(\GD,\xd)$ is also degenerate in $(\GD,\yd)$. Choosing the relative orderings $\preceq_z$ (cf. \cref{rmk:MCE:near_emb_combin}) to be the same for $\xd$ and $\yd$, we see that $\yT$ is a \wemb. Thus, $(\GD,\yd)$ satisfies~\itemref{MCE1:emb}. Suppose that $\{\ff,\f\}$ is a clique in $(\GD,\xd)$. Let 
$\preKmaxfl$ consist of all $\fff\in\Faces$ such that $\Gfl$ contains a directed path from either $\ff$ or $\f$ to $\fff$. 
Iterating \cref{lemma:MCE:clique_union}, we see that $\preKmaxfl$ is a clique in $(\GD,\xd)$. 
Thus, $\preKmaxfl$ is also a clique in $(\GD,\yd)$ since for each $\fff\in\preKmaxfl$, we have $\yd(\fff)\in\Conv\xd(\preKmaxfl)$. In particular, for all $\ff,\f\in\Faces$, either $(\xd(\ff)-\xd(\f))^2=(\yd(\ff)-\yd(\f))^2=0$ or $(\xd(\ff)-\xd(\f))^2>0$, in which case $(\yd(\ff)-\yd(\f))^2>0$ for small $|\bt'-\bt|$. 
It follows that $(\GD,\yd)$ satisfies~\itemref{MCE2:null_edges}--\itemref{MCE3:M-tnn}. 
Since the sets of ambiguous corners and their colorings coincide for $(\GD,\xd)$ and $(\GD,\yd)$, and since $\sumwT(\f),\sumbT(\f)$ are locally constant, we see that~\itemref{MCE4:properly_colored} is satisfied for $(\GD,\yd)$. 
 Since $\Gloop$ is \terminal and $(\GD,\yd)$ has the same boundary as $(\GD,\xd)$,~\itemref{MCE5:Mpos_chords} and~\eqref{eq:MCMS_angles_vs_k} are vacuously satisfied for $(\GD,\yd)$. We have shown that $(\GD,\yd)\in\MCM(\Gloop)$ for small $|\bt'-\bt|$. 
By \cref{lemma:MCE:non_inj_edge,lemma:MCE:sticky}, for any distinct $\ff,\f\in\Faces$ such that $\xT(\ff)=\xT(\f)$, we have $\ff,\f\in\Fflmid$. Thus, for generic $\bt'$, we have $\yT(\ff)\neq\yT(\f)$, so $\yT:\Faces\to\C$ is injective. 

If $\Gloop$ is not \terminal, we apply the \ORA to $(\GD,\xd)\in\MCM(\Gloop)$ to obtain a \terminal \MCE $(\GDout,\xdout)$. 
As we showed above, for generic $(\GDout,\ydout)\in\MCM(\Glout)$, $\yTout:\Fout\to\C$ is injective. 
 By \cref{lemma:Res_of_MCE_is_MCE}, $\ResG\MCM(\Glout)\subset\MCM(\Gloop)$. 
 Thus, for generic points $(\GD,\yd)\in\MCM(\Gloop)$, $\yT:\Faces\to\C$ is injective. 
\end{proof}

\section{BCFW tilings of \MCMSsTITLE}\label{sec:BCFW}

We
 formulate the \emph{loop BCFW recursion} of~\cite[Section~4.2]{AHBC_all_loop} (see also~\cite{BCFW,abcgpt}) and prove that the corresponding \emph{BCFW tiles} form a tiling of the underlying \MCMS; see \cref{thm:BCFW:triang_MCE}.
We will deduce the BCFW tiling results for loop amplituhedra in \cref{sec:finish}.

\subsection{BCFW steps}\label{sec:BCFW_steps}

Let $\Gloop$ be \algg satisfying \cref{ass:lgg}. 
 Each step of the BCFW recursion depends on a fixed choice of \aBCFWid $\BCin=(\cc)$ for $\GD$ as in \cref{dfn:OR_move}. 
 Recall that for a corner $\corner$ of $\GD$, the set $\Pivots(\corner)\subset\partF\corv$ was defined in~\eqref{eq:Pivots_dfn}. 

\begin{definition}[BCFW step]\label{dfn:BCFW:step}
For a fixed \BCFWid $\BCin=(\corner,\conven)$ for $\GD$, let $\BCmvs(\Gloop)$ be the set of all \lggs $\Glout$ satisfying conditions~\itemref{BCFWstep1}--\itemref{BCFWstep4} below.
\setlength{\leftmargini}{60pt}
\begin{enumerate}[label=(BCFW\arabic*)]
\item\label{BCFWstep1} The planar dual $\GDlout$ of $\Glout$ is obtained from $\GD$ by adding one extra vertex $\ffout$ of degree~$4$, connected to $\corfp$, $\corf$, $\corfm$, and exactly one other vertex $\fpiv\in\Pivots(\corner)$. 
\end{enumerate}
 The faces of $\GDlout$ are denoted as in \cref{fig:BCFW-vertices}. We refer to the case where $\fpiv$ 
is isolated (resp., not isolated) in $\GD$ as the \emph{non-splitting} (resp., \emph{splitting}) case, shown in \figref{fig:BCFW-vertices}(left) (resp., \figref{fig:BCFW-vertices}(right)).
We impose the following conditions on the functions $\gelWout,\gelBout:\Vintout\to\Z_{\geq1}$.
\begin{enumerate}[label=(BCFW\arabic*)]\setcounter{enumi}{1}
\item\label{BCFWstep2} $\gelWout(\vv) =\gelW(\vv)$ and $\gelBout(\vv) = \gelB(\vv)$ for all $\vv\in\Vint\setminus\{\corv\}$.
\item\label{BCFWstep3} %
$\gelcpmout(\uLR) = 1$. (In other words, if $\conven=\BWcon$ then $\uL$ is black and $\uR$ is white, and vice versa.)
\item\label{BCFWstep4} In the non-splitting case, $\gelWout(\vout) = \gelW(\corv) + 1$ and $\gelBout(\vout) = \gelB(\corv) + 1$. In the splitting case, 
\begin{equation}\label{eq:BCFW:step_split_hel}
 \gelWout(\vL) + \gelWout(\vR) = \gelW(\corv) + 1,\quad 
 \gelBout(\vL) + \gelBout(\vR) = \gelB(\corv) + 1, \quad\text{and}\quad 
 \gelcpmout(\vLR)\geq2.
\end{equation}
\end{enumerate}
\end{definition}

\begin{remark}
The distinguished vertices $\Puncs:=(\ploc_1,\ploc_2,\dots,\ploc_\nL)$ of $\GD$ are also the distinguished vertices of $\GDlout$, so the loop data of $\Glout$ can be read off directly from $\GDlout$ following \cref{rmk:BCFW:dual_loop_data_extraction}. In the non-splitting case with $\fpiv = \ploc_{\rhopiv}$, we have $\brnLcorv = \brnLvout\sqcup\{\rhopiv\}$, so that the \Lfuncture $\rhopiv$ is floating in $\Gloop$ but fixed in $\Glout$. In the splitting case, we have $\brnLcorv = \brnLvL\sqcup\brnLvR$.
\end{remark}

Our next goal is to relate BCFW steps 
 to \ORsts introduced in \cref{sec:ORA_overview}.
\begin{definition}\label{dfn:BCin-generic}
Let $\BCin$ be an input datum for $\GD$. 
We say that $(\GD,\xd)\in\MCM(\Gloop)$ is \emph{$\BCin$-generic} if $\BCin$ is a valid input datum for $(\GD,\xd)$ and $(\GDout,\xdout):=\ORmv(\GD,\xd)$ is obtained from $(\GD,\xd)$ by adding exactly $|\Eastout| - |\East|=4$ edges; cf.~\eqref{eq:gdim_diff}. 
We denote $\MCMgen(\Gloop):=\{(\GD,\xd)\in\MCM(\Gloop)\mid (\GD,\xd)\text{ is $\BCin$-generic}\}$.
\end{definition}

\begin{definition}
For \lggs $\Gloop,\Glout$ and an input datum $\BCin$ for $\GD$, let
\begin{equation}\label{eq:MCMgen_Gl->Glout_dfn}
 \MCMgen(\Gloop\to\Glout):=\{(\GD,\xd)\in\MCMgen(\Gloop)\mid \ORmv(\GD,\xd)\in\MCM(\Glout)\}.
\end{equation}
\end{definition}

\begin{proposition}\label{lemma:BCFW:BCFW_vs_OR_step_forward}
Let $(\GD,\xd)\in\MCM(\Gloop)$, and let $\BCin$ be \aBCFWid for $\GD$. Then 
\begin{equation}\label{eq:BCFW_vs_OR_step_forward}
 \MCMgen(\Gloop) = \bigsqcup_{\Glout\in\BCmvs(\Gloop)} \MCMgen(\Gloop\to\Glout).
\end{equation}
\end{proposition}

\begin{proof}
By~\eqref{eq:MCMgen_Gl->Glout_dfn}, 
the left-hand side of~\eqref{eq:BCFW_vs_OR_step_forward} contains the right-hand side. 
Conversely, fix $(\GD,\xd)\in\MCMgen(\Gloop)$ and let $(\GDout,\xdout) := \ORmv(\GD,\xd)$. 
By \cref{rmk:BCFW:ass_propagates}, $\GDout$ satisfies both conditions in \cref{lemma:BCFW:ass_on_GD}. 
By \cref{lemma:BCFW:from_MCM_to_Glout}, there exists a unique \lgg $\Glout$ compatible with $(\GDout,\xdout)$, i.e., such that $(\GD,\xd)\in\MCMgen(\Gloop\to\Glout)$. Thus, the union on the right-hand side of~\eqref{eq:BCFW_vs_OR_step_forward} is disjoint. 
It remains to show that $\Glout\in\BCmvs(\Gloop)$, i.e., that $\Glout$ satisfies \itemref{BCFWstep1}--\itemref{BCFWstep4}. 

\itemref{BCFWstep1} is satisfied by \cref{dfn:BCin-generic}. \itemref{BCFWstep2}--\itemref{BCFWstep3} are satisfied by construction. 
Let $\Rgout$ be the set of faces of $\GDout$ contained inside $\corv$. 
In view of~\eqref{eq:MCMS_angles_vs_k}, we compare $\sumbT(\corv)=(\gelW(\corv)-1)\pi$ to 
$\sumbT(\Rgout):=\sum_{\vvout\in\Rgout}\sumbT(\vvout) = \sum_{\vvout\in\Rgout}(\gelWout(\vvout)-1)\pi$. 
The corners contributing to each sum are related by~\eqref{eq:induced_angle_sums}. 
In the splitting (resp., non-splitting) case, the corners incident to $\ffout$ (resp., to $\ffout$ and $\fout$) contribute to $\sumbT(\Rgout)$ but not to $\sumbT(\corv)$, so by~\eqref{eq:MCE:sumwT_sumbT=pi}, $\sumbT(\Rgout)-\sumbT(\corv)$ equals $\pi$ (resp., $2\pi$). 
Since the vertices $\uLR$ are trivalent and of opposite color, 
 $(\gelWout(\uL)-1) + (\gelWout(\uR)-1) = 1$. Substituting this into $\sumbT(\Rgout)-\sumbT(\corv)$, we find $\gelW(\corv) = \gelWout(\vL) + \gelWout(\vR) - 1$ (resp., $\gelW(\corv) = \gelWout(\vout) - 1$). 

To complete the proof of~\itemref{BCFWstep4}, it remains to check that in the splitting case, $\gelcpmout(\vLR)\geq2$. 
We first prove it in the case when $\BCin$ is \flex. 
 Suppose that we have, say, $\conven=\BWcon$. Thus, $\uL$ is black, $\uR$ is white. We show that $\gelBout(\vL)\geq2$. Suppose otherwise that $\gelBout(\vL) = 1$, i.e., $\vL\in\BVintout$. The faces $\uL$ and $\vL$ of $\GDout$ are both black and share an edge, so $\xdout(\{\corfp,\corf,\ffout,\fpiv\})$ is a clique. Let $\Kbig\in\Kbigs$ be a maximal clique containing it. 
 By \cref{lemma:ORA:ORA_rigid_faces_descr}, $\vL$ must be an external degenerate triangular face, so by \cref{dfn:ORA:coloring}, the colors of $\vL,\uL$ must be different, a contradiction. 
In the case of rigid $\BCin$, $\gelcpmout(\vLR)\geq2$ follows immediately since $\colop(\vL)\neq\colop(\uL)$ and $\colop(\vR)\neq\colop(\uR)$ by \cref{dfn:ORA:rigid_GDout_descr}. 
\end{proof}

\subsection{Loop BCFW recursion and \oraTITLE}
\label{sec:loop_BCFW_rec}

 Let $\Glinit$ be an \lgg of type $\knl$ containing a single interior vertex $\vo$ connected to $n$ boundary vertices, with $\gelW(\vo)=k$, $\gelB(\vo) = n-k$, and $\brnLvo = \brnL$. 
Let $\GD_0$ be the planar dual of $\Glinit$, i.e., an $n$-cycle with $\nL$ isolated vertices inside. 

\begin{algorithm}[Loop BCFW recursion]\label{dfn:BCFW:rec}
Fix $n\geq4$, $2\leq k\leq n-2$, and $\nL\geq0$. 
Let $\BCGs$ be a collection of \terminal \lggs of type $\knl$ obtained as follows.
\begin{enumerate}[label=(\arabic*)]
\item\label{dfn:BCFW1} Set $\Gbf_0:=\{\Glinit\}$.
\item\label{dfn:BCFW2} For $t=0,1,\dots$, add each \terminal graph $\Gloop\in\Gbf_{t}$ to $\BCGs$. 
For each non-\terminal graph $\Gloop\in\Gbf_t$, choose \aBCFWid $\BCin_{\Gloop}$. Set $\Gbf_{t+1}:=\bigcup_{\Gloop} \BCmvsx_{\BCin_{\Gloop}}(\Gloop)$, where the union is taken over non-\terminal graphs $\Gloop\in\Gbf_t$.
\item\label{dfn:BCFW3} Repeat~\itemref{dfn:BCFW2} until $\Gbf_{t+1}$ becomes empty.
\end{enumerate}
\end{algorithm}

Note that each BCFW step (cf. \cref{dfn:BCFW:step}) preserves the type $\knl$ of \algg. Furthermore, one can show similarly to the proof of \cref{lemma:ORA:ORA_termination} that the loop BCFW recursion always terminates after a finite number of steps.

\begin{remark}
The above description of the loop BCFW recursion matches the one given in~\cite[Section~2.6]{abcgpt}; see also~\cite[Section~4.2]{AHBC_all_loop}.
\end{remark}

 From now on, we fix one collection $\BCGs$ produced by \cref{dfn:BCFW:rec}. In particular, for each non-\terminal graph $\Gloop$ appearing in \cref{dfn:BCFW:rec}, we fix a choice of \aBCFWid $\BCin_{\Gloop}$. 

\begin{definition}[Branch of the \ORA]\label{dfn:BCFW:exc}
Fix a choice of \aBCFWid $\BCin_{\Gloop}$ for \emph{any} non-\terminal \lgg $\Gloop$, extending the choices made in \cref{dfn:BCFW:rec}. 
For $(\GD_0,\xd_0)\in \MCM(\Glinit)$, we apply the \ORA according to these choices. 
Explicitly, for $t=0,1,\dots$, let $(\GD_{t+1},\xd_{t+1}):=\ORmvt(\GD_t,\xd_t)$, where $\BCin_t:=\BCin_{\Gloopsub_t}$ and $\Gloopsub_t$ is obtained from $(\GD_t,\xd_t)$ via \cref{lemma:BCFW:from_MCM_to_Glout}. 
(Note that $\BCin_t$ is automatically valid for $(\GD_t,\xd_t)$ by \cref{lemma:ORA:iccar_valid}.) 
Continue this process for $t=0,1,\dots,\Tmax-1$ inductively until the graph $\Gfin:=\Gloopsub_\Tmax$ is \terminal. Thus, each $(\GD_0,\xd_0)\in\MCM(\Glinit)$ gives rise to a \emph{branch}
\begin{equation}\label{eq:BCFW:BCFW_branch}
 \Gloopsub_0\xrightarrow{\BCin_0} \Gloopsub_1 \xrightarrow{\BCin_1}\cdots\xrightarrow{\BCin_{\Tmax-1}} \Gloopsub_\Tmax=\Gfin,
\quad
(\GD_0,\xd_0)\xrightarrow{\ORmvx_{\BCin_0}}(\GD_1,\xd_1)\xrightarrow{\ORmvx_{\BCin_1}}\cdots
\xrightarrow{\ORmvx_{\BCin_{\Tmax-1}}}(\GD_\Tmax,\xd_\Tmax)
\end{equation}
of the \ORA. 
For a \terminal graph $\Gfin$, we let $\MCMto(\Gfin)$ be the set of $(\GD_0,\xd_0)\in\MCM(\Glinit)$ such that the above process terminates in $\Gfin=\Gloopsub_\Tmax$. We set
\begin{equation}\label{eq:MCMo_Zexc_dfn}
 \MCMo(\Glinit):=\bigsqcup_{\Gfin\in\BCGs} \MCMto(\Gfin)
 \quad\text{and}\quad
 \Zexc:=\MCM(\Glinit)\setminus\MCMo(\Glinit).
\end{equation}
\noindent 
In other words, $\Zexc = \bigsqcup_{\Gfin'\notin\BCGs} \MCMto(\Gfin')$, where the (finite) union is taken over all realizable \terminal graphs $\Gfin'\notin\BCGs$. Thus, we have $(\GD_0,\xd_0)\in \MCMo(\Glinit)$ if for each $0\leq t<\Tmax$, $(\GD_t,\xd_t)\in\MCMgenx_{\BCin_t}(\Gloopsub_t)$ is $\BCin_t$-generic. 
\end{definition}

\begin{definition}
A \terminal graph $\Gfin$ is called \emph{\realizable} if $\MCMto(\Gfin)\neq\emptyset$. We let 
\begin{equation*}%
\rBCGs:=\{\Gfin\in\BCGs\mid \Gfin\text{ is realizable}\}. 
\end{equation*}
\end{definition}

\begin{proposition}\label{lemma:MCMto=Res}
For each \realizable \terminal \lgg $\Gfin$, 
\begin{equation}\label{eq:MCMto=Res}
 \MCMto(\Gfin) = \ResGlinit \MCM(\Gfin).
\end{equation}
\end{proposition}
\begin{proof}
By construction, $\MCMcellint(\Gfin)\subset \ResGlinit \MCM(\Gfin)$. Conversely, consider a branch~\eqref{eq:BCFW:BCFW_branch} of the \ora terminating in some $(\GD_\Tmax,\xd_\Tmax)\in\MCM(\Gfin)$. Consider any other point $(\GD_\Tmax,\yd_\Tmax)\in\MCM(\Gfin)$ and for each $t$, set $(\GD_t,\yd_t):=\ResGt(\GD_\Tmax,\yd_\Tmax)$. Since $(\GD_t,\xd_t)\in\MCM(\Gloopsub_t)$, by \cref{lemma:Res_of_MCE_is_MCE}, we have $(\GD_t,\yd_t)\in\MCM(\Gloopsub_t)$ 
for each $0\leq t\leq\Tmax$. 

We claim that $(\GD_{t+1},\yd_{t+1}) = \ORmvt(\GD_t,\yd_t)$ for each $0\leq t<\Tmax$. Let us denote $(\GD,\xd):=(\GD_t,\yd_t)$, $\BCin:=\BCin_t=(\cc)$, and let $\ffout$, $\GDr$, $\Rcc$ be as in \cref{sec:ORA}. Note that the triangular faces $\uLR$ created during the \ORst $(\GD,\xd)\mapsto\ORmv(\GD,\xd)$ are present in $\GD_\Tmax$, and their colors in $\Gfin$ agree with the coloring convention $\conven$. Thus, $\yd_{t+1}(\ffout)\in\FRay$. 
Since $(\GD_{t+1},\yd_{t+1})$ is an \MCE, it is \finj, so $\yd_{t+1}(\ffout)\notin\yd_{t+1}(\partF\corv)$. By \cref{lemma:MCE:Res}, $(\GDr,\yd_{t+1})=\Res_{\GDr}(\GD_{t+1},\yd_{t+1})$ is a \pMNE. It is not an \MCE because $(\GD_{t+1},\yd_{t+1})$ contains at least one extra edge incident to $\ffout$. 
By \cref{lemma:ORA:uniqueness}, we get $(\GD_{t+1},\yd_{t+1}) = \ORmvt(\GD_t,\yd_t)$. Thus, $\ResGlinit(\GD_\Tmax,\yd_\Tmax)=(\GD_0,\yd_0)\in\MCMto(\Gfin)$. 
\end{proof}

\begin{proposition}\label{lemma:MCMto_open}
For each $\Gfin\in\rBCGs$, $\MCMto(\Gfin)\subset\MCM(\Glinit)$ is a (nonempty) open subset. 
\end{proposition}
\begin{proof}
By \cref{dfn:BCFW:exc}, $\MCM(\Glinit)=\bigsqcup_{\Gfin'} \MCMto(\Gfin')$, where the (finite) union is taken over all realizable \terminal graphs $\Gfin'$. Thus, if $\MCMto(\Gfin)$ is not open then there exists some other \terminal graph $\Gfin'$ such that some point $(\GD_0,\xd_0)\in\MCMto(\Gfin)$ is a limit of points $(\GD_0,\xdtls_0)\in\MCMto(\Gfin')$. Let $0\leq \tdiv<\Tmax$ be the first step where the branches~\eqref{eq:BCFW:BCFW_branch} for $\Gfin$ and $\Gfin'$ diverge. Thus, we have $\Gloopsub_{\tdiv}=\Gloopsubp_{\tdiv}$ and $\BCin_{\tdiv}=\BCin'_{\tdiv}$ but $\Gloopsub_{\tdiv+1}\neq\Gloopsubp_{\tdiv+1}$.
Note that $(\GD_{\tdiv},\xd_{\tdiv}) \in\MCMgenx_{\BCin_{\tdiv}}(\Gloopsub_{\tdiv})$ is $\BCin_{\tdiv}$-generic while $(\GDp_{\tdiv},\xdtls_{\tdiv})$ may or may not be $\BCin_{\tdiv}$-generic.
Fix $0\leq t\leq \tdiv$. 
To match notation to \cref{sec:ORA}, denote
 $\GD:=\GD_t=\GDp_t$, $\BCin=(\cc):=\BCin_t=\BCin'_t$, $\xd:=\xd_t$, $\xdtls:=\xdtls_t$, and let $\GDr$ be obtained from $\GD$ via \cref{dfn:GDr}. 
Recall from \cref{lemma:BCFW:clique_vs_Rg} that $\corv$ is \flex in $(\GD,\xd)$ if and only if it is \flex in $(\GD,\xdtls)$. 
Let $\rcrit$ (resp., $\rcrittls$) be given by~\eqref{eq:rcrit_dfn} for $(\GD,\xd)$ (resp., $(\GD,\xdtls)$).
 Let $(\GDout,\xdout):=\ORmv(\GD,\xd)$ (resp., $(\GDpout,\xdouttls):=\ORmv(\GD,\xdtls)$) be the result of applying the corresponding \ORst. 
Since $\rcrittls$ is bounded by \cref{lemma:ORA:r_large_not_MCE}, after possibly passing to a subsequence, we assume that there exists a limit $\rcritzls=\lim_{\tlim\to0}\rcrittls \in [0,\infty)$. We denote $\xdoutzls:=\lim_{\tlim\to0}\xdouttls$. 

First, we prove by induction on $t=0,1,\dots,\tdiv$ that $\xdoutzls=\xdout$. 
The base case holds since $\lim_{\tlim\to0} \xdtls_0 = \xd_0$. 
By the induction hypothesis, we may assume that $\xdoutzls|_{\Faces} = \lim_{\tlim\to0} \xdtls = \xd$.
Suppose that $\xdoutzls(\ffout)=\xdoutzls(\ff)$ for some $\ff\in\partF\corv$, and let $\fpiv\in\Pivots(\corner)$ be such that $(\GDpout,\xdouttls)$ contains an outgoing edge connecting $\ffout$ to $\fpiv$. 
Then $(\GD,\xd)$ admits \pchords inside $\corv$ connecting $\ff$ to each of $\{\corfp,\corf,\corfm,\fpiv\}$.
By~\itemref{MCE5:Mpos_chords}, each of these \pchords must be isotopic to an edge of $(\GD,\xd)$. 
If $\ff=\corf$ or $\ff\in\Pivots(\cor)$ then $(\GD,\xd)$ contains an edge located weakly inside $\cor$ connecting $\corf$ to a vertex in $\Pivots(\cor)$, a contradiction. 
Otherwise, $\ff\in\{\corfm,\corfp\}$; say, $\ff=\corfp$. 
If $\corfm=\corfp$, then $\sumwT(\cor)=\sumbT(\cor)=\pi$, so the folding ray $\RccT$ points away from $\xT(\corfm)$, a contradiction. 
Otherwise, $\cor$ is a corner of a triangle with vertices $\{\corfm,\corf,\corfp\}$, contradicting $\corv\in\Vsact$. 
We have shown that $\xToutzls$ is injective on the faces of $\GDpout$ and $\GDr$.

 By \cref{lemma:limit_of_pcMNE_is_pMNE}, $(\GDpout,\xdoutzls)$ is a \pMNE. By \cref{lemma:MCE:Res}, so is its restriction $(\GDr,\xdoutzls)$. 
By \cref{lemma:ORA:new_edge_at_least_one,lemma:ORA:rigid_Chords}, $\GDpout$ contains at least one outgoing edge, so $(\GDr,\xdoutzls)$ is not an \MCE. 
Since $\cor$ is not ambiguous in $(\GD,\xd)$ by \cref{lemma:ambig=>triangular}, we get $\sumbT_{\xdtls}(\cor)\to\sumbT_{\xd}(\cor)$ and $\sumwT_{\xdtls}(\cor)\to\sumwT_{\xd}(\cor)$ as $\tlim\to0$. Thus, 
$\xdoutzls(\ffout)$ lies on the folding ray $\FRay$. By \cref{lemma:ORA:uniqueness}, we find $\rcrit = \rcritzls$ and so $\xdoutzls = \xdout$. 

Assume now that $t=\tdiv$. 
As we showed above, $\xdoutzls=\xdout$ and $(\GDpout,\xdout)$ is a \pMNE. 
By \cref{lemma:ORA:ORA_Cext_chords,lemma:ORA:rigid_Chords}, $(\GDout,\xdout)$ contains all \pchords of $(\GDo,\xdout)$ incident to $\ffout$ (of which there are exactly four since $(\GD,\xd)$ is $\BCin$-generic). In particular, $(\GDout,\xdout)$ contains every edge of $(\GDpout,\xdout)$. Since $\deg_{\GDpout}(\ffout)\geq4$, we get $\GDout = \GDpout$. 
 Since $(\GDout,\xdouttls)$ and $(\GDout,\xdout)$ are \MCEs, each of their ambiguous corners is contained in a degenerate triangular face by \cref{lemma:ambig=>triangular}. Given $\vout\in\Vintout$ such that $\xdout(\vout)$ is not a degenerate triangle, since~\eqref{eq:MCMS_angles_vs_k} holds for 
$(\GDout,\xdout)\in\MCM(\Gloopsub_{t+1})$ and $(\GDout,\xdouttls)\in\MCM(\Gloopsubp_{t+1})$, 
we see by taking the $\tlim\to0$ limit that $\gelWsub_{\Gloopsub_{t+1}}(\vout) = \gelWsub_{\Gloopsubp_{t+1}}(\vout)$ and $\gelBsub_{\Gloopsub_{t+1}}(\vout) = \gelBsub_{\Gloopsubp_{t+1}}(\vout)$. 
Finally, assume that $\xdout(\vout)$ is a degenerate triangle and let $\f$ be its middle vertex. If $\f$ is not doubly ambiguous then the color of $\vout$ is the same in $\Gloopsub_{t+1}$ and $\Gloopsubp_{t+1}$ in view of \cref{lemma:MCE:coloring_improper_verts}. The only doubly ambiguous vertex that may arise during a $\BCin$-generic \ORst is $\corf$; see \figref{fig:fout-bend}(d). In this case, the color of $\vout$ is determined by $\conven$. We conclude that $\Gloopsub_{t+1} = \Gloopsubp_{t+1}$ as \lggs, a contradiction.
\end{proof}

\subsection{\Kinsupp}
Let $\gdimo:=\gdim(\GD_0) = 3n+4\nL$; cf.~\eqref{eq:gdim_dfn}. 

\begin{definition}\label{dfn:BCFW:kinsupp}
We say that $\Gfin\in\BCGs$ has \emph{\kinsupp} if $\dimZ\ResGlinit \MCM(\Gfin) = \gdimo$. 
\end{definition}

\begin{lemma}\ \label{lemma:BCFW:dim_M0_irr}
$\MCMalg(\GDinit)$ is an irreducible algebraic variety of dimension $\gdimo$.%
\end{lemma}
\begin{proof}
It is well known that $\lalatsMAT$ is an irreducible variety of dimension $4n-4$. 
Denote $\GDinit=\BCGDinitx_{k,n;L}$ to make the dependence on $\knL$ explicit. 
We have a dominant map $\lalatsMAT\to\MCMalg(\BCGDinitx_{k,n;L=0})/\Rdd$ sending $\lalat\mapsto\Pll$, where
 $\Rdd$ acts on $\MCMalg(\GDinit)$ by translations. 
A generic fiber of this map is the little group $\LG\cong(\Rast)^n$. 
Thus, $\MCMalg(\BCGDinitx_{k,n;L=0})/\Rdd$ is irreducible of dimension $3n-4$. 
It follows that $\MCMalg(\BCGDinitx_{k,n;L})$ is irreducible of dimension $3n+4\nL=\gdimo$.
\end{proof}

\begin{corollary}%
\label{lemma:MCE:fullysep<=kinsupp}
If $\Gfin\in\BCGs$ \haskinsupp then $\Gfin$ is \fullysepMCE. Furthermore, it is \fullysep.
\end{corollary}
\begin{proof}
If $\ploc_\seps,\ploc_\sept$ are not \twosepMCE for some $\sepst\in\brnLbdsep$ then $\ResGlinit \MCM(\Gfin)$ is contained in a proper subvariety of $\MCMalg(\GDinit)$ given by $(\xd(\ploc_\seps)-\xd(\ploc_\sept))^2=0$, contradicting \cref{lemma:BCFW:dim_M0_irr}. Thus, $\Gfin$ is \fullysepMCE. Since $\Gfin$ \haskinsupp, $\MCM(\Gfin)\neq\emptyset$. By \cref{lemma:fullysepMCE_vs_fullysep}, $\Gfin$ is \fullysep.
\end{proof}

\begin{proposition}%
\label{lemma:kinsupp=realizable}
Let $\Gfin\in\BCGs$. Then $\Gfin$ \haskinsupp if and only if it is \realizable.
\end{proposition}

\begin{example}\label{ex:fullysep_BCFW_check}
The graph $\Gfin\in\BCGs$ in \figref{fig:BCFW-full}(e) is \realizable. 
By \cref{lemma:kinsupp=realizable}, $\Gfin$ \haskinsupp.
By \cref{lemma:MCE:fullysep<=kinsupp}, $\Gfin$ must be \fullysep. This agrees with \cref{ex:BCFW-T-dual}.
\end{example}

\begin{proof}[Proof of \cref{lemma:kinsupp=realizable}]
If $\Gfin\in\rBCGs$ is \realizable then by 
\crefrange{lemma:MCMto=Res}{lemma:MCMto_open} and \cref{lemma:BCFW:dim_M0_irr}, 
$\MCMto(\Gfin)=\ResGlinit \MCM(\Gfin)$ is Zariski dense in $\MCM(\Glinit)$, so $\Gfin$ \haskinsupp. 

Conversely, suppose that $\Gfin\in\BCGs$ \haskinsupp. First, we claim that $\ResGlinit \MCM(\Gfin)\subset\MCMp(\Glinit)$. 
Let $(\GD,\xd)\in\MCM(\Gfin)$ and $(\GDinit,\xdinit):=\ResGlinit(\GD,\xd)$. 
By \cref{lemma:MCE:fullysep<=kinsupp}, $\Gfin$ is \fullysepMCE. 
Thus, for all $\sepst\in\brnLbdsep$, we have $(\xd(\ploc_\seps)-\xd(\ploc_\sept))^2>0$, so $(\GDinit,\xdinit)$ satisfies~\itemref{MCE3:M-tnn}. \itemref{MCE2:null_edges} and~\itemref{MCE4:properly_colored}--\itemref{MCE5:Mpos_chords} trivially hold for $(\GDinit,\xdinit)$. 
\itemref{MCE1:emb} is satisfied by \cref{lemma:TOP:Mpos=>simple}. 
Finally,~\eqref{eq:MCMS_angles_vs_k} is deduced for $\vo$ by applying~\eqref{eq:BCFW:k(G)_dfn} and~\eqref{eq:Rg_sum_bd_vs_chi} to $(\GD,\xd)$ with $\Rg=\Vint$.

Suppose for contradiction that $\Gfin\in\BCGs\setminus\rBCGs$ \haskinsupp but is not \realizable. 
Since $\Gfin$ \haskinsupp, $\MCM(\Gfin)\neq\emptyset$.
Let $(\GD,\xd)\in\MCM(\Gfin)$ be generic so that $\xT:\Faces\to\C$ is injective (cf. \cref{lemma:injective_MCE_exists}). 
Consider the branch~\eqref{eq:BCFW:BCFW_branch} of the \ORA starting with $(\GD_0,\xd_0):=\ResGlinit(\GD,\xd)$ (which belongs to $\MCMp(\Glinit)$ as shown above). Since $\MCMto(\Gfin)=\emptyset$, we have $\Gloopsub_{\Tmax}\neq\Gfin$. 
Consider the branch 
$\Glinit= \Gloopsubp_0\xrightarrow{\BCin'_0} \Gloopsubp_1 \xrightarrow{\BCin'_1}\cdots\xrightarrow{\BCin'_{\Tmax'-1}} \Gloopsubp_{\Tmax'}=\Gfin$ 
of the loop BCFW recursion (\cref{dfn:BCFW:rec}) terminating in $\Gfin$. 
Denote $(\GDp_{t'},\yd_{t'}):=\Res_{\GDp_{t'}}(\GD,\xd)$ for $0\leq t'\leq \Tmax'$. 

We claim that $\Tmax'=\Tmax$ and for all $0\leq t\leq \Tmax$, $(\GD_t,\xd_t) = (\GDp_t,\yd_t)$. 
By construction, $(\GD_0,\xd_0) = (\GDp_0,\yd_0)$.
Observe that if, say, $\Tmax\leq\Tmax'$ and $(\GD_\Tmax,\xd_\Tmax) = (\GDp_\Tmax,\yd_\Tmax)$ then we must have 
$\Tmax = \Tmax'$ since the property of being \terminal depends only on whether each face of $\GD_\Tmax=\GDp_\Tmax$ is \btrar. Thus, it suffices to consider the case where $(\GD_{t+1},\xd_{t+1}) \neq (\GDp_{t+1},\yd_{t+1})$ for some $0\leq t<\min(\Tmax,\Tmax')$, and we let $t$ be the minimal such index. 
Since $\xT$ is injective on $\Faces$, by \cref{lemma:MCE:Res}, $(\GDp_{t+1},\yd_{t+1})$ is a \pMNE. 
(Note that $(\GDp_{t+1},\yd_{t+1})$ is proper while $(\GD_{t+1},\xd_{t+1})$ is properly colored.)
Similarly to the proof of \cref{lemma:MCMto=Res}, we deduce from \cref{lemma:ORA:uniqueness} that $\yd_{t+1} = \xd_{t+1}$. Thus, we must have $\GD_{t+1}\neq\GDp_{t+1}$. Since $(\GD_{t+1},\xd_{t+1})$ contains all \pchords 
of $(\GDo,\xd_{t+1})=(\GDo,\yd_{t+1})$ 
incident to $\ffout$, it contains all edges of $(\GDp_{t+1},\xd_{t+1})$. 
Since $\GD_{t+1}\neq\GDp_{t+1}$ and $\deg_{\GDp_{t+1}}(\ffout)=4$, we must have $\deg_{\GD_{t+1}}(\ffout)>4$.
Thus, $(\GD_0,\xd_0)\in\Zexc$. 
By~\eqref{eq:gdim_diff}, $\gdim(\GD_{t+1})<\gdim(\GD_t)\leq \gdimo$. 
By~\eqref{eq:OCP:gdim_Mmce}, $\dimZ\MCM(\Gloopsub_{t+1})<\gdimo$, so
\begin{equation}\label{eq:dimZ_Zexc}
 \dimZ\Zexc < \gdimo.
\end{equation}
We have shown that for a generic point $(\GD,\xd)\in\MCM(\Gfin)$, the restriction $(\GD_0,\xd_0)=\ResGlinit(\GD,\xd)$ belongs to $\Zexc$. By~\eqref{eq:dimZ_Zexc}, 
$\dimZ\ResGlinit\MCM(\Gfin)<\gdimo$, a contradiction. 

Thus, we indeed have $\Tmax'=\Tmax$ and $(\GD_t,\xd_t) = (\GDp_t,\yd_t)$ for all $0\leq t\leq \Tmax$. 
Recall that the only way a doubly ambiguous vertex could arise in $(\GD_\Tmax,\xd_{\Tmax})$ is shown in \figref{fig:fout-bend}(d), in which case the coloring of the two ambiguous corners incident to it is determined by $\conven$. Thus, this coloring choice coincides with the corresponding vertex coloring choice in $\Gloop = \Gloopsubp_\Tmax$. 
Since $(\GDp_\Tmax,\yd_\Tmax)\in\MCM(\Gloopsubp_\Tmax)$ and $(\GD_\Tmax,\xd_\Tmax)\in\MCM(\Gloopsub_\Tmax)$, we see from \cref{lemma:MCE:coloring_improper_verts,lemma:BCFW:from_MCM_to_Glout} that the \lggs $\Gloop = \Gloopsubp_\Tmax$ and $\Gloopsub_\Tmax$ coincide, a contradiction.
\end{proof}

\begin{figure}
 \def\inputscale{1.5}
\begin{tabular}{cc}
\includegraphics[scale=\inputscale]{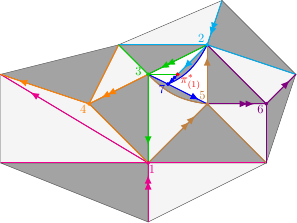}
&
\includegraphics[scale=\inputscale]{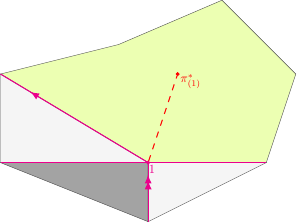}
\end{tabular}

 \caption{\label{fig:counter-kinsupp} A terminal \MCE $(\GDfin,\xdfin)\in\MCM(\Gfin)$ of $\Gfin\in\BCGs$ (left) whose restriction $\ResGx_1(\GDfin,\xdfin)\notin\MCM(\Gloopsub_1)$ is not an \MCE of $\Gloopsub_1$ (right). The red dashed line indicates the \pchord violating~\itemref{MCE5:Mpos_chords} for $\ResGx_1(\GDfin,\xdfin)$.
See \cref{ex:counter}. 
}
\end{figure}

\begin{example}[\Fullysepon is not sufficient for \kinsupp]
\label{ex:counter}
Consider a \terminal \MCE $(\GDfin,\xd)\in\MCM(\Gfin)$ %
 shown in \figref{fig:counter-kinsupp}(left).
 We have $\Gfin\in\BCGs$ for $\knL = (3,7;1)$. 
Similarly to \cref{fig:BCFW-full}, for each $t=1,2,\dots,7$, the interior vertex $\ffout_t$ that appeared during the $t$-th \ORst is labeled by $t$ in \figref{fig:counter-kinsupp}(left). 
 Since the white clique 
$\{\ffout_t\mid t=1,2,3,5,7\}\sqcup\{\ploc_1\}$ 
of $(\GDfin,\xdfin)$ does not contain any boundary vertices, and since no clique of $(\GDfin,\xdfin)$ contains non-adjacent boundary vertices, we see that $\Gfin$ is \fullysepMCE. 
By \cref{lemma:fullysepMCE_vs_fullysep}, $\Gfin$ is \fullysep.

However, we claim that $\Gfin$ does not \havekinsupp. The culprit is that the face $\ffout_1$ is not \twosepMCE from $\ploc_1$ or $\ffout_2$. Consider the branch $(\Gloopsub_0,\Gloopsub_1,\dots,\Gloopsub_\Tmax=\Gfin)$ of the loop BCFW recursion terminating at $\Gfin$. 
We see from \cref{fig:counter-kinsupp} that the faces $\ploc_1$ and $\ffout_1$ 
 are \twosepMCE in $\Gloopsub_t$ for $t=1,2,3$ but not for $t\geq4$. 
It follows from \cref{lemma:MCMto=Res,lemma:kinsupp=realizable} that $\Gfin$ does not \havekinsupp. 
In other words, even though $(\GDfin,\xdfin)$ is an \MCE, it is not realizable as the output of the \ORA as some of the \pchords incident to vertices $\ffout_1,\ffout_2$ were not added to the set of outgoing edges during the first two \ORsts.
\end{example}

\begin{conjecture}[Combinatorial criterion for \kinsupp]\label{conj:kinsupp}
A graph $\Gfin\in\BCGs$ \haskinsupp if and only if for all $\ff,\f\in\Faces$ and all $t=0,1,\dots,\Tmax$ such that $\ff,\f$ are faces of $\Gloopsub_t$, $\ff$ and $\f$ are \twosepMCE in $\Gloopsub_t$ if and only if they are \twosepMCE in $\Gloopsub_\Tmax=\Gfin$.
\end{conjecture}
\noindent 
This criterion is combinatorial in view of \cref{lemma:BCFW:NullE_vs_NullED}. 
By \cref{lemma:MCMto=Res,lemma:kinsupp=realizable}, the condition in \cref{conj:kinsupp} is necessary in order for $\Gfin$ to \havekinsupp. 
 The $t=0$ case of this condition recovers the necessity of \fullysepon shown in \cref{lemma:MCE:fullysep<=kinsupp}.

\subsection{BCFW tilings of \MCMSs} 
Recall the notion of a \emph{tiling} (\cref{dfn:BCFW:tiling}). 

\begin{theorem}[BCFW tilings of \MCMSs]\label{thm:BCFW:triang_MCE}
The tiles $\{\MCMcellint(\Gfin)\mid \Gfin\in\oBCGs\}$ form a tiling of $\MCM(\Glinit)$.
\end{theorem}

\begin{proof}%

In the notation of \cref{dfn:BCFW:tiling}, we set 
$\MCX:=\bigsqcup_{\Gfin\in\oBCGs} \MCM(\Gfin)$,
$\MCW:=\MCM(\Glinit)$, $\MCPhi:=\ResGlinit$, and for $\Gfin\in\oBCGs$, we set $\MCXo_{\Gfin}:=\MCM(\Gfin)$ and $\MCWo_{\Gfin}:=\MCMcellint(\Gfin)$. 
By~\eqref{eq:MCMto=Res}, $\MCPhi:\MCXo_{\Gfin}\to\MCW$ has image $\MCWo_{\Gfin}$, 
and the \ORA yields a continuous inverse to the map $\ResGlinit:\MCM(\Gfin)\to\MCMcellint(\Gfin)$. This verifies condition~\itemref{tiling1} in \cref{dfn:BCFW:tiling}. 
Condition~\itemref{tiling2} follows from~\eqref{eq:MCMo_Zexc_dfn}. 
Finally, since $\MCMcellint(\Gfin)=\emptyset$ for $\Gfin\in\BCGs\setminus\oBCGs$, 
condition~\itemref{tiling3} follows from~\eqref{eq:MCMo_Zexc_dfn},~\eqref{eq:dimZ_Zexc} and \cref{lemma:BCFW:dim_M0_irr}. 
\end{proof}

\section{Finishing the proof}\label{sec:finish}
Our final goal is to deduce \cref{thm:main_triangulation} from \cref{thm:BCFW:triang_MCE}. Recall from \cref{lemma:tMtiling} that we mostly focus on \tmtilings.

\subsection{BCFW tilings of ambient loop momentum amplituhedra}\label{ssec:BCFW_triang_final1}
\begin{theorem}[BCFW tilings of ambient loop momentum amplituhedra]
\label{lemma:BCLOOP:MPknL_tiling}
In the notation of \cref{dfn:RX_triple}, the tiles 
$\{\MPGfin\mid \Gfin\in\oBCGs\}$
 form \atmtilingA of $\MPknL$.
\end{theorem}
\begin{proof}
Let $\Gfin\in\oBCGs$.
We write $\Meas$ as a shorthand for $\Measknk(\Gfin,\cdot)$. 
We claim that we have a commutative diagram 
\begin{equation}\label{eq:Gfin_cd}
 \begin{tikzcd}[column sep=2em,row sep=1.5em]
\MCMp(\Gfin) \dar["\ResGlinit",twoheadrightarrow] 
& 
\MCMpdec(\Gfin) \lar[twoheadrightarrow,"\RpiLG"'] \rar[twoheadrightarrow,"\RpiGGsh"] 
\dlar[phantom, "\square",pos=0.43] 
\drar[phantom, "\square"] 
\dar["\dResGlinit",twoheadrightarrow]
&
\MCMpdec(\Gfin)/\GGsh \arrow[r,"\sim"',"\RMOphi"] \dar["\dResGlinit/\GGsh",twoheadrightarrow] &
\ReloO_{\Gfin} \dar["\RprojoO_{\Gfin}",twoheadrightarrow]
\\
\MCMcellint(\Gfin) & 
\dMCMcellint(\Gfin) \lar[twoheadrightarrow,"\RpiLG"'] \rar[twoheadrightarrow,"\RpiGGsh"] & 
\dMCMcellint(\Gfin)/\GGsh \arrow[r,"\sim"',"\RMOpsi"] & 
\RYoO_{\Gfin}
\end{tikzcd} 
\end{equation}
obtained as the restriction of another commutative diagram 
\def\Gfincupsub{\Gfin}
\def\shortpt{-8pt}
\def\shortpos{0.3}
\begin{equation}\label{eq:Gfin_cd_sqcup}
 \begin{tikzcd}[column sep=1.5em,row sep=1.5em]
\bigsqcup\limits_{\Gfincupsub}\MCMp(\Gfin) 
\dar["\ResGlinit",shorten <=\shortpt,pos=\shortpos] 
& 
\bigsqcup\limits_{\Gfincupsub}\MCMpdec(\Gfin) \lar[twoheadrightarrow,"\RpiLG"'] \rar[twoheadrightarrow,"\RpiGGsh"] 
\dlar[phantom, "\square",pos=0.25] 
\dar["\dResGlinit",shorten <=\shortpt,pos=\shortpos]
&
\bigsqcup\limits_{\Gfincupsub}\MCMpdec(\Gfin)/\GGsh \arrow[r,"\sim"',"\RMOphi"] 
\dar["\dResGlinit/\GGsh",shorten <=\shortpt,pos=\shortpos] 
\dlar[phantom, "\square",pos=0.27] 
&
\bigsqcup\limits_{\Gfincupsub}\ReloO_{\Gfin} 
\dar["\RprojO",shorten <=\shortpt,pos=\shortpos]
\\
\MCMp(\Glinit) & 
\dMCMp(\Glinit) \lar[twoheadrightarrow,"\RpiLG"'] \rar[twoheadrightarrow,"\RpiGGsh"] & 
\dMCMp(\Glinit)/\GGsh \arrow[r,"\sim"',"\RMOpsi"] & 
\RYO,
\end{tikzcd} 
\end{equation}
where the disjoint unions are taken over all $\Gfin\in\oBCGs$ and Cartesian squares are marked with $\square$ similarly to~\eqref{eq:TREE:comm_diag}. 
The spaces $\MCMp(\Gfin)$ and $\MCMpdec(\Gfin)$ were introduced in \cref{dfn:BCFW:MCE_compatible,dfn:MCMp_MCMpdec} and the subset $\MCMcellint(\Gfin)\subset\MCMp(\Glinit)$ was introduced in \cref{dfn:BCFW:exc}; see also~\eqref{eq:MCMto=Res}. The \emph{decorated restriction operator} $\dResGlinit:\MCMpdec(\Gfin)\to\MCMpdec(\Glinit)$ sends $(\GDfin,\xdfin,\la,\lat)\mapsto(\GDinit,\xdfin|_{\Faces_0},\la,\lat)$. 
 We denote its image by $\dMCMcellint(\Gfin):=\dResGlinit(\MCMpdec(\Gfin)) \subset \MCMpdec(\Glinit)$. 

We show that the $\GGsh$-action (\cref{dfn:BACKGR:GGsh}) preserves the subsets $\MCMpdec(\Glinit)$ and $\MCMpdec(\Gfin)$ and is free on each subset. (See \cref{ssec:APP:RP3} for related discussion.) 
Recall that $\GDinit$ is an $n$-cycle with $\nL$ isolated vertices $(\ploc_1,\dots,\ploc_{\nL})$ inside. Comparing \cref{dfn:LPUNC:amb_loop_ampl,dfn:MCMp_MCMpdec} and applying~\eqref{eq:MCMS_angles_vs_k}, we see that any point $(\GD_0,\xd_0,\la,\lat)\in\MCMpdec(\Glinit)$ gives rise to a point $\llPllL\in\MPknL$ with $\byL=(\xd_0(\ploc_1),\dots,\xd_0(\ploc_\nL))$, assuming $\Pbdxo=(\bdx_1=0,\bdx_2,\dots,\bdx_n)$ is in normal form. Conversely, for any point $\llPllL\in\MPknL$, choosing matrix representatives for $\lalat\in\Gror(2,n)^2$ yields a point in $\MCMpdec(\Glinit)$. By \cref{rmk:GGsh_action_preserves_MPknL}, $\GGsh$ preserves the space $\MCMpdec(\Glinit)$ (and clearly acts freely on it), 
so we obtain a homeomorphism $\RMOpsi:\MCMpdec(\Glinit)/\GGsh \xrasim \RYO = \MPknL$ in~\eqref{eq:Gfin_cd}--\eqref{eq:Gfin_cd_sqcup}.

For $\Gfin\in\oBCGs$, $\MCMpdec(\Gfin)$ is ($\GGsh$-equivariantly) homeomorphic to $\MdteoMP(\Gbip)$ by \cref{lemma:TE:TE_to_MCE_and_back}. 
By definition, the $\GGsh$-action preserves $\MdteoMP(\Gbip)$ and is free, and thus the same is true for $\MCMpdec(\Gfin)$. Since restriction commutes with the $\GGsh$-action, the square in~\eqref{eq:Gfin_cd} involving the quotient map $\RpiGGsh$ commutes, is Cartesian, and all four maps involved in it are surjective. 

Let $(\GD,\xd,\la,\lat)\in\MCMpdec(\Gfin)$ with $\Pbdx=\Pll$ in normal form. Set $\byL=(\xd(\ploc_1),\dots,\xd(\ploc_\nL))$. 
Let $\mceTOwte(\GD,\xd,\la,\lat)=\datrQ\in\MdteoMP(\Gfin)$; cf.~\eqref{eq:ORA::mceTOwte_dfn_iso}. 
 We claim that $\RMOphi(\GD,\xd,\la,\lat):=(\wt,\llPllL)$ belongs to $\ReloO_{\Gfin}$. We have $\wt\in\RXoO_{\Gfin}$.
By \cref{dfn:MCMp_MCMpdec}, $\lalat\in\MPkntreeMAT$. 
 By \cref{lemma:MCE:fullysep<=kinsupp}, $\Gfin$ is \fullysepMCE, so $\llPllL$ satisfies condition~\itemref{MPknL2} in \cref{dfn:LPUNC:amb_loop_ampl}. By \cref{lemma:TOP:Mpos=>simple}, each $\yTloc_\rho$ is disjoint from the polygon $\PllT$, and since $\xT$ is a \wemb of $\GDfin$ by~\itemref{MCE1:emb}, $\llPllL$ satisfies condition~\itemref{MPknL4}. Thus, $\llPllL\in\MPknL$.
Since~\eqref{eq:LOOP:Rel_dfn} holds by construction, we get $\RMOphi(\GD,\xd,\la,\lat)\in\ReloO_{\Gfin}$. 

 By~\eqref{eq:ORA::mceTOwte_dfn_iso}, $\mceTOwte:\MCMpdec(\Gfin) \xrasim \MdteoMP(\Gfin)$ is a homeomorphism. 
As we show in \cref{ssec:collapsed_MCE}, 
\begin{equation}\label{eq:kinsupp_einj_automatic}
\MdteoMP(\Gfin)=\MdteMP(\Gfin) \quad\text{for $\Gfin\in\oBCGs$,}
\end{equation} 
where $\MdteMP(\Gfin)$ is the set of \wtembs of $\Gfin$ with \Mbd similarly to~\eqref{eq:MdteoMP_dfn}. 
By \cref{thm:TE:OAC,lemma:TE:hasMbd=>simple_and_MCE3}, $\MdteMP(\Gfin)/\GGsh\xrasim\ReloO_{\Gfin}$ is a homeomorphism with inverse $(\wt,\llPllL)\mapsto(\GD,\xll,\la,\lat)$.
Composing these maps, we obtain a homeomorphism $\RMOphi:\MCMpdec(\Gfin)/\GGsh\xrasim\ReloO_{\Gfin}$ in~\eqref{eq:Gfin_cd}--\eqref{eq:Gfin_cd_sqcup}. 
 It is clear that the squares involving $\RMOphi$ and $\RMOpsi$ commute. In particular, the homeomorphism $\RMOpsi:\MCMpdec(\Glinit)/\GGsh \xrasim \RYO$ restricts to a homeomorphism $\RMOpsi:\dMCMcellint(\Gfin)/\GGsh \xrasim \RYoO_{\Gfin}$ for each $\Gfin\in\oBCGs$. 

By \cref{rmk:MCMp->MCM_surj}, each point in $\MCMp(\Gfin)$ lifts to a point in $\MCMpdec(\Gfin)$, unique up to $\LGpm$-action. For $(\GDfin,\xdfin)\in\MCMp(\Gfin)$, the boundary polygon $\Pbdx$ is \einj, so the $\LGpm$-action is free. We denote the corresponding quotient map by $\RpiLG:\MCMpdec(\Gfin)\to\MCMp(\Gfin)$. 
We see that the squares in~\eqref{eq:Gfin_cd}--\eqref{eq:Gfin_cd_sqcup} involving $\RpiLG$ commute and are Cartesian, and that the double-headed arrows indeed correspond to surjective maps in each case. 

By \cref{thm:BCFW:triang_MCE}, the tiles $\{\MCMcellint(\Gfin)\mid\Gfin\in\oBCGs\}$ form a tiling of $\MCMp(\Glinit)$. We propagate this result through each square in~\eqref{eq:Gfin_cd}--\eqref{eq:Gfin_cd_sqcup} from left to right. Since $\RpiLG$ and $\RpiGGsh$ are quotient maps, it follows that the tiles $\{\dMCMcellint(\Gfin)/\GGsh\mid\Gfin\in\oBCGs\}$ form a tiling of $\dMCMp(\Glinit)/\GGsh$; cf. \cref{lemma:RX_triple}. Since $\RMOphi$ and $\RMOpsi$ are homeomorphisms in both~\eqref{eq:Gfin_cd} and~\eqref{eq:Gfin_cd_sqcup}, the tiles $\{\RYoO_{\Gfin}\mid\Gfin\in\oBCGs\}$ form a tiling of $\RYO$. 
In particular, $\RprojoO_{\Gfin}:\ReloO_{\Gfin}\xrasim \RYoO_{\Gfin}$ is a homeomorphism.
\end{proof}%

\begin{remark}\label{rmk:Gfin_reduced}
By \cref{lemma:tMtiling}, 
the map
$\Measknk(\Gfin,\cdot):\Rtpgauge\xrasim\PtpLfunc_{\Gfin}$ is a homeomorphism for each $\Gfin\in\oBCGs$,
where for the continuous section $(\id_{\RXoA_{\Gfin}},\PhiLLL)$ we use the \emph{loop momentum amplituhedron map} $\PhiLLL:\PtpLfunc_{\Gfin}\to\RYO$ introduced in \cref{ssec:flip_proj} below; see also \cref{lemma:LOOP:proj_subset_amb}.
Thus, $\Gfin$ is reduced in the sense of \cref{dfn:Lfunc_positroid_cell_and_reduced}.
\end{remark}

\subsection{Collapsed \MCEsnoacr}\label{ssec:collapsed_MCE}
The goal of this subsection is to prove~\eqref{eq:kinsupp_einj_automatic}. 

\begin{figure}
 \def\inputscale{1.2}
 \def\inputscalex{1.5}
 \def\resc#1{\scalebox{0.9}{\setlength{\tabcolsep}{2pt}#1}}
 \setlength{\tabcolsep}{-3pt}
 \def\endln{\\[5pt]}
\begin{tabular}{ccccc}
\includegraphics[scale=\inputscale]{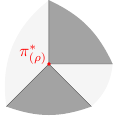}
\hspace{4pt}
&
\includegraphics[scale=\inputscale]{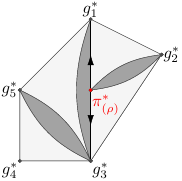}
&
\includegraphics[scale=\inputscale]{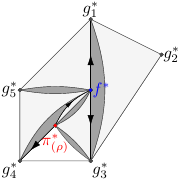}
&
\includegraphics[scale=\inputscale]{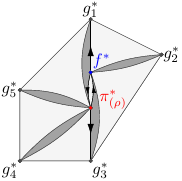}
&
\includegraphics[scale=\inputscalex]{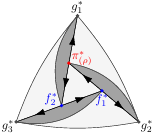}
\\
 \resc{\begin{tabular}{rl}
 (a) & $\Sloc_\rho=\{\ploc_\rho\}$\endln
 & $\Tloc_\rho=\emptyset$
 \end{tabular}}
&
 \resc{\begin{tabular}{rl}
 (b) & $\Sloc_\rho=\{\f_1,\f_3\}$\endln
 & $\Tloc_\rho=\{\ploc_\rho\}$
 \end{tabular}}
&
 \resc{\begin{tabular}{rl}
 (c) & $\Sloc_\rho=\{\f_1,\f_3,\f_4\}$\endln
 & $\Tloc_\rho=\{\ploc_\rho,\ff\}$
 \end{tabular}}
&
 \resc{\begin{tabular}{rl}
 (d) & $\Sloc_\rho=\{\f_1,\f_3\}$\endln
 & $\Tloc_\rho=\{\ploc_\rho,\ff\}$
 \end{tabular}}
&
 \resc{\begin{tabular}{rl}
 (e) & $\Sloc_\rho=\{\f_1,\f_2,\f_3\}$\endln
 & $\Tloc_\rho=\{\ploc_\rho,\ff_1,\ff_2\}$
 \end{tabular}}
\end{tabular}

 \caption{\label{fig:S-rho} Computing $\protect\Sloc_\rho$ and $\protect\Tloc_\rho$ in the proof of \cref{lemma:BCFW:counting_stars}.}
\end{figure}

Let $\Gfin\in\oBCGs$ and $(\GDfin,\xdfin)\in\MCM(\Gfin)$. 
Similarly to the proof of \cref{lemma:injective_MCE_exists}, consider a directed graph $\GflW$ with vertex set $\Faces$ and arrows $\ff_1\leftarrow\ff_2\to\ff_3$ for each degenerate \emph{black} triangle $\b\in\BVint$, where $\ff_2$ is the middle vertex of $\xT(\b)$. We denote by $\FflWsink$ (resp., $\FflWmid$) the set of sinks (resp., non-sinks) of $\GflW$. 
For $\rho\in\brnL$, let $\preKmaxloc_\rho:=\{\ff\in\Faces\mid \GflW\text{ contains a directed path from $\ploc_\rho$ to $\ff$}\}$, and set $\Sloc_\rho:=\preKmaxloc_\rho\cap \FflWsink$ and $\Tloc_\rho:=\preKmaxloc_\rho\cap \FflWmid$. 
See \cref{fig:S-rho} for some examples. 
Iterating \cref{lemma:MCE:clique_union}, we see that $\preKmaxloc_\rho$ is a clique in $(\GDfin,\xdfin)$. 
Let $\tloc_\rho:=|\Tloc_\rho|$ and $\dloc_\rho:=\dim\Conv\KTmaxloc_\rho$. 
 Let $\tbloc:=\sum_{\rho\in\brnL}\tloc_\rho$ and $\dbloc:=\sum_{\rho\in\brnL}\dloc_\rho$. 
We will compare the following result to \cref{lemma:counting_stars_type_1_n}. 

\begin{lemma}\label{lemma:BCFW:counting_stars}
For $\Gfin\in\oBCGs$ and $(\GDfin,\xdfin)\in\MCM(\Gfin)$, the sets $(\Tloc_\rho)_{\rho\in\brnL}$ are pairwise disjoint,
\begin{equation}\label{eq:counting_stars}
 \FflWmid = \bigsqcup_{\rho\in\brnL} \Tloc_\rho,
 \quad\text{and}\quad
 \dloc_\rho = \tloc_\rho = |\Sloc_\rho|-1 \quad\text{for each $\rho\in\brnL$}. 
\end{equation}
\end{lemma}

\begin{proof}

Following \cref{lemma:TOP:restr}, let $\CollWFaces$ denote the set of faces of the \wdash collapsed graph $\CollGW$. We claim that 
\begin{equation}\label{eq:CollW=FflWsink}
 \CollWFaces = \FflWsink.
\end{equation}
Let $\ff\in\Faces$. Suppose first that $\ff\notin\FflWsink$; thus, $\ff\in\FflWmid$ is the middle vertex of some degenerate black triangle $\xT(\b)$, $\b\in\BVint$. Setting $\Rg_0:=\{\b\}$, we see that $\xd(\RgFaclx_0)$ is a white clique. Consider the simply connected subset $\Rg\subset\Vint$ satisfying $\helW(\Rg)=1$ provided by \cref{lemma:BCFW:clique_vs_Rg}. By~\eqref{eq:BCFW:sumbTcond=0}, $\sumbT\cond(\f_i|\Rg)=0$ for each $\f_i\in\RgFbd$ (cf. \cref{notn:RgFbd}). Since $\b\in\Rg$, we have $\sumbT\cond(\ff|\Rg)=\pi$, so $\ff\in\RgFint$. Thus, $\ff\notin\CollWFaces$. Conversely, suppose that $\ff\notin\CollWFaces$. Then $\ff\in\RgFint$ for some maximal \wdash collapsible subset $\Rg\subset\Verts$. By~\eqref{eq:MCE:sumwT_sumbT=pi}, $\sumbT\cond(\ff|\Rg) = \pi$. Since $\xd(\RgFacl)$ is a white clique, every black face $\xT(\b)$, $\b\in\RB$, must be degenerate. Thus, in order to have $\sumbT\cond(\ff|\Rg) = \pi$, $\ff$ must be adjacent to a black corner $\cor$ with $\sumT(\cor)=\pi$. Since $\G$ satisfies~\eqref{eq:ass_lgg}, $\ff$ is the middle vertex of the degenerate black triangle $\xT(\corv)$, so $\ff\in\FflWmid$. 

Since $\Gfin\in\BCGs$, iterating~\eqref{eq:gdim_diff}, we get $\gdim(\GDfin) = \gdimo = 3n+4\nL$.
By~\eqref{eq:OCP:dim_atr}--\eqref{eq:ORA:terminal_vs_atr_dim_only}, $\gdim(\GDfin) = \gdimatr(\G) = |\Faces| + n + 3$. Thus, $|\Faces| = 2n + 4\nL - 3$. For example, the graph $\Gfin\in\oBCGs$ with $(k,n;\nL)=(2,4;1)$ shown in \figref{fig:BCFW-full}(e) has $9$ faces.

Consider the restriction $(\CollGDW,\CollWxd)$ of $(\GD,\xd)$ to the faces of the \wdash collapsed graph $\CollGW$. 
By \cref{lemma:TE:TE_to_MCE_and_back,lemma:TOP:restr}, $\CollWxd$ is a \wtemb of $\CollGDW$. 
For $\rho\in\brnL$, 
 the point $\xd(\ploc_\rho)$ may be recovered from $(\CollGDW,\CollWxd)$ by additionally specifying the location of $\xd(\ploc_\rho)$ inside the $\dloc_\rho$-dimensional clique $\Conv\Kmaxloc_\rho$. In particular, the restriction $\ResGlinit(\GD,\xd)$ may be encoded by 
$\dimZ(\Mdte(\CollGW)/\LGpm) + \dbloc \leq \gdimatr(\CollGW) + \dbloc$ parameters; cf.~\eqref{eq:OCP:dim_atr} and the proof of \cref{lemma:ORA:terminal_vs_atr_dim_only}. Thus, 
$\dimZ\MCMcellint(\Gfin) \leq \gdimatr(\CollGW) + \dbloc$. 
Substituting $\gdimatr(\CollGW)=|\CollWFaces|+n+3$ and 
$\dimZ\MCMcellint(\Gfin) = \gdimo=3n+4\nL$ (since $\Gfin\in\oBCGs$ \haskinsupp), we get
$|\CollWFaces| + \dbloc \geq 2n + 4\nL - 3$ which equals $|\Faces|$ as we showed above. By~\eqref{eq:CollW=FflWsink}, we get
\begin{equation}\label{eq:dbloc>=|FflWmid|}
 \dbloc \geq |\Faces| - |\CollWFaces| = |\FflWmid|.
\end{equation}

We show that the sets $(\Tloc_\rho)_{\rho\in\brnL}$ are pairwise disjoint. Suppose otherwise that $\ff_2\in\FflWmid$ with outgoing edges $\ff_1\leftarrow\ff_2\to\ff_3$ and that $\ff_2\in\Tloc_\rho\cap\Tloc_\gamma$ for distinct $\rho,\gamma\in\brnL$. The cliques $\Conv\KTmaxloc_\rho$ and $\Conv\KTmaxloc_\gamma$ both contain the line segment $[\xT(\ff_1),\xT(\ff_3)]$. By \cref{lemma:MCE:fullysep<=kinsupp}, $\Gfin$ is \fullysepMCE, so the union $\preKmaxloc_\rho\cup\preKmaxloc_\gamma$ cannot be a clique. 
Similarly to \cref{lemma:MCE:two_cliques_intersection}, since the cliques $\Conv\KTmaxloc_\rho$ and $\Conv\KTmaxloc_\gamma$ are both white and share a line segment, their union must be a white clique, a contradiction.
Thus, the sets $(\Tloc_\rho)_{\rho\in\brnL}$ are indeed pairwise disjoint. Therefore, $|\FflWmid|\geq \tbloc$.
By construction, $\dloc_\rho\leq|\Sloc_\rho|-1 \leq \tloc_\rho$ for each $\rho\in\brnL$, 
 so $\dbloc\leq\tbloc$. Combining this with~\eqref{eq:dbloc>=|FflWmid|}, 
 we arrive at~\eqref{eq:counting_stars}. 
\end{proof}

\begin{proof}[Proof of~\eqref{eq:kinsupp_einj_automatic}]
Let $\datrQL=\datrQ\in\MdteMP(\Gfin)$ and let $\lalat$ be as in~\eqref{eq:TE:y_to_lalat}--\eqref{eq:TE:lalat_vs_pFw_pFb}. 
Since $\Gfin\in\oBCGs$, by \cref{rmk:MCMp->MCM_surj,lemma:kinsupp=realizable}, $\MCMpdec(\Gfin)\neq\emptyset$, so let $(\GDfin,\yd,\la',\lat')\in\MCMpdec(\Gfin)$ and let $\mceTOwte(\GDfin,\yd,\la',\lat')=\datrQp\in\MdteoMP(\Gfin)$; cf. \cref{lemma:TE:TE_to_MCE_and_back}.

In view of \cref{rmk:TE:einj_nonvanishing}, suppose for contradiction that we have, say, $\Fw(\w)=0$ for some $\w\in\WVint$. 
Let $\ff\in\partF\w$. Suppose first that $\ff$ is incident to some black corner $\cor\in\cornersb(\ff)$ with $\sumT_{\yd}(\cor)\in(0,\pi)$. Then we can find another white vertex $\w'$ incident to $\ff$ such that the sum $\sumbT_{\yd}(\w\to\w')$ of black angles incident to $\ff$ located between $\w$ and $\w'$ belongs to $(0,\pi)$. 
Similarly to \cref{lemma:TE:Kawangle_alg}, we see that $\sumbT_{\yd}(\w\to\w')$ equals $\arg(\pm\Fwp(\w')/\Fwp(\w))$ modulo $\pi$. It follows that $\det(\Fwp(\w')|\Fwp(\w))\neq0$. 
Since $\w,\w'$ share the face $\ff$ of $\Gfingr$, by \crefi{lemma:TE:brla_nonzero_if_apm}{lak_implies1}, $\Gfingr\rem\{\w,\w'\}$ admits an \APM. 
By \crefi{lemma:TE:brla_nonzero_if_apm}{lak_implies1} again, $\det(\Fw(\w)|\Fw(\w'))\neq0$, so $\Fw(\w)\neq0$, a contradiction.

Thus, for each $\ff\in\partF\w$ and $\cor\in\cornersb(\ff)$, we have $\sumT_{\yd}(\cor)\in\{0,\pi\}$. By~\itemref{TE:bdry_angle}, we must have $\ff\in\Fint$. By~\eqref{eq:TE:angle_cond}, $\sumbT_{\yd}(\ff)=\pi$, so $\sumT_{\yd}(\cor)=\pi$ for some $\cor\in\cornersb(\ff)$. Thus, $\partF\w\subset\FflWmid$, where $\FflWmid$ is defined with respect to $(\GD,\yd)$. 
Let $\preKmaxnoloc_\w$ be the set of $\f\in\Faces$ such that $\GflW$ contains a directed path from some $\ff\in\partF\w$ to $\f$. 
By \cref{lemma:MCE:clique_union}, $\preKmaxnoloc_\w$ 
 is a clique. 
By \cref{lemma:MCE:fullysep<=kinsupp}, $\Gfin$ is \fullysepMCE, so 
by \cref{lemma:BCFW:counting_stars}, $\preKmaxnoloc_\w\cap\FflWmid$ is contained in $\Tloc_\rho$ for a unique $\rho\in\brnL$. In particular, by~\eqref{eq:counting_stars}, $|\Tloc_\rho| = \dloc_\rho\leq2$. Since $\partF\w\subset\FflWmid$, we have $\partF\w\subset\Tloc_\rho$ and thus $|\partF\w|\leq2$. Therefore, $\w$ is a bigonal face of $\GD$ and $\Tloc_\rho=\partF\w$. Denote $\partF\w=\{\ff_1,\ff_2\}$.

If $\w$ shares a degree-$2$ black neighbor with some other white vertex $\w'$ then by~\eqref{eq:OCP:holom}, $\Fw(\w)=0$ implies $\Fw(\w')=0$. Repeating the above argument, we see that $\w'$ is also degree-$2$. Continuing in this fashion, we can find two trivalent (cf.~\eqref{eq:ass_lgg}) black vertices $\b_1,\b_2\in\BV$ connected by a path consisting of degree-$2$ vertices one of which is $\w$. Thus, $\ff_1,\ff_2\in\partF\b_1\cap\partF\b_2$. Since $\sumT_{\yd}(\cor)\in\{0,\pi\}$ for all $\cor\in\cornersb(\ff_1)$ and $\cor\in\cornersb(\ff_2)$, it follows that $\yT(\b_1),\yT(\b_2)$ are degenerate black triangles. Their middle vertices must belong to $\Tloc_\rho=\{\ff_1,\ff_2\}$. By~\eqref{eq:MCE:sumwT_sumbT=pi}, $\ff_1$ (resp., $\ff_2$) must be the middle vertex of $\b_1$ (resp., $\b_2$) or vice versa. Thus, the graph $\GflW$ contains a directed $2$-cycle, contradicting $\dloc_\rho=\tloc_\rho$; see \figref{fig:S-rho}(d) for an example. 
We have shown that $\Fw$ takes nonzero values. Similarly, $\Fb$ also takes nonzero values. By \cref{rmk:TE:einj_nonvanishing}, $\xT$ is \einj. 
\end{proof}

\subsection{BCFW tilings of ambient loop \mtas}\label{ssec:BCFW_triang_final2}
A major complication when translating the BCFW recursion from momentum space to momentum-twistor space at loop level is that the graphs $\{\Gfin\in\oBCGs\}$ do not usually satisfy $\helWmin(\Gfingr)\geq2$ as the \terminal \MCEs produced by the \ora may contain degenerate \btrar faces; see \figref{fig:BCFW-full}(e). 
 (By \cref{lemma:BCFW:MCM_nonempty=>hel>=1,lemma:BCFW:gelWmin_vs_helWmin}, the graphs $\Gfin\in\oBCGs$ do satisfy $\helmin(\Gfingr)\geq1$.) 
 Thus, the graphs $\Gfin\in\oBCGs$ are in general not T-dualizable (\cref{dfn:G:T_dualizable}). 
For each $\Gfin\in\oBCGs$, we consider the \wdash collapsed planar bipartite graph $\CollGW$. By \cref{lemma:DIM:CollGW_geq_2}, $\CollGW$ satisfies $\helWmin(\CollGW)\geq2$ and $\helBmin(\CollGW)\geq1$. 
By \cref{lemma:MCE:fullysep<=kinsupp}, the next-to-boundary white vertices of $\CollGW$ are pairwise distinct, so $\CollGW$ is T-dualizable. We let $\ddCollGW$ be the T-dual planar bipartite graph of $\CollGW$. 
See \cref{ex:BCFW-T-dual}.

For each $\rho\in\brnL$ such that $\ploc_\rho\in\Fint\ind[\Rgloc_\rho]$ for some maximal \wdash collapsible subset $\Rgloc_\rho\subset\Verts$, by \cref{prop:ccc}, 
 we obtain a tuple $\Sloc_\rho$ of faces of $\CollGW$ incident to the collapsed white vertex $\wloc_\rho$ of $\CollGW$. 
Let $\CollGWfunc=(\CollGW,\Slocs)$ be the resulting generalized \Lfunc graph (\cref{dfn:gen_Lfunc_graph}).
 We denote the collection of such collapsed BCFW graphs $\CollGWfunc$ by $\CollBCGs$. 
Applying generalized T-duality (\cref{dfn:generalized_T-duality}) to each $\CollGWfunc\in\CollBCGs$, we obtain a collection $\ddoBCGs$ of generalized \Lvunc graphs.
For $\ddCollGWfin\in\ddoBCGs$, we define the tile $\AA_{\ddCollGWfin}\subset\AAknL$ similarly to~\eqref{eq:LPUNC:AAddG_dfn} using the generalized \Lvunc boundary measurement map introduced in~\eqref{eq:LPUNC:Measbiv_dfn}. 

\begin{theorem}[BCFW tilings of ambient loop \mtas]
\label{lemma:BCLOOP:AAknL_tiling}
The tiles $\{\AA_{\ddCollGWfin}\mid \ddCollGWfin\in\ddoBCGs\}$ form \atddmtiling of $\AAknL$.
\end{theorem}
\begin{proof}
Let $\Gfin\in\oBCGs$ and suppose that $\ploc_\rho\notin\CollFaces$ for some $\rho\in\brnL$. Let $\wloc_\rho$ and $\Rgloc_\rho$ be as above. 
Let $\CycRgloc_\rho$ be the cycle in $\GDfin$ enclosing $\Rgloc_\rho$ as in \cref{lemma:BCFW:dualizable_holess_sconn}. Then $\G\ind[\CycRgloc_\rho]$ is of type $(1,\nloc_\rho)$ for some $\nloc_\rho\geq\mloc_\rho$, where $\mloc_\rho$ was introduced in \cref{dfn:ddGlocs}. As explained in the proof of \cref{lemma:TOP:restr}, since $\helmin(\Gfingr)\geq1$, the boundary measurements of $\G\ind[\CycRgloc_\rho]$ belong to $\Grtp(1,\nloc_\rho)$. Let $\bcccloc_\rho\in\Simplexr$ be the convex combination coefficients expressing $\Hknkloc_\rho$ in terms of the values of $\Hknk$ at the boundary faces of $\G\ind[\CycRgloc_\rho]$ given by \cref{prop:ccc}. 
It follows from \cref{lemma:BCFW:counting_stars} and its proof that $\G\ind[\CycRgloc_\rho]$ (with $|\Tloc_\rho|\leq2$ interior faces) is \easyred and satisfies~\eqref{eq:counting_stars_type_1_n}. 
By \cref{lemma:counting_stars_type_1_n}, the resulting map $\Rtpgauge \to \CollWRtpgauge \times \SimplexbL$ sending $\wt\mapsto (\CollWwt,\bcccnL)$ is a homeomorphism. We have a commutative square
\begin{equation}\label{eq:Coll_square}
\begin{tikzcd}[column sep=3em, row sep=1.5em]
\Rtpgauge \arrow[r,"\sim"] \arrow[d,"{\Meas(\Gfin,\cdot)}","\sim"'] 
& \CollWRtpgauge \times \SimplexbL \arrow[d,"{\Meas(\CollGWfin,\cdot)}",twoheadrightarrow] 
\\
\PtpLfunc_{\Gfin}\arrow[r,"{\id}"]%
& \PtpLfunc_{\CollGWfin},
\end{tikzcd} 
\end{equation} 
where the vertical map $\Meas(\Gfin,\cdot):\Rtpgauge\xrasim\PtpLfunc_{\Gfin}$ is a homeomorphism by \cref{rmk:Gfin_reduced}. Therefore, all four maps in~\eqref{eq:Coll_square} are homeomorphisms. 
Thus, by \cref{lemma:BCLOOP:MPknL_tiling}, the tiles 
$\{\MPCollGWfunc\mid \CollGWfunc\in\CollBCGs\}$
 form \atmtilingA of $\MPknL$. 
\Cref{lemma:GENPUNC:T_duality_D_vs_H} extends \cref{lemma:LPUNC:T_duality_D_vs_H} to the case of generalized \Lpunc graphs. This leads to a generalization of T-duality for \tmtilings of ambient loop amplituhedra (\cref{lemma:TREE:amb_tiling_equivalence}). Combining it with \cref{lemma:LPUNC:twosep_vs_twoind,lemma:MCE:fullysep<=kinsupp}, 
 we obtain the result. 
\end{proof}

Combining \cref{lemma:BCLOOP:MPknL_tiling,lemma:BCLOOP:AAknL_tiling} with \cref{lemma:tMtiling}, we obtain the following.
\begin{corollary}\ \label{cor:from_tMtiing_to_Mtiling}
\begin{enumerate}[label=(\arabic*)]
\item\label{Mtil1} The tiles $\{\MPGfin\mid \Gfin\in\oBCGs\}$ form \amtilingA of $\MPknL$.
\item\label{Mtil2} The tiles $\{\AA_{\ddCollGWfin}\mid \ddCollGWfin\in\ddoBCGs\}$ form \addmtiling of $\AAknL$.
\end{enumerate}
\end{corollary}

\subsection{Sign flip and linear projection loop amplituhedra}\label{ssec:flip_proj}
Let $\LaLat\in\LaLaimmnn$. Similarly to the momentum amplituhedron map $\PhiLL:\Grtnn(k,n)\to\overline{\MPkntree}$ defined in~\eqref{eq:intro:PhiLL_dfn}, we introduce a map $\PhiLLL:\GrtnnLfuncx_2(k,n)\to\MPknL$. For $\CHL\in\GrtnnLfuncx_2(k,n)$, let $\lalat:=\PhiLL(C)$. Similarly to~\eqref{eq:LOOP:Rel_dfn}, let $\Amat\in\Mator_{2,k}$ and $\Atmat\in\Mator_{2,n-k}$ be such that $\la = \Amat\cdot C$ and $\lat = \Atmat\cdot C^\perp$. We define $\Hdndloc_\rho:=\Atmat\cdot \Hknkloc_\rho\cdot \Amat^T$ and let $\yloc_\rho\in\Rdd$ be given by $\yMloc_\rho = \Hdndloc_\rho$ for $\rho\in\brnL$. We set $\byL:=(\yloc_1,\dots,\yloc_\nL)$ and $\PhiLLL\CHL:=\llPllL$. 
\begin{definition}
The \emph{linear projection} and \emph{sign flip} loop momentum amplituhedra are 
given by
\begin{equation*}%
 \MomLLLprojo:=\PhiLLL(\GrtnnLfuncx_2(k,n))
\ \ \ \text{and}\ \ \ 
\MomLLLamb:=\{\llPllL\in\MPknL\mid \la\subset\La,\ \lat\subset\Lat\}.
\end{equation*}
\end{definition}

\begin{remark}\label{rmk:FL_FGLS_compare}
The ``sign flip'' ambient loop momentum amplituhedron $\MPknLFGLS$ of~\cite{FGLS} 
 is defined as the set of points $\llPllL$ satisfying 
parts~\itemref{MPknL1}--\itemref{MPknL2} of \cref{dfn:LPUNC:amb_loop_ampl}, 
and part~\itemref{MPknL4} is replaced with $n$ sign variation conditions listed in~\cite[Equation~(3.8)]{FGLS}.
 One can check that the element $\ellastij\in\Rdd$ introduced in~\cite[Equation~(3.4)]{FGLS} satisfies 
$\frac14(\ellastij - \yloc_\rho)^2 = \brVL[i,j,\rho] \cdot \brla<i,j>^{-1}$ 
for all $i,j\in\brn$ such that $\brla<i,j>\neq0$ and all $\rho\in\brnL$. 
Thus,~\cite[Equation~(3.8)]{FGLS} records the sign variation of the sequence 
$\left(\brVL[i,i+1,\rho],\brVL[i,i+2,\rho],\dots,\brVL[i,i+n-1,\rho]\right)$;
cf.~\eqref{eq:varVL_dfn}. By \cref{thm:AAshift}, these conditions are equivalent to part~\itemref{MPknL4} of \cref{dfn:LPUNC:amb_loop_ampl}, so $\MPknL = \MPknLFGLS$.

A ``linear projection'' loop momentum amplituhedron $\MomLLLprojoFL$ was introduced in~\cite{FL}. 
It is stated in~\cite{FGLS} that $\MomLLLprojoFL\subset\MPknLFGLS$, so by~\eqref{eq:LOOP:proj=amb} below,
$\MomLLLprojoFL\subset \clMomLLLprojo$. However, the precise relationship between $\MomLLLprojoFL$ and $\MomLLLprojo$ is unclear to us. 

One advantage of our definitions of $\MPknL$ and $\MomLLLprojo$ is that they do not involve any objects on the ``momentum-twistor'' side of T-duality. By contrast, the definitions in~\cite{FGLS} (resp.,~\cite{FL}) rely
on the T-dual $4\times 4$ minors $\brVL[i,i+2,\rho]$ (resp., on $\Qla$ and $\GrtnnLvuncambx_2(k-2,n)$). 
\end{remark}

Next, given
 $Z\in\Grtp(k+2,n)$, the \emph{\mta map} $\PsiZtree:\Grtnn(k-2,n)\to\Gror(4,n)$ sends
$\ddC\mapsto V:=\ddC^\perp\cap Z$. We extend $\PsiZtree$ to a map $\PsiZL:\GrtnnLvuncambx_2(k-2,n)\to\GrL(4,n)$. For $\ddCDL\in\GrtnnLvuncambx_2(k-2,n)$, let $V:=\PsiZtree(\ddC)=\ddC^\perp\cap Z$ and $\Vloc_\rho:=\Dbivloc_\rho^\perp\cap Z$ for $\rho\in\brnL$. 
The following result is well known~\cite{AHT}; see e.g.~\cite[Definition~3.8]{KW} for a proof.
\begin{lemma}[{\cite[Definition~3.8]{KW}}]
\label{lemma:KW_intersect_Z_dim}
Let $\ddC\in\Grtnn(k-2,n)$, $\Dbivloc_\rho\in\Grtnn(k,n)$, $\Dbivlocs_{\sepst}\in\Grtnn(k+2,n)$, and $Z\in\Grtp(k+2,n)$. Set $V:=\ddC^\perp\cap Z$, $\Vloc_\rho:=\Dbivloc_\rho^\perp\cap Z$, and $\Vlocs_{\sepst}:=\Dbivlocs_{\sepst}^\perp\cap Z$. Then $\dim(V) = 4$, $\dim(\Vloc_\rho)=2$, and $\dim(\Vlocs_{\sepst})=0$. 
\end{lemma}

Similarly to \cref{rmk:orientation_inherits}, we orient $V$ (resp., $\Vloc_\rho$) so that $\brV[i,i+1,i+2,i+3]>0$ (resp., $[i\,i+1]_{\Vloc_\rho}>0$) for all $i\in\brn$. This is possible when $\ddCDL$ is \twoind in view of~\eqref{eq:KW_minors} below. 
For $\rho\in\brnL$, let $\pLr\in\Gror(2,4)$ be such that $\pLr\cdot V = \Vloc_\rho$ and let $\Liner\in\Mator_{4,2}$ be a matrix representing the (oriented) orthogonal complement of $\pLr$. We set $\Linelocs:=(\Line_1,\dots,\Line_\nL)$ and $\PsiZL\ddCDL := \VLpunc$.

\begin{definition}
The \emph{linear projection} and \emph{sign flip} loop \mtas are 
given by
\begin{equation*}%
 \AZprojLo:=\PsiZL(\GrtnnLvuncambx_2(k-2,n)) \quad\text{and}\quad
 \AZambL:=\{\VLpunc\in\AAknL\mid V\subset Z\}.
\end{equation*}
\end{definition}
\noindent We will show below that the closure of $\AZprojLo$ coincides with that of $\PsiZL(\GrtnnLvuncprojx_2(k-2,n))$.

\begin{lemma}
\label{lemma:LOOP:proj_subset_amb}
For all $\LaLat\in\LaLaimmnn$ and $Z\in\Grtp(k+2,n)$, we have inclusions
\begin{equation}\label{eq:TREE:proj_subset_cl_amb}
\MomLLLprojo\subset\MomLLLamb
 \quad\text{and}\quad 
 \AZprojLo\subset\AZambL.
\end{equation}
\end{lemma}

\begin{proof}%
 Let $\llPllL=\PhiLLL\CHL\in\MomLLLprojo$ for some $\CHL\in\GrtnnLfuncx_2(k,n)$. We have $\lalat=\PhiLL(C)\in\lalak$. 
Let $(\Gfunc,\wt)$ be such that $\Gfunc$ is \fullysep and $\CHL=\Measknk(\Gfunc,\wt)$. Observe that $\yMloc_\rho=\Hdndloc_\rho$ for $\rho\in\brnL$ and $\bdx_i=\Hdnd(\bdf_i)$ for $i\in\brn$ by construction of $\PhiLLL$. 
Since $\Gfunc$ is \fullysep, by \cref{lemma:PROP:Mpos}, we obtain a proof of parts~\itemref{MPknL1}--\itemref{MPknL2} of \cref{dfn:LPUNC:amb_loop_ampl}. 
 By \cref{thm:TE:OAC} and \cref{rmk:TE:Mpos_existence}, $\lalat$ gives rise to a \wtemb of $(\G,\wt)$. 
Thus, each $\yTloc_\rho$ is located strictly inside (cf. \cref{lemma:TOP:Mpos=>simple}) the simple polygon $\PbdxT$.
 This verifies part~\itemref{MPknL4}.
Thus, we obtain the first inclusion in~\eqref{eq:TREE:proj_subset_cl_amb}. 

Suppose now that $\VLpunc=\PsiZL\ddCDL$ for some $\ddCDL\in\GrtnnLvuncambx_2(k-2,n)$. 
Since $\ddCDL$ is \twoind, we have $\Dbivlocs_{\sepst}\in\Grtnn(k+2,n)$. 
Since $V=\ddC^\perp\cap Z$, $\Vloc_\rho = \Dbivloc_\rho^\perp\cap Z$,
 and $\Dbivlocs_{\sepst}^\perp\cap Z=\{0\}$ by \cref{lemma:KW_intersect_Z_dim}, we have~\cite[Equation~(3.11)]{KW}
\begin{equation}\label{eq:KW_minors}
 \brLL[\seps,\sept] = \sum_{I\in{\brn\choose k+2}}\Delta_I(\Dbivlocs_{\sepst}) \Delta_{I}(Z) \quad\text{for all $\sepst\in\brnLbdsep$}.
\end{equation}
Since $Z\in\Grtp(k+2,n)$ and $\Dbivlocs_{\sepst}\in\Grtnn(k+2,n)$ with $\Dbivlocs_{\sepst}^\perp\cap Z=\{0\}$, 
 we get $\brLL[\seps,\sept]>0$ for all $\sepst\in\brnLbdsep$. 
This shows part~\itemref{AAknL2} of \cref{dfn:LSH:AAknL}. 
We have $V=\PsiZtree(\ddC)\in\AZprojtree:=\PsiZtree(\Grtnn(k-2,n))$. It is well known that $\AZprojtree\subset\overline{\AAkntree}$; see~\cite[Section~5.4]{AHTT}. Thus, $\varxxx(V)=k-2$. 
This verifies part~\itemref{AAknL1}.
Finally, for $\rho\in\brnL$, $\Vloc_\rho = Z\cap \Dbivloc_\rho^\perp$ belongs to the \emph{$m=2$ \mta}; cf.~\cite[Theorem~5.1]{PSBW}. In particular, $\varx(\Vloc_\rho) = k$. Since $\brVL[1,i,\rho]=[1\,i]_{\Vloc_\rho}$ for all $i\in\brn$, part~\itemref{AAknL4} follows. 
\end{proof}

For our final result, denote 
$\MomLLprojG:=\PhiLLL(\PtpLfunc_{\Gfunc})$ and $\AZprojddG:=\PsiZL(\PtpLvunc_{\ddCollGWvunc})$.

\begin{theorem}[BCFW tilings of loop amplituhedra]\ \label{lemma:proj_tiling}
\begin{enumerate}[label=(\arabic*)]
\item\label{proj_tiling1} The tiles $\{\MomLLprojG\mid \Gfin\in\oBCGs\}$ form a tiling of $\MomLLLprojo$ for all $\LaLat\in\LaLaimmnn$.
\item\label{proj_tiling2} The tiles 
$\{\AZprojddG\mid \ddCollGWfin\in\ddoBCGs\}$ 
 form a tiling of $\AZprojLo$ for all $Z\in\Grtp(k+2,n)$. 
\item\label{proj_tiling3} The linear projection and sign flip definitions of the loop amplituhedron agree: we have
\begin{equation}\label{eq:LOOP:proj=amb}
\clMomLLLprojo = \clMomLLLamb,
 \quad\text{resp.,}\quad 
 \clAZprojLo = \clAZambL
\end{equation}
for all $\LaLat\in\LaLaimmnn$, resp., $Z\in\Grtp(k+2,n)$.
\end{enumerate}
\end{theorem}
\begin{proof}
Let $\LaLat\in\LaLaimmnn$ and $Z\in\Grtp(k+2,n)$. 
Let $\RXA,\RYA$ and $\RXC,\RYC$ be as in \cref{dfn:RX_triple}. 
Our goal is to apply \cref{lemma:Rtiling_to_tiling} to $\WA:=\MomLLLamb$ and $\WC:=\AZambL$. 

Let $\RXAcl:=\GrtnnLfuncx_0(k,n)$, 
$\RYAcl:=\lalats\times(\Rdd)^{\nL}$ (cf. \cref{rmk:GGsh_action_preserves_MPknL}),
 and $\WAcl:=\{\llPllL\in\RYAcl\mid \la\subset\La\text{ and }\lat\subset\Lat\}$. Thus, $\RXA\subset\RXAcl$, $\RYA\subset\RYAcl$, and $\WA=\WAcl\cap \RYA$.
Let $\RelAcl\subset\RXAcl\times\RYAcl$ be the (closed) subset of points satisfying~\eqref{eq:LOOP:Rel_dfn}.
Thus, 
 $\RelA = \RelAcl\cap(\RXA\times\RYA)$. 
 By \crefi{cor:from_tMtiing_to_Mtiling}{Mtil1},
 $\{\RYoA_{\Gfin}\mid \Gfin\in\oBCGs\}$ is \amtilingA of $\RYA$. 
The map $\RPhiA=\PhiLLL:\RXA\to\WA$ (cf. \cref{lemma:LOOP:proj_subset_amb}) extends to a map $\RPhiAcl:\RXAcl\to\WAcl$ defined in the same way. Both maps are continuous by \cref{lemma:OAC:PhiLL_in_lalak}. 

We check~\eqref{eq:RGraph}. By construction, the output of the map $\RPhiAcl$ satisfies~\eqref{eq:LOOP:Rel_dfn}; thus, $\RGraph_{\RPhiAcl} \subset \RelAcl\cap (\RXAcl\times \WAcl)$. Conversely, a point $(\CHL,\llPllL)\in\RelAcl\cap (\RXAcl\times \WAcl)$ satisfies~\eqref{eq:LOOP:Rel_dfn} for some $\CHL\in\RXAcl$ and $\la\subset\La$, $\lat\subset\Lat$. By \cref{lemma:OAC:PhiLL_in_lalak}, both intersections $C\cap\La$ and $C^\perp\cap\Lat$ are $2$-dimensional. Thus, we must have $\lalat=\PhiLL(C)$, and therefore $\llPllL=\PhiLLL\CHL$. 

It remains to check that the closure $\RXoAcl_{\Gfin}$ of $\RXoA_{\Gfin}=\PtpLfunc_{\Gfin}$ is compact for each $\Gfin\in\oBCGs$. This would follow immediately if one could show that $\RXAcl=\GrtnnLfuncx_0(k,n)$ is compact; see \cref{problem:GrtnnLfunc_by_ineq}. Instead, we realize $\RXoAcl_{\Gfin}$ explicitly as a continuous map image of a compact set.
Consider the \emph{fractional matching polytope} 
$P_{\G}:=\{h\in[0,1]^{\Edges}\mid \sum_{\e\sim\v}h(\e)=1\text{ for each $\v\in\Vint$}\}$. 
Here, the summation is taken over all edges incident to $\v$. Similarly to~\cite[Lemma~3.1]{PSW}, it follows that $\Measknk:\Rtpgauge\to\PtpLfunc_{\Gfin}$ extends to a continuous map from the nonnegative part of the toric variety associated to $P_{\G}$ to $\RXAcl=\GrtnnLfuncx_0(k,n)$, with image $\PtnnLfunc_{\Gfin}:=\bigcup_{\Gfin'}\PtpLfunc_{\Gfin'}$, where the union is taken over all graphs $\Gfin'$ obtained from $\Gfin$ by deleting a subset of edges so that the resulting graph still admits an \APM. Thus, $\PtnnLfunc_{\Gfin}\subset\RXAcl$ is indeed an image of a compact set under a continuous map, so $\RXoAcl_{\Gfin}$ is compact. 
We conclude that the tiles $\{\WoA_{\Gfin}\mid \Gfin\in\oBCGs\}$ form a tiling of $\WA=\MomLLLamb$. By~\eqref{eq:TREE:proj_subset_cl_amb}, since each tile is contained inside $\MomLLLprojo$, they also form a tiling of $\MomLLLprojo$.

For $\WC:=\AZambL$, we can choose $\RXCcl:=\Grtnnflamb$ itself to be compact; cf. \cref{dfn:Grtnnflamb_CDind,que:LPUNC:GrtnnLvuncproj=GrtnnLvuncamb}. Let $\RYCcl:=\GrL(4,n)$ and $\WCcl:=\{\VLpunc\in\RYCcl\mid V\subset Z\}$. 
We have $\RXC\subset\RXCcl$, $\RYC\subset\RYCcl$, and $\WC=\RYC\cap\WCcl$. 
Let $\RelCcl\subset\RXCcl\times\RYCcl$ be the (closed) subset of points satisfying~\eqref{eq:TREE:ddRel_dfn}.
Thus, 
 $\RelC = \RelCcl\cap(\RXC\times\RYC)$.
 By \crefi{cor:from_tMtiing_to_Mtiling}{Mtil2},
 $\{\RYoC_{\ddCollGWfin}\mid \ddCollGWfin\in\ddoBCGs\}$ is \amtilingC of $\RYC$.
The map $\RPhiC=\PsiZL:\GrtnnLvuncambx_2(k-2,n)\to\AZprojLo$ and its extension $\RPhiCcl:\RXCcl\to\WCcl$ (defined in the same way) are continuous by \cref{lemma:KW_intersect_Z_dim}. The proof of~\eqref{eq:RGraph} for $\RPhiCcl$ is deduced from \cref{lemma:KW_intersect_Z_dim} similarly to the above. Finally, the closure $\RXoCcl_{\ddCollGWfin}$ of $\RXoC_{\ddCollGWfin}$ in $\RXCcl$ is compact since $\RXCcl$ is compact. 
Thus, the tiles $\{\WoC_{\ddCollGWfin}\mid \ddCollGWfin\in\ddoBCGs\}$ form a tiling of $\WC=\AZambL$. 
By~\eqref{eq:TREE:proj_subset_cl_amb}, since each tile is contained inside $\AZprojLo$, these tiles also form a tiling of $\AZprojLo$.

This concludes the proof of parts~\itemref{proj_tiling1}--\itemref{proj_tiling2} of the theorem.
Since $\bigsqcup_{\Gfin\in\oBCGs}\WoA_{\Gfin}\subset\MomLLLprojo$
 is dense in $\WA=\MomLLLamb$ 
and 
$\bigsqcup_{\ddCollGWfin\in\ddoBCGs}\WoC_{\ddCollGWfin}\subset\AZprojLo$ is dense in $\WC=\AZambL$, part~\itemref{proj_tiling3} follows. 
\end{proof}

\appendix

\section{Loop BCFW recursion in momentum-twistor space}\label{ssec:APP:RP3}
We briefly explain how to run the BCFW recursion directly in momentum-twistor space while still working with \lggs $\Gfunc$ as opposed to their T-duals $\ddCollGWfin$.

\begin{definition}\label{dfn:APP:mtconf}
Let $\Gloop$ be \algg satisfying \cref{ass:lgg}. 
A \emph{\mtconf} $(\GD,\bmtV,\bmtL)$ on $\GD$ is a choice of a point $\mtV_{\east}\in\RP^3$ for each edge $\east\in\East$ of $\GD$ and an oriented line $\mtL_{\ff}\subset\RP^3$ for each vertex $\ff\in\Faces$ of $\GD$, satisfying the following conditions.\footnote{We thank D.~Chelkak for explaining an equivalent Pl\"ucker-quadric viewpoint to us in 2024.}
\begin{enumerate}[label=(\arabic*)]
\item\label{mtconf1} For every edge $\east$ incident to a vertex $\ff$, the line $\mtL_{\ff}$ passes through the point $\mtV_{\east}$.
\item\label{mtconf2} 
For every white face $\w\in\WVint$ of $\GD$, there exists a point $\mtV_\w\in\RP^3$ such that $\mtV_{\east} = \mtV_\w$ for each $\east\in\partEast\w$, and such that each line $\mtL_{\ff}$, $\ff\in\partF\w$, passes through $\mtV_\w$.
\item\label{mtconf3} For every black face $\b\in\BVint$ of $\GD$, there exists a plane $\mtP_\b\subset\RP^3$ such that all points $\{\mtV_{\east}\mid\east\in\partEast\b\}$ and all lines $\{\mtL_{\ff}\mid\ff\in\partF\b\}$ are contained in $\mtP_\b$.
\end{enumerate}
\end{definition}

Following \cref{ssec:LOOP:ambient_ampl_T_dual}, we explain how to convert $(\GD,\xd)\in\MCM(\Gloop)$ into a \mtconf $\mtwphi(\GD,\xd):=(\GD,\bmtV,\bmtL)$ on $\GD$. By~\itemref{MCE2:null_edges}, for each edge $\ebarast=\{\ff,\f\}\in\Ebarast$ of $\GD$, we have a (nonzero) null vector $\Pmom_{\east}:=\xd(\ff)-\xd(\f)$. 
We choose a bispinor representation $\xM_{\Pmom_{\east}} = \lat_{\east}\cdot \la_{\east}^T$ as in~\eqref{eq:TE:decor_dfn}. Thus, $\xM_{\Pmom_{\east}}\cdot \CMI\cdot \la_{\east} = \bzero_{2\times1}$ by~\eqref{eq:dual_spinor}. 
 We set 
\begin{equation}\label{eq:mtwphi_dfn}
\mu_{\east}:=\xM_{\xd(\ff)}\cdot \CMI\cdot \la_{\east} = \xM_{\xd(\f)}\cdot \CMI\cdot \la_{\east},
\quad
\mtV_{\east}:=\begin{pmatrix}
\la_{\east} \\ \mu_{\east}
\end{pmatrix},\quad
\mtL_{\ff}:=\begin{pmatrix}
 \Id_2 \\ \xM_{\xd(\ff)}\cdot \CMI
 \end{pmatrix}.
\end{equation}

For an oriented line $\mtL_{\ff}$ in $\RP^3$ with Pl\"ucker coordinates $(\Delta_{ij})_{1\leq i<j\leq 4}$ such that 
$\Delta_{12}\neq0$, we may recover $\xT(\ff)$ from~\eqref{eq:mtwphi_dfn} and~\eqref{eq:x_vs_ap_vs_am} via
\begin{equation}\label{eq:APP:mtmapT_dfn}
 \mtmapT(\mtL_{\ff}) := \left(\frac{\Delta_{13}+\Delta_{24}}{\Delta_{12}},\frac{\Delta_{23} - \Delta_{14}}{\Delta_{12}}\right) = \xT(\ff).
\end{equation}

\begin{remark}\label{rmk:APP:cliques_vs_lines}
 Let $\preK\subset\Faces$ be a white (resp., black) clique in $(\GD,\xd)$. It follows from the above construction that the lines $\{\mtL_{\ff}\mid\ff\in\preK\}$ all pass through a single point (resp., are all contained in a single plane) in $\RP^3$. Observe that the variety of all oriented lines in $\RP^3$ passing through a given point (resp., contained in a given plane) and satisfying $\Delta_{12}\neq0$ is isomorphic to the affine plane $\R^2$, 
and the map $\mtmapT$ provides a specific affine isomorphism, identifying each line $\mtL_{\ff}$ with the corresponding point $\xT(\ff)\in\Conv\KT$. In other words, the oriented line configuration $\{\mtL_{\ff}\mid\ff\in\preK\}$ in $\RP^3$ is 
affinely isomorphic to the point configuration $\{\xT(\ff)\mid\ff\in\preK\}$ in $\R^2$. Under this isomorphism, the t-immersion condition~\itemref{TE:t_imm} translates into orientation conditions on the lines $\mtL_{\ff}$: e.g., for a nondegenerate white triangle $\w\in\WVint$ in $(\GD,\xd)$ with vertices $\ff_1,\ff_2,\ff_3$ in counterclockwise order, the oriented lines $\mtL_{\ff_1},\mtL_{\ff_2},\mtL_{\ff_3}$ passing through the point $\mtV_\w$ form a positively oriented basis of the tangent space $\R^3=T_{\mtV_\w}\RP^3$, and similarly for black triangles. 

\end{remark}

\noindent One can use this observation to formulate the \ORA purely in terms of the \mtconf $\mtwphi(\GD,\xd)=(\GD,\bmtV,\bmtL)$, thus showing that the \ORA is \emph{conformally invariant}, i.e., invariant under the action of $\SL_4(\R)$ on $(\bmtV,\bmtL)$. 

Explicitly, since rigid \ORsts (\cref{sec:ORA_TORA}) occur inside $\Conv\KTmax$ for a single clique $\Kmax\in\Kmaxs$, they are translated into the momentum-twistor space using the affine isomorphism $\mtmapT$ above. The \flex \ORsts are translated slightly differently for the different cases in \cref{fig:fout-bend}. In each case, the conditions in \cref{dfn:APP:mtconf} applied to the newly created triangular faces $\uLR$ yield a $1$-parameter family $\mtLr_{\ffout}$ of oriented lines depending on $\r\in\Rtp$. 

For example, suppose that, say, $\conven=\WBcon$. Let $\eastout_\pm$ be the edge connecting $\ffout$ to $\corfpm$. 
If $\sumbT(\cor)<\pi$ as in \figref{fig:fout-bend}(a,b) then we have $\mtV_{\coream}\neq\mtV_{\coreap}$. In this case, $\mtLr_{\ffout}$ is the line passing through the point $\mtV_{\coreap}=\mtV_{\eastout_+}=\mtV_{\uL}$ and the point $\mtVrm$ that is moving along the oriented line $\mtL_{\corfm}$ in the positive direction as $\r$ increases, starting from $\mtVom=\mtV_{\coream}$. The line $\mtLr_{\ffout}$ is oriented so that in the $\r\to0$ limit, its orientation coincides with that of $\mtL_{\corf}$. 
 On the other hand, if $\sumbT(\cor)=\pi$ as in \figref{fig:fout-bend}(c,d) then
$\{\corfp,\corf,\corfm,\ffout\}$ is a white clique with $\mtV_{\coream}=\mtV_{\coreap}=\mtV_{\eastout_-}=\mtV_{\eastout_+}=\mtV_{\uL}$. The oriented line $\mtLr_{\ffout}$ is obtained by rotating $\mtL_{\corf}$ around the point $\mtV_{\uL}$ inside the plane $\mtP_{\uR}$, starting from $\mtLro_{\ffout}=\mtL_{\corf}$ and rotating away from $\mtL_{\corfm}$. 
The ``output'' line $\mtLrcrit_{\ffout}$ is described similarly to~\eqref{eq:rcrit_dfn}: 
a violation of each condition~\axrangeMCEall can be translated into the language of \mtconfs, and $\rcrit$ is the minimal positive value for which such a violation occurs.
Similarly, the compatibility condition~\eqref{eq:MCMS_angles_vs_k} for arbitrary $\v\in\Vint$ may be translated into a sign variation condition (cf.~\eqref{eq:TREE:AAkn_dfn}) on the points $\{\mtV_{\east}\mid\east\in\partEast\v\}$. We leave the details for future work.

\apptocmd{\thebibliography}{\setlength{\itemsep}{-1pt}}{}{}
\bibliographystyle{alpha}
\bibliography{loop_ampl}

\end{document}